\documentclass[a4paper,fleqn]{cas-sc}
\usepackage{amsmath}
\usepackage{amssymb}
\usepackage{graphicx}
\usepackage{mathrsfs}
\usepackage{amsfonts}
\usepackage{amsthm}
\usepackage{color}
\usepackage{titletoc}

\titlecontents{section}
              [1.1cm]
              { \normalsize}%
              {\contentslabel{2em}}%
              {}%
              {\titlerule*[0.3pc]{$\cdot$}\normalsize\contentspage\hspace*{0cm}}%
\titlecontents{subsection}
              [2cm]
              {\normalsize}%
              {\contentslabel{2em}}%
              {}%
              {\titlerule*[0.3pc]{$\cdot$}\normalsize\contentspage\hspace*{0cm}}%
\titlecontents{subsubsection}
              [2.9cm]
              {\normalsize}%
              {\contentslabel{2.5em}}%
              {}%
              {\titlerule*[0.3pc]{$\cdot$}\normalsize\contentspage\hspace*{0cm}}%

\usepackage{lipsum}
\usepackage{gensymb}
\usepackage[numbers,sort&compress]{natbib}
\usepackage[numbers]{natbib} 
\def\tsc#1{\csdef{#1}{\textsc{\lowercase{#1}}\xspace}}
\tsc{WGM}
\tsc{QE}
\tsc{EP}
\tsc{PMS}
\tsc{BEC}
\tsc{DE}

\begin{document}
\let\WriteBookmarks\relax
\def\floatpagepagefraction{1}
\def\textpagefraction{.001}
\shorttitle{Tao Yu, Zhaochu Luo, and  Gerrit E. W. Bauer}
\shortauthors{Tao Yu et~al.}

\title [mode = title]{Chirality as Generalized Spin-Orbit Interaction in Spintronics}                      

\author[a]{Tao Yu}[orcid=https://orcid.org/0000-0001-7020-2204]

\address[a]{School of Physics, Huazhong University of Science and Technology, Wuhan 430074, China}
\cormark[1]
\ead{taoyuphy@hust.edu.cn}

\author[b,c]{Zhaochu Luo}[orcid=https://orcid.org/0000-0002-7543-3244]
\address[b]{State Key Laboratory of Artificial Microstructure and Mesoscopic Physics, School of Physics, Peking University, 100871 Beijing, China}
\address[c]{Beijing Key Laboratory for Magnetoelectric Materials and Devices, Beijing, 100871, China}

\author[d,e]{Gerrit E. W. Bauer}[orcid=https://orcid.org/0000-0002-3615-8673]
\address[d]{WPI-AIMR and Institute for Materials Research and CSRN, Tohoku University, Sendai 980-8577, Japan}
\address[e] {Kavli Institute for Theoretical Sciences, University of  the Chinese Academy of Sciences, Beijing 100190, China}

\cortext[cor1]{Corresponding author.}

\begin{abstract}
Chirality or handedness  distinguishes an object from its mirror images, such as the spread thumb, index finger, and middle finger of the right and left hand. In mathematics, it is described by the outer product of three vectors that obey a right-hand \textit{vs.} left-hand rule.
The chirality of \textit{ground state} magnetic textures defined by the vectors of magnetization, its gradient, and an electric field from broken inversion symmetry can be fixed by a strong relativistic spin-orbit interaction. This review focuses on the chirality observed in the \textit{excited states} of the magnetic order, dielectrics, and conductors that hold transverse spins when they are evanescent. Even without any relativistic effect, the transverse spin of the evanescent waves is locked to the momentum and the surface normal of their propagation plane. This chirality thereby acts as a generalized spin-orbit interaction, which leads to the discovery of various chiral interactions between magnetic, phononic, electronic, photonic, and plasmonic excitations in spintronics that mediate the excitation of quasiparticles into a single direction, leading to phenomena such as chiral spin and phonon pumping, chiral spin Seebeck, spin skin, magnonic trap, magnon Doppler, and spin diode effects. Intriguing analogies with electric counterparts in the nano-optics and plasmonics exist.
	After a brief review of the concepts of chirality that characterize the ground state chiral magnetic textures and chirally coupled magnets in spintronics, we turn to the chiral phenomena of excited states.  We present a unified electrodynamic picture for dynamical chirality in spintronics in terms of generalized spin-orbit interaction and compare it with that in nano-optics and plasmonics. Based on the general theory, we subsequently review the theoretical progress and experimental evidence of chiral interaction, as well as the near-field transfer of the transverse spins, between various excitations in magnetic, photonic, electronic and phononic nanostructures at GHz time scales. We provide a perspective for future research before concluding this article. 
\end{abstract}


\maketitle

\tableofcontents

\section{Introduction}
\label{section1}

\subsection{Chirality}
\label{Sec_chirality}
To start, let us define the concept of ``chirality" or handedness.  A phenomenon that is governed by a vector product that obeys a right-hand \textit{vs.} left-hand rule is characterized by a binary, such as a positive or negative chirality index. Two objects with opposite chirality are each other's mirror image, such as the spread fingers of the right and left hand. Chirality may be also defined as property under inversion and/or time-reversal symmetry operation that depends on the nature of the objects (waves, particles, molecules, knots \textit{etc.}) and spatial dimension \cite{Barron,magnetic_chirality}. Throughout this review, we stick to the simple geometrical convention in terms of a relation between three vectors. 

An object is chiral when distinguishable from its mirror image, \textit{viz.}, the two cannot be physically superimposed. Chirality of molecular enantiomers, as well as chiral continuous \cite{PR_soliton_1,PR_soliton_2,PR_soliton_3,PR_soliton_4} or  discrete \cite{Ref_DMIChiral1,Ref_DMIChiral7} magnetic textures fulfill this condition. The reason for the chiral symmetry breaking in chemistry is still under debate.   
\textcolor{blue}{The chirality of magnetic textures emerges by either a sufficiently strong relativistic spin-orbit coupling in the presence of a broken inversion symmetry or the dipolar interaction that may also stabilize spirals and local textures such as skyrmions and vortices.}
Ground state textures such as helices and spirals can be characterized by the local magnetization direction, its gradient tensor, and  a symmetry-breaking electric field direction \cite{PR_soliton_1,PR_soliton_4,Ref_DMIChiral1,Ref_DMIChiral7}. Recent reviews on chirality in magnetism and spintronics \cite{PR_soliton_2,PR_soliton_3,PR_soliton_4} address domain walls, vortices, and skyrmions under the spin-orbit interaction, referred to as anti-symmetric exchange or Dzyaloshinskii-Moriya interaction (DMI) \cite{DMI_1,DMI_2}. Their robustness and fast dynamics raise the hope for application in memories or for non-conventional computing.  We refer readers with an interest in chiral  domain walls and their interaction with spin currents to Refs.~\cite{chiral_spintronics_1,chiral_spintronics_2}.

Chirality is also a dynamic property.  Consider for example the precession of a classical spinning top under the torque of the gravity field  \textbf{F}.  In terms of the mechanical angular momentum \textbf{L} the torque and precession direction follows the right-hand rule \(\dot{\textbf{L}} \sim \textbf{L} \times \textbf{F}\).  Analogously, a magnetic moment  \textbf{M} in a magnetic field \textbf{H} obeys the Landau-Lifshitz equation \(\dot{\textbf{M}} = - \gamma\mu_0 \textbf{M} \times \textbf{H}\), where \(-\gamma \) is the gyromagnetic ratio of the electron and $\mu_{0}$ is the vacuum permeability, with handedness in the precession. The electric or magnetic vector fields of electromagnetic waves in the vacuum are purely transverse, \textit{i.e.}, the field direction is normal to the wave propagation direction. The two circular polarizations are degenerate and carry angular momentum, photon spin quantum \(\pm \hbar\), in the direction of its wave vector. The ``helicity" of a polarized photon (beam) is the sign of the inner product of wave vector and spin. \textcolor{blue}{For photons in a vacuum, polarization spin and wave vector are parallel, so helicity and chirality are the same. In condensed matter systems, excitations may propagate normally to their polarization spin vector, however, i.e. with zero helicity, but still finite chirality in the presence of spin-momentum locking.}

While electromagnetic waves in the vacuum are purely transverse, they develop longitudinal components when in contact with condensed matter, \textit{e.g.}, at metallic interfaces and in wave guides. Elastic media harbor both longitudinal (pressure) and transverse sound waves. Spin waves are transverse excitations of a magnet, which is intrinsically polarized already. In all these cases the simple helicity index of a photon in the vacuum is not sufficient to characterize a given polarization state in which the polarization spin and propagation vectors may appear at any angle.

\begin{figure}[htp]
    \centering
    \includegraphics[width=16 cm]{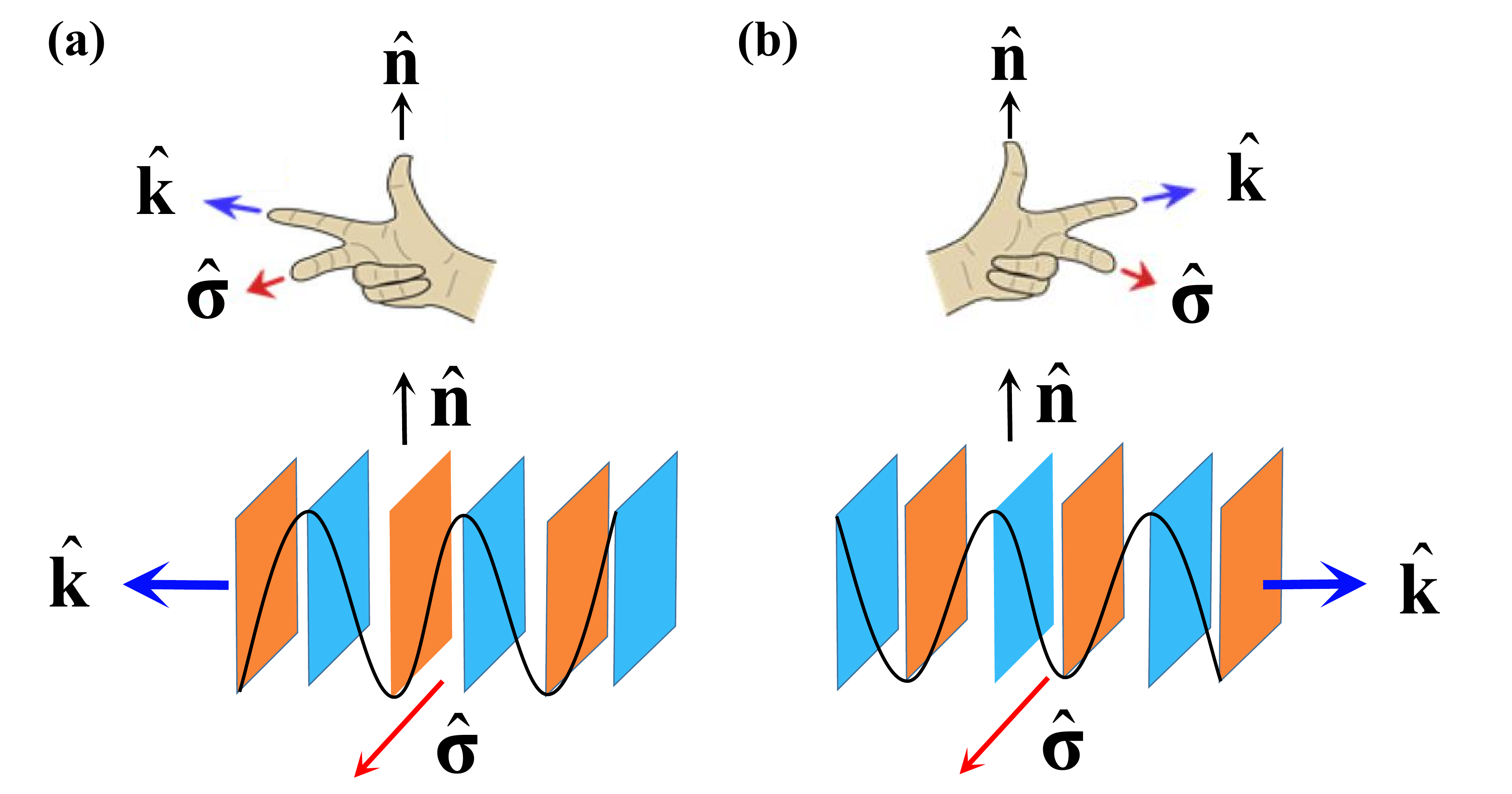}
    \caption{Plane spin waves with opposite chirality index. They propagate with wave vector ${\bf k}$ normal to $\hat{\bf n}$, which lies in the planes  formed by the wave fronts. In this example the polarization spins ${{\pmb \sigma}}$ are transverse to the propagation since ${\hat{\pmb \sigma}}\cdot \hat{\bf k}=0$. The chirality index $Z=\hat{\bf k}\cdot({\hat{\pmb \sigma}\times \hat{\bf n}})=1$ in (a) is ``right-handed", but becomes ``left-handed" in (b) since $Z=-1$.}
    \label{spin_momentum_locking}
\end{figure}

The central topic of this review is a special type of waves in which the polarization spin is \textit{normal} to the propagation direction. Such a ``transverse spin" is common to many wave forms. Here we focus on the magnetism and spintronics in wave phenomena that have only recently been fully appreciated and harbor promising functionalities for various applications. The chirality is here a relation between the wave vector \textbf{k},  polarization spin \(\pmb{\sigma}\), and a symmetry-breaking electric field \textbf{E}.  The chirality index 
$ Z=  \hat{\mathbf{k}} \cdot ( \hat{\pmb{\sigma}} \times \hat{\mathbf{E}}) $ or commutations of it, where the hat indicates unit vectors, measures the direction of the precession of the polarization in an effective magnetic field. Such vector combinations typically appear in  spin-orbit interaction Hamiltonians. The inversion symmetry is always broken at interfaces at which  we may replace   \textbf{E} by the surface normal \textbf{n} in  the chirality index Z. \textcolor{blue}{Figure~\ref{spin_momentum_locking} illustrates chiral plane spin waves with opposite chirality index.}

When the polarization and surface normal are fixed by the system, a wave with a given chirality can propagate only in one direction.  ``Unidirectionality" is the extreme form of non-reciprocity, \textit{i.e.}, a difference between waves propagating with positive and negative wave vectors. 
The classical example for a perfect chiral wave with \(Z=1\) are the Damon-Eshbach spin waves \cite{DE,Walker_sphere} at the surface of a ferromagnet that propagate normal to the magnetization in one direction. Numerous other types of chiral waves with non-reciprocal propagation and even unidirectionality will be reviewed in this article.

The restriction of a fixed transverse spin direction is completely released in the presence of the time-reversal symmetry, with which the spin can be reversed such that the waves are allowed to propagate along opposite directions.
The propagation direction of surface acoustic waves with fixed chirality enforces the sign of the spin  \cite{Kino1987,Viktorov1967}, which is referred to as spin-momentum ``locking".  Similar phenomena are observed in plasmonics (``spin-orbit interaction of light" \cite{plasmonics_1,plasmonics_2,Nori,nano_optics,chiral_optics}).  Chiral  electromagnetic and acoustic waves are  evanescent \cite{plasmonics_1,plasmonics_2,Nori}, \textit{i.e.}, their amplitudes  decay exponentially as a function of distance from surface or interface that is characterized by the normal vector $\hat{\bf n}$. The chirality of all these evanescent waves is well characterized by a chirality index $ Z=  \hat{\mathbf{k}} \cdot ( \hat{\pmb{\sigma}} \times \hat{\mathbf{n}}) $.

 Although chiral waves may exist in bulk systems, the examples discussed in this review are evanescent. Chiral surface states exist at the interfaces of topologically non-trivial materials to vacuum in the gap of the bulk dispersion \cite{quantum_Hall_1,quantum_Hall_2,quantum_anomalous_Hall,topological_insulator_1,topological_insulator_2,Chaoxing_Liu,xiaogang_wen,chiral_superconductor_1,chiral_superconductor_2,Majorana_1,Majorana_2}. The question whether all chiral waves have a topological origin has to best of our knowledge not been answered. Yamamoto \textit{et al.}  \cite{Kei_topology} only recently demonstrated that the Damon-Eshbach modes are topological, even though they do not lie in a gap, in contrast to the edge modes in magnonic crystals  \cite{topology_magnonic_crystal}. Here we do not concern ourselves with the intricate mathematics of topological materials and their boundaries but focus on the microscopic physics, phenomenology, and comparison with experiments.

\subsection{Chiral interaction}

The physical origin of the chirality of a given state originates from the underlying Hamiltonian which contains chiral interactions, such as the spin-orbit coupling. The examples above are the eigenstates of the system Hamiltonians. Often the chiral interactions are weak and can be treated by perturbation theory. A typical example is the electric or magnetic dipole-dipole interaction, purely electromagnetic origin without relativistic effect. In this case, the unperturbed state can be a normal metal or magnet. For the emergence of chiral interaction, consider for example the two coupled systems, one being the source and the other being the medium. Their evanescent electric (or magnetic) dipolar fields have transverse spins that are governed by a chirality index. When the coupling matrix elements conserve the transverse spin, as it often does, the excited states have to be chiral as well.  If the  transverse spin from the source is fixed in such as a magnet or circularly-polarized electric dipoles, the direction of the pumped spin carried by the chiral waves is fixed as well, which can propagate along only one direction due to the spin-momentum locking. Spin torques emerge when the transverse spin has additional sinks. So the states excited by a chiral perturbation are unidirectional or non-reciprocal. A typical example is the chiral light–matter interaction widely studied in nano-optics \cite{nano_optics,chiral_optics,Tang_Cohen} and plasmonics \cite{plasmonics_1,plasmonics_2,Nori,chiral_speed,chiral_loss}. 
We will review various unidirectional pumping in the magnetism and spintronics based on this chiral interaction.

A prototype of chiral quantum optics \cite{nano_optics,chiral_optics} is a specially designed rotating electric dipole. In the proximity of an optical fiber, it leads to a unidirectional energy flow. Rotating dipoles  close to a metallic waveguide excites unidirectional surface plasmon polaritons  \cite{plasmonics_1,plasmonics_2,Nori}. Here the surface plasmon polariton's electric field is spin-momentum locked, so when its electric field only interacts with the electric dipole of one circular polarization, the excitation is unidirectional. The equivalent viewpoint is that the electric field of the circularly-polarized electric dipole is unidirectional such that it only excites the surface plasmon polariton in one direction. Unidirectional excitation reduces reaction-induced quantum noise, thereby enhancing the capacity of communication channels, and protecting the transmission against possible signal instabilities. 
Chiral interaction also governs optical valley selection rules in monolayer transition-metal dichalcogenides that may be useful in ``valleytronics" which employs the  pseudo-spin index of the graphene-like conduction band valleys of atomic  monolayers \cite{valleytronics}.

While the chiral index can be an integer, the chirality of the interaction can be weak or strong, so the physical consequences can vary from subtle to dramatic, and different names are in use to describe the effects.  ``Non-reciprocities" is transport properties in opposite direction that can be caused by chiral interactions but also asymmetries in the dispersion relation \(\omega_\textbf{k} \ne \omega_{-\textbf{k}} \) that are caused by chiral terms in the system Hamiltonian. ``Unidirectional" refers to a very strong non-reciprocity that implies that one direction is completely suppressed. In the case of the Damon-Eshbach mode, this is caused not by an asymmetry in the dispersion, but a complete suppression of modes in one direction. Unidirectional transport by chiral interaction terms is never complete but easily comes close to it. So while ``chirality" is the property of the Hamiltonian, ``non-reciprocity" and unidirectionality are its consequence in observations.  However, unidirectional or non-reciprocal phenomena do not necessarily imply chirality, but may also be caused by wave interference or simple symmetry breaking, \textcolor{blue}{such as the electric diode, unidirectional magnetoresistance, and nonreciprocal transport in noncentrosymmetric
superconductors} \cite{non_reciprocity_Nagaosa}.

\subsection{Chirality as generalized spin-orbit interaction}

Non-reciprocities addressed above can be traced to different microscopic phenomena, \textit{viz.}, the relativistic spin-orbit interaction in different disguises, and the electric and magnetic dipole-dipole interactions, so one might ask what these have in common.

\textcolor{blue}{The chirality index is binary in the sense that it can have either sign, i.e. ``$+$" or ``$-$", or ``left" or ``right", but it becomes an integer only under ideal conditions.}
The chirality index is maximized when the transverse spin is perpendicular to the momentum and the surface normal.  In analogy with electrons, the polarized quantum or classical wave can be  characterized as ``spin"-$\uparrow$ and ``spin"-$\downarrow$. The relativistic spin-orbit interaction (Rashba or DMI) can be visualized as the torque on the spin  \(\pmb{\sigma}\) by a magnetic field in the rest frame of the electron  \(\sim\)\textbf{k} that moves normal to an electric field \(\sim\)\textbf{n}. The dipolar field when evanescent is similar. Indeed, in the long wavelength limit, the self-consistent dipolar interaction that governs the energy of the Damon-Eshbach modes is proportional to the magnetization gradient, \textit{i.e.}, magnon wave vector. This field can act on the spin, here parallel to the equilibrium magnetization, only when normal to it and when in the plane of the interface, leading to a Hamiltonian term that looks the same as a spin-orbit interaction. This analogy turns out to be quite universal that governs the chirality of many evanescent fields of electromagnetic origin and can be also extended to other wave physics such as acoustics.

\begin{figure}[ptbh]
	\begin{centering}
		\includegraphics[width=1.0\textwidth]{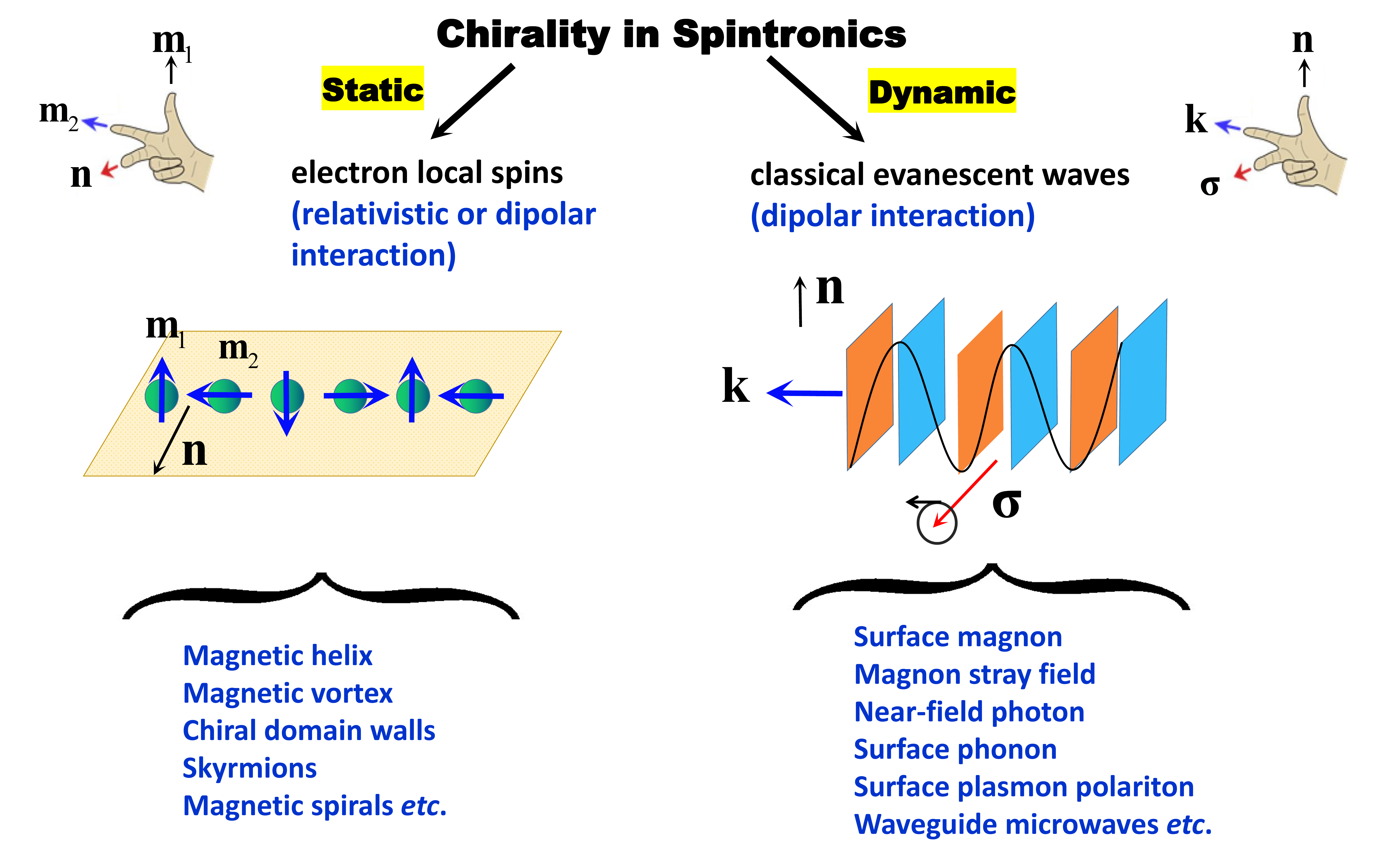}
		\par\end{centering}
	\caption{Overview of chirality in spintronics in the presence of a generalized spin-orbit interaction. The effects can be measured in terms of the chirality index $ Z=  \hat{\mathbf{k}} \cdot ( \hat{\pmb{\sigma}} \times \hat{\mathbf{n}}) $, which is maximized when wave vector, circular polarization or ``spin", and surface normal are all at right angles. The sources of chirality are divided into two categories with the relativistic origin and dipolar interaction, respectively, resulting in a similar spin-momentum locking phenomenology. The existence of such chirality in various waves appears in many physical systems including evanescent waves such as surface magnon, magnon stray field, near-field photon, surface phonon, surface plasmon polariton, and waveguide microwaves, or in perturbations such as light-matter interactions to systems that are in itself not chiral.  The conservation of transverse spin and chirality index when transferred leads to many novel functionalities in magnetism and spintronics.}
	\label{issues}
\end{figure}

A non-conserved transverse spin angular momentum density of electromagnetic waves, dominated by the electric field component in optics \cite{Nori,Nori_PRL_1,Nori_PRL_2} but the magnetic field in the microwave regime implies the existences of spin transfer torques that may rotate an object. Similarly, the acoustics defines the transverse spin for the acoustic waves \cite{phonon_spin_1962,Zhang_phonon_spin,phonon_angular_momentum_2015}. The chirality index \textit{Z} reflects per definition a generalized spin-orbit interaction in surface magnon, magnon stray field, near-field photon, surface phonon, surface plasmon polariton, and waveguide microwaves, as illustrated in Fig.~\ref{issues} for an overview. So relativistic and dipolar spin-orbit interaction cause a very similar phenomenology, while the physics governs the magnitude of the prefactors in the Hamiltonians that can of course be very different. 

The relativistic spin-orbit interaction is essential for the control and detection of the spin current \cite{spintronics_1,spintronics_2}, and drives the topological phases \cite{quantum_anomalous_Hall,topological_insulator_1,topological_insulator_2,Chaoxing_Liu}. The dipolar interaction has been studied in magnetism for many decades, but the analogy with the spin-orbit interaction in creating chirality was emphasized only recently.  Recently, many theoretical predictions and experimental discoveries of chiral interactions among magnetic \cite{Chiral_pumping_Yu,Chiral_pumping_grating,magnon_trap,Haiming_exp_grating,Haiming_exp_wire,Hanchen_damping,Au_first,DMI_circulator,bilayer_dipolar_1,bilayer_dipolar_2,Dirk_transducer,Chuanpu_NC}, photonic \cite{waveguide_Yu_1,waveguide_Yu_2,circulating_polariton,Xufeng_exp,Canming_exp,Teono_NV,Yu_Springer,Doppler_Yu,circulator_Tang,stripline_poineering_1,stripline_poineering_2,Yuxiang_subradiance,chiral_waveguide1,chiral_waveguide2,chiral_waveguide3}, phononic \cite{phonon_Yu_1,Xu,DMI_phonon_exp,Onose_exp,Nozaki_exp,phonon_Yu_2,Otani_exp,phonon_Kei,Page_exp,Page_exp_2}, electronic \cite{electron_spin_Yu}, and plasmonic \cite{plasmonics_spin_NC,plasmonics_spin_APL,plasmonics_spin_PRB,plasmonics_spin_NJP,plasmonics_spin_PRL,plasmonics_spin_OOP} excitations in spintronics are revealed, which are found to mediate the excitation of quasiparticles into a single direction, leading to phenomena such as chiral spin pumping \cite{Chiral_pumping_Yu}, chiral spin Seebeck \cite{Chiral_pumping_Yu}, chiral phonon pumping by magnetization dynamics \cite{phonon_Yu_1}, magnon diodes \cite{Chiral_pumping_Yu}, photon diodes \cite{waveguide_Yu_1}, photon diodes \cite{waveguide_Yu_1}, phonon diodes \cite{phonon_Yu_2},  magnonic non-Hermitian skin \cite{waveguide_Yu_1}, magnon trap \cite{magnon_trap}, and magnon Doppler \cite{Doppler_Yu} effects, to illustrate a few. Since an overview and unified understanding over these seemingly very different systems in different sub-fields of spintronics are still lacking, we focus on this class of chirality in this review and pragmatically unify the spin of classical waves with the quantum spin of electrons \cite{electron_spin_Yu}. We summarize these two important issues in Fig.~\ref{issues} that are addressed in this review.

Chirality has profound and universal implications 
on various phenomena ranging from quantum information science, optics and plasmonics, and condensed matter physics.
This review focuses on the class of chirality in the interaction between excitations by their electrodynamics that includes but does not require relativistic spin-orbit interaction. It is rooted in the coupling strength between quasiparticles that depends on the chirality index that parameterizes the handedness of transverse spin ${\pmb \sigma}$, momentum ${\bf k}$, and surface normal ${\bf n}$ vectors. To be familiar with the definition of chirality, we briefly revisit the chirality of magnetic textures  and introduce the static chiral interaction between two nanomagnets that is forced by an energy-favorable magnetization alignment (Sec.~\ref{section2}).
We then present the theoretical basis for the chirality of various waves that are employed in spintronics (Sec.~\ref{chiral_spin_waves} to \ref{Sec_chiral_phonon}), address the unified electromagnetic origin of chirality with intriguing analogies with electric counterparts in the nano-optics and plasmonics (Sec.~\ref{unification}), and review the theoretical progress and experimental evidence of chiral interaction between various excitations in magnetic, photonic, electronic and phononic nanostructures at GHz time scales (Sec.~\ref{Chiral_interaction}). In Sec.~\ref{Near_field_spintronics}, we review the theoretical demonstration and experimental progress of the transfer of transverse spin of classical electromagnetic and surface acoustic waves to the electron spins.
We provide a perspective for future research before concluding this article (Sec.~\ref{outlook}).

\section{Chiral magnetic textures and static chiral interaction}

\label{section2}

  A recurrent theme of this review is the difference between static and dynamic chiralities in magnetism, spintronics, and magnonics. In this Chapter, we briefly address magnetic textures  that arise from static chiral interactions. The chiral magnetic textures are a counterpart of the dynamic chiral phenomena as discussed in subsequent chapters. A chirality in real space can  be grasped intuitively in terms of the handedness of spiral spin structures such as extended helices or domain walls that separate domains with constant magnetizations.

\subsection{Chiral magnetic textures}
Magnetism is caused by the Coulomb exchange interaction between electrons. In magnets with local magnetic moments \(\textbf{M}_i\) and spins \(\textbf{S}_i =-\textbf{M}_i / \gamma \) on a lattice site \(\textbf{r}_i\), where  $-\gamma<0$ is the gyromagnetic ratio, the Heisenberg model for the  exchange energy reads
\begin{align}
 	H_{\rm ex}=- J \sum_{ij}^{n.n.} {\bf S}_{i}\cdot{\bf S}_{j},
 	\label{Eexchange}
\end{align}
where \(J\) is the exchange integral between nearest neighbors (\textit{n.n}). Exchange interaction favors a spatially constant ferromagnetic \( (J>0)\) or antiferromagnetic \( (<0)\) order. Textures can develop under   the long-range magneto-dipolar interaction
\begin{align}
	H_{\rm dip}=-\sum_{i\ne j}\frac{\mu_0\gamma^2\hbar^2}{4\pi |{\bf r}_{ij}|^5}
	\left(3({\bf S}_i\cdot{\bf r}_{ij})({\bf S}_j\cdot {\bf r}_{ij})-{\bf S}_i\cdot{\bf S}_j{\bf r}^2_{ij}\right),
	\label{Edipolar}
\end{align}
where $\mu_0$ is the permeability of the free space and ${\bf r}_{ij}={\bf r}_i-{\bf r}_j$. Another ingredient is the relativistic  spin-orbit interaction that in local spin systems has the form of the Dzyaloshinskii-Moriya interaction (DMI) by breaking inversion symmetry 
inherent to the crystal structure,  interfaces between different materials or larger scale inhomogeneities. This interaction has the form of an asymmetric exchange interaction with Hamiltonian \cite{DMI_1,DMI_2}
 \begin{align}
 	H_{\rm DMI}=-\sum_{ij}^{n.n.}{\bf D}_{ij}\cdot({\bf S}_i\times{\bf S}_j),
 	\label{DMI}
\end{align}
where ${\bf D}_{ij}$ is a vector that depends on the local electric field and ${\bf r}_{ij}$. The source of the DMI interaction at an interface is often a heavy metal atom in a non-magnetic metal which affects the hopping of electrons between spins  ${\bf S}_{i}$ and ${\bf S}_{j}$ by its spin-orbit interaction.

The ground state magnetic texture minimizes the total magnetic energy, i.e., the sum of the symmetric and asymmetric exchange interactions, dipolar interaction (demagnetization energy), the magnetic anisotropy energy, and the Zeeman energy due to an external magnetic field: 
\begin{align}
 	H_{\rm total}={H}_{\rm ex}+H_{
 	\rm DMI}+{H}_{\rm dip}+{H}_{\rm an}+{H}_{\rm Z}.
 	\label{Etotal}
\end{align}
A chiral texture is non-collinear with fixed handedness. Such handedness is governed by the relation between one geometry-related vector ${\bf n}$, \textcolor{blue}{which is the unit vector of an effective magnetic field that defines the magnetization rotation direction of the spatial texture}, and two neighboring magnetization directions ${\bf m}_1$ and ${\bf m}_2$ as indicated in each example in Table~\ref{table_ChiralTexture}. \textcolor{blue}{In typical domain walls  ${\bf n}$ is perpendicular to the other two vectors.} \textcolor{blue}{The chirality or handedness is indicated by the sign of the cross product of three vectors $Z={\bf n}\cdot({\bf m}_1\times {\bf m}_2)$,} \textit{i.e.}, a specific locking between these three vectors ${\bf m}_1$, ${\bf m}_2$ and ${\bf n}$ by the handed rule. This geometric definition of chirality is employed throughout this review in both static and dynamic situations.

   \begin{table}[htb]
   	\caption{Typical chiral magnetic textures with the chirality index $Z={\bf n}\cdot({{\bf m}_1\times{\bf m}_2})$ governed by the locking of two neighboring magnetization directions ${\bf m}_{1,2}$ and the geometric vector ${\bf n}$, which is the unit vector of an effective ``magnetic field" that defines the magnetization rotation direction of the spatial texture. } \label{table_ChiralTexture}
   	\centering
   	\begin{tabular}{ccc}
   		\toprule
   	\hspace{-0.6cm}	Texture Type (\textbf{mechanisms})&  \hspace{-1.9cm} Magnetic texture & \hspace{-1.4cm}Applications \\
   		\toprule
   		\hspace{-0.13cm}\begin{minipage}[m]{.3\textwidth}
   		 Magnetic helix\\
   		 (\textbf{Magnetostatics} or \textbf{DMI}) 
   		 \end{minipage}
   		 &\hspace{-1.5cm}\begin{minipage}[m]{.5\textwidth}
   			\centering\vspace*{4pt}
   		\includegraphics[width=2.75cm]{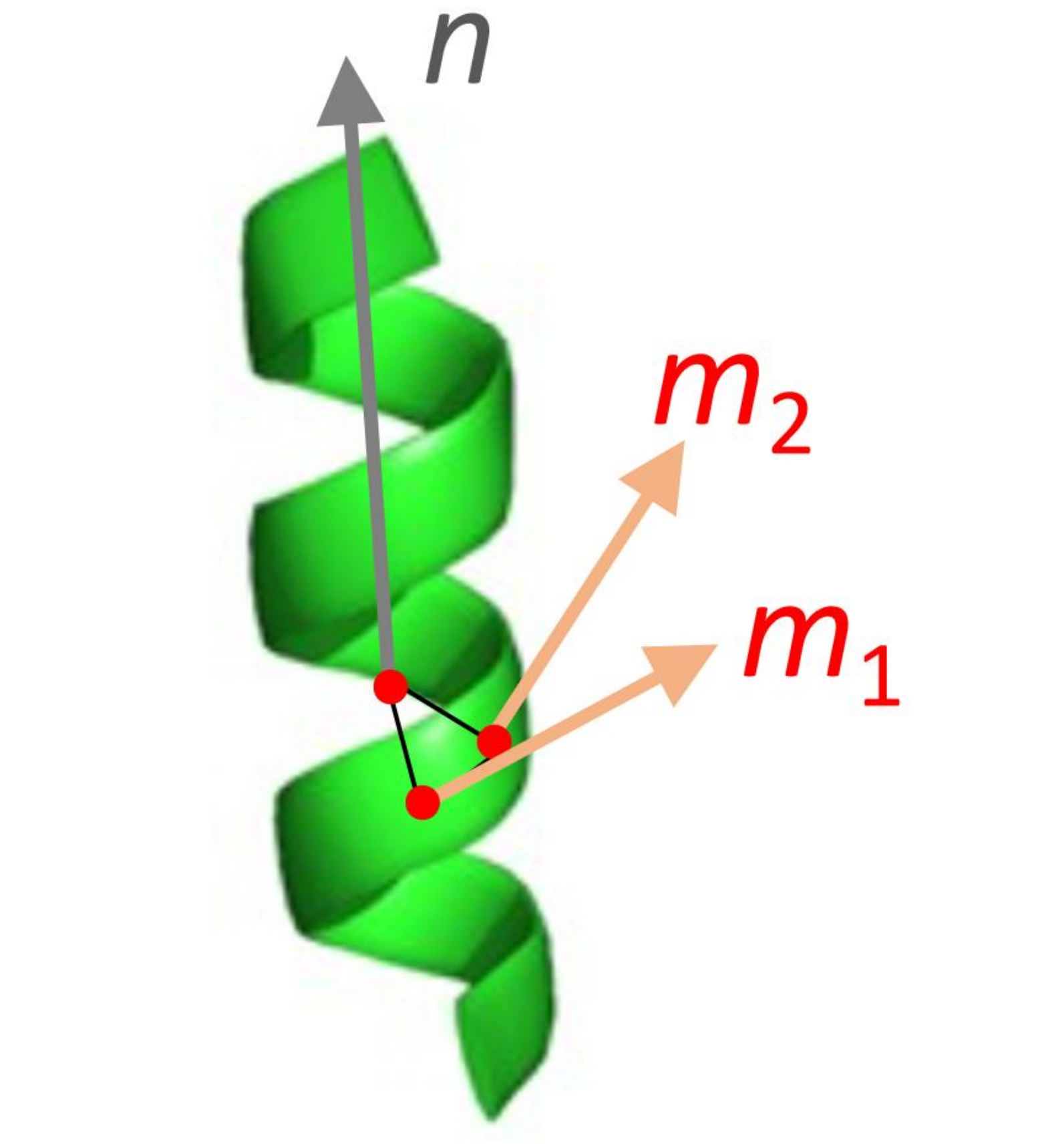}\vspace*{4pt}
   		\end{minipage} &
   		\hspace{-1.5cm}\begin{minipage}[m]{4.8cm}
   			\begin{itemize}
   				\item Optical modulator \cite{Ref_Geochiral3}
   				\item Magnetic robotic \cite{Ref_Geochiral6, Ref_Geochiral7}
   			\end{itemize}
   		\end{minipage} 
   		\\
   \toprule
   \hspace{-0.13cm}\begin{minipage}[m]{.3\textwidth}
   		Magnetic vortex\\ 
   		(\textbf{Magnetostatics} or \textbf{DMI}) 
   		\end{minipage}&\hspace{-1.5cm}\begin{minipage}[m]{.5\textwidth}
   			\centering\vspace*{4pt}
   			\includegraphics[width=3.8cm]{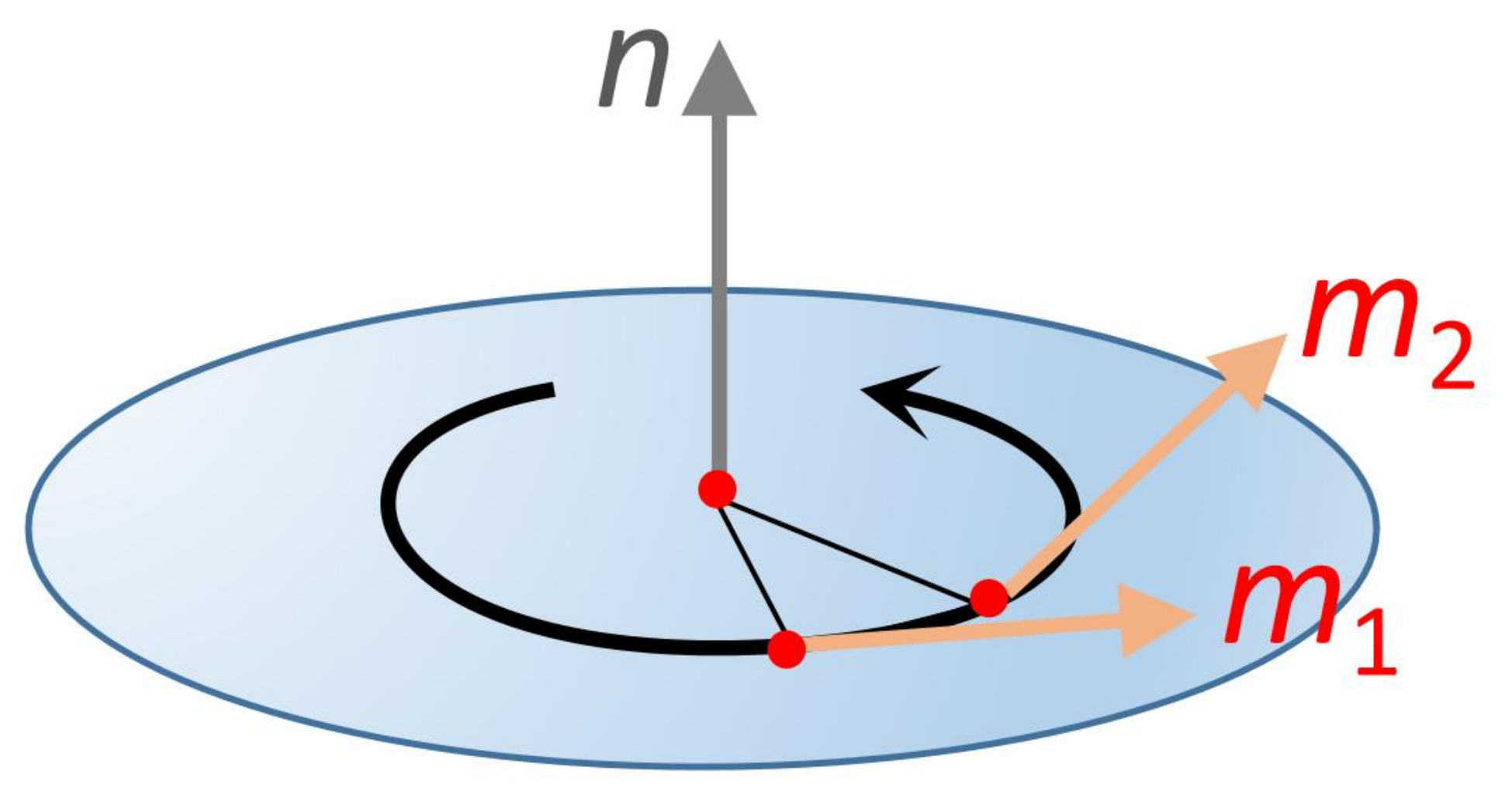}\vspace*{4pt}
   		\end{minipage} &
   		\hspace{-1.5cm}\begin{minipage}[m]{4.8cm}
   			\begin{itemize}
   				\item Data storage \cite{Ref_Geochiral12}
   				\item Oscillator \cite{Ref_Geochiral13}
				\item Magnetic sensor \cite{Ref_Geochiral14}
   			\end{itemize}
   		\end{minipage} 
   		\\
   		\toprule
   		\hspace{-0.13cm}\begin{minipage}[m]{.3\textwidth}
   		Chiral domain walls \\
   		(\textbf{Magnetostatics} or \textbf{DMI}) 
   		\end{minipage}&\hspace{-1.5cm}\begin{minipage}[m]{.5\textwidth}
   			\centering\vspace*{4pt}
   	\includegraphics[width=4.3cm]{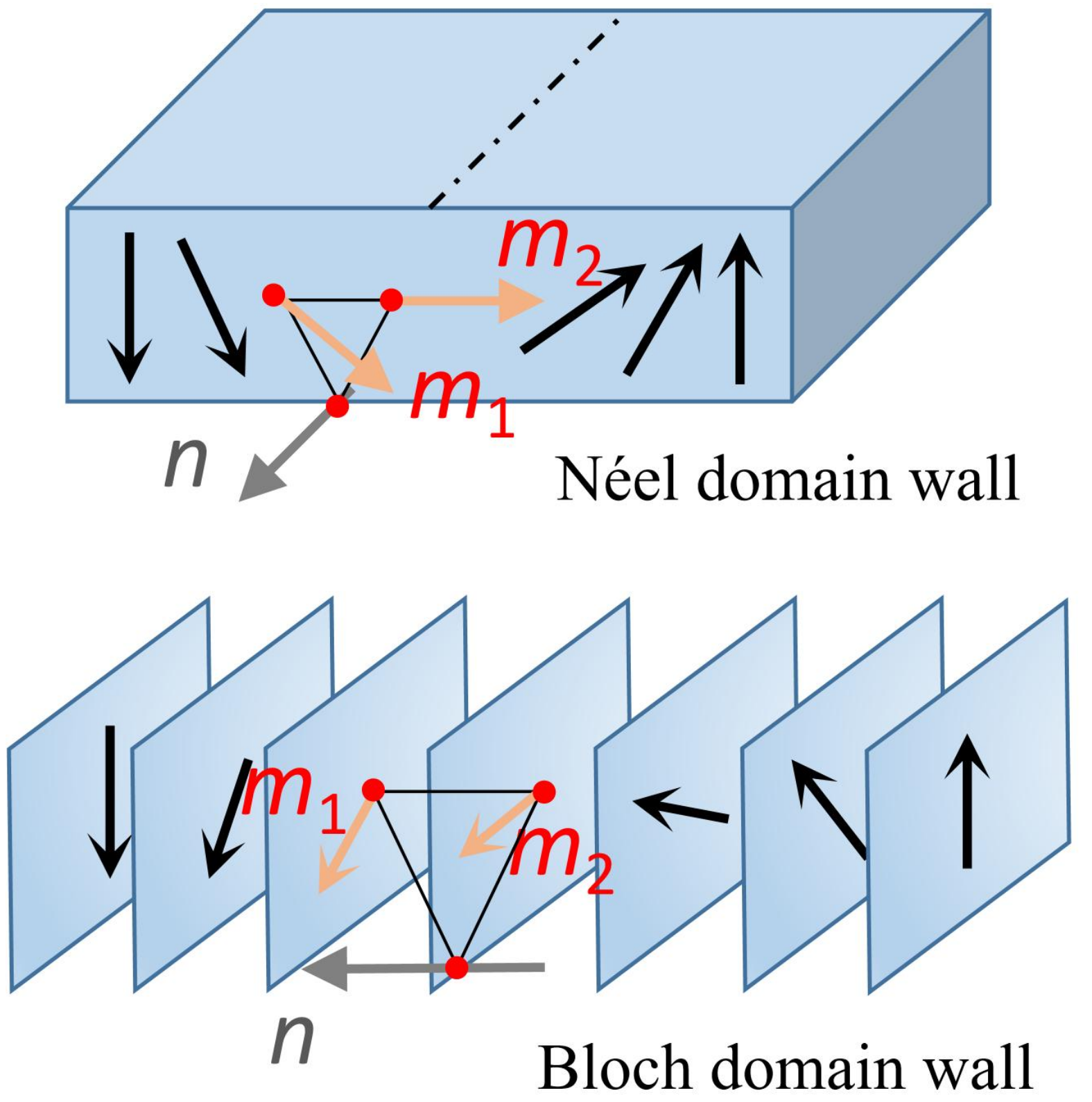}\vspace*{4pt}
   		\end{minipage} &
   		\hspace{-1.5cm}\begin{minipage}[m]{4.8cm}
   			\begin{itemize}
   				\item Data storage\\
   				(racetrack memory) \cite{Ref_DMIDW8, Ref_DMIDW9}
   				\item Magnetic synapse \cite{Ref_DWSynapse1, Ref_DWSynapse2}
   				\item Boolean logic gate \cite{Ref_DWLogic1,Ref_DWLogic2,Ref_DWLogic_1,Ref_DWLogic_2,Ref_DMIChiral7}
   			\end{itemize}
   		\end{minipage} 
   		\\
\toprule
\hspace{-0.13cm}\begin{minipage}[m]{.3\textwidth}
   		Skyrmions\\
   		(\textbf{DMI} or \textbf{magnetostatics}) 
   		\end{minipage}
   		&\hspace{-1.5cm}\begin{minipage}[m]{.5\textwidth}
   			\centering\vspace*{4pt}
   			\includegraphics[width=5.1cm]{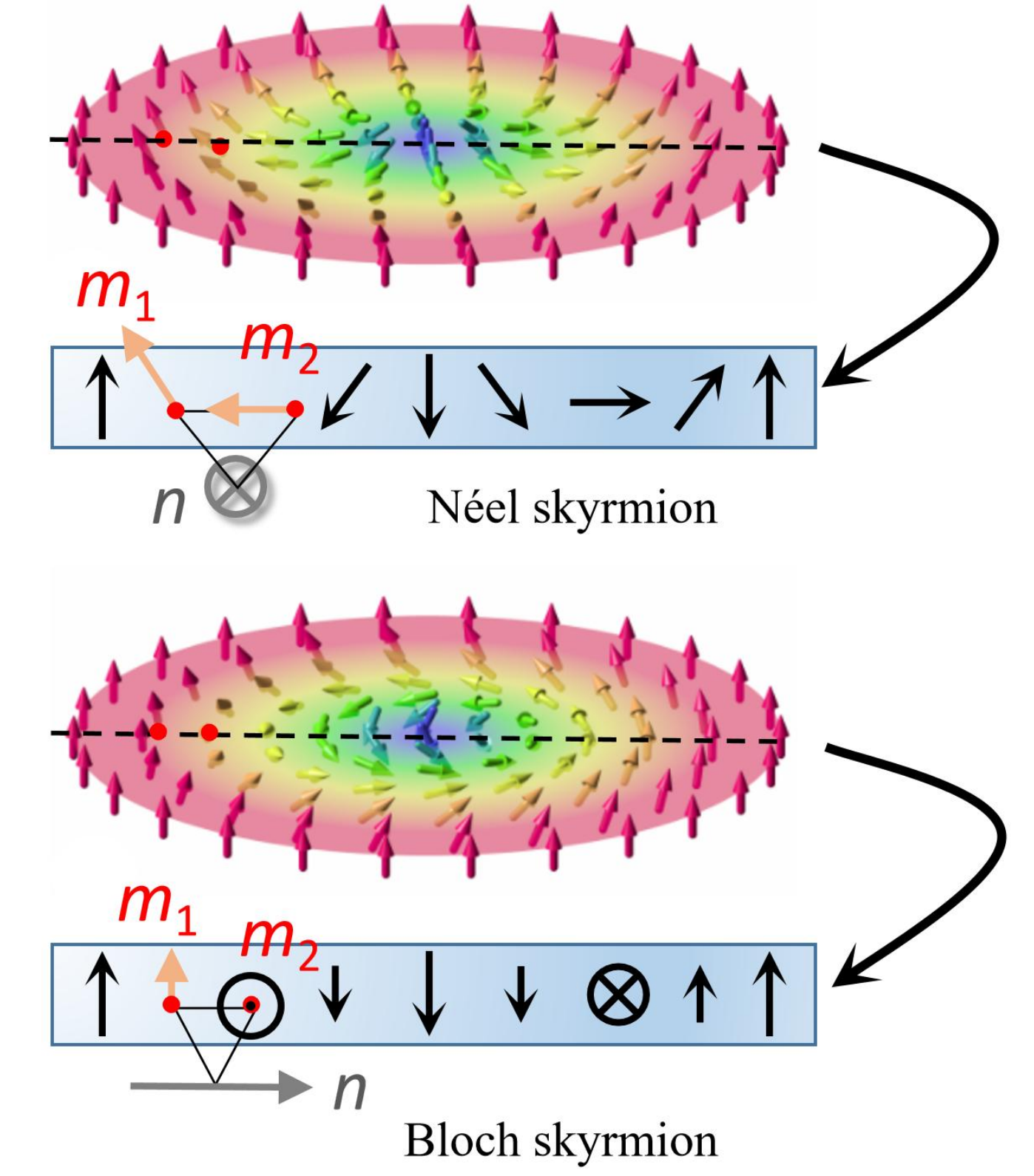}\vspace*{4pt}
   		\end{minipage} &
   		\hspace{-1.5cm}\begin{minipage}[m]{4.8cm}
   			\begin{itemize}
   				\item Data storage
   				(skyrmion racetrack memory) \cite{Ref_DMIskr4, Ref_DMIskr5}
   				\item Boolean logic gate \cite{Ref_SkrCompute1}
   				\item Probablistic computing device \cite{Ref_SkrCompute2}
   				\item Neuromorphic computing device \cite{Ref_SkrCompute3}
   			\end{itemize}
   		\end{minipage} 
   		\\
   		\hline
   	\end{tabular}
   \end{table}

\begin{figure}[ht]
	\begin{centering}
		\includegraphics[width=0.99\textwidth]{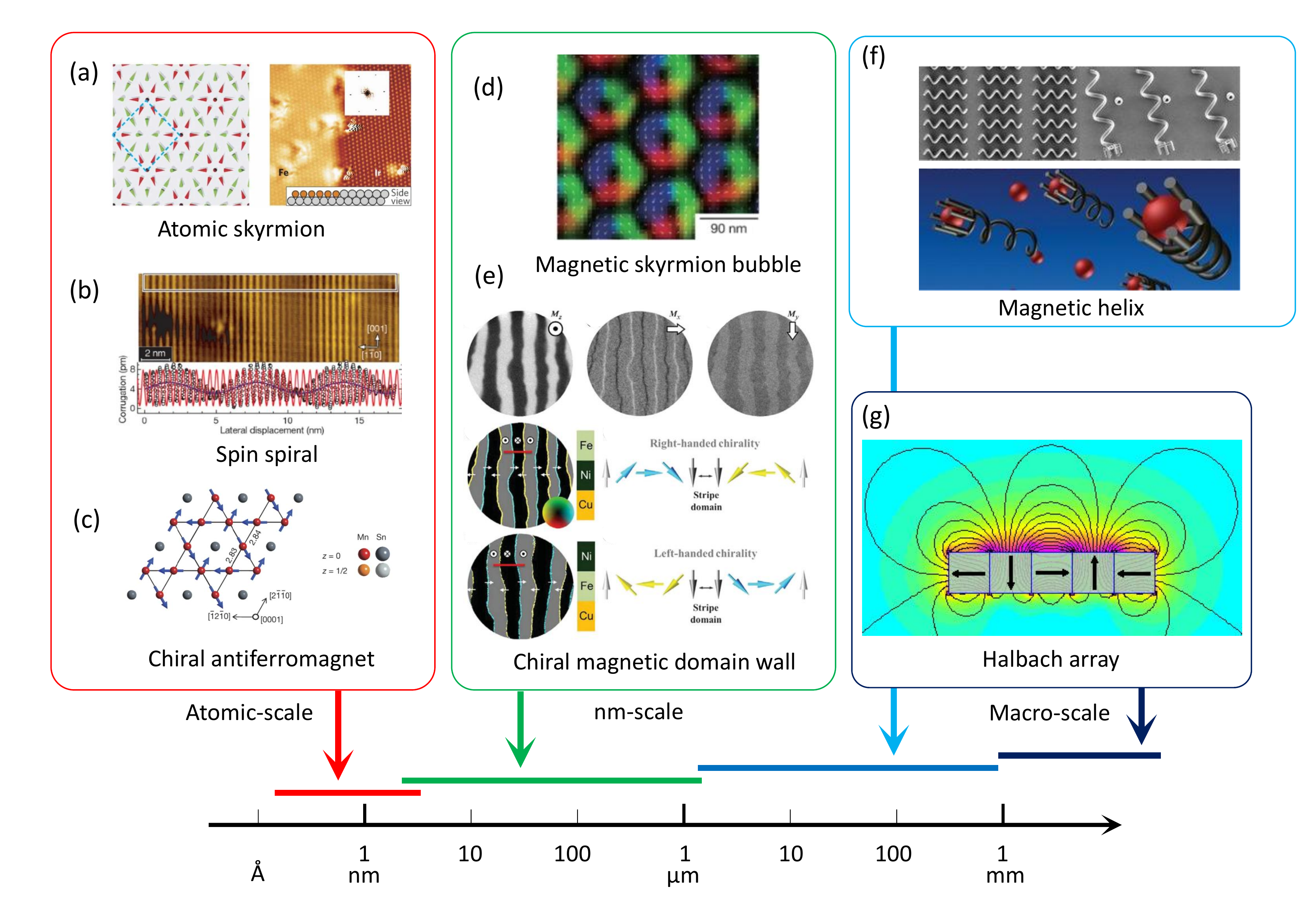}
		\par\end{centering}
	\caption{Chiral magnetic textures of different length scales. At atomic scale: (a) atomic skyrmion \cite{Ref_AtomSkyrmion}, (b) spin helix \cite{Ref_SpinHelix} and (c) non-collinear antiferromagnetism \cite{Ref_NAFM}; at nano-scale: (d) skyrmion bubble \cite{Ref_SkyrmionBubble} and (e) chiral N\'eel domain wall \cite{Ref_ChiralNeelDW}; at micro-scale, (f) helical structure \cite{Ref_Geochiral6}; at macro-scale, (g) magnetic Halbach array. The figures are reproduced with permission from \cite{Ref_AtomSkyrmion, Ref_SpinHelix, Ref_NAFM, Ref_SkyrmionBubble, Ref_ChiralNeelDW, Ref_Geochiral6}.}
	\label{Chiraltexture1}
\end{figure}

The scale of magnetic texture may vary from interatomic distances to millimeters (mm), as depicted in  Fig.~\ref{Chiraltexture1}. The DMI causes chiralities at an atomic and nanometer scale, while that at the micro- and macro-scale depends on the sample shape and is imprinted by the magnetostatic interaction. 
The relative importance of exchange interaction, dipolar interaction, anisotropy energy, DMI, and  sample size and shape generate many chiral magnetic textures ranging from an atomic scale, such as atomic skyrmions \cite{Ref_AtomSkyrmion} [Fig.~\ref{Chiraltexture1}(a)], spin helices \cite{Ref_SpinHelix} [Fig.~\ref{Chiraltexture1}(b)] and non-collinear antiferromagnetism \cite{Ref_NAFM} [Fig.~\ref{Chiraltexture1}(c)], to the micro-scale, such as skyrmionic ``bubbles" \cite{Ref_SkyrmionBubble} [Fig.~\ref{Chiraltexture1}(d)] and chiral N\'eel domain walls \cite{Ref_ChiralNeelDW} [Fig.~\ref{Chiraltexture1}(e)].  \textcolor{blue}{As an example shown in Fig.~\ref{Chiraltexture1}(c), the chirality index of the antiferromagnet Mn$_3$Sn can be defined by any two neighboring magnetization at the hexagonal lattice and the normal vector. For the magnetic textures, the chirality can be either left-handed or right-handed, given by positive or negative chirality index (refer to Table~\ref{table_ChiralTexture} for details).} At an even larger scale, magnetic chirality may follow that of the sample shape such as a helically wound film \cite{Ref_Geochiral6} [Fig.~\ref{Chiraltexture1}(f)]. At the macro-scale, a chiral magnetic texture can be enforced by judiciously arranging permanent magnets. An example is the ``Halbach array" \cite{Halbach} that focuses the stray magnetic field on one side of the array while nearly eliminating it on the other side [Fig.~\ref{Chiraltexture1}(g)]. Such textures are not only of fundamental interest but exhibit many attractive functionalities.

\subsubsection{Geometry-induced chiral magnetic textures}\label{section2.1.1} 

\begin{figure}[ht]
	\begin{centering}
		\includegraphics[width=0.88\textwidth]{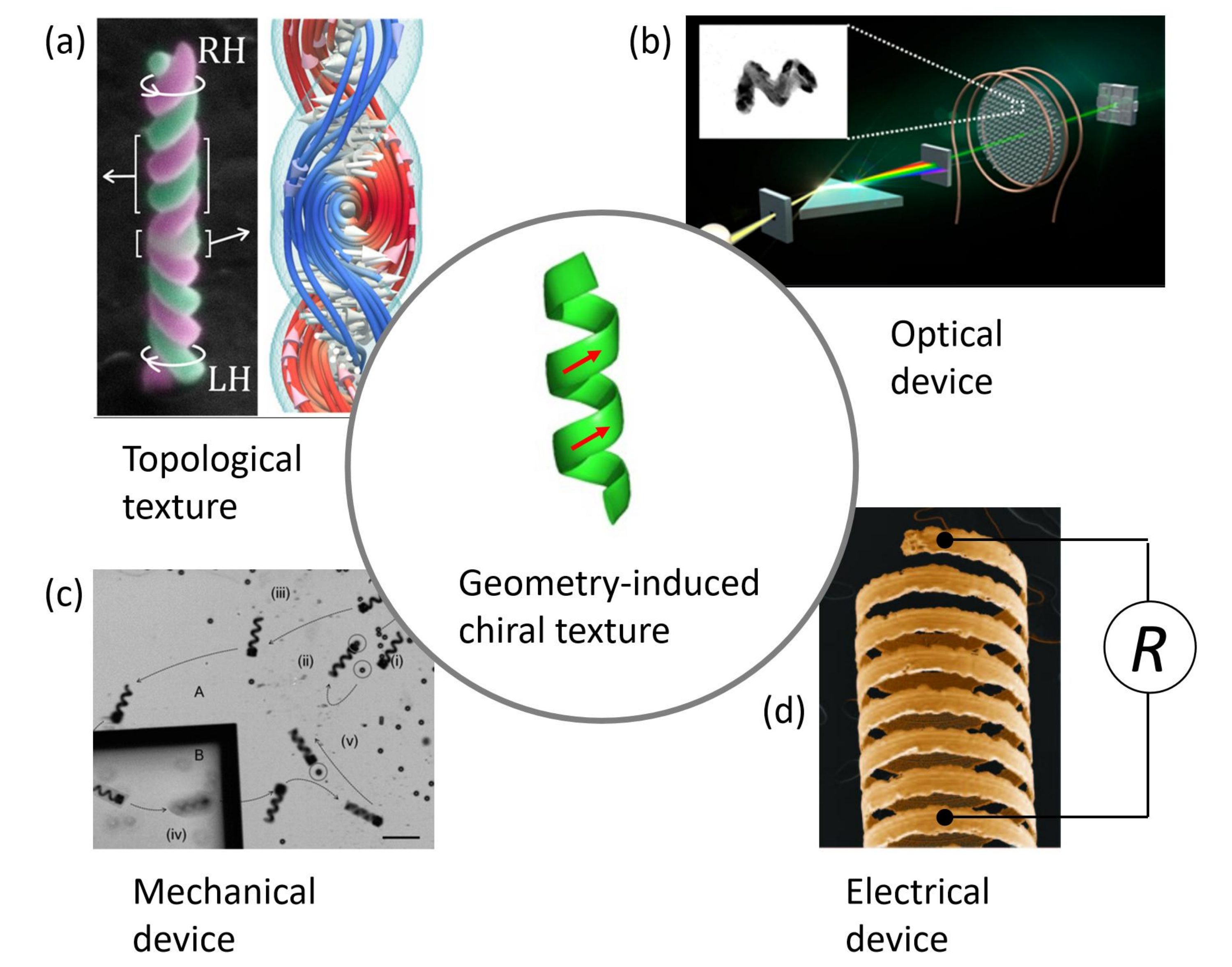}
		\par\end{centering}
	\caption{Chiral magnetic texture in helix-shaped magnet and its applications in nontrivial topological physics \cite{Ref_Geochiral2} and in optical \cite{Ref_Geochiral3}, electrical \cite{Ref_Geochiral5} and mechanical \cite{Ref_Geochiral6} devices. The figure is reproduced with permission from \cite{ Ref_Geochiral2, Ref_Geochiral3, Ref_Geochiral5, Ref_Geochiral6}.}
	\label{Chiraltexture2}
\end{figure}

We first introduce the micrometer-sized chiral texture induced by the magnetostatic interaction and the shape of the magnet. In films, a magnetization parallel to the surface minimizes the demagnetization energy. Hence, a chiral magnetic texture can be fabricated by winding a magnetic strip into a chiral helix with a given handedness, as shown in  Fig.~\ref{Chiraltexture2}. Intricate topological magnetic textures can be realized by nanofabricating magnetic materials into single and double helices \cite{Ref_Geochiral1,Ref_Geochiral2} [Fig.~\ref{Chiraltexture2}(a)]. Chiral magnetic textures can exhibit unique optical properties such as magnetochiral dichroism, \textit{i.e.}, a magnetic-field-tunable, polarization-independent absorption effect \cite{Ref_Geochiral3} [Fig.~\ref{Chiraltexture2}(b)]. Electric fields may affect chiral magnetizations by the magnetoelectric coupling \cite{Ref_Geochiral4}. Ferromagnetic CoNi micro helices fabricated by electrodeposition and laser printing showed an anisotropic magnetoresistance (AMR)  with a contribution that was attributed to the self-magnetic field induced by the current \cite{Ref_Geochiral5} [Fig.~\ref{Chiraltexture2}(c)]. Rotating magnetic helices in liquids propel themselves under a magnetic field and at low Reynolds numbers  may transport cargo \cite{Ref_Geochiral6, Ref_Geochiral7}  [Fig.~\ref{Chiraltexture2}(d)].

Magnetic thin-films patterned into a ring or disk shape, a two-dimensional closed-loop chiral vortex texture minimizes the dipolar energy \cite{Ref_Geochiral8, Ref_Geochiral10, Ref_Geochiral12, Ref_Geochiral9, Ref_Geochiral11}. The absence of magnetic stray fields and therefore cross-talk between neighboring elements is attractive for applications in magnetic memories \cite{Ref_Geochiral12}. The vortex core in thin disks cannot accommodate the in-plane magnetization, which leads to a small up or down out-of-plane magnetization in its center, which offers a second digital degree of freedom. Its motion under DC and AC magnetic field constitutes a magnetic oscillator with low and tunable frequency \cite{Ref_Geochiral13} and sensitive magnetic field detector \cite{Ref_Geochiral14}. A local magnetization gradient in samples patterned into curvilinear shapes or defects generates an effective magnetic field that is generated solely by the exchange interaction without the need for either dipolar interaction or DMI \cite{Ref_Geochiral15, Ref_Geochiral16, Ref_Geochiral17, Ref_Geochiral18}. 
 
\subsubsection{DMI-induced chiral magnetic textures}\label{section2.1.2}  
The magnetodipolar interaction is long-ranged, which means that it dominates the texture in bigger samples since a large number of spins contributes, but in nanoscale samples, it becomes very small. The microscopic DMI, on the other hand, dominates on the atomic unit nanoscale and offers a new way to construct chiral magnetic textures. It favors a non-collinear alignment of neighboring spins, striving to twist a collinear magnetization. The  competition between the (symmetric) exchange interaction, dipolar interaction, DMI, and magnetic anisotropy can generate a rich zoo of ground state textures,  such as chiral N\'eel domain walls,  spin helices, cycloids, or skyrmions (Table~\ref{table_ChiralTexture}) that offer a platform to realize high-performance spintronic devices.

\begin{figure}[ht]
	\begin{centering}
		\includegraphics[width=0.96\textwidth]{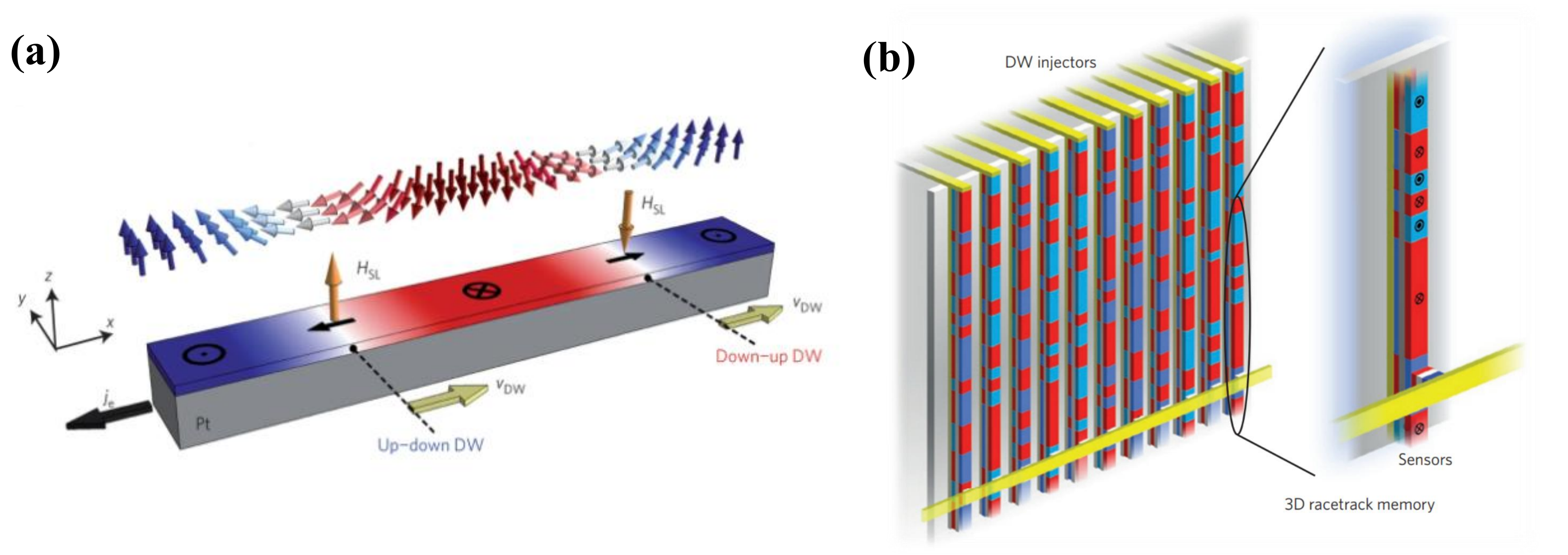}
		\par\end{centering}
	\caption{Current-driven chiral domain wall motion and domain wall racetrack memory. (a) is the illustration of left-handed chiral N\'eel domain walls in Pt/CoFe/MgO \cite{Ref_DMIDW4}. As the magnetization in the middle of the $\otimes|\odot$ and $\odot|\otimes$ domain walls is opposite, the effective field ${\mathbf{H}}_{\rm SOT}\sim ({\bf M}\times \pmb{ \sigma})\parallel z$ from the damping-like torque moves adjacent $\otimes|\odot$ and $\odot|\otimes$ domain walls with velocity ${\bf v}_{\rm 
	DW}$ in the same direction against electron flow ${\bf j}_{\rm e}$. The effect of damping-like \(\tau_{SOT}\) looks like an out-of-plane field only for the magnetization in the middle of N\'eel domain walls, which drives both $\otimes|\odot$ and $\odot|\otimes$ domain walls in the same direction. (b) is the schematics of 3D domain wall racetrack memory device \cite{Ref_DMIDW9}. The figure is reproduced with permission from \cite{Ref_DMIDW4, Ref_DMIDW7}.}
	\label{Chiraltexture3}
\end{figure}

Perpendicularly magnetized films support two types of domain walls, \textit{viz.}, N\'eel and Bloch domain wall. The magnetization rotates in one and the same plane in a N\'eel wall, but parallel to the domain wall (normal to the domain magnetizations) in a Bloch wall. The trade-off between the exchange, magnetostatic, and perpendicular anisotropy energies governs their relative stability. Usually, the N\'eel domain walls appear in thin and narrow magnetic wire \cite{Ref_DMIDW1}. In a perpendicularly magnetized film, the DMI not only favors the N\'eel configuration but also selects its chirality. The spin-orbit torques (SOTs) enable current-induced motion of N\'eel-type domain walls \cite{Ref_DMIDW2, Ref_DMIDW3, Ref_DMIDW4, Ref_DMIDW5, Ref_DMIDW6, Ref_DMIDW7} [Fig.~\ref{Chiraltexture3}(a)]. SOTs may originate from the anomalous Hall effect and in bilayers with heavy metals,  spin Hall and Rashba-Edelstein  effects. Phenomenologically, SOTs vectors normal to the magnetization with two components that act as (anti) damping and effective magnetic field when added to the Landau-Lifshitz-Gilbert (LLG) equation: 
 \begin{align}
 	\pmb{\tau}_{\rm SOT}=a{\bf M}\times \pmb{\sigma}+b{\bf M}\times({\bf M}\times \pmb{ \sigma}),
 	\label{SOT}
\end{align}
where $a$ and $b$ are parameters depending on the electric current, magnetization, device geometry, and materials. $\pmb{\sigma}$ is the direction of the injected spin moment of electrons. For the current-induced motion  of N\'eel domain walls  the SOT is more efficient than that of the conventional spin-transfer torque (STT) \cite{Ref_STT1} because of the interplay of the chiral magnetic texture and SOT. Moreover, the damping-like SOTs move  $\otimes|\odot$ and $\odot|\otimes$ domain walls  in the same direction, such that a chain of domain walls moves in a rigid fashion [see the case of left-handed N\'eel domain wall in Fig.~\ref{Chiraltexture3}(a)]. This is an important feature of the domain wall shift register or ``racetrack" memory \cite{Ref_DMIDW8, Ref_DMIDW9}, as illustrated in Fig.~\ref{Chiraltexture3}(b), in which digital information of \textquotedblleft 1\textquotedblright\ and \textquotedblleft 0\textquotedblright\ is encoded by a pattern of $\odot$ and $\otimes$ magnetization in a nanowire. A domain wall injector writes the data by transferring an electric signal into magnetization. An electric current flowing in the magnetic nanowire then shifts the domain-wall pattern under a magnetic tunnel junction that detects the magnetization direction by its electric resistance. This memory scheme allows for ultra-high storage density that can be far superior to magnetic hard disk drives. It can achieve a higher density by vertical integration, has a faster read/write response, lower energy consumption, and is more reliable because there are on moving parts. The efficiency of current-driven domain wall motion can be further improved by using magnetic insulators \cite{Ref_DMIDW10, Ref_DMIDW11}, ferromagnetic materials \cite{Ref_DMIDW12, Ref_DMIDW13, Ref_DMIDW14, Ref_DMIDW15}, and synthetic antiferromagnetic multilayers \cite{Ref_DMIDW16, Ref_DMIDW17}. In the case of current-driven DW motion, STT originates from the noncollinear orientation of magnetization inside the domain wall. The STT efficiency depends on the gradient of the magnetization. The domain wall width is usually several tens to hundreds of nanometers, so the current-driven domain wall motion via STT is slow and inefficient. In contrast, the orientation of spin polarization is usually independent on the magnetization and perpendicular to the magnetization in the middle of the N\'eel domain wall. So the magnitude of SOT torque on the domain is stronger than that of STT torque at the same intensity of the spin current. 

\begin{figure}[ht]
	\begin{centering}
		\includegraphics[width=0.99\textwidth]{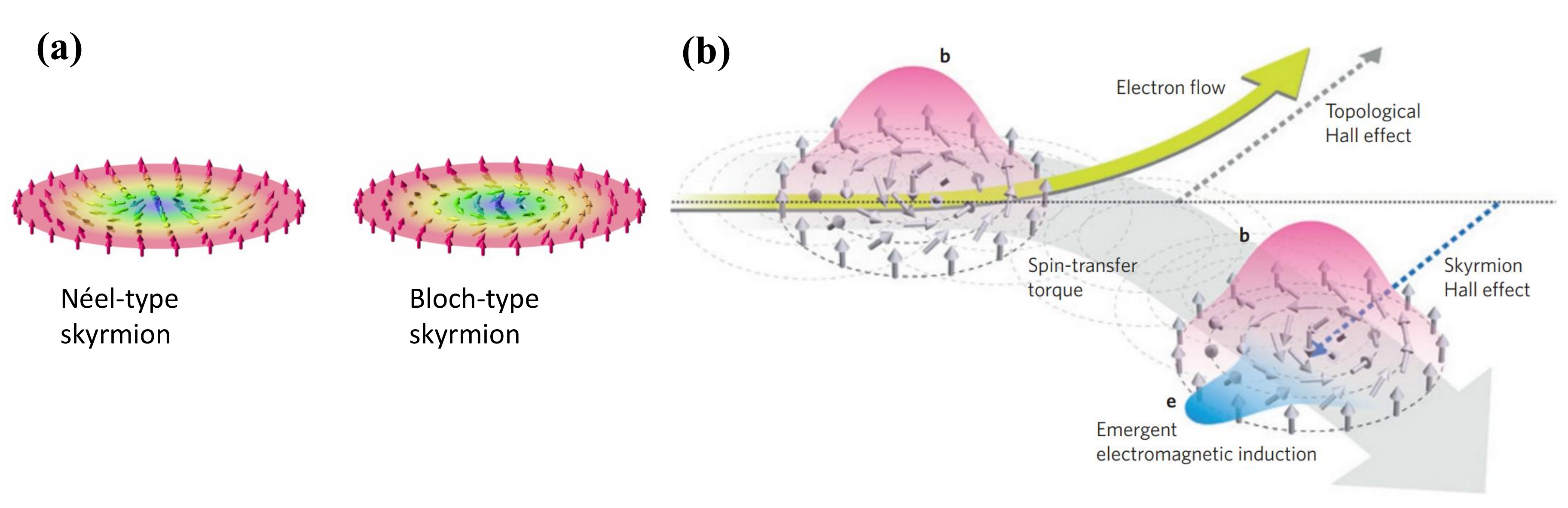}
		\par\end{centering}
	\caption{Skyrmion structure and associates topological phenomena. (a) N\'eel-type (left) and Bloch-type (right) skyrmions with magnetization rotating from the down  direction in the center to the up (external uniform) magnetization at the edge, analogous to N\'eel or in Bloch domain walls. (b) Schematic picture of current-driven skyrmion motion and associated physical phenomena. The figure is reproduced with permission from \cite{Ref_DMIskr18}.}
	\label{Chiraltexture4}
\end{figure}

A chiral magnetic texture of relatively recent interest is the magnetic skyrmion, a local ``whirl" in the magnetic texture that is quite different from the vortex \cite{Ref_DMIskr1, Ref_DMIskr2, Ref_DMIskr3}. Typically, it is a topological defect in which the magnetization rotates in a tube with a fixed chirality from the up direction at the edges to the down direction at the center. As in domain walls, skyrmions exist as both N\'eel and Bloch-type [Fig.~\ref{Chiraltexture4}(a)]. Bloch skyrmions are predominantly found in bulk materials, while  N\'eel skyrmions are favored by the DMI in interfacial DMI systems. Skyrmions are labeled by their topological number $S$, which is a measure of the winding of the unit vectors of the local magnetization direction ${\bf m}({\pmb \rho},z)$ with ${\pmb \rho}=x\hat{\bf x}+y\hat{\bf y}$. In a thin film, we may drop any dependence on $z$ normal to the interfaces and 
\begin{align}
	S=\frac{1}{4\pi}\int d^{2}{\pmb \rho}\left[\mathbf{m}\cdot(\partial_{x}\mathbf{m}\times\partial_{y}\mathbf{m})\right]=\pm1.
	\label{Skr_Winding}
\end{align}

Skyrmions can be very small, with diameters in the nanometer range, and behave like mobile particles. While topologically protected against weak perturbations, they can be created and annihilated, which makes them suitable for applications in information storage and computation technologies \cite{Ref_DMIskr4,Ref_DMIskr5,Ref_SkrCompute1,Ref_SkrCompute2,Ref_SkrCompute3}. Skyrmions may replace domain walls in the racetrack memory \cite{Ref_DMIskr4, Ref_DMIskr5}. The skyrmion texture can be considered as a closed loop of a chiral domain wall and hence the current-induced SOTs can drive the entire texture without distorting it. However, current-driven skyrmions move at a finite angle  $\theta_{sk}$ with respect to the current direction due to gyrotropic (Magnus) forces depending on the skyrmion number \textit{S} \cite{Ref_DMIskr6, Ref_DMIskr7, Ref_DMIskr8} [Fig.~\ref{Chiraltexture4}(b)].
\textcolor{blue}{For the isolated circular skyrmions in the magnetic film normal to $z$ that for weak driving may be treated as a rigid
point-like object, the skyrmion Hall angle }
\begin{align}
\tan \theta_{sk}=-{S}/({\alpha \widetilde{D}}),
	\label{Skr_Hall}
\end{align}
where $\alpha$ represents the damping factor and the dissipation factor $\widetilde{D}$ is isotropic, given by
\begin{align}
\widetilde{D}=({1}/{4\pi})\int(\partial_{x}\mathbf{M})\cdot(\partial_{x}\mathbf{M})\mathrm{d}x\mathrm{d}y=({1}/{4\pi})\int(\partial_{y}\mathbf{M})\cdot(\partial_{y}\mathbf{M})\mathrm{d}x\mathrm{d}y.
\end{align}
For the other shape of skyrmions, the dissipation factor is generally a tensor with components $\tilde{D}_{ij}=1/(4\pi)\int (\partial_i{\bf m})\cdot(\partial_j{\bf m})dxdy$, however. This skyrmion Hall effect can be detrimental for device applications since skyrmions may be trapped and annihilated at the edges.
\textcolor{blue}{The skyrmion Hall angle dependence in Eq.~(\ref{Skr_Hall}) on the damping
and skyrmion geometry changes for temperatures and driving amplitudes at which the ``rigid" approximation breaks down \cite{Ref_DMIskr8}. Its dynamics may also be
affected by the local energy landscape including structural and magnetic disorder, leading to a Hall angle that does not depend on
skyrmion diameter \cite{skyrmion_Marrows}. More studies may be needed to fully clarify this issue. }
The skyrmion Hall angle vanishes in antiferromagnets \cite{Ref_AntiSkr1, Ref_AntiSkr2, Ref_AntiSkr3, Ref_AntiSkr4}, at the compensation point of ferrimagnets \cite{Ref_DMIskr9, Ref_DMIskr10, Ref_DMIskr11}, and in synthetic antiferromagnetic stacks \cite{Ref_DMIskr12, Ref_DMIskr13,Ref_DMIskr14}.

The carriers can drive the motion of magnetic textures as mentioned above. Inversely, the chirality of the magnetic textures can influence the motion of carriers such as electrons and magnons.
The ``topological Hall effect" is an additional contribution to the transverse voltage in metallic ferromagnets in the presence of a topologically non-trivial spin texture \cite{Ref_DMIskr15, Ref_DMIskr16,  Ref_DMIskr17, Ref_DMIskr18} [Fig.~\ref{Chiraltexture4}(b)]. The magnitude of the topological Hall resistance can be written as 
\begin{align}
    \rho^{\mathrm{THE}}_{xy}=R_{0}PB^{z}_{\rm eff},
	\label{Skr_THE}
\end{align}
where $R_{0}$ is the ordinary Hall coefficient, $P$ is the spin polarization of the conduction electrons, and $B^{z}_{\rm eff}$ is an effective field experienced by the conduction electrons due to Berry phase of adiabatically traversing the skyrmion texture. The topological Hall voltage is sensitive enough to detect individual skyrmions \cite{Ref_DMIskr19, Ref_DMIskr20,  Ref_DMIskr21}. 

\subsection{Static chiral interaction between nanomagnets}

In this review, we will focus on the chirality of the  interactions that cause among others unidirectional propagation of magnons, photons, and phonons. 
As mentioned in the introduction and above, chirality affects both equilibrium and non-equilibrium properties of magnetic structures and devices. We call a coupling ``chiral" in this static situation by the locking of three vectors defined in the chirality of magnetic textures, implying that changing one of the three vectors will change another that acts as an efficient way to locally control spintronic devices. Here we first introduce the static chiral coupling phenomena in spintronics that affect the equilibrium magnetization texture, which is very useful for some magnetic devices that operate by current-induced spin transfer and spin-orbit torques. 
For example, we can engineer chiral magnetic textures with deliberate patterns by shaping the magnetic elements or locally modulating their magnetic properties through interface engineering. The 
measure of a magnetic texture is the chirality and the pitch between neighboring magnetic moments that in the ground state rotate as a function of position. Here, we first discuss the geometry-induced chiral coupling depending on the magnetic configuration and show how it brings about uniquely ordered textures that are potentially interesting for applications (Sec.~\ref{geometry_static_chiral}). We then discuss DMI-induced chiral coupling in both lateral and vertical magnetic systems (Sec.~\ref{geometry_DMI_chiral}). This is a predominantly interface effect with high coupling strength that can be engineered by the choice of materials and nanofabrication conditions.

\subsubsection{Geometry-induced chiral coupling}
\label{geometry_static_chiral}

\begin{figure}[ht]
	\begin{centering}
		\includegraphics[width=0.76\textwidth]{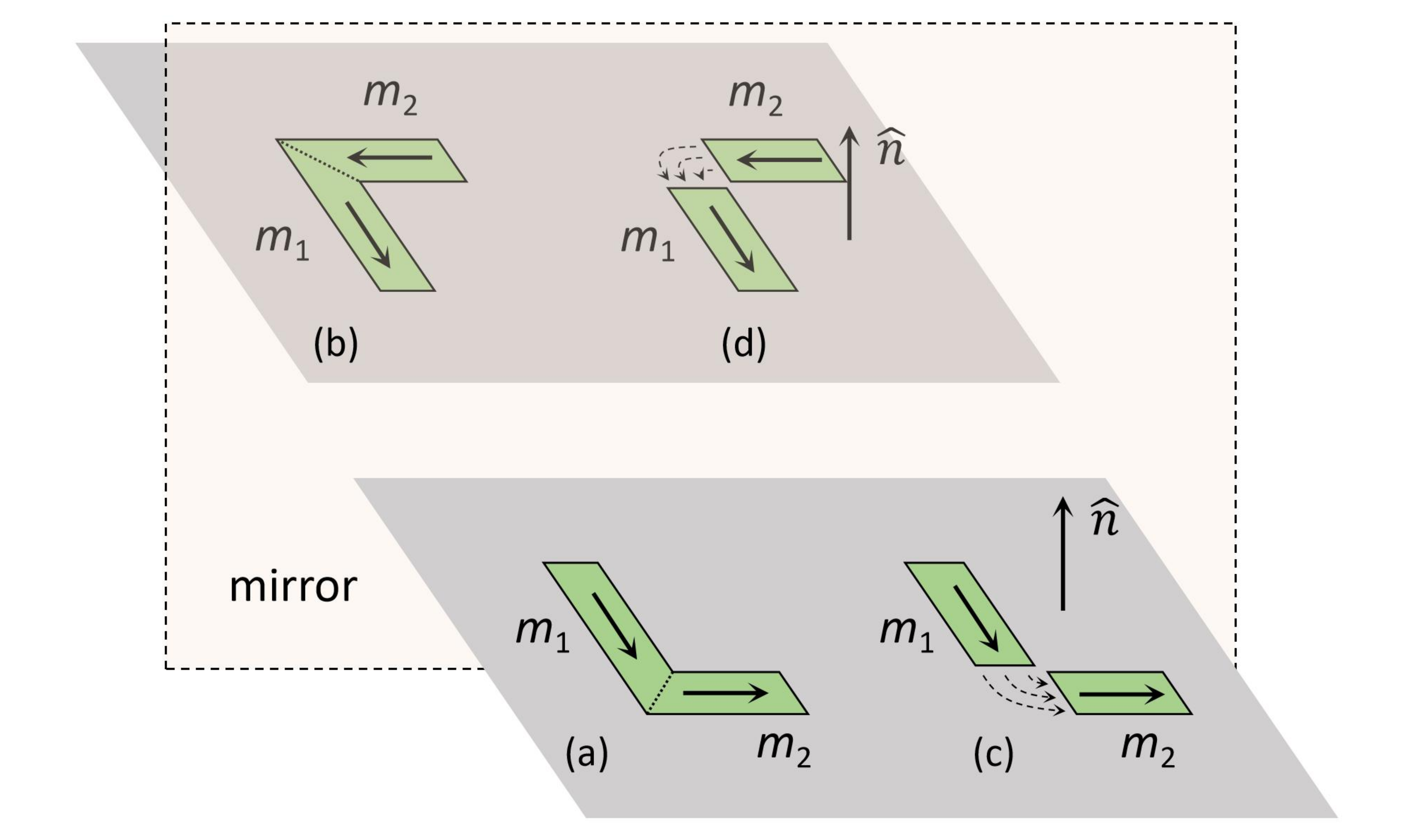}
		\par\end{centering}
	\caption{Geometry-induced chiral coupling \textcolor{blue}{by the magnetostatics. (a) and (b) are the two L-shaped magnetic conduit with low energy, in which the magnetization direction is fixed by the chirality. They are mirrors of each other by recalling that magnetic moments ${\bf m}_{1,2}$ are pseudo-vectors.} The chirality $({\bf m}_{1} \times {\bf m}_{2}) \cdot \bf{n}$  of the mirror structure is opposite. In (c) and (d), the dashed lines illustrate the approximate flux closure when the particles do not touch at heads and tails, where the chirality remains due to the long-range nature of the dipolar coupling.}
	\label{GeoChiralCoupling1}
\end{figure}

The geometry-induced chiral coupling emerges in certain magnetic configurations via the magneto-dipolar energy. For example, consider the  L-shaped magnetic conduit consisting of two strips with magnetization directions ${\bf m}_{1}$ and ${\bf m}_{2}$ lying in a plane with normal unit vector  ${\bf n}$  in Fig.~\ref{GeoChiralCoupling1}. The dominant shape anisotropies enforce the magnetization to be at right angles. The possible magnetic configurations with low anisotropy energy are  (${\bf m}_{1}=\downarrow, {\bf m}_{2}=\leftarrow$), $\downarrow\rightarrow$, $\uparrow\leftarrow$, and $\uparrow\rightarrow$. The head-to-tail or tail-to-head orders ($\downarrow\rightarrow$ and $\uparrow\leftarrow$)  generate less stray magnetic fields and have reduced dipolar energy
\begin{align}
	H^{\rm Chiral}_{\rm Dipole}=-\frac{\mu_{0}}{4\pi}
	\left(3({\bf m}_1\cdot{\bf r}_{12})({\bf m}_2\cdot {\bf r}_{12})\right).
	\label{ChiralDipole}
\end{align}
\textcolor{blue}{Figure~\ref{GeoChiralCoupling1}(a) and (b) show plots of these low-energy magnetization configurations, which are mirror images of each other.}
When the magnetic particles are disconnected, we can still have an approximate flux closure, as shown in Figs.~\ref{GeoChiralCoupling1}(c) and (d). The two configurations are degenerate in energy and with opposite chirality defined by as  $({\bf m}_{1} \times {\bf m}_{2}) \cdot \bf{n}$ and each-others  mirror, as illustrated in Fig.~\ref{GeoChiralCoupling1}. Hence, the geometry-induced chiral coupling can be controlled by the sample design. 

\begin{figure}[ht]
	\begin{centering}
		\includegraphics[width=0.98\textwidth]{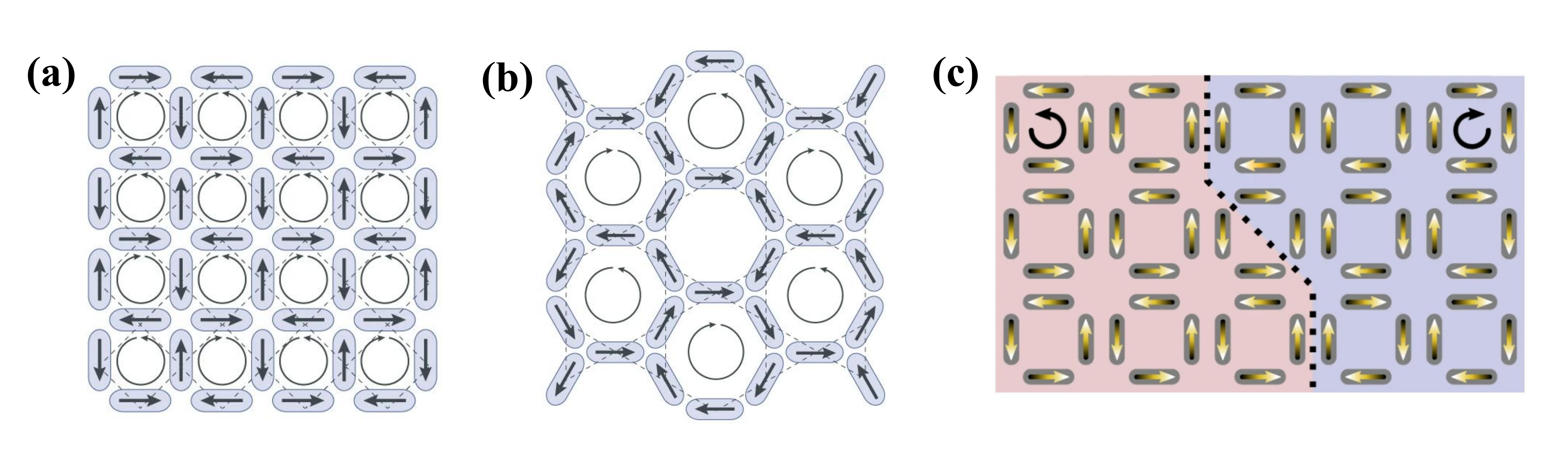}
		\par\end{centering}
	\caption{Achiral alignment in artificial spin ices on the square  (a) and kagome (b) lattices \textit{vs}. chiral alignment in the toroidal (c) lattice \cite{Ref_ASI12}. The arrows indicate the head-to-tail (closed loop) alignments of the magnetizations.  The figures are reproduced with permission from \cite{ Ref_ASI12}.}
	\label{GeoChiralCoupling2}
\end{figure}

The chiral coupling between nanomagnets leads to emergent phenomena in nanomagnetic arrays. Artificial spin ice is a  metamaterial made from magnetic nanorods (single-domain elongated islands) on periodic and aperiodic lattices \cite{Ref_ASI1, Ref_ASI2, Ref_ASI3, Ref_ASI4}, with \textit{e.g.},  square or kagome unit cells illustrated by Figs.~\ref{GeoChiralCoupling2}(a) and (b), \textcolor{blue}{which are here achiral, however, since in every unit cell
opposite circulations of magnetization of equal magnitude exist}. Also here a head-to-tail or tail-to-head magnetic configuration minimizes the total dipolar energy  and leads to a ground state with vortex-like magnetic alignments \cite{Ref_ASI5, Ref_ASI6}. The handedness of the magnetic order \textcolor{blue}{can emerge} in artificial magneto-toroidal crystals, as shown in Fig.~\ref{GeoChiralCoupling2}(c), in which the unit cell consists of 4 nanomagnets arranged in a square \cite{Ref_ASI11, Ref_ASI12}. In every domain there is an ordered state that corresponds to equal chirality, between which there is a domain interface [Fig.~\ref{GeoChiralCoupling2}(c)]. This chiral order can be locally manipulated by an effective toroidal field around a close-by magnetic tip. This artificial magneto-toroidal crystal is a model system for a novel ferroic phase without stray fields and a route to non-reciprocal physical phenomena that are based on the simultaneous breaking of inversion and time reversal symmetries. 

Additional phenomena arise when the artificial spin ice lattice itself has a spatial chirality such as the pinwheel (or windmill) system, referred to as ``chiral ice"  \cite{Ref_ASI8, Ref_ASI9, Ref_ASI10}. Here the unit cell of the square lattice consists of 4 magnets that rotate through 45\textdegree. Minimization of the dipolar energy at the edges of the array results in a ratchet effect, in which thermal relaxation leads to a uni-directional rotation of the entire magnetization that is governed by the edge structure. This uni-directional rotation in chiral ice exemplifies how magnetostatically coupled nanomagnet arrays can be designed to harness thermal fluctuations of the magnetization and convert heat into motion, opening an alternative possibility to devise mesoscopic motors and rotors.

\begin{figure}[ht]
	\begin{centering}
		\includegraphics[width=0.99\textwidth]{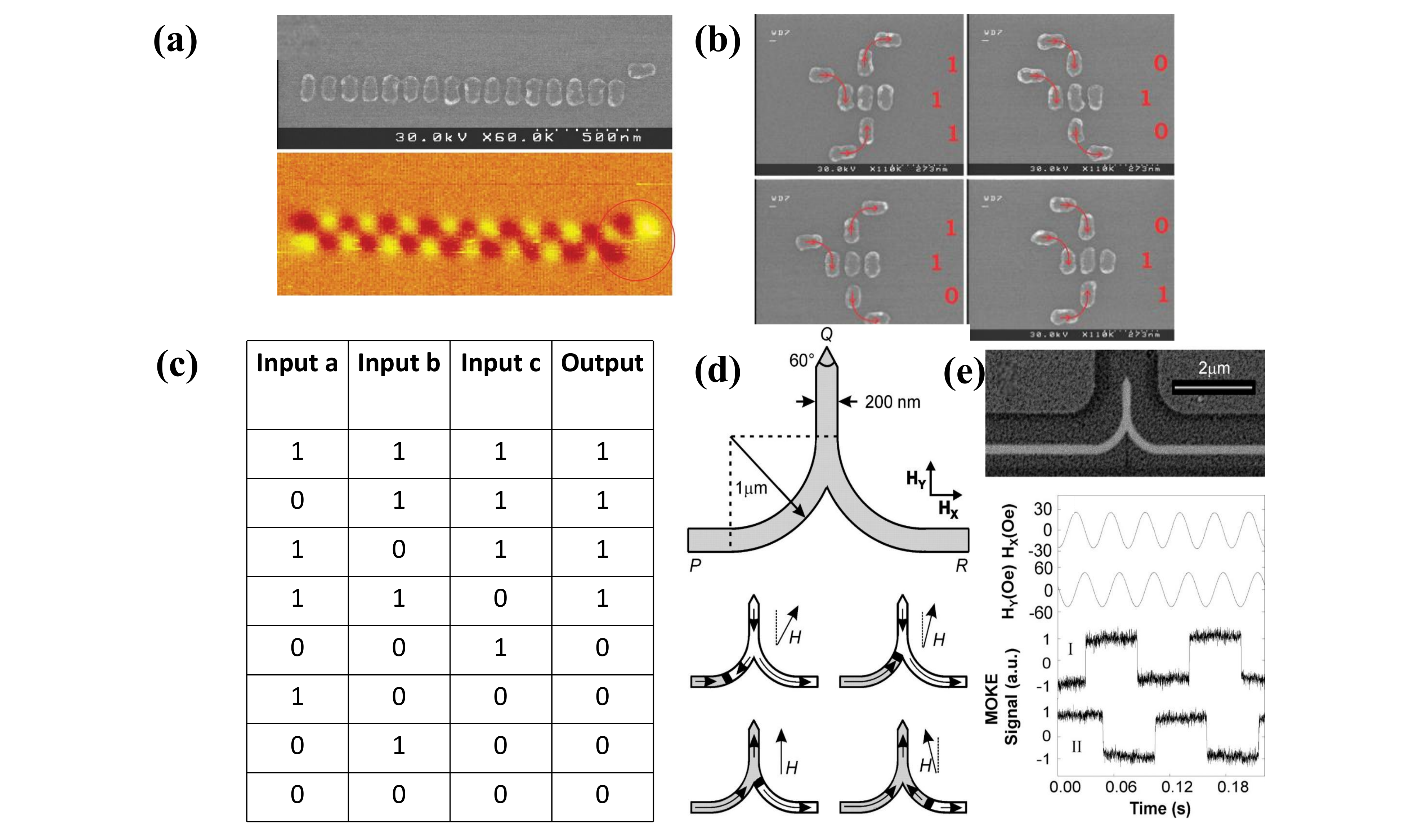}
		\par\end{centering}
	\caption{Logic devices that employ the geometry-induced chiral coupling. (a) Switchable magnetic chain \cite{Ref_NMLogic2}. Top panel: SEM image of closely spaced magnetic nanoislands. Bottom panel: the corresponding MFM image shows the anti-parallel coupling between nearest-neighbor islands. The most-right horizontal island selects one of the two degenerate ground states of the anti-parallel chain via the chiral coupling. (b) SEM images of a majority gate \cite{Ref_NMLogic2}. The red arrows indicate the logical input  that affects the magnetic configuration  by the chiral coupling. (c) The truth table is that of a majority gate. (d) Schematics of a magnetic NOT gate in which a rotating magnetic field inverts an injected domain wall (DW)  in a flipped T-shape magnetic structure. The figure illustrates the successive magnetization directions (arrows) and domain wall positions (black squares) \cite{Ref_DWLogic1}. (e) SEM image of such a microfabricated DW NOT gate (top panel) and the MOKE signals from the left branch (input, trace I) and the right branch (output, trace II) under a counterclockwise rotating magnetic field (bottom panel). The figure is reproduced with permission from \cite{Ref_NMLogic2, Ref_DWLogic1}.}
	\label{GeoChiralCoupling4}
\end{figure}

The geometry-induced chiral coupling may lead to many applications. Carefully designed nanomagnetic islands define a dipole energy landscape that provides the functionality needed  for certain computational architectures \cite{Ref_NMLogic1, Ref_NMLogic2}, in which the direction of a magnetic element encodes a single bit (\textit{e.g.}, up=``1" and down=``0" or vice versa).  As shown in Fig.~\ref{GeoChiralCoupling4}(a), an array of magnetic nanoislands orders antiferromagnetically. The degeneracy of the two ground states is broken by a control island at right angles that may select and switch the configuration of the entire array by the chiral coupling introduced above. An arrangement of  horizontal and vertical magnetic islands as in Fig.~\ref{GeoChiralCoupling4}(b) acts as a majority gate with the truth table given in Fig.~\ref{GeoChiralCoupling4}(c). Here the directions of the horizontal control magnets serve as the 3 logic inputs while the magnetization of the rightmost vertical magnet islands is the logic output. 

In the domain wall (DW) logic, the information is stored in the magnetization direction in nano-sized magnetic conduits, while logic operations are accomplished by propagating domain walls that switch the magnetizations.  Figures~\ref{GeoChiralCoupling4}(d) and (e) \cite{Ref_DWLogic1, Ref_DWLogic2} illustrate an elementary structure that operates as a DW inverter that switches between the head-to-head and tail-to-tail DW polarities. This is a T-shaped structure operating under a rotating in-plane magnetic field. The alignment  favored by the dipolar energy is $\rightarrow\uparrow\leftarrow$ and $\leftarrow\downarrow\rightarrow$. A magnetic field with $+\hat{\bf x}$ component pushes a DW of the head-to-head polarity from the left to the T-shape. When the magnetic field rotates adiabatically to  $-\hat{\bf x}$ component leads to a switch of the vertical magnetization from $-\hat{\bf y}$ to $+\hat{\bf y}$, while the DW moves to the right side with tail-to-tail  polarity.

The  magnetodipolar interaction causes the chirality in the static geometric coupling (as well as in the dynamic coupling discussed in later Chapters). In principle, we can model its effects easily and optimize the functionality of devices. However, the geometry-induced chiral coupling  is also prone to disorder and imprecise fabrication issues that become more serious for smaller structures. The long range of the dipolar interaction can also render the modeling cumbersome for complex structures.

\subsubsection{Chiral coupling by spin-orbit interaction}
\label{geometry_DMI_chiral}

\begin{figure}[ht]
	\begin{centering}
		\includegraphics[width=0.98\textwidth]{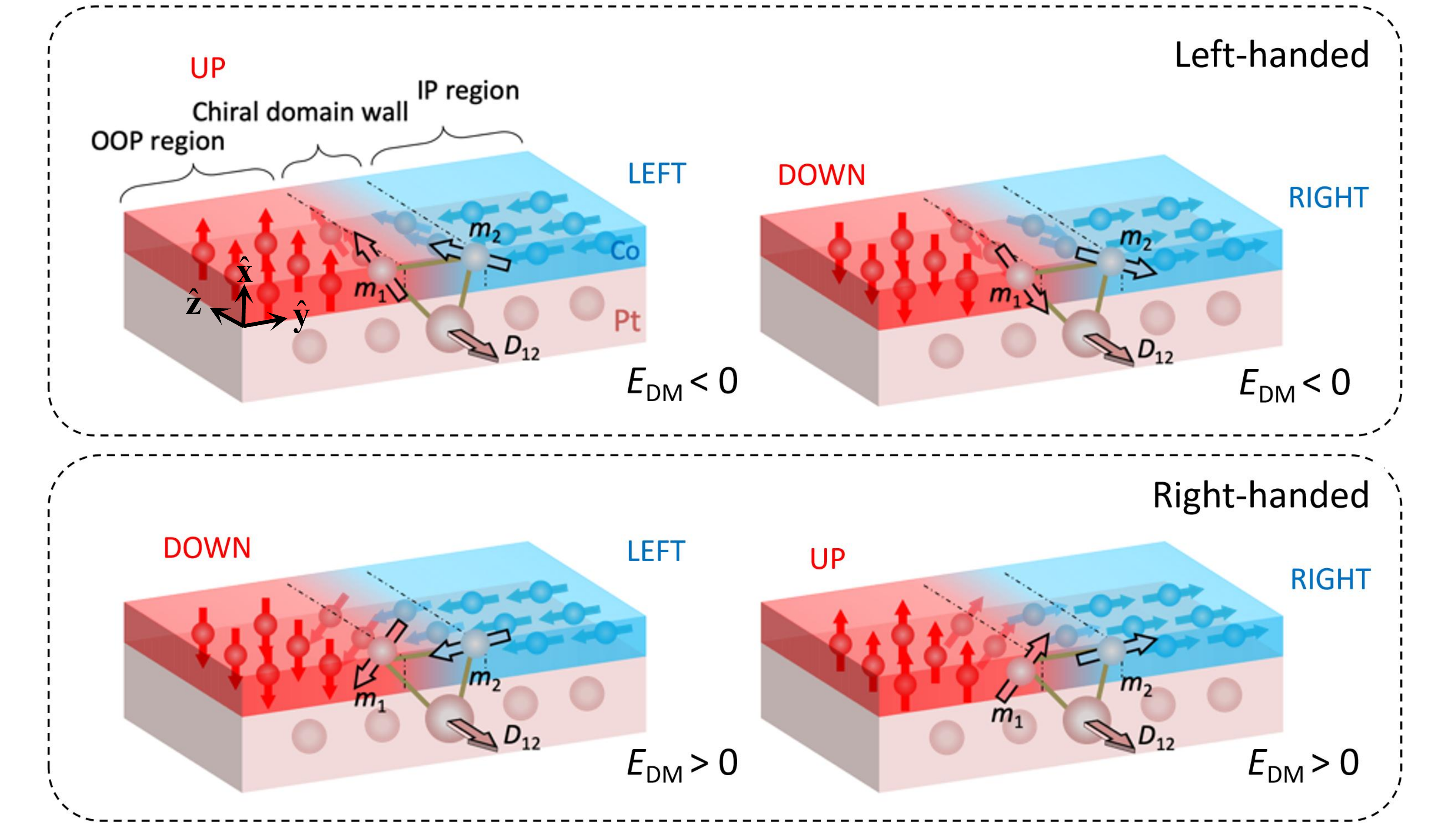}
		\par\end{centering}
	\caption{\textcolor{blue}{Chirally coupled Co magnetic domains with out-of-plane (red) and in-plane (blue) magnetizations in proximity
to Pt. The chirality or handedness of the domain wall is fixed by the interfacial DMI that favors the configurations in the upper panel indicated by $E_{\rm DM}<0$, but not those in the lower panel with $E_{\rm DM}>0$.} The figure is reproduced with permission from \cite{Ref_DMIChiral1}.}
	\label{DMIChiralCoupling1}
\end{figure}

The spin-orbit interaction is another important mechanism responsible for chiral magnetic ground states that again can be employed as a chiral coupling between nanomagnets. The Dzyaloshinskii-Moriya interaction (DMI) can induce chirality in noncollinear magnetic textures such as domain walls as reviewed in Table~\ref{table_ChiralTexture}. In order to assess the effect of the DMI we have to control the orientation of local magnetizations. Let us consider   a magnetic element that connects a left side with  out-of-plane magnetic anisotropy with a right side with an in-plane uniaxial anisotropy that is depicted in Fig.~\ref{DMIChiralCoupling1}, which has four distinct magnetic configurations of $\uparrow\rightarrow$, $\downarrow\rightarrow$, $\uparrow\leftarrow$ and $\downarrow\leftarrow$ with identical exchange and dipolar energy. 
To be specific, here we consider a specific configuration shown in Fig.~\ref{DMIChiralCoupling1} with interface normal along $\hat{\bf x}$ and magnetization direction ${\bf m}$ varies in the $x$-$y$ plane.
The DMI energy in the continuum model reads 
\begin{align}
\nonumber
    E^{\rm Chiral}_{\rm DMI}&=-\frac{D}{2\mu_0}\int d{\bf r} {\bf m}\cdot[(\hat{\bf x}\times\nabla)\times{\bf m}]=-\frac{D}{2\mu_0}\int d{\bf r} \hat{\bf z}\cdot[({\bf m}\times\partial_y{\bf m}]\\
    &=-\frac{D}{2\mu_0}\int d{\bf r} \left(m_x\partial_ym_y-m_y\partial_ym_x\right)=\pm\frac{\pi}{2}\frac{D}{2\mu_0}lt,
\end{align}
\textcolor{blue}{where $D$ is the DMI strength, $t$ and $l$ are the thickness and length (along $\hat{\bf z}$) of the magnetic film, and the sign ``$\pm$" is determined by the chirality of the domain wall.}
 The magnetization direction ${\bf m}$ varies by $\pm \pi/2$ across the domain, with which we can perform the integration. The sign of \(D\) defines the chirality; it is negative for a left-handed texture ($\uparrow\leftarrow$ and $\downarrow\rightarrow$) and positive for the right-handed one ($\uparrow\rightarrow$ and $\downarrow\leftarrow$). Hence, the four configurations fall into two categories with left-handed (top) and right-handed (bottom) chiralities (Fig.~\ref{DMIChiralCoupling1}).  
 
 The DM vector direction governs the DMI energy difference. In Pt/Co the \(D\)  is negative. The domain wall between the out-of-plane and in-plane magnetizations is then left-handed: when the out-of-plane magnetization is $\uparrow$  $(\downarrow)$, the adjacent in-plane magnetization prefers to be $\leftarrow$ $(\rightarrow)$, \textcolor{blue}{as shown in the upper panel of Fig.~\ref{DMIChiralCoupling1}. The chiral domain wall between two single-domain nanomagnets locks their magnetization, i.e. changing the magnetization direction in one nanomagnet flips that of its neighbor \cite{Ref_DMIChiral1}, a mechanism promising many functionalities as summarized in the following.}

The relative importance of the DMI \textit{vs.} dipolar coupling depends on the geometry and size of the magnetic devices. In the nanomagnetic structures of ultra-thin films, the dipolar coupling becomes very weak, which leads to reduced performance of dipolar-coupled logic devices \cite{Ref_NMLogic1, Ref_NMLogic2}. On the other hand, the DMI coupling is an interfacial effect that becomes relatively important for thin films. Moreover, the chiral coupling happens close to the boundary of the out-of-plane/in-plane magnetized regions on the scale of a domain wall width, which becomes increasingly important when scaling down the  magnetized regions. Finally, in contrast to the dipolar interaction, the DM coupling is local and avoids the complexity caused by the non-locality of the dipolar coupling.

\begin{figure}[ht]
	\begin{centering}
		\includegraphics[width=0.95\textwidth]{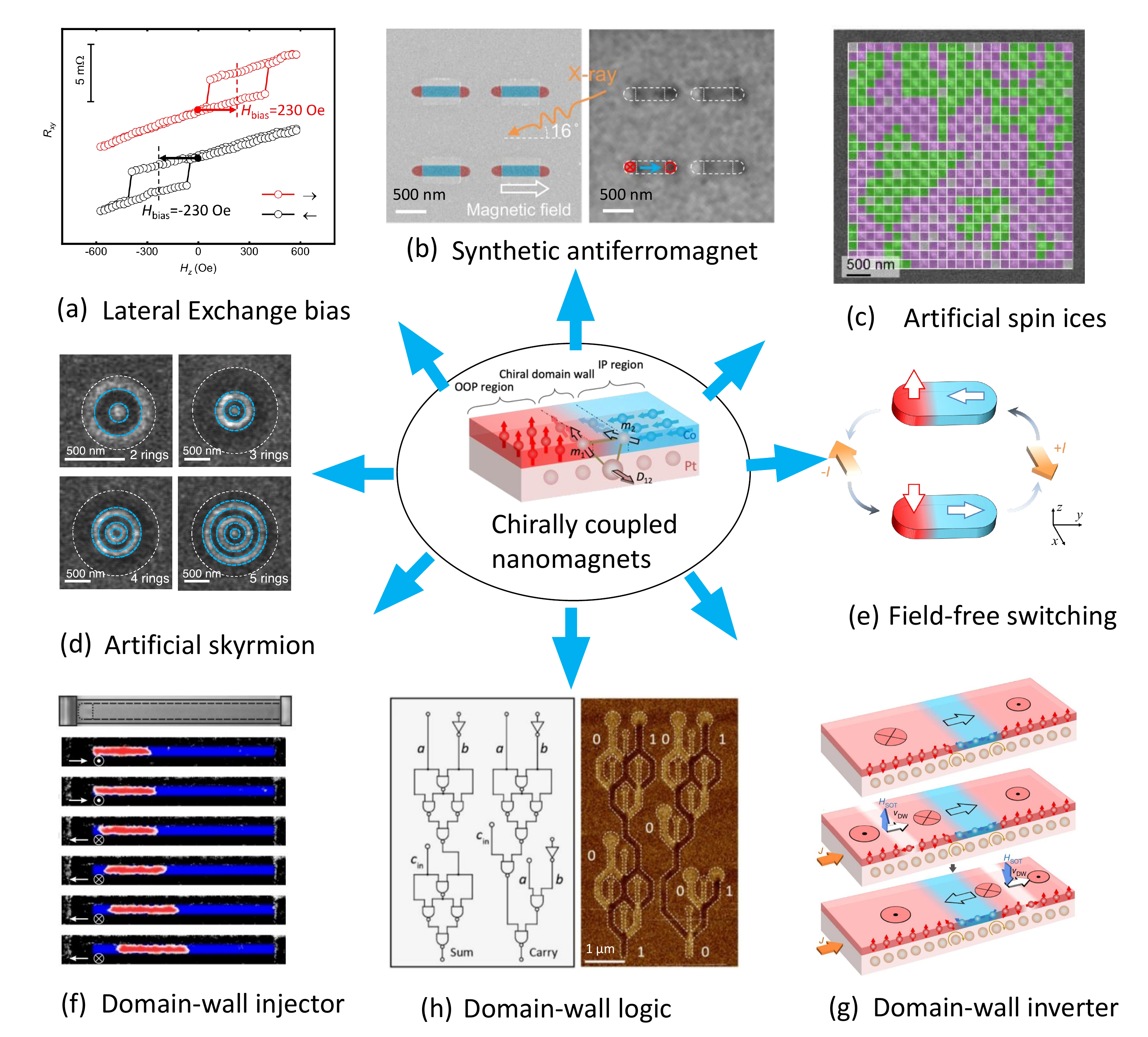}
		\par\end{centering}
	\caption{Collective phenomena based on the concept of chirally coupled nanomagnets: (a) lateral exchange bias, (b) synthetic antiferromagnet, (c) artificial spin ices, (d) artificial skyrmions, (e) spin-orbit torque induced field-free switching \cite{Ref_DMIChiral1}, (f) current-driven domain wall injector \cite{Ref_DMIChiral2}, (g) domain-wall inverter, and (h) domain-wall logic device \cite{Ref_DMIChiral7}. The figures are reproduced with permission from \cite{Ref_DMIChiral1, Ref_DMIChiral2, Ref_DMIChiral7}.}
	\label{DMIChiralCoupling2}
\end{figure}

\textbf{Chirally coupled nanomagnets}.---Chirally coupled magnetic structures induced by DMI display many novel and potentially useful phenomena \cite{Ref_DMIChiral1, Ref_DMIChiral2, Ref_DMIChiral3, Ref_DMIChiral4, Ref_DMIChiral5, Ref_DMIChiral6, Ref_DMIChiral7} that is summarized in  Fig.~\ref{DMIChiralCoupling2} for such as lateral exchange bias, synthetic antiferromagnet, artificial spin ices, artificial skyrmion, field-free switching \cite{Ref_DMIChiral1}, domain-wall injector \cite{Ref_DMIChiral2}, domain-wall logic and domain-wall inverter \cite{Ref_DMIChiral7}. 

In the non-collinear junctions discussed above the out-of-plane magnetization experiences an effective out-of-plane magnetic field that depends on the direction of the in-plane magnetization, which causes an effect that is reminiscent of the exchange bias, as shown in Fig.~\ref{DMIChiralCoupling2}(a) for a ferromagnet/antiferromagnet bilayer \cite{Ref_ExchangeBias1, Ref_ExchangeBias2, Ref_ExchangeBias3}. The chiral coupling is not governed by the exchange interaction but nevertheless enables a ``lateral exchange bias" effect with coupling strength that can easily be tuned by a magnetic field at room temperature.

Non-collinear planar junctions can become the elements of  complex magnetic circuits. That can be combined to form synthetic antiferromagnets in which the two out-of-plane magnetizations are anti-parallel [Fig.~\ref{DMIChiralCoupling2}(b)]. When arranged on periodic lattices with frustrated couplings the collective textures behave like an Ising-like artificial spin ice as shown in Fig.~\ref{DMIChiralCoupling2}(c). Ring-shaped magnetic structures, on the other hand, arrange themselves into artificial skyrmionic  textures [Fig.~\ref{DMIChiralCoupling2}(d)] \cite{Ref_DMIChiral1}. Advanced lithography will allow the fabrication of wafer-scale chiral magnetic textures and the study of artificial chiral and topological magnetism in the thermodynamic limit.

The spin-orbit DMI coupling is large at the interface between a magnet and a heavy metal such as platinum. This not only enhances the chiral coupling but also leads to large current-induced spin-orbit torques that can be used to control the magnetization [Fig.~\ref{DMIChiralCoupling2}(e)]. Magnetization switching by spin-orbit torques is a promising technique for next-generation magnetic random access memories (MRAM) promising fast and low-power data storage \cite{Ref_SOT1, Ref_SOT2, Ref_SOT3}. However, the switching of a perpendicular magnetization by the conventional spin-orbit torques is not deterministic unless an applied magnetic field breaks the symmetry, which is unattractive for applications. Proposals for field-free switching include the introduction of anisotropy gradients \cite{Ref_FFSwitch1}, exchange bias \cite{Ref_FFSwitch2}, out-of-plane spin polarization \cite{Ref_FFSwitch3} and built-in magnetic fields \cite{Ref_FFSwitch4}. Here we point out that the chirality of the out-of-plane/in-plane magnetic junction naturally breaks the symmetry without the need for a magnetic field  \cite{Ref_DMIChiral1}: the  spin Hall spin current in the heavy metal film switches the in-plane magnetization \cite{Ref_FFSwitch5} that by the chiral coupling forces the reversal of the out-of-plane magnetization. The entire element reversibly switches between two DMI-favored configurations by changing the direction of the electric current, eventually facilitated by current-induced domain-wall motion from the edges \cite{Ref_FFSwitch6, Ref_FFSwitch7}. The chiral coupling is also essential for the field-free switching under a gradient in the perpendicular magnetic anisotropy \cite{Ref_FFSwitch1, Ref_FFSwitch8}.

The static chiral coupling enables versatile  manipulation of domain walls, with potential applications in information processing and storage \cite{Ref_DMIChiral2, Ref_DMIChiral4, Ref_DMIChiral7}. Patterning the end of an out-of-plane magnetic conduit generates a small in-plane region that serves as a current-driven domain-wall injector [Fig.~\ref{DMIChiralCoupling2}(f)] \cite{Ref_DMIChiral2}. When the length of the perpendicularly magnetized region exceeds the critical single-domain size, the DMI affects only the in-plane/out-of-plane domain wall that moves under an electric current. An in-plane magnetic field reverses the in-plane magnetization and nucleates a  domain wall at the in-plane/out-of-plane transition region, which when repeated generates a domain-wall array in the out-of-plane magnetic conduit.  The time sequence of switching the magnetic field thus controls the domain wall spacings.  Electric current pulses normal to the in-plane magnetization may fulfill the same role by switching  magnetizations by the spin-orbit torques.

A narrow in-plane magnetized region inverts a domain-wall propagating through a perpendicular magnetic conduit  [Fig.~\ref{DMIChiralCoupling2}(g)] \cite{Ref_DMIChiral4, Ref_DMIChiral7}. Here the chiral coupling operates at the edges of the in-plane magnetized islands. We start from a configuration with the uniform magnetization of out-of-plane and in-plane regions [see the top panel in Fig.~\ref{DMIChiralCoupling2}(g)]. A domain-wall incident from the left can propagate through the in-plane island  only at the cost of a high-energy texture that is topologically protected. It can be unwound only by reversing the magnetization of the in-plane region, which is accompanied by annihilation of the domain wall at the left side and nucleation of a new on the right with inverted polarity, as shown by the bottom schematic in Fig.~\ref{DMIChiralCoupling2}(g). This inverter is equivalent to a Boolean NOT gate. Chirally coupled structures can operate as well as a NAND/NOR gate which completes the domain-wall logic since arbitrary Boolean operations can be carried out by a cascade of NAND gates. The 15 NAND gates complemented by a fan-out and NOT gate as depicted in Fig.~\ref{DMIChiralCoupling2}(h) operates as a binary full adder.

\begin{figure}[ht]
	\begin{centering}
		\includegraphics[width=0.99\textwidth]{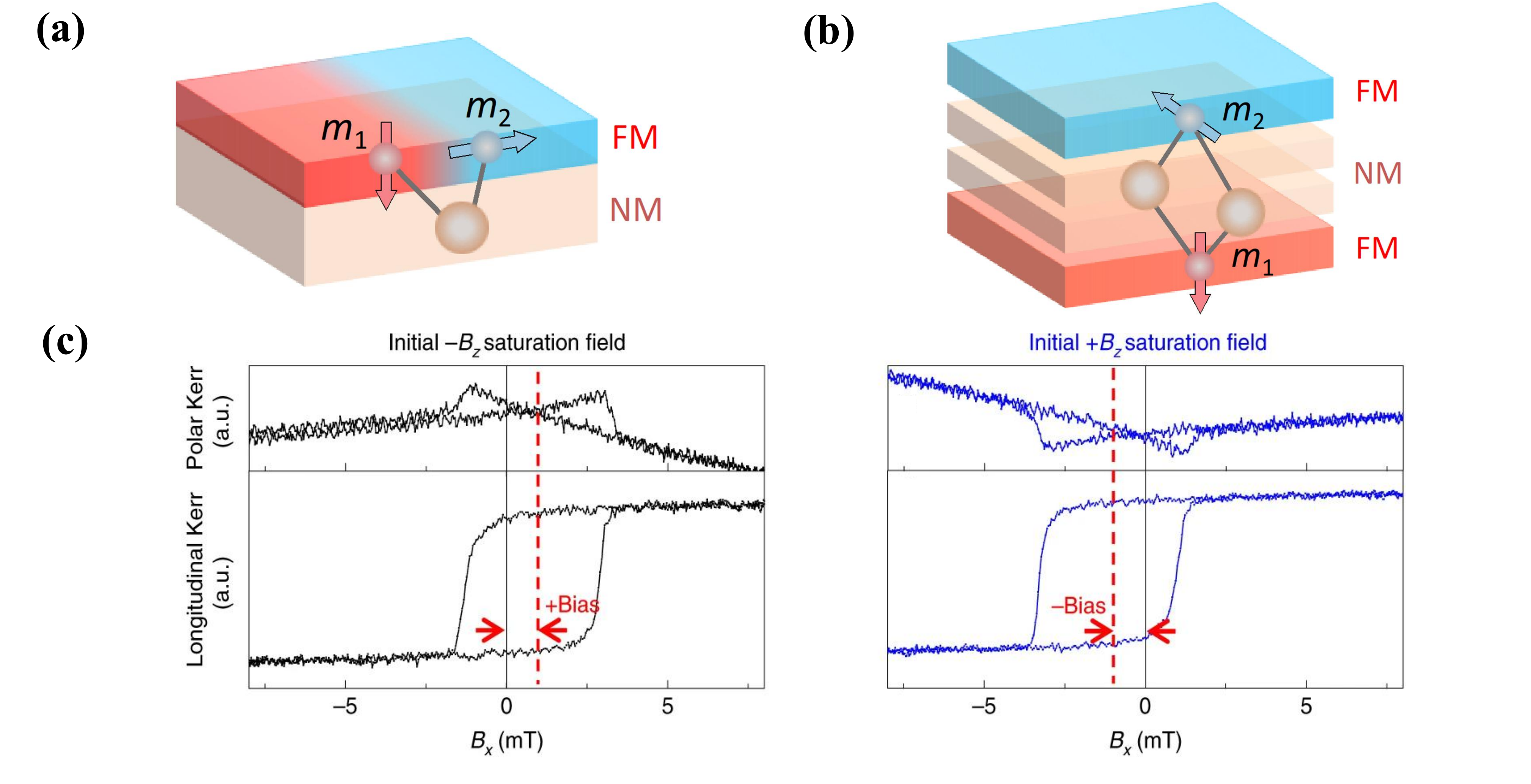}
		\par\end{centering}
	\caption{Intralayer (a) and interlayer (b) chiral coupling. (c) Exchange bias caused by the interlayer chiral coupling. The out-of-plane and in-plane magnetization components of the CoFeB film are measured by the polar and longitudinal Kerr effect under ${\bf B}_{x}$ magnetic fields, respectively, after negative (left) and positive (right) initial saturating orthogonal ${\bf B}_{z}$ fields that define the out-of-plane magnetic state. The favorable magnetic configurations of the negative and positive out-of-plane initialization are shown in the inset. The figures are reproduced with permission from \cite{Ref_InterDMI1}.} 
	\label{DMIChiralCoupling4}
\end{figure}

\textbf{Interlayer chiral coupling}.---The chiral coupling also exists in multilayers normal to the interfaces in the form of the asymmetric part of the non-local exchange (RKKY) interaction \cite{Ref_InterDMI5, Ref_InterDMI1, Ref_InterDMI2, Ref_InterDMI4} (Fig.~\ref{DMIChiralCoupling4}). The basic structure comprises a spin valve with two magnetic layers separated by a non-magnetic spacer [Fig.~\ref{DMIChiralCoupling4}(b)] chiral coupling requires that the magnetizations are non-symmetric, \textit{e.g.}, with different magnetic anisotropies. The conventional  RKKY interaction is symmetric and favors either  anti-parallel or parallel alignment as a function of the thickness of the non-magnetic layer. The asymmetric interaction, on the other hand, strives to twist the two magnetizations into a chiral manner. It exists again by means of the spin-orbit interaction and originates from the interface DMI. The coupling can be observed by sweeping an in-plane magnetic field in the form of a bias field in the hysteresis loops [Fig.~\ref{DMIChiralCoupling4}(c)]. In contrast to the intralayer DMI that is governed by the exchange between two neighboring magnetic sites via the non-magnetic metal, the interlayer chiral coupling is between sites that are separated by the non-magnetic spacer  of typically a few nanometers and therefore much weaker, unless the composition and thickness of the nonmagnetic spacer are further optimized. A chiral orientation of in-plane and out-of-plane magnetic layers in the vertical structure can also be obtained by carefully tilting the out-of-plane anisotropy, which enables field-free switching by spin-orbit torques \cite{Ref_InterDMI3}. The interlayer chiral coupling opens up an additional dimension to engineer magnetic structures and may control novel three-dimensional chiral magnetic devices.

In the static case, as reviewed, both the dipolar coupling and DMI play important roles in the stabilization of various chiral structures.
The chirality of dipolar coupling and DMI in the static situation is inherited in the dynamical case that will be discussed in the later Chapters.

\section{Chiral waves with transverse spins in spintronics}
\label{section3}

\begin{figure}[ht]
	\begin{centering}
		\includegraphics[width=0.996\textwidth]{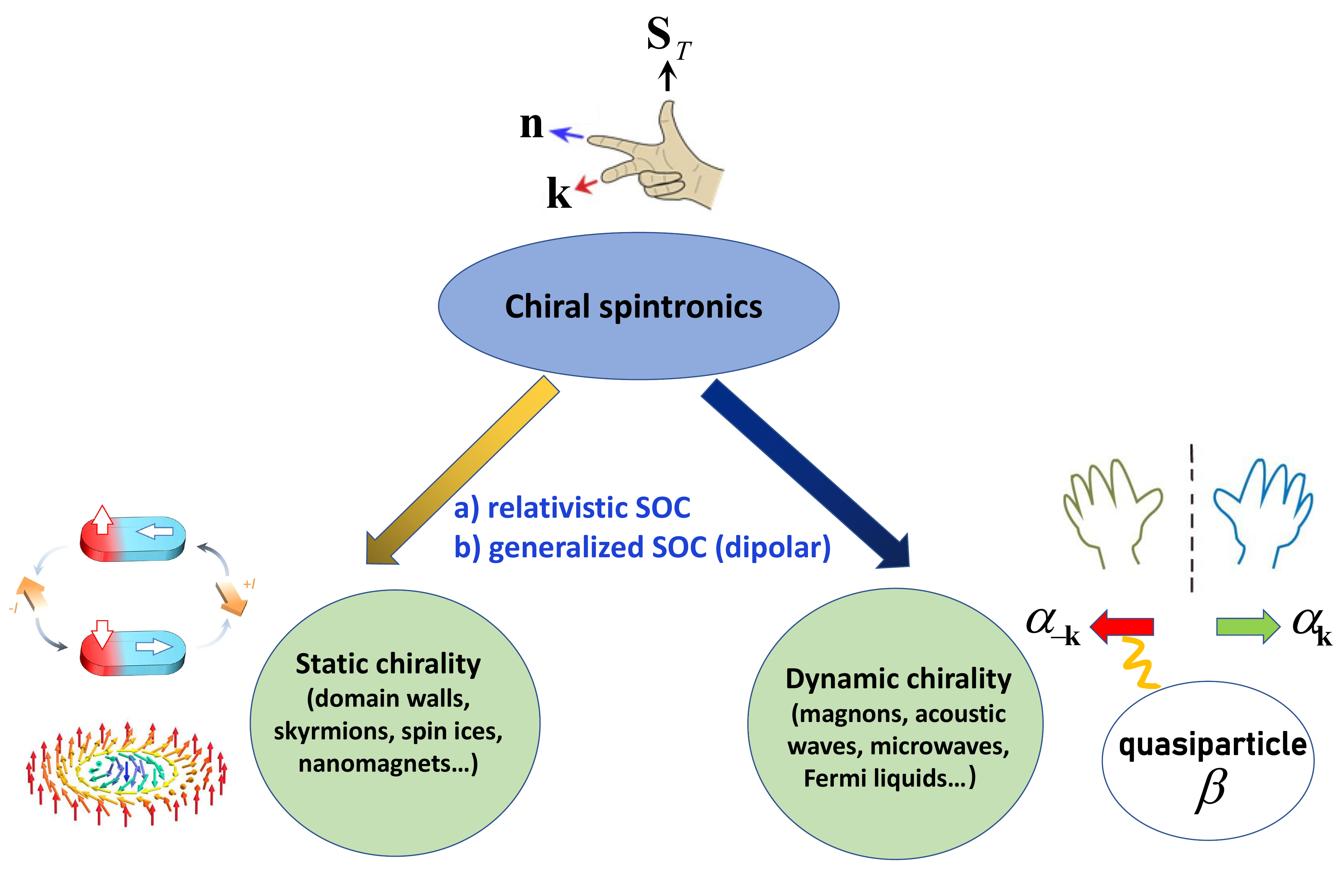}
		\par\end{centering}
	\caption{Static chirality \textit{vs.} dynamical chirality in chiral spintronics with analogous physical origin but for different physical objects. The branch termed ``static chirality" focuses on the chiral spin textures \textit{etc.} that can control the local magnetization in a chiral manner. The other branch, referred to as ``dynamical chirality", exploits various chiral waves (or quasiparticles) whose motion is governed by the chirality index that provides generalized spin-orbit interaction (SOC) for various new functionalities. 
	\textcolor{blue}{The red and green arrows on the right  represent the left- and right-moving $\alpha$ modes, while $\beta$ represents another quasiparticle that couples with $\alpha_{-{\bf k}}$ but not $\alpha_{\bf k}$ due to a chiral interaction.}}
	\label{Chiral_spintronics}
\end{figure}

The above Chapter addresses the \textit{ground-state} chirality in the spin textures that serve as a static chiral interaction between nanomagnets. This kind of chirality, originating from both the relativistic spin-orbit interaction for the small spin objects and dipolar coupling for the large spin objects, is an important branch in chiral spintronics that has been extensively investigated (Sec.~\ref{section2}), which can be referred to as the ``static chirality" in the sense that they are time-independent as depicted in Fig.~\ref{Chiral_spintronics} with some examples.

As addressed and defined in the introduction Sec.~\ref{section1}, a different branch that describes the chiral (or directional, or non-reciprocal) interaction between \textit{excitations} including various quasiparticles (Fig.~\ref{issues}) can originate from both the relativistic effect and generalized spin-orbit interaction, the latter often being of electromagnetic dipolar origin. The spin concept  is thereby extended to the classical evanescent waves (or quasiparticles in their quanta) in the context of a cross-product locking of transverse spin ${\bf S}_T$ and momentum ${\bf k}$ with a fixed surface normal ${\bf n}$ of the waves propagation plane, or the locking between ${\bf S}_T$ and ${\bf n}$ with fixed ${\bf k}$, which is thereby maximal in the chirality index 
\begin{equation}
Z=\hat{\bf n}\cdot({\hat{\bf S}_T\times \hat{\bf k}}),
\label{chirality_index}
\end{equation}
or their commutations. 
Those waves with a fixed $Z=+1$ or $-1$ are termed as the chiral waves that act as spin carriers as in Fig.~\ref{issues}, as reviewed from Sec.~\ref{chiral_spin_waves} to \ref{Sec_chiral_phonon} below for various waves. In particular, when the transverse spin ${\bf S}_T$ is fixed by time-reversal breaking in such as a ferromagnet, the chiral waves are unidirectional. Otherwise, they are allowed to propagate in both directions but may have different amplitudes with opposite momentum, as reviewed in Sec.~\ref{unification} in this Chapter. Since this kind of chirality  involves dynamics, we refer it to as ``dynamic chirality" to distinguish it from its static counterpart in Fig.~\ref{Chiral_spintronics}.
The direct physical consequence of this chirality of wave propagation is the chiral interaction between quasiparticles and near-field spin pumping by transferring the transverse spin of waves to electrons which will be reviewed in the later Chapters.

We turn subsequently to the dynamic counterparts, \textit{i.e.}, the elementary chiral waves summarized in Table~\ref{table_chiral_waves}, focusing on their key features and potential functionalities. We introduce chiral spatiotemporal eigenmodes of  Hamiltonians of various subsystems  that form the basis of their interaction with other excitations or objects in the later sections. Let us introduce again for convenience a terminology to be used here and later. \textbf{Transverse spin} is a unique property of evanescent waves, which are waves decaying in particular directions with complex momentum, or confined waves in cavity and waveguide. Different from the free traveling waves with spin direction usually parallel to the propagation direction \cite{Jackson}, the spin of evanescent waves is perpendicular to the momentum \cite{Nori} and thus ``transverse". Excitations are {\bf spin-momentum locked} when their transverse spin is locked to the propagation direction, which in principle is arbitrary, however.
The {\bf spatial chirality} refers to surface waves propagating in opposite directions on opposite surfaces of film or on extremal orbits of finite ellipsoids such as the Damon Eshbach modes and the electric stray field by rotating electric dipoles. Their spin direction is not necessarily locked to the momentum. 

 \begin{table}[htp]
   	\caption{Typical chiral waves with features and functionalities in spintronics. } \label{table_chiral_waves}
   	\centering
   	\begin{tabular}{ccc}
   		\toprule
   		Chiral waves & \hspace{-0.0cm}Features  & \hspace{-0.0cm} Functionalities \\
   		\toprule
   		\hspace{-0.8cm}\begin{minipage}[m]{.45\textwidth}
   			\centering\vspace*{5pt}
   			\textbf{Damon-Eshbach spin waves of magnetic slabs and spheres}
   			\includegraphics[width=6.3cm]{DE_figure}\vspace*{2pt}
   		\end{minipage} &
   		\hspace{-0.7cm}\begin{minipage}[m]{5.1cm}
   			\begin{itemize}
   				\item Spatial chirality (unidirectional at one surface) \cite{DE,Walker_sphere}
   				\item Small group velocity and sensitive to disorder \cite{surface_roughness_Yu}
   				\item Topological origin \cite{Kei_topology}
   			\end{itemize}
   		\end{minipage} &
   		\hspace{-0.5cm}\begin{minipage}[m]{5.1cm}
   			\begin{itemize}
   				\item Unidirectional heat conveyer \cite{heatconveyer1,heatconveyer2,heatconveyer3,heatconveyer4}
   				\item Optical magnon cooling \cite{optical_cooling}
   			\end{itemize}
   		\end{minipage}
   		\\
   		\toprule
   		\hspace{-0.8cm}\begin{minipage}{.45\textwidth}
   			\centering\vspace*{5pt}
   			\textbf{Dipolar stray fields of spin waves in ultrathin films}
   		\includegraphics[width=6.3cm]{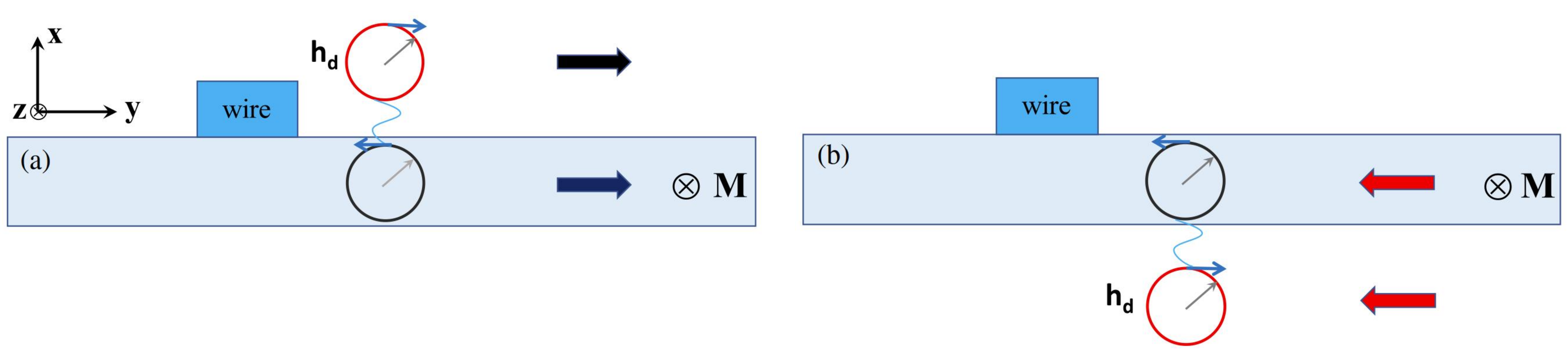}\vspace*{2pt}
   		\end{minipage} &
   		\hspace{-0.7cm}\begin{minipage}[m]{5.1cm}
   			\begin{itemize}
   			    \item Evanescent waves
   				\item Transverse spin opposite to the spin waves
   				\item Spatial chirality \cite{Chiral_pumping_Yu}
   				\item Robust against disorder
   			\end{itemize}
   		\end{minipage} &
   		\hspace{-0.5cm}\begin{minipage}[m]{5.1cm}
   			\begin{itemize}
   				\item Chiral coupling between nanomagnets
   				\item Chiral sensing \cite{chiral_sensing}
   				\item Chiral damping \cite{chiral_damping}
   			\end{itemize}
   		\end{minipage}
   		\\
   		\toprule
   		\hspace{-0.8cm}\begin{minipage}{.45\textwidth}
   			\centering\vspace*{5pt}
   			\textbf{Near field microwaves of striplines}\\
   			\includegraphics[width=4.6cm]{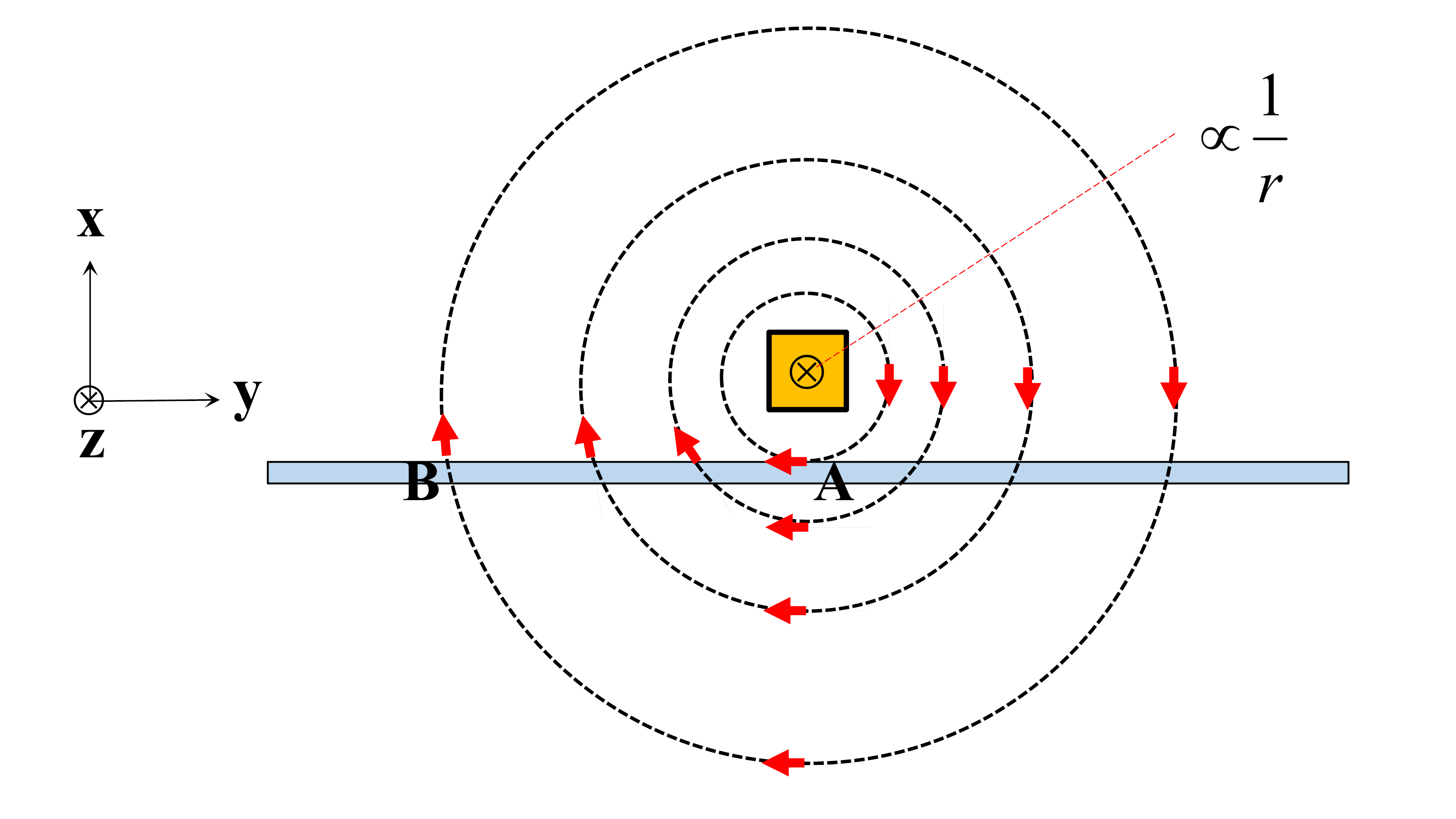}\vspace*{2pt}
   		\end{minipage} &
   		\hspace{-0.7cm}\begin{minipage}[m]{5.1cm}
   			\begin{itemize}
   			    \item Evanescent waves
   				\item Spin-momentum locking \cite{stripline_poineering_1,stripline_poineering_2,Yu_Springer}
   				\item GHz regime, wavelength governed by strip width 
   			\end{itemize}
   		\end{minipage} &
   		\hspace{-0.5cm}\begin{minipage}[m]{5.1cm}
   			\begin{itemize}
   				\item Chiral excitation of non-chiral spin waves
   				\item Spin-wave focusing/caustics
   			\end{itemize}
   		\end{minipage}
   		\\
   		\toprule
   		\hspace{-0.8cm}\begin{minipage}{.45\textwidth}
   			\centering\vspace*{2pt}
   			\textbf{Microwave waveguides and cavities} 
         \includegraphics[width=6.0cm]{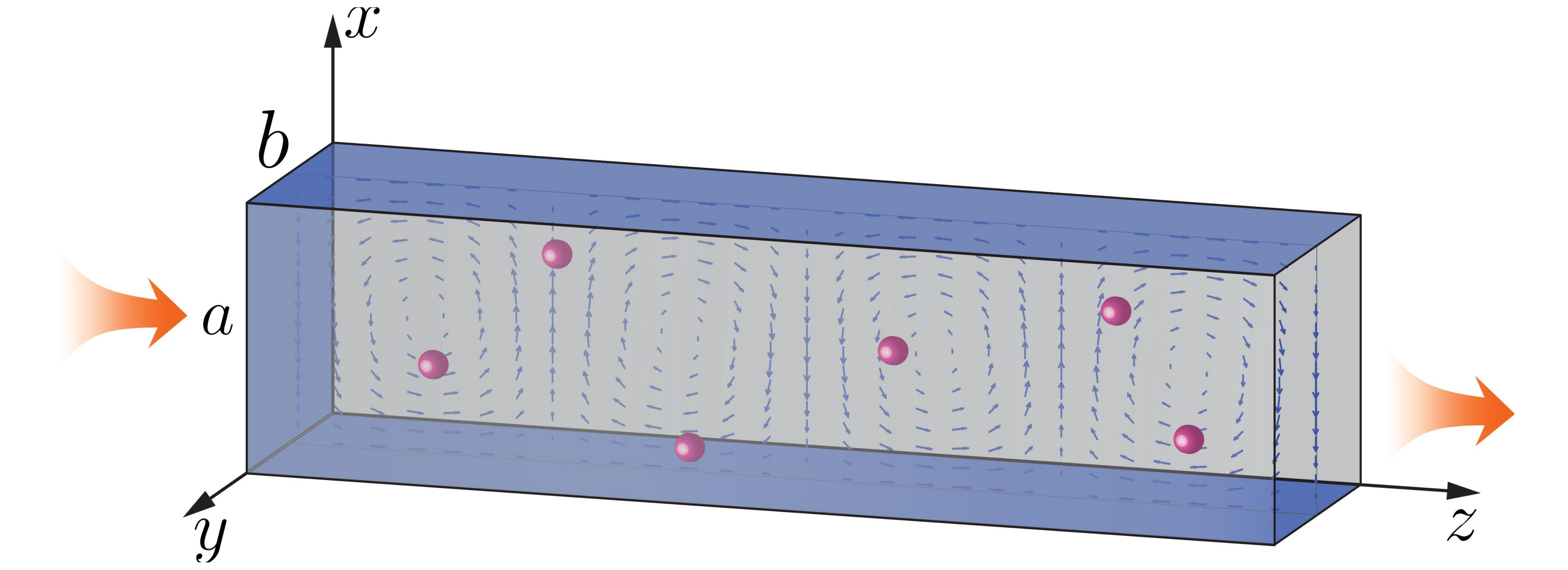}\vspace*{5pt}
   		\end{minipage} &
   		\hspace{-0.7cm}\begin{minipage}[m]{5.1cm}
   			\begin{itemize}
   				\item Chiral surface/line \cite{waveguide_Yu_1,waveguide_Yu_2,circulating_polariton}
   				\item Small damping \cite{chiral_optics}
   				\item Strong chiral coupling to magnetic loads
   			\end{itemize}
   		\end{minipage} &
   		\hspace{-0.5cm}\begin{minipage}[m]{5.1cm}
   			\begin{itemize}
   				\item Microwave circulator \cite{circulator_Tang,circulating_polariton}
   				\item Microwave diodes \cite{waveguide_Yu_1,waveguide_Yu_2}
   				\item Broadband microwave non-reciprocity \cite{Xufeng_exp,Zhong}
   				\item Classical or quantum transducers  \cite{chiral_optics}
   			\end{itemize}
   		\end{minipage} 
   		\\
   		\toprule
   		\hspace{-0.8cm}\begin{minipage}{.45\textwidth}
   			\centering\vspace*{2pt}
   			\textbf{Optical waveguides, fibers and resonators}
   			\includegraphics[width=6.3cm]{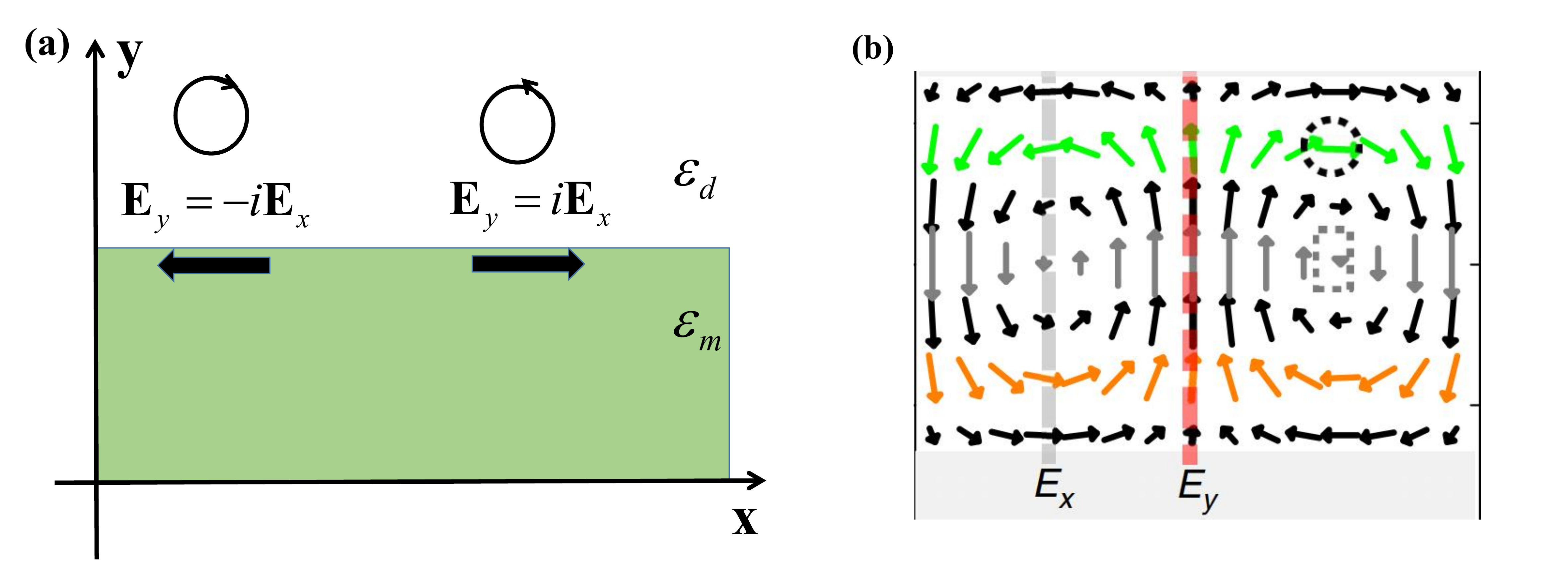}\vspace*{5pt}
   		\end{minipage} &
   		\hspace{-0.7cm}\begin{minipage}[m]{5.1cm}
   			\begin{itemize}
   			\item Transverse spin  \cite{Nori}
   				\item Spin-momentum locking (bidirectional at one surface) \cite{SSP_review} 
   				\item Chiral coupling to electric dipoles \cite{plasmonics_1,plasmonics_2}
   			\end{itemize}
   		\end{minipage} &
   		\hspace{-0.5cm}\begin{minipage}[m]{5.1cm}
   			\begin{itemize}
   				\item Field localization \cite{nano_optics}
   				\item Field enhancement
   				\item Spin transfer to electrons \cite{plasmonics_spin_NC,plasmonics_spin_APL,plasmonics_spin_PRB,plasmonics_spin_NJP,plasmonics_spin_PRL}
   			\end{itemize}
   		\end{minipage}
   		\\
   		\toprule
   		\hspace{-0.8cm}\begin{minipage}{.45\textwidth}
   			\centering\vspace*{5pt}
   			\textbf{Rayleigh surface acoustic waves}
   			\includegraphics[width=5.1cm]{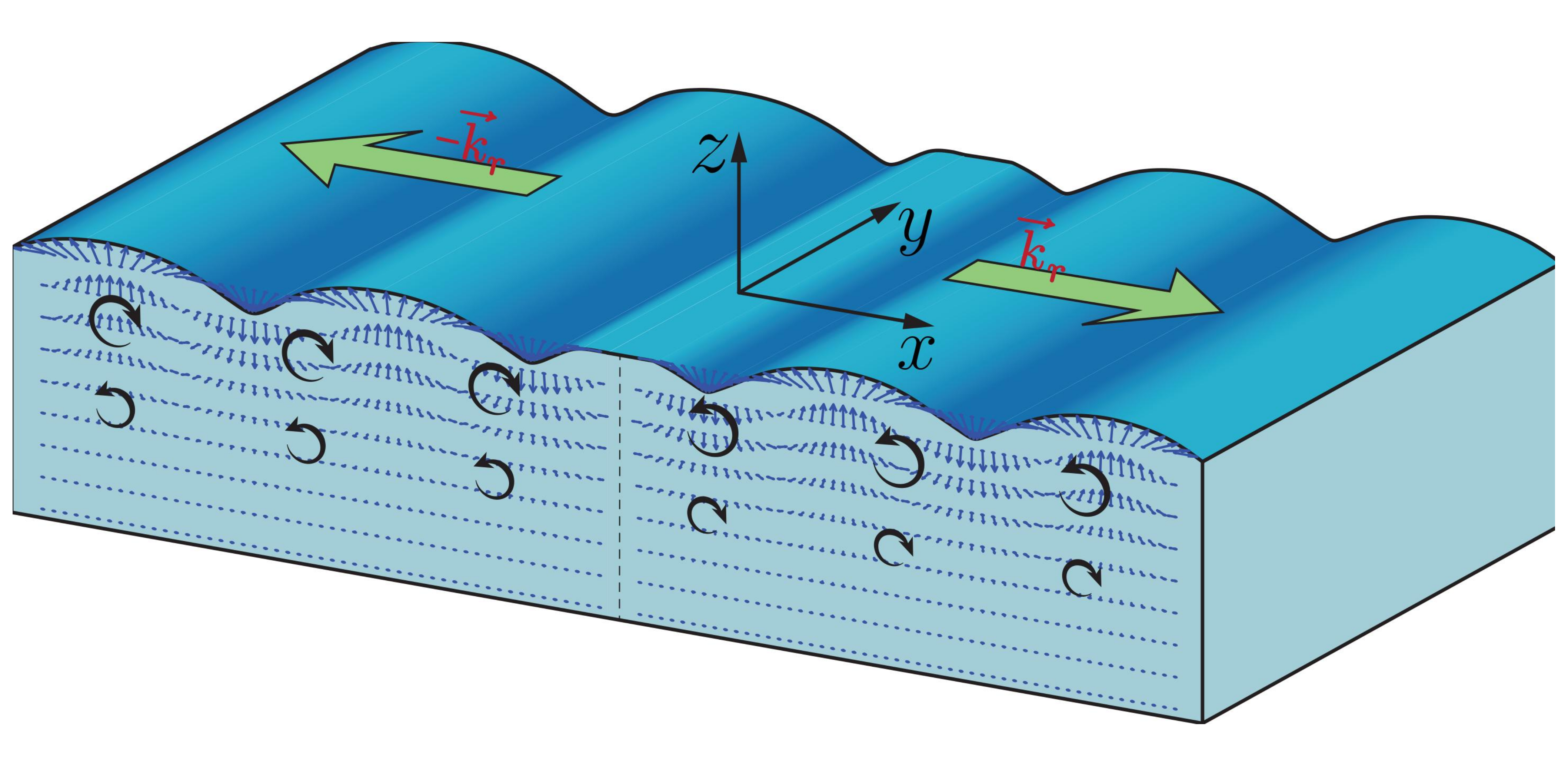}\vspace*{5pt}
   		\end{minipage} &
   		\hspace{-0.7cm}\begin{minipage}[m]{5.1cm}
   			\begin{itemize}
   			     \item Weak damping/scattering
   				\item Transverse spin
   				\item Spin-momentum locking \cite{Kino1987,Viktorov1967} 
   			\end{itemize}
   		\end{minipage} &
   		\hspace{-0.5cm}\begin{minipage}[m]{5.1cm}
   			\begin{itemize}
   				\item Phonon cavity \cite{phonon_Yu_1} 
                \item Phonon diode \cite{Xu,Onose_exp,Nozaki_exp,phonon_Yu_2,phonon_Yu_1,Otani_exp,Page_exp,Page_exp_2,phonon_Kei,DMI_phonon_exp}
                \item Chiral pumping of surface acoustic waves \cite{phonon_Yu_1,phonon_Kei,phonon_Yu_2}
                \item Classical or quantum transducer
   			\end{itemize}
   		\end{minipage}
   		\\
   		\hline
   	\end{tabular}
   \end{table}

\subsection{Chiral spin waves}
\label{chiral_spin_waves}

In the following, we first focus on the excitations of the collinear ferromagnetic order, \textit{i.e.}, the spin waves. The Hamiltonian of a ferromagnet is governed by the exchange [Eq.~(\ref{Eexchange})], dipolar [Eq.~(\ref{Edipolar})], and DM [Eq.~(\ref{DMI})] interaction, as summarized in Eq.~(\ref{Etotal}). Magnetic order breaks time-reversal symmetry and the spins in an excited state precess around the (effective) magnetic field in an anticlockwise fashion. In finite size magnets such as films and spheres, they display spatial chirality determined by magnetization, propagation, and surface normal directions.  This is the essence of the Damon-Eshbach (DE) spin waves \cite{DE,Walker_sphere} in the magnetostatic approximation (Sec.~\ref{DE_review}).   The vector product in the DMI  Eq.~(\ref{DMI}) obviously causes spatial chirality (Sec.~\ref{DMI_spin_waves}). Dipolar interaction and DMI are therefore two key tunable factors to realize non-reciprocal functionality in magnetic devices and  metamaterials such as multilayers.

\subsubsection{Damon-Eshbach spin waves} 
\label{DE_review} 
The excitations of the magnetic order at the surfaces of slabs or extremals of particles with wave numbers normal to the magnetic order are the archetypal chiral waves that were predicted more than 70 years ago. Here we sketch the essentials of the derivation and the properties of these modes and introduce a Hamiltonian formulation that will be useful in later Chapters.

\textbf{Basis}.---Magnetostatic waves of thick in-plane magnetized films are governed by the dipolar interaction that dominates over exchange interaction in the long-wavelength limit \cite{Walker_sphere,DE,YIG_magnonics}.  Among them, the DE spin waves are those propagating nearly perpendicular to the magnetization and exponentially localized at the equator of spheres  \cite{Walker_sphere} or the surfaces of slabs \cite{DE} and are chiral \cite{Walker_sphere,DE}. The chirality can be traced to the interaction of the waves with their own stray magnetic fields caused by surface and bulk magnetic charges that must  self-consistently be taken into account. Spin waves with in-plane wave vectors nearly perpendicular to the magnetization can then propagate only in one direction that is fixed by the outer product of surface normal $\hat{\bf n}$ and magnetization direction $\hat{\bf M}$, \textit{i.e.}, characterized by the chirality index $Z=\hat{\bf n}\cdot(\hat{\bf M}\times\hat{\bf k})$.  
 
 \begin{figure}[ptbh]
 	\begin{centering}
 		\includegraphics[width=0.96\textwidth]{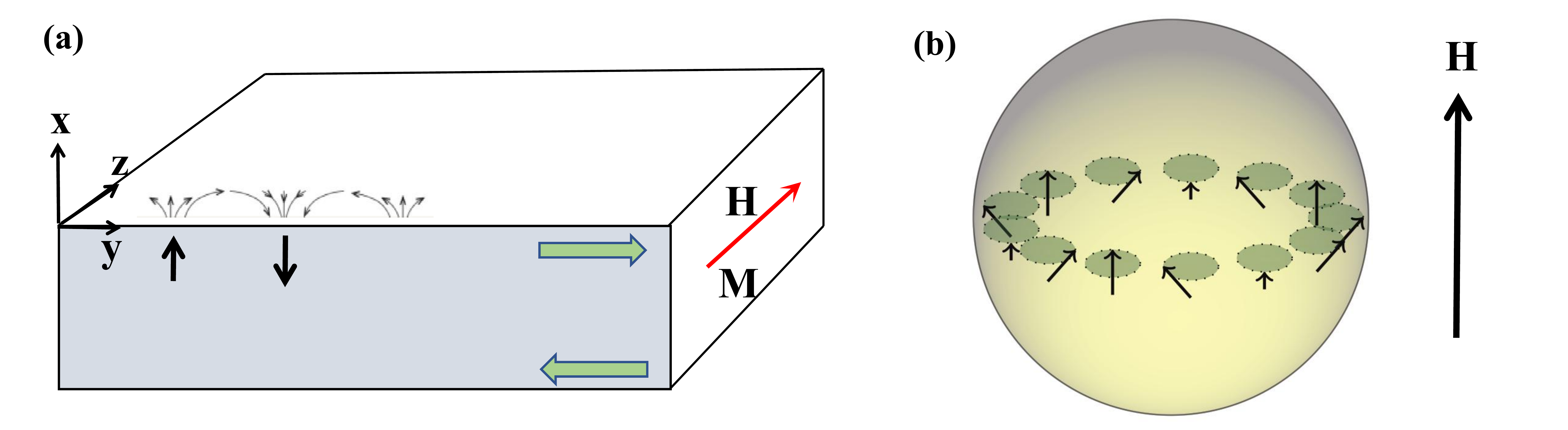}
 		\par\end{centering}
 	\caption{DE surface modes with a unidirectional propagation direction perpendicular to the magnetization in the in-plane magnetized films [(a)] and magnetic spheres [(b)].}
 	\label{DE_figure}
 \end{figure}

We focus on the configuration in Fig.~\ref{DE_figure}(a), where the film surface is perpendicular to the $\hat{\mathbf{x}}$-direction, the equilibrium magnetization ${\bf M}_s=M_{s}\hat{\mathbf{z}}$ points along the $\hat{\mathbf{z}}$-direction parallel to an applied magnetic field $H_z\hat{\bf z}$, and we address the spin waves that propagate along the $\hat{\mathbf{y}}$-direction. The film thickness is $d$. We disregard here the crystalline anisotropy and exchange interaction, which is an excellent approximation for the long-wavelength spin waves in magnets such as yttrium iron garnet (YIG). Then, the  magnetic free energy functional of magnetization ${\bf M}$
     \begin{equation}
     	F = -{\mu_{0}}\int d\mathbf{r}\left[  M_{z}H_{z} + \frac{\mathbf{M}%
     		(\mathbf{r})}{8\pi}\cdot\nabla\int d\mathbf{r}^{\prime}\frac{\nabla^{\prime
     		}\cdot\mathbf{M} (\mathbf{r}^{\prime})}{|\mathbf{r} - \mathbf{r}^{\prime}%
     		|}\right]
     	\label{free_energy_DE}
 \end{equation} 
governs the effective magnetic field 
$\mathbf{H}_{\mathrm{eff}}(\mathbf{r}) = - (1/\mu_{0}){\delta F}\left[\mathbf{M} \right]  /{\delta\mathbf{M}}(\mathbf{r})$ in the Landau-Lifshitz equation \cite{Landau}
\begin{equation}
	{\partial\mathbf{M}(\mathbf{r})}/{\partial t} = - \gamma\mu_{0}\mathbf{M}%
	(\mathbf{r})\times\mathbf{H}_{\mathrm{eff}}(\mathbf{r}).
	\label{Heisenberg}
\end{equation}
Here, $\gamma$ denotes the modulus of the gyromagnetic ratio, and $\mu_{0}$ is the vacuum permeability.
Small  amplitude fluctuations are transverse to the equilibrium magnetization ($\mathbf{M}_s = M_{s}\hat{\mathbf{z}}$). They propagate with in-plane momenta $\mathbf{k} = k_{y}\hat{\mathbf{y}} + k_{z}\hat{\mathbf{z}}$ and amplitudes   
\begin{equation}
	\mathbf{M}_{\gamma= x,y}^{j\mathbf{k}} = m_{\gamma}^{j\mathbf{k}}(x)e^{ik_{y}y}
e^{ik_{z}z}.
\end{equation}
 Into the film:
     \begin{align}
         m_{x}^{j\mathbf{k}}(x)  = a_{j\mathbf{k}}e^{i\kappa_{j}x} + b_{j\mathbf{k}}
     e^{- i\kappa_{j}x},\hspace{1cm}
         m_{y}^{j\mathbf{k}}(x)  = c_{j\mathbf{k}}e^{i\kappa_{j}x} + d_{j\mathbf{k}}
         e^{- i\kappa_{j}x},
     \end{align}
where $j$ is an integer when \(d\) is finite. The coefficients and wave numbers depend on the boundary conditions from the Maxwell equations (the tangential ${\bf H}_{S}$ at the surface is continuous, and the normal ${\bf B}_S$ at the surface is continuous).
To leading order in small $\textbf{m}$, the Landau-Lifshitz equation leads to the dispersion relation \cite{DE}
     \begin{equation}
       \omega_{j\mathbf{k}} = \sqrt{\omega_{H}^{2} + \omega_{H} \omega_{M}\frac{\kappa_{j}^{2} + k_{y}^{2}}{\kappa_{j}^{2} + |\mathbf{k}|^{2}} },
        \label{energy_spectra}
     \end{equation}
     where $\kappa_{j}$ solves the characteristic equation \begin{equation}
     	(\beta k_{y})^{2} + \kappa_{j}^{2}(\alpha+ 1)^{2} - |\mathbf{k}|^{2} -
     	2\kappa_{j}| \mathbf{k}|(\alpha+ 1)\cot(\kappa_{j}d) = 0.
     	\label{characteristic}%
     \end{equation}
Here, $\omega_{H} = \gamma\mu_{0} H_{z}$, $\omega_{M} = \gamma\mu_{0} M_{s}$, $\alpha= \omega_{H} \omega_{M}/(\omega_{H}^{2} - \omega_{j\mathbf{k}%
     }^{2})$, and $\beta= \omega_{j\mathbf{k}} \omega_{M}/(\omega_{H}^{2} -
     \omega_{j\mathbf{k}}^{2})$.
         
We refer to the solutions with purely imaginary $\kappa_{j} = iq_{x}$ as DE modes \cite{DE}. For ${\bf k}=k_y\hat{\bf y}$ they are highly degenerate because their frequency $\omega_{\mathbf{k}}$  approaches a constant $\sqrt{\omega_{H}^{2} + \omega_{H} \omega_{M} }$ in the limit of large momenta $|k_y| \gtrsim 1/d$. In the film, 
\begin{subequations}
    \begin{align}
    \label{wavefunctions_DE_x}
     m_{x}^{\mathbf{k}}(x)  & = C\left[  e^{- q_{x}x}(- \alpha q_{x} + \beta k_{y})
     + De^{q_{x}(x+d)}(\alpha q_{x} + \beta k_{y})\right],\\
     m_{y}^{\mathbf{k}}(x)  & = iC\left[  e^{- q_{x}x}(- \beta q_{x} + \alpha
     k_{y}) + De^{q_{x}(x+d)}(\beta q_{x} + \alpha k_{y})\right],
     \label{wavefunctions_DE_y}
     \end{align}
 \end{subequations}
 in which $C$ is a normalization constant, and $D = [{q_{x}(\alpha+ 1) - \beta k_{y} + \left\vert \mathbf{k} \right\vert}]/[{q_{x}(\alpha+ 1) + \beta k_{y} - \left\vert \mathbf{k}\right\vert }]$.
  $q_x$ solves the characteristic equation \cite{DE},
     \begin{equation}
       (\beta k_{y})^{2} - q_{x}^{2}(\alpha+ 1)^{2} - \left\vert \mathbf{k}%
       \right\vert ^{2} - 2q_{x}|\mathbf{k}|(\alpha+ 1)\coth(q_{x}d) = 0.
       \label{ce_de}%
      \end{equation}
We observe that the DE modes are exponentially localized at the upper and lower surfaces depending on their propagation direction. Note that $x<0$ in the coordinate system of Fig.~\ref{DE_figure}(a) and $\beta<0$ for the DE modes. When $e^{q_{x}d}\gg1$, $\coth(q_xd)\rightarrow 1$, and Eq.~(\ref{ce_de}) becomes
     	\begin{equation}
     		q_{x}(\alpha+1)+\left\vert \mathbf{k}\right\vert \approx|\beta k_{y}|.
     \end{equation} 
     When $k_y<0$, $D\rightarrow 0$ and $m_{x,y}^{\bf k}\propto e^{-q_xx}$ is localized at the lower surface, for $k_y>0$ at the upper surface. Moreover, the DE mode is (almost) circularly polarized. For example,  for surface magnons with $k_{y}<0$ and $e^{q_{x}d}\gg1$, $im_{x}^{\mathbf{k}}+({k_{y}}/{\left\vert \mathbf{k}\right\vert })m_{y}^{\mathbf{k}}\approx0$, 
 \textit{i.e.}, when $\mathbf{k}=k_{y}\hat{\mathbf{y}}$ the DE magnons are circularly polarized $m_{y}^{\mathbf{k}}\approx im_{x}^{\mathbf{k}}$. The DE-like modes with $k_{z}\neq0$ acquire a finite ellipticity. Equation~(\ref{ce_de}) implies that  these modes preserve their surface character and chirality as long as $\left\vert k_{z}\right\vert <\left\vert k_{y}\right\vert \sqrt{M_{s}/H_{z}}$ \cite{DE}. 
         
The eigenmodes (\ref{wavefunctions_DE_x}) and (\ref{wavefunctions_DE_y}) are suitable basis functions into which to expand the free energy Eq.~(\ref{free_energy_DE}) and formulate a Hamiltonian function. The latter can be easily quantized \cite{squeezed_magnon,Kittel_book,HP,Sanchar_PRB} by replacing the magnetization amplitudes by spin operators $\mathbf{M}\rightarrow- \gamma\hbar\hat{\mathbf{S}}$ (and $M_{s}=\gamma\hbar S$). In extended systems where the latter can be expressed in terms of magnons with wave vector \(\mathbf{k}\) and operators	$\hat{\alpha}_{\mathbf{k}},\hat{\alpha}^{\dagger}_{\mathbf{k}}$ by the leading order term of the Holstein-Primakoff expansion \cite{squeezed_magnon,Kittel_book,HP,Sanchar_PRB}:
     \begin{subequations}
     	\begin{align}
     		\hat{S}_{x,y}(\mathbf{r})  & = \sqrt{2S}\sum_{\mathbf{k}}\left(m_{x,y}%
     		^{\mathbf{k}}(\mathbf{r})\hat{\alpha}_{\mathbf{k}} + m_{x,y}^{\mathbf{k}%
     		}(\mathbf{r})^{\ast}\hat{\alpha}_{\mathbf{k}}^{\dagger}\right),
     		\label{Bogoliubov_a}\\
     		\hat{S}_{z}(\mathbf{r})  & = - S + (\hat{S}_{x}^{2} + \hat{S}_{y}^{2})/(2S)
     		.\label{Bogoliubov}%
     	\end{align}
     \end{subequations}
     With the following normalization of the eigenmodes \cite{Walker_sphere,magnetic_nanodots},
      \begin{equation}
      	\int d\mathbf{r}\left(m_{x}^{\mathbf{k}}(\mathbf{r})m_{y}^{\mathbf{k}}(\mathbf{r})^{\ast} - m_{x}^{\mathbf{k}}(\mathbf{r})^{\ast}m_{y}%
      	^{\mathbf{k}}(\mathbf{r})\right) = - i/2,
      	\label{normalization2}
      \end{equation} 
we arrive at a magnon Hamiltonian as a sum of harmonic oscillators $\hat{H}_m=\sum_{\bf k}\hbar \omega_{\bf k}\hat{\alpha}_{\bf k}^{\dagger}\hat{\alpha}_{\bf k}$.

   In finite magnets, the wave vector is not conserved and the eigenmodes can be expressed on a symmetry-adapted basis. Walker \cite{Walker_sphere} solved the magnetostatic problem for  magnetic spheres before Damon and Eshbach \cite{DE} addressed the slab geometry. The surface modes in the sphere with magnetization pointing to the north pole propagate along the equator  with angular momenta $l\gg 1$ in only one direction, as illustrated in  Fig.~{\ref{DE_figure}}(b). They are strongly localized and circularly polarized with amplitudes \cite{Walker_sphere}
     \begin{align}
     	m_{\phi}=im_{\rho}=\left(\frac{l}{\pi}\right)^{3/4}
     	\sqrt{\frac{\hbar \gamma M_s}{R^3}}\left(\frac{\rho}{R}\right)^{l-1}. 
     \end{align}
   Their wavelength $\lambda\sim R/l$ and frequency
     \begin{align}
     	\omega_{\rm DE}=\gamma\mu_0\left(H_z-\frac{M_s}{3}+\frac{lM_s}{2l+1}\right)
     \end{align}
saturates at $\omega_{\rm DE}\rightarrow \gamma\mu_0\left(H_z+{M_s}/{6}\right)$ when $l\gg 1$, very similar to the DE modes for larger \(k\).

\begin{figure}[ptbh]
     	\begin{centering}
     		\includegraphics[width=0.98\textwidth]{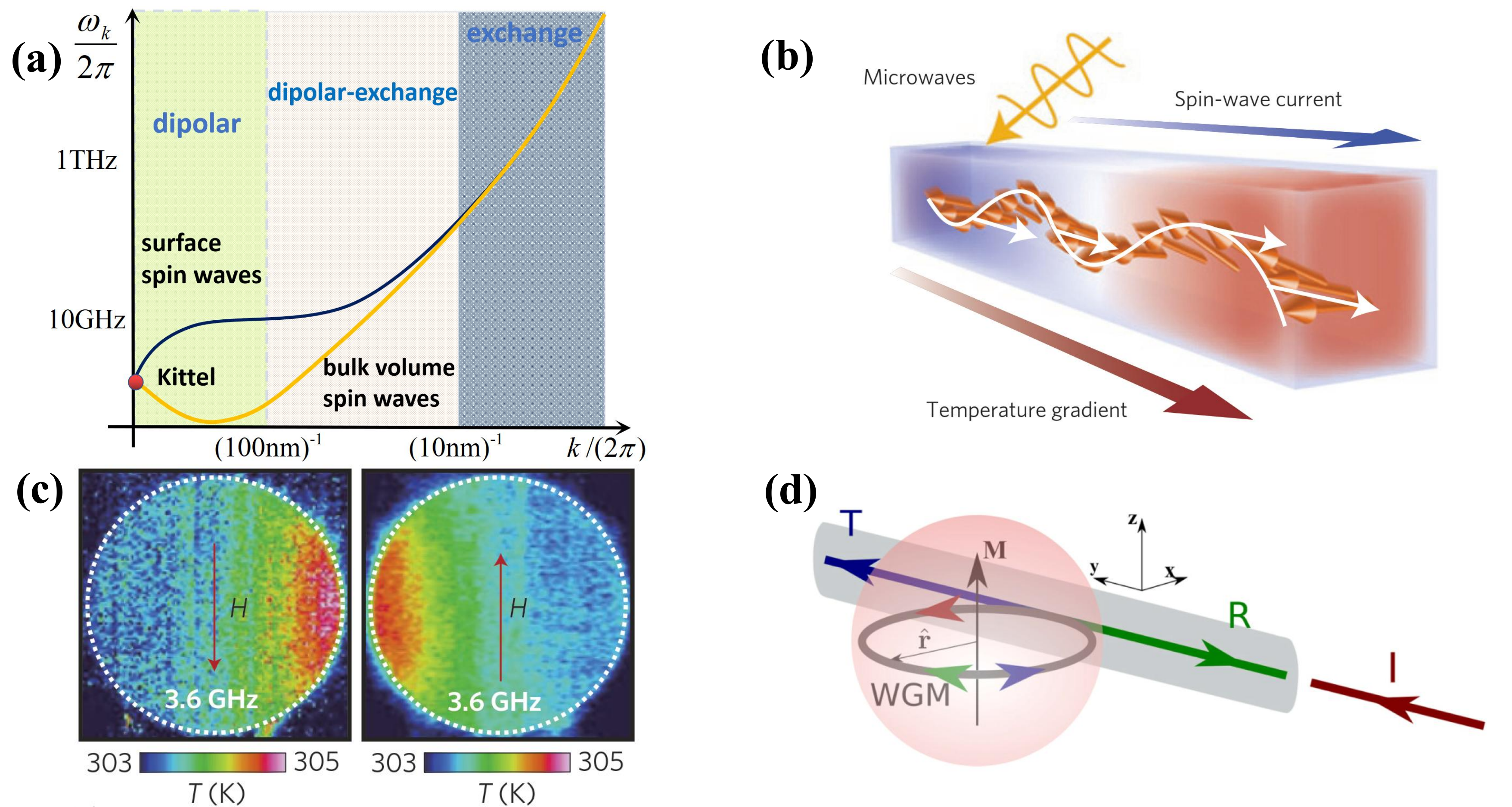}
     		\par\end{centering}
     	\caption{Damon-Eshbach configurations. (a) illustrates the spin waves propagating perpendicular to the in-plane magnetization of thick YIG films with the film thickness $\sim 100$~nm. For small wave numbers $k$ the modes are governed by the dipolar interaction and the Damon-Eshbach treatment holds \cite{DE}. At large \(k\) the exchange interaction dominates and dispersion is parabolic. The dipolar-exchange waves are those with intermediate \(k\). The frequencies range from gigahertz to many terahertz. (b)-(c) describes the heat conveyer effect by the Damon-Eshbach modes that can unidirectionally transmit angular momentum even against the temperature gradient. (d) is the spherical magnetic resonator in which the optical whispering gallery
     		modes are excited by proximity optical fiber. The figures (b)-(d) are taken from Refs.~\cite{heatconveyer1,Sanchar_PRB}.}
     	\label{DE_summary}
     \end{figure}

In spite of the long history, the topology of  Damon-Eshbach modes was addressed only recently by Yamamoto {\it et al.}  \cite{Kei_topology}. In contrast to the conventional bulk-edge correspondence, their frequency band does not overlap with the bulk modes and therefore does not require gap formation, as shown in Fig.~\ref{DE_summary}(a) for the dispersion of spin waves in YIG slabs. The topology of Damon-Eshbach modes of the infinite half-space is characterized by vortex lines and corresponding winding numbers in reciprocal space. This theory also provides topological characterizations of Hamiltonians for other classical waves.

The Damon-Eshbach surface modes were also theoretically predicted in antiferromagnet long ago \cite{DE_AFM_1,DE_AFM_2,DE_AFM_3,DE_AFM_4,DE_AFM_5,DE_AFM_6}, but still without experimental confirmation.  Recently, Liu {\it et al}.  showed that dipolar spin waves in uniaxial easy-axis antiferromagnets are naturally in a topological nodal-line semimetal phase \cite{DE_AFM_7}. The topological surface modes reside in the dipolar-induced gap between the bulk bands, different from its ferromagnet counterpart \cite{Kei_topology}, and display a chirality-momentum locking \cite{DE_AFM_7}. For bulk antiferromagnets, Shen predicted that the dipole-dipole interaction acts as an effective spin-orbit coupling between magnons with opposite polarization (effective spin) that induces such as the
 	D'yakonov-Perel'-type magnon spin relaxation
 	and the intrinsic magnon (inverse) spin Hall effect \cite{shen_dipolar}.

     \textbf{Functionalities by chirality}.---Unidirectional propagation of waves provides unique functionalities that can be useful in  magnetic, optic, mechanical, and electronic devices. As indicated in Fig.~\ref{DE_summary}(b),  the surface modes of sufficiently thick magnetic films transport heat in a particular direction, even against a temperature gradient, \textit{i.e.}, they act like a ``heat conveyer-belt" \cite{heatconveyer1,heatconveyer2,heatconveyer3,heatconveyer4}. Figure~\ref{DE_summary}(c) shows that the  magnetization direction  controls the temperature gradients of the heat generated by microwave absorption \cite{heatconveyer1}.

     YIG magnetic spheres are also excellent optical resonators that support whispering gallery photon modes as sketched in Fig.~\ref{DE_summary}(d). Sharma {\it et al}. predicted that the surface magnons can strongly interact with optical whispering gallery modes \cite{WGM1,WGM2,WGM3} due to the surface nature \cite{Sanchar_PRB,Sanchar_optimal}.
     By angular momentum conservation in the inelastic scattering of a whispering-gallery photon by magnons, surface magnons can contribute only by reflection when tuned to the anti-Stokes line with full suppression of the Stokes scattering, which improves the efficiency of optical cooling of magnons \cite{optical_cooling}. The efficiency also depends on the spatial overlap between  photon and magnon modes and can be enhanced by tuning the systems to magnon modes that are still chiral but slightly separated from the surface, which is possible in a regime in which the exchange interaction contributes \cite{Sanchar_optimal}. These modes are also less prone to dephasing by surface roughness scattering \cite{surface_roughness_Yu} discussed below.

Surface waves should be sensitive to surface roughness. The topological protection that implies a complete absence of backscattering for sufficiently thick films \cite{Kei_topology,Pirro} does not mean that surface disorder has no effect  \cite{surface_roughness_Yu}  because 
 forward scattering is still allowed. Typically, the Gilbert damping is enhanced hundreds of times for the Damon Eshbach modes in YIG films of micrometers in thickness and surface roughness of several nanometers \cite{surface_roughness_Yu}. Thus, for the long-range disorder, the transport is much less affected by the suppressed backscattering (vertex correction). The wavelength of the Damon-Eshbach modes $\sim 2\pi d$ should be much larger than the exchange wavelength $\lambda_{\rm ex}$, leading to $d\gg \lambda_{\rm ex}/(2\pi)$. Thus, the surface modes do not exist in sufficiently thin films.
     Their group velocity tends to be zero when the linear momentum is larger than the inverse of film thickness, leading
     to a small spin conductivity.   These may limit the applications of Damon–Eshbach spin waves in magnonic devices.

     The polarization of the DE spin waves at one surface and the dipolar field of the DE modes of the other surface does not match (see also in Sec.~\ref{dipolar_fields_1} below), so the DE modes at the two opposite surfaces do not really interact via direct dipolar interaction due to the surface disorder \cite{surface_roughness_Yu}.
     In films with finite thickness, the disorder can induce back scattering between the surface modes at both surfaces, which break topological protection and reduce the conductivity of the surface modes. Nevertheless, Mohseni \textit{et al.}  \cite{Pirro} reported strongly suppressed backscattering of dipole-exchange magnetostatic surface spin waves even in thin films.

 \subsubsection{Chiral spin waves by spin-orbit interaction}
\label{DMI_spin_waves}

Besides the dipolar origin addressed above, another source of chirality of spin waves originates from the relativistic spin-orbit interaction.
The broken inversion symmetry at the interfaces of ferromagnets gives rise to the interface Dzyaloshinskii-Moriya (DMI) interaction  Eq. (\ref{DMI}). The coefficient \(D\) is a spin-orbit interaction that can be enhanced at the interface to a heavy metal such as Pt that mediates the hopping between neighboring spins ${\bf S}_i$ and ${\bf S}_j$. Interfacial DMI causes non-reciprocal spin-wave propagation, such as different group velocities for spin waves of the same frequency that propagate in opposite directions  \cite{DMI_shift}.

\begin{figure}[ptbh]
	\begin{centering}
		\includegraphics[width=0.96\textwidth]{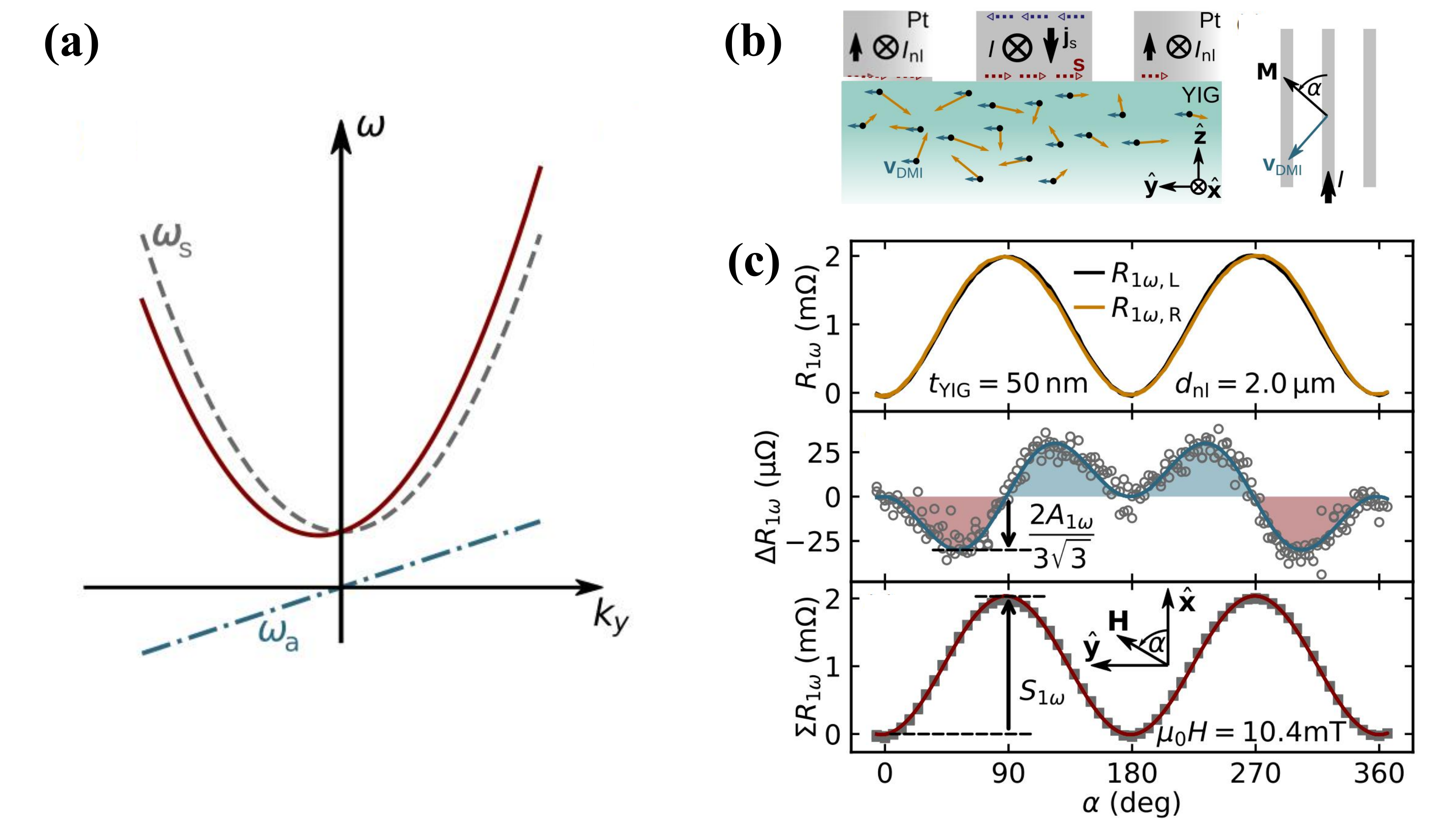}
		\par\end{centering}
	\caption{Non-reciprocity  observed experimentally in electrical injection and detection of incoherent magnons of YIG films. (a) illustrates the shift of dispersion by the DMI. In (b), the left and right platinum electrodes detect the diffusive magnon injected by the middle platinum electrode via the spin Hall effect. The difference of the signals detected by the left and right platinum depends on the magnetization direction $\alpha$ (relative to the electrode wire $\hat{\bf x}$-direction) by $\propto \sin^2\alpha\cos\alpha$, as shown in (c). The figures are taken from Ref.~\cite{DMI_shift_2}.}
	\label{DMI_shifts}
\end{figure}

Now consider a magnetic film with the surface normal along the $\hat{\bf x}$-direction and the saturated magnetization along the $\hat{\bf z}$-direction. 
The associated interfacial DMI \textcolor{blue}{in the continuum model} reads
\cite{DMI_shift}
\begin{align}
\nonumber
	\hat{H}_{\mathrm{DMI}}&=-\frac{D}{2\mu_{0}M_{s}^{2}}\int d\mathbf{r}{\bf M}\cdot[(\hat{\bf x}\times\nabla)\times{\bf M}]\\
	&=-\frac{D}{2\mu_{0}M_{s}^{2}}\int d\mathbf{r}\left[
	\hat{\mathbf{y}}\cdot(\mathbf{M}\times\partial_{z}\mathbf{M})-\hat{\mathbf{z}}\cdot(\mathbf{M}\times\partial_{y}\mathbf{M})\right],
	\label{Hamiltonian_DMI}
\end{align}
where $D$ is the DMI constant. For the spin waves propagating perpendicular to the magnetization, the magnon dispersion limited by the exchange and DM interactions to leading order reads 
\begin{align}
	\omega_{k_{y}}=\mu_{0}\gamma H_{\mathrm{z}}+\omega_M\alpha_{\mathrm{ex}%
	}k_{y}^{2}-\gamma Dk_{y}/M_{s},
\end{align}
where $\alpha_{\rm ex}$ is the exchange stiffness. The DMI then causes a rigid shift of the parabolic band, as illustrated in Fig.~\ref{DMI_shifts}(a). For propagation along the general direction with momentum ${\bf k}$, the dispersion is governed by the surface normal $\hat{\bf n}$, the magnetization direction $\hat{\bf M}_s$ and the momentum that constructs a chirality by the right-hand rule,
\begin{align}
	\omega_{{\bf k}}=\mu_{0}\gamma H_{\mathrm{z}}+\omega_M\alpha_{\mathrm{ex}%
	}{\bf k}^{2}-\gamma D/M_{s}(\hat{\bf n}\times \hat{\bf M}_s)\cdot{\bf k},
\end{align}
\textit{i.e.}, there is no non-reciprocity for spin waves propagating along the magnetization direction. The chirality index $(\hat{\bf n}\times \hat{\bf M}_s)\cdot{\bf k}$ naturally appears in the dispersion relation.
Additional numerical calculations showed that the non-reciprocity affects not only
 the frequency but also the amplitude and the attenuation length \cite{DMI_chiral_amplitude,DMI_chiral_attenuation}.

The impact of the interfacial DMI $|D|$ on all but the surface spin waves decreases with the magnetic film thickness \cite{DMI_2019,DMI_2017}. For thin ferromagnetic films in contact with heavy metals typically $D \sim 1~{\rm mJ/m^2}$ (techniques for measuring $D$ and its values for different materials have been reviewed in \cite{interface_DMI_review}). The DMI also favors out-of-plane magnetization since the DMI vector ${\bf D}$ is perpendicular to the film normal and when strong enough may pull the magnetization out of the film plane (refer to Sec.~\ref{section2.1.2}). The system which exhibits large interfacial DMI usually contains heavy metal adjacent to the ferromagnet layer. The heavy metal may be notorious for spin wave application as it will introduce additional damping in magnetization dynamics and curb the long-distance propagation of the spin waves. Recently, it has been reported that DMI can also arise in the absence of heavy metal. All-oxide interface can provide a sizable DMI, which extends the material set of chiral magnetism \cite{Ref_NewDMI1,Ref_NewDMI3}.

Recently, a significant DMI was reported for the interface between GGG and rare earth iron garnets \cite{DMI_shift_3}. Chiral velocities were observed in thin YIG film of thickness $7$~nm on GGG substrate in the measurement via the microwave transmission between two strip lines on YIG, which were interpreted by the above DMI-shift with $D=-16~\mu{\rm J}/{\rm m^2}$ \cite{DMI_shift_1}. A competitive mechanism such as the Doppler shift by the stripline chirality (introduced in Sec.~\ref{Stripline} below) \cite{Doppler_Yu} in the nonlinear regime and chiral damping by the proximity metallic contact in the linear response (reviewed in Sec.~\ref{Sec_chiral_damping}) \cite{chiral_gate} provides the same phenomenology. A similar mechanism was employed to interpret the experimentally observed non-reciprocity in electrical injection and detection of incoherent magnons of YIG films with thickness $50$~nm in Fig.~\ref{DMI_shifts}(b), where $D=-2~\mu{\rm J}/{\rm m^2}$ was fitted \cite{DMI_shift_2}. However, the DMI can be eliminated by a global gauge transformation in terms of a phase that has no physical consequences on transport in that the group velocity of the spin waves is the same at the same frequency. A similar argument applies to the absence of equilibrium spin current under the Rashba spin-orbit coupling. The possibility of the DMI between YIG and GGG awaits first-principle calculation for the origin of the DMI by closed-shell magnetic moments with small spin-orbit interaction.

\subsection{Chiral magnetic fields  }
\label{evanescent_magnetic_field}
A plane wave with (real) frequency \(\omega\) and a three-dimensional complex wave vector \textbf{k} is called propagating when \textbf{k} is real and evanescent when imaginary.  Surface or interface wave propagates in a plane but is evanescent normal to it, therefore natural conduits of guided transport in a plane. In wave guides only the wave vector component along its axes is real and propagation is in one dimension only. Here we focus on the electrodynamics of magnetic fields with evanescent components that are chiral.

 \subsubsection{Stripline}
   \label{Stripline}
   
  A stripline is a metallic wire biased by an ac electric current. It generates Amp\'{e}re magnetic fields, \textit{e.g.}, Ref.~\cite{Jackson} with interesting chiralities that have been overlooked in the past.

\begin{figure}[ptbh]
   	\begin{centering}
   		\includegraphics[width=1.0\textwidth]{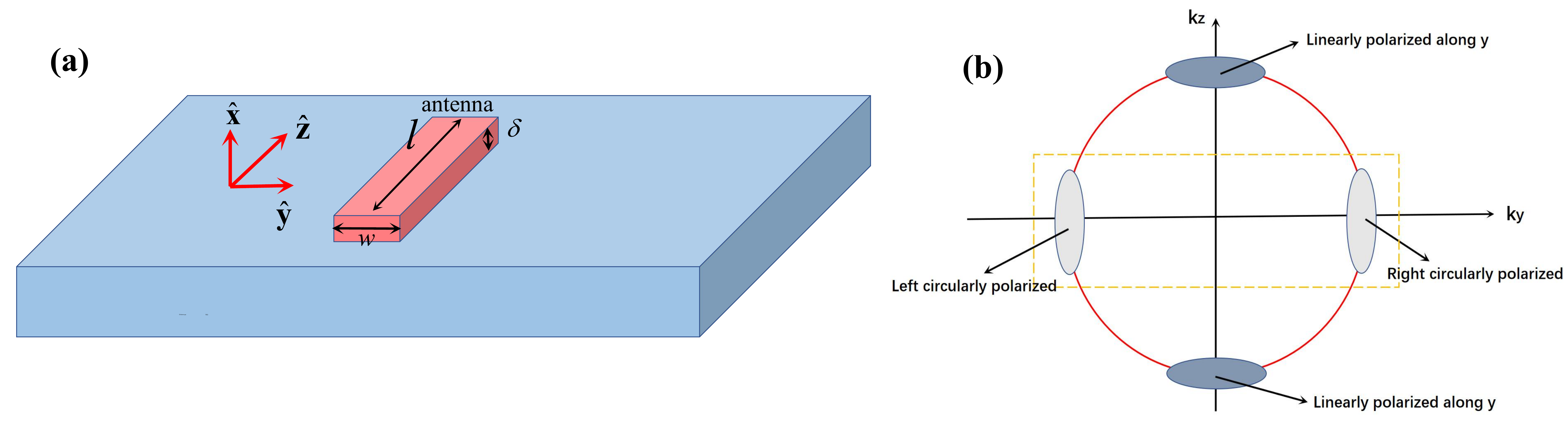}
   		\par\end{centering}
   	\caption{A stripline of rectangular section with length $l$, width $w$ and thickness $\delta$ supported by a planar substrate [(a)] and biased by an ac current.  (b) illustrates the polarization of the Amp\'{e}re magnetic field in the momentum space.}
   	\label{polarization}
   \end{figure}

  Let us focus on the short stripline in Fig.~\ref{polarization}(a) of length $l$ along  \(\hat{\mathbf{z}}\) and rectangular cross section \(w\delta\), with width $w$ and thickness $\delta$, supported by a planar substrate. An ac current  density bias $\mathbf{J}(\mathbf{r},\omega) \parallel \hat{\bf z}$ with frequency $\omega$ generates the vector potential \cite{Jackson}
   \begin{equation}
   	\mathbf{A}(\mathbf{r},\omega)=\frac{\mu_{0}}{4\pi}\int d\mathbf{r}^{\prime}\mathbf{J}(\mathbf{r}^{\prime},\omega)\frac{e^{ik|\mathbf{r}-\mathbf{r}^{\prime}|}}{|\mathbf{r}-\mathbf{r}^{\prime}|},
   	\label{vector_potential}
   \end{equation}
 where $\mu_{0}$ is the vacuum permeability and $k=\sqrt{k_{x}^{2}+k_{y}^{2}+k_{z}^{2}}=\omega/c$. The magnetic field $(H_{x},H_{y})=(1/\mu_0)(\partial A_{z}/\partial
   y,\\-\partial A_{z}/\partial x)$ decays as a function of distance \(r\) from the stripline by $\sim 1/r$ and $H_{z}=0$. To see that this magnetic field is evanescent and chiral we substitute the Weyl identity \cite{nano_optics},
   \begin{equation}
   	\frac{e^{ik\sqrt{\left(x-x^{\prime}\right)^{2}+\left(  y-y^{\prime}\right)^{2}+\left(z-z^{\prime}\right)  ^{2}}}}{\sqrt{\left(x-x^{\prime}\right)^{2}+\left(  y-y^{\prime}\right)^{2}+\left(z-z^{\prime}\right)^{2}}}
   	=\frac{i}{2\pi}\int dk_{y}dk_{z}\frac{e^{ik_{x}|x-x^{\prime}|+ik_{y}(y-y^{\prime})+ik_{z}(z-z^{\prime})}}{k_{x}},
   	\label{Weyl_identity}
   \end{equation}
   where  $k_{x}=\sqrt{(\omega/c)^{2}-k_{y}^{2}-k_{z}^{2}}$. 
   
   In the mixed real and Fourier space
   \begin{subequations}
   \begin{align}
   	H_{x}(x;k_{y},k_{z}) &  =2iJ(\omega)\frac
   	{e^{-ik_{x}x}}{k_{x}}\frac{e^{ik_{x}\delta}-1}{k_{x}}\sin\left(  k_{y}\frac
   	{w}{2}\right)  \frac{\sin(k_{z}l/2)}{k_{z}}e^{-ik_{z}z},\\
   	H_{y}(x;k_{y},k_{z}) &  =2iJ(\omega)\frac
   	{e^{-ik_{x}x}}{k_{x}}\frac{e^{ik_{x}\delta}-1}{k_{y}}\sin\left(  k_{y}\frac
   	{w}{2}\right)  \frac{\sin(k_{z}l/2)}{k_{z}}e^{-ik_{z}z}.
   \end{align}
\end{subequations}
Hence
\begin{align}
	k_{x}H_{x}(x;k_{y},k_{z})=k_{y}H_{y}(x;k_{y},k_{z})
\end{align}
obeys a polarization (spin)-momentum locking relation. The propagation direction, \textit{i.e.}, the rotation direction of the fields, is along the $\hat{\bf y}$-direction, but the polarization is along $\hat{\bf z}$, so the polarization (spin) of $H_{x,y}$ is transverse. In Sec.~\ref{unification}, we will strictly define the transverse spin density of the electromagnetic waves, where we find this implies the spin-momentum locking and chirality of the evanescent magnetic field.
For a long wire $l\gg w$, the magnetic field oscillates as function of $k_{z}$ with
   a short period of $4\pi/l$, while the dependence on $k_{y}$ is slow, \textit{i.e.}, with a periodicity of 
   $4\pi/w$.

   The evanescent limit $\sqrt{k_{y}^{2}+k_{z}^{2}}\gg\omega/c$ is typical for magnonics: Microwaves with frequencies of $\omega/(2\pi)\sim 1$~GHz, $k_{0}\equiv\omega/c=2.1$~m$^{-1}$, and a characteristic wavelength $\lambda_{0}=2\pi/k_{0}=0.3$~m. As long as we focus on wave lengths of interest that are much shorter than this scale, $k_{x}%
   \rightarrow i\sqrt{k_{y}^{2}+k_{z}^{2}}=i\kappa$. When additionally $\kappa\delta\ll 1$, the magnetic near-fields read 
   \begin{subequations}
   \begin{align}
   	H_{x}(x;k_{y},k_{z})  &  =-2iJ(\omega)e^{\kappa
   		x}\frac{e^{-\kappa\delta}-1}{\kappa^{2}}
        \sin\left(  k_{y}\frac{w}{2}\right)
   	\frac{\sin(k_{z}l/2)}{k_{z}}e^{-ik_{z}z},\\
   	H_{y}(x;k_{y},k_{z})  &  =2J(\omega)e^{\kappa x}%
   	\frac{e^{-\kappa\delta}-1}{\kappa k_{y}}\sin\left(  k_{y}\frac{w}{2}\right)
   	\frac{\sin(k_{z}l/2)}{k_{z}}e^{-ik_{z}z}, \label{magnetic_fields}%
   \end{align}
   \end{subequations}
   and thereby
   \begin{equation}
   	H_{x}(x;k_{y},k_{z})=-i(k_{y}/\kappa)H_{y}(x;k_{y},k_{z}).
   \end{equation}
When $\left\vert k_{y}\right\vert \ll|k_{z}|$ the radiation is nearly
   linearly-polarized along the $\hat{\mathbf{y}}$-direction $(\left\vert
   H_{x}\right\vert \ll\left\vert H_{y}\right\vert) $.
   On the other hand, when $\left\vert k_{y}\right\vert \gg|k_{z}|$,
   $H_{x}(x;k_{y},k_{z})\rightarrow-i\mathrm{sgn}(k_{y})H_{y}(x;k_{y},k_{z})$. The field is then
   nearly right- (left-) circularly polarized for positive (negative) $k_{y}$, \textit{i.e.}, propagating in  positive and (negative) directions.
   Figure~\ref{polarization}(b) summarizes the field polarization (spin) as a function of wave vector direction for 
 a short stripline. For a long stripline, $l\rightarrow \infty$ and $k_z\rightarrow 0$, on the other hand, the polarization and momentum are locked. The chirality of the stripline magnet field causes the chiral pumping of circularly-polarized spin waves \cite{stripline_poineering_1,stripline_poineering_2,Yu_Springer,Teono_NV} as reviewed in the next Chapter (Sec.~\ref{stripline_excitation}).

\begin{figure}[ptbh]
	\begin{centering}
		\includegraphics[width=0.68\textwidth]{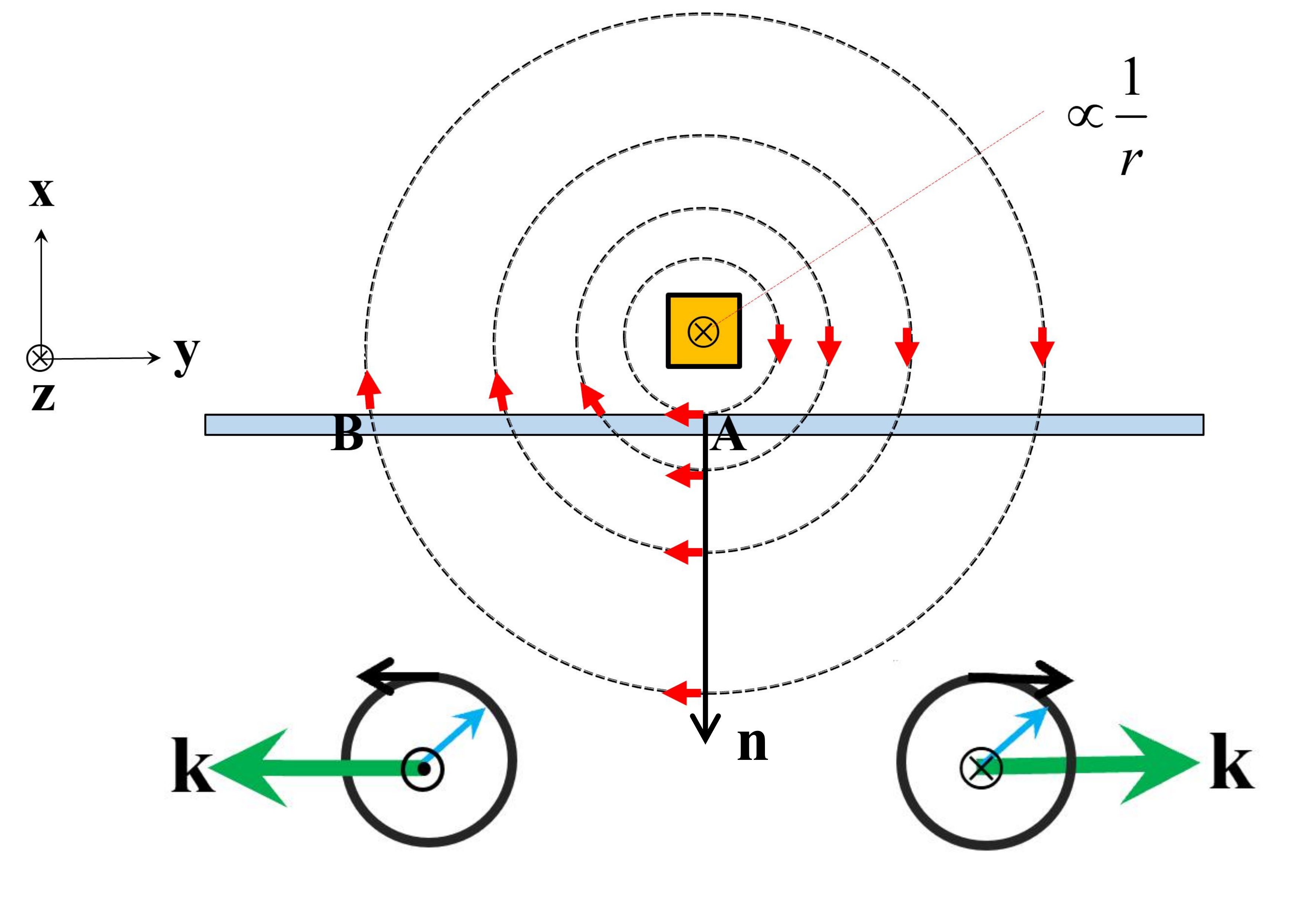}
		\par\end{centering}
	\caption{The chirality of an ac magnetic field emitted by an infinitely long stripline on a plane defined by $AB$ with surface normal ${\bf n}$. The polarization-momentum locking is illustrated by the thick black circles with red (green) arrows for  the field direction (momentum ${\bf k}$) at the bottom of the figure.}
	\label{stripline_field}
\end{figure}

We can understand the chirality best for an infinitely long wire, \textit{e.g.}, along the $\hat{\bf z}$-direction, as depicted in the snapshot Fig.~\ref{stripline_field} and a plane $AB$ with the surface normal along the $\hat{\bf x}$-direction.  The magnetic field is indicated by its direction (red arrows) while its amplitude decays slowly like $1/r$. The polarization of the radial waves periodically changes sign, but not direction. On the line from $A$ to $B$ (to the left) the instantaneous vector rotates in a  clockwise fashion, but anti-clockwise on the line from $B$ to $A$. In the time domain, the spherical waves emitted by the stripline move outward, so the wave on the AB (BA) line has a wave vector to the left (right). Therefore, in a particular plane (that does not cross the stripline), the magnetic field is polarization-momentum locked, with a propagation direction governed by the outer product of the polarization and surface normal of the chosen plane, \textit{i.e.}, the ac magnetic field is chiral. As discussed above the magnetic field wave on the AB plane is not periodic and is composed of a linear combination of evanescent plane waves.

The chirality of the ac magnetic field implies a unidirectional launch of spin waves in the in-plane magnetized magnetic field below the stripline \cite{stripline_poineering_1,stripline_poineering_2,Yu_Springer}, as we will review in Sec.~\ref{stripline_excitation}.

 \subsubsection{Dipolar stray magnetic fields of spin waves}
 \label{dipolar_fields_1}
 
   Another example of evanescent ac magnetic fields is the stray field generated by spin waves. The field of a magnetic dipole decays faster than that of the stripline (monopole), \textit{i.e.}, as $1/r^3$ . We discuss in the following the chirality of the  fields emitted by coherent spin waves in a magnetic film and a nanowire.
   
\begin{figure}[ptbh]
   	\begin{centering}
   		\includegraphics[width=0.99\textwidth]{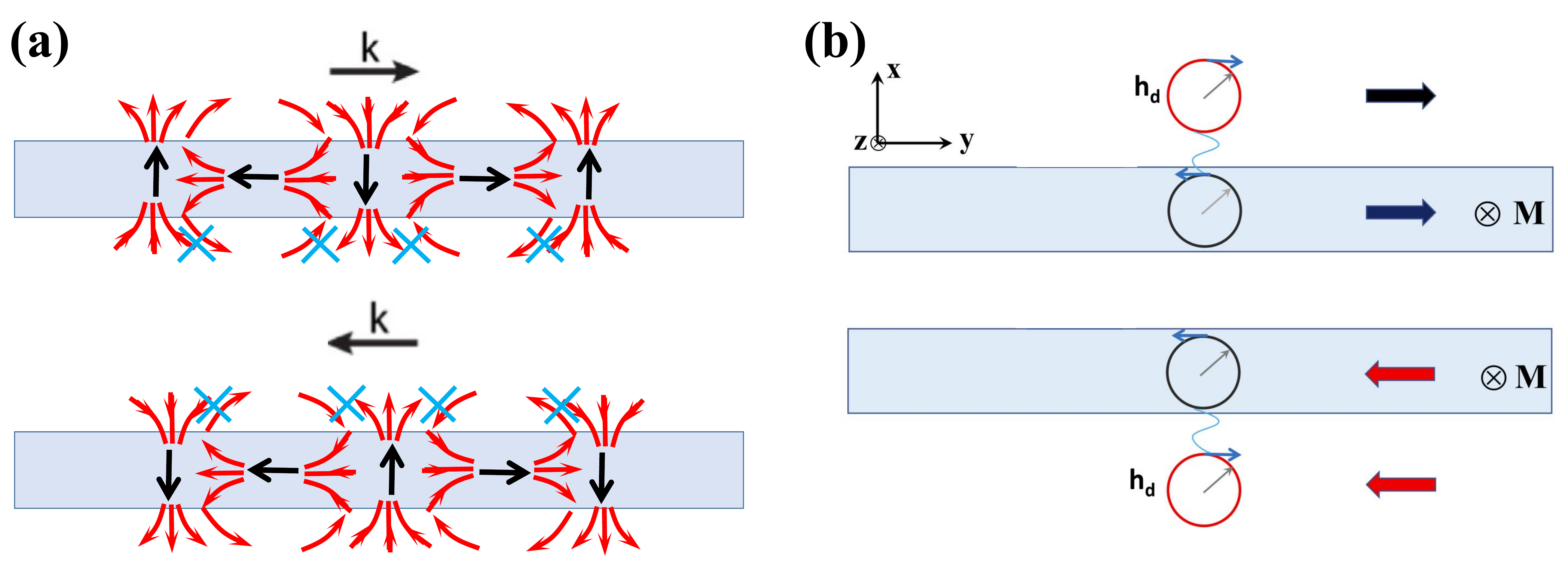}
   		\par\end{centering}
   	\caption{Dipolar fields generated by spin waves propagating perpendicular to the in-plane magnetization of a thin film. (a) illustrates the particular nature of the stray fields of surface and bulk magnetic charges that vanish on one side of the film depending on the wave vector of the spin wave. (b) summarizes the polarization (spin), propagation, and spatial distribution of the stray fields. The figure [(b)] is taken from Ref.~\cite{Chiral_pumping_Yu}.}
   	\label{spin_wave_field}
   \end{figure}

   \textbf{In-plane magnetized film}.---Spin waves cause periodic perturbations of the equilibrium magnetization and associated stray magnetic fields. Here we focus on the excitations of homogeneously in-plane magnetized magnetic films. Waves at GHz are in the  (quasi-)static regime, which allows a treatment close to conventional electrostatics. A non-uniform magnetization in which the divergence of the magnetic vector field generates a magnetic charge density $\rho_M$:\begin{equation}
   \nabla\cdot{\bf M}\equiv \rho_M.
   \end{equation}
It is convenient to express the stray field in terms of the magnetic potential
\begin{equation}
	{\bf H}=-\nabla\phi_M, 
\end{equation} 
which leads to the Poisson equation:
\begin{equation}
	\nabla^2\phi_M=\nabla\cdot{\bf M}. 
\end{equation} 
The stray field is the negative gradient of the magnetic potential 
\begin{equation}
	\phi_M({\bf r})=-\frac{1}{4\pi}\int_{\rm volume}\frac{\nabla'\cdot{\bf M}({\bf r}')}{|{\bf r}-{\bf r}'|}dV'+\frac{1}{4\pi}\int_{\rm surface}\frac{{\bf n}\cdot{\bf M}({\bf r}')}{|{\bf r}-{\bf r}'|}dS',
	\label{potential_general}
\end{equation}
where we separated the contributions from  ``volume" and ``surface" charges. 
 Figure~\ref{spin_wave_field} illustrates the physical origin of the chirality of the dipolar field \cite{Chiral_pumping_Yu,Chiral_pumping_grating}  for a magnetization that is uniform across a (sufficiently) thin film of thickness $\delta$. Spin waves that are right circularly polarized and propagate perpendicular to the in-plane magnetization, generate both surface and bulk magnetic charges. Their contributions generated by plane waves that move to the right add up to generate a finite stray field above the film, but cancel below the film and vice versa for left-moving spin waves.  The stray fields are therefore unidirectional with a handedness governed by the outer vector product of the surface normal and equilibrium magnetization directions. This mechanism is responsible for the chiral pumping of spin waves by nearby nanomagnets, as discussed in the following Sec.~\ref{dipolar_pumping}.  
   
In general, spin waves are elliptically polarized and may propagate in any direction in the plane.  We fix the  equilibrium magnetization along the $\hat{\mathbf{z}}%
   $-direction. A spin wave consists of a norm-conserving transverse magnetization amplitude \textbf{M}. Expressed as a superposition of right ($m_{R}$) and left ($m_{L}$) circularly polarized components,
   \begin{align}
   \left(
   \begin{array}
   	[c]{c}%
   	M_{x}(\mathbf{r},t)\\
   	M_{y}(\mathbf{r},t)
   \end{array}
   \right)  &=\left(
   \begin{array}
   	[c]{c}%
   	m_{x}^{\mathbf{k}}(x)\cos\left(  \mathbf{k}\cdot\pmb{\rho}-\omega t\right) \\
   	-m_{y}^{\mathbf{k}}(x)\sin\left(  \mathbf{k}\cdot\pmb{\rho}-\omega t\right)
   \end{array}
   \right)\nonumber\\
   &=m_{R}^{\mathbf{k}}(x)\left(
   \begin{array}
   	[c]{c}%
   	\cos\left(  \mathbf{k}\cdot\pmb{\rho}-\omega t\right) \\
   	-\sin\left(  \mathbf{k}\cdot\pmb{\rho}-\omega t\right)
   \end{array}
   \right)  +m_{L}^{\mathbf{k}}(x)\left(
   \begin{array}
   	[c]{c}%
   	\cos\left(  \mathbf{k}\cdot\pmb{\rho}-\omega t\right) \\
   	\sin\left(  \mathbf{k}\cdot\pmb{\rho}-\omega t\right)
   \end{array}
   \right),
   \end{align}
   where $m_{R}^{\mathbf{k}}(x)=[m_{x}^{\mathbf{k}}(x)+m_{y}^{\mathbf{k}}(x)]/2$, 
    $m_{L}^{\mathbf{k}}(x)=[m_{x}^{\mathbf{k}}(x)-m_{y}^{\mathbf{k}}(x)]/2$,  \(\boldsymbol{\rho}=(y,z)\), and \(\textbf{k}=(k_y,k_z\)).
   This magnetization generates a dynamical dipolar field $\left(  \alpha,\beta
   \in\{x,y,z\}\right)  $
   \begin{equation}
   	h_{\beta}(\mathbf{r},t)=\frac{1}{4\pi}\partial_{\beta}\partial_{\alpha}\int d\mathbf{r}^{\prime}\frac{M_{\alpha}(\mathbf{r}^{\prime},t)}{|\mathbf{r}-\mathbf{r}^{\prime}|}. \label{hdip}
   \end{equation}
   By substituting into Eq.~(\ref{hdip}) and using the
   Coulomb integral
   \begin{equation}
   	I=\int d\mathbf{r}^{\prime}\frac{e^{i{\bf k}\cdot{\pmb{\rho}}^{\prime}}f\left(  x^{\prime
   		}\right)  }{|\mathbf{r}-\mathbf{r}^{\prime}|}=\frac{2\pi}{|{\bf k}|}e^{i{\bf k}\cdot{\pmb \rho}}\int dx^{\prime}e^{-\left\vert x-x^{\prime}\right\vert |{\bf k}|}f\left(
   	x^{\prime}\right)  ,
   \end{equation} 
   their dipolar field outside the film reads
   \begin{align}
\textbf{h}(\mathbf{r},t) & =\frac{1}{2}\left(
   	\begin{array}
   		[c]{c}%
   		\left(  \left\vert \mathbf{k}\right\vert +\mathrm{sgn}(x)k_{y}\right)
   		\cos\left(  \mathbf{k}\cdot\pmb{\rho}-\omega t\right) \\
   		k_{y}\left(  \frac{k_{y}}{\left\vert \mathbf{k}\right\vert }+\mathrm{sgn}%
   		(x)\right)  \sin\left(  \mathbf{k}\cdot\pmb{\rho}-\omega t\right) \\
   		k_{z}\left(  \frac{k_{y}}{\left\vert \mathbf{k}\right\vert }+\mathrm{sgn}
   		(x)\right)  \sin\left(  \mathbf{k}\cdot\pmb{\rho}-\omega t\right)
   	\end{array}
   	\right)  e^{-\left\vert \mathbf{k}\right\vert \left\vert x\right\vert }\int_0^{\delta}
   	dx^{\prime}m_{R}^{\mathbf{k}}\left(  x^{\prime}\right)  e^{\left\vert
   		\mathbf{k}\right\vert \mathrm{sgn}(x)x^{\prime}}\nonumber\\
   	&  +\frac{1}{2}\left(
   	\begin{array}
   		[c]{c}%
   		(\left\vert \mathbf{k}\right\vert -\mathrm{sgn}(x)k_{y})\cos\left(
   		\mathbf{k}\cdot\pmb{\rho}-\omega t\right) \\
   		k_{y}\left(  -\frac{k_{y}}{\left\vert \mathbf{k}\right\vert }+\mathrm{sgn}%
   		(x)\right)  \sin\left(  \mathbf{k}\cdot\pmb{\rho}-\omega t\right) \\
   		k_{z}\left(  -\frac{k_{y}}{\left\vert \mathbf{k}\right\vert }+\mathrm{sgn}%
   		(x)\right)  \sin\left(  \mathbf{k}\cdot\pmb{\rho}-\omega t\right)
   	\end{array}
   	\right)  e^{-\left\vert \mathbf{k}\right\vert \left\vert x\right\vert }\int_0^{\delta}
   	dx^{\prime}m_{L}^{\mathbf{k}}\left(  x^{\prime}\right)  e^{\left\vert
   		\mathbf{k}\right\vert \mathrm{sgn}(x)x^{\prime}},
   \end{align}
where $\mathrm{sgn}(x)$ is the sign function.  A spin wave propagating normal to the magnetization $\left(  k_{z}\rightarrow0\right)  $ generates a dipolar field above the film with \(h_{z}=0\) and
   \cite{Chiral_pumping_grating,Chiral_pumping_Yu}
   \begin{align}
   	\left(
   	\begin{array}
   		[c]{c}%
   		h_{x}\\
   		h_{y}
   	\end{array}
   	\right)_{x>\delta} (\mathbf{r},t)  &  =\frac{\left\vert k_{y}\right\vert +k_{y}}{2}\left(
   	\begin{array}
   		[c]{c}%
   		\cos(k_{y}y-\omega t)\\
   		\sin(k_{y}y-\omega t)
   	\end{array}
   	\right)  e^{-|k_{y}|x}\int_0^{\delta} dx^{\prime}m_{R}^{k_{y}}\left(  x^{\prime}\right)
   	e^{|k_{y}|x^{\prime}}\nonumber\\
   	&  +\frac{\left\vert k_{y}\right\vert -k_{y}}{2}\left(
   	\begin{array}
   		[c]{c}%
   		\cos(k_{y}y-\omega t)\\
   		-\sin(k_{y}y-\omega t)
   	\end{array}
   	\right)  e^{-|k_{y}|x}\int_0^{\delta} dx^{\prime}m_{L}^{k_{y}}\left(  x^{\prime}\right)
   	e^{|k_{y}|x^{\prime}},
   	\label{above_film}
   \end{align}
   and below the film
   \begin{align}
   	\left(
   	\begin{array}
   		[c]{c}%
   		h_{x}\\
   		h_{y}
   	\end{array}
   	\right)_{x<0} (\mathbf{r},t)   &  =\frac{\left\vert k_{y}\right\vert -k_{y}}{2}\left(
   	\begin{array}
   		[c]{c}%
   		\cos(k_{y}y-\omega t)\\
   		\sin(k_{y}y-\omega t)
   	\end{array}
   	\right)  e^{|k_{y}|x}\int_0^{\delta} dx^{\prime}m_{R}^{k_{y}}\left(  x^{\prime}\right)
   	e^{-\left\vert k_{y}\right\vert x^{\prime}}\nonumber\\
   	&  +\frac{\left\vert k_{y}\right\vert +k_{y}}{2}\left(
   	\begin{array}
   		[c]{c}%
   		\cos(k_{y}y-\omega t)\\
   		-\sin(k_{y}y-\omega t)
   	\end{array}
   	\right)  e^{|k_{y}|x}\int_0^{\delta} dx^{\prime}m_{L}^{k_{y}}\left(  x^{\prime}\right)
   	e^{-\left\vert k_{y}\right\vert x^{\prime}}.
   \end{align}
 The polarization of the dipolar field of circularly polarized spin waves is opposite to that of magnetization. Right (left) propagating spin waves with $k_{y}>0$  ($k_{y}<0$) only generate a dipolar field above (below) the film. Spin waves propagating \textit{parallel} to the equilibrium magnetization $\left(  k_{y}%
   \rightarrow0\right)  $ generate the fields
   \begin{align}
   \left(
   \begin{array}
   	[c]{c}%
     		h_{x}\\
   		h_{y}
   	\end{array}
   	\right) (\mathbf{r},t)    =\frac{1}{2}\left(
   \begin{array}
   	[c]{c}%
   	|{k}_{z}|\cos\left(  \mathbf{k}\cdot\pmb{\rho}-\omega t\right) \\
   	\mathrm{sgn}\left(  x\right)  k_{z}\sin\left(  \mathbf{k}\cdot
   	\pmb{\rho}-\omega t\right)
   \end{array}
   \right)  e^{-\left\vert {k}_{z}\right\vert \left\vert x\right\vert }\int_0^{\delta}
   dx^{\prime}\left(  m_{R}^{{k}_{z}}\left(  x^{\prime}\right)  +m_{L}^{{k}_{z}%
   }\left(  x^{\prime}\right)  \right)  e^{|{k}_{z}|\mathrm{sgn}\left(  x\right)
   	x^{\prime}}.
   \end{align}
Above the film, the dipolar field of spin waves with positive (negative) $k_{z}$, is always left (right) circularly polarized, \textit{viz.},
polarization-momentum locked. Below the film, the polarization is reversed.

\begin{figure}[ptbh]
   	\begin{centering}
   		\includegraphics[width=1.0\textwidth]{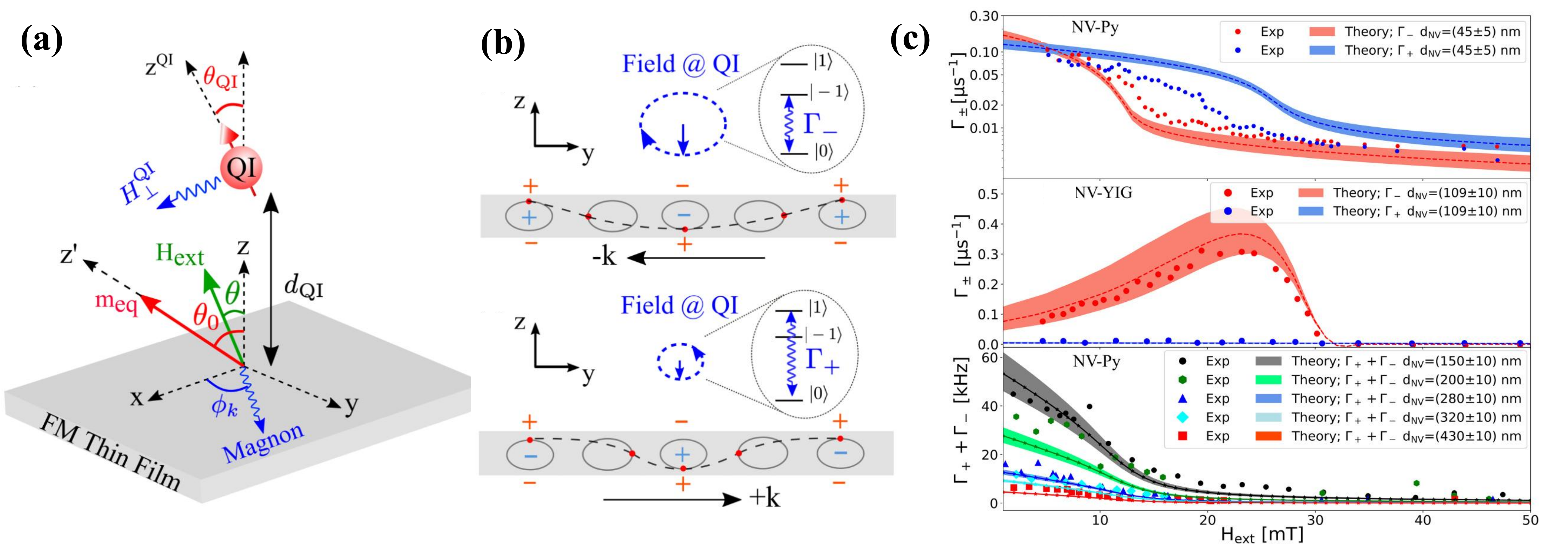}
   		\par\end{centering}
   	\caption{Sensing of the chirality of magnetic stray field by quantum impurity (diamond NV center) relaxometry. In (a), NV centers (QI) are placed above the Py (and YIG) magnetic thin films, whose stray-field noise triggers the transitions between states under the selection rules sketched in (b). The chirality caused the difference in the observed transition rates in panels (c). The figures are taken from Ref.~\cite{chiral_sensing}.}
   	\label{chiral_sensing}
   \end{figure}

  This dipolar field can be detected directly by local probes such as nitrogen-vacancy (NV) centers in diamonds that are sensitive to weak magnetic fields and can be read out optically.  Quantum impurity relaxometry \cite{quantum_impurity_1,quantum_impurity_2} as  illustrated in Fig.~\ref{chiral_sensing}(a) measures the relaxation rate of an NV center spin by its coupling with a magnetic noise source \cite{magnetic_noise_1,magnetic_noise_2,magnetic_noise_3,magnetic_noise_4}. The chirality of the stray dipolar fields plays a role because of the selection rules sketched in Fig.~\ref{chiral_sensing}(b), the transitions from state $|0\rangle\rightarrow|-1\rangle$ and $|0\rangle\rightarrow|+1\rangle$ are triggered by left and right circularly polarized stray fields with different and observable rates $\Gamma_{-}$ and $\Gamma_+$ because the amplitudes of the two differ \cite{Chiral_pumping_Yu}. The modeling based on this mechanism successfully explains the recent experimental observations as a function of applied field \cite{chiral_sensing} in Fig.~\ref{chiral_sensing}(c).

We review other physical consequences of the chiral stray field in Sec.~\ref{nanomagnets}.
   
    \textbf{Magnetic wire}.---The dipolar fields of an excited magnetic wire combine features of the ac current-biased stripline with the stray field from magnetic films. Here we focus on the chirality of the dipolar stray field generated by the ferromagnetic resonance (FMR) of a magnetic nanowire. We consider a nanowire and its equilibrium magnetization along the $\hat
   {\mathbf{z}}$ direction. The magnetic oscillations are the real part of
   \begin{equation}
   	\tilde{M}_{x,y}(\mathbf{r},t)=\tilde{m}_{x,y}\Theta(x)\Theta(-x+d)\Theta
   	(y+w/2)\Theta(-y+w/2)e^{-i\omega t},
   \end{equation}
   where $d$ and $w$ are the thickness and width of the nanowire and \(\omega\) is the FMR frequency. The
   corresponding dipolar magnetic field
   \begin{equation}
   	\tilde{h}_{\beta}(\mathbf{r},t)=\frac{1}{4\pi}\partial_{\beta}\partial
   	_{\alpha}\int\frac{\tilde{M}_{\alpha}(\mathbf{r}^{\prime},t)}{|\mathbf{r}%
   		-\mathbf{r}^{\prime}|}d\mathbf{r}^{\prime}=\frac{1}{4\pi}\partial_{\beta
   	}\partial_{\alpha}\int dz^{\prime}\int_{0}^{d}dx^{\prime}\int_{-w/2}%
   	^{w/2}dy^{\prime}\frac{\tilde{m}_{\alpha}e^{-i\omega t}}{\sqrt{z^{\prime
   				2}+(x-x^{\prime})^{2}+(y-y^{\prime})^{2}}}.
   \end{equation}
   By substituting the Coulomb integral \cite{Chiral_pumping_grating,nano_optics},
   \begin{equation}
   	\frac{1}{\sqrt{z^{\prime2}+(x-x^{\prime})^{2}+(y-y^{\prime})^{2}}}=\frac
   	{1}{2\pi}\int dk_{x}dk_{y}\frac{e^{-|z^{\prime}|\sqrt{k_{x}^{2}+k_{y}^{2}}}%
   	}{\sqrt{k_{x}^{2}+k_{y}^{2}}}e^{ik_{x}(x-x^{\prime})+ik_{y}(y-y^{\prime})},
   	\label{Coulomb_integral}
   \end{equation}
   the magnetic field with Fourier component $k_{y}$ below the nanowire ($x<0$) 
   \begin{align}
   	\tilde{h}_{\beta}(k_{y},x,t)  &  =\int h_{\beta}(\mathbf{r},t)e^{-ik_{y}%
   		y}dy\nonumber\\
   	&  =\frac{1}{2\pi}\int dk_{x}(k_{x}\tilde{m}_{x}+k_{y}\tilde{m}_{y})k_{\beta
   	}e^{ik_{x}x-i\omega t}\frac{1}{k_{x}^{2}+k_{y}^{2}}\frac{1}{ik_{x}%
   	}(1-e^{-ik_{x}d})\frac{2\sin(k_{y}w/2)}{k_{y}}.
   \end{align}
   Closing the contour of the $k_{x}$ integral in the lower half complex plane
   \begin{equation}
   	\left(
   	\begin{array}
   		[c]{c}%
   		\tilde{h}_{x}\\
   		\tilde{h}_{y}
   	\end{array}
   	\right)  (k_{y},x,t) =-\frac{i}{4\pi}e^{\left\vert k_{y}\right\vert x}(1-e^{-\left\vert
   		k_{y}\right\vert d})\frac{2\sin(k_{y}w/2)}{k_{y}\left\vert k_{y}\right\vert
   	}\left(
   	\begin{array}
   		[c]{cc}%
   		\left\vert k_{y}\right\vert  & ik_{y}\\
   		ik_{y} & -\left\vert k_{y}\right\vert
   	\end{array}
   	\right)  \left(
   	\begin{array}
   		[c]{c}%
   		\tilde{m}_{x}\\
   		\tilde{m}_{y}%
   	\end{array}
   	\right)  e^{-i\omega t}.
   	\label{wire_dipolar_field}
   \end{equation}
The dynamics of a cylindrical nanowire  is right circularly polarized $\left(  \tilde{m}%
   _{y}=i\tilde{m}_{x}\right)  $, which implies that the Fourier components of  $\tilde{\mathbf{h}}$ with $k_{y}>0$ vanish, \textit{i.e.}, being unidirectional, while that with  $k_{y}<0$ is left circularly polarized $\left(  \tilde{h}%
   _{y}=-i\tilde{h}_{x}\right)  $.

  \subsubsection{Waveguide and cavity microwaves}
   \label{waveguide_fields}
Here we review chiral electromagnetic waves that emerge due to partial confinement which leads to waves that are partly evanescent and propagating only along particular directions. We discuss two examples in more detail, \textit{i.e.}, a waveguide and a torus cavity \cite{Toroidal_waveguide_1,Toroidal_waveguide_2}. We focus on the microwave regime, but the formalism can be used as well for THz and optical frequencies. 
    
\textbf{Waveguide}.---A waveguide confines electromagnetic fields in two directions, but allows propagation along a third one. Examples are co-planar waveguides that in their simplest form are three parallel strip lines on a dielectric surface. Here we focus on the confinement caused by hollow tubes of highly conducting metals. The eigenmodes are the solutions of Maxwell's equations with appropriate boundary conditions.  
  
  The free-space Maxwell equations for the electric \textbf{E} and magnetic \textbf{H} fields in the frequency space
\cite{Jackson}
\begin{subequations}
\begin{align}
	\boldsymbol{\nabla }\times \mathbf{E}& =i\mu_0\omega {\bf H},\\ \boldsymbol{\nabla }\times \mathbf{H}&=-i\varepsilon_0 \omega \mathbf{E},   \\
	\boldsymbol{\nabla }\cdot \mathbf{H}& =0,\\ 
	\boldsymbol{\nabla }\cdot \mathbf{E}&=0,
	\label{Maxwell_free}
\end{align}
\end{subequations}
lead to the wave equations \cite{Jackson}
\begin{equation}
	\left(  \nabla^{2}-\frac{1}{c^{2}}\frac{\partial^{2}}{\partial t^{2}}\right)
	\{\mathbf{E},\mathbf{ B}\}=0. 
	\label{Maxwell_waves}
\end{equation}
The conducting boundary conditions at thick metallic walls are
\begin{equation}
	\mathbf{n}\times\mathbf{E}|_{S}=0,~~~~~~\mathbf{n}\cdot\mathbf{H}|_{S}=0.
	\label{boundary_condition}
\end{equation}
For a waveguide with translational symmetry along $\hat{\bf z}$, we separate the transverse and longitudinal components of the fields to be $\{{\bf E}_t,{\bf H}_t\}$ and $\{{\bf E}_z,{\bf H}_z\}$, respectively.
The modes are divided into two classes, \textit{viz.}, transverse magnetic (TM) modes with ${\bf B}_{z}=0$ and ${\bf E}_{z}%
|_{S}=0$ and transverse electric (TE) modes with ${\bf E}_{z}=0$ and $\partial {\bf B}_{z}/\partial n|_{S}=0$.
In the TE modes, the electric fields are transverse to the propagation $\hat{\bf z}$-direction with finite field components $\{E_x,E_y,H_z\}$. The magnetic field of the TM modes is transverse to the propagation direction and the field components are $\{H_x,H_y,E_z\}$.

We  may separate the magnetic field of a mode $\lambda$ with momentum $k>0$ in the $\hat{\bf z}$-direction as
\begin{subequations}
\begin{align}
	&{\bf H}_{k}^{(\lambda)}(x,y,z)=\big[{\pmb {\cal H}}_k^{(\mathrm{t},\lambda)}(x,y)+{\bf {\cal H}}_{kz}^{(\lambda)}(x,y)\hat{\bf z}\big]e^{ikz}/\sqrt{L},\\
	&{\bf H}_{-k}^{(\lambda)}(x,y,z)=\big[-{\pmb {\cal H}}_k^{(\mathrm{t},\lambda)}(x,y)+{\bf {\cal H}}_{kz}^{(\lambda)}(x,y)\hat{\bf z}\big]e^{-ikz}/\sqrt{L},
\end{align}
\end{subequations}
where ``$t$" means ``transverse", and $L \rightarrow \infty$ is the length of the waveguide. When there are no charges in the waveguide $\nabla\cdot {\bf H}_{\pm k}^{(\lambda)}=0$. Analogous relations hold for the electric fields $\boldsymbol{\cal E}^{\lambda}_{\pm k}$. These modes satisfy the orthonormal relations \cite{Jackson},
\begin{subequations}
\begin{align}
	&\int \left( {\mbox{\boldmath$\cal H$\unboldmath}}_{k,x}^{\left( \lambda
		\right) \ast }{\mbox{\boldmath$\cal H$\unboldmath}}_{k,x}^{\left( \lambda
		^{\prime }\right) }+{\mbox{\boldmath$\cal H$\unboldmath}}_{k,y}^{\left(
		\lambda \right) \ast }{\mbox{\boldmath$\cal H$\unboldmath}}_{k,y}^{\left(
		\lambda ^{\prime }\right) }\right) d\pmb{\rho}=\frac{A_{k}^{\left( \lambda
			\right) }}{(Z_{k}^{\left( \lambda \right) })^{2}}\delta _{\lambda \lambda
		^{\prime }},  \\
	&\int \mathcal{H}_{k,z}^{\left( \lambda \right) \ast }\mathcal{H}%
	_{k,z}^{\left( \lambda ^{\prime }\right) }d\pmb{\rho}=\frac{\gamma
		_{\lambda }^{2}A_{k}^{\left( \lambda \right) }}{k^{2}(Z_{k}^{\left( \lambda
			\right) })^{2}}\delta _{\lambda \lambda ^{\prime }},\hspace{0.4cm}(\mathrm{%
		TE)}  \\
	&\int \left( {\mbox{\boldmath$\cal E$\unboldmath}}_{k,x}^{\left( \lambda
		\right) \ast }{\mbox{\boldmath$\cal E$\unboldmath}}_{k,x}^{\left( \lambda
		^{\prime }\right) }+{\mbox{\boldmath$\cal E$\unboldmath}}_{k,y}^{\left(
		\lambda \right) \ast }{\mbox{\boldmath$\cal E$\unboldmath}}_{k,y}^{\left(
		\lambda ^{\prime }\right) }\right) d\pmb{\rho}=A_{k}^{\left( \lambda
		\right) }\delta _{\lambda \lambda ^{\prime }},  \\
	&\int \mathcal{E}_{k,z}^{\left( \lambda \right) \ast }\mathcal{E}%
	_{k,z}^{\left( \lambda ^{\prime }\right) }d\pmb{\rho}=\frac{\gamma
		_{\lambda }^{2}A_{k}^{\left( \lambda \right) }}{k^{2}}\delta _{\lambda
		\lambda ^{\prime }},\hspace{0.8cm}(\mathrm{TM)}  \label{orthogonal}
\end{align}
\end{subequations}
where $\pmb{\rho}=(x,y)$ and we introduced the impedances $Z_k^{(\lambda)}=\mu_0\omega_k^{(\lambda)}/k$  and $k/(\varepsilon_0\omega_k^{(\lambda)})$ as well as $A_k^{(\lambda)}=\hbar \omega_k^{(\lambda)}/(2\varepsilon_0)$ and $\hbar/(2\varepsilon_0\omega_k^{(\lambda)})$ for the TE and TM modes with frequency $\omega_k^{(\lambda)}$, respectively, while
\begin{align}
\gamma_{\lambda}^2=\mu_0\varepsilon_0(\omega_k^{(\lambda)})^2-k^2=(\omega_k^{(\lambda)})^2/c^2-k^2.
	\label{spectra}
\end{align} 
 $\gamma_{\lambda}$ in $\omega_k^{(\lambda)}=c\sqrt{k^2+\gamma_{\lambda}^2}$ only depends on the band index but not the momentum. 

The electromagnetic fields can be expanded in the waveguide-photon operators $\hat{p}_k^{\lambda}$, 
	\begin{equation}
		\mathbf{H}(\mathbf{r})=\sum_{\lambda }\int_{-\infty }^{\infty }\left( %
		\pmb{\cal H}_{k}^{\left( \lambda \right) }(\pmb{\rho})e^{ikz}\hat{p}%
		_{k}^{\left( \lambda \right) }+\mathrm{H.c.}\right) \frac{dk}{\sqrt{2\pi }},
		\label{HExp}
	\end{equation}%
	and similarly for the electric field with $\mathbf{H}\rightarrow \mathbf{E}$ and 
	amplitudes $\pmb{\cal H}\rightarrow \pmb{\cal E}$. The orthonormal relations (\ref{orthogonal}) lead to a compact expansion of the field energy 
\begin{align}
	\frac{\varepsilon_0}{2}\int d{\bf r}{\bf E}({\bf r})\cdot{\bf E}({\bf r})+\frac{\mu_0}{2}\int d{\bf r}{\bf H}({\bf r})\cdot{\bf H}({\bf r})=\sum_{k,\lambda}\hbar\omega_{k}^{(\lambda)}(\hat{p}_{k,\lambda}^{\dagger}\hat{p}_{k,\lambda}
	+\hat{p}_{-k,\lambda}^{\dagger}\hat{p}_{-k,\lambda}).
\end{align} 

\begin{figure}[ptbh]
	\begin{centering}
		\includegraphics[width=0.98\textwidth]{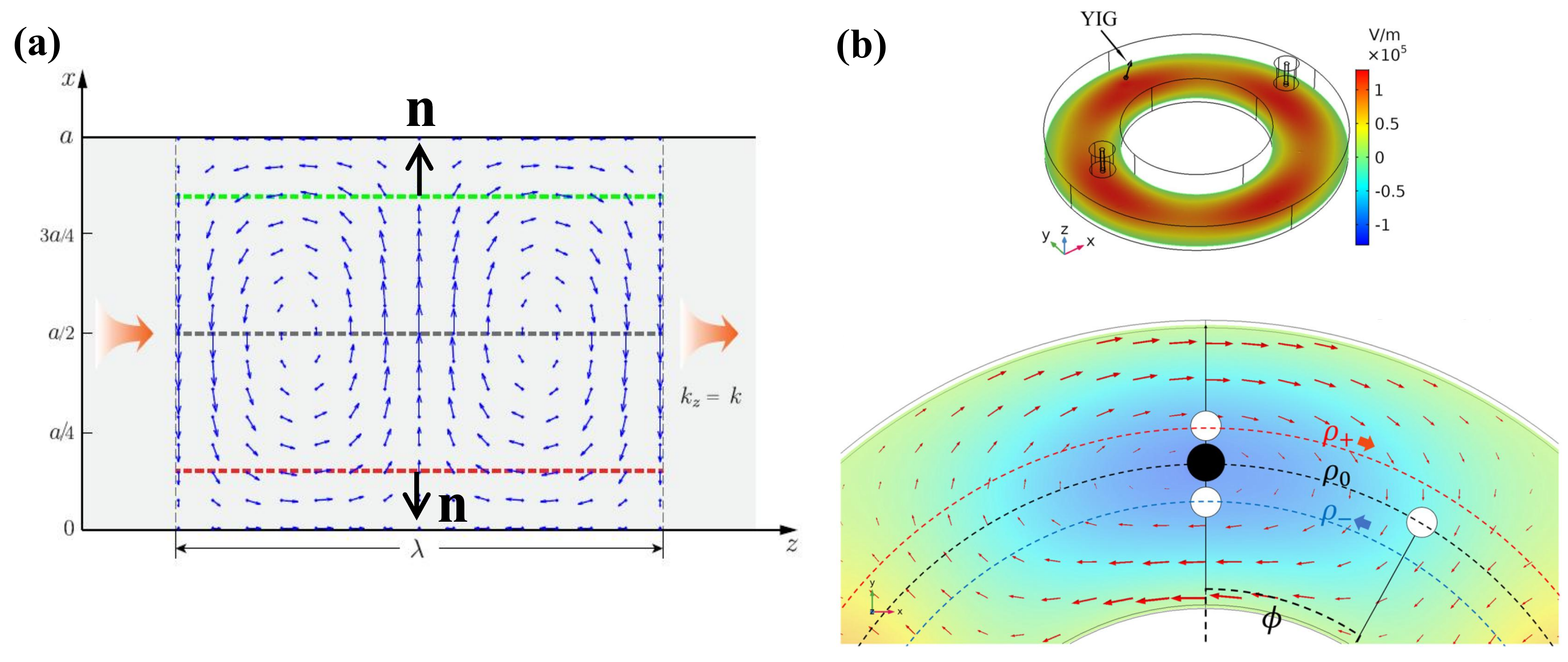}
		\par\end{centering}
	\caption{Chirality of microwaves in a rectangular waveguide and a torus-shaped cavity. (a) A snapshot of the magnetic-field vector distribution of the TE$_{10}$ waveguide mode over one wavelength \(\lambda\) in the open $\hat{\bf z}$-direction. The green and red dashed lines in the side view indicate lines (planes) at which the magnetic field rotates with constant modulus, \textit{i.e.}, is (oppositely) circularly polarized.  (b) sketches the geometry of a torus cavity and the snapshot of the magnetic-field vector distribution of the lowest TE mode.   In the enlarged cavity section indicating the polar angle $\phi$, $\rho_{+}$ and $\rho_{-}$ indicate the chiral lines (planes) of the magnetic field  that are calculated by Eq.~(\ref{chiral_lines}), while $\rho_0$ is for the position of a linearly polarized magnetic field. The figures are taken from Refs.~\cite{waveguide_Yu_1,circulating_polariton}.}
	\label{waveguide_waves}
\end{figure}

At ideal metallic walls [Eq.~(\ref{boundary_condition})] the magnetic field is parallel to the boundary.
Therefore, for most traveling magnetic fields, it must have at least one vortex in the field lines.
We illustrate the physics for a rectangular waveguide with dimensions $a>b$,
focusing on the lowest TE$_{10}$ mode, where the subscripts 1 and 0 are indices of the two transverse Cartesian modes.  Then $\gamma_{{\rm TE}_{10}}=\pi/a$, $Z_k=\mu_0\omega_k/k$ and $A_k=\hbar \omega_k/(2\epsilon_0)$,
and
\begin{subequations}
\begin{align}
	&{\cal H}_{\pm k,x}=\mp\sqrt{\frac{2}{ab}}\frac{\sqrt{A_k}}{\mu_0\omega_k}\sqrt{\Big(\frac{\omega_k}{c}\Big)^2-\Big(\frac{\pi}{a}\Big)^2}\sin\Big(\frac{\pi x}{a}\Big),\\
	&{\cal H}_{\pm k,y}=0,\\
	&{\cal H}_{\pm k,z}=-i\sqrt{\frac{2}{ab}}\frac{\sqrt{A_k}}{\mu_0\omega_k}\frac{\pi}{a}\cos\Big(\frac{\pi x}{a}\Big).
	\label{rec}
\end{align}
\end{subequations}
Of special interest is the existence of 2 ``chiral planes (lines in side view)"  along the waveguide at which the magnetic fields are circularly polarized in the $(x,z)$ plane. Its precise location depends on the waveguide geometry and the polarization on the 2 planes is opposite.  The sign of ${\cal H}_{\pm k,x}$ depends on the propagation direction which means that the polarization and momentum are locked. 
Figure~\ref{waveguide_waves}(a) shows a snapshot of the magnetic-field vector distribution over one wavelength, in which the chiral planes are indicated by the green and red dashed lines, and we see that on these lines the magnetic field vector of constant modulus rotates clockwise and anticlockwise.

\textbf{Torus cavity}.---Bending a straight waveguide into a circle generates a closed, torus-shaped cavity \cite{Toroidal_waveguide_1,Toroidal_waveguide_2}. It is easily seen that in the limit of a large aspect ratio the chiral lines (planes) should at which the polarization is locked to the ``momentum" form closed circles at approximately equal distance from the boundaries. The electric field of the TM modes of the torus-like cavity with inner
and outer radii $R_{1}$ and $R_{2}$ and height $b$ in the coordinate system of Fig.~\ref{waveguide_waves}(b) 
with ${\bf E}_{t}={\bf E}-{E}_{z}\hat{\mathbf{z}}=0$ at $z=0$ and $z=b$ read in cylindrical
coordinates $\{\rho,\phi\}$
\begin{equation}
	{\bf E}_{z}=\psi(\rho,\phi)\cos\left(  \frac{p\pi z}{b}\right)  ,~~~~~p=0,1,2,\cdots.
\end{equation}
with the radial wave equation 
\begin{equation}
	(\nabla_{t}^{2}+\gamma_{p}^{2})\psi_{p}=0,
\end{equation}
where $\gamma_{p}^{2}=\omega^{2}/c^{2}-(p\pi/b)^{2}$ and the index $p$  labels the standing waves in the $\hat{\bf z}$-direction.  With the cylindrical coordinates, 
\begin{equation}
	\frac{1}{\rho}\frac{\partial}{\partial\rho}\left(  \rho\frac{\partial\psi
	}{\partial\rho}\right)  +\frac{1}{\rho^{2}}\frac{\partial^{2}\psi}{\partial\phi^{2}}%
	+\gamma_{p}^{2}\psi=0.
\end{equation}
With $\psi(\rho,\phi)=R_{m}(\rho)e^{im\phi}$ and $m\in\mathbb{Z}_{0}$, we arrive at a Bessel equation of order $m$: 
\begin{equation}
	\frac{1}{\rho}\frac{\partial}{\partial\rho}\left(  \rho\frac{\partial
		R_{m}(\rho)}{\partial\rho}\right)  +\left(  \gamma_{p,m}^{2}-\frac{m^{2}}{\rho^{2}}%
	\right)  R_{m}(\rho)=0,
\end{equation}
with a general solution
\begin{equation}
	R_{m}(\rho)=J_{m}(\gamma_{p,m}\rho)+CN_{m}(\gamma_{p,m}\rho),
\end{equation}
where $J_{m}(x)$ and $N_{m}(x)$ are the Bessel functions of the first and second
kind, respectively. The boundary conditions with zero tangential $E_z$ at the inner and outer walls 
\begin{subequations}
\begin{align}
	J_{m}(\gamma_{p,m}R_{2})+CN_{m}(\gamma_{p,m}R_{2})  &  =0,\\
	J_{m}(\gamma_{p,m}R_{1})+CN_{m}(\gamma_{p,m}R_{1})  &  =0,
\end{align}
\end{subequations}
lead to $C=-J_{m}(\gamma_{p,m}R_{1})/N_{m}(\gamma_{p,m}R_{1}),$
\begin{equation}
	J_{m}(\gamma_{p,m}R_{2})N_{m}(\gamma_{p,m}R_{1})=J_{m}(\gamma_{p,m}R_{1}%
	)N_{m}(\gamma_{p,m}R_{2}),
	\label{characteristic_Bessel}%
\end{equation} and an electric field along \(z\)
\begin{equation}
	{\cal E}_{z}(\rho,\phi,z)=\left(  J_{m}(\gamma_{p,m}\rho)-\frac{J_{m}(\gamma_{p,m} R_{1})}%
	{N_{m}(\gamma_{p,m} R_{1})}N_{m}(\gamma_{p,m}\rho)\right)  e^{im\phi}\cos\left(
	\frac{p\pi z}{b}\right),
\end{equation}
which still has to be normalized. According to the Maxwell equations $(\nabla\times{\bf B})=(1/c^{2})\partial
_{t}{\bf E}$ and $\nabla\times{\bf E}=-\partial_{t}{\bf B}$, the magnetic field of the TM mode 
\begin{subequations}
\begin{align}
	{\cal H}_{x}(\rho,\phi,z)  &  =-i\frac{\omega}{\mu_{0}\gamma_{p,m}^{2}c^{2}}\partial
	_{y}{\cal E}_{z}=-i\frac{\omega}{\mu_{0}\gamma_{p,m}^{2}c^{2}}\left(  \sin\phi
	\frac{\partial}{\partial\rho}+\frac{1}{\rho}\cos\phi\frac{\partial}%
	{\partial\phi}\right)  {\cal E}_{z},\\
	{\cal H}_{y}(\rho,\phi,z)  &  =i\frac{\omega}{\mu_{0}\gamma_{p,m}^{2}c^{2}}\partial
	_{x}{\cal E}_{z}=i\frac{\omega}{\mu_{0}\gamma_{p,m}^{2}c^{2}}\left(  \cos\phi\frac
	{\partial}{\partial\rho}-\frac{1}{\rho}\sin\phi\frac{\partial}{\partial\phi
	}\right)  {\cal E}_{z}.
\end{align}
\end{subequations}

Figure~\ref{waveguide_waves}(b) shows the lowest standing mode ($p=0$).
In cylindrical coordinates  $\pmb{\cal H}(\rho,\phi)={\cal H}_{\rho}{\bf e}_{\rho}+{\cal H}_{\phi}{\bf e}_{\phi}$  with components
\begin{subequations}
\begin{align}
	{\cal H}_{\rho}(\rho,\phi)&=\frac{1}{\mu_0\gamma_mc}\frac{m}{\rho}{\cal E}_z,\\
	{\cal H}_{\phi}(\rho,\phi)&=-i\frac{1}{\mu_0\gamma_mc}\frac{\partial {\cal E}_z}{\partial \rho}.
\end{align}
\end{subequations}
We are now interested in the locking between the polarization and momentum, which is now not governed by the wave vector but by the orbital angular momentum that is governed by the integer $m$ that indicates clockwise (CW, $m>0$)  or counter-clockwise (CCW, $m<0$ ) motion. The rotation direction is ``locked" to the momentum by $\pm|m|$ at the ``chiral lines" $\rho_{\pm}$ in Fig.~\ref{waveguide_waves}(b) that solve
\begin{equation}
\frac{m}{\rho_{\pm}}{\cal E}_{z}(\rho_{\pm})+\left.  \frac{\partial {\cal E}_{z}(\rho
)}{\partial\rho}\right\vert _{\rho=\rho_{\pm}}=0,
\label{chiral_lines}
\end{equation}
since then magnetic field is circularly polarized $H_{\phi}=i\mathrm{sgn}%
(m)H_{\rho}$. For dimensions of tens of millimeters, the cavity is in GHz regime, \textit{e.g.}, 
with  $R_{1}=15$~mm and $R_{2}=30$~mm, by solving
Eq.~(\ref{characteristic}) to find $\gamma_{p,m}$,  we find the
frequency 
\begin{align}
	\omega_m=c\gamma_{p=0,m}=\{10.84,11.87,13.16,14.63,16.22,17.88,
	19.59,21.33,23.08\}~{\rm GHz}
\end{align}
when
$m=\{2,3,...,10\}$.

The energy of the electromagnetic field is
\begin{equation}
	\hat{H}_p=\int d{\bf r}\left[\frac{\varepsilon_0}{2}{\bf E}({\bf r})\cdot{\bf E}({\bf r})+\frac{\mu_0}{2}{\bf H}({\bf r})\cdot{\bf H}({\bf r})\right].
	\label{photon_Hamiltonian}
\end{equation}
We may quantize the fields for our TM mode in terms of the photon operator $\hat{\alpha}_{p,m}$,
\begin{subequations}
\begin{align}
	{\bf H}({\bf r})&=\sum_p\sum_m\left[\pmb{\cal H}_m^p(\rho,z)e^{im\phi}\hat{\alpha}_{p,m}+\pmb{\cal H}_m^{p*}(\rho,z)e^{-im\phi}\hat{\alpha}^{\dagger}_{p,m}\right],\\
	{\bf E}({\bf r})&=\sum_p\sum_m\left[\pmb{\cal E}_m^p(\rho,z)e^{im\phi}\hat{\alpha}_{p,m}+\pmb{\cal E}_m^{p*}(\rho,z)e^{-im\phi}\hat{\alpha}^{\dagger}_{p,m}\right],
	\label{quantization_photon}
\end{align}
\end{subequations}
such that Eq.~(\ref{photon_Hamiltonian}) assumes the form of a collection of harmonic oscillators $\hat{H}_{p}=\sum_{p,m}\hbar\omega_{p,m}\hat{\alpha}^{\dagger}_{p,m}\hat{\alpha}_{p,m}$. The normalization conditions
\begin{align}
	\int d{\bf r}\left(\frac{\varepsilon_0}{2}|\pmb{\cal E}_m^p(\rho,z)|^2+\frac{\mu_0}{2}|\pmb{\cal H}_m^p(\rho,z)|^2\right)=\frac{\hbar \omega_{p,m}}{2}
\end{align}
read in cylindrical coordinates
\begin{align}
	\pi\int_{R_1}^{R_2} \rho d\rho\int_0^b dz\left(\varepsilon_0|\pmb{\cal E}_m^p(\rho,z)|^2+\mu_0|\pmb{\cal H}_m^p(\rho,z)|^2\right)=\frac{\hbar \omega_{p,m}}{2}.
\end{align}

The field distribution of small torus cavities and therefore large curvature differs from straight open waveguides. The maxima of the magnetic field and the chiral lines shift to the outer edges. Increasing the circumference of the torus at constant cross-section shifts the maximum field into its center, but also reduces the local amplitude of a given cavity mode at equal power. The straight waveguide limit is reached when the discrete \(m\)-modes merge into a one-dimensional continuum.

\subsection{Chiral electric fields}
\label{Electric_field}

In the microwave regime, the magnetic field component of electromagnetic radiation dominates the interaction with matter. At optical frequencies, the magnetic field can be disregarded and the electric field response takes over. Evanescent electric fields play a central role in (nano)optics of structures with sub-wavelength dimensions. As discussed above evanescent electric fields can be expanded into plane waves of the form ${\bf E}e^{i{\bf k}\cdot{\bf r}-\omega t}$ in which at least one component of the wave vector \textbf{k} is imaginary and the field amplitude decays exponentially. 

\subsubsection{Surface plasmon polaritons }
\label{Sec_SPP}

The surface plasmon polariton (SPP) is a hybrid of  the electromagnetic field and collective motion of electrons at the surfaces of metals, insulators with a large dielectric constant, or mismatched material interfaces \cite{SSP_1,SSP_textbook_1,SSP_textbook_2,SSP_review}. Its amplitude decays exponentially as a function of distance from the surface or interface, while the propagation direction is governed by the outer product of the surface normal and polarization, \textit{i.e.}, exhibiting chirality, as explained in more detail below.

\textbf{Bulk plasmon}.---We first recall the concepts of bulk plasmon and its interaction with electromagnetic waves. Maxwell equations read
\begin{align}
	\nonumber
	\nabla\cdot {\bf D}&=\rho_f,\nonumber\\
	\nabla\cdot {\bf B}&=0,\nonumber\\
	\nabla\times{\bf E}&=-{\partial {\bf B}}/{\partial t},\nonumber\\
	\nabla\times{\bf H}&={\bf J}_f+{\partial {\bf D}}/{\partial t},
	\label{Maxwell_materials}
\end{align}
where $\rho_f$ is the density of net free charges that contribute to a free charge current ${\bf J}_f$ via the conservation law $\partial \rho_f/\partial t=-\nabla\cdot {\bf J}_f$, the electric displacement field ${\bf D}=\varepsilon_0{\bf E}+{\bf P}$, and ${\bf H}={\bf B}/\mu_0-{\bf M}$.  On the other hand, the electric polarization ${\bf P}$ is related to the bounded charge accumulation $\rho_b$ according to
\begin{align}
	\nabla\cdot{\bf P}=-\rho_b.
\end{align}
The charge conservation $\nabla\cdot {\bf J}_b=-\partial\rho_b/\partial t$ defines the associated bounded charge current ${\bf J}_b$ that leads to the displacement current 
\begin{align}
	{\bf J}_b={\partial {\bf P}}/{\partial t}.
	\label{current_P}
\end{align}
Here we focus on the electrodynamics of isotropic materials in the linear response regime. In the frequency and momentum space
	\begin{subequations}
\begin{align}
	\label{DD}
	&{\bf D}({\bf k},\omega)=\varepsilon_0\varepsilon({\bf k},\omega){\bf E}({\bf k},\omega),\\
	&{\bf J}({\bf k},\omega)=\sigma({\bf k},\omega){\bf E}({\bf k},\omega),
\end{align} 
\end{subequations}
where  $\varepsilon_0\varepsilon({\bf k},\omega)$ is the dielectric constant and $\sigma({\bf k},\omega)$ is the dynamic conductivity. The Maxwell equations (\ref{Maxwell_materials}) with Eq.~(\ref{current_P}) lead to 
\begin{align}
	\varepsilon({\bf k},\omega)=1+\frac{i\sigma({\bf k},\omega)}{\varepsilon_0\omega}.
\end{align}
The imaginary part of $\sigma({\bf k},\omega)$ contributes to the real part of $\varepsilon({\bf k},\omega)$, \textit{i.e.}, the ability for a material to be polarized by the electric field; while the real part of $\sigma({\bf k},\omega)$ corresponds to the imaginary part of $\varepsilon({\bf k},\omega)$, \textit{i.e.}, the ability of a material to absorb electromagnetic waves.

At the zeros of the dielectric constant $\varepsilon({\bf k},\omega)={\bf D}=0$, the polarization ${\bf P}$ exactly cancels the electric field ${\bf E}$, which implies that the motion of the electrons becomes collective. The root of  $\varepsilon({\bf k},\omega)=0$ defines the dispersion of this collective mode or ``plasmon".
The Lindhard formula for the dielectric constant of free electrons in bulk metals reads \cite{Mahan}
\begin{align}
	\varepsilon({\bf q},\omega)=1+\frac{8\pi e^2}{ q^2}\sum_{{\bf k}}n_{\bf k}\frac{E_{{\bf k}+{\bf q}}-E_{\bf k}}{(E_{{\bf k}+{\bf q}}-E_{\bf k})^2-\omega^2},
\end{align}
where $n_{\bf k}$ is the Fermi distribution and $E_{\bf k}$ is the electronic band structure. In the long-wavelength limit with ${\bf q}\rightarrow 0$, the plasmon dispersion is the root of
\begin{align}
	\varepsilon({\bf q},\omega_q)=1-\frac{\omega_{\rm P}^2}{\omega_q^2}-\frac{3\omega_{\rm P}^2}{5\omega_q^4}q^2v_F^2=0,
\end{align}
where $\omega_{\rm P}=\sqrt{{4\pi ne^2}/{m}}$ is the plasmon frequency from the Drude model, and $v_F$ is the Fermi velocity. The plasmon frequency in bulk metal is then obtained to be
\begin{align}
	\omega_q=\omega_{\rm P}\sqrt{1+\frac{3v_F^2}{5\omega_{\rm P}^2}q^2},
\end{align}
with a frequency gap $\omega_{\rm P}$.

\textbf{Plasmon polariton}.---The bulk plasmon couples strongly with the optical electric field to form a hybrid plasmon polariton. The wave equation for the electric field in metals follows from the Maxwell equations (\ref{Maxwell_materials}) and Eq.~(\ref{DD}):
\begin{align}
	{\bf k}({\bf k}\cdot{\bf E})-k^2{\bf E}=-\varepsilon({\bf k},\omega)\frac{\omega^2}{c^2}{\bf E}.
\end{align}
The longitudinal mode with ${\bf k}\cdot{\bf E}=kE$, leads to $\varepsilon({\bf k},\omega)=0$, \textit{i.e.}, corresponding to the bare plasmon excitation. The plasmon polariton is therefore associated with the transverse mode with ${\bf k}\cdot{\bf E}=0$ that satisfies
\begin{align}
	k^2=\varepsilon({\bf k},\omega)\omega^2/c^2,
	\label{PP}
\end{align}
which leads to its dispersion relation.

 In the long-wave length limit of a metal \(\varepsilon_m(\omega)\rightarrow 1-\omega_{\rm P}^2/\omega^2\), Eq.~(\ref{PP}) leads to the frequency of plasmon polariton 
 \begin{equation}
 \omega_{\rm PP}=\sqrt{\omega_{\rm P}^2+k^2c^2}.
 \label{PP_dispersion}
 \end{equation}
 The light is then not allowed to propagate below $\omega_P$ in the bulk because ${\bf k}$ thereby becomes imaginary. For an optical field of $\omega>\omega_P$, only the transverse plasmon couples to the optical field which leads to a hybridization.

\begin{figure}[ptbh]
	\begin{centering}
		\includegraphics[width=0.99\textwidth]{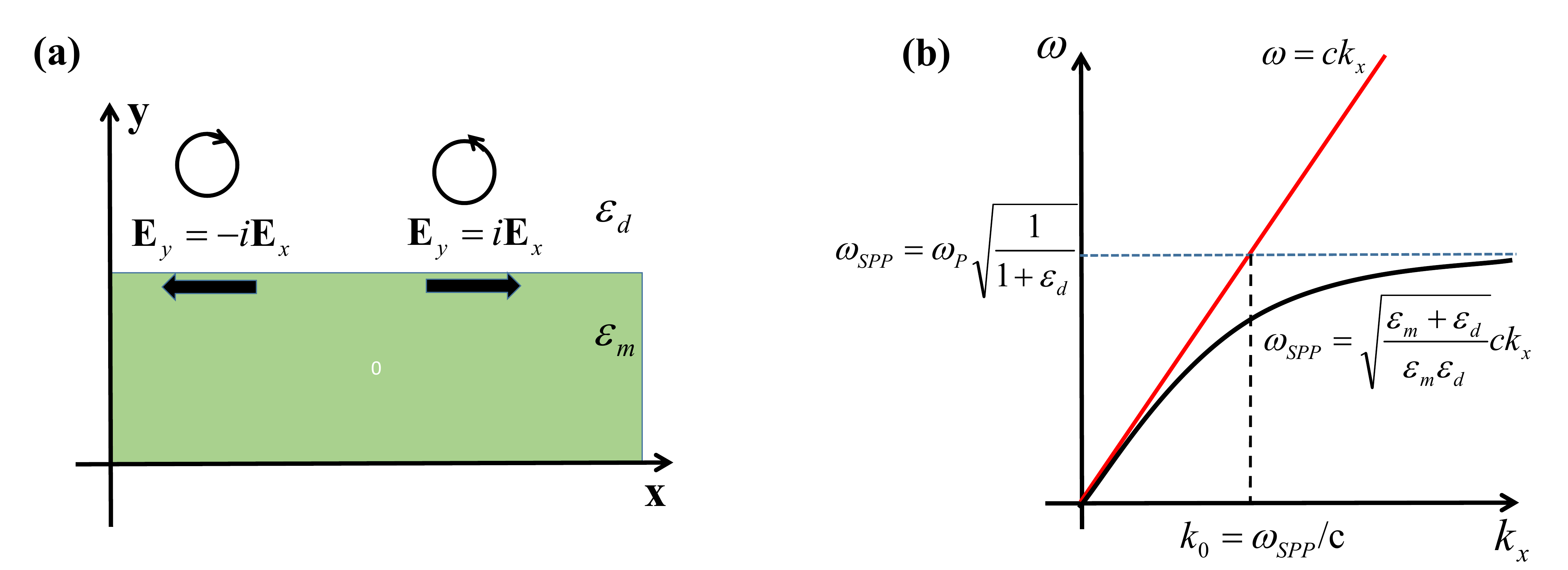}
		\par\end{centering}
	\caption{ Chirality and dispersion of surface plasmon polaritons. (a) illustrates the polarization-momentum locking of the electric field of surface plasmon polaritons: the electric field is right (left) circularly polarized for the right (left) moving excitations. (b) plots the dispersion of surface plasmon polaritons. $k_0=\omega_{\rm SPP}/c$ is indicated.}
	\label{SPP}
\end{figure}

\textbf{Surface plasmon polariton}.---Surface plasmon polariton emerges at the interface between a metal and a material with dielectric constant $\varepsilon_d$ (or surface to the vacuum with $\varepsilon_d= 1$). For
a flat interface between two semi-infinite spaces of a conductor ($y\le 0$) and a dielectric ($y>0$), as in Fig.~\ref{SPP}(a), the Maxwell’s equations in frequency space read
\begin{subequations}
 	\begin{align}
 		\nabla\times{\bf E}&=i\omega\mu_0{\bf H},\label{Maxwell_4}\\
 		\nabla\times{\bf H}&=-i\omega\varepsilon_0\varepsilon_r(y){\bf E},
 		\label{Maxwell_3}
 	\end{align}
 	\end{subequations}
 where $\varepsilon_r(y\le 0)=\varepsilon_m$ at the metal and $\varepsilon_r(y>0)=\varepsilon_d$ at the dielectric side. The solutions are  planes waves that we assume to propagate along the $\hat{\bf x}$-direction, \textit{i.e.}, $\pmb{\cal H}(y)e^{ik_xx}$ and $\pmb{\cal E}(y)e^{ik_xx}$. The TE mode with ${\bf E}=E_z\hat{\bf z}$ and ${\bf H}=H_x\hat{\bf x}+H_y\hat{\bf y}$ does not solve Eqs.~(\ref{Maxwell_3}) and (\ref{Maxwell_4})  \cite{SSP_review}, so the SSP must be a TM mode with ${\bf E}=E_x\hat{\bf x}+E_y\hat{\bf y}$ and ${\bf H}=H_z\hat{\bf z}$ with wave equation 
  \begin{align}
  	\frac{\partial^2H_z}{\partial y^2}+(\varepsilon_rk_0^2-k_x^2)H_z=0,
\end{align}
where $k_0^2=\omega^2/c^2$. Since the in-plane magnetic field is continuous across the interface, 
\begin{align}
	H_z(x,y)=A e^{ik_xx}e^{-\sqrt{k_x^2-\varepsilon_rk_0^2}y}.
	\label{magnetic_solutions}
\end{align}
The electric field components
\begin{subequations}
\begin{align}
	E_x&=\frac{i}{\omega\varepsilon_0\varepsilon_r}\frac{\partial H_z}{\partial y}=-\frac{i}{\omega\varepsilon_0\varepsilon_r}\sqrt{k_x^2-\varepsilon_rk_0^2}H_z,\\
	E_y&=-\frac{i}{\omega\varepsilon_0\varepsilon_r}\frac{\partial H_z}{\partial x}=\frac{k_x}{\omega\varepsilon_0\varepsilon_r}H_z,
\end{align}
\end{subequations}
are related as $E_x=-iE_y\sqrt{k_x^2-\varepsilon_rk_0^2}/k_x$. When $k_x\gg k_0$, $E_y=i {\rm sgn}(k_x)E_x$, \textit{i.e.}, a polarization-momentum locking of the electric field as illustrated in Fig.~\ref{SPP}(a).

We can derive the dispersion relation \(\omega_{\rm SPP}(k_x)\) from the boundary conditions, \textit{i.e.}, the continuity of the normal components of ${\bf D}$ (and ${\bf B}$) and that of the tangential components of ${\bf E}$ (and ${\bf H}$). From $E_x^{(m)}=E_x^{(d)}$,
\begin{align}
	\omega_{\rm SPP}=\sqrt{\frac{\varepsilon_d+\varepsilon_m(\omega_{\rm SPP})}{\varepsilon_d\varepsilon_m(\omega_{\rm SPP})}}c|k_x|.
	\label{dispersion_SPP}
\end{align}
Equation~(\ref{magnetic_solutions}) is a surface mode if $\varepsilon_rk_0^2<k_x^2$. Therefore  $\varepsilon_m(\omega_{\rm SPP})<0$ and $\omega_{\rm SPP}<\omega_{\rm P}$. Moreover, \(\omega_{\rm SPP}\) must be real and positive, which requires $\varepsilon(\omega_{\rm SPP})+\varepsilon_d<0$ and $\omega_{\rm SPP}<\omega_{\rm P}/\sqrt{1+\varepsilon_d}$. These features are summarized in Fig.~\ref{SPP}(b). The upper branch, if we wonder, is just the dispersion of plasmon polariton Eq.~(\ref{PP_dispersion}), which is bulk in nature. 

SPPs allow  sub-wavelength confinement of optical fields. Plasmonic structures act as antennae of electromagnetic radiation that strongly enhance the amplitude of incident signals. As reviewed in Sec.~\ref{SPP_spin_pumping}, SPPs can generate electronic spin currents in normal metals \cite{plasmonics_spin_NC,plasmonics_spin_APL,plasmonics_spin_PRB,plasmonics_spin_NJP,plasmonics_spin_PRL}.

 \subsubsection{Stray fields by dynamic electric dipoles}
\label{Stray_field_electric_dipole}
	
Analogous to the magnetic dipoles discussed in the previous Sec.~\ref{dipolar_fields_1}, electric dipoles emit stray electric fields that become chiral when evanescent. They are important in the  near-field regime of nano-optics \cite{nano_optics}. Here we discuss the waves emitted by  electric dipoles  that oscillate or rotate in a plane normal to an interface, as  formulated  by Fortu\~no {\em et. al.} \cite{plasmonics_1}.

According to non-relativistic radiation theory \cite{Jackson}, Eq.~(\ref{vector_potential}), an electric free or bounded current \textbf{J}  generates a vector potential ${\bf A}({\bf r},t)=({\mu_0}/4\pi)\int d{\bf r}'{\bf J}({\bf r}',t){e^{ik|{\bf r}-{\bf r}'|}}/{|{\bf r}-{\bf r}'|}$. A uniform line of dipoles ${\bf p}=p_x\hat{\bf x}+p_z\hat{\bf z}$ along the $\hat{\bf y}$-direction that oscillates with frequency \(\omega\) generates a displacement current 
\begin{equation}
	{\bf J}(x',z')=-i\omega\delta(x',z'){\bf p}.
\end{equation}
Its vector potential 
\begin{align}
	{\bf A}({\bf r},t)=\frac{\mu_0}{4\pi}\int dy'\frac{-i\omega{\bf p}}{\sqrt{x^2+(y-y')^2+z^2}}e^{ik\sqrt{x^2+(y-y')^2+z^2}}
\end{align}
does not depend on \(y\) as can be made explicit by using the Weyl identity (\ref{Weyl_identity}), 
\begin{equation}
	{\bf A}(x,z)=\frac{\mu_0\omega}{4\pi}{\bf p}\int dk_x\frac{e^{ik_xx+ik_z|z|}}{k_z}.
\end{equation}
The vector potential is parallel to ${\bf p}$,
and the magnetic field ${\bf H}({\bf r})=({1}/{\mu_0})\nabla\times {\bf A}({\bf r})=(0,H_y,0)$
becomes for $z<0$
\begin{equation}
	H_y(x,z)=\int {dk_x}e^{ik_xx}H_y(k_x,z),
\end{equation} 
where $k_z=\sqrt{(\omega/c)^2-k_x^2}$ and
\begin{equation}
	H_y(k_x,z)=-\frac{i\omega}{4\pi}\Big(p_x+\frac{k_x}{k_z}p_z\Big)e^{-ik_zz}.
\end{equation}
The magnetic field generated by an in-plane dipole is always linearly polarized. When $k_x\gg \omega/c\equiv k_0$, it becomes evanescent with
\begin{equation}
	H_y(k_x,z)=-\frac{i\omega}{4\pi}\Big(p_x-i{\rm sgn}({k_x})p_z\Big)e^{-|k_x||z|}.
\end{equation}
When the dipole \textit{rotates} with $p_z=ip_x$, this Fourier component
\begin{equation}
	H_y(k_x,z)=-\frac{i\omega}{4\pi}p_x\Big(1+{\rm sgn}({k_x})\Big)
	e^{-|k_x||z|}
\end{equation} 
is unidirectional, \textit{viz}., it vanishes when $k_x<0$ and $H_y(k_x,z)e^{ik_xx}$ is a linear polarized traveling wave with a chirality governed by the rotation direction of the dipole.

According to Eq.~(\ref{Maxwell_3}), the associated electric field in the evanescent limit ($z<0$)
\begin{subequations}
\begin{align}
	&E_x(k_x,z)=-\frac{1}{\varepsilon_0\varepsilon_r}\frac{1}{4\pi}\left(p_x-i{\rm sgn}(k_x)p_z\right)|k_x|e^{-|k_x||z|},\\
	&E_z(k_x,z)=\frac{i}{\varepsilon_0\varepsilon_r}\frac{1}{4\pi}\left(p_x-i{\rm sgn}(k_x)p_z\right)k_xe^{-|k_x||z|},
\end{align}
\end{subequations}
\textit{i.e.}, $E_z(k_x,z)=i{\rm sgn}(k_x)E_x(k_x,z)$. A linearly polarized electric dipole generates a circularly polarized electric field wave that propagates in both directions, with polarization (spin) locked to the momentum. On the other hand, a circularly polarized electric dipole emits a (circularly polarized) electric field in only one direction that is locked to the  rotation direction of the dipole, i.e., being unidirectional.

\begin{figure}[ptbh]
\begin{centering}
\includegraphics[width=0.98\textwidth]{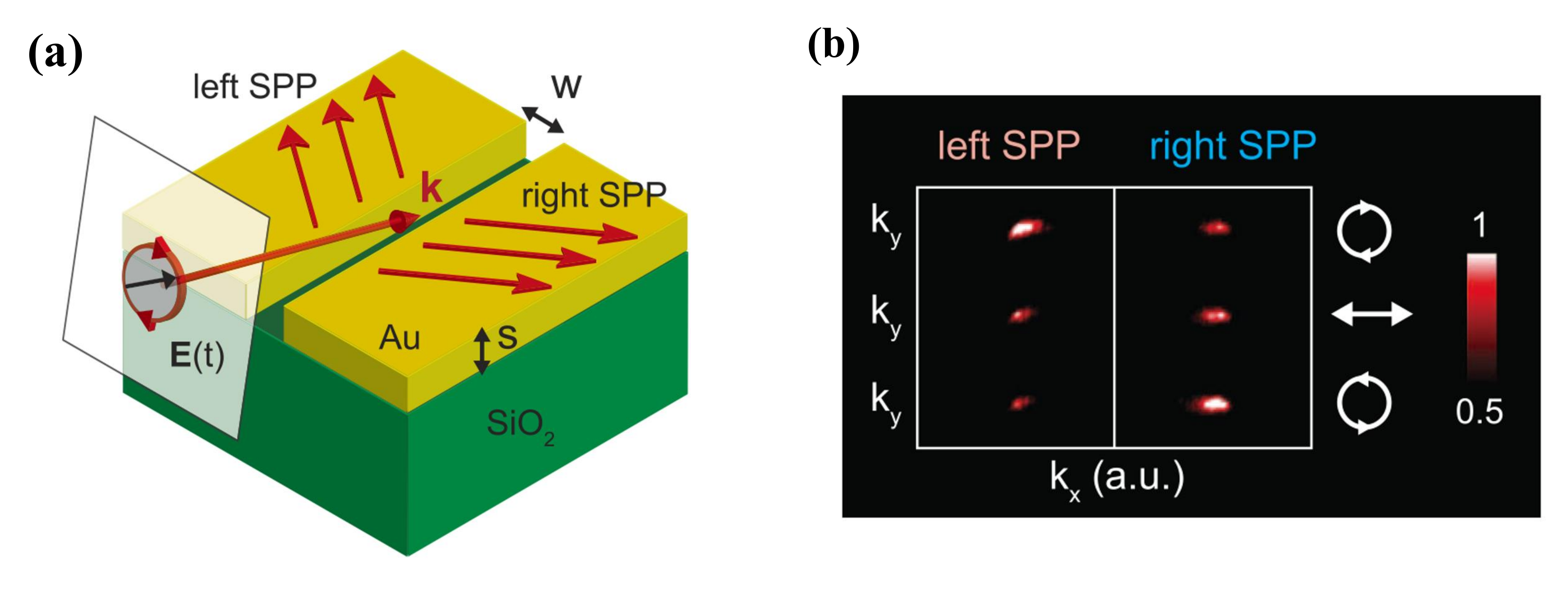}
	\par\end{centering}
	\caption{Directional excitation of surface plasmon polaritons by local electric-dipole ``wire". (a) is the schematics of the experiment, where a plane wave is an incident at a nearly grazing angle ($70^{\degree}$) onto a 130 nm slit in a thin gold film. The excitation of the surface plasmon polariton is recorded by the leakage radiation from either side of the slit.  The results in Fourier space in (b) show the dependence of the excitation direction on the light polarization. The figures are taken from Ref.~\cite{plasmonics_1}.}
	\label{electric_dipole}
\end{figure}

From the above discussion and Sec.~{\ref{Sec_SPP}} we understand that the dynamics of an electric dipole close to a metal surface may chirally excite surface plasmon polaritons, as reported by  Rodr\'iguez-Fortu\~no \textit{et al.}   \cite{plasmonics_1}. As illustrated in Fig.~\ref{electric_dipole}(a), a line of in-plane dipoles can be created by illuminating a narrow slit (130 nm) in a thin gold film by circularly or linearly polarized light. At grazing incidence, the circular polarization of the light's electric field is approximately perpendicular to the metal interface. The excited SPPs are recorded by their leakage radiation on either side of the slit. As shown in Fig.~\ref{electric_dipole}(b),  the plasmons escape symmetrically to both sides for a linearly polarized dipole excitation, but the excitation becomes unidirectional with increasing polarization. 

\subsection{Chiral phonons}

\label{Sec_chiral_phonon}

The excitations of elastic media such as ultrasound obey wave equations that are similar, but also characteristically different from those of  electromagnetic radiation or the magnetic order. It should come as no surprise by now that also sound and lattice vibrations display chiral properties when evanescent. Continuum mechanics is appropriate to describe pressure and shear waves at long wavelengths, while at short wavelengths the atomic dynamics should be taken into account. The quanta of mechanical excitations are phonons. Their eigenmodes are found by solving the lattice dynamical problem, \textit{i.e.}, the classical equation of motion of the atomic displacement field. We consider a lattice periodic system with $r$ atoms per unit cell located at ${\bf R}_l+{\bf d}_s$, where ${\bf R}_l$ is the origin of the $l$-th unit cell and ${\bf d}_s$ is the position of the $s$-th atoms in this unit cell. Their displacement into the $\alpha$-direction ${u}_l^{\alpha}(s,t)$ so obeys the (linearized) equation of motion \cite{Callaway}
	\begin{align}
		M_a u_l^{\alpha}(s)=-\sum_{l's'\beta}\Phi_{\alpha\beta}
		\left(
		\begin{matrix}
			l&l'\\
			s&s'
		\end{matrix}\right) u_{l'}^{\beta}(s'),
		\label{EOM_atoms}
	\end{align}
 where $M_a$ is the atomic mass and the $3r\times 3r$ matrix $\Phi_{\alpha\beta}
\left(
\begin{matrix}
	l&l'\\
	s&s'
\end{matrix}\right)$ contains the force constants between different atoms. With an ansatz of a plane wave solution with wave vector \textbf{k}, the $3r$ degrees of freedom in the unit cell lead to the same number of eigenvalues or Eq.~(\ref{EOM_atoms}) that we label as $\omega_{j=\{1,2,\cdots,3r\}}({\bf k})$. These form $3r$ branches of phonons in the reciprocal space, including 3 acoustic and $(3r-3)$ optical phonon bands, and the $3r$ eigenvectors ${\pmb \xi}^{(j)}_{s}({\bf k})$ indicate the phonon polarization. In second quantization, the bosonic phonon operators $\hat{b}_{\bf k}$ and $\hat{b}_{\bf k}^{\dagger}$ govern the displacement field for atom $s$ in unit cell \(l\)
\begin{align}
	\hat{\bf u}_l(s)=\frac{1}{\sqrt{NM_a}}\sum_{{\bf k}}\sum_{j=1}^{3r}\sqrt{\frac{\hbar}{2\omega_{j}({\bf k})}}\left(\hat{b}_{{\bf k}j}+\hat{b}_{-{\bf k}j}^{\dagger}\right)\pmb{\xi}_{s}^{(j)}({\bf k})e^{i{\bf k}\cdot{\bf R}_l},
	\label{displacement_s}
\end{align}
where $N$ is the number of unit cells. The phonon Hamiltonian is a distributed set of quantum harmonic oscillators $\hat{H}=\sum_{{\bf k}}\sum_j\hbar\omega_{j}({\bf k})(\hat{b}^{\dagger}_{{\bf k}j}\hat{b}_{{\bf k}j}+1/2)$.

Phonons have been called chiral \cite{Lifa_chiral_phonon} when i) their polarization is circular, \textit{i.e.}, the displacement field rotates around the equilibrium position, and ii) the direction of the polarization rotation (spin) is locked to the linear momentum under the time-reversal symmetry. We focus here on two types of chiral phonons, \textit{i.e.}, bulk phonons (Sec.~\ref{chiral_bulk_phonon}) and surface acoustic waves (Sec.~\ref{surface_acoustic_waves}).

\subsubsection{Chiral bulk phonons}
\label{chiral_bulk_phonon}

A chirality of phonon modes in a bulk crystal affects the electron scattering \cite{chiral_phonon_electron} and the interaction with light \cite{chiral_phonon_light}. In the reduced zone scheme the phonon crystal momenta lie in the first Brillouin zone and the phonon eigenmodes of the $j$-th band are the Bloch waves
\begin{align}
{\pmb \psi}^{(j)}_{l}({\bf k})=\frac{1}{\sqrt{NM_a}}
\sqrt{\frac{\hbar}{2\omega_{j}({\bf k})}}
\pmb{\xi}^{(j)}({\bf k})e^{i{\bf k}\cdot{\bf R}_l}.
\end{align}
These basis functions transform according to the point group symmetry of the lattice spanned by a number of symmetry operations $\hat{O}_g$ such that 
$\hat{O}_g{\pmb \psi}^{(i)}_l({\bf k})=R_{ji}(g){\pmb \psi}_l^{(j)}(g^{-1}{\bf k})$ in terms of the irreducible representation matrix $R_{ji}(g)$. At high-symmetry points of the Brillouin zone,  eigenmodes tend to have degenerate eigenvalues and higher-dimensional irreducible representations of the point group. The phonon polarization is then not well-defined as an arbitrary superposition of the polarization of the degenerate states. A definite polarization of the phonons at symmetry points can be generated by breaking a symmetry.  Zhang \textit{et al}. showed that breaking the inversion symmetry generates a ``pseudoangular" momentum of phonons at special points in the Brillouin zone of  a two-dimensional honeycomb lattice such as monolayer transition-metal dichalcogenide \cite{Lifa_chiral_phonon}. The ``pseudo-angular" momentum should not be confused with the ``phonon spin", \textit{i.e.}, the rotational motion of a single atom  that we discuss below.  The ``pseudo-angular" momentum is defined by the transformation law of the displacement field under the rotation operation, but not that defined via Noether's theorem below. Throughout this review, we stick to the latter description as addressed in Sec.~\ref{phonon_angular_momentum} for the phonon spins and orbital angular momentum.

\begin{figure}[ptbh]
	\begin{centering}
		\includegraphics[width=1.0\textwidth]{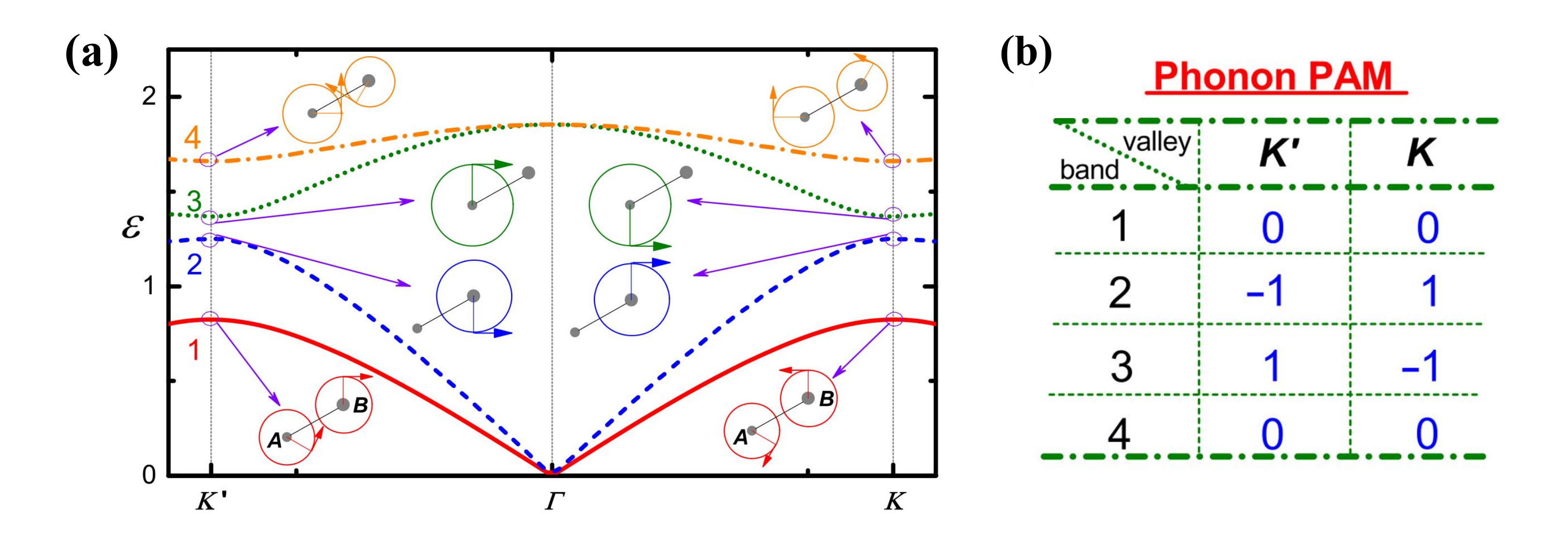}
		\par\end{centering}
	\caption{Dispersion relation [(a)] of phonon modes with in-plane dynamics and their pseudoangular momentum [(b)] at the $K$ and $K'$ valleys of a two-dimensional honeycomb lattice of monolayer transition-metal dichalcogenides. The insets in (a) show the motions of the $A$ and $B$ atoms in every band at $K$ and $K'$ valleys. These figures are taken from Ref.~\cite{Lifa_chiral_phonon}.}
	\label{chiral_phonon}
\end{figure}

We illustrate these abstract notions for  a monolayer of a hexagonal lattice \cite{Lifa_chiral_phonon}. With $r=2$ atoms, labeled by $A$ and $B$ in the unit cell, we have $3r=6$ phonon modes of every momentum in total. Limiting attention to the in-plane motion, 2 acoustic and 2 optical phonon branches remain. At the Brillouin-zone corners points ${\bf K}$ and ${\bf K}'$, the phonon amplitudes transform according to the $C_{3h}$ symmetry group that contains six elements including, in particular, the threefold rotation $C_3$, as shown in Table~\ref{MX2_symmetry} together with their basis functions. All six representations $\{A_1,A_2,A_3,A_4,A_5,A_6\}$ are one-dimensional and non-degenerate. The basis of the representations $A_3$ and $A_4$ are right-handed and left-handed, respectively, thus corresponding to circular polarization and pseudo-angular momentum. They are not degenerate because of the lack of inversion symmetry.  The threefold rotation of the basis at ${\bf K}$,
 $\hat{O}_{C_3}{\pmb \psi}_l^{(j)}({\bf K})={\pmb \psi}_l^{(j)}(C_3^{-1}{\bf K})=e^{-2\pi l_s^{(j)} i/3}{\pmb \psi}_l^{(j)}({\bf K})$ therefore implies  a pseudoangular momentum $l_s^{(j)}$of the band with index  $j$. For the representations $A_3$ and $A_4$,  $l_s^{(A_3)}=1$ and $l_s^{(A_4)}=-1$, while $l_s^{(j)}=0$ for all others. These pseudoangular momenta are governed by the crystal symmetry and generate selection rules in the matrix elements of electron-phonon scattering  \cite{chiral_phonon_electron} and photon-phonon \cite{chiral_phonon_light} scattering.  Figure~\ref{chiral_phonon}(a) shows the dispersion of the four phonon bands of the in-plane motion along the $K'$-$\Gamma$-$K$ lines \cite{Lifa_chiral_phonon}. The motion of the $A$ and $B$ atoms in each band at $K$ and $K'$ valleys are illustrated in the insets. Here bands 2 and 3 at the ${\bf K}$-valley has pseudoangular momentum $+1$ and $-1$, respectively, with opposite sign at the ${\bf K}'$-valley, as shown in Fig.~\ref{chiral_phonon}(b). There are more complicated cases in three dimensions and other symmetry groups \cite{phonon_chirality_PRR}.

\begin{table}[hp]
\caption{Character table for $C_{3h}$ group \cite{32_point_groups}. $\omega=e^{2\pi i/3}$.}
\begin{center}
\begin{tabular}{l|l|l|l|l|l|l|l}
  \hline 
    & $e$  & $C_3$   & $C^2_3$ & $\sigma_h$ & $S_3$ & $\sigma_h C^2_3$ & $~~~~~~~~~$ basis function  \\ 
  \hline
  $A_1$    &1  &1    &1 &1 &1 &1 &$x^2+y^2$;$\sigma_z$;$(x+iy)^3$;$(x-iy)^3$\\
  \hline
  $A_2$    &1 &1 &1  &-1 &-1 &-1 &\\
  \hline
  $A_3$    &1   &$\omega$ &$\omega^*$  &1   &$\omega$ &$\omega^*$  &$x+iy$;$(x-iy)^2$ \\
  \hline
  $A_4$  &1 &$\omega^*$ &$\omega$ &1 &$\omega^*$ &$\omega$
  &$x-iy$;$(x+iy)^2$  \\ 
\hline
$A_5$    &1   &$\omega$ &$\omega^*$  &-1   &-$\omega$ &-$\omega^*$  &$\sigma_x+i\sigma_y$\\
  \hline
$A_6$  &1 &$\omega^*$ &$\omega$ &-1 &-$\omega^*$ &-$\omega$  &$\sigma_x-i\sigma_y$  \\ 
\hline
\end{tabular}
\end{center}
\label{MX2_symmetry}
\end{table}

\subsubsection{Surface acoustic waves}
\label{surface_acoustic_waves}

The phonon modes localized at surfaces and interfaces have a lot in common with the surface plasmons discussed above (Sec.~\ref{Sec_SPP}), but with idiosyncrasies that induce subtle differences as well. 
Surface acoustic waves (SAWs) are evanescent normal to the plane and propagating in the plane, with  chirality as discussed below.

\begin{figure}[ptbh]
	\begin{centering}
	\includegraphics[width=0.72\textwidth]{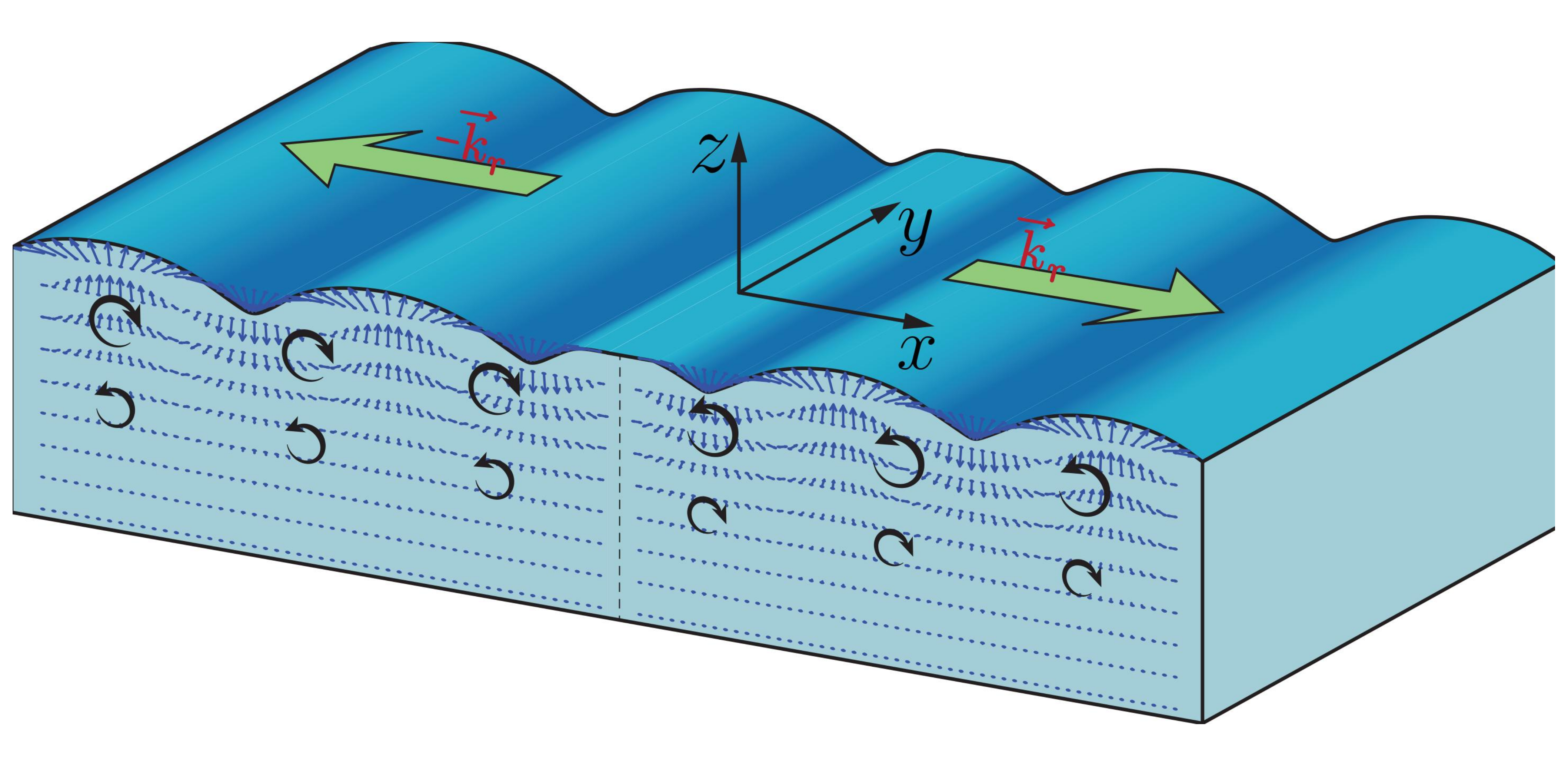}
	\par\end{centering}
	\caption{Rayleigh SAWs with opposite momentum propagate and rotate in opposite directions, as indicated by the green and black arrows, respectively.}
	\label{SAW}
\end{figure}

The deformation vector field \textbf{u}(\textbf{r}) in a homogeneous and isotropic elastic continuum obeys the equation of motion 
\cite{Kino1987,Viktorov1967}
\begin{equation}
	\rho\frac{\partial^{2}u_{i}}{\partial t^{2}}=\sum_{j}\frac{\partial\sigma
		_{ij}}{\partial x_{j}},
	\label{EOM_phonon}
\end{equation}
where \(\rho \)  is the mass density and $\sigma_{ij}$ the stress tensor. The boundary at the surface or interface $S$ to air or vacuum is free, \textit{i.e.},
\begin{align}
	\sigma_{ij}|_{S}=0.
\end{align}
For small deformations Hooke's Law holds:
\begin{align}
\sigma_{ij}=\sum_{kl}C_{ijkl}\epsilon_{kl},
\end{align}
with stiffness coefficient $C_{ijkl}$ and strain tensor $\epsilon
_{ij}=({\partial u_{i}}/{\partial x_{j}}+{\partial u_{j}}/{\partial x_{i}})/2$
\cite{Kino1987}. For the surface acoustic waves propagating along $\hat{\textbf{x}}$ in Fig.~\ref{SAW}, we have to consider the elements $u_x(x,z)$ and $u_z(x,z)$ that do not depend on $y$.

The non-zero strain tensor elements are $\epsilon_{xx}$, $\epsilon_{zz}$ and $\epsilon_{xz}$, and Hooke's Law can be simplified to
\begin{align}
	\left(\begin{matrix}
		\sigma_{xx}\\
		\sigma_{zz}\\
		\sigma_{xz}
	\end{matrix}\right)=
\left(\begin{matrix}
	\lambda+2\mu&\lambda&0\\
	\lambda&\lambda+2\mu&0\\
	0&0&2\mu
\end{matrix}\right)\left(\begin{matrix}
\epsilon_{xx}\\
\epsilon_{zz}\\
\epsilon_{xz}
\end{matrix}\right),
\end{align} 
where $\mu$ and $\lambda$ are the elastic Lam\'{e} force constants,
The displacement field can be expressed in terms of the elastic scalar and vector potentials $V$ and $\mathbf{A}$ as $\boldsymbol{u}=\nabla
V+\nabla\times\mathbf{A}$, with $\mathbf{A}=A\hat{\textbf{y}}$ because the displacement
field of an acoustic wave propagating in the $\hat{\textbf{x}}$ direction does not
depend on $y$. With $u_{x}=\partial V/\partial x-\partial A/\partial z$ and
$u_{z}=\partial V/\partial z+\partial A/\partial x$ we arrive at the wave equations
\begin{subequations}
\begin{align}
&\rho\frac{\partial^{2}V}{\partial t^{2}}=(\lambda+2\mu)\left(\frac{\partial^{2}V}{\partial x^{2}}+\frac{\partial^{2}V}{\partial z^{2}}\right),\\
&\rho\frac{\partial^{2}A}{\partial t^{2}}=\mu\left(\frac{\partial^{2}A}{\partial x^{2}}+\frac{\partial^{2}A}{\partial z^{2}}\right).
\label{elastic_EOM}
\end{align}
\end{subequations}
The evanescent plane wave ansatz 
\begin{subequations}
	\begin{align}
		V({\bf r},t)&=V_0e^{-q|z|}e^{i(kx-\omega_k t)},\\
		A({\bf r},t)&=A_0e^{-s|z|}e^{i(kx-\omega_k t)},
	\end{align}
\end{subequations}
solves the wave equations of a surface at  $z=0$, where $\{V_0,A_0\}$ are the wave amplitudes, $q\equiv \sqrt{k^2-k_l^2}$, $s=\sqrt{k^2-k_t^2}$, and $k_{l}=\omega_{k}\sqrt{\rho/(\lambda+2\mu)}$ and $k_{t}=\omega_{k}\sqrt
{\rho/\mu}$ are the wave vectors of the longitudinal and transverse bulk waves,
respectively. The surface acoustic waves exist only when $k>\{k_l,k_t\}$. With boundary condition of vanishing stress along the surface normal $\hat{\textbf{z}}$, \textit{i.e.}, $\sigma_{zz}|_{z=0}=0$ and $\sigma_{xz}|_{z=0}=0$,
\begin{align}
	\left(\begin{matrix}
		2\mu k^2-(\lambda+2\mu)k_l^2&-2i\mu k\sqrt{k^2-k_t^2}\\
		2ik\sqrt{k^2-k_l^2}&-k_t^2+2k^2
	\end{matrix}\right)
\left(\begin{matrix}
	V_0\\A_0
\end{matrix}\right)=0,
\end{align}
and hence,
\begin{align}
	4k^2qs-(k^2+s^2)^2=0.
	\label{roots}
\end{align}
With $\eta=k_l/k<1$ and $\xi=k_l/k_t=\sqrt{\mu/(\lambda+2\mu)}$, Eq.~(\ref{roots}) leads to the characteristic equation
\cite{Viktorov1967} 
\begin{align}
\eta^6-8\eta^4+8(3-2\xi^2)\eta^2+16(\xi^2-1)=0.
\end{align}
Its roots are the eigenfrequencies
\begin{align}
	\omega_{k}= \eta\sqrt{\frac{\mu}{\rho}}\left\vert k\right\vert=c_{r}|k|,
\end{align} 
 with sound velocity \(c_r\) that depends only on the Lam\'{e} constants. The amplitudes of these ``Rayleigh" surface acoustic waves are entirely transverse 
\begin{equation}%
	\begin{split}
		\psi_{x}  &  =ik\varphi_{k}\left(  e^{qz}-\frac{2qs}{k^{2}+s^{2}}%
		e^{sz}\right)  e^{ikx},\\
		\psi_{z}  &  =q\varphi_{k}\left(  e^{qz}-\frac{2k^{2}}{k^{2}+s^{2}}%
		e^{sz}\right)  e^{ikx},
	\end{split}
	\label{eqn:SAW_profile}%
\end{equation}
where $\varphi_{k}$ is a normalization constant. The relative
phase of the displacement field ${\mathrm{Arg}}(\psi_{z}/\psi_{x})|_{z=0}=\pm
i$ is opposite for right- and left-propagating waves, which reflects the rotation-momentum locking \cite{Viktorov1967}, as illustrated in Fig.~\ref{SAW}. With a proper definition of phonon spin density (Sec.~\ref{phonon_angular_momentum}) below, this implies the spin-momentum locking of SAWs, which together with the surface normal is well characterized by the chirality index defined in Eq.~(\ref{chirality_index}).

The quantized displacement field $(\hat{u}_{x},\,\hat{u}_{z})$ can be expanded
in terms of these eigenmodes $\boldsymbol{\psi}(k)$ and bosonic phonon operators $\hat{b}%
_{k}(t)$ \cite{PRX}
\begin{equation}
	\hat{\mathbf{u}}(x,z,t)=\sum_{k}\left[  \boldsymbol{\psi}(x,z,k)\hat{b}%
	_{k}(t)+\boldsymbol{\psi}^{\ast}(x,z,k)\hat{b}_{k}^{\dagger}(t)\right]  .
	\label{eqn:phonon_operator}%
\end{equation}
We normalize the mode amplitudes $\boldsymbol{\psi}$ 
\begin{equation}
	\int_{-\infty}^{0}dz\left(  |\psi_{x}|^{2}+|\psi_{z}|^{2}\right)  =\frac
	{\hbar}{2\rho L\omega_{k}},
\end{equation}
to recover the quantum harmonic oscillator form of the
Hamiltonian for Rayleigh SAWs
\begin{equation}
	\hat{H}_{\mathrm{e}}=\rho\int d\boldsymbol{r}\,\dot{\hat{\boldsymbol{u}}}%
	^{2}(x,z,t)=\sum_{k}\hbar\omega_{k}\hat{b}_{k}^{\dagger}\hat{b}_{k}.
	\label{eqn:Hamiltonian_E}%
\end{equation}
The normalization factor
\begin{equation}
	\varphi_{k}=\frac{1}{2|k|}
	\sqrt{\frac{2\hbar}{\rho
			Lc_{r}}}\left(  \frac{1+a^{2}}{2a}%
		+\frac{2a(a-2b)}{b(1+b^{2})}\right)^{-1/2},
		\label{eqn:SAW_normalization}%
\end{equation}
with dimensionless material constants $a=q/\left\vert k\right\vert =\sqrt{1-(c_{r}/c_{l})^{2}}$ and $b=s/\left\vert k\right\vert =\sqrt{1-\eta^{2}}$.
Here $c_{l}=\sqrt{(\lambda+2\mu)/\rho}$ is
the sound velocity of the longitudinal bulk waves.

\subsubsection{Phonon orbital and spin angular momentum}

\label{phonon_angular_momentum}

The angular momentum of a photon and its separation into a photon spin and orbital angular momentum is well established \cite{Allen,Jackson}.
In spite of the similarities of the respective wave equations, there are differences that show up when the phonon wavelength approaches either the lattice constant or the size of the dielectric particle under consideration.  
The concept of the phonon spin can be helpful in understanding microscopic processes such as the spin-transfer processes between magnons, electrons, phonons, and rigid mechanical rotation. The chirality implies the unidirectional spin current of phonons (Sec.~\ref{Sec_chiral_phonon_pumping}). 

\begin{figure}[ptbh]
	\begin{centering}
		\includegraphics[width=0.99\textwidth]{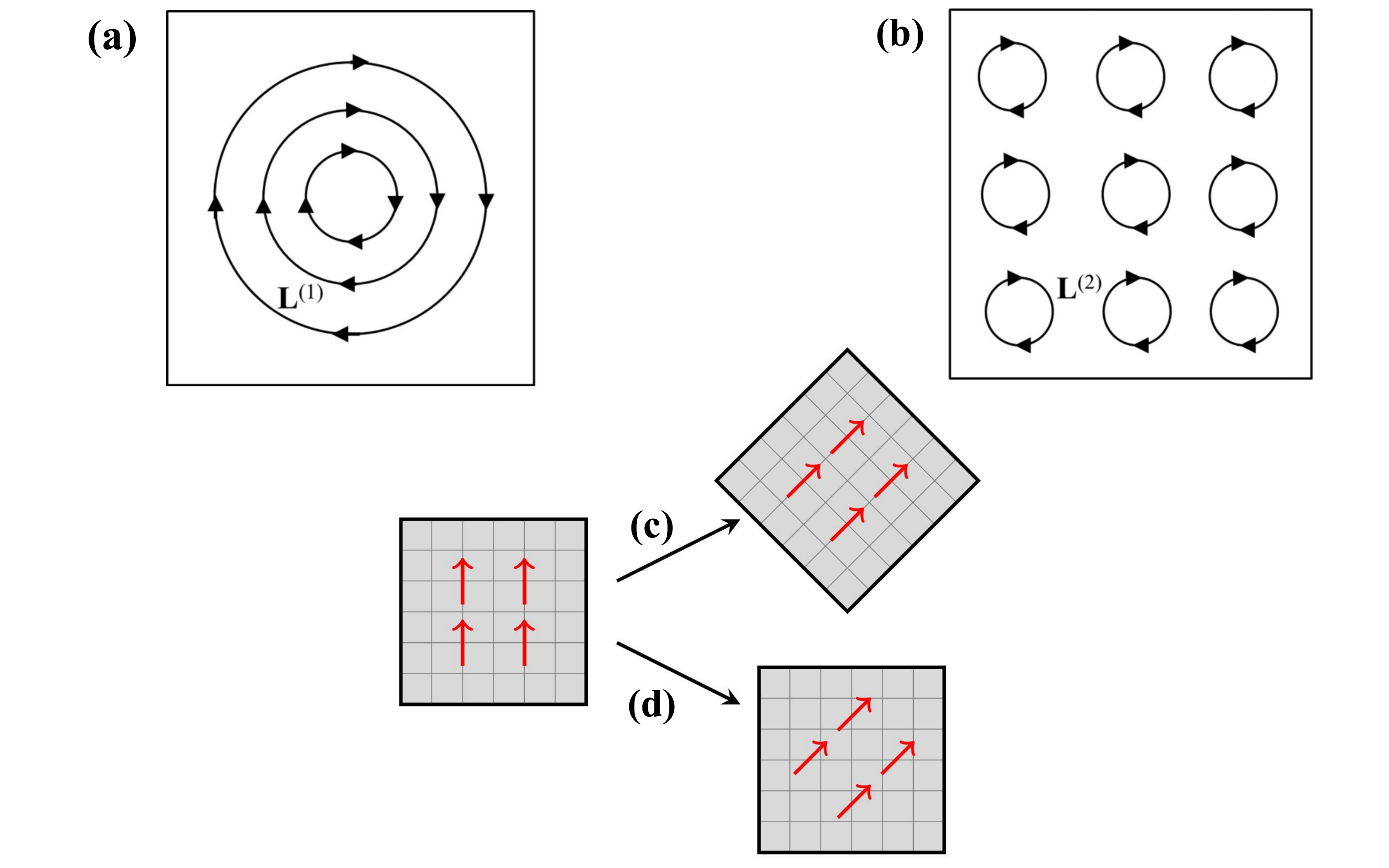}
		\par\end{centering}
	\caption{Orbital and spin angular momentum of phonons. (a) and (b) distinguish the mechanical angular momentum of the rotation of a finite particle that forms the atomic angular momentum in which the local deformation field rotates around the equilibrium position. Panels (c) and (d) illustrate the ambiguity of the phonon angular momentum for large bulk systems. Here the square lattice sketches a homogeneous elastic continuum, while the red arrows indicate a deformation field. Process (c) illustrates the invariance with respect to a global rotation of the whole levitated particle, while the invariance under rotations of the vector field [(d)] should hold in the thermodynamic limit of the particle size. The figures are taken from Ref.~\cite{phonon_angular_momentum_2015} [(a) and (b)] and Ref.~\cite{Simon_Noether} [(c) and (d)].}
	\label{phonon_spin}
\end{figure}

Vonsovskii and Svirskii \cite{phonon_spin_1962} introduced the concept of phonon spin, while  Zhang and Niu \cite{Zhang_phonon_spin} brought the concept to recent attention. 
The total angular momentum of a homogeneous elastic body is defined by \cite{Zhang_phonon_spin,phonon_angular_momentum_2015}
\begin{align}
{\bf L}=\int d{\bf r}({\bf r}+{\bf u})\times {\bf p},
\end{align}
to leading order in a small displacement ${\bf u}({\bf r},t)$ at a position  ${\bf r}$ of the body at rest, and ${\bf p}({\bf r},t)=\rho\dot{\bf u}({\bf r},t)$ is the momentum density. \(\bf L\) is the sum of two contributions, as shown in Fig.~\ref{phonon_spin} (a) and (b) for the motion of media. 
\begin{align}
    {\bf L}^{(1)}=\int d{\bf r}\rho({\bf r}\times \dot{\bf u})
    \label{L1}
\end{align}
is the angular momentum due to the rotation of the elastic medium relative to the origin; while 
\begin{align}
    {\bf L}^{(2)}=\int d{\bf r}\rho({\bf u}\times \dot{\bf u})
    \label{L2}
\end{align}
is the phonon spin,  which is finite for circular shear displacements of  volume elements around their equilibrium positions. Accordingly, we can define $\boldsymbol{l}_{\mathrm{DC}}\left({\bf r}\right)  =\rho\left\langle
\mathbf{u}\times\mathbf{\dot{u}}\right\rangle _{t}$ as the time averaged mechanical angular momentum density. This phonon spin density is constant for the SAWs discussed above
\cite{Long2018,Holanda2018,Shi2019,phonon_Yu_1}. The spin of water (surface) waves is defined by the water velocity in a similar way \cite{water_spin}. The equation of motion for the displacement field [Eq.~(\ref{EOM_phonon})]  leads to the conservation law of total phonon angular momentum ${\dot{\bf L}}={\dot{\bf L}}^{(1)}+{\dot{\bf L}}^{(2)}=0$  only when including non-linear terms beyond the harmonic oscillator Hamiltonian~\cite{phonon_angular_momentum_2015}, which indicates the subtlety of the concept.

Using Noether's theorem, according to which the conservation of angular momentum  is a consequence of rotational symmetry, Nakane and Kohno derived two inequivalent expressions for the phonon angular momentum \cite{two_expressions}. The invariance of a (very large or spherical) body under a rotation operator $\cal{R}$, \textit{i.e.}, ${\bf u}({\bf r})+{\bf r}\rightarrow {\cal R}[{\bf u}({\bf r})+{\bf r}]$  as in Fig.~\ref{phonon_spin}(c), leads to a phonon angular momentum via Eqs.~(\ref{L1}) and (\ref{L2}). In Figs.~\ref{phonon_spin}(c) and (d), the square lattice indicates the elastic continuum, while the red arrow denotes a rotation of a pinned and free vector field. On the other hand, when the medium is very large,  the invariance of the phonon displacement field under rotation, \textit{i.e.}, ${\bf u}({\bf r})\rightarrow {\cal R}{\bf u}({\cal R}^{-1}{\bf r})$ as illustrated in Fig.~\ref{phonon_spin}(d) appears by the natural symmetry. The phonon ``orbital" and spin angular momentum then read
\begin{subequations}
\begin{align}
\tilde{{\bf L}}^{(1)}&=\sum_{\alpha=x,y,z}\int d{\bf r}\rho\dot{u}_{\alpha}{\bf r}\times (-\nabla)u_{\alpha},\label{L1_tilde}\\
\tilde{\bf L}^{(2)}&=\int d{\bf r}\rho({\bf u}\times \dot{\bf u}).
\label{L2_tilde}
\end{align}    
\end{subequations}

Streib  \cite{Simon_Noether} referred to the two definitions as the (real) angular momentum [Eqs.~(\ref{L1}) and (\ref{L2})] and the pseudo-angular momentum [Eqs.~(\ref{L1_tilde}) and (\ref{L2_tilde})] by the analogy with linear momentum in free space and pseudo-momentum in crystals. Both definitions lead to the same expression for the phonon spin [Eqs.~(\ref{L2}) and (\ref{L2_tilde})] but different angular momentum conservation laws. When considering an embedded spin such as a magnetic moment in an elastic medium \cite{two_expressions},  the spin transfer should be invariant to a combined rotation of spin directions and phonon fields as in Fig.~\ref{phonon_spin}(c), \textit{i.e.}, pseudo-angular momentum should be conserved. On the other hand, when taking into account the angular momentum transfer of the intrinsic spins of magnons, photons, phonons, and the global lattice rotation \cite{angular_momentum_global}, the conserved quantity is the real rather than the pseudo-angular momentum. The relaxation of magnetization dynamics in a small particle, therefore, leads to a rigid rotation of the entire particle \cite{angular_momentum_global}. Chaplain \textit{et al.} demonstrated that the flexural guided elastic waves in elastic pipes carry a well-defined orbital angular momentum that enables the transfer of elastic
orbital angular momentum to a surrounding fluid in contact with the pipe \cite{OAM_phonon_beams}.

\subsection{Universal features of chirality of evanescent waves}
\label{unification}

After reviewing various chiral waves in the condensed matter as summarized in the left panel of Table~\ref{table_chiral_waves}, we here address some unifying features in terms of fixed handedness of three vectors as defined in the introduction \ref{section1}. This unification is achieved in terms of the transverse spin ${\bf S}_T$ (its direction is denoted by $\hat{\pmb \sigma}$), the surface normal ${\bf n}$ of the propagation plane of interest, and the momentum ${\bf k}$ in this plane of evanescent waves, which defines the chirality index $Z=\hat{\bf n}\cdot(\hat{\pmb \sigma}\times\hat{\bf k})$, where the hat indicates unit vectors. We demonstrate the following ``theorem":
\begin{itemize}
    \item Chirality index $Z=\hat{\bf n}\cdot(\hat{\pmb \sigma}\times\hat{\bf k})$ is fixed for the evanescent waves;
    \item In the presence of time-reversal symmetry, their transverse spin is locked to the momentum with fixed ${\bf n}$, \textit{i.e.}, the spin-momentum locking, leading to an opposite spin of opposite propagation;
    \item The breaking of time-reversal symmetry leads to the non-reciprocal and even unidirectional propagation of waves with fixed ${\bf n}$, \textit{i.e.}, the spatial chirality. The wave propagation is reversed with opposite ${\bf n}$.
\end{itemize}

\subsubsection{The zoo of chiral evanescent waves}

Before showing the unifying features, we first address the diverse description of the chirality of these waves in the above sections (Sec.~\ref{chiral_spin_waves} to \ref{Sec_chiral_phonon}).
\begin{itemize}
    \item In Sec.~\ref{DE_review}, when the time-reversal symmetry is broken by the magnetization, the propagation direction ${\hat{\bf k}}$ of  Damon-Eshbach (DE) modes at the surface of a ferromagnet is determined by the cross product of the surface normal ${\bf n}$ and the magnetization ${\bf M}_s$ [Fig.~\ref{DE_figure}(a)]. The chirality of the DE modes in the magnetic films can be understood in terms of the self-interaction of the magnetic dipoles that interact with the magnetic charges generated by the spin waves themselves.
    \item In Sec.~\ref{Stripline}, the stripline magnetic field with the surface normal ${\bf n}$ that is perpendicular  to the plane of propagation of interest as in Fig.~\ref{stripline_field}, is polarized with a transverse spin ${\bf S}_T$ along the cross product of its momentum ${\bf k}$ and the surface normal. 
    \item In Sec.~\ref{Stray_field_electric_dipole}, the electric field emitted by the linearly polarized  electric dipoles along a surface normal  shares a similar chirality to that of the stripline magnetic field by being spin-momentum locking at a plane with the fixed surface normal ${\bf n}$.
    \item In Sec.~\ref{Stray_field_electric_dipole}, when the time-reversal symmetry is broken with circularly polarized dipoles, the electric field emitted by it is unidirectional, similar to that of DE modes in the magnetism.
    \item In Sec.~\ref{dipolar_fields_1}, the Kittel mode of a magnetic nanowire causes a rotating two-dimensional magnetic dipole with in general elliptical polarization dynamics. When being linearly polarized, the spin of its dipolar field shares the same chirality with the electric counterpart by being spin-momentum locking. 
    \item In Sec.~\ref{dipolar_fields_1}, while when being circularly polarized for the Kittel mode of a magnetic nanowire, the time-reversal symmetry is broken that renders the generated dipolar field to be unidirectional in propagation, similar to the electric stray field by the circularly-polarized electric-dipole wires.
    \item In Sec.~\ref{dipolar_fields_1}, in sufficiently thin films, spin waves are uniform across the film and therefore not chiral anymore, but their stray magnetic fields are. The dipolar field of these spin waves propagating normal to the magnetization in a direction ${\bf k}$ that is governed by the cross product of the surface normal ${\bf n}$ and the in-plane magnetization ${\bf M}_s$ [Fig.~\ref{spin_wave_field}(b)]; the spin of the dipolar field is also locked by the momentum and surface normal. 
    \item In Sec.~\ref{waveguide_fields}, for the chirality of a waveguide magnetic field on special lines or planes, we can also identify a surface normal by which locking effects between polarization and propagation rotation (spin) direction can be formulated in terms of vector products, as shown in Fig.~\ref{waveguide_waves}(a), with similar physics for the electric field in optical fibers.
    \item In Sec.~\ref{Sec_SPP}, the electric field of the surface plasmon polariton is circularly polarized, with polarization rotation direction determined by the chiral relation with the surface normal and the propagation direction [refer to Fig.~\ref{SPP}(a)].
    \item In Sec.~\ref{surface_acoustic_waves}, the surface acoustic waves are evanescent and hold chirality in that their polarization, momentum, and surface normal direction obey the handed rule (refer to Fig.~\ref{SAW}).
\end{itemize}

Blow we unify these features with a strict definition of the spin density of these waves (Sec.~\ref{spin_density_waves}) and highlight the role of evanescence for the chirality (Sec.~\ref{evanescence}), with the unifying features summarized in Tables~\ref{table_chiral_universality} and \ref{table_chiral_universality_continued}, which will be explained in detail in the following.

   \begin{table}[htbp]
	\caption{Universal origin of typical chiral waves exploited in Spintronics.} \label{table_chiral_universality}
	\centering
	\begin{tabular}{ccc}
		\toprule
		\hspace{-2.5cm}Chiral Waves & \hspace{-3.5cm} Wave Equations& \hspace{-0.1cm}Spin $vs$  Momentum $vs$ Surface\\
		\toprule
		\hspace{-2.5cm}\begin{minipage}[m]{.53\textwidth}
			\centering\vspace*{5pt}
			Damon-Eshbach spin waves
			\[
			\hspace{1.8cm}\boxed{\nabla \cdot {\bf M}\rightarrow 0~{\rm for}~{\rm large}~k}
			\]
			\vspace*{2pt}
		\end{minipage} &
		\hspace{-2.7cm}\begin{minipage}[m]{6.9cm}
			\begin{eqnarray}
				\nonumber
				&&\partial {\bf M}/\partial t=-\gamma\mu_0{\bf M}\times {\bf H}_{\rm eff}\\
				\nonumber
				&&H^{\beta}_{\rm eff}({\bf r})=H_z\hat{\bf z}\delta_{\beta z}+\frac{1}{4\pi}\partial_{\beta}\sum_{\alpha}\partial_{\alpha}\int \frac{M_{\alpha}({\bf r}')}{|{\bf r}-{\bf r}'|}d{\bf r}'
			\end{eqnarray}
		\end{minipage} &
		\hspace{-0.55cm}\begin{minipage}[m]{5.3cm}
			\centering\vspace*{5pt}
   		\includegraphics[width=5.35cm]{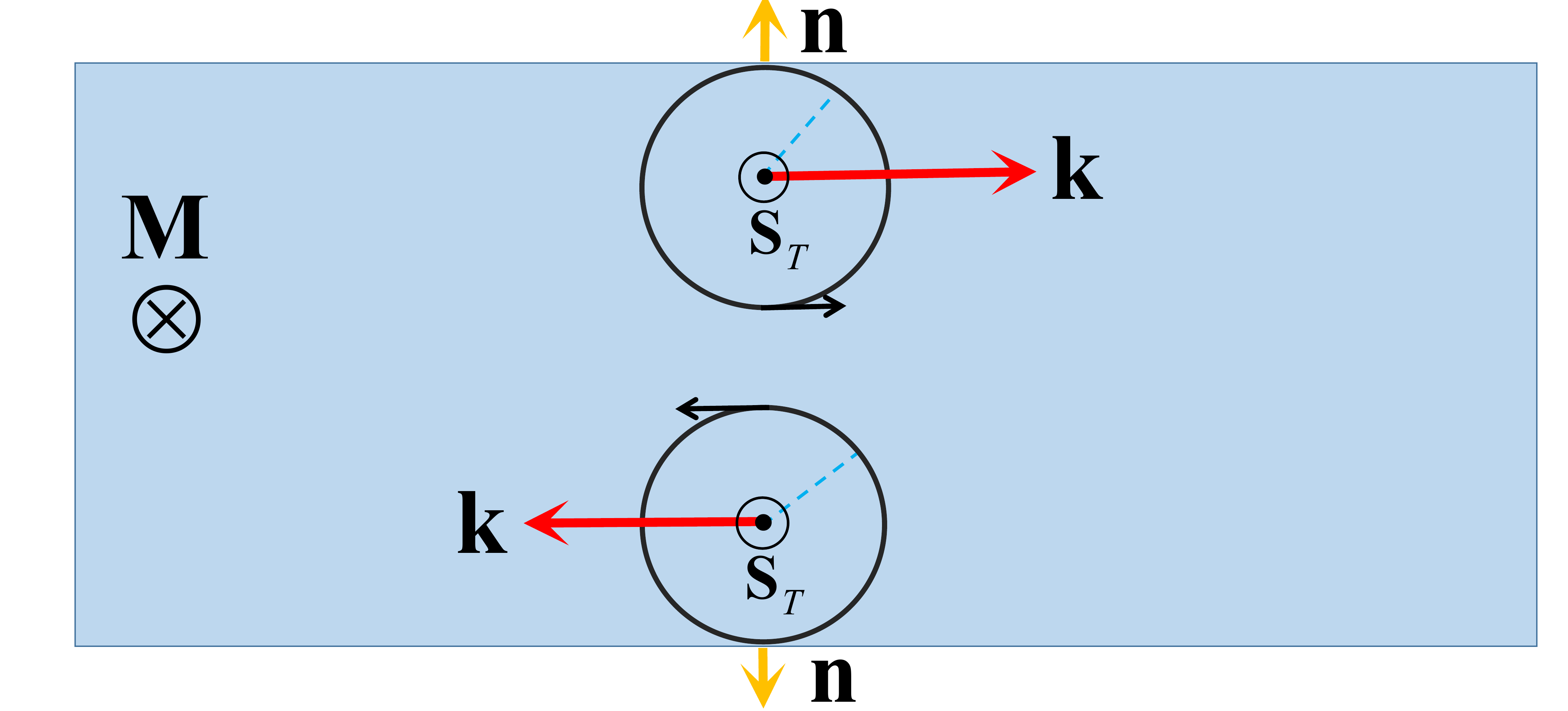}\vspace*{5pt}
		\end{minipage}\\
		\toprule
		\hspace{-2.5cm}\begin{minipage}{.53\textwidth}
			\centering\vspace*{5pt}
			Dipolar field of spin waves\\
			of film
			\[
			\hspace{1.7cm}\boxed{\nabla \cdot {\bf h}= 0~{\rm outside}~{\rm the}~{\rm magnet}}
			\]
			\vspace*{2pt}
		\end{minipage} &
		\hspace{-2.7cm}\begin{minipage}[m]{6.9cm}
			\begin{eqnarray}
				\nonumber
				~~~~h_{\beta}({\bf r})=\frac{1}{4\pi}\partial_{\beta}\sum_{\alpha}\partial_{\alpha}\int \frac{M_{\alpha}({\bf r}')}{|{\bf r}-{\bf r}'|}d{\bf r}'
			\end{eqnarray}
		\end{minipage} &
		\hspace{-0.62cm}
		\begin{minipage}[m]{5.3cm}
			\centering\vspace*{5pt}
   			\includegraphics[width=5.4cm]{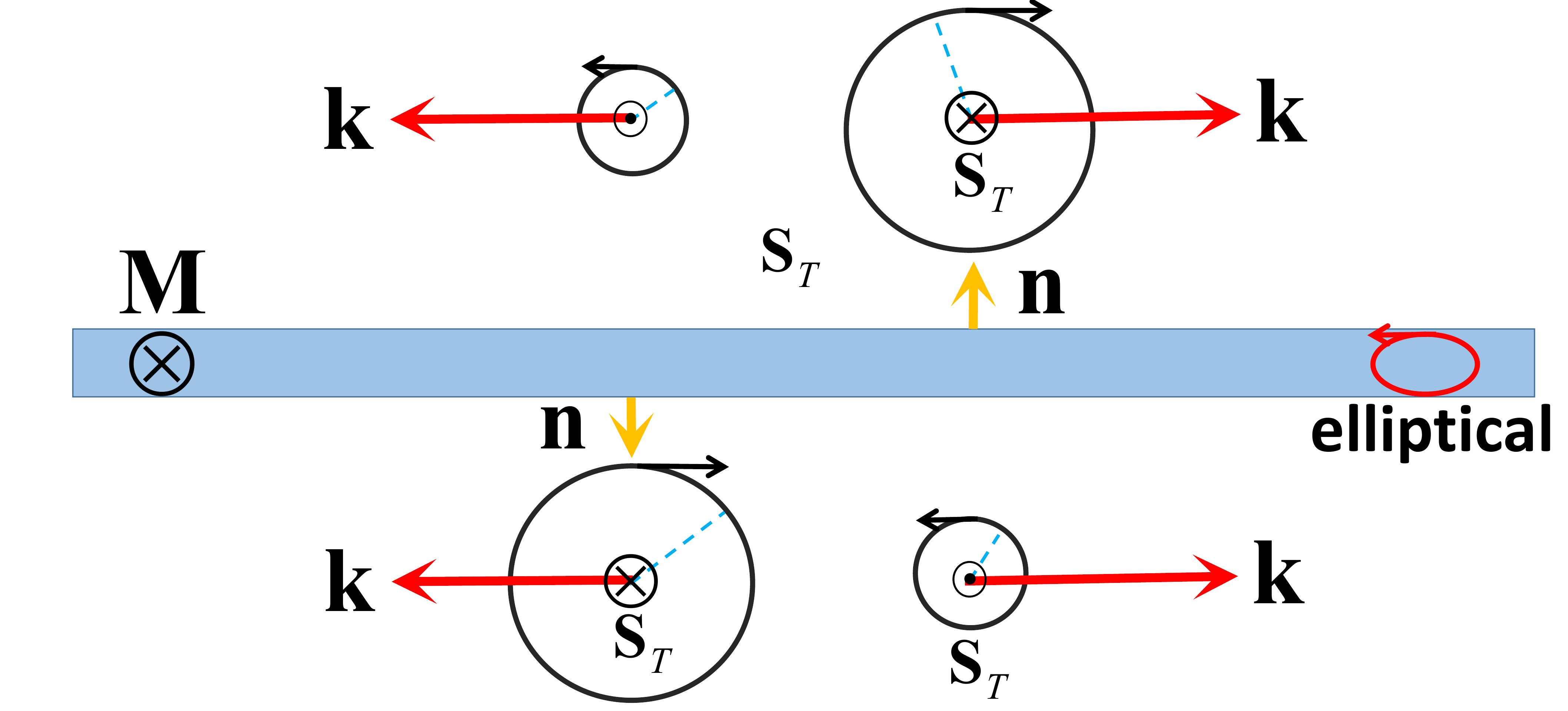}\vspace*{5pt}
		\end{minipage}\\
		\toprule
		\hspace{-2.5cm}\begin{minipage}{.53\textwidth}
			\centering\vspace*{5pt}
			Dipolar field of spin waves\\
			of nanowire
			\[
			\hspace{1.7cm}\boxed{\nabla \cdot {\bf h}= 0~{\rm outside}~{\rm the}~{\rm magnet}}
			\]
			\vspace*{2pt}
		\end{minipage} &
		\hspace{-2.7cm}\begin{minipage}[m]{6.9cm}
			\begin{eqnarray}
				\nonumber
				~~~~h_{\beta}({\bf r})=\frac{1}{4\pi}\partial_{\beta}\sum_{\alpha}\partial_{\alpha}\int \frac{M_{\alpha}({\bf r}')}{|{\bf r}-{\bf r}'|}d{\bf r}'
			\end{eqnarray}
		\end{minipage} &
		\hspace{-0.62cm}
		\begin{minipage}[m]{5.3cm}
			\centering\vspace*{5pt}
   			\includegraphics[width=5.5cm]{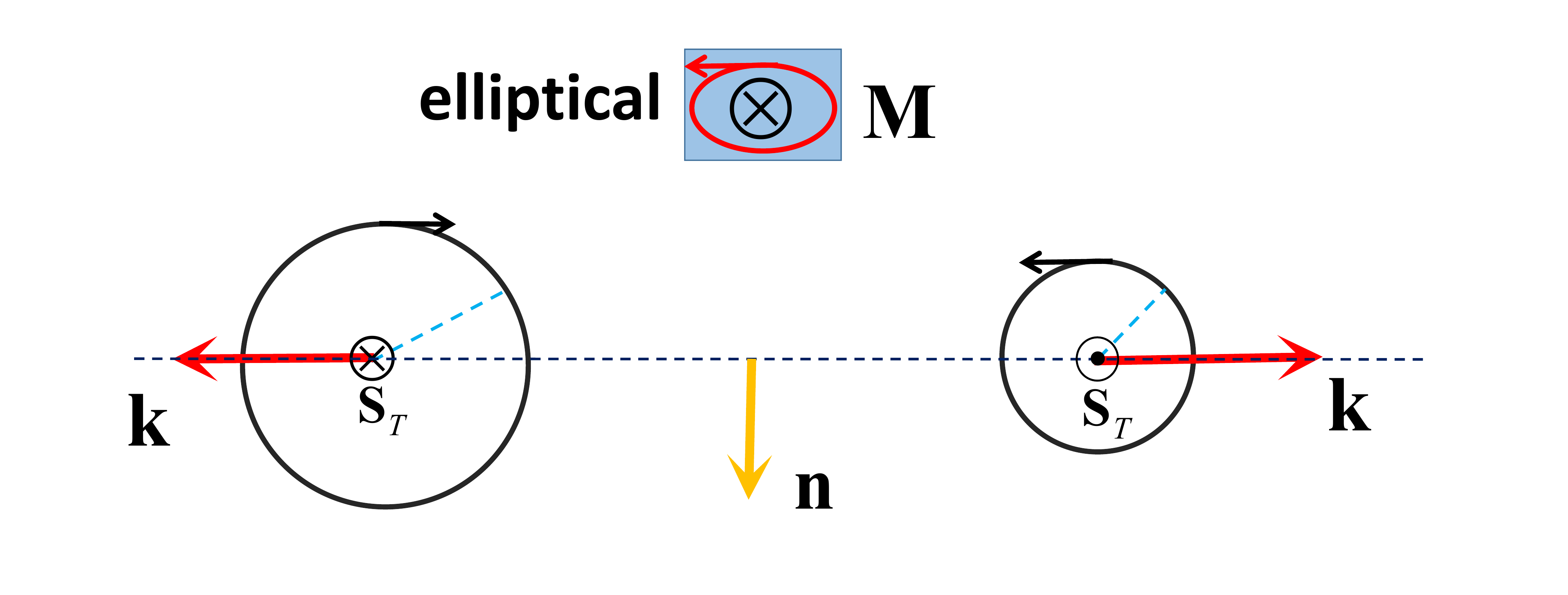}\vspace*{5pt}
		\end{minipage}\\
		\toprule
		\hspace{-2.5cm}\begin{minipage}{.53\textwidth}
			\centering\vspace*{5pt}
			Stripline magnetic field
			\[
			\hspace{1.7cm}\boxed{\nabla \cdot {\bf H}= 0~{\rm outside}~{\rm the}~{\rm magnet}}
			\]
			\vspace*{2pt}
		\end{minipage} &
		\hspace{-2.7cm}\begin{minipage}[m]{6.9cm}
			\begin{eqnarray}
				\nonumber
				&&{A}_z({\bf r},\omega)=\frac{\mu_0}{4\pi}\int d{\bf r}'{J}_z({\bf r}',\omega)\frac{e^{ik|{\bf r}-{\bf r}'|}}{|{\bf r}-{\bf r}'|}\\
				\nonumber
				&&{\bf H}=(1/\mu_0)\nabla\times {\bf A}
			\end{eqnarray}
		\end{minipage} &
		\hspace{-0.55cm}\begin{minipage}[m]{5.3cm}
		     \centering\vspace*{5pt}
   			\includegraphics[width=5.5cm]{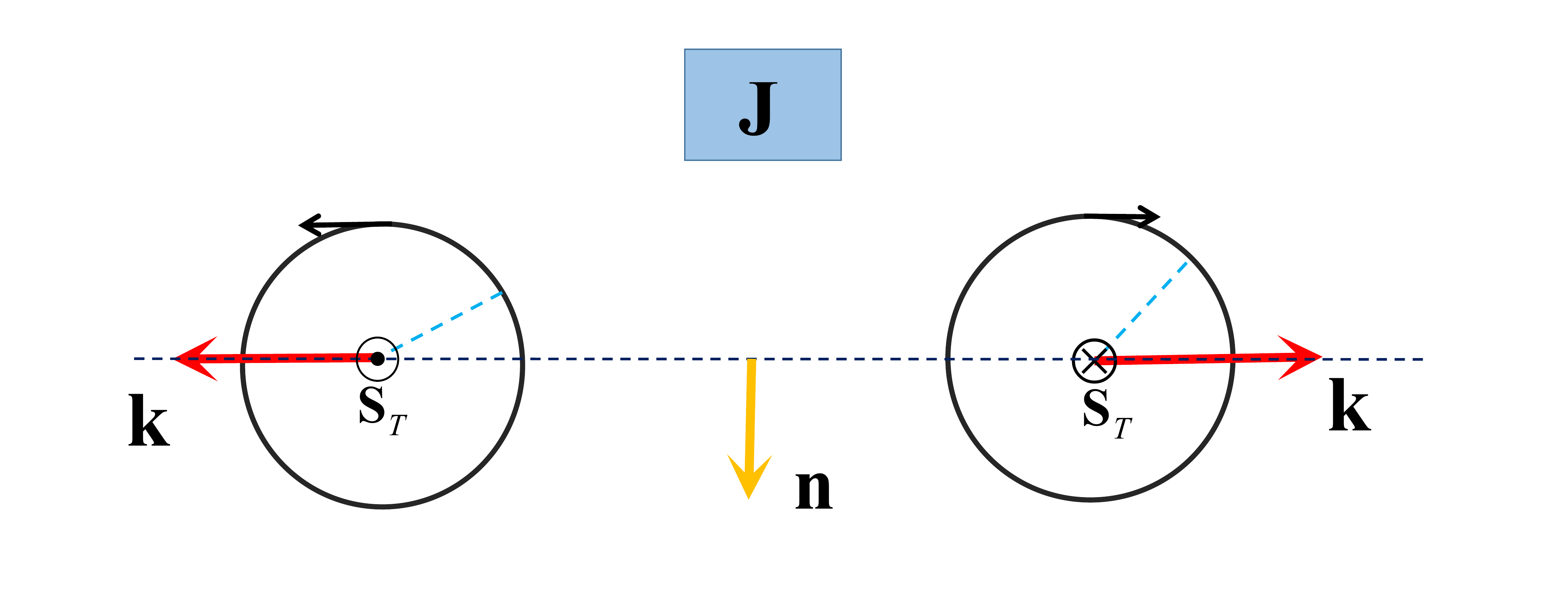}\vspace*{5pt}
		\end{minipage}\\
	\toprule
	\hspace{-2.5cm}\begin{minipage}{.53\textwidth}
		\centering\vspace*{5pt}
		\hspace{0.9cm}Electric field of two-dimensional, \\
		\hspace{0.9cm}linearly polarized electric dipoles
		\[
		\hspace{1.7cm}\boxed{\nabla \cdot {\bf H}= 0~{\rm and}~\nabla \cdot {\bf E}= 0}
		\]
		\vspace*{2pt}
	\end{minipage} &
	\hspace{-2.7cm}\begin{minipage}[m]{6.9cm}
		\begin{eqnarray}
			\nonumber
			&&{\bf A}({\bf r},t)=\frac{\mu_0}{4\pi}\int d{\bf r}'\frac{-i\omega{\bf p}\delta(x',z')}{|{\bf r}-{\bf r}'|}e^{ik|{\bf r}-{\bf r}'|}\\
			\nonumber
			&&{\bf H}=(1/\mu_0)\nabla\times {\bf A}\\
			\nonumber
			&&{\bf E}=\frac{i}{\omega\varepsilon_0\varepsilon_r}\nabla\times{\bf H}
		\end{eqnarray}
	\end{minipage} &
	\hspace{-0.55cm}\begin{minipage}[m]{5.3cm}
		\centering\vspace*{5pt}
   			\includegraphics[width=5.5cm]{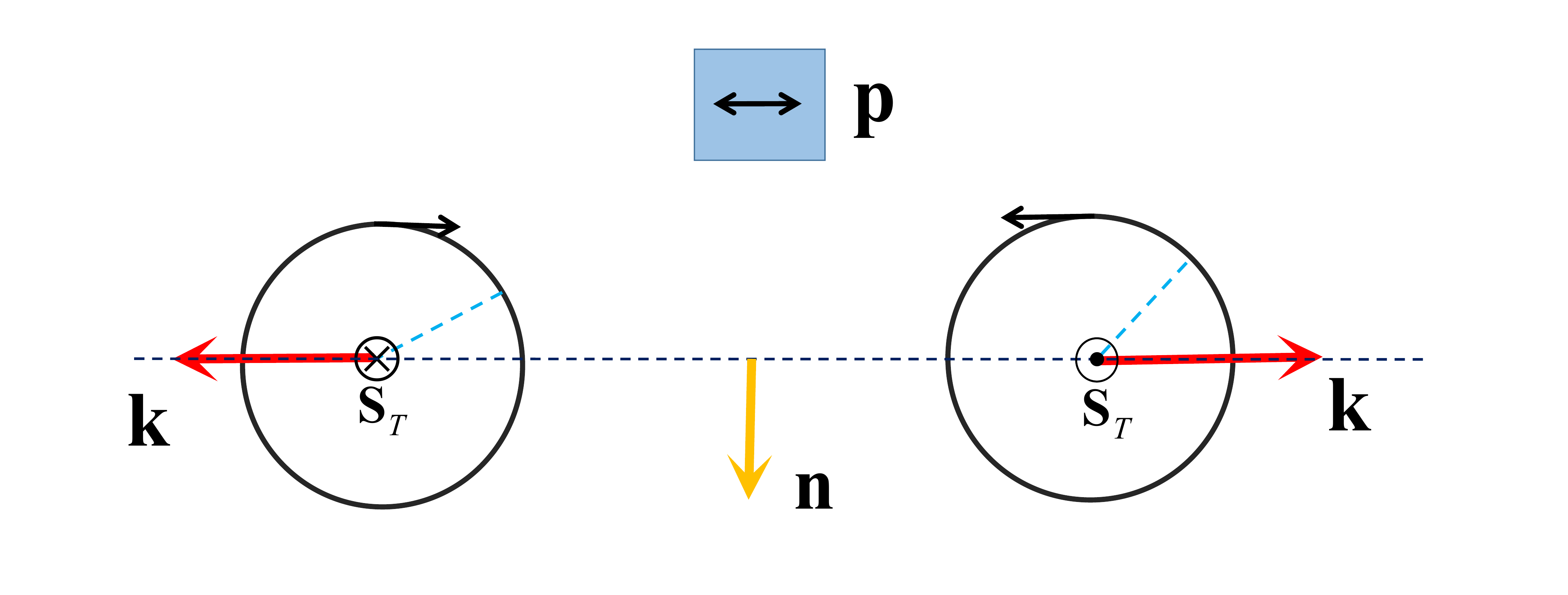}\vspace*{5pt}
	\end{minipage}\\
	\toprule
	\hspace{-2.5cm}\begin{minipage}{.53\textwidth}
		\centering\vspace*{5pt}
		\hspace{0.9cm}Electric field of two-dimensional, \\
		\hspace{0.9cm}circularly polarized electric dipoles
		\[
		\hspace{1.7cm}\boxed{\nabla \cdot {\bf H}= 0~{\rm and}~\nabla \cdot {\bf E}= 0}
		\]
		\vspace*{2pt}
	\end{minipage} &
	\hspace{-2.7cm}\begin{minipage}[m]{6.9cm}
		\begin{eqnarray}
			\nonumber
			&&{\bf A}({\bf r},t)=\frac{\mu_0}{4\pi}\int d{\bf r}'\frac{-i\omega{\bf p}\delta(x',z')}{|{\bf r}-{\bf r}'|}e^{ik|{\bf r}-{\bf r}'|}\\
			\nonumber
			&&{\bf H}=(1/\mu_0)\nabla\times {\bf A}\\
			\nonumber
			&&{\bf E}=\frac{i}{\omega\varepsilon_0\varepsilon_r}\nabla\times{\bf H}
		\end{eqnarray}
	\end{minipage} &
	\hspace{-0.55cm}\begin{minipage}[m]{5.3cm}
		\centering\vspace*{5pt}
   			\includegraphics[width=5.5cm]{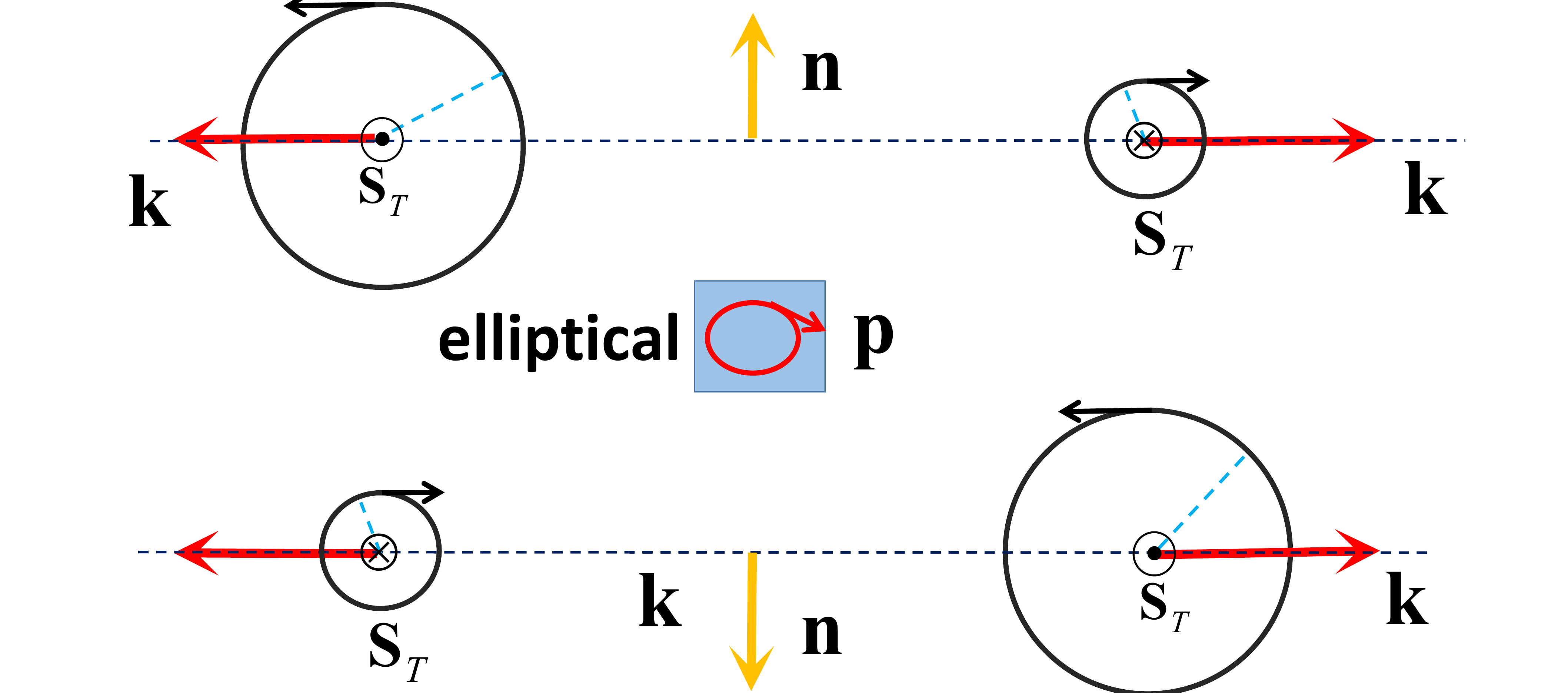}\vspace*{5pt}
	\end{minipage}\\
		\hline
		\hline
	\end{tabular}
\end{table}

\begin{table}[htbp]
	\caption{Universal origin of typical chiral waves exploited in Spintronics (continued).} \label{table_chiral_universality_continued}
	\centering
	\begin{tabular}{ccc}
		\toprule
		\hspace{-2.5cm}\begin{minipage}{.53\textwidth}
			\centering\vspace*{5pt}
			Waveguide and cavity\\ microwaves
			\[
		\hspace{2.8cm}\boxed{\nabla \cdot {\bf H}= 0}
		\]
			\vspace*{2pt}
		\end{minipage} &
		\hspace{-2.7cm}\begin{minipage}[m]{6.9cm}
			\begin{eqnarray}
				\nonumber
				&&\left(\nabla^2-({1}/{c^2}){\partial^2}/{\partial t^2}\right){\bf E}=0\\
				\nonumber
				&&{\bf n}\times {\bf E}|_S=0\\
				\nonumber
				&&{\bf H}=-{i}/({\mu_0\omega})\nabla\times{\bf E}\\
				\nonumber
				&&{\rm \textit{e.g.},}~E_z\propto \cos(p \pi z/b)
			\end{eqnarray}
		\end{minipage} &
		\hspace{-0.55cm}\begin{minipage}[m]{5.3cm}
			\centering\vspace*{5pt}
   			\includegraphics[width=5.5cm]{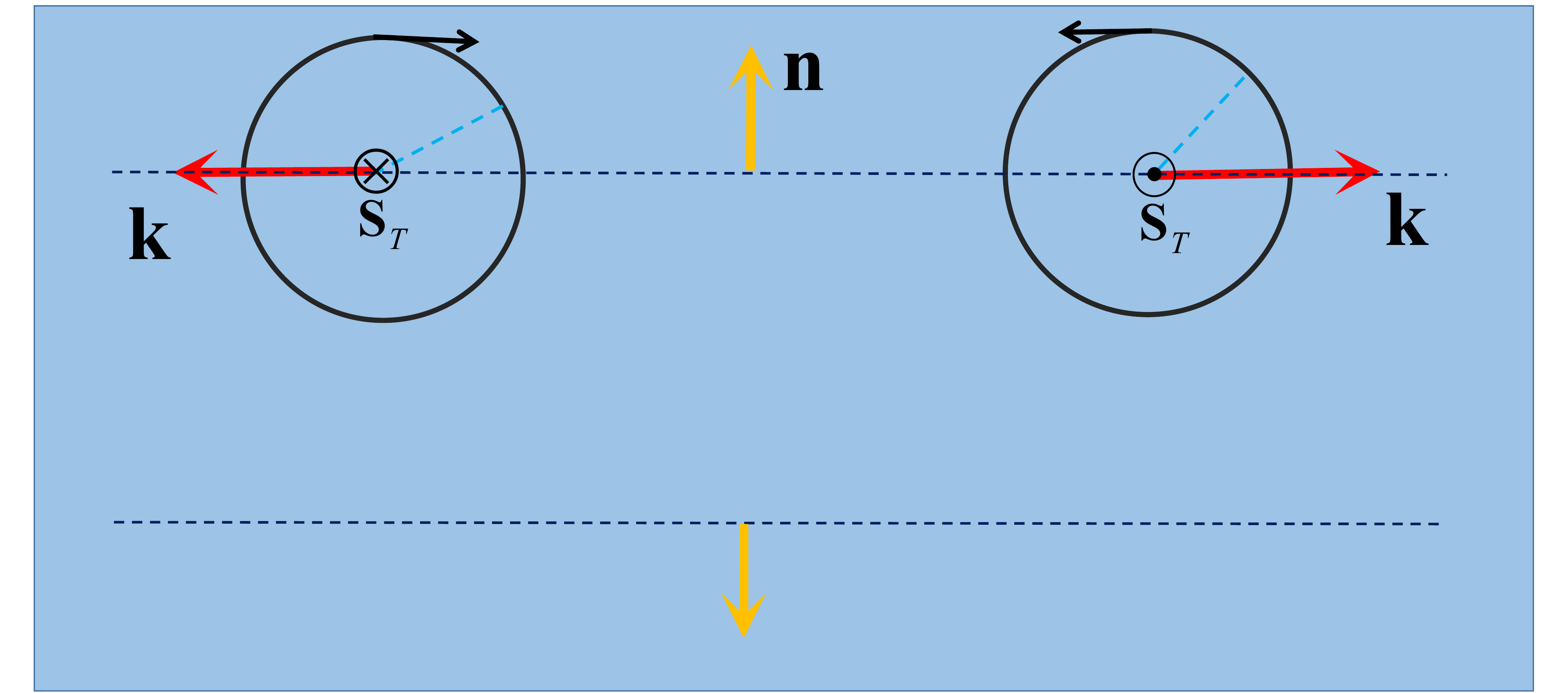}\vspace*{5pt}
		\end{minipage}\\
		\toprule
		\hspace{-2.5cm}\begin{minipage}{.53\textwidth}
			\centering\vspace*{5pt}
			Surface plasmon polaritons
			\[
			\hspace{2.8cm}\boxed{\nabla \cdot {\bf E}= 0}
			\]
			\vspace*{2pt}
		\end{minipage} &
		\hspace{-2.7cm}\begin{minipage}[m]{6.9cm}
			\begin{eqnarray}
				\nonumber
				&&\partial^2H_z/\partial y^2+(\varepsilon_rk_0^2-k_x^2)H_z=0\\
				\nonumber
				&&H_z\propto e^{ik_xx}e^{-\sqrt{k_x^2-\varepsilon_rk_0^2}y}\\
				\nonumber
				&&{\bf E}=\frac{i}{\omega\varepsilon_0\varepsilon_r}\nabla\times{\bf H}
			\end{eqnarray}
		\end{minipage} &
		\hspace{-0.55cm}\begin{minipage}[m]{5.3cm}
			\centering\vspace*{5pt}
   			\includegraphics[width=5.5cm]{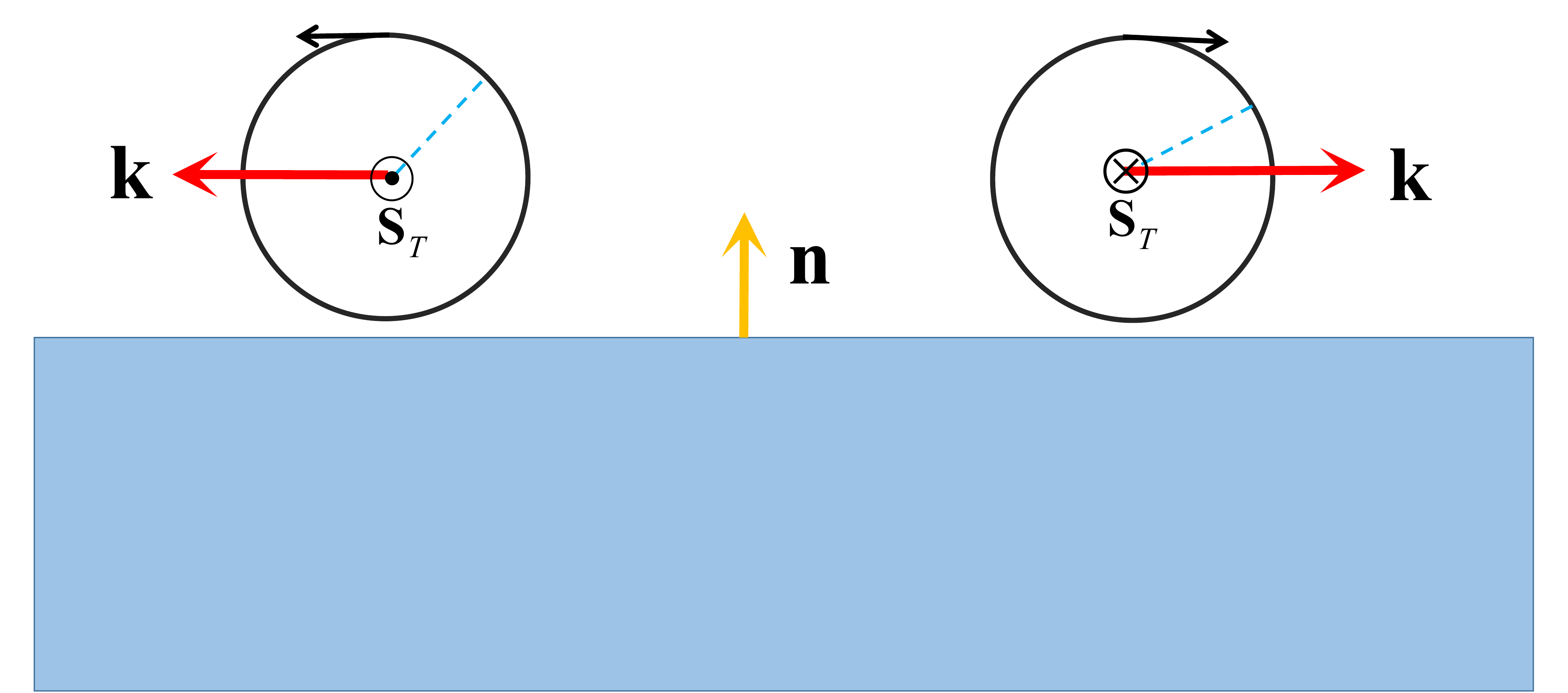}\vspace*{5pt}
		\end{minipage}\\
		\toprule
		\hspace{-2.5cm}\begin{minipage}{.53\textwidth}
			\centering\vspace*{5pt}
			Surface acoustic waves
			\[
			\hspace{1.7cm}\boxed{\nabla \cdot {\bf u}_T= 0,~{\rm where}~{\bf u}_T=\nabla\times {\bf A}}
			\]
			\vspace*{2pt}
		\end{minipage} &
		\hspace{-2.7cm}\begin{minipage}[m]{6.5cm}
			\begin{eqnarray}
				\nonumber
				&&\rho\frac{\partial^2 V}{\partial t^2}=(\lambda+2\mu)\left(\frac{\partial^2V}{\partial x^2}+\frac{\partial^2V}{\partial z^2}\right)\\
				\nonumber
				&&\rho\frac{\partial^2 {\bf A}}{\partial t^2}=\lambda\left(\frac{\partial^2{\bf A}}{\partial x^2}+\frac{\partial^2{\bf A}}{\partial z^2}\right)\\
				\nonumber
				&&{\bf u}=\nabla V+\nabla\times {\bf A}\\
				\nonumber
				&&{\bf A}\propto e^{-s|z|}e^{i(kx-\omega_kt)}
			\end{eqnarray}\\
		\end{minipage} &
		\hspace{-0.55cm}\begin{minipage}[m]{5.3cm}
			\centering\vspace*{5pt}
   			\includegraphics[width=5.5cm]{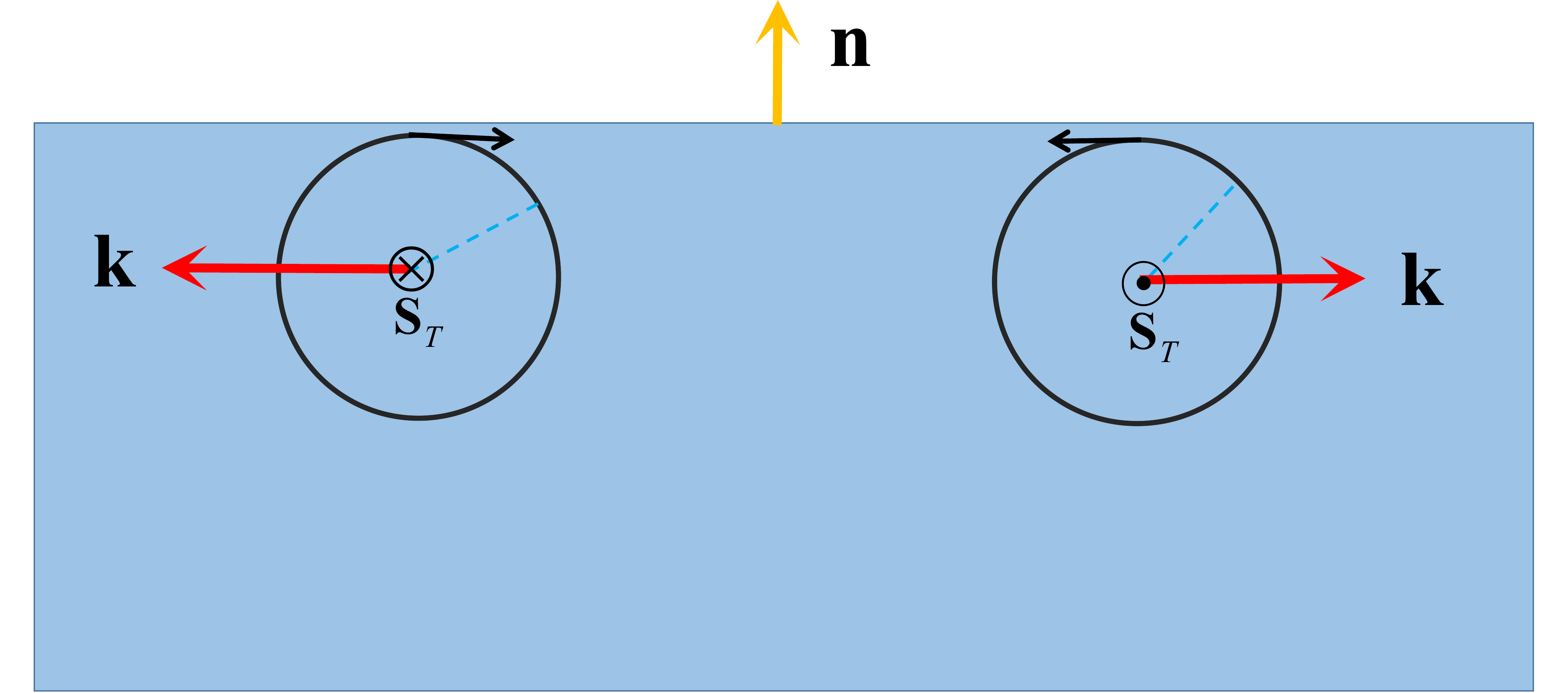}\vspace*{5pt}
		\end{minipage}\\
		\hline
		\hline
	\end{tabular}
\end{table}

\subsubsection{Spin density of classical waves}
\label{spin_density_waves}

To unify the chirality of these waves, including the magnetization ${\bf M}$, magnetic field ${\bf H}$, electric field ${\bf E}$, and displacement field ${\bf u}$, we may employ the generalization of spin concept from the particles, of relativistic origin, to waves \cite{Nori}, of dipolar origin. The electric polarization ${\bf P}$ may also be contained in this generalization. 

The spin operator $\hat{\bf S}$ has elements $(\hat{S}_j)_{ab}=-i\epsilon_{abj}$ for the spin-1 representation, which in matrix form reads 
\begin{align}
    \hat{S}_x=-i\left(\begin{array}{ccc}
        0 & 0 & 0 \\
        0 & 0 & 1 \\
        0 &-1 & 0
    \end{array}\right),~~~~~
    \hat{S}_y=-i\left(\begin{array}{ccc}
        0 & 0 & -1 \\
        0 & 0 & 0 \\
        1 & 0 & 0
    \end{array}\right),~~~~~
    \hat{S}_z=-i\left(\begin{array}{ccc}
        0 & 1 & 0 \\
        -1 & 0 & 0 \\
        0 & 0 & 0
    \end{array}\right).
\end{align}
The definition of spin density of waves is given for the monochromatic waves, which for our purpose is sufficient to identify the chirality. This definition, however, cannot be simply applied to wavepacket. For the electromagnetic waves propagating in the dispersive medium, the spin density is defined as the average value between $\langle \tilde{{\pmb \psi}}|=(\tilde{\varepsilon}{\bf E}^*,-i\tilde{\mu}{\bf H}^*)/2$ and $|{\pmb \psi}\rangle=({\bf E},i{\bf H})^T/2$, \textit{i.e.} \cite{Nori,Nori_PRL_1,Nori_PRL_2},
\begin{align}
{\bf S}_{\rm EM}=\langle \tilde{{\pmb \psi}}|\hat{\bf S}|{\pmb \psi}\rangle=(1/4)\tilde{\varepsilon}{\bf E}^*\hat{S}{\bf E}+(1/4)\tilde{\mu}{\bf H}^*\hat{S}{\bf H}=(1/4)\tilde{\varepsilon}{\rm Im}({\bf E}^*\times {\bf E})+(1/4)\tilde{\mu}{\rm Im}({\bf H}^*\times {\bf H}).
\label{transverse_spin_EM}
\end{align}
Here $\tilde{\varepsilon}=\partial[\omega\varepsilon]/\partial \omega$ is the group permittivity and $\tilde{\mu}=\partial[\omega\mu]/\partial \omega$ is the group permeability \cite{Nori_PRL_2}.
Because these are defined for density, spatial variation is allowed which is convenient for addressing the physical process that happens locally.

Making an analogy to this definition with keeping the same unit, we may define the spin of magnetization (and electric polarization) of the monochromatic oscillation as
\begin{align}
{\bf S}_{\rm M}&=(1/4)\mu_r{\mu}_0{\bf M}^*\hat{S}{\bf M}=(1/4)\tilde{\mu}{\rm Im}({\bf M}^*\times {\bf M}),
\end{align}
a quantity not defined before, which, however, can auxiliarily help to unify the description of this review article.

Similarly, in terms of $|{\bf u}\rangle=\sqrt{\rho \omega/2}{\bf u}$ and $\langle {\bf u}|=\sqrt{\rho \omega/2}{\bf u}^*$,  the spin angular momentum density is defined for the displacement field of phonon as \cite{Long2018,Holanda2018,Shi2019,phonon_Yu_1}
\begin{align}
    {\bf S}_{\bf u}=\langle {\bf u}|\hat{\bf S}|{\bf u}\rangle=(\rho \omega/2){\rm Im}({\bf u}^*\times {\bf u}). 
\end{align}
The phonon displacement field of surface acoustic waves contains the transverse component ${\bf u}_T=\nabla \times {\bf A}$ and longitudinal component ${\bf u}_L=\nabla V$, respectively (refer to Sec.~\ref{surface_acoustic_waves}).

\begin{figure}[ptbh]
	\begin{centering}
		\includegraphics[width=0.98\textwidth]{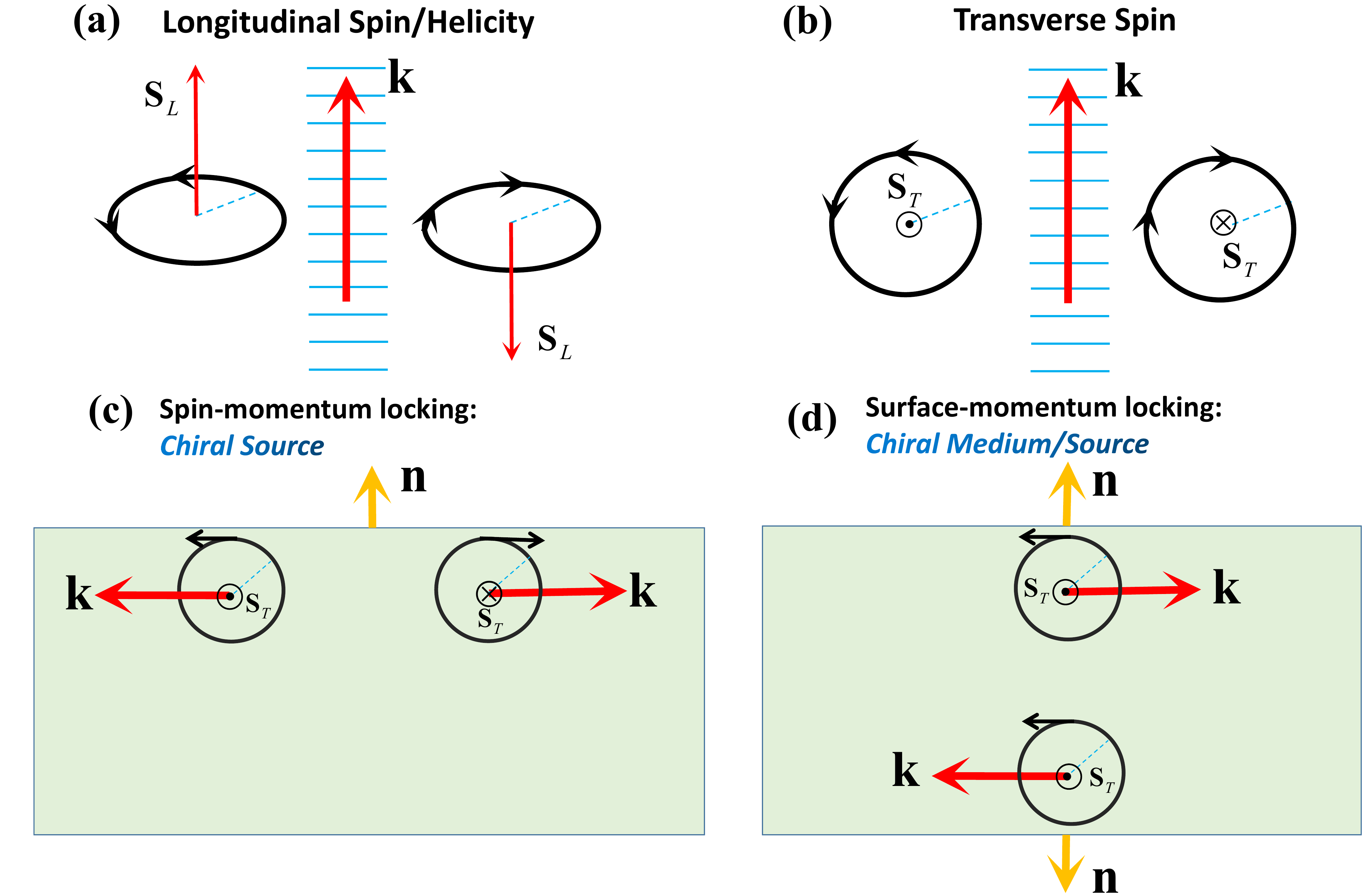}
		\par\end{centering}
	\caption{Chirality of evanescent waves. (a) and (b) are a comparison between the longitudinal and transverse spins of evanescent waves. The longitudinal spin ${\bf S}_L$ in (a) is parallel or anti-parallel to the wave propagation direction, implied by the red thick arrow, while the transverse spin ${\bf S}_T$ in (b) is perpendicular to it. (c) and (d) are two key chiral relations that lead to the chiral interactions and excitations in spintronics, where (c) retains the time-reversal symmetry, while (d) breaks the time-reversal symmetry. (c) describes the spin-momentum locking of waves that often acts as sources for chiral excitation, while (d) describes the surface-momentum locking of waves that can both act as a chiral medium and chiral sources in spintronics devices.}
	\label{polarizations}
\end{figure}

The spin (density) of waves is longitudinal when it is parallel to the wave vector. The longitudinal spin ${\bf S}_L$ is also known as the helicity of waves, as discussed in the introduction (Sec.~\ref{Sec_chirality}). While it is transverse when it is normal to the propagation direction. 
For a comparison, Figs.~\ref{polarizations}(a) and (b) define waves with longitudinal ${\bf S}_L$ and transverse  ${\bf S}_T$ polarizations, respectively, according to their direction relative to the momentum (thick red arrow). 

\subsubsection{Transverse spin and origin of chirality}
\label{evanescence}

We recall from Helmholtz's theorem, also known as the fundamental theorem of vector calculus, that any sufficiently smooth, rapidly decaying vector field in three dimensions can be resolved into the sum of a curl-free vector field and a divergence-free vector field. Here, as listed in the left panel of Tables~\ref{table_chiral_universality} and \ref{table_chiral_universality_continued}, we are allowed to focus on the divergence-free vector field because the waves we are interested in turn out to be either exactly or approximately divergence-free (\textit{e.g.}, $\nabla\cdot{\bf M}\approx 0$, $\nabla\cdot {\bf H}=0$, $\nabla\cdot {\bf E}=0$, and $\nabla\cdot {\bf u}_T=0$).

We now demonstrate that for this kind of elliptically polarized evanescent waves, the spin density is transverse, \textit{viz.}, the spin direction is perpendicular to the propagation direction; they contain a locking between the transverse spin and momentum when fixing a surface normal of the propagation plane. We consider generally a monochromatic wave of vector field propagating along the $\hat{\bf y}$-direction with wave number $k_y$, decaying along the surface normal $\hat{\bf x}$-direction by rate $\kappa$, and is homogeneous along the $\hat{\bf z}$-direction, \textit{i.e.}, 
\begin{align}
{\bf C}({\bf r},t)=\left(C_x\hat{\bf x}+C_y\hat{\bf y}+C_z\hat{\bf z}\right)e^{ik_y y-\kappa x-i\omega t},
\end{align}
where $C_{x,y,z}$ are amplitudes and we do not require decaying $\kappa=|k_y|$.
The divergence-free condition $\nabla\cdot {\bf C}=0$ renders that $C_x=(ik_y/\kappa)C_y$ and an arbitrary $C_z$. We are interested in those fields with $C_z=0$ and we obtain ${\bf C}=C_y(ik_y/\kappa,1,0)^Te^{ik_y y-\kappa x}$ and the spin angular momentum density [refer to above Sec.~\ref{spin_density_waves}]
\begin{align}
    {\bf S}_T\propto {\rm Im}({\bf C}^*\times{\bf C})=-2(k_y/\kappa)|C_y|^2e^{-2\kappa x}\hat{\bf z},
\end{align}
which is time independent, decays along the surface normal with the rate $2\kappa$, perpendicular to the propagation direction (transverse), and, more importantly, is locked to its momentum $k_y$. This spin is known as transverse spin. This analysis demonstrates, quite generally, that an elliptical, evanescent, and divergence-free vector field contains a transverse spin that is locked to the momentum when propagating within a particular plane fixed by a surface normal ${\bf n}$. Reversing the direction of the surface normal ${\bf n}$ retains this relation. So the chirality is well captured by the chirality index Eq.~(\ref{chirality_index}) defined in this review article.

These monochromatic waves can be real evanescent waves of particular momentum or the Fourier components of decaying electric and magnetic dipolar fields that we decompose with Weyl or Coulomb identities [Eqs.~(\ref{Weyl_identity}) and (\ref{Coulomb_integral})]. So there exists not a few but many such waves that are governed by various wave equations, such as those listed in the middle panel of Tables~\ref{table_chiral_universality} and \ref{table_chiral_universality_continued}. The locking between the transverse spin, momentum and the surface normal of the propagation plane of various chiral waves are depicted in the right panel of Tables~\ref{table_chiral_universality} and \ref{table_chiral_universality_continued}. This locking may be treated as a generalization of the locking of spin and momentum of electrons where the relativistic effect overrules, or \textit{generalized spin-orbit interaction}, a new resource that inspires spintronics functionality and devices. The chiralities in the propagation as summarized above all hold for near-field evanescent wave fields with polarization perpendicular to the wave vector and surfaces or interfaces. We do not find chirality in waves that are longitudinal in spins or fully propagating.  For Rashba-like spin-orbit interaction of surface states of a topological insulator \cite{topological_insulator_1,topological_insulator_2}, the direction of the spin is also perpendicular to the momentum, so maybe also a transverse spin that is well characterized by the chirality index. They propagate in one direction at a given boundary and are immune to  (weak) backscattering.

With a fixed maximal chirality index $Z=+1$ (or $-1$), there are two categories for the wave propagation as classified in Figs.~\ref{polarizations}(c) and (d), respectively. When there exists time-reversal symmetry, the transverse spin is allowed to be reversed. So with a fixed surface normal ${\bf n}$ for the wave propagation, the propagation is bidirectional, but the transverse spin is reversed with the opposite wave vector, as shown in Fig.~\ref{polarizations}(c). Many examples in Tables~\ref{table_chiral_universality} and \ref{table_chiral_universality_continued} belong to this category. On the other hand, additional features arise when the sources of waves break the time-reversal symmetry. Particularly, interesting phenomena appear when the sources become circularly polarized such as magnetic and electric dipoles, in which the transverse spin is fixed by the source. Accordingly, the direction of the wave vector is fixed as well by the restriction of the chirality index, so being unidirectional, as shown in Fig.~\ref{polarizations}(d). Sometimes, the sources hold both spin components but of different amplitudes, in which case the propagation of their evanescent waves is non-reciprocal with different amplitudes of opposite directions. Examples vary from the stray fields of elliptical magnetic dipoles to that of the elliptical electron dipoles as listed in Tables~\ref{table_chiral_universality} and \ref{table_chiral_universality_continued}.

The chirality index thereby provides an intuitive explanation of the unidirectionality of well-known examples. There are unidirectional waves propagating within one plane such as the Damon-Eshbach spin waves, magnetic stray fields generated by circularly polarized spin waves, and electric dipolar fields generated by the circularly polarized electric dipoles. The broken time-reversal symmetry forces that when close to one surface normal there is no counter-propagating modes that otherwise break the transverse-spin--momentum locking effect. The counter-propagating modes are near the opposite surface or plane that renders a momentum-surface locking for waves. The transverse spins of these waves close to the opposite surface are the same since fixed by the equilibrium magnetic or electric dipoles.
Figure~\ref{polarizations}(d) depicts that for the same transverse spin at opposite surfaces, the propagation direction must be opposite.

\subsubsection{An example}
\label{An_example} 

\begin{figure}[htp]
    \centering
    \includegraphics[width=10.5 cm]{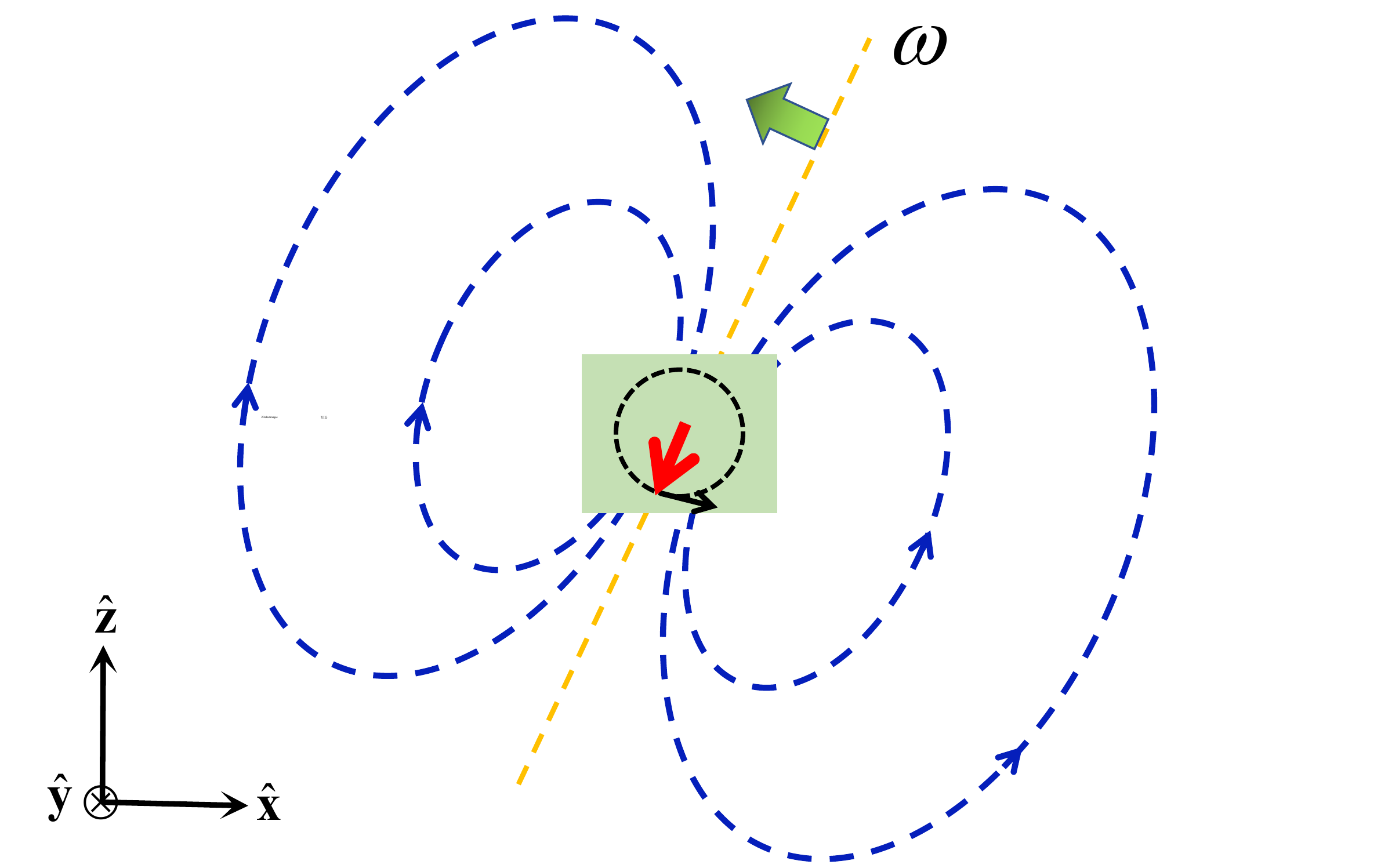}
    \caption{The transverse spin of the magnetic stray field emitted by a magnetization (Kittel mode) of a magnetic nanowire that precesses in frequency $\omega$ and an anti-clockwise direction around its equilibrium $\hat{\bf y}$-direction. The stray field precesses at the same frequency $\omega$ but in opposite direction, as illustrated by the black arrow for the magnetization precession and the green arrow for the stray-field rotation.
    }
    \label{transverse_spin_example}
\end{figure}

\textcolor{blue}{We may intuitively associate the ``spin density" of classical waves with a vector field that locally rotates around a fixed axis. This can be the magnetization vector of spin waves that precess around the saturation magnetization in a ferromagnet or circularly polarized electric, magnetic, or displacement fields. When divergence free and evanescent, the precession axis, or spin, of polarization vectors of these waves are transverse to their propagation direction.  In the previous sections we reviewed this concept for several types of classical waves and its appeal for spintronics. Here}
we illustrate nature of the classical spin density by explicitly calculating the transverse spin of a particular chiral wave, \textcolor{blue}{i.e., the magnetic stray field emitted by the ferromagnetic magnetic resonance (Kittel)  mode of a magnetic nanowire  \cite{electron_spin_Yu}, as sketched in Fig.~\ref{transverse_spin_example}.} We also refer to the review of the theoretical formulation and experimental evidence of \textcolor{blue}{chirality} in Chapter~\ref{Near_field_spintronics}.

Here we analyze in detail the  transverse
angular momentum or \textquotedblleft spin\textquotedblright\ \cite{electron_spin_Yu}
of the evanescent magnetodipolar field emitted by an excited magnetic nanowire (Sec.~\ref{dipolar_fields_1})
with thickness $d$, width $w$, and equilibrium magnetization $\mathbf{M}_{s}$
along the wire $\hat{\bf y}$-direction (see Fig.~\ref{transverse_spin_example}).  \textcolor{blue}{The magnetization and magnetic stray field of the fundamental mode precess at the same frequency $\omega$ but in opposite direction.} For later convenience  (Sec.~\ref{Sec_evanescent_pumping}), we adopt a quantum mechanical notation, but in
the classical limit operators can be simply replaced by the field amplitudes.

The magnetization dynamics \textcolor{blue}{of the wire} as expressed by the spin operator $\hat{\mathbf{S}%
}(\mathbf{r},t)$ generates a magnetic \textcolor{blue}{stray} field that obeys Coulomb's Law \cite{Landau,Jackson},
\begin{equation}
	\mathbf{h}_{\beta}(\mathbf{r},t)=-\frac{\gamma\hbar}{4\pi}\partial_{\beta
	}\partial_{\alpha}\int d\mathbf{r}^{\prime}\frac{\langle\hat{\mathbf{S}%
		}_{\alpha}(\mathbf{r}^{\prime},t)\rangle}{|\mathbf{r}-\mathbf{r}^{\prime}
		|},
		\label{dipolar}
\end{equation}
where we adopted the summation convention over repeated indices
$\{\alpha,\beta\}=\{x,y,z\}$ and $-\gamma$ is the gyromagnetic ratio. The magnon modes in the wire move with wave number $k_y$ along the wire $\hat{\bf y}$-direction. For sufficiently weak excitation the spin operator can
be expanded into magnon field operators $\hat{\alpha}_{k_{y}}$ with
amplitudes $m_{x,z}^{k_{y}}(x,z)$  [refer to Eq.~(\ref{Bogoliubov_a}) for magnetic films]:
\begin{equation}
	\hat{\mathbf{S}}_{x,z}(\mathbf{r})=\sqrt{2S}\sum_{k_{y}}\left(  m_{x,z}%
	^{k_{y}}(x,z)e^{ik_{y}y}\hat{\alpha}_{k_{y}}+\mathrm{H.c.}\right),
	\label{magnon_operator}
\end{equation}
where $S=M_{s}/(\gamma\hbar)$ in terms of the saturation magnetization \(M_{s}\). The static stray field vanishes for long wires. The  dynamic stray field $\mathbf{h}$ is a response to $\langle\hat{\alpha}_{k_{y}}\rangle$, the coherent amplitude of
magnons with momentum $k_{y}\hat{\mathbf{y}}$  that are excited by external microwaves with frequency \(\omega=\omega_{k_y}\). \textcolor{blue}{We are interested in the Fourier components \(\mathbf{h}_{\beta}(z,\mathbf{k},t)\) of the magnetic stray field Eq.~(\ref{dipolar}) with momenta $\mathbf{k}=k_{x}\hat{\bf x}+k_{y}\hat{\bf y}$ that represent chiral waves propagating in the $x$-$y$ plane.} At frequency $\omega$
\begin{align}
    \mathbf{h}_{\beta}(z,\mathbf{k},t)=\tilde{\mathbf{h}}_{\beta}(z,\mathbf{k})e^{-i\omega t}+\tilde{\mathbf{h}}_{\beta
}^{\ast}(z,-\mathbf{k})e^{i\omega t}
\end{align}
is evanescent, i.e. decays exponentially with distance from the $x$-$y$ plane. Below the nanowire ($z<0$) \cite{Chiral_pumping_Yu,Chiral_pumping_grating} Eq.~(\ref{dipolar}) leads to
\begin{align}
	\left(
	\begin{array}
		[c]{c}%
		\tilde{\mathbf{h}}_{x}(z,\mathbf{k})\\
		\tilde{\mathbf{h}}_{y}(z,\mathbf{k})\\
		\tilde{\mathbf{h}}_{z}(z,\mathbf{k})
	\end{array}
	\right)   &  =F_{\mathbf{k}}\left(  m_{z}^{k_{y}}+\frac{ik_{x}}{k}m_{x}%
	^{k_{y}}\right)  \left(
	\begin{array}
		[c]{c}%
		ik_{x}/k\\
		ik_{y}/k\\
		1
	\end{array}
	\right)  e^{kz}\left\vert \left\langle \hat{\alpha}_{k_{y}}\right\rangle
	\right\vert ,
\label{near_dipolar_field}%
\end{align}
where $F_{\mathbf{k}}=-\gamma\hbar\sqrt{2S}(1-e^{-kd})\sin(k_{x}w/2)/k_{x}$ is
the form factor of the rectangular wire, which decays exponentially $\sim e^{-k|z|}$ on a
scale governed by the complex momentum $k_{x}\hat{\mathbf{x}}+k_{y}\hat
{\mathbf{y}}-ik\hat{\mathbf{z}}$ \cite{Jackson}.

Spatially homogeneous microwaves excite the Kittel magnon \cite{Kittel_mode} with $k_{y}=0$ by ferromagnetic resonance (FMR).  \textcolor{blue}{Then, according to Eq.~(\ref{near_dipolar_field}),} $\tilde{h}%
_{y}(\mathbf{k})$ vanishes and $\tilde{h}_{x}(\mathbf{k})=i\mathrm{sgn}%
(k_{x})\tilde{h}_{z}(\mathbf{k})$. The polarization of the Kittel mode is
governed by the shape anisotropy and applied magnetic field. In nanowires with circular/square cross sections or
sufficiently large magnetic field the spin waves are circularly polarized with
$m_{x}\rightarrow im_{z}$ (refer to Sec.~\ref{magnon_film_wire} below for details). The Fourier components $\tilde{\mathbf{h}%
}(\mathbf{k})=0$ for $k_{x}>0$, i.e.  the stray field is unidirectional.

\textcolor{blue}{Electromagnetic fields  in vacuum are divergence-free, so when $k_y=0$ Eq.~(\ref{near_dipolar_field}) tells us that $ik_x\tilde{h}_x(z,k_x)+|k_x|\tilde{h}_z(z,k_x)=0$.}
The photon spin density is at GHz frequency strongly dominated by the magnetic component. Under the nanowire in vacuum  Eq.~(\ref{transverse_spin_EM}) leads to \cite{Nori,Nori_PRL_1,Nori_PRL_2}:
\begin{align}
    {\pmb{\mathcal{D}}}(z,k_x)&=\mu
_{0}\mathrm{\mathrm{\operatorname{Im}}}\left[  \tilde{\mathbf{h}}^{\ast
}(z,k_x)\times\tilde{\mathbf{h}}(z,k_x)\right]  /(4\omega)\nonumber\\
&=\mu_0{\rm Im}\left[\tilde{h}_x(z,k_x)\tilde{h}_z^*(z,k_x)\right]/(2\omega)\hat{\bf y}
\label{transverse_spin_density}
\end{align}
is purely transverse since ${\pmb {\cal D}}\cdot\mathbf{h}=0$ and normal to the propagation $x$-$y$ plane ${\pmb {\cal D}}\cdot\hat{\bf z}=0$.
\textcolor{blue}{The vector $\hat{\pmb \sigma}$ in the chirality index  $Z=\hat{\bf n}\cdot(\hat{\pmb \sigma}\times\hat{\bf k})$ defined in the beginning of this review is here the unit vector in the direction of the spin density ${\pmb{\cal D}}$. }

The photon magnetic field may couple to the electron spin by the Zeeman interaction non-locally over distances limited by the evanescent decay length  \cite{electron_spin_Yu} (refer to Sec.~\ref{Sec_evanescent_pumping}). This can be compared  with conventional spin pumping by interface exchange, which is a contact interaction.

Spintronics is concerned with the information carried by the spin degree of freedom either by electron particle transport or by spin waves. Here we showed that spin dynamics generates chiral magnetic stray fields that transfer and share this information  with other waves or quasi-particles in a chiral manner.  Section~\ref{Chiral_interaction} reviews the associated dynamical chiral interaction (Fig.~\ref{Chiral_spintronics}).

\section{Dynamic chiral interactions: unidirectional transport with transverse spins}
\label{Chiral_interaction}

As reviewed in Sec.~\ref{section2} and \ref{section3}, the physical origin of the chirality of given states either comes from the ground-state Hamiltonian that contains strong chiral interactions, such as the dipolar and relativistic spin-orbit coupling or is captured by the chirality of the excitations  (Fig.~\ref{Chiral_spintronics}). For the latter, the interaction between the excitations or quasiparticles is usually weak and can be treated by perturbation theory. Nevertheless, the interaction can be chiral that is governed by the generalized spin-orbit interaction that renders the dynamics in nanostructures to be totally determined by the chirality but not the trivial ground state that itself does not have any chirality. In this Chapter, we first address a universal quantum description of chiral interaction between excitations (Sec.~\ref{universal_chiral_interaction}) and then address their numerous realizations in magnetism, spintronics, and magnonics (Sec.~\ref{nanomagnets}, \ref{magnon_photon}, and \ref{magnon_phonon}).

\subsection{Chiral interaction between excitations}

\label{universal_chiral_interaction}

Much of condensed matter physics is based on the concept of a quasiparticle, \textit{i.e.}, a collective excitation with a long lifetime due to weak interaction with other degrees of freedom. The quasiparticle picture for electrons works well because charges in metals are very efficiently screened and fields can be mapped on a quadratic Hamiltonian when amplitudes are small. Similarly, the excitations of ground states  such as photons, phonons, and magnons,  are usually weakly interacting (Bosonic) quasiparticles. Nevertheless, interactions are essential for many phenomena, such as superconductivity and magnetism. While electron-electron, phonon-phonon, and magnon-magnon interactions are of course important and interesting, we focus here on interactions between different quasiparticles, say $\hat{\alpha}$ and $\hat{\beta}$, that can be Fermions or Bosons. The Hamiltonian can be written
\begin{equation}
\hat{H}=\hat{H}_{\alpha}+\hat{H}_{\beta}+\hat{H}_{\alpha\beta}.
    \end{equation}
When the system is translational invariant in a spatial direction the conjugate  linear momentum \(k\) is a good quantum number. The non-interacting quasiparticle Hamiltonian then read in the second quantization
\begin{subequations}
\begin{align}
    \hat{H}_{\alpha}&= \sum_k \epsilon_k \hat{\alpha}_k^{\dagger} \hat{\alpha}_k,\\
    	\hat{H}_{\beta}&= \sum_n \hbar\omega_n \hat{\beta}_n^{\dagger} \hat{\beta}_n,
    \end{align}
    \end{subequations}
where \(\epsilon_k=\hbar \omega_k\) is the energy of a quasiparticle \(\hat{\alpha}\) with continuous momentum variable $k$ created by field operator $\hat{\alpha}_k^{\dagger} $, while $n$ labels the modes of quasiparticle $\hat{\beta}$ that we assume here to be discrete.
    The leading term in the interaction between $\hat{\alpha}$ and $\hat{\beta}$ then reads  
    \begin{equation}
    	\hat{H}_{\alpha\beta}= \sum_{k,n} g_{k,n} \hat{\alpha}_{-k}^{\dagger} \hat{\beta}_n + \mathrm{H.c.}
    \end{equation}
    \textcolor{blue}{We note that we do not address interactions involving the electron charge that cannot be expressed in a bilinear form of field operators.}
    We call the interaction ``chiral'' when the interaction vanishes for one sign of \(k\) but is finite for the other sign. It is ``non-reciprocal'' when $|g_{k,n}| \ne |g_{-{k,n}}|$.

The coupling between quasiparticles \textcolor{blue}{or excitations} such as electron \textcolor{blue}{spins}, phonons, photons, and magnons often leads to undesired effects such as dephasing and reduced lifetimes. However, when coherently controlled, it can also be a boon, \textit{e.g.}, in transducers for telecommunication or resulting in an attractive effective interaction that causes superconductivity. Chiral interaction between different excitations may help to build key components for future communication technologies.

Here we focus on the phenomena in spintronics, which is rich in chiral and non-reciprocal effects.  Table~\ref{table_chiral_interaction} summarizes a few of its chiral interactions, illustrating the theories and typical experimental observations. We here define some key functionalities of chirality that are addressed in different circumstances in this review. 
\begin{itemize}
\item ``Diodes" are two-terminal (electronic) components that conduct current primarily in one direction with an asymmetric conductance. 
\item ``Isolators" are wave devices that are based on chirality. They are two-port diodes that transmit a coherent wave signal only in one direction, stabilizing device performance by suppressing back reflections.
\item ``Circulators" are three-port devices that route a signal between ports in a clockwise, or counter-clockwise, direction.
\end{itemize}

This Chapter focuses on the (linear) chiral interaction between quasi-particles. Table~\ref{table_chiral_interaction} summarizes the key effects, approaches, and experimental evidence of chiral interactions in spintronics as discussed in detail in the sections below.

\begin{table}[htb]
	\caption{Chiral interactions in spintronics. } \label{table_chiral_interaction}
	\centering
	\begin{tabular}{ccc}
		\toprule
		\hspace{-0.7cm}Systems & \hspace{-0.6cm} Theory & ~~~Samples and phenomena \\
		\toprule
		\hspace{-1.2cm}\begin{minipage}[m]{.45\textwidth}
			\centering\vspace*{5pt}
			\textbf{Propagating spin wave spectroscopy}
	    \includegraphics[width=5.5cm]{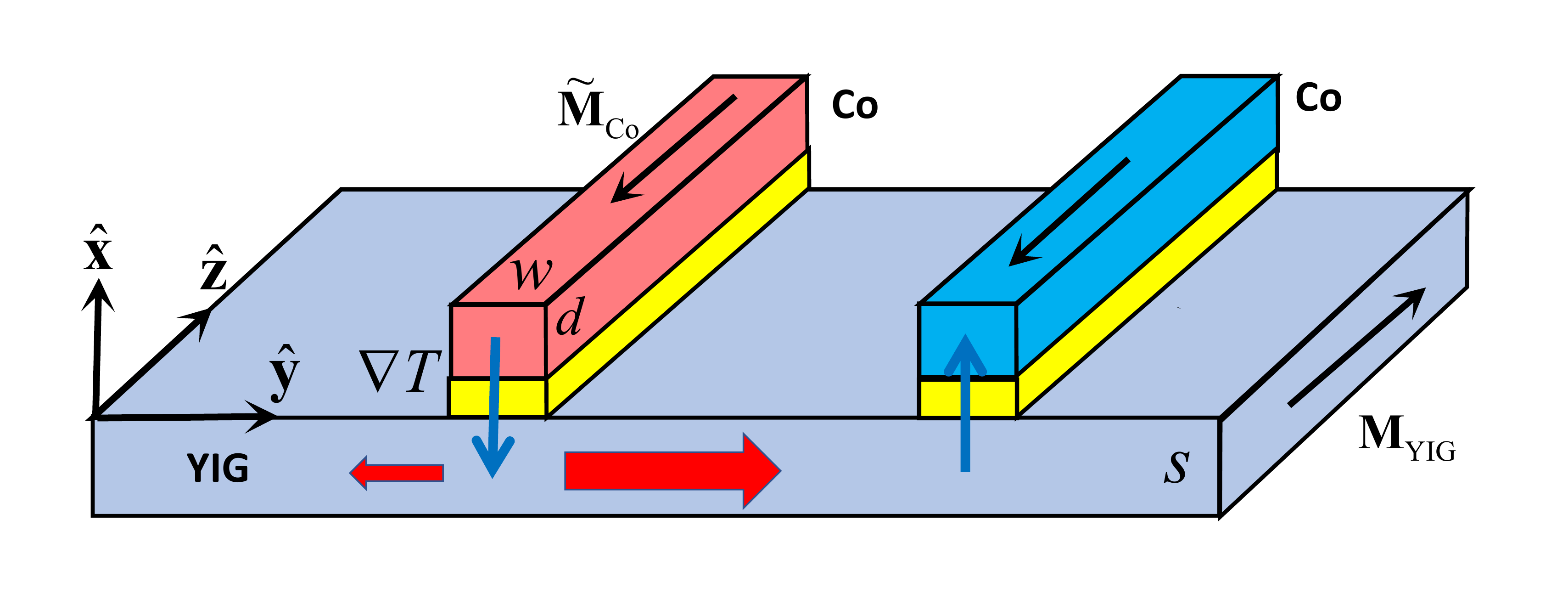}\vspace*{2pt}
		\end{minipage} &
		\hspace{-1.1cm}\begin{minipage}[m]{5.8cm}
			\begin{itemize}
				\item Chiral spin pumping (input-output \cite{Yu_Springer,Chiral_pumping_Yu,Chiral_pumping_grating}, Green function \cite{Chiral_pumping_Yu}, micromagnetic simulation \cite{Au_first,Jilei_skyrmion})
				\item Chiral spin Seebeck effect 
	\cite{Chiral_pumping_Yu,Yu_Springer}
				\item Additional damping due to spin and magnon pumping \cite{Chiral_pumping_Yu,magnon_trap}
				\item Magnon trapping \cite{magnon_trap} and unidirectional blocking \cite{chiral_gate}
			\end{itemize}
		\end{minipage} &
		\hspace{-0.5cm}\begin{minipage}[m]{4.5cm}
			\begin{itemize}
				\item Magnetic nanowire arrays on thin YIG films \cite{Haiming_exp_grating}
				\item Antiparallel spin valves \cite{bilayer_dipolar_1,bilayer_dipolar_2}
				\item Two magnetic nanowires on thin YIG films \cite{Haiming_exp_wire}
				\item Enhanced damping and magnon trap  \cite{Hanchen_damping}
			\end{itemize}
		\end{minipage}
		\\
		\toprule
		\hspace{-0.7cm}\begin{minipage}{.45\textwidth}
			\centering\vspace*{5pt}
			\textbf{Magnons and photons}
			\includegraphics[width=5.0cm]{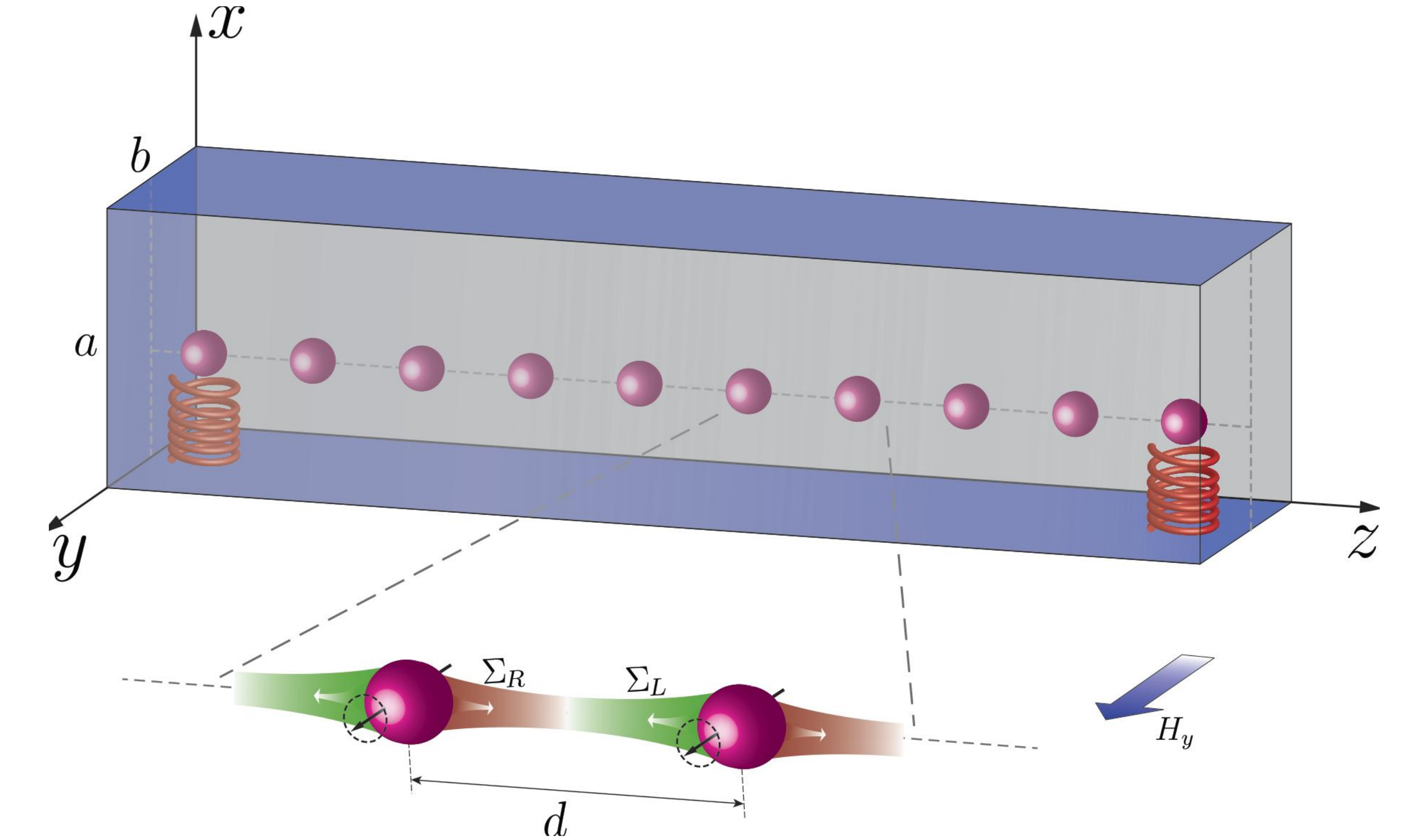}\vspace*{2pt}
		\end{minipage} &
		\hspace{-1.1cm}\begin{minipage}[m]{5.8cm}
			\begin{itemize}
				\item Chiral magnon pumping by microwave striplines 
				 \cite{Yu_Springer}
				\item Chiral magnon polaritons (master equation \cite{waveguide_Yu_2}, input-output \cite{magnon_trap})
				\item Spin skin effect 
				 \cite{waveguide_Yu_1,waveguide_Yu_2})
				\item Cavity microwave circulators (input-output theory \cite{circulating_polariton,circulator_Tang}, finite element simulations \cite{circulating_polariton})
				\item Spin-wave Doppler shift \cite{Doppler_Yu}
			\end{itemize}
		\end{minipage} &
		\hspace{-0.5cm}
		\begin{minipage}[m]{4.5cm}
			\begin{itemize}
			\item Radiative damping in waveguides \cite{radiative_damping_exp}
			\item Chiral pumping of propagating magnons  \cite{stripline_poineering_1,stripline_poineering_2,Teono_NV}
				\item Chiral damping of spin waves \cite{chiral_damping}
				\item Broadband non-reciprocity of magnon polaritons \cite{Xufeng_exp}
			\end{itemize}
		\end{minipage}
		\\
		\toprule
		\hspace{-0.7cm}\begin{minipage}{.45\textwidth}
			\centering\vspace*{5pt}
			\textbf{Magnons and surface phonons}
			\includegraphics[width=4.6cm]{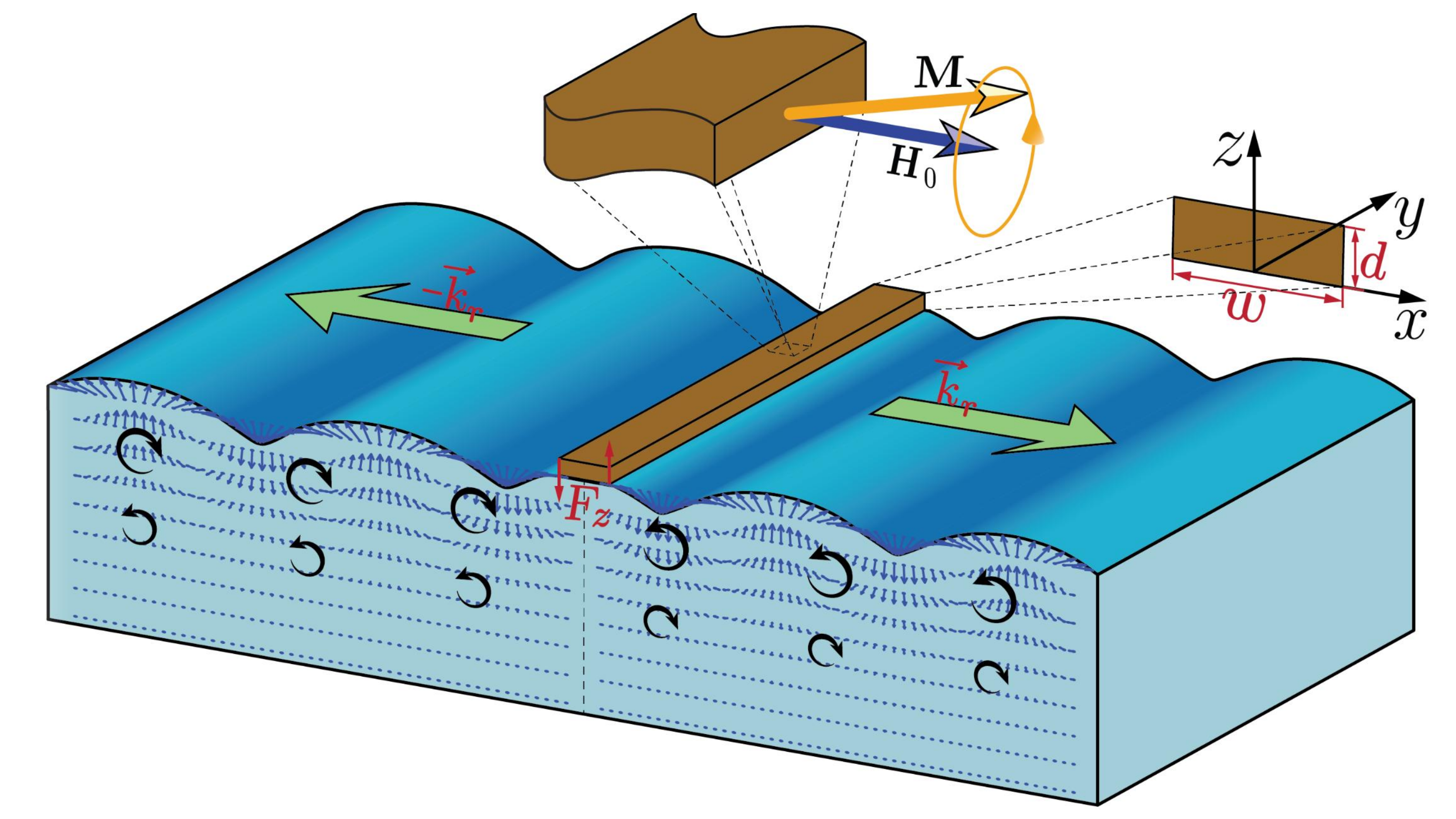}\vspace*{2pt}
		\end{minipage} &
		\hspace{-1.1cm}\begin{minipage}[m]{5.8cm}
			\begin{itemize}
				\item Phonon diodes by magnetoelasticity \cite{Xu} (input-output \cite{phonon_Yu_1} and scattering \cite{phonon_Yu_2} theory)
				\item Chiral pumping of surface phonons (input-output theory \cite{phonon_Yu_1}, magnetoelasticity \cite{phonon_Kei}, simulation \cite{phonon_Kei}, and Landauer-B\"uttiker \cite{phonon_Yu_2})
				\item Spin Seebeck effect by chiral bulk phonon (Boltzmann equations \cite{Lifa_Seebeck}) 
			\end{itemize}
		\end{minipage} &
		\hspace{-0.6cm}\begin{minipage}[m]{4.5cm}
			\begin{itemize}
				\item Chiral attenuation of surface phonon
				\item AC spin current generation in copper 
				\item Chiral phonon transmission \cite{Xu,Onose_exp,Otani_exp,Page_exp,Page_exp_2}
				\item Non-reciprocal magnetoelastic waves
			\end{itemize}
		\end{minipage}
		\\
		\toprule
		\hspace{-0.7cm}\begin{minipage}{.45\textwidth}
			\centering\vspace*{5pt}
			\textbf{Chiral photons and electron spin}
			\includegraphics[width=4.7cm]{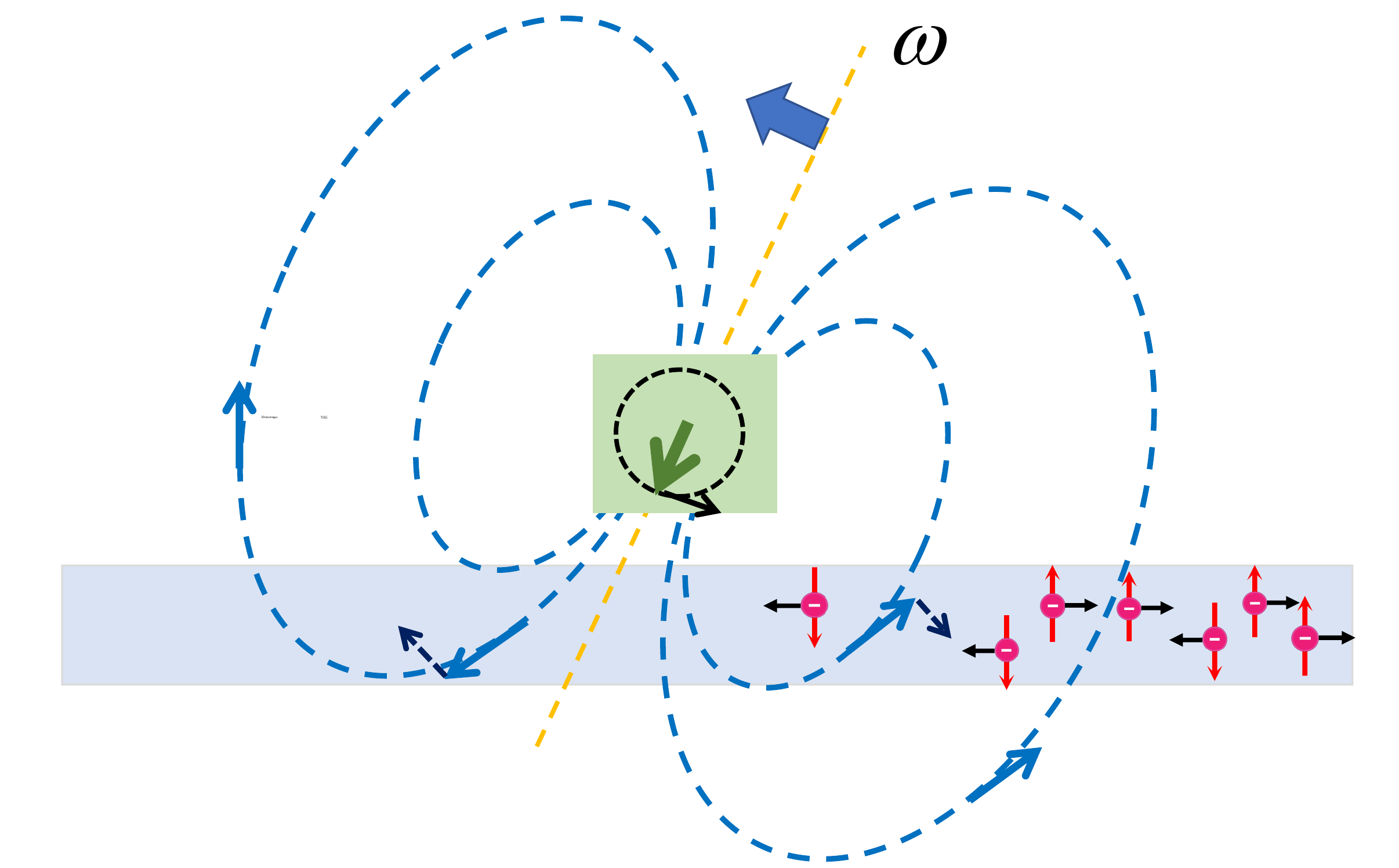}\vspace*{2pt}
		\end{minipage} &
		\hspace{-1.1cm}\begin{minipage}[m]{5.8cm}
			\begin{itemize}
				\item Transverse spin of magnetic near-field 
				 (Maxwell equations \cite{Nori,electron_spin_Yu})
				\item Contactless spin pumping\\
				 (nonlinear response theory \cite{electron_spin_Yu})
				\item Chiral spin pumping in Luttinger\\
				 liquid (linear response theory \cite{electron_spin_Yu})
			\end{itemize}
		\end{minipage} &
		\hspace{-0.5cm}\begin{minipage}[m]{4.5cm}
			\begin{itemize}
				\item No conclusive experiment available
			\end{itemize}
		\end{minipage}
		\\
		\toprule
		\hspace{-0.7cm}\begin{minipage}{.45\textwidth}
			\centering\vspace*{5pt}
			\textbf{Electron spin and SPP}
			\includegraphics[width=5.1cm]{SSP.pdf}\vspace*{2pt}
		\end{minipage} &
		\hspace{-1.1cm}\begin{minipage}[m]{5.8cm}
			      \begin{itemize}
			      	\item Transverse spin of electric near-field 
			      	(Maxwell equations \cite{Nori,plasmonics_spin_NJP,plasmonics_spin_PRB})
                  \item Spin-current generation in metals (spin-diffusion equation \cite{plasmonics_spin_PRB,plasmonics_spin_NJP})
                  \end{itemize}
		\end{minipage} &
	\hspace{-0.6cm}\begin{minipage}[m]{4.5cm}
		\begin{itemize}
			\item Enhanced spin current signals by surface plasmon resonance \cite{plasmonics_spin_APL,plasmonics_spin_NC,plasmonics_spin_PRL}
		\end{itemize}
		\end{minipage}
		\\
		\hline
	\end{tabular}
\end{table}

\subsection{Ferromagnets}
\label{nanomagnets}

The physical properties and weak excitations of ferromagnets are  affected by the dipolar self-interaction, \textit{i.e.}, the Zeeman energy of local spins in the stray field of all other spins, which introduces strong anisotropies in the magnon dispersion that depend on the shape of the magnets such as small particles, thin or thick films, or wires (Sec.~\ref{chiral_spin_waves}). The dipolar interaction causes the intrinsic chiralities in the Damon-Eshbach (DE) surface modes of ferromagnetic spheres and films (Sec.~\ref{chiral_spin_waves}). This chirality ceases to exist when the film thickness \(s\) is much smaller than the exchange length $\lambda_{\rm ex}=2\pi\sqrt{\alpha_{\rm ex}}$, which is of the order of 100~nm in YIG. The bulk and surface modes then merge into perpendicular standing spin waves (PSSW)  with $\omega_{l\bf k}= \omega_{-{l\bf k}}$, $l$ being the band index. However, a chiral interaction emerges in the form of the dipolar coupling to microwaves as generated by excited nanoscale magnetic wires and striplines. Here we set the stage by discussing the magnon modes in thin films and nanowires, without chirality themselves, in Sec.~\ref{magnon_film_wire} before introducing the chiral coupling in Sec.~\ref{dipolar_pumping},  below.

    \subsubsection{Magnons in thin films and nanowires}
    \label{magnon_film_wire}
In thin films, the exchange interaction dominates the dipolar one in the magnon energy when \(ks \gg 1\), where \(k\) is the wave number and \(s\) the magnetic film thickness. So in that limit, the dispersion relation is \(\omega_k \sim k^2\)  and the magnetization undergoes circular precession.  
When the dipolar interaction kicks in at longer wavelengths, the precession  becomes elliptical, as shown below.

\textbf{Ultrathin magnetic films}.---The magnetic mode amplitude $\mathbf{M}$ is the solution of the Landau-Lifshitz equation \cite{Landau}
\begin{equation}
{d}\mathbf{M}/{dt}=-\mu_{0}\gamma\mathbf{M}\times
\left(\mathbf{H}_{\mathrm{app}}+\mathbf{H}_{d}+\mathbf{H}_{\mathrm{ex}}\right), 
\label{EOM_film}%
\end{equation}
where $\mathbf{H}_{\mathrm{app}}=H_{\mathrm{app}}\mathbf{\hat{z}}$ is
the applied magnetic field that saturates the film magnetization $M_s\hat{\bf z}$, $\mathbf{{H}}_{d}$ is the dipolar field
[Eq.~(\ref{Edipolar})] \cite{Kalinikos}, and the exchange field
$\mathbf{H}_{\mathrm{ex}}=\alpha_{\mathrm{ex}}\left(  \partial_{x}%
^{2}\mathbf{M}+\partial_{y}^{2}\mathbf{M}+\partial_{z}%
^{2}\mathbf{M}\right)$. By translational symmetry in the film ($\hat{\mathbf{y}}$-$\hat{\mathbf{z}}$) plane,
${M}_{x,y}(\mathbf{r})={m}_{x,y}^{\mathbf{k}}(x)e^{ik_{y}
y+ik_{z}z}e^{-i\omega_{\bf k} t}$ with $\mathbf{k}\equiv k_{y}\hat{\mathbf{y}}%
+k_{z}\hat{\mathbf{z}}$. We
choose free boundary conditions $d\mathbf{M}(\mathbf{r})/dx|_{x=0,-d}%
=0$ \cite{inplane1,inplane2,inplane3}, since the lowest mode in
sufficiently thin films is not affected by dipolar pinning
\cite{exchange_1969,exchange_six_order,Stamps_0,Stamps}. 
Thus, by writing ${\bf m}^{{\bf k}}_{\pm}(x)={\bf m}^{{\bf k}}_x(x)\pm i{\bf m}^{{\bf k}}_{{y}}(x)$, we have the Fourier series \cite{Kalinikos,Kostylev},
\begin{equation}
{\bf m}^{{\bf k}}_{\pm}(x)=\sum_{l=0}^{\infty}\frac{\sqrt{2}}{\sqrt{1+\delta_{l0}}}\cos\big(l\pi x/s\big){\bf m}^{{\bf k}}_{l,\pm}.
\end{equation}
When the film
is sufficiently thin such that
$k_{y}^{2}\ll\left(  {l\pi}%
/{s}\right)  ^{2}$ for $l\geq1$, we may
confine our attention to the spin waves in the lowest branch ${l=0}$ with
amplitude governed by \cite{Kalinikos,Kostylev,Chiral_pumping_grating} 
\begin{align}
\hspace{-0.6cm}\omega_{\bf k}\left(
\begin{array}{cc}
{\bf m}^{{\bf k}}_{0,+}\\
{\bf m}^{{\bf k}}_{0,-}\\
\end{array}
\right)=
\mu_0\gamma {M}_s\left(
\begin{array}{cc}
-\Omega_H-\alpha_{\rm ex} {k}^2-\frac{1}{2}+\frac{1}{2}(1-\frac{{k}^2_y}{{k}^2})f({k})&-\frac{1}{2}+\frac{1}{2}(1+\frac{{k}^2_y}{{k}^2})f({k})\\
\frac{1}{2}-\frac{1}{2}(1+\frac{{k}^2_y}{{k}^2})f({k})&\Omega_H+\alpha_{\rm ex} {k}^2+\frac{1}{2}-\frac{1}{2}(1-\frac{{k}^2_y}{{k}^2})f({k})
\end{array}
\right)\left(
\begin{array}{cc}
{\bf m}^{{\bf k}}_{0,+}\\
{\bf m}^{{\bf k}}_{0,-}\\
\end{array}
\right),
\label{homogeneous}
\end{align}
where $f({k})\equiv 1-{1}/{(s|k_y|)}+{1}/{(s|k_y|)}\exp({-s|k_y|})$ and $\Omega_{H}\equiv H_{\mathrm{app}}/{M}_{s}$.
The dispersion \cite{Kalinikos,inplane1,inplane2,inplane3} 
\begin{equation}
\omega_{\bf k}=\mu_0\gamma{M}_s\sqrt{\left(\Omega_H+\alpha_{\rm ex} {k}^2+1-f({k})\right)\left(\Omega_H+\alpha_{\rm ex} {k}^2+\frac{{k}_y^2}{{k}^2}f({k})\right)}
\end{equation}
is reciprocal with $\omega_{{\bf k}}=\omega_{-{\bf k}}$, but is not isotropic.

    The amplitudes $m_x$ and $m_y$ are constant and when normalized by Eq.~(\ref{normalization2}) read
    \begin{align}
    	m_x({\bf k})=\sqrt{\frac{F({\bf k})+1}{4d\left(F({\bf k})-1\right)}},~~~~~m_y({\bf k})=i\sqrt{\frac{F({\bf k})-1}{4d\left(F({\bf k})+1\right)}},
    \end{align}
    in which 
    \begin{align}
    	F({\bf k})=\frac{1/2-(1/2)\left(1+k_y^2/k^2\right)f(|k_y|)}{\omega_{\bf k}/(\mu_0\gamma M_s)-\left(\Omega_H+\alpha_{\rm ex}k_y^2+1/2\right)+(1/2)\left(1-k_y^2/k^2\right)f(|k_y|)}.
    \end{align}
   The local magnetization precesses in an orbit with ${\bf k}$-dependent ellipticity. When the wavelengths are shorter than the exchange length, $|F|\gg 1$, and the precession becomes circular with ${m}_{y}^{{\bf k}}=i{m}_{x}^{{\bf k}}=i\sqrt{1/(4s)}$. When ${\bf k}\rightarrow 0$, $f(|{k_y}|)\rightarrow 0$, $\omega_{\bf k}\rightarrow \mu_0\gamma M_s\sqrt{\Omega_H(\Omega_H+1)}$, and 
   \[
   F\rightarrow \frac{1}{2}\left(\sqrt{\Omega_H^2+\Omega_H}-2\Omega_H-1\right)^{-1}=-1-2\Omega_H-2\sqrt{\Omega_H(\Omega_H+1)}.
   \]
   When $\Omega_H\rightarrow 0$ with a small static magnetic field, $F\rightarrow -1-2\sqrt{\Omega_H}$, $|m_y|\gg |m_x|$, leading to a (nearly) linearly polarized Kittel modes. 
   
\textbf{Magnetic nanowires}.---The stray fields of excited magnetic nanowires  are excellent pumps of exchange spin waves in magnetic films. For an equilibrium magnetization, $\tilde{M}_s\hat{\bf z}$ parallel to the wire and applied magnetic field the dipolar field Eq.~(\ref{EOM_film}) becomes
\begin{align}
{\bf H}_d=-N_{xx}\tilde{M}_x\hat{\bf x}-N_{yy}\tilde{M}_y\hat{\bf y},
\end{align}
where the  $N_{xx}\simeq w/(d+w)$ and $N_{yy}\simeq d/(d+w)$ \cite{Chiral_pumping_grating,magnon_trap} are demagnetization factors for thickness $d$ and width $w$. With ansatz $\tilde{M}_{x,y}\propto e^{i\kappa z}e^{-i\omega_{\kappa}t}$, where $\kappa$ and $\omega_{\kappa}$ are the momentum and frequency of the spin waves, the equation of motion reads
\begin{subequations}
\begin{align}
&i\omega_{\kappa}\tilde{M}_x=\left(\mu_0\gamma H_{\rm app}+N_{yy}\mu_0\gamma\tilde{M}_s+\lambda_{\rm ex}\kappa^2\mu_0\gamma\tilde{M}_s\right)\tilde{M}_y,\\
&i\omega_{\kappa}\tilde{M}_y=-\left(\mu_0\gamma H_{\rm app}+N_{xx}\mu_0\gamma\tilde{M}_s+\lambda_{\rm ex}\kappa^2\mu_0\gamma \tilde{M}_s\right)\tilde{M}_x.
\end{align}
\end{subequations}
 Here $\lambda_{\mathrm{ex}}$ is the exchange length in the nanowire. 
 The dispersion of the lowest mode
 \begin{equation}
    	\omega_{\kappa}=\mu_{0}\gamma\sqrt{\left(H_{\mathrm{app}}+N_{xx}\tilde{M}_{s}+\lambda_{\rm ex}\kappa^2\tilde{M}_s\right)\left(H_{\mathrm{app}}+N_{yy}\tilde{M}_{s}+\alpha_{\rm ex}\kappa^2\tilde{M}_s\right)}, \label{omegaK}%
    \end{equation}
reduces to the Kittel formula \cite{Kittel_mode} when $\kappa=0$.
The normalized amplitudes of the dipolar-exchange spin waves 
    \begin{equation}
    	\tilde{m}_{x}^{\kappa}=\sqrt{\frac{1}{4\mathcal{D}
    	(\kappa)wd}},~~~~\tilde{m}_{y}^{\kappa}=i\sqrt{\frac{\mathcal{D}(\kappa)}{4wd}}, 
    	\label{nanowire_waves}%
    \end{equation}
    where
    \begin{equation}
    	\mathcal{D}(\kappa)=\sqrt{\frac{H_{\mathrm{app}}+N_{xx}\tilde{M}_{s}+{\lambda}_{\mathrm{ex}}\kappa^{2}\tilde{M}_{s}}{H_{\mathrm{app}}+N_{yy}\tilde{M}_{s}+{\lambda}_{\mathrm{ex}}\kappa^{2}\tilde{M}_{s}}}.
    \end{equation}
precess elliptically unless the wire cross section is rectangular or cylindrical with $w=d$ and $\mathcal{D}= 1$.

An analytic treatment of the anisotropic magnon dispersion of finite systems in the dipolar-exchange regime is complicated. However, when the films are much thinner than the magnon wave length, the splitting between the perpendicular modes becomes large and the intrinsic pinning is suppressed \cite{Pirro}. The amplitude of the lowest mode is then constant over the thickness, which drastically simplifies the problem.  This regime is relevant for high-quality ultra-thin YIG films that can be grown down to nanometer thickness \cite{DMI_shift_1,ultrathin_1,ultrathin_2}.

   \subsubsection{Chiral pumping of spin waves}\label{dipolar_pumping}
    
\begin{figure}[ptbh]
    	\begin{centering}
    		\includegraphics[width=0.98\textwidth]{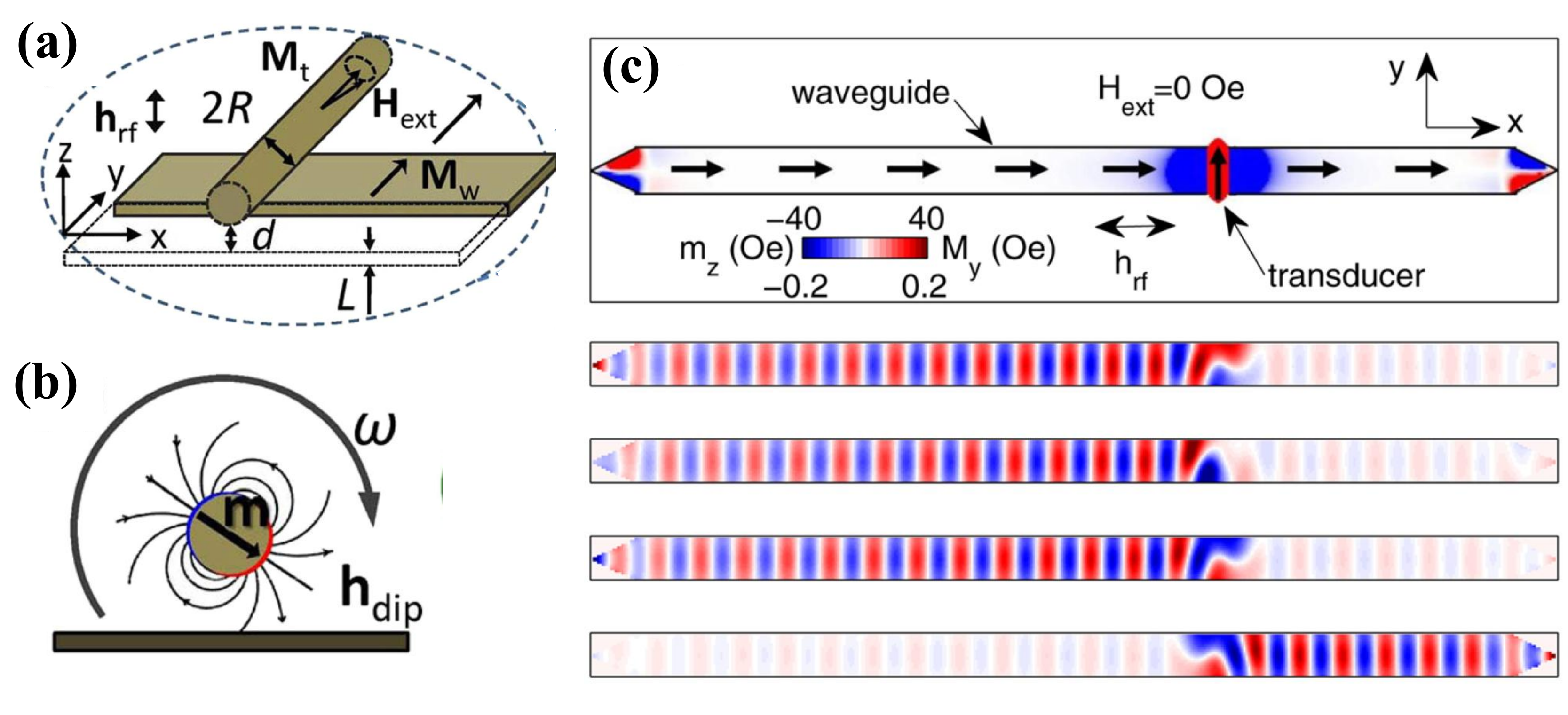}
    		\par\end{centering}
    	\caption{Micromagnetic simulations of the unidirectional excitation of  spin waves in a Py magnetic ribbon by the dynamical dipolar field of a nearby cylindrical Py wire  resonantly excited by a microwave stripline (not shown). Here $L$ is the thickness of the ribbon, $R$ is the radius of the wire, and $d$ is the distance between the bottom of the wire and the upper surface of the ribbon. (a) and (b) are sketches of the configuration and the dipolar field of the wire, respectively. (c) shows calculated results for 100 nm wide and 10 nm thick ribbon, crossed by 50 nm wide, 150 nm long, and 30 nm thick cylindrical wires with the shortest distance of 10 nm. Py's saturation magnetization is 800 kA/m and the Gilbert damping coefficient $\alpha=0.005$. The stray field of the wire excites the magnetization component $m_ y$ of half of the waveguide that can be chosen by the direction of the wire magnetization. \textcolor{blue}{In the last panel of (c), the excited magnetization appears at the opposite half space compared with the other panels with reversed saturated magnetization direction.} The figures are taken from Ref.~\cite{Au_first}.}
    	\label{Au_simulation}
    \end{figure}

Polushkin \textit{et al.}  \cite{transducer_Polushkin}  proposed an all-metallic magnetic microwave-to-spin-wave  transducer. They pointed out that the evanescent dipolar stray field of a magnetic wire antenna enhances the  incident microwave field by an order of magnitude.  Au {\it et al.} \cite{Au_first}  studied the coupling between an excited magnetic wire and a  magnetic ribbon mediated by the local dynamical dipolar field.  Figures~\ref{Au_simulation}(a) and (b) show the resonantly excited spin waves in the waveguide.  Micromagnetic simulations reveal that a wire can  unidirectionally pump exchange spin waves, see Fig.~\ref{Au_simulation}(c) with the direction that can be switched by the wire or film magnetization. These authors only address spin wave vectors normal to the equilibrium magnetizations, \textit{i.e.}, the Damon-Eshbach configuration. In the thick film limit, this chirality is therefore expected, as explained in Sec.~\ref{DE_review}, but it was surprising  for  thin films. 

  Yu $\textit{et al.}$ \cite{Chiral_pumping_Yu,Chiral_pumping_grating}  addressed the dipolar fields emitted by the Kittel mode of a magnetic wire with rectangular cross section. The analytic form Eq.~(\ref{wire_dipolar_field}) clearly reveals the spin-momentum locking of the stray magnetic field. The Kittel mode's orbit is elliptic, \textit{i.e.}, a linear combination of right and left circularly polarized components that  strongly depends on the shape of the wire cross section [refer to Eq.~(\ref{nanowire_waves})]. Pure  exchange spin waves in an underlying film magnetized parallel to the wire as in Fig.~\ref{Au_simulation}(a) are circularly polarized and couple only to the matching component of the wire dipolar field, which fixes the propagation direction of the excitation. Equivalently, we may start from  the dipolar fields of spin waves magnetic films that possess the spatial chirality illustrated by Fig.~\ref{spin_wave_field}: the dipolar field of right (left) moving spin waves vanishes below (above) the film, \textit{i.e.}, being unidirectional (also refer to Sec.~\ref{unification}). This implies that spin waves propagating along a particular direction normal to the magnetization can couple only to nanomagnets on one side of the film.

  In the following, we review the properties of the magnetic wire$\&$film device for non-collinear magnetic configurations  and elliptic precessions. When the (exchange and/or dipolar) coupling between the wire and film exceeds the lifetime broadening caused by magnetization and radiation damping, we are in the strong coupling regime and dynamics that can be accessed by solving the coupled Maxwell and LLG equations. Here we focus on the limit of weak dipolar and contact exchange interactions that are relevant for most experiments. We may then use perturbation/linear-response theory to capture the physics by transparent analytic results.
    
\begin{figure}[ptbh]
    	\begin{centering}
    		\includegraphics[width=0.99\textwidth]{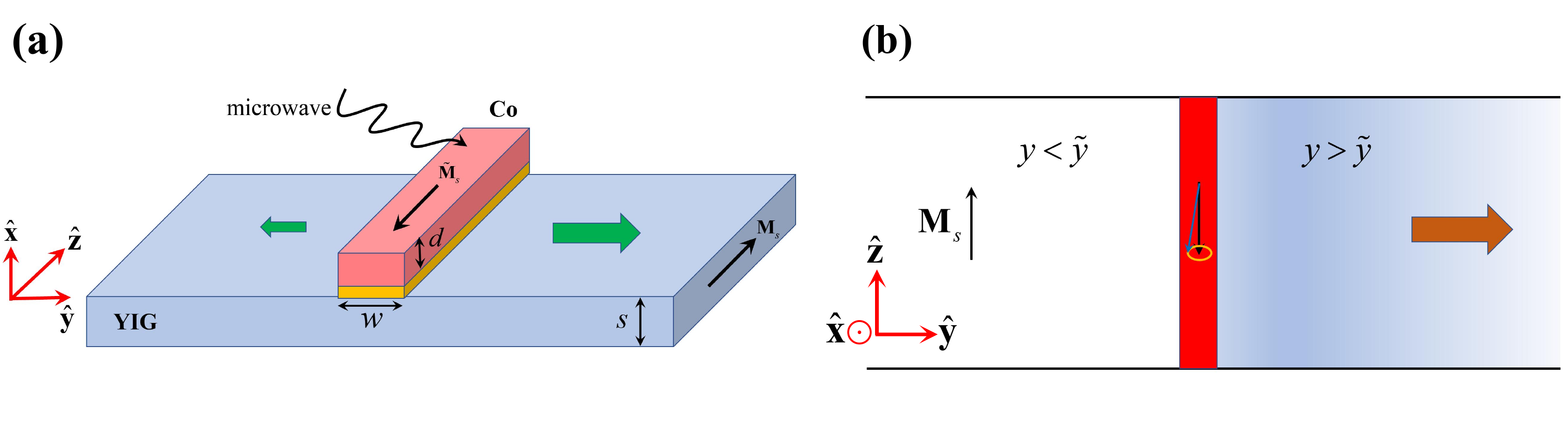}
    		\par\end{centering}
    	\caption{Chiral pumping of spin waves by a magnetic wire under the ferromagnetic resonance. (a) depicts a magnetic (Co) nanowire transducer separated by a non-magnetic spacer (optional) from a YIG film. The  $\pm \hat{\bf y}$-directions of the magnon spin currents are indicated by the green arrows, whose size indicates their magnitude. (b) illustrates the top view of ``spatial chirality'' of the excited magnetization, \textit{viz.}, the injection of magnetization, energy and momentum into half-space with $y>\tilde{y}$. The figure [(a)] is taken from Ref.~\cite{Yu_Springer}.}
    	\label{Yu_theory}
    \end{figure}

    \textbf{Linear response theory.}---Here we address the excitation of magnetization in a thin film by a proximity wire that acts as a microwave antenna, taking into account arbitrary magnetization direction \cite{Chiral_pumping_Yu,Chiral_pumping_grating}. We use perturbation theory
    \cite{Mahan,pumping_linear,spin_pumping1,Maekawa_linear} combined  with the input-output theory of the microwave scattering matrix (see below). Linear-response/perturbation theory holds in the ``weak coupling" regime, disregarding the back-action of the film dynamics on the wire. We consider an infinite long magnetic wire of rectangular cross section with thickness $d$ and width $w$ above an in-plane magnetized magnetic film of thickness $s$, as illustrated in Fig.~\ref{Yu_theory}(a) for a Co wire on top of YIG film. The Zeeman interaction Hamiltonian
    \begin{equation}
    \hat{H}_{\mathrm{int}}=-\mu_{0}\int\mathbf{\tilde{M}}(\mathbf{r},t)\cdot\mathbf{h}(\mathbf{r},t)d\mathbf{r}=-\mu_{0}\int\mathbf{M}(\mathbf{r},t)\cdot\mathbf{\tilde{h}}(\mathbf{r},t)d\mathbf{r},
    \label{H_int}
    \end{equation}
between the magnetization $\mathbf{\tilde{M}}$ of the wire and that of the film $\mathbf{M}$ via the stray dipolar magnetic fields $\mathbf{h}$ and $\mathbf{\tilde{h}}$ \cite{Landau}, respectively. Both expressions are equivalent but lead to quite different physical pictures, as briefly mentioned above. 
We introduce the  \(3 \times 3\) Cartesian Green function tensor    \begin{equation}
    \mathcal{G}(\pmb{k},x-x^{\prime })=e^{-|x-x^{\prime}||\mathbf{k}|}\left(
    	\begin{array}{ccc}
    	\frac{|\mathbf{k}|}{2} & -\frac{ik_{y}}{2}\mathrm{sgn}\left(
    	x-x^{\prime }\right) & -\frac{ik_{z}}{2}\mathrm{sgn}\left( x-x^{\prime}\right) \\
    	-\frac{ik_{y}}{2}\mathrm{sgn}\left( x-x^{\prime }\right) & -\frac{k_{y}^{2}}{2|\mathbf{k}|} & -\frac{k_{y}k_{z}}{2|\mathbf{k}|}\\
    	-\frac{ik_{z}}{2}\mathrm{sgn}\left( x-x^{\prime }\right) & -\frac{k_{y}k_{z}}{2|\mathbf{k}|} & -\frac{k_{z}^{2}}{2|\mathbf{k}|}%
    	\end{array}
    	\right) -\delta _{\mathrm{in}}(x-x^{\prime })\mathcal{I},
    	\label{Green_function_tensor}
    	\end{equation}
where $\mathcal{I}$ is the \(3 \times 3\) unity tensor and $\delta
    _{\mathrm{in}}(x)$ vanishes when $x$ lies outside the magnetic film, as a convenient instrument to formulate the dipolar fields.  A spin wave in the film of the form ${\bf m}_{\alpha}(\mathbf{k},x)e^{i\mathbf{k}\cdot\pmb{\rho}}$, for example, emits the magnetic field
    \begin{equation}
    	\left(
    	\begin{array}
    		[c]{c}%
    		{{h}}_{x}(\mathbf{r})\\
    		{{h}}_{y}(\mathbf{r})\\
    		{{h}}_{z}(\mathbf{r})
    	\end{array}
    	\right)  =e^{i\mathbf{k}\cdot\pmb{\rho}}\int_{-s}%
    	^{0}dx^{\prime}\mathcal{G}(\mathbf{k},x-x^{\prime})\left(
    	\begin{array}
    		[c]{c}%
    		{{m}}_{x}(\mathbf{k},x^{\prime})\\
    		{{m}}_{y}(\mathbf{k},x^{\prime})\\
    		{{m}}_{z}(\mathbf{k},x^{\prime})
    	\end{array}
    	\right)  . \label{Green}%
    \end{equation}
${\mathbf{h}}$ is a demagnetization field that naturally satisfies the electromagnetic boundary condition, \textit{i.e.}, continuity of the electromagnetic fields and currents at the surfaces of the film \cite{Kalinikos}.
    
    The film magnetization excited by the dynamical stray field follows from the time-dependent perturbation theory \cite{Mahan,Vignale} as:
    \begin{align}
    	\nonumber
    	\hspace{-0.7cm}{M}_{\alpha}(x,\pmb{\rho},t)&=-i\int_{-\infty}^{t}dt^{\prime}\left\langle
    	\left[  \hat{M}_{\alpha}(x,\pmb{\rho},t),\hat{H}_{\mathrm{int}}(t^{\prime
    	})\right]  \right\rangle \\
    &=\mu_{0}(\gamma\hbar)^{2}\sum_{\mathbf{k}}%
    \int_{-\infty}^{\infty}dt^{\prime}\int_{0}^{d}d\tilde{x}d\tilde{\pmb{\rho}}%
    \int_{-s}^{0}dx^{\prime}\tilde{M}_{\beta}(\tilde{x},\tilde{\pmb{\rho}}%
    ,t^{\prime})\mathcal{G}_{\beta\xi}(-\mathbf{k},\tilde{x}-x^{\prime})
    \chi_{\alpha\xi
    }\left(  x,x^{\prime};\mathbf{k};t-t^{\prime}\right)  e^{i\mathbf{k}%
    	\cdot(\pmb{\rho}-\tilde{\pmb{\rho}})},
    \label{film_magnetization}
    \end{align}
where the spin operator $\hat{S}_{\alpha}=-\hat{M}_{\alpha}/(\gamma\hbar)$ and we expanded the retarded spin susceptibility $\chi_{\alpha\delta}(x,x^{\prime};\pmb{\rho}-\pmb{\rho}^{\prime};t-t^{\prime
    })=i\Theta(t-t^{\prime})\left\langle \left[  \hat{S}_{\alpha}%
    (x,\pmb{\rho},t),\hat{S}_{\delta}(x^{\prime},\pmb{\rho}^{\prime},t^{\prime
    })\right]  \right\rangle$, 
 in the spin wave modes $m_{\alpha}^{\mathbf{k}}(x)e^{i\mathbf{k}%
	\cdot\pmb{\rho}}$ with frequency $\omega_{\mathbf{k}}$. The coefficients in the momentum-frequency space read
\begin{equation}
	\chi_{\alpha\xi}(x,x^{\prime};\mathbf{k};\omega)=-\frac{2M_{s}}{\gamma\hbar
	} \frac{m_{\alpha}^{\mathbf{k}}(x){m_{\xi}^{\mathbf{k}*}(x')}} 
	{\omega-\omega_{\mathbf{k}}+i0_{+}}.
	\label{spin_susceptibility}
\end{equation}
In the presence of Gilbert damping, the infinitesimal  $0_+$ should be replaced by \(2\alpha_G |\omega|\). Under FMR of the Kittel frequency $\omega_{\rm K}$ the wire magnetization precesses uniformly (Kittel mode). In the perturbation regime, it is safe to employ the rotating wave approximation
    \begin{equation}
    	\tilde{M}_{\beta}(\tilde{x},\tilde{\pmb{\rho}},t^{\prime})\approx\tilde
    	{M}_{\beta}(\tilde{x},\tilde{\pmb{\rho}},t)e^{i\omega_{\mathrm{K}}%
    		(t-t^{\prime})},
    	\label{wire_magnetization}
    \end{equation}
where $\mathbf{}{M}(t)$ varies slowly on the \(1/\omega_K\) time scale. We address the effects of  finite damping and back-action by the film  in the input-output theory section below.
    With Eqs.~(\ref{spin_susceptibility}) and (\ref{wire_magnetization}), the film magnetization (\ref{film_magnetization}) becomes
    \begin{align}
    M_{\alpha}(\mathbf{r},t) & =-2\mu_{0}M_{s}\gamma\hbar\sum_{\mathbf{k}}\int%
    _{0}^{d}d\tilde{x}d\tilde{\pmb{\rho}}\int_{-s}^{0}dx^{\prime}e^{i\mathbf{k}%
    	\cdot(\pmb{\rho}-\tilde{\pmb{\rho}})}\frac{m_{\alpha}(\mathbf{k},x)\tilde{M}_{\beta}%
    (\tilde{x},\tilde{\pmb{\rho}},t)}{\omega_{\mathrm{K}}%
    	-\omega_{\mathbf{k}}+i0_{+}}\nonumber\\
	&  \times  \mathcal{G}_{\beta\xi}(-\mathbf{k},\tilde{x}-x^{\prime
    }){m_{\xi}^*(\mathbf{k},x)}.
    \end{align}

When the nanowire and its equilibrium magnetizations are parallel to $\mathbf{\hat{z}}$, the excited magnetization does not depend on $z$ and the momentum integral in
    \begin{align}
    	M_{\alpha}(x,y,t)  &  =-2\mu_{0}M_{s}\gamma\hbar\int\frac{dk_{y}}{2\pi}%
    	\int_{0}^{d}d\tilde{x}d\tilde{y}\int_{-s}^{0}dx^{\prime}e^{ik_{y}(y-\tilde
    		{y})}\frac{1}{\omega_{\mathrm{K}}-\omega_{k_{y}}+i0_{+}}\nonumber\\
    	&  \times m_{\alpha}(k_{y},x)M_{\beta}(\tilde{x},\tilde{y},t)
    	\mathcal{G}_{\beta\xi
    	}(-k_{y},\tilde{x}-x^{\prime}){m_{\xi}^{\ast}(k_{y},x)}%
    \end{align}can be evaluated by contours in the complex plane. The zeros of the denominator $\omega_{\mathrm{K}}-\omega_{k_{y}}+i0_{+}$ generate two singularities at $k_{\pm}=\pm(\kappa+i0_{+})$ where $\kappa$ is the positive root of $\omega_{k_y}=\omega_{\rm K}$, so $k_{+}$ and $k_{-}$ lie in the upper and lower half planes, respectively.   When $y>\tilde{y}$ the contour should be closed in the upper half plane and
    \begin{align}
     M_{\alpha}^{>}(x,y,t)  =2i\mu_{0}M_{s}\gamma\hbar\frac{1}{v_{\kappa}}  	    	\int_{0}^{d}d\tilde{x}d\tilde{y}\int_{-s}^{0}dx^{\prime}e^{i\kappa(y-\tilde{y}%
    		)}m_{\alpha}(\kappa,x)M_{\beta}(\tilde{x},\tilde{y},t)\mathcal{G}_{\beta\xi
    	}(-\kappa,\tilde{x}-x^{\prime}){m_{\xi}^{\ast}(\kappa,x)},
 \end{align}
A small spin-wave group velocity $v_{\kappa}=\left.  \partial\omega_{k}/\partial k\right\vert
    _{k=\kappa}$ corresponds to a large density of states and excitation efficiency. When $y<\tilde
    {y}$,
    \begin{equation}
    	M_{\alpha}^{<}(x,y,t)=2i\mu_{0}M_{s}\gamma\hbar\frac{1}{v_{\kappa}}\int%
    	_{0}^{d}d\tilde{x}d\tilde{y}\int_{-s}^{0}dx^{\prime}e^{-i\kappa(y-\tilde
    		{y})}m_{\alpha}(-\kappa,x)M_{\beta}(\tilde{x},\tilde{y},t){\cal G}_{\beta\xi
    	}(\kappa,\tilde{x}-x^{\prime}){m_{\xi}^{\ast}(-\kappa,x)}.
    \end{equation}
    When the spin waves in the film are circularly polarized with $m_{y}=im_{x}$,
    \begin{equation}
    	{\cal G}_{\beta\xi}(-\kappa,\tilde{x}-x^{\prime})
    	{m_{\xi}^{\ast}%
    		(\kappa,x)}\longrightarrow\frac{e^{-|\tilde{x}-x^{\prime}||\kappa|}}{2}\left(
    	\begin{array}
    		[c]{cc}%
    		\kappa & i\kappa\\
    		i\kappa & -\kappa
    	\end{array}
    	\right)  \left(
    	\begin{array}
    		[c]{c}%
    		m_{x}\\
    		-im_{x}%
    	\end{array}
    	\right)  =\left(
    	\begin{array}
    		[c]{c}%
    		0\\
    		0
    	\end{array}
    	\right),
    \end{equation}
leading to zero $M_{\alpha}^{<}(x,y,t)$, but finite $M_{\alpha}^{>}(x,y,t)$, \textit{i.e.}, perfect chirality. When the magnetic nanowire excites only spin waves with positive momentum, energy and momentum are injected into the half-space $y>\tilde{y}$, as sketched in Fig.~\ref{Yu_theory}(b). This ``spatial chirality'' is a consequence of the causality or retardation in electrodynamics and persists in the presence of magnetic damping.
    
   \textbf{Input-output theory.}---References \cite{input_output_Cardiner,input_output_Clerk} addressed the scattering matrix of waves in finite systems such as cavities that includes (quantum) noise and dissipation. Here we employ this formalism to calculate the transmission and reflection amplitudes of microwave photons between emitters and detectors formed by striplines or co-planar wave guides, which are modulated by the interaction with the spin waves \cite{Chiral_pumping_Yu,magnon_trap}. 
     As above, we adopt the configuration of a weakly excited magnetic nanowire on a thin magnetic film and use the language of second quantization. The Cartesian components $\beta\in\left\{
    x,y\right\}  $ of the magnetization dynamics of film ($\mathbf{\hat{M}}$) and nanowire ($\mathbf{\hat{\tilde{M}}}$) can be expanded into magnon creation and annihilation operators \cite{Kittel_book,HP},
    \begin{subequations}
    \begin{align}
    	\hat{M}_{\beta}(\mathbf{r})  &  =-\sqrt{2M_{s}\gamma\hbar}\sum_{\mathbf{k}%
    	}\left[  m_{\beta} (\mathbf{k},x)e^{i{\mathbf{k}}%
    		\cdot\pmb{\rho}}\hat{\alpha}_{\mathbf{k}}+\mathrm{h.c.}\right],\\
    	\hat{\tilde{M}}_{\beta}(\mathbf{r})  &  =-\sqrt{2\tilde{M}_{s}\gamma\hbar}%
    	\sum_{k_{z}}\left[  \tilde{m}_{\beta}(k_{z},x,y) e^{ik_{z}z}\hat{\beta
    	}_{k_{z}}+\mathrm{h.c.}\right],
    	\label{expansion}
    \end{align}
    \end{subequations}
where $M_{s}$ and $\tilde{M}_{s}$ are the saturation magnetizations, $m_{\beta} ( \mathbf{k},x)$ and $\tilde{m}_{\beta} (k_{z},x,y)$ are the normalized spin wave mode amplitudes, while $\hat{\alpha}_{\mathbf{k}}$ and $\hat{\beta}_{k_{z}}$denote the magnon (annihilation) operators in the film and nanowire, respectively. 

We focus, from now on, on ultrathin films with typical dimensions $s,d=\mathcal{O}\left(  10~\mathrm{nm}\right)  $ and nanowires of width $w=\mathcal{O}\left(  50~\mathrm{nm}\right)$. These structures can be fabricated with high quality only recently \cite{ultrathin_1,ultrathin_2,Haiming_exp_grating,Jilei_PRL}.  The  weakly excited magnetizations of film and nanowire (centered at $y_{0}\hat{\mathbf{y}%
    }$) are then nearly constant over a cross section of film and wire (Sec.~\ref{magnon_film_wire}) \cite{Chiral_pumping_Yu,Chiral_pumping_grating}, \textit{viz.}:
\begin{subequations}
\begin{align}
	&m_{\beta}^{\left(  \mathbf{k}\right)  }(x)\approx
	m_{\beta}^{\left(  \mathbf{k}\right)  }\Theta(-x)\Theta(x+s),\\
	&\tilde{m}_{\beta}^{\left(  k_{z}\right)  }(x,y)\approx\tilde{m}_{\beta}^{\left(
		k_{z}\right)  }\Theta(x)\Theta(-x+d)\Theta(y-y_{0}+w/2)\Theta(-y+y_{0}+w/2),
\end{align}
\end{subequations}
 where $\Theta(x)$ is the Heaviside step function. The mode selectivity of the dipolar coupling allows us to disregard higher perpendicular standing spin waves, even when they are thermally occupied and important for diffuse magnon transport. 
    The system Hamiltonian then simplifies to
    \begin{align}
    \hat{H}/\hbar  =\sum_{\mathbf{k}}\omega_{\mathbf{k}}\hat{\alpha}%
    _{\mathbf{k}}^{\dagger}\hat{\alpha}_{\mathbf{k}}+\sum_{k_{z}}\tilde{\omega}_{k_{z}}\hat{\beta}_{k_{z}}^{\dagger}\hat{\beta}_{k_{z}}
    +\sum_{\mathbf{k}}\left(  g_{\mathbf{k}}e^{-ik_{y}y_{0}}\hat{\alpha
    }_{\mathbf{k}}^{\dagger}\hat{\beta}_{k_{z}}+g_{\mathbf{k}}^{\ast}
    e^{ik_{y}y_{0}}\hat{\beta}_{k_{z}}^{\dagger}\hat{\alpha}_{\mathbf{k}}\right),
    \end{align}
where $\omega_{\mathbf{k}}$ and $\tilde{\omega}_{k_{z}}$ are the frequencies of spin waves in the film and nanowire and the Zeeman coupling constant follows from the results of the previous Sec.~\ref{magnon_film_wire}
    \begin{equation}
    	g_{\mathbf{k}}=F(\mathbf{k})\left(  m_{x}^{\ast}(\mathbf{k}),m_y ^{\ast}(\mathbf{k})%
    	\right)  \left(
    	\begin{array}
    		[c]{cc}%
    		|\mathbf{k}| & ik_{y}\\
    		ik_{y} & -k_{y}^{2}/|\mathbf{k}|
    	\end{array}
    	\right)  \left(
    	\begin{array}
    		[c]{c}%
    		\tilde{m}_x (k_{z})\\
    		\tilde{m}_{y}(k_{z}) %
    	\end{array}
    	\right)  ,\label{coupling_constant}%
    \end{equation}
   where $F(\mathbf{k})=-\mu_{0}\gamma \phi\left(
    \mathbf{k}\sqrt{M_{s}\tilde{M}_{s}/L}\right)  $. The form factor 
     \begin{equation}
     \phi\left(  \mathbf{k}\right)
    =2\sin(k_{y}w/2) \frac{(1-e^{-kd})(1-e^{-ks})}{k_{y}{k^{2}}}
     \end{equation}
is significant when the wavelengths are of the order of the nanowire width. Right-circularly polarized exchange waves with $m_{y}^{(k_{y})}=im_{x}^{(k_{y})}$ couple with perfect chirality $g_{-|k_{y}|}=0$.
    
   The microwave photon operator $\hat{p}_{k_{z}}(t)=\int\hat{p}_{k_{z}}(\omega)e^{-i\omega t}d\omega/(2\pi)$  represents a coherent input centered at the frequency $\tilde{\omega}_{k_{z}}$ with expectation value $\langle\hat{p}_{k_{z}}(\omega)\rangle\rightarrow2\pi
    \mathit{\mathcal{D}}\delta(\omega-\tilde{\omega}_{k_{z}})$ and amplitude  $\mathcal{D}$. The Heisenberg-like operator equations of motion of the input-output formulation are then
    \begin{subequations}
    \begin{align}
    	\frac{d\hat{\beta}_{k_{z}}}{dt}  &  =-i\tilde{\omega}_{k_{z}}\hat{\beta
    	}_{k_{z}}(t)-\sum_{k_{y}}ig_{\mathbf{k}}^{\ast}e^{ik_{y}y_{0}}\hat{\alpha
    	}_{\mathbf{k}}(t)-\left(  \frac{\tilde{\kappa}_{k_{z}}}{2}+\frac{\zeta_{k_{z}}}{2}\right)
    	\hat{\beta}_{k_{z}}(t)-\sqrt{\tilde{\kappa}_{k_{z}}}\hat{\tilde{N}}_{k_{z}%
    	}(t)-\sqrt{\zeta_{k_{z}}}\hat{p}_{k_{z}}(t),\\
    	\frac{d\hat{\alpha}_{\mathbf{k}}}{dt}  &  =-i\omega_{\mathbf{k}}\hat{\alpha
    	}_{\mathbf{k}}(t)-ig_{\mathbf{k}}e^{-ik_{y}y_{0}}\hat{\beta}_{k_{z}}%
    	(t)-\frac{\kappa_{\mathbf{k}}}{2}\hat{\alpha}_{\mathbf{k}}(t)-\sqrt{\kappa_{\mathbf{k}}}\hat{N}_{\mathbf{k}}(t),
    \end{align}
\end{subequations}
 where $\tilde{\kappa}_{k_{z}}\equiv2\tilde{\alpha}_G\tilde{\omega}_{k_{z}}%
    \ $($\kappa_{\mathbf{k}}\equiv2\alpha_G\omega_{\mathbf{k}}$) is the magnon dissipation rate in terms of the Gilbert damping constant $\tilde{\alpha}_G$ $(\alpha_G)$ in the nanowire (film) and $\zeta_{k_{z}}$ is the photon dissipation rate. The thermal environment causes random fluctuations in the magnon amplitudes $\hat
    {N}_{\mathbf{k}}$ \cite{input_output_Cardiner,input_output_Clerk}. Their correlation functions can be calculated assuming Markovian statistics that obeys the (quantum) fluctuation-dissipation theorem with $\langle{\hat
    	{N}_{\mathbf{k}}}\rangle=0$ and $\langle{\hat{N}_{\mathbf{k}}^{\dagger}%
    	(t)\hat{N}_{\mathbf{k}^{\prime}}(t^{\prime})}\rangle=n_{\mathbf{k}}%
    \delta(t-t^{\prime})\delta_{\mathbf{k}\mathbf{k}^{\prime}}$. $n_{\mathbf{k}%
    }=1/\left\{  \exp\left[  {\hbar\omega_{\mathbf{k}}}/({k_{B}T_{2}})\right]
    -1\right\}  $ is the magnon population at temperature $T_{2}$ of the film, and $k_{B}T_{2}$ should be larger than the spectral width of the mode \(\mathbf{k}\) (narrow band white noise).  The thermal fluctuations $\hat{\tilde{N}}_{k_{z}}$ in the nanowire are governed by a different temperature $T_{1}$ and distribution $\tilde
    {n}_{k_{z}}=1/\left\{  \exp\left[  {\hbar\tilde{\omega}_{k_{z}}}/({k_{B}T_{1}%
    })\right]  -1\right\}  $. The solutions to the equations of motion are then
    \begin{subequations}
    \begin{align}
    	\hat{\beta}_{k_{z}}(\omega)  &  =\frac{i\sum\limits_{k_{y}}\gamma_{\mathbf{k}%
    		}G_{\mathbf{k}}\hat{N}_{\mathbf{k}}(\omega)-\sqrt{\tilde{\kappa}_{k_{z}}}%
    		\hat{\tilde{N}}_{k_{z}}(\omega)-\sqrt{\zeta_{k_{z}}}\hat{p}_{k_{z}}(\omega
    		)}{-i(\omega-\tilde{\omega}_{k_{z}})+\frac{\tilde{\kappa}_{k_{z}}}{2}%
    		+\frac{\zeta_{k_{z}}}{2}+i\sum_{k_{y}}|g_{\mathbf{k}}|^{2}G_{\mathbf{k}%
    		}\left(  \omega\right)  },
    		\label{shift}\\
    	\hat{\alpha}_{\mathbf{k}}(\omega)  &  =G_{\mathbf{k}}\left(  \omega\right)
    	\left(  g_{\mathbf{k}}e^{-ik_{y}y_{0}}\hat{\beta}_{k_{z}}(\omega
    	)-i\sqrt{\kappa_{\mathbf{k}}}\hat{N}_{\mathbf{k}}(\omega)\right)  ,
    	\label{Kittel_excitation}%
    \end{align}
\end{subequations}
 where the Green function $G_{\mathbf{k}}\left(  \omega\right)  =\left(
    (\omega-\omega_{\mathbf{k}})+i\kappa_{\mathbf{k}}/2\right)  ^{-1}$ and
    $\gamma_{\mathbf{k}}=ig_{\mathbf{k}}^{\ast}e^{ik_{y}y_{0}}\sqrt{\kappa
    	_{\mathbf{k}}}$ reveal the effect of thermal fluctuations in both wire
    ($\hat{\tilde{N}}_{k_{z}}$) and film ($\hat{N}_{\mathbf{k}}$) on  $\hat{\beta}_{k_{z}}(\omega)$. Moreover, the magnon pumping enhances the damping in the nanowire by
    \begin{align}
    \delta\tilde{\kappa}_{k_{z}}=2\pi\sum_{k_{y}
    }|g_{\mathbf{k}}|^{2}\delta(\tilde{\omega}_{k_{z}}-\omega_{\mathbf{k}})
    \end{align}
    to
    $\tilde{\kappa}_{k_{z}}^{\prime}$and the frequency shift to $\tilde{\omega
    }_{k_{z}}^{\prime}$ follows from the real part of $\sum_{k_{y}}|g_{\mathbf{k}}|^{2}G_{\mathbf{k}}$ .
   The (plasmonically enhanced) uniform microwave field drives the Kittel mode ($k_{z}=0$)  at the ferromagnetic resonance of the nanowire, which can be pushed to much higher frequencies than the Kittel mode of the film, \textit{e.g.}, by the high form anisotropy  associated with  a large magnetization. Spin waves  with finite $k_{y}\equiv q$ in the film are excited directly only by the inhomogeneous components of the  microwave field, which is dominated by the stray fields emitted by the magnetization dynamics of the wire. 
   
The equations of motion  can now be solved for the magnon operators. With $T_{1}=T_{2}\equiv T_{0}$ 
    \begin{equation}
    	\hat{\alpha}_{q}(t)=\int\frac{d\omega}{2\pi}\frac{e^{-i\omega t}%
    		ig_{q}e^{-iqy_{0}}}{-i(\omega-\omega_{q})+\frac{\kappa_{q}}{2}}\frac
    	{\sqrt{\zeta_{0}}\hat{p}_{0}(\omega)}{-i(\omega-\tilde{\omega}_{0}^{\prime
    		})+\frac{\tilde{\kappa}_{0}^{\prime}}{2}+\frac{\zeta_{0}}{2}}.
    \end{equation}
Under coherent excitation of the wire $\langle\hat{\alpha
    }_{q}^{\dagger}(t)\hat{\alpha}_{q}(t)\rangle=\langle\hat{\alpha}_{q}^{\dagger
    }(t)\rangle\langle\hat{\alpha}_{q}(t)\rangle$ and in the absence of damping, $\kappa_{q}$ is a positive infinitesimal that safeguards causality. The magnetization in position and time domain reads
    \begin{align}
    	\delta{M}_{\beta}(\mathbf{r},t)  =\sqrt{2M_{s}\gamma\hbar}\mathcal{D}\frac
    	{1}{v_{q_{\ast}}}\frac{e^{-i\tilde{\omega}_{0}t}\sqrt{\zeta_{0}}}
    	{(\tilde{\omega}_{0}-\tilde{\omega}_{0}^{\prime})+i(\tilde{\kappa}_{0}^{\prime}+\zeta_{0})/2}
    	\left\{
    	\begin{array}
    		[c]{c}%
    		m_{\beta}^{(q_{\ast})}(x)g_{q_{\ast}}e^{iq_{+}(y-y_{0})}+\mathrm{h.c.}\\
    		m_{\beta}^{(-q_{\ast})}(x)g_{-q_{\ast}}e^{iq_{-}(y-y_{0})}+\mathrm{h.c.}%
    	\end{array}
    	\text{ for }%
    	\begin{array}
    		[c]{c}%
    		y>y_{0}\\
    		y<y_{0}%
    	\end{array}
    	\right.  . \label{excited_M}%
    \end{align}
  $\tilde{\omega}_{0}-\omega_{q}+i\kappa_{q}/2$ vanishes for  $q_{\pm}=\pm(q_{\ast}+i\delta_{\Gamma})$ where $q_{\ast
    }>0$, $\delta_{\Gamma}$ is the inverse propagation length, and $v_{q_{\ast}}=\left.  \partial\omega_{q}/\partial q\right\vert
    _{q_{\ast}}$ is the magnon group velocity. Perfect chiral coupling  $g_{-q_{\ast}}=0$ implies actuation of the magnetization only in the half space $y>y_{0}$.
    
  \textbf{Microwave transmission
    spectra}.---Coherent chiral pumping can be observed in microwave transmission
    spectra \cite{Dirk_transducer,Chuanpu_NC,Jilei_PRL} with two nanowires at
    $\mathbf{r}_{1}=R_{1}\hat{\mathbf{y}}$ and $\mathbf{r}_{2}=R_{2}%
    \hat{\mathbf{y}}$ that act as inductive actuators and detectors. 
    When representing the magnons in the wires $R_{1}$ and $R_{2}$  by the
    operators $\hat{m}_{L}$ and $\hat{m}_{R}$, respectively, the equations of motion in the time domain become
    \begin{subequations}
    \begin{align}
    	\frac{d\hat{m}_{L}}{dt} &  =-i\omega_{\mathrm{K}}\hat{m}_{L}(t)-i\sum_{q}%
    	g_{q}e^{iqR_{1}}\hat{\alpha}_{q}(t)-\left(  \frac{\kappa_{L}}{2}+\frac
    	{\kappa_{p,L}}{2}\right)  \hat{m}_{L}(t)-\sqrt{\kappa_{p,L}}\hat
    	{p}_{\mathrm{in}}^{(L)}(t),\\
    	\frac{d\hat{m}_{R}}{dt} &  =-i\omega_{\mathrm{K}}\hat{m}_{R}(t)-i\sum_{q}%
    	g_{q}e^{iqR_{2}}\hat{\alpha}_{q}(t)-\frac{\kappa_{R}}{2}\hat{m}_{R}%
    	(t),\\
    	\frac{d\hat{\alpha}_{q}}{dt} &  =-i\omega_{q}\hat{\alpha}_{q}(t)-ig_{q}%
    	e^{-iqR_{1}}\hat{m}_{L}(t)-ig_{q}e^{-iqR_{2}}\hat{m}_{R}(t)-\frac{\kappa_{q}%
    	}{2}\hat{\alpha}_{q}(t),
    \end{align}
\end{subequations}
where $\kappa_{L}$ and $\kappa_{R}$ are the intrinsic magnetization damping rates of the Kittel
modes in the left and right nanowires, respectively, $\kappa_{p,L}$ represents the radiative damping of the microwave photons $\hat
    {p}_{\mathrm{in}}^{(L)}$ by the coupling of the left nanowire with the microwave source, and $\kappa_{q}$ denotes the intrinsic damping of
the spin waves in the film. In frequency space:
    \begin{subequations}
    \begin{align}
    	\hat{\alpha}_{q}(\omega) &  =g_{q}G_{q}\left(  \omega\right)  \left[
    	e^{-iqR_{1}}\hat{m}_{L}(\omega)+e^{-iqR_{2}}\hat{m}_{R}(\omega)\right],\\
    	\hat{m}_{R}(\omega) &  =\frac{-i\sum_{q}g_{q}^{2}G_{q}\left(  \omega\right)
    		e^{iq(R_{2}-R_{1})}}{-i(\omega-\omega_{\mathrm{K}})+\kappa_{R}/2+i\sum
    		_{q}g_{q}^{2}G_{q}\left(  \omega\right)  }\hat{m}_{L}(\omega),\\
    	\hat{m}_{L}(\omega) &  =\frac{-\sqrt{\kappa_{p,L}}}{-i(\omega-\omega
    		_{\mathrm{K}})+(\kappa_{L}+\kappa_{p,L})/2+i\sum_{q}g_{q}^{2}G_{q}\left(
    		\omega\right)  -f(\omega)}\hat{p}_{\mathrm{in}}^{(L)}(\omega),
    \end{align}
\end{subequations}
    where $G_{q}\left(  \omega\right)  =\left[  (\omega-\omega_{q})+i\kappa
    _{q}/2\right]  ^{-1}$ and
    \begin{equation}
    	f(\omega)\equiv-\frac{\left(  \sum_{q}g_{q}^{2}G_{q}\left(  \omega\right)
    		e^{iq(R_{1}-R_{2})}\right)  \left(  \sum_{q}g_{q}^{2}G_{q}\left(
    		\omega\right)  e^{iq(R_{2}-R_{1})}\right)  }{-i(\omega-\omega_{\mathrm{K}%
    		})+\kappa_{R}/2+i\sum_{q}g_{q}^{2}G_{q}\left(  \omega\right)  }.
    \end{equation}
   When the chiral coupling is perfect $f(\omega)$ vanishes without the back-action. The spin waves excited by the left nanowire in the film propagate to the right nanowire, where microwave striplines or wave guides detect their stray fields  inductively. The relation between the output 
    $\hat{p}_{\mathrm{out}}^{(L)}(\omega)$ and $\hat{p}_{\mathrm{out}}%
    ^{(R)}(\omega)$ and the input  $\hat{p}_{\mathrm{in}}^{(L)}(\omega)$ amplitudes  is
    \cite{input_output_Cardiner,input_output_Clerk}
    \begin{subequations}
    \begin{align}
    	\hat{p}_{\mathrm{out}}^{(L)}(\omega) &  =p_{\mathrm{in}}^{(L)}(\omega
    	)+\sqrt{\kappa_{p,L}}\hat{m}_{L}(\omega),\\
    	\hat{p}_{\mathrm{out}}^{(R)}(\omega) &  =\sqrt{\kappa_{p,R}}\hat{m}_{R}%
    	(\omega),
    \end{align}
\end{subequations}
    where $\kappa_{p,R}$ is the radiative damping rate induced by the
    detector. The elements in the microwave scattering matrix, \textit{i.e.},
    microwave reflection $\left(  S_{11}\right)  $ and transmission $\left(
    S_{21}\right)  $ amplitudes are then
    \begin{subequations}
    \begin{align}
    	S_{11}(\omega) &  \equiv\frac{\hat{p}_{\mathrm{out}}^{(L)}}{\hat
    		{p}_{\mathrm{in}}^{(L)}}=1-\frac{\kappa_{p,L}}{-i(\omega-\omega_{\mathrm{K}%
    		})+(\kappa_{L}+\kappa_{p,L})/2+i\sum_{q}g_{q}^{2}G_{q}\left(  \omega\right)
    		-f(\omega)},\\
    	S_{21}(\omega) &  \equiv\frac{\hat{p}_{\mathrm{out}}^{(R)}}{\hat
    		{p}_{\mathrm{in}}^{(L)}}=\left[  1-S_{11}(\omega)\right]  \sqrt{\frac
    		{\kappa_{p,R}}{\kappa_{p,L}}}\frac{i\sum_{q}g_{q}^{2}G_{q}\left(
    		\omega\right)  e^{iq(R_{2}-R_{1})}}{-i(\omega-\omega_{\mathrm{K}})+\kappa
    		_{R}/2+i\sum_{q}g_{q}^{2}G_{q}\left(  \omega\right)  }.\label{S}
    \end{align}
\end{subequations}

\begin{figure}[th]
    	\begin{center}
    	{\includegraphics[width=16.6cm]{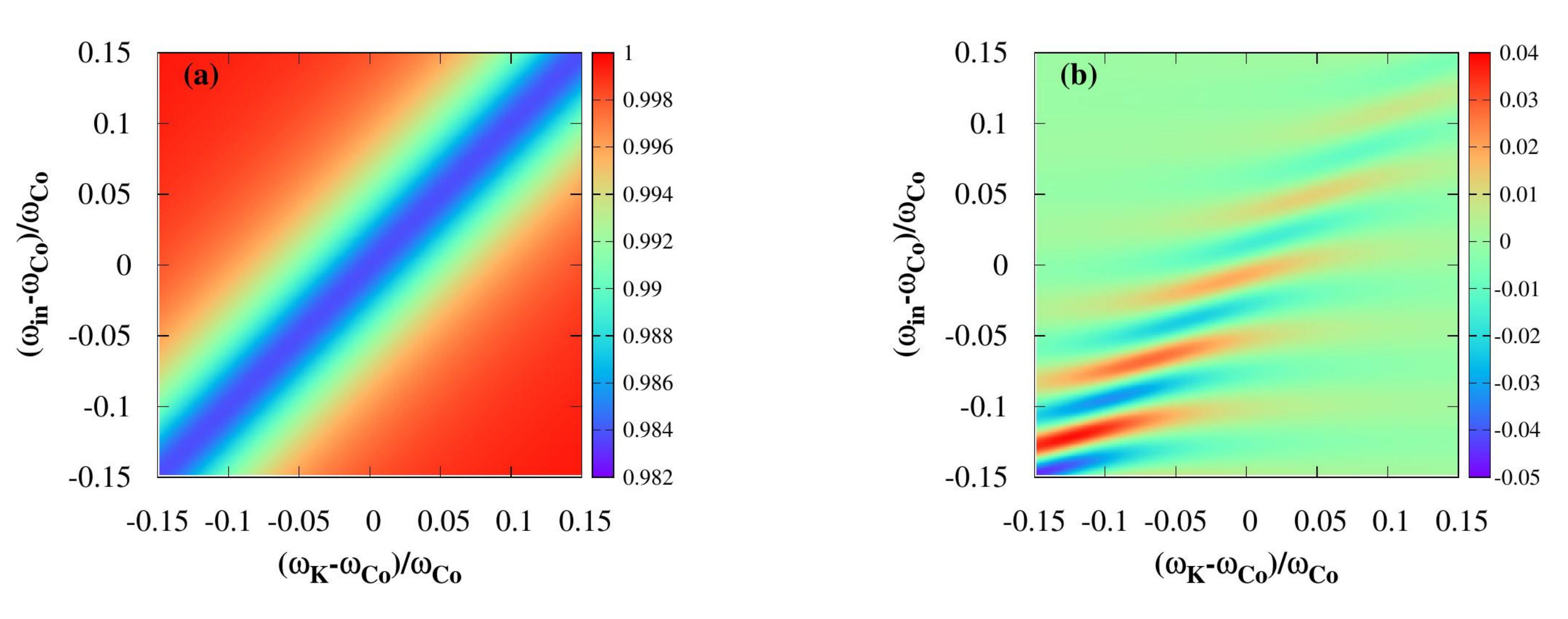}} 
    	\end{center}
    	\caption{Reflection $\operatorname{Re}(S_{11})$ [(a)] and transmission $\operatorname{Re}(S_{12})$ [(b)] amplitudes of microwaves, Eq.~(\ref{S}), between two parallel magnetic nanowires on a magnetic film. $\omega
    		_{\mathrm{Co}}$ is the Kittel mode frequency Eq.~(\ref{omegaK}) of the cobalt nanowire at a  fixed applied magnetic field ($H^{0}_{\mathrm{app}}=0.05$~T) that forces the magnetization of the soft YIG film to be antiparallel to that of the wires. $\omega_{\mathrm{in}}$ is the microwave frequency and $\omega_{\mathrm{K}}$ is the  Co Kittel mode frequency as a function of $H_{\mathrm{app}}$ close to $H^{0}_{\mathrm{app}}$. The radiative damping rate for both nanowires is chosen as $\kappa_{p}/(2\pi)=10$~MHz. The Co nanowire with magnetization $\mu_{0}\tilde{M}_{s}=1.1$~T is $w=70$~nm wide and $d=20$~nm thick. We adopt an exchange stiffness $\alpha_{\rm ex}^{({\rm Co})}=3.1\times10^{-13}$~cm$^{2}$, and the Gilbert damping coefficient $\alpha_G^{\mathrm{(Co)}}=2.4\times10^{-3}$. The YIG film thickness $s=20$~nm with magnetization $\mu_{0}M_{s}=0.177$~T and exchange stiffness $\alpha^{(\rm YIG)}_{\rm ex}=3.0\times 10^{-12}$~cm$^2$. The figures are adopted from Ref.~\cite{Yu_Springer}.}
    	\label{transmission}%
    \end{figure}

Figure~\ref{transmission} shows the real components of $S_{11}$ and $S_{12}$ when the magnetizations of Co nanowire and YIG film are anti-parallel, which can be realized in experiments because the magnetic form anisotropy of the wire is  much larger than that of the film. The absorption of the microwaves at the FMR of the  Co wire causes the pronounced dip in the reflection amplitude in Fig.~\ref{transmission}(a). The fringes in the transmission spectra in Fig.~\ref{transmission}(b) reflect the spin wave propagation in the film from the left to the right nanowire, caused by the propagation phase factor $e^{iq(R_{1}-R_{2})}$ in Eq.~(\ref{S})
that oscillates with the momentum $q$ selected by the frequency \(\omega_{in}\). This is not an interference pattern since spin waves are not reflected at the right nanowire. The chirality is implicit: When $g_{-|q|}=0$ and  $R_{2}>R_{1}$, spin and energy are transmitted from 1 to 2 and \(S_{21} \ne 0\),  while no waves can travel in the opposite direction, \textit{i.e.},  \(S_{21} = 0\).
    
\begin{figure}[ptbh]
    	\begin{centering}
    		\includegraphics[width=0.99\textwidth]{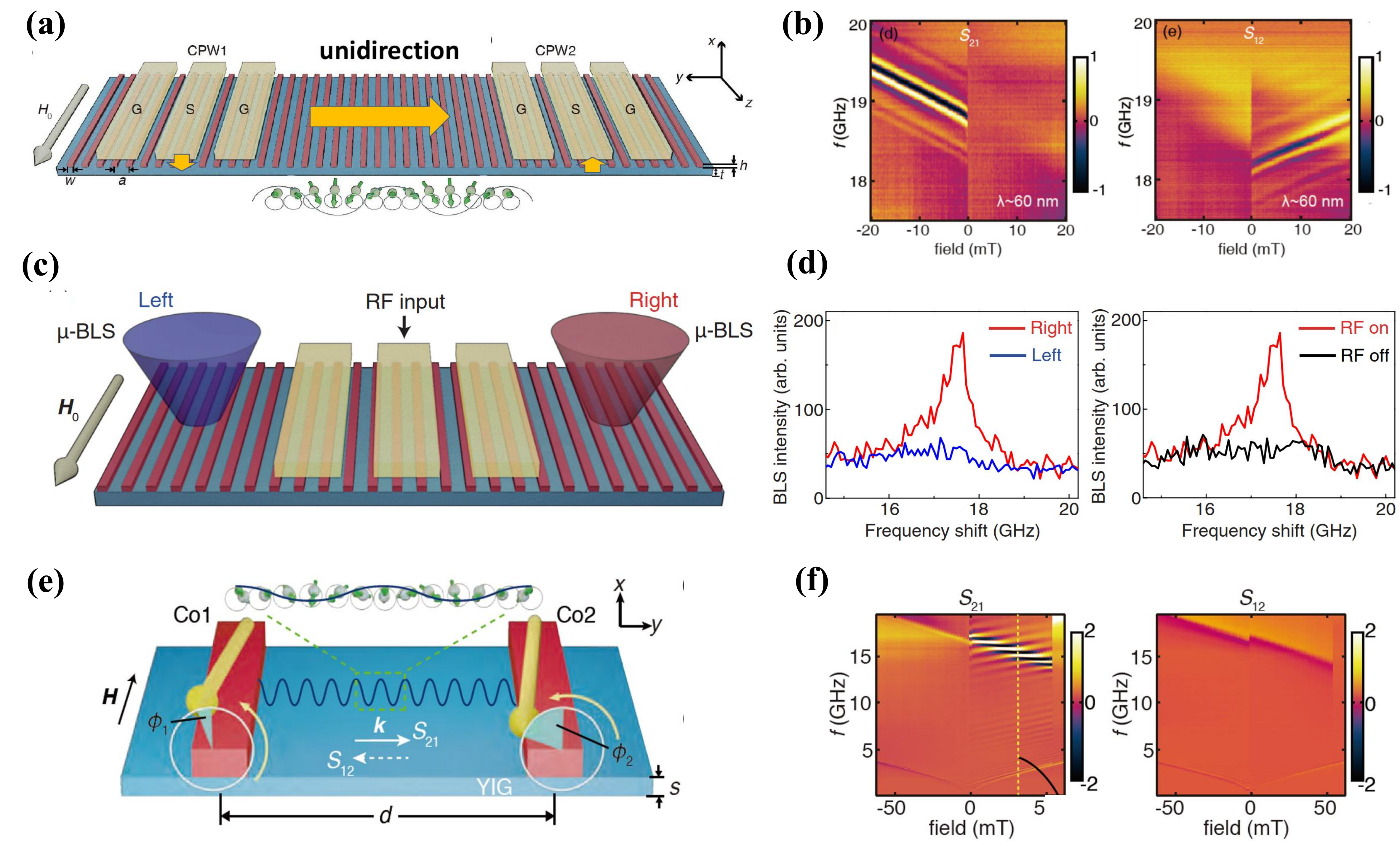}
    		\par\end{centering}
    	\caption{Observation of chiral pumping of spin waves with magnetic wire array [(a)] or two magnetic wires [(b)] on top of thin YIG films. The Co wire magnetizations are either parallel or anti-parallel to that of YIG film in these observations. (b) and (f) are the microwave transmission amplitudes from port 1 to 2 ($S_{21}$) and vice versa ($S_{12}$). (c) illustrates the setup of  the Brillouin light scattering experiments of the magnon distribution in the magnetic wires with results shown in (d). The figures are taken from Ref.~\cite{Haiming_exp_grating,Haiming_exp_wire}.}
    	\label{Haiming_exp}
    \end{figure}

    \textbf{Experimental evidence.}---Chiral microwave transmission was reported in two different systems. Chen \textit{et al}. \cite{Haiming_exp_grating} first reported unidirectional pumping of dipolar-exchange spin waves in a Co nanowire array on top of ultrathin (\(s=\)20 nm)  YIG films, as illustrated in Fig.~\ref{Haiming_exp}(a).  Subsequently, they simplified the sample to only two magnetic nanowires \cite{Haiming_exp_wire}, as shown in Fig.~\ref{Haiming_exp}(e), which is the configuration discussed above.
   In the earlier experiments, a co-planar waveguide on top of the device generates  nearly uniform microwaves that hardly excite propagation spin waves in the YIG film at low resonant frequencies.  When reaching the resonance of the Co nanowire array at much higher frequencies, the dipolar  or interface exchange interaction  with the precessing Co magnetization excites spin waves in the YIG film with momenta  up to the order of the reciprocal wire width. The spin waves travel to and inductively excite another coplanar waveguide. Figure~\ref{Haiming_exp}(b) shows that the device functions indeed as an almost perfect and switchable microwave diode.  For the negative applied field (anti-parallel magnetizations) the transmission $S_{21}$ from waveguide 1 to 2  is finite, but it vanishes for the opposite direction $S_{12}$.  Brillouin light scattering detects an imbalance in the magnon numbers in the Co wires on both sides of a waveguide, as illustrated in Figs.~\ref{Haiming_exp}(c) and (d) \cite{Hanchen_damping}. Since magnetizations are anti-parallel, excess magnons are detected on the right side. 
The effects observed for the nanowire array can be interpreted in terms of the chiral pumping mechanism via the dipolar interaction as explained above with minor modifications for the array configuration \cite{Chiral_pumping_grating}. The unidirectionality of the spin-wave excitation switches with the YIG magnetization from the parallel to anti-parallel configurations as shown in Fig.~\ref{transmission}(b) with symmetry $S_{12}(\textbf{M})=S_{21}(-\textbf{M})$ because of the chirality of dipolar field of the wire that is locked to its momentum of the spin wave in the film (refer to Fig.~\ref{spin_wave_field}) \cite{Chiral_pumping_Yu,Yu_Springer,Chiral_pumping_grating}. The spin wave polarization switches sign with the film magnetization and couples to the dipolar field of the wire with opposite momentum components. The theory \cite{Chiral_pumping_Yu} also predicts a critical angle $\arccos(\sqrt{d/w})$ (for $w>d$) between the magnetizations of film and wire at which the chirality vanishes.

The transmission of the spin waves between the microwave source and detector is complicated by the nanowire configuration. Subsequent  experiments \cite{Haiming_exp_wire} retain only two Co nanowires on the YIG films, as illustrated in Fig.~\ref{Haiming_exp}(e). They are capped by metallic striplines that act as source and detector of wire magnons (not shown). The microwave transmission in Fig.~\ref{Haiming_exp}(f) is also non-reciprocal, but now the phase coherence of the  magnons is resolved in terms of fringes that reflect the propagation phase in Fig.~\ref{transmission}(b) \cite{Chiral_pumping_Yu,Yu_Springer} as explained above. Recently, Temdie \textit{et al.} further downscaled the chiral
coupling technique implementing permalloy nanowires arrays, with which they observed the non-reciprocal transmission of exchange spin waves with shorter wavelengths down to $\lambda\approx 80$~nm \cite{ Vlaminck_array}.

\textbf{DMI}.---The chirality from the dipolar interaction may coexist with that caused by DMI. Szulc \textit{et al.}  \cite{DMI_circulator} demonstrated  by micromagnetic simulations that the interfacial DMI can induce a non-reciprocity of the magnon propagation in the Damon-Eshbach configuration in metallic magnetic spin valves. The DMI induced by the interface to heavy metal on one side shifts the spin-wave dispersion in one layer such that its stray fields couple in-phase with spin waves in the other layer that propagate in one direction, but out-of-phase with those moving into another direction. The authors suggested that spin-wave diode and  circulator functionalities may be useful in magnonic logic circuits.

  \subsubsection{Chiral spin Seebeck effect}\label{section4.1.3}
    
The chirality of the dipolar spin-photon interaction is ubiquitous and can significantly affect  observables other than the microwave scattering matrix for ferromagnets. 

    \textbf{Chiral spin Seebeck effect}.---The generation of a spin current by a temperature gradient is called spin Seebeck effect \cite{Spin_seebeck_exp,Spin_seebeck_theory1,Spin_seebeck_theory2,Spin_caloritronics}. It can be detected by the perturbation of the equilibrium balance that cancels spin pumping and torque at the interface of the magnet to a normal metal. \textquotedblleft Thermal spin pumping\textquotedblright\ can be generated by the Joule heating caused by an electric current in a metallic contact on top of a magnetic film, while the magnon accumulation can be detected electrically in a second contact via the inverse spin Hall effect  \cite{Ludo}. In contrast to the single-mode propagating microwave spectroscopy discussed above, the magnon transport is diffuse. In principle, all thermal magnons contribute, at room temperature up to THz \cite{Ludo2}.

   The chiral spin Seebeck effect refers to the directional generation of a spin current in a ferromagnet under a temperature gradient. Here we illustrate its input-output theory by focusing on the Kittel modes of two \textit{magnetic} nanowires at 
    $\mathbf{r}_{2}=R_{2}\hat{\mathbf{y}}$ (the detectors) and at $\mathbf{r}_{1}=R_{1}\hat{\mathbf{y}}$ with $R_{1}<R_{2}$  on top a magnetic insulator. We focus on the configuration in which the magnetization of the wires is opposite to that of the film. We disregard thermal occupation of higher standing magnon modes in the wires and the film,  which is allowed at low temperatures, fine wires, and thin films.  We assume that the magnon  mean-free path is much larger than the distance between the wires. The propagating magnons in the long nanowires then couple coherently only to the magnons in the film that propagate perpendicular to the wires.  We assume that the non-local dipolar interaction between the wires and film dominates the contact exchange interaction and therefore the conventional spin Seebeck effect.

    The coupled equations of motion of the Kittel modes  in the nanowires \(\hat{m}_{L}\), \(\hat{m}_{R}\) and spin waves in the film \(\hat{\alpha}_{q}\) then read
    \begin{subequations}
    \begin{align}
    	\frac{d\hat{m}_{L}}{dt}  &  =-i\omega_{\mathrm{K}}\hat{m}_{L}-\sum_{q}%
    	ig_{q}^{\ast}e^{iqR_{1}}\hat{\alpha}_{q}-\frac{\kappa}{2}\hat{m}_{L}%
    	-\sqrt{\kappa}\hat{N}_{L},\\
    	\frac{d\hat{m}_{R}}{dt}  &  =-i\omega_{\mathrm{K}}\hat{m}_{R}-\sum_{q}%
    	ig_{q}^{\ast}e^{iqR_{2}}\hat{\alpha}_{q}-\frac{\kappa}{2}\hat{m}_{R}%
    	-\sqrt{\kappa}\hat{N}_{R},\\
    	\frac{d\hat{\alpha}_{q}}{dt}  &  =-i\omega_{q}\hat{\alpha}_{q}-ig_{q}%
    	e^{-iqR_{1}}\hat{m}_{L}-ig_{q}e^{-iqR_{2}}\hat{m}_{R}-\frac{\kappa_{q}}{2}%
    	\hat{\alpha}_{q}-\sqrt{\kappa_{q}}\hat{N}_{q},
    \end{align}
    \end{subequations}
where $\kappa= \alpha_G \omega_\mathrm{K}$ represents the Gilbert damping in the contacts and \(\kappa_q\) that of the film. We model the equilibrium thermal fluctuations $\hat{N}_{L}$ and $\hat{N}_{R}$ by the white noise time-correlation  functions $\langle{\hat{N}_{\eta}^{\dagger}(t)\hat{N}_{\eta^{\prime}}(t^{\prime})}\rangle=n_{\eta}\delta(t-t^{\prime})\delta
    _{\eta\eta^{\prime}}$, where $\eta\in\{L,R\}$, $n_{\eta}=1/\left\{
    \exp\left[  \hbar\omega_{\mathrm{K}}/(k_{B}T_{\eta})\right]  -1\right\}  $. By integrating out the spin-wave modes in the film, we obtain equations for the dynamics of the two wires that are dissipatively coupled by the incoherent thermal fluctuations in the film. 
    The wires couple differently to right  $\left(\Gamma_{1}=|g_{q_{\ast}}|^{2}/v_{q_{\ast}}\right)$ and left-moving $\left(\Gamma_{2}%
    =|g_{-q_{\ast}}|^{2}/v_{q_{\ast}}\right)$ modes, where $q_{\ast}$ is the positive root of $\omega_{q_{\ast}}=\omega
    _{\mathrm{K}}$. The wire-film interaction is fully chiral when $\Gamma_{2}=0$. With $\hat{m}_{L,R}(t)=\int e^{-i\omega t}\hat{m}_{L,R}(\omega)d\omega/(2\pi
    )$ and $G_q(\omega)=1/[(\omega-\omega_q)+i\kappa_q/2]$,
    \begin{subequations}
    \begin{align}
    	\hat{m}_{L}(\omega)  &  =\frac{\sum_{q}ig_{q}^{\ast}e^{iqR_{1}}\sqrt
    		{\kappa_{q}}G_{q}(\omega)\hat{N}_{q}(\omega)-\sqrt{\kappa}\hat{N}_{L}(\omega
    		)}{-i(\omega-\omega_{\mathrm{K}})+\frac{\kappa}{2}+\frac{\Gamma_{1}}{2}%
    	},\\
    	\hat{m}_{R}(\omega)  &  =\frac{\sum_{q}ig_{q}^{\ast}e^{iqR_{2}}\sqrt
    		{\kappa_{q}}G_{q}(\omega)\hat{N}_{q}(\omega)-\sqrt{\kappa}\hat{N}_{R}%
    		(\omega)-\Gamma_{1}e^{q_{\ast}(R_{2}-R_{1})}\hat{m}_{L}(\omega)}%
    	{-i(\omega-\omega_{\mathrm{K}})+\frac{\kappa}{2}+\frac{\Gamma_{1}}{2}}.
    \end{align}
    \end{subequations}
Disregarding the damping in the film $\left(  \kappa
    _{q}\rightarrow0\right)$ is an excellent approximation when the contacts distance is much smaller than the magnon propagation length. The Kittel mode occupation number then becomes
    \begin{subequations}
    \begin{align}
    	\rho_{L}  &  \equiv\langle\hat{m}_{L}^{\dagger}(t)\hat{m}_{L}(t)\rangle
    	=n_{L}+\int\frac{d\omega}{2\pi}\frac{\kappa}{(\omega-\omega_{\mathrm{K}}%
    		)^{2}+(\kappa/2+\Gamma_{1}/2)^{2}}(n_{q_{\ast}}-n_{L}),\\
    	\rho_{R}  &  \equiv\langle\hat{m}_{R}^{\dagger}(t)\hat{m}_{R}(t)\rangle
    	=n_{R}+\int\frac{d\omega}{2\pi}\frac{\Gamma_{1}^{2}\kappa}{\left[
    		(\omega-\omega_{\mathrm{K}})^{2}+(\kappa/2+\Gamma_{1}/2)^{2}\right]  ^{2}%
    	}(n_{L}-n_{q_{\ast}}).
    	\label{right_wire}
    \end{align}
    \end{subequations}
We can now compute the effect of a temperature change $T_L$ in $n_L$ in the left wire  with $n_R=n_{q_*}$ is at the same thermal equilibrium temperature $T_0$. To leading order in the temperature difference, the non-local thermal injection of magnons into the passive right nanowire by the active left one then reads from Eq.~(\ref{right_wire})
    \begin{subequations}
    \begin{align}
    \label{no_Maxwell_Demon}
    	\delta\rho_{R}  &  =
    	\mathcal{S}_{\mathrm{CSSE}}(T_{L}-T_0),\\
    	\mathcal{S}_{_{\mathrm{CSSE}}}  &  =\int\frac{d\omega}{2\pi}\frac{\Gamma
    		_{1}^{2}\kappa}{\left[  (\omega-\omega_{\mathrm{K}})^{2}+(\kappa/2+\Gamma
    		_{1}/2)^{2}\right]  ^{2}}\left.  \frac{dn_{L}}{dT}\right\vert _{T_0}.
    \end{align}
\end{subequations}
On the other hand, when we only change the temperature of the right wire \(\delta\rho_{L}=0\). This defines the chiral (or dipolar) spin Seebeck effect with coefficient  $\mathcal{S}_{_{\mathrm{CSSE}}}.$

\textcolor{blue}{We note that the chirality does not cause a rectification of the heat current that in the linear response would require a ``Maxwell Demon". When $T_L>T_0$, magnons are injected from the left wire to the right one, while when $T_L<T_0$ the current is induced by ``holes" in the thermal distribution, which is equivalent to a spin and heat current in the opposite direction. We hereby resolve an issue raised in the Supplemental Material of Ref.~\cite{Chiral_pumping_Yu}. Instead, the chirality is exposed in the form of a non-local heat current switch.}

\begin{figure}[ptbh]
    	\begin{centering}
    		\includegraphics[width=0.99\textwidth]{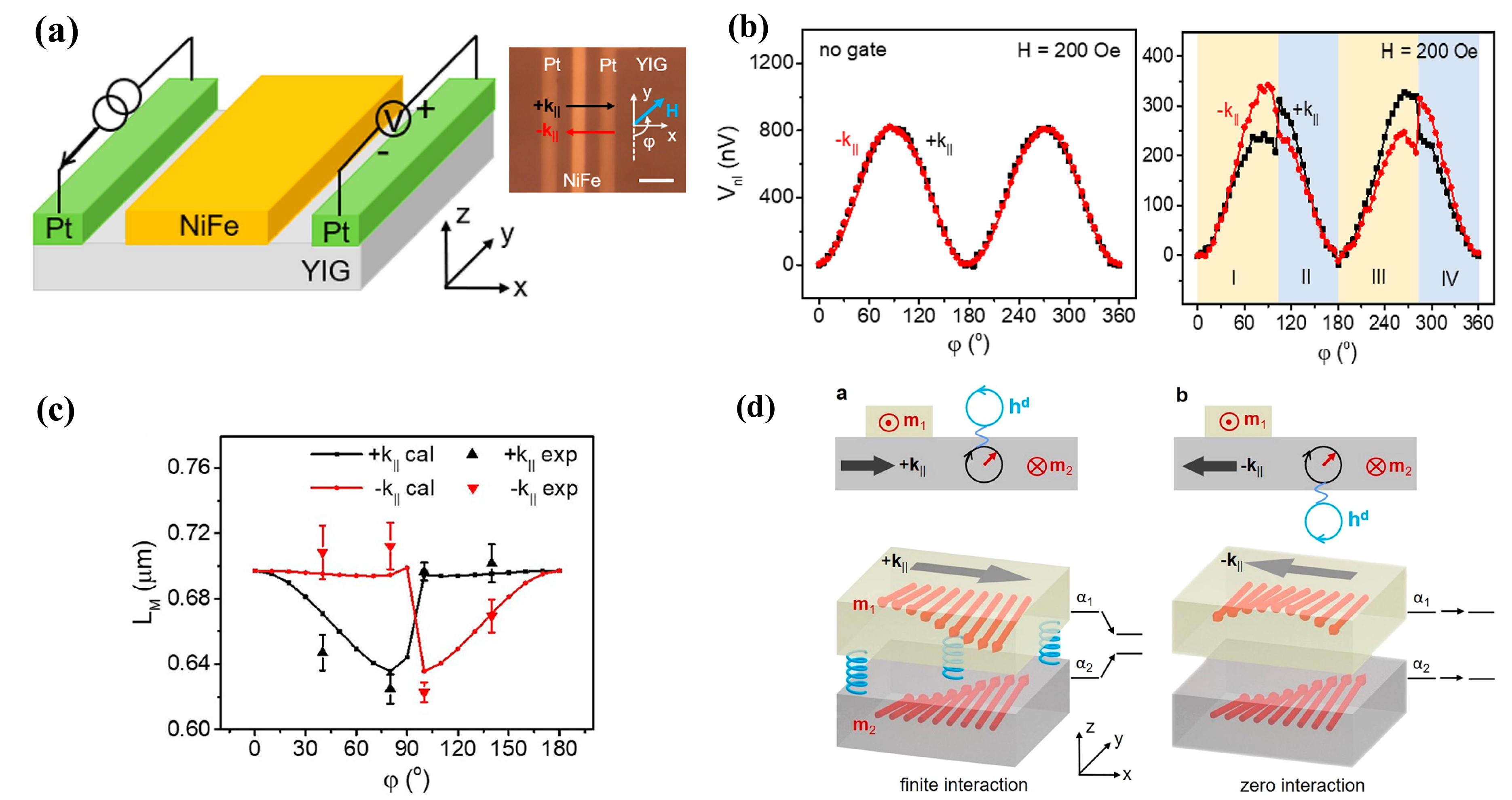}
    		\par\end{centering}
    	\caption{Observation of  asymmetric diffusion of thermal magnons in YIG capped by a film of the magnetic alloy NiFe. (a) is the configuration with two Pt wires that excite and detect the magnon transport. The dependence of the detected voltage in one Pt wire on the magnetization direction is plotted in (b) without and with the NiFe overlayer. The magnon diffusion length  depends on the width of the NiFe strip as shown in (c) for different magnetization configurations. (d) illustrates the chiral interaction mechanism. The figures are taken from Ref.~\cite{Luqiao_exp}.}
    	\label{incoherent_chirality}
    \end{figure}

   \textbf{Experiments in diffuse regime}.--- As sketched above, chiral coupling can affect magnon transport in the diffuse regime even without any spin-orbit coupling.  Han \textit{et al.} \cite{Luqiao_exp}  studied the non-local magnon transport in a YIG film (thickness \(s=\)50~nm, Gilbert damping $\alpha_G=10^{-4}$) under a gate of the magnetic metal alloy NiFe (40~nm thick) as sketched in Fig.~\ref{incoherent_chirality}(a). They reported tunable non-reciprocal propagation  of magnons between Pt wires at room temperature. Here a large number of thermally excited magnons up to several THz contribute to the signal. The NiFe layer has a strong in-plane uniaxial form anisotropy with an easy axis along the $\hat{\bf y}$ direction, while the soft YIG magnetization direction can be rotated freely in the film plane by an applied field \(\mathbf{H}\), as illustrated in Fig.~\ref{incoherent_chirality}(a). In the bare YIG film the non-local spin-Hall voltage depends as $\sin^2\varphi$ on the field angle \(\varphi\) [the left panel of Fig.~\ref{incoherent_chirality}(b)]   \cite{Ludo}. The NiFe gate distorts the signal line shape and introduces a dependence on the magnon current direction, \textit{i.e.}, when the role of source and detector are exchanged [the right panel of Fig.~\ref{incoherent_chirality}(b)].  Figure~\ref{incoherent_chirality}(c) shows the magnon diffusion lengths extracted from experiments with different gate widths. The difference on current direction is the largest when $\varphi$ is close to $90^{\degree}$ and $270^{\degree}$, but tends to vanish when $\varphi=0^{\degree}$ and $180^{\degree}$.  The sudden jump of the signal when $\varphi\sim 100^{\degree}$ and $\sim 280^{\degree}$ is attributed to an abrupt switch of the $y$-component of NiFe with a form anisotropy of the order of the applied magnetic field.

Han \textit{et al.} \cite{Luqiao_exp} attributed this non-reciprocity in terms of a chiral dipolar interaction as sketched in Fig.~\ref{incoherent_chirality}(d), arguing that the dipolar interaction modifies the damping of right and left moving magnons in the film differently \cite{Luqiao_exp}. Whether the additional damping by chiral interactions \cite{Yu_Springer} can really explain these experiments requires additional theoretical efforts.

      \subsubsection{Magnon trap}
      \label{Sec_magnon_trap}
      
      In a field-effect transistor, an electronic gate controls the flow of electrons, amplifying a gate signal or switching the source-drain current on and off. \textquotedblleft Spin transistors\textquotedblright\ are three-terminal spintronic devices that manipulate a source-drain spin current by a gate that couples to the spin degree of freedom. Manipulation of magnon currents in an on-chip circuit is a key functionality in magnonics \cite{magnonics_roadmap,spintronics_2,magnonics_2,magnonics_3,magnonics_4,magnonics_5}. Electric gates easily control electrons,  but manipulating magnons on a small length scale is challenging. Chumak \textit{et al.} \cite{Chumak_transistor} etched a magnonic crystal of 20 parallel grooves into the surface of a YIG strip. The interaction with magnons injected into this region suppresses the source-drain magnon currents by three orders of magnitude. Wu \textit{et al}. reported a 32$\%$  modulation of magnon current in a YIG|Au|YIG perpendicular spin valve when comparing a parallel and  anti-parallel alignment \cite{Xiufeng_valve}, while Cramer \textit{et al.} \cite{Klaui_valve} reported a 290$\%$ difference in a cobalt|cobalt oxide|YIG trilayer. By injecting magnons by a Pt modulator into the source-drain channel of YIG films,  the magnon conductivity can be strongly enhanced \cite{Bart_transistor,Huebl_transistor,Xiangyang_transistor}. 
  
\begin{figure}[ptbh]
    	\begin{centering}
    		\includegraphics[width=0.99\textwidth]{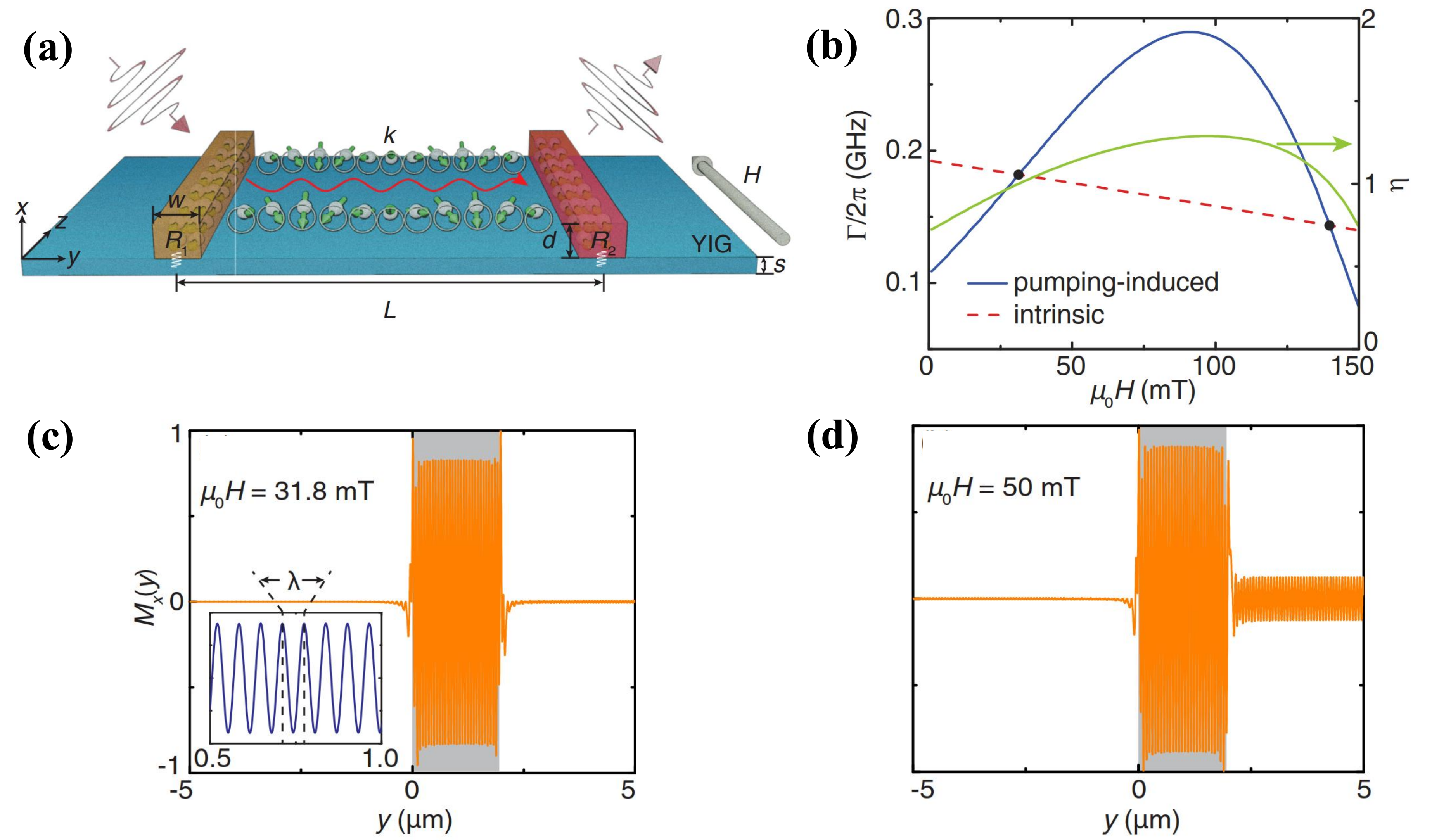}
    		\par\end{centering}
    	\caption{Chirality-induced trapping of magnons. (a) The device consists of two parallel magnetic wires on top of a  thin YIG film: microwaves excite only the left one. (b) compares the intrinsic (Gilbert) and pumping-induced damping (\(\Gamma\)) and the transmission amplitude (\(\eta\)) as a function of magnetic field. (c) shows the localized magnetization amplitude in the film around the second contact under \textquotedblleft trapping\textquotedblright\ conditions.  (d) illustrates a partial trapping in which some spin wave energy escapes to the right. The figures are taken from Ref.~\cite{magnon_trap}.}
    	\label{magnon_trapping}
    \end{figure}

In spite of the recent progress,  magnon transistors still manage only small on-off ratios of the source-drain current.  We believe that chiral spin pumping in multi-terminal magnonic devices may offer more efficient control mechanisms. An example is the trapping and release of spin wave excitations \cite{magnon_trap} as realized between long  magnetic nanowires on top of  a thin film of YIG  as in Fig.~\ref{magnon_trapping}(a). We can treat this case by  extending the theory of coherent chiral excitation by a single wire (Sec.~\ref{dipolar_pumping}) to two or more wires. Here we address the effect of passive wires that act as gates on the coherent spin waves underneath. A Kittel magnon mode $\hat{m}_{l}$ with frequency $\omega_{\mathrm{K},l}$ in the $l$-th magnetic wire at $R_{l}\hat{\mathbf{y}}$ interacts with the propagating spin waves in the magnetic film  $\hat{\alpha}_{k}$ of frequency $\omega
    _{k}$ \cite{Au_first,Chiral_pumping_Yu}. By the chiral coupling, the precession in the wires interacts preferentially with the traveling waves propagating in one direction when either one of the coupling constants $g_{|k|}$ or $g_{-|k|}$ vanishes. Even more than that in the diffuse situation discussed in the previous Sec.~\ref{section4.1.3}, the passive wire  exerts strong effects in the coherent case.  The spin waves that connect magnetic wires can be integrated out of the problem mediates an effective interaction that couples their dynamics.
    
Let us put the wires on   $\mathbf{r}_{1}=R_{1}%
    \hat{\mathbf{y}}$ and $\mathbf{r}_{2}=R_{2}\hat{\mathbf{y}}$ and assume couplings $g_{k,1}=g_{k,2}=g_{k}$ and frequencies $\omega_{\mathrm{K},1}=\omega
    _{\mathrm{K},2}=\omega_{\mathrm{K}}$, the equation of motion becomes
    \begin{align}
    \frac{d}{dt}\left(
    \begin{matrix}
    	\hat{m}_{L}\\
    	\hat{m}_{R}%
    \end{matrix}
    \right)  +i\left(
    \begin{matrix}
    	\tilde{\omega}_{\mathrm{K}}-i\Gamma(\omega) & -i\Gamma_{12}(\omega)\\
    	-i\Gamma_{21}(\omega) & \tilde{\omega}_{\mathrm{K}}-i\Gamma(\omega)
    \end{matrix}
    \right)  \left(
    \begin{matrix}
    	\hat{m}_{L}\\
    	\hat{m}_{R}%
    \end{matrix}
    \right)  =\left(
    \begin{matrix}
    	\hat{P}_{1}\\
    	\hat{P}_{2}%
    \end{matrix}
    \right)  ,
    \end{align}
where $\tilde{\omega}_{\mathrm{K}}=\omega_{\mathrm{K}}-i\kappa/2$ and $\kappa=2\alpha_{\mathrm{G}}\omega_{\mathrm{K}}$ is the material damping rate of the Kittel modes in terms of  the Gilbert constant $\alpha_{\mathrm{G}}$, while $\hat{P}_{l}$ is the input power of the local antennas. The self-interaction
    \begin{equation}
    	\Gamma(\omega)=\frac{1}{2v(k_{\omega})}\left(  |g_{k_{\omega}}|^{2}%
    	+|g_{-k_{\omega}}|^{2}\right)  \label{addition_damping1}%
    \end{equation}
is a pumping-induced loss rate for a single nanowire \cite{Chiral_pumping_Yu,Yu_Springer}. With $R_{2}>R_{1}$, the directional coupling rate reads
\begin{subequations}
    \begin{align}
    	\Gamma_{12}(\omega)   &=\frac{1}{v(k_{\omega})}|g_{-k_{\omega}}%
    	|^{2}e^{ik_{\omega}(R_{2}-R_{1})},\\
    	\Gamma_{21}(\omega)   &=\frac{1}{v(k_{\omega})}|g_{k_{\omega}}|^{2}e^{ik_{\omega}(R_{2}-R_{1})},
    	\label{dissipative_1}
    \end{align}
    \end{subequations}
where  $v(k)$ is the group velocity of the traveling spin  waves and $k_{\omega}$ is the positive root of $\omega_{k}=\omega=\omega_{\mathrm{K}}$ at the ferromagnetic resonance (FMR). When $g_{-k}=0$, $|\Gamma_{21}(\omega)|=2\Gamma(\omega)$, \textit{i.e.}, twice as large as the  magnon broadening Eq.~(\ref{addition_damping1}) induced by the local spin pumping. In the perfect chiral limit, one magnet affects the other without any back action. The input-output theory introduced in Sec.~\ref{dipolar_pumping} leads to expressions for the transmission amplitude of microwaves that excite one of the magnetic wires and inductively escape from the other  \cite{magnon_trap,phonon_Yu_1}. The two \textit{magnetic} amplitudes are related as
    \begin{equation}
    	\hat{m}_{R}(\omega_{\mathrm{K}})=\eta(\omega_{\mathrm{K}})e^{i\pi
    		+ik_{\omega}(R_{2}-R_{1})}\hat{m}_{L}(\omega_{\mathrm{K}}),
    	\label{phase_relation2}%
    \end{equation}
where $\eta(\omega_{\mathrm{K}})={2\Gamma(\omega_{\mathrm{K}}) }/({\kappa
    		/2+\Gamma(\omega_{\mathrm{K}}) })$ is the spin wave transmission amplitude. We observe that the magnon propagation from \(R_{1}\) to \(R_{2}\) contributes a phase factor 
    \begin{align}
    	\Delta\phi=\pi+k_{r}(R_{2}-R_{1}),
    	\label{phase_shift_magnon}
    \end{align}
 where  $k_{r}(R_{2}-R_{1})$ is the conventional plane wave phase delay, while $\pi$
is twice the dissipative phase shift $\pi/2$ between the Kittel modes and the propagating waves. Remarkably, when $\kappa/2\ll\Gamma$, $\eta\rightarrow2$, it corresponds to an enhancement of the magnon amplitude by a factor of 2. The damping $\Gamma$ and coupling $\eta$ can be tuned by the external magnetic field as shown in Fig.~\ref{magnon_trapping}(b). When  $\kappa/2=\Gamma$, $\eta=1$ and we observe a perfect ``trapping'' of magnons in and under the passive magnet, as illustrated in  Fig.~\ref{magnon_trapping}(c).
   
We can express the excited magnetization in the real space by the Green function method introduced in Sec.~\ref{dipolar_pumping}. For $y<R_{1}<R_{2}$,
    \begin{align}
    	\hat{M}^{L}_{\alpha}(x)   =\frac{2i}{v_{k_{\omega}}}\sqrt{2M_{s}\gamma\hbar}%
    	\hat{m}_{L}(\omega_{\mathrm{K}})m_{\alpha}^{(k_{\omega})}(x)g_{-k_{\omega}%
    	}\left(  e^{ik_{\omega}(R_{1}-y)}-\eta(\omega_{\mathrm{K}})e^{ik_{\omega}%
    		(2R_{2}-R_{1}-y)}\right)  
    \end{align}
    vanishes for chiral coupling $g_{-k}=0$. The magnetization amplitude to the right of both wires
    \begin{align}
    	\hat{M}_{\alpha}^{R}(x)   =\frac{2i}{v_{k_{\omega}}}\sqrt{2M_{s}\gamma\hbar}%
    	\hat{m}_{L}(\omega_{\mathrm{K}})m_{\alpha}^{(k_{\omega})}(x)g_{k_{\omega}%
    	} e^{-ik_{+}^{\ast}(R_{1}-y)}\left(  1-\eta(\omega_{\mathrm{K}%
    	})\right)  
    \end{align}
vanishes for $\eta(\omega_{\mathrm{K}})\rightarrow1$ as in Fig.~\ref{magnon_trapping}(c). Here, $k_{+}^{\ast}=(k_{\omega}+i\epsilon)$ in terms of the inverse extinction length $\epsilon$. Between the wires ($R_{1}<y<R_{2}$),
    \begin{equation}
    	\hat{M}_{\alpha}^{M}(x)=\frac{2i}{v_{k_{\omega}}}\sqrt{2M_{s}\gamma\hbar}\hat
    	{m}_{L}(\omega_{\mathrm{K}})m_{\alpha}^{(k_{\omega})}(x)g_{k_{\omega}}%
    	e^{-ik_{+}^{\ast}(R_{1}-y)}
    \end{equation}
is a right-propagating wave for $g_{-k}=0$. The excited magnetization decays on the length scale of the inverse of  \(1/\)Im$k_{+}^{\ast}$, which is not significant in the region plotted here.
    The trapped magnetization is not a standing wave that requires the interference of back and forth reflections. The magnetic energy that is injected from the left accumulates in a small region under the contact and for weak damping reaches a large steady-state density. Tuning $\eta$ allows a controlled escape of magnons to the right as  illustrated by Fig.~\ref{magnon_trapping}(d).
    
\begin{figure}[ptbh]
    	\begin{centering}
    		\includegraphics[width=0.99\textwidth]{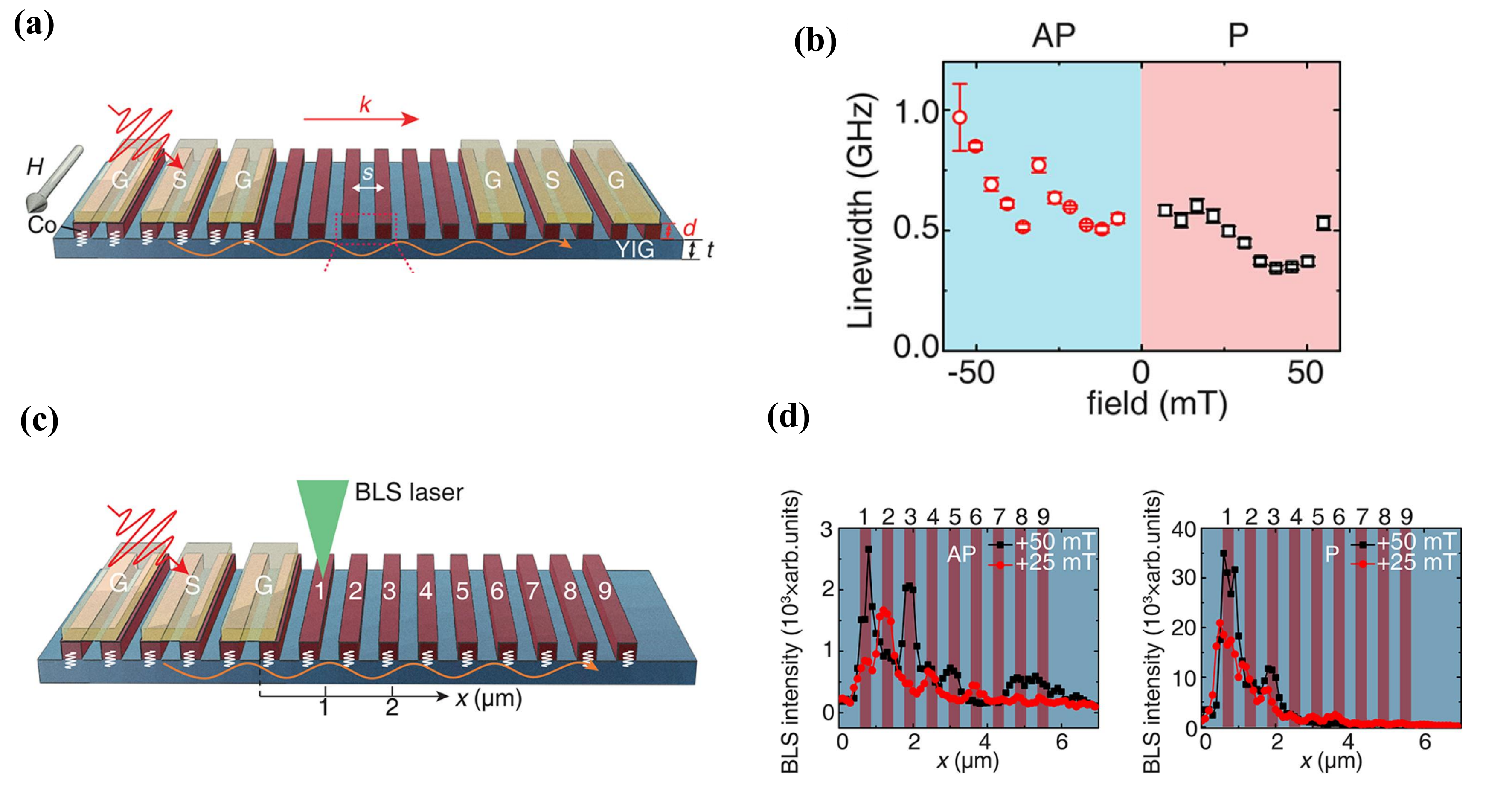}
    		\par\end{centering}
    	\caption{Tunable damping of magnons in an array of Cobalt nanowires on top of a thin YIG film. (a) shows the microwave transmission in the Damon-Eshbach configuration with parallel and anti-parallel magnetization. The extracted line width in the  parallel (\textquotedblleft P\textquotedblright) and anti-parallel (\textquotedblleft AP\textquotedblright) configurations of Co and YIG magnetizations are shown in (b). The Brillouin light scattering microscopy sketched in (c) measures the spatial distribution of the excitation over the individual cobalt wires as shown in (d). The figures are taken from Ref.~\cite{Hanchen_damping}.}
    	\label{Hanchen_damping}
    \end{figure}

\textbf{Experiment}.---The experiments carried out by the Wang \textit{et al.} on the chiral magnon pumping by an array of Co wires  Ref.~\cite{Hanchen_damping} into YIG thin films provides  additional evidence on the magnon trapping. They measured microwave transmission, see Fig.~\ref{Hanchen_damping}(a,b), and Brillouin light scattering (BLS), see Fig.~\ref{Hanchen_damping}(c,d) when the wire and YIG film magnetizations are parallel and anti-parallel. The broadening of the FMR of Co wires in Fig.~\ref{Hanchen_damping}(b) is larger in the anti-parallel configuration. The BLS signals in Fig.~\ref{Hanchen_damping}(c) image the magnon density in the Co wires that decays  with distance from the coplanar wave guide on a scale much shorter than the spin wave decay length in YIG, see Fig.~\ref{Hanchen_damping}(d). Larger coupling strengths $g_k$ for the parallel alignment  \cite{Chiral_pumping_Yu,Haiming_exp_grating}, \textit{i.e.},  Eq.~(\ref{addition_damping1}) explains the faster decay in that configuration.  
The authors interpreted the strong enhancement of the BLS signal in the third wire in Fig.~\ref{Hanchen_damping}(d) in terms of the magnon trapping effect. Wang \textit{et al.} \cite{Haiming_exp_wire} observed the phase shift Eq.~(\ref{phase_shift_magnon}) with two
magnetic nanowires on top of the magnetic film by microwave spectroscopy.

 \textbf{Miscellaneous}.--- A not yet solved issue is the relative importance of the dipolar coupling compared to the interface exchange interaction between a magnetic insulator film and a metallic magnetic wire. This is not universal, but depends on the material and interface quality. The short-range exchange interaction can be suppressed by atomically thin insulating spacers. Reference \cite{Spin_torque_excitation} reported a large antiferromagnetic interface exchange effect in the FMR of Co|YIG bilayers with direct contact \cite{Spin_torque_excitation}.  Model calculations pointed to an important exchange contribution to the strong coupling observed between propagating magnons in YIG films and Ni wires  \cite{Chiral_pumping_grating} that agree with observations \cite{Jilei_PRL}. Reference \cite{Hanchen_damping,Haiming_exp_grating}  found evidence for a dominant dipolar interaction over the exchange ones for Co wires. First principle calculations \cite{first_principle_exchange} can help  in assessing the importance of the interface exchange for different interfaces.

 Equation~(\ref{addition_damping1})  in Refs.~\cite{Chiral_pumping_Yu,magnon_trap,Yu_Springer} holds for a single magnetic wire  on top of magnetic film \cite{Chiral_pumping_Yu,magnon_trap}  and weak spin wave  damping. The magnon trapping effect was originally formulated for two-wire configuration \cite{magnon_trap}. On the other hand, the dynamics of magnonics crystal formed by a closely spaced  array of nanomagnets \cite{Hanchen_damping,Haiming_exp_grating,Jilei_PRL} is collective \cite{phonon_Yu_2,waveguide_Yu_1,waveguide_Yu_2,subradiance1,Yuxiang_subradiance}, which deserves more attention in the future.

\subsubsection{Synthetic antiferromagnets}
    \label{synthetic_antiferromagnet}
    The chiral dipolar field emitted by a magnon can drive other magnets. Here we address a spin valve of two magnets that are separated by a thin normal metal spacer that non-locally couples  into an anti-parallel configuration, \textit{i.e.}, a ``synthetic antiferromagnet'' (SAFM).  References \cite{bilayer_dipolar_1,bilayer_dipolar_2,slow_wave,bilayer_2019} reported non-reciprocal spin wave propagation in SAFMs. We argue now that the dipolar coupling discovered in the nanowire on film geometry \cite{Hanchen_damping,Haiming_exp_grating,Jilei_PRL} is also at work here, which may come as a surprise since the dipolar coupling does not affect the equilibrium configuration in extended thin films. To this end, we proceed from the quantum formalism introduced above \cite{Chiral_pumping_Yu,slow_wave,bilayer_2019} and discuss the relevant publications.
    	
\begin{figure}[ht]
     \begin{center}
    {\includegraphics[width=15.6cm]{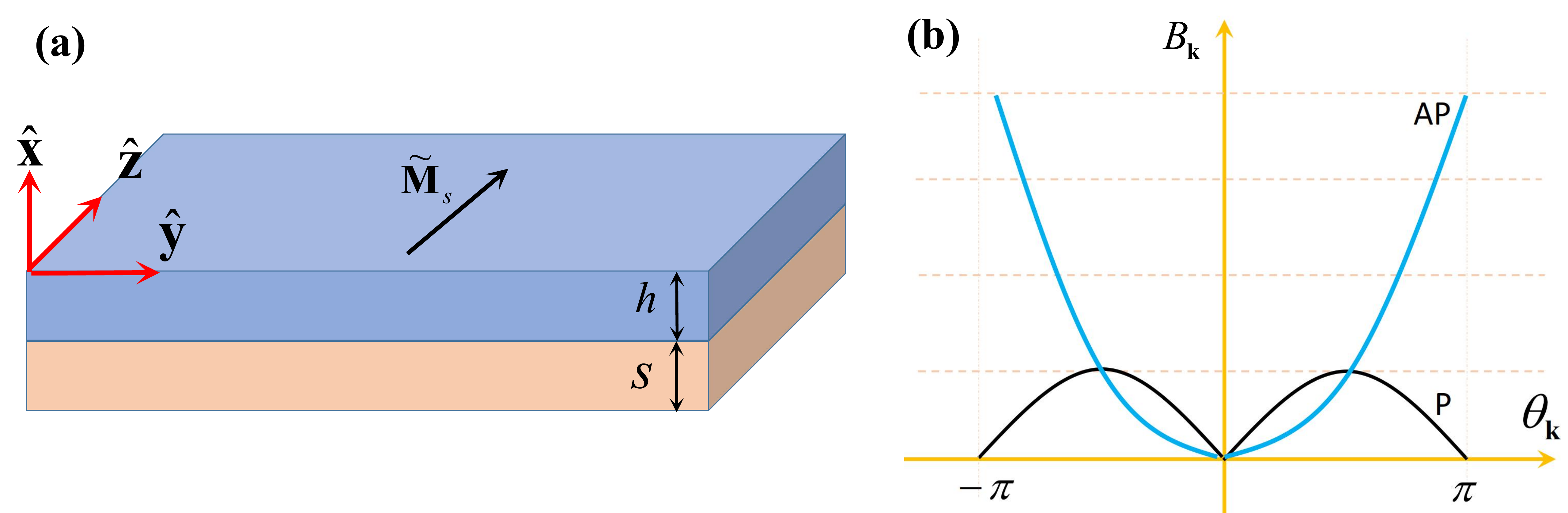}}
    \caption{(a) Bilayer of two magnetic films with thicknesses \textit{h} and \textit{s}. The upper magnetization $\tilde{\mathbf{M}}_s $ is fixed along $\hat{\mathbf{z}}$, while the lower one may rotate in the plane by an angle \(\phi\). Here we disregard a thin spacer layer that suppresses the exchange but not the dynamical dipolar coupling. (b) Angle dependence of the interlayer dipolar couplings for spin waves in the lowest subbands \(i,j=1\) when the magnetizations of the two layers are parallel [P, Eq.~(\ref{P})] and anti-parallel [AP, Eq.~(\ref{AP})]. The unit of $B_{\bf k}/\hbar$ is in GHz.
    }
    \label{bilayer_dipolar_model}
    \end{center}
  \end{figure}

    \textbf{Bilayer}.---Figure~\ref{bilayer_dipolar_model}(a) shows a bilayer of two magnetic films  with thicknesses $h$ and $s$. The magnetodipolar interlayer coupling reads 
    \begin{equation}
    	\hat{H}_d=-\mu_0\gamma^2\frac{1}{4\pi}\int d{\bf r}\hat{\tilde{\bf S}}_{\beta}({\bf r})\partial_{\beta}\partial_{\alpha}\int d{\bf r}'\frac{\hat{S}_{\alpha}({\bf r}')}{|{\bf r}-{\bf r}'|},
    	\label{dipolar_bilayer}
    \end{equation}
where $\alpha,\beta=\{x,y\}$, while $\hat{\tilde{\bf S}}_{\beta}({\bf r})$ and $\hat{\bf S}_{\alpha}({\bf r})$ are spin fields operators in, respectively, the upper and lower layer. The spacer layer communicates the equilibrium coupling into the anti-parallel configuration, but strongly suppresses exchange effects on the dynamics that we disregard in the following. We take the saturated magnetization $\tilde{\bf M}_s$ in the upper layer to be  parallel to the in-plane $\hat{\bf z}$-direction, while ${\bf M}_s$ of the lower layer may rotate in the plane with ${\bf M}_s\cdot \tilde{\bf M}_s = M_s \tilde{M_s} \cos\varphi$. The leading term of the Holstein-Primakoff expansion of spin field operator of the upper layer in terms of the magnon operators $\hat{\alpha}_{j{\bf k}}$ of subband $j$ and wave vector \textbf{k} with mode amplitudes \(\tilde{m}_{\beta}^{j{\bf k}}\) reads:
    \begin{equation}
    	\hat{\tilde{{\bf S}}}_{\beta}({\bf r},t)=\sqrt{2\tilde{S}}\sum_{j{\bf k}}\left(\tilde{m}_{\beta}^{j{\bf k}}(x)e^{i{\bf k}\cdot\pmb{\rho}}\hat{\alpha}_{j{\bf k}}+\tilde{m}_{\beta}^{j{\bf k}*}(x)e^{-i{\bf k}\cdot\pmb{\rho}}\hat{\alpha}^{\dagger}_{j{\bf k}}\right).\\
    \end{equation}
For the lower layer, ${\bf S}_x={\bf S}_{x'}$ and ${\bf S}_y={\bf S}_{y'}\cos\varphi$, while
    \begin{equation}
    	\hat{{\bf S}}_{\delta=\{x',y'\}}({\bf r},t)=\sqrt{2S}\sum_{i{\bf q}}\left(\tilde{m}_{\delta}^{i{\bf q}}(x)e^{i{\bf q}\cdot\pmb{\rho}}\hat{\beta}_{i{\bf q}}+\tilde{m}_{\delta}^{i{\bf q}*}(x)e^{-i{\bf q}\cdot\pmb{\rho}}\hat{\beta}^{\dagger}_{i{\bf q}}\right)
    \end{equation}
with magnon operators $\hat{\beta}_{i{\bf q}}$. Inserting these expansions, Eq.~(\ref{dipolar}) reads:
    \begin{equation}
    	\hat{H}_d=\sum_{ij,{\bf k}}\left({{\cal A}_{ij,{\bf k}}}\hat{\alpha}_{j{\bf k}}\hat{\beta}_{i{\bf k}}+{{\cal B}_{ij,{\bf k}}}\hat{\alpha}^{\dagger}_{j{\bf k}}\hat{\beta}_{i{\bf k}}+{\rm H.c.}\right).
    \end{equation}
In the rotating wave approximation we disregard  ${{\cal A}_{ij,{\bf k}}}$, while
    \begin{equation}
    	{\cal B}_{ij,{\bf k}}=-\mu_0\gamma\sqrt{\tilde{M}_sM_s}\int_0^h dx \int_{-s}^0 dx'e^{-|{\bf k}|(x-x')}(\tilde{m}_x^{j{\bf k}*}(x),\tilde{m}_y^{j{\bf k}*}(x))
    	\left(
    	\begin{array}{cc}
    		|{\bf k}|&-ik_y\\
    		-ik_y&-\frac{k_y^2}{|{\bf k}|}
    	\end{array}
    	\right)
    	\left(
    	\begin{array}
    		[c]{c}
    		m_{x'}^{i{\bf k}}(x')\\
    		m_{y'}^{i{\bf k}}(x')\cos\varphi
    	\end{array}
    	\right).
    \end{equation}
Assuming for simplicity circular precession with $\tilde{m}_y=i \tilde{m}_x$, and for \(\varphi=0\)  (parallel configuration, P) $m_y=im_x$, while for  \(\varphi=\pi\) (anti-parallel configuration, AP) $m_y=-im_x$, and with  $\cos\theta_{\bf k}=k_y/|{\bf k}|$
    \begin{subequations}
    \begin{align}
    	&{\cal B}^{\rm P}_{ij,{\bf k}}=-\mu_0\gamma\sqrt{\tilde{M}_sM_s}|{\bf k}|\sin^2\theta_{\bf k}\int_0^h dx \int_{-s}^0 dx'e^{-|{\bf k}|(x-x')}
    	\tilde{m}_x^{j{\bf k}*}(x)
    	m_{x'}^{i{\bf k}}(x'), \label{P}\\
    	&{\cal B}^{\rm AP}_{ij,{\bf k}}=-\mu_0\gamma\sqrt{\tilde{M}_sM_s}|{\bf k}|(1-\cos\theta_{\bf k})^2\int_0^h dx \int_{-s}^0 dx'e^{-|{\bf k}|(x-x')}
    	\tilde{m}_x^{j{\bf k}*}(x)
    	m_{x'}^{i{\bf k}}(x').\label{AP}
    \end{align}
   \end{subequations}
Figure~\ref{bilayer_dipolar_model}(b) illustrates the anisotropy of the interlayer dipolar couplings of the lowest subband in the P and AP configurations, with maximum coupling four times larger for the latter. Switching the AP ground state of a SAFM into a P state by a magnetic field strongly suppresses the magnetodipolar coupling between the films. Calculation of the spin wave dispersion and propagation including exchange and dipolar interaction as a function of magnitude and direction of the magnetic field is now straightforward.

\begin{figure}[ptbh]
    	\begin{centering}
    	\includegraphics[width=1\textwidth]{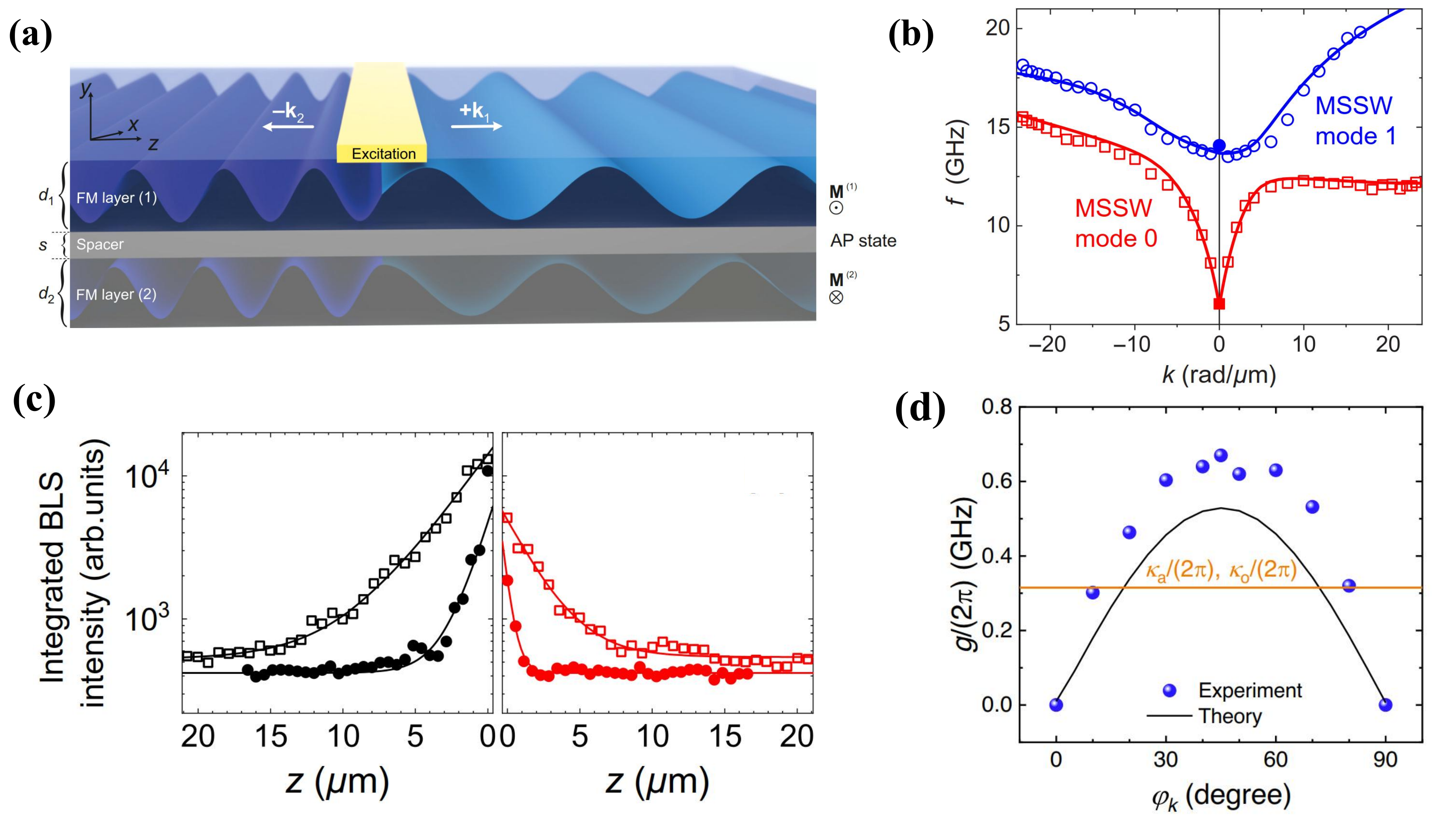}
    		\par\end{centering}
    	\caption{Non-reciprocal propagating spin waves in synthetic antiferromagnets (SAFMs) caused by interlayer dipolar coupling. (a) illustrates the different wave lengths of magnons  with the same frequency but propagating in opposite directions.  (b) is the measured and modeled dispersion relation of spin waves in a SAFM that become ``slow" for positive momentum. Brillouin light scattering spectra reproduced in (c) find the magnon diffusion in the positive suppressed compared to that in the negative direction.   (d) shows the coupling strength between optical and acoustic modes in a SAFM as a function of canting angle. 
    	The figures are taken from Refs.~\cite{bilayer_2019}[(a)], \cite{slow_wave} [(b) and (c)], and \cite{bilayer_dipolar_2} [(d)].}
    	\label{bilayer_dipolar}
    \end{figure}

 \textbf{Experiments}.---The non-reciprocity of Brillouin light scattering by magnons provided the first evidence for  the non-local exchange coupling in magnetic multilayers decades ago \cite{bilayer_BLS_1,bilayer_BLS_2,bilayer_BLS_3,bilayer_BLS_4,bilayer_BLS_5}. 
   More recently, Gallardo \textit{et al.} attributed the non-reciprocity when the wave-vector is perpendicular to the magnetization (Damon-Eshbach configuration) observed by Brillouin light scattering in Permalloy/Ir/CoFeB SAFMs to an interlayer dipolar coupling \cite{bilayer_2019}, as illustrated in Fig.~\ref{bilayer_dipolar}(a). An analytical formulation with arbitrary angles between the magnetizations of two magnetic layers concludes that the non-reciprocity increases linearly for small wave vectors but decreases again for large wave vectors. 
    Microwave spectroscopy of propagating spin waves between two co-planar wave guides is a new and effective instrument to study these effects   \cite{bilayer_dipolar_1,bilayer_dipolar_2,slow_wave}.
    Grassi \textit{et al.} demonstrated non-reciprocity (``diode effect") in metallic SAFMs by Brillouin light scattering and propagating-spin-wave spectroscopy \cite{slow_wave}.  They engineered the spin-wave dispersion relation of the bilayer system such that the magnon group velocities in opposite directions differ, as seen in Fig.~\ref{bilayer_dipolar}(b). Brillouin light scattering demonstrates that the diffusion length of excited spin waves with opposite momenta  differs in the Damon-Eshbach configuration, as shown in Fig.~\ref{bilayer_dipolar}(c).  Ishibashi \textit{et al.} \cite{bilayer_dipolar_1} and Shiota \textit{et al.} \cite{bilayer_dipolar_2} observed similar features such as a switchable non-reciprocal frequency shift of propagating spin waves in the SAFM FeCoB/Ru/FeCoB. Here the observed dipolar+exchange coupling strength between optical and acoustic magnon modes under an applied magnetic field that tilts the upper layer magnetization relative to the spin wave propagation direction and that of the magnetization of the lower layer in Fig.~\ref{bilayer_dipolar}(d) agrees with the model expectations.

\begin{figure}[ptbh]
    	\begin{centering}
    		\includegraphics[width=0.97\textwidth]{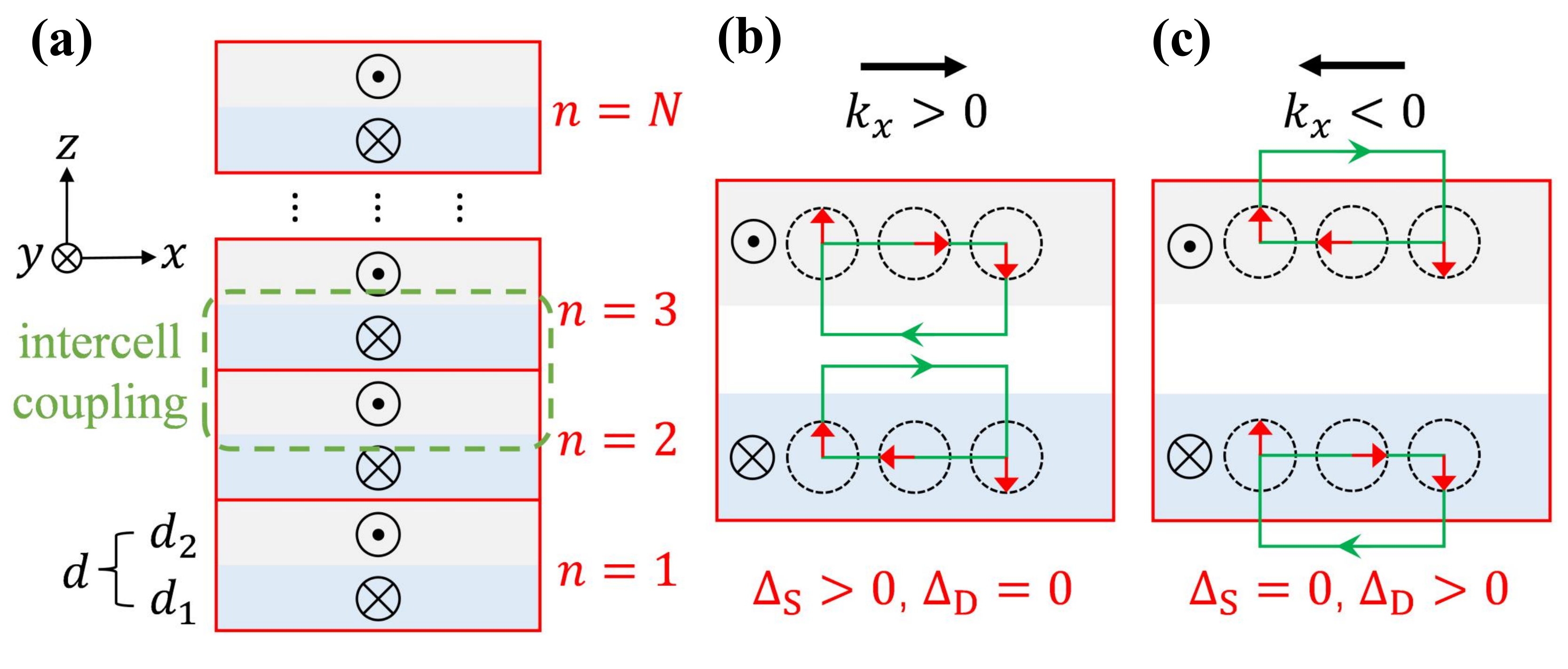}
    		\par\end{centering}
    	\caption{(a) Magnetic multilayer which at equilibrium is coupled into an anti-parallel in-plane configuration (SAFM) and the dipolar coupling of magnons that propagate normal to the magnetization. In (a), the red, solid frames correspond to unit cells indexed by $n = \{1, 2, ..., N\}$. The blue (grey) blocks with thicknesses of $d_1$ and $d_2$ form the 1st and 2nd sublattice. The equilibrium magnetizations in the 1st (2nd) sublattice are parallel to the $+\hat{\bf y}$ and $-\hat{\bf y}$ directions, denoted by $\otimes$ and $\odot$, respectively. (b) depicts the dipolar fields (green lines and arrows) in two layers within a unit cell, generated by propagating magnons with $k_x > 0$ for the precessing magnetization (red arrow) propagating from left to right.  The magnons of the two layers experience finite coupling constant $\Delta_S>0$. Meanwhile, the generated dipolar fields from magnons are zero between the layers of neighboring cells, corresponding to $\Delta_D=0$. (c) depicts the dipolar fields generated by propagating magnons with $k_x<0$ for two layers within a unit cell, corresponding to $\Delta_S=0$ and $\Delta_D>0$. The figures are taken from Ref.~\cite{multilayer_topology}.}
    	\label{multilayer_dipolar}
    \end{figure}

\textbf{Multilayer SAFM and topological edge modes}.---The chirality itself, as pointed out in Sec.~\ref{unification}, indicates the close analogy of dipolar interaction and relativistic spin-orbit interaction. He \textit{et al.} proposed that in magnetic multilayers with anti-parallel alignment that are coupled by the chiral dipolar field as illustrated in Fig.~\ref{multilayer_dipolar}(a), the magnon modes are characterized by non-zero Chern integers and ultra-localized magnonic surface states that carry chiral spin currents \cite{multilayer_topology}.

As explained above, the coupling of magnons that propagate normally to the in-plane magnetization of a bilayer depends on the propagation direction, \textit{viz.}, parallel with no coupling and anti-parallel with net coupling in configuration Fig.~\ref{bilayer_dipolar}(a). So in  anti-parallel multilayers (SAFMs) with two layers of different magnetic films in a unit cell $n$ as shown in Fig.~\ref{multilayer_dipolar}(a), the coupling of magnons in the intra- and inter-unit cell differs depending on the propagation direction $k_x$ normal to the in-plane magnetization. Here $n=\{1,2,\cdots,N\}$ indicate the unit cells, and the lower and upper layers in each cell are labeled by ``1" and ``2", respectively. The intra- and inter-unit cell coupling constants are  $\Delta_S(k_x)$ and $\Delta_D(k_x)$, respectively, where $\Delta_S(k_x<0)=0$ and $\Delta_D(k_x>0)=0$, as sketched in Fig.~\ref{multilayer_dipolar}(b) and (c), respectively. Hence, when $k_x=0$, $\Delta_S=0$ and the two layers in the unit cell $n=1$ and $n=N$ decouple from the rest and the magnons are confined in the 1st layer of the 1st unit cell and the 2nd layer of the $N$th unit cell,  that become discrete edge modes. He \textit{et al.} demonstrated that this phenomenon can be mapped on the Su-Schrieffer-Heeger model \cite{SSH} with a non-trivial topology and non-vanishing Chern number \cite{multilayer_topology}. The non-trivial topology of the bulk magnons implies the emergence of chiral edge states.

\subsection{Magnons and photons}

\label{magnon_photon}

Due to the broken time-reversal symmetry of ferro- or ferrimagnets their excitations are anticlockwise precessions around the equilibrium magnetization. A photon magnetic field \(\mathbf{H}_{ac}\) interacts with the precessing magnetic moment \(\mathbf{M}\) via the Zeeman coupling $-\mu_0 \mathbf{H}_{ac} \cdot \mathbf{M}$, which implies that in the absence of magnetic anisotropies that deform the circular precession only one photon polarization can interact with a magnon. A locking between photon polarization and its wave vector at interfaces is a feature of the Maxwell equation that explains the chirality of the magnon-photon interaction and hybridization. The reported electromagnetic non-reciprocities 
 range from microwave \cite{microwave_nonreciprocity_1,microwave_nonreciprocity_2,microwave_nonreciprocity_3,microwave_nonreciprocity_4,waveguide_Yu_1,waveguide_Yu_2,Canming_exp,chiral_waveguide1,chiral_waveguide2,chiral_waveguide3,Xufeng_exp}, 
to terahertz \cite{terahertz_nonreciprocity}, to optical frequencies \cite{Faraday_rotation_1,Faraday_rotation_2,Faraday_rotation_3,Faraday_rotation_4}. Related phenomena are found in (nonlinear) optomechanical  interactions \cite{microwave_nonreciprocity_2,microwave_nonreciprocity_3,optics_nonreciprocity_1,optics_nonreciprocity_5}, and the tuning of the interaction between a system and its reservoir \cite{optics_nonreciprocity_4,reservior_engineering}.

Optical chiralities are not limited to the interaction with magnets and the associated broken time-reversal symmetry. In the optical regime, photon electric fields $\mathbf{E}_{ac}$ interact with matter  via $- \mathbf{E}_{ac} \cdot \mathbf{P}$, where \textbf{P} is the electric polarization, which is much stronger than the Zeeman interaction at microwave frequencies. However, chiral interaction in the optical regime may require polarized light  \cite{chiral_optics,nano_optics,optics_nonreciprocity_1,optics_nonreciprocity_2,optics_nonreciprocity_3,optics_nonreciprocity_4,optics_nonreciprocity_5,chiral_emitter_light_1,chiral_emitter_light_2,chiral_emitter_light_3,chiral_emitter_light_4,chiral_emitter_light_5}.

The advantage of using magnets to realize microwave non-reciprocity is its flexible control and broadband access in the linear response regime \cite{Xufeng_exp,Canming_exp,waveguide_Yu_1,waveguide_Yu_2}. The photon is an excellent long-range information carrier  of spin information, and the chirality brings about additional functionalities. In the following, we review the theory and experiments of chiral interaction between magnons and microwaves that lead to chiral pumping of magnons (Sec.~\ref{stripline_excitation}), chiral damping and blocking of magnons (Sec.~\ref{Sec_chiral_damping}), spin Doppler effect (\ref{Spin_Doppler}), broadband microwave diodes (\ref{Microwave_diodes}), and a magnetic skin effect (\ref{spin_skin_effect}).

\subsubsection{Chiral excitation of spin waves by striplines}
    \label{stripline_excitation}

    \textbf{Chiral pumping}.---In Sec.~\ref{Stripline}, we showed that the polarization of the Oersted magnetic field generated by an AC current in a long stripline is locked to its momentum, a phenomenon with generalized spin-orbit interaction (Sec.~\ref{unification}). The interaction with nearby magnets then depends on the angle  with the local magnetization which governs the direction into which spin waves can be emitted. The general problem of the coupled LLG and Maxwell equations becomes analytically tractable in the linear response, which is the regime appropriate for many experiments \cite{Chiral_pumping_Yu,Chiral_pumping_grating,Haiming_exp_grating,Teono_NV,Huajun}. We emphasize the similarities and differences with the magnon pumping by the dipolar field of magnetic nanowires discussed in Sec.~\ref{dipolar_pumping}.

\begin{figure}[ptbh]
    	\begin{centering}
    		\includegraphics[width=0.99\textwidth]{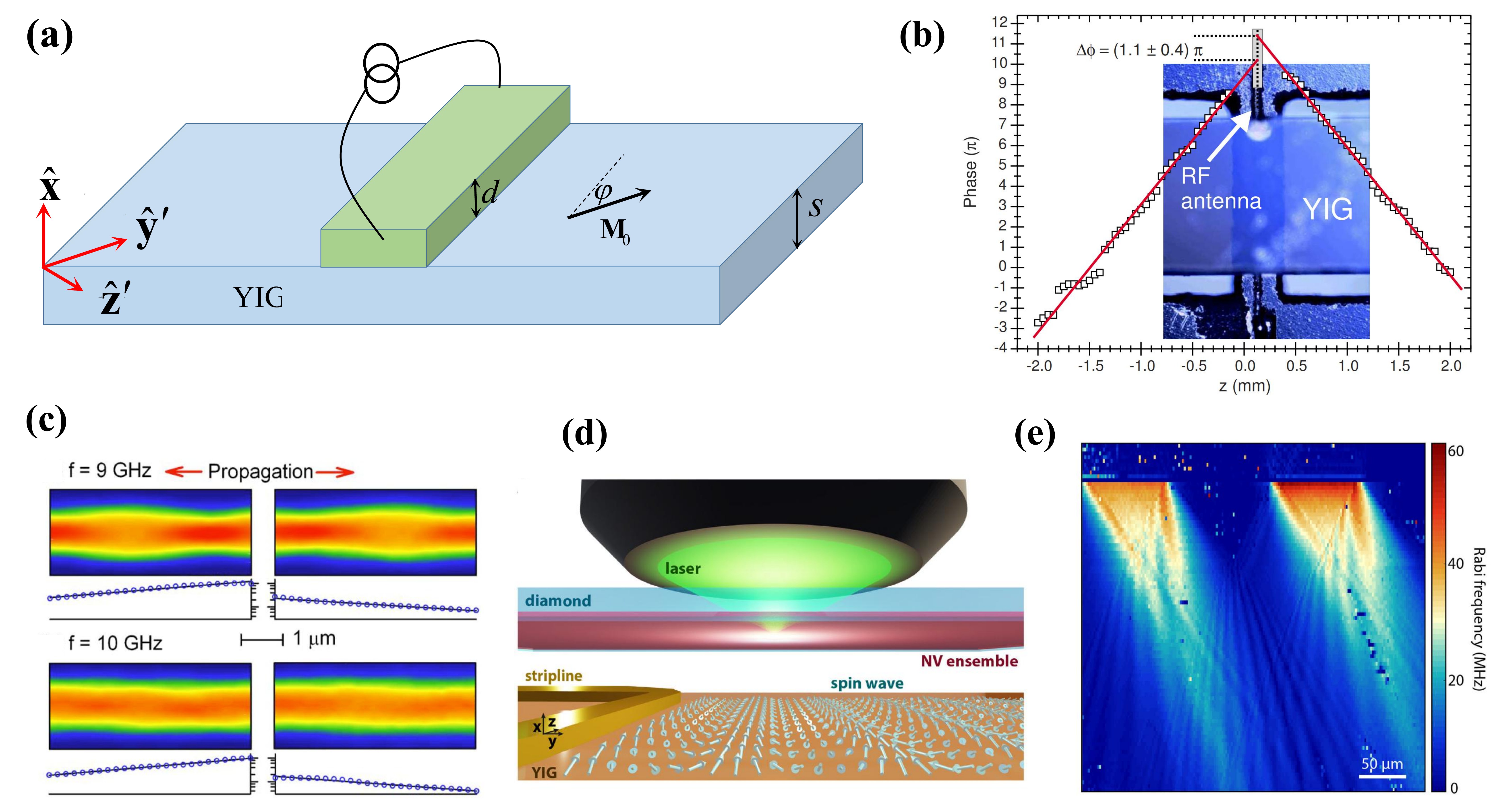}
    		\par\end{centering}
    	\caption{Non-reciprocal and chiral excitation of spin waves by microwave striplines. (a) sketches the typical configuration of a long and narrow metal wire with an ac current bias. The stray magnetic field of this stripline excites the spin waves in an in-plane magnetized film below. Panel (b) shows a  non-reciprocity in the phase of spin wave  propagation parallel to the magnetization and normal to the stripline in opposite direction. (c)-(e) demonstrate an amplitude non-reciprocity when the magnetization is parallel to the stripline. The figures are taken from Refs.~\cite{stripline_poineering_1} [(b)], \cite{stripline_poineering_2} [(c)], and \cite{Teono_NV} [(d) and (e)].}
    	\label{Toeno_exp}
    \end{figure}

We focus on the geometry in Fig.~\ref{Toeno_exp}(a), in which the surface normal of the magnetic film is along $\hat{\bf x}$. The magnetization along  $\hat{\bf z}'$-direction can be rotated by an applied magnetic field. A current-biased stripline along $\hat{\bf z}$ excites the spin waves into the $\hat{\bf y}$-direction and  the relative angle between the saturated magnetization and the stripline direction is $\varphi=\arccos ( \hat{\bf z} \cdot\hat{\bf z}' )$. The Oersted magnetic fields from the stripline interact with the spin waves by the Zeeman interaction \cite{Landau}
     \begin{equation}
     	\hat{H}_{\mathrm{int}}=-\mu_{0}\int\mathbf{M}(\mathbf{r})\cdot\mathbf{H}%
     	(\mathbf{r})dV.
     \end{equation}
     The linear response of the magnetization reads
    \begin{align}
    	M_{\alpha}(x,{\bf k},\omega)=\mu_0(\gamma\hbar)^2\sum_{\beta=\{x,y'\}}\int_{-s}^0dx'\chi_{\alpha\beta}(x,x',{\bf k},\omega)H_{\beta}(x',{\bf k},\omega),
    \end{align}
    where
    \begin{eqnarray}
    	\chi_{\alpha\beta}(x,x',{\bf k},\omega)=-\frac{2M_s}{\gamma\hbar}m_{\alpha}^{{\bf k}}(x)m_{\beta}^{{\bf k}*}(x')\frac{1}{\omega-\omega_{\bf k}+i\Gamma_{\bf k}}
    \end{eqnarray}
    is the spin susceptibility, $m^{\bf k}_{\alpha}(x)$ is the amplitude profile of a spin wave with in-plane momentum ${\bf k}$ into the film and $\Gamma_{\bf k}=2\alpha\omega_{\bf k}$ is the Gilbert damping of the spin waves with frequency $\omega_{\bf k}$. The Zeeman coupling energy is   $-\mu_{0}\sum_{\beta}{m_{\beta}^{{\bf k}*}(x')}H_{\beta}(x',{\bf k},\omega)$.

We limit our theoretical treatment to ultrathin films with $ks\ll1$ in which $m^{\bf k}_{\alpha}(x)$ does not depend on $x$ (Sec.~\ref{magnon_film_wire}).  When the saturated magnetization is along $\hat{\bf z}$ or $\varphi=0$, the excited magnetization in the time domain and position space is the real part of the inverse Fourier transform ($q\equiv k_{y}$),
  \begin{align}
  	\nonumber
  	\mathbf{M}_{\alpha}(x,y,t) &=\sum_{q}e^{iqy-i\omega t}\mathbf{M}_{\alpha
  	}(x,q)\\
  &\approx2i\mu_{0}\gamma\hbar dM_{s}m_{\alpha}^{\left(  q_{\omega}\right)
  	}\sum_{\beta}m_{\beta}^{\left(  q_{\omega}\right)  \ast}\frac{1}{v_{q_{\omega}}%
  	}e^{-i\omega t}\left\{
  	\begin{array}
  		[c]{c}%
  		e^{iq_{\omega}y-\delta_{\omega}y}H_{\beta}(q_{\omega},\omega)\\
  		e^{-iq_{\omega}y+\delta_{\omega}y}H_{\beta}(-q_{\omega},\omega)
  	\end{array}
  	\text{ for }%
  	\begin{array}
  		[c]{c}%
  		y>0\\
  		y<0
  	\end{array}
  	\right.  ,
  	\label{excitations}%
  \end{align}
  where $q_{\omega}+i\delta_{\omega}$ is the positive root of $\omega_{q}%
  =\omega+i\Gamma_{q}$, and $v_{q_{\omega}}$ is the modulus of the group velocity $|\partial\omega_{q}/\partial q|_{q_{\omega}}$. We see that circularly polarized spin waves  propagate in one direction, thereby exciting only half of the film. For the arbitrary directions $\varphi\ne 0$ 
  \[
  m_{x}^{(k_{y})}\rightarrow m_{x}^{(\mathbf{p})},~~~~m_{y}^{(k_{y})}%
  \rightarrow\cos\varphi m_{y}^{(\mathbf{p})}%
  \]
in Eq.~(\ref{excitations}) \cite{Yu_Springer},  where $\mathbf{p}=(0,q\cos\varphi,q\sin\varphi)$ and $q$ solves
  $\omega_{\mathbf{p}}+i2\alpha\omega_{\mathbf{p}}=\omega$. When $\varphi=\pi/2$, only the $x$-component of the magnetic field participates in the spin-wave excitation, and the chirality vanishes. The phases on opposite sides of the stripline are not the same, reflecting  the phase shift $\pi$ between $H_{x}(q_{\omega})$ and $H_{x}(-q_{\omega})$, see Eq.~(\ref{magnetic_fields}).

   \textbf{Experiments}.---Schneider \textit{et al.}  \cite{stripline_poineering_1} observed phase shifts between  spin waves  excited in thick YIG films by striplines when counter-propagating parallel to the  magnetization (volume waves) by space-, time-, and phase-resolved Brillouin light-scattering spectroscopy, as reproduced in Fig.~\ref{Toeno_exp}(b) and called it ``phase non-reciprocity''. Demidov \textit{et al.} \cite{stripline_poineering_2} reported a non-reciprocity in the \textit{amplitude} of spin waves in thin Permalloy films in the Damon-Eshbach configuration, also by Brillouin light scattering microscopy.  Figure~\ref{Toeno_exp}(c) shows the different amplitudes of the magnetization on both sides of the stripline. The chirality is not perfect because the spin waves are elliptically polarized by the strong dipolar interaction. More recently, Bertelli \textit{et al}. \cite{Teono_NV} demonstrated chiral pumping in YIG films by imaging the spin-wave stray fields using diamond NV centers. Figure~\ref{Toeno_exp}(d) and (e) show nearly perfectly chiral magnon pumping into half space.
    
    \subsubsection{Chiral damping and blocking of spin waves}
    \label{Sec_chiral_damping}
    
    \textbf{Chiral damping of magnon by normal metal}.---The interaction between the spin waves of the magnetic films with the proximity metallic gate turns out to be chiral as well because of the chirality of the spin-wave dipolar field (Sec.~\ref{dipolar_fields_1}). The dispersion of long-wavelength magnetostatic spin waves in thick magnetic films or magnonic crystals can be modulated in a non-reciprocal fashion $\omega_{\bf k}\ne \omega_{-{\bf k}}$ by the electromagnetic
interaction with metallic gates, even when they are reciprocal without the gate. The latter can be modelled as thick perfect \cite{perfect_conductor_1} or Ohmic conductors \cite{conductor_1,conductor_2,conductor_3,conductor_4,conductor_5}. Seshadri addressed the
screening of the stray field by a ``perfect" conductor with infinite conductivity, such that the stray fields cannot penetrate the metal cap because of the zero skin depth for electromagnetic waves, in which case the role of the perfect conductor is reduced to the interface boundary condition ${\bf B}_{\perp}=0$ \cite{perfect_conductor_1}. Mruczkiewicz \textit{et al.} reported the effects of thick metal films or focused on the magnonic crystals, where they found that the dispersion of magnetostatic spin
waves is strongly modulated by the conductors \cite{conductor_1,conductor_2,conductor_3,conductor_4,conductor_5}, but the damping is less affected. These studies focused on thick magnetic films.  Recent research on ultrathin films accumulates
evidence that the physics changes drastically when thicknesses are reduced down to nanometer scale \cite{Xiangyang_record}.

Recent experiments showed very significant effects of a metallic environment on the spin-wave dynamics of the stray fields. The stray fields interact with the conduction electrons in metallic contacts including the stripline to generate ``eddy'' currents  \cite{chiral_damping,eddy_damping_1,eddy_damping_2,eddy_damping_3,eddy_damping_4,eddy_damping_5,eddy_damping_6,eddy_damping_7}.  The model  in Ref.~\cite{chiral_damping} considered spin waves in a YIG film of thickness $s$ propagating along the $\hat{\bf y}$-direction under a wide and long metal strip relative to the wavelength \(\lambda\), with thickness $0<x<d$  and magnetization along the $\hat{\bf z}$. The eddy currents in the conductor generate in turn Oersted magnetic fields that affect the spin waves. When $d\ll 1~{\mathrm{\mu}}$m, the eddy currents are uniform across the metal film thickness. The dipolar fields generated by spin waves $\{m_x(y,t),m_y(y,t)\}\propto e^{ik_yy-i\omega t}$ for  $x>0$ are, according to Eq.~(\ref{above_film}), given by the real part of \cite{Chiral_pumping_Yu}
     \begin{subequations}
     \begin{align}
     	&{h}_x(x,y,t)=\frac{1}{2}e^{-|k_y|x}(1-e^{-|k_y|s})\left(m_x(y,t)-i\frac{k_y}{|k_y|}m_y(y,t)\right),\\
     	&{h}_y(x,y,t)=\frac{1}{2}e^{-|k_y|x}(1-e^{-|k_y|s}) \left(-i\frac{k_y}{|k_y|}m_x(y,t)-m_y(y,t)\right).
     \end{align}
     \end{subequations}
 By Faraday's Law, this magnetic field generates a vortex electric field with ${\partial E_z}/{\partial y}=-\mu_0{\partial h_x}/{\partial t}$ and $-{\partial E_z}/{\partial x}=-\mu_0{\partial h_y}/{\partial t}$, where
     	\begin{align}
     		E_z(x,y,t)=(\mu_0\omega/k_y) {h}_x(x,k_y,t)
     	\end{align}
     	drives a current by Ohm's Law 
     	\begin{align}
     		J_{\rm edd}(x,y,t)=\sigma_c(\omega\rightarrow 0)E_z=(\mu_0\omega \sigma_c/k_y){h}_x(x,y,t),
     		\label{eddy_current}
     	\end{align}
with a conductivity $\sigma_c \approx 10^7~{\rm \Omega^{-1}\cdot m^{-1}}$ for good metals. For thin metal films we may set $x\rightarrow d/2$ in Eq.~(\ref{eddy_current}). The Oersted field generated this  current below the conductor ($x\le 0$) reads
    \begin{subequations}
    \begin{align}
    	{\cal H}_x(x,y,t)&=\frac{i\mu_0\omega\sigma_cd}{2}h_x\left(x'=\frac{d}{2},y,t\right)\frac{e^{|k_y|(x-d/2)}}{|k_y|}
    	=\frac{\mu_0\sigma_cds}{4}g\left(-\frac{dm_x}{dt}+i{\rm sgn}(k_y)\frac{dm_y}{dt}\right),\\
    	{\cal H}_y(x,y,t)&=-\frac{\mu_0\omega\sigma_cd}{2} h_x\left(x'=\frac{d}{2},y,t\right)\frac{e^{|k_y|(x-d/2)}}{k_y}=\frac{\mu_0\sigma_cds}{4}g\left(-\frac{dm_y}{dt}-i{\rm sgn}(k_y)\frac{dm_x}{dt}\right),
    \end{align}
\end{subequations}
where $g=e^{-|k_y|d/2}/(|k_y|s)(1-e^{-|k_y|s})\approx 1$ is a form factor for thin films $s,d\ll 1/|k_y|$. When the spin waves are circularly polarized, the Landau-Lifshitz equation (\ref{Heisenberg}) leads to the dispersion
\begin{align}
	\tilde{\omega}_{k_y}=\omega_{k_y}-i\alpha_m\tilde{\omega}_{k_y}\left({\rm sgn}(k_y)+1\right),
\end{align}
with current-induced damping coefficient \cite{chiral_damping}
\begin{align}
	\alpha_m=\mu_0^2\gamma\sigma_cdsM_sg/4,
\end{align}
which should be added to the intrinsic Gilbert damping constant.  This eddy current or radiative damping is chiral since only  spin waves with positive wave number $k_y>0$ generate a dipolar field above the film and suffer from the back action. 

\begin{figure}[ptbh]
    	\begin{centering}
    		\includegraphics[width=0.99\textwidth]{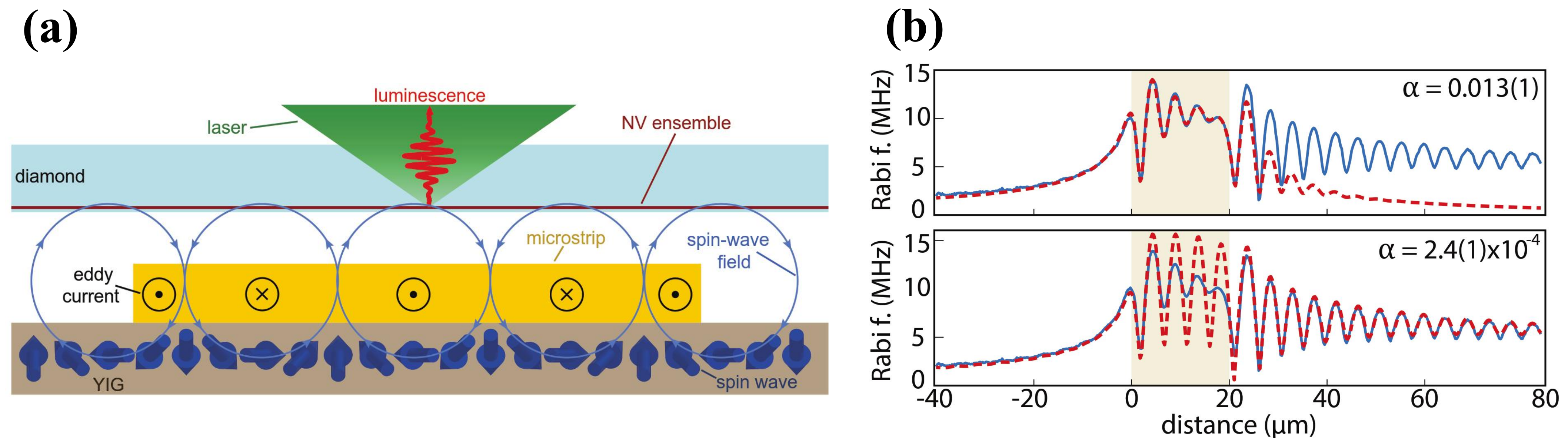}
    		\par\end{centering}
    	\caption{Observation of significant additional damping of the spin wave propagation induced by the eddy-current in the metal contact above the YIG film. In (a), the NV center in diamonds is used to detect the stray field of the spin waves that can penetrate the metal films of sub-skin-depth thickness. (b) The red dashed lines obtained for large (upper panel) and small (lower panel) damping constants \(\alpha\) demonstrate that the damping of spin waves beneath the metallic microstrip is about 50 times larger in magnitude than that outside the microstrip region. The figures are taken from Ref.~\cite{chiral_damping}.}
    	\label{chiral_damping}
    \end{figure}

 The NV-center in diamonds can detect stray fields emitted by spin waves through metal films of sub-skin-depth thickness as sketched in Fig.~\ref{chiral_damping}(a)  by the modulation of their Rabi frequencies. The observation of a significant additional damping of  propagating spin waves  in a YIG film under a metal gate in Fig.~\ref{chiral_damping}(b) agrees with the above estimates for eddy currents \cite{chiral_damping}. The damping parameter under the metallic strip is enhanced up to two orders of  magnitude. Our prediction that the eddy-current induced damping of spin waves propagating in the opposite direction is much smaller still awaits direct experimental observation.
   
\textbf{Unidirectional blocking by proximity superconductor}.---When replacing the normal-metal layer by a superconducting cap, the superconductor has profound effects on magnon transport as well. Golovchanskiy \textit{et al.} reported the effects of thick superconducting films that fully screen the stray magnetic fields, in which limit the method of images is a suitable approach \cite{magnon_superconductor_1,magnon_superconductor_2,magnon_superconductor_3,magnon_superconductor_4}. But the assumption of a
vanishing London’s penetration depth obviously does not apply to the case of a thin superconducting film that is placed on top of ultrathin magnetic films.

\begin{figure}[ptbh]
	\begin{centering}
		\includegraphics[width=1\textwidth]{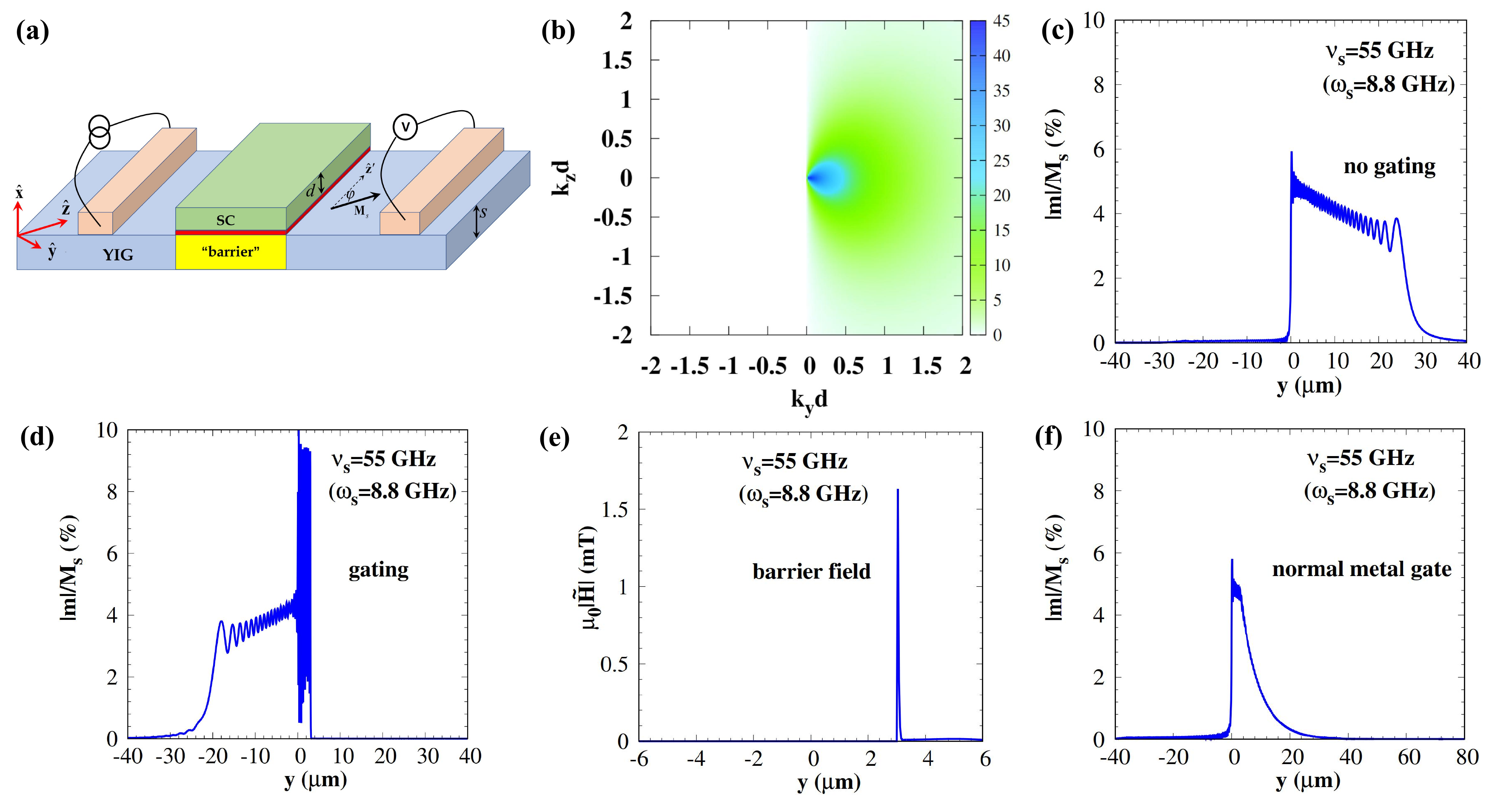}
		\par\end{centering}
	\caption{Unidirectional blocking of spin waves by thin superconducting layer on top of a thin YIG film. (a) illustrates the configuration, in which the equilibrium magnetization is allowed to be tunned in the film plane by an angle $\varphi$ with respect to the stripline direction. In (b), we plot the frequency shift in the momentum space by an extended superconducting layer. When the superconductor is of finite width, this provides a barrier for the spin waves launching the superconductor, such as those propagating along the $\hat{\bf y}$ direction, resulting in contrast behaviors without [(c)] and with [(d)] the superconducting gate located at $y\ge 3~\mu$m. This is because the launching spin waves induce an Oersted field at the edge of the superconductor, as shown in (e), which reflects the spin waves. For comparison, we plot the propagation of spin waves through a normal metal in (f). The figures are taken from Ref.~\cite{chiral_gate}.}
	\label{magnon_gate}
\end{figure}

We recently \cite{chiral_gate} found the chiral gating effect of floating superconducting
films on top of ultrathin films of a magnetic insulator such
as YIG in the configuration of Fig.~\ref{magnon_gate}(a). The thicknesses  of the superconducting layer and thin YIG film are $d$ and $s$, respectively. When the superconducting layer is thinner than London's penetration depth, the stray field, which is chiral as reviewed in Sec.~\ref{dipolar_fields_1}, can penetrate through the superconductor. It generates the electric field that induces the eddy current in the superconducting layer with components given by \cite{chiral_gate}
\begin{subequations}
\begin{align}
J_{x}(x,\pmb{\rho})&=0,\\
J_{y}(x,\pmb{\rho})&=-\left(  {i\sigma_{c}\omega\mu_{0}}/\left\vert
\mathbf{k}\right\vert \right)  {H}_{z}^{(0)}(x,\pmb{\rho}),\\
J_{z}(x,\pmb{\rho})&=\left(  {i\sigma_{c}\omega\mu_{0}}/\left\vert
\mathbf{k}\right\vert \right)  {H}_{y}^{(0)}(x,\pmb{\rho}),
\end{align}
\end{subequations}
where $\pmb{\rho}=y\hat{\bf y}+z\hat{\bf z}$ lies in the film plane and  $\sigma_{c}=i{n_{s}e^{2}}/(m_{e}\tilde{\omega}_{\mathbf{k}})$ is the conductivity of superconductor at magnon frequency $\tilde{\omega}_{\mathbf{k}}$ that is
purely imaginary, with $n_s$ being the superfluid density and $m_e$ being the effective electron mass. $J_x=0$ because the normal component of the electric field is immediately screened by the conductor. By the Maxwell equation $\nabla\cdot {\bf H}=0$, the eddy current is normal to the spin wave propagation ${\bf k}\cdot {\bf J}=0$. This eddy current, as reviewed in Sec.~\ref{Stripline}, in turn, generates Oersted magnetic fields with chirality that oppose the original ones, i.e., the Lenz effect. For $x<0$, 
the transverse ($x$- and $y$-) components
of the Oersted magnetic field read 
\begin{subequations}
\begin{align}
\tilde{H}_{x}(\mathbf{r},\omega)&=ie^{|\mathbf{k}|(x-d)}{d}\sigma
_{c}\omega\mu_{0}({1-e^{-|\mathbf{k}|s}})/(4{|\mathbf{k}|})\left(  m_{x}^{\mathbf{k}}(\pmb{\rho},\omega)-im_{y}^{\mathbf{k}%
}(\pmb{\rho},\omega){k_{y}}/{|\mathbf{k}|}\right)  ,\\
\tilde{H}_{y}(\mathbf{r},\omega)&=(ik_{y}/|\mathbf{k}|)\tilde{H}_{x}(\mathbf{r},\omega).
\end{align}
\end{subequations}
The spin of the Oersted fields contains a transverse component that, again, is locked with the momentum (Sec.~\ref{unification}).
The effect on the gate itself may be disregarded
since we consider only films much thinner than London's penetration. However, they affect the spin wave dynamics by a field-like torque that causes a frequency shift.

By self-consistently solving the linearized LLG equation, we find the superconducting gate shifts the frequency of the spin waves in half momentum space by tens of GHz, given by
\begin{equation}
\delta\omega_{\mathbf{k}}=\frac{d}{4}e^{-|\mathbf{k}|\left(  \frac{s}%
{2}+d\right)  }\frac{1-e^{-\left\vert \mathbf{k}\right\vert s}}{\left\vert
\mathbf{k}\right\vert }\mu_{0}^{2}\gamma M_{s}\frac{n_{s}e^{2}}{m_e
}\left(  \frac{k_{y}}{\left\vert \mathbf{k}\right\vert }+\frac{|k_{y}%
|}{\left\vert \mathbf{k}\right\vert }\right)  ,
\end{equation}
which is real and positive definite. Figure~\ref{magnon_gate}(b) plots the chirality of the frequency shift with $d=40$~nm thick NbN
superconducting film with electron density $n_{s}=10^{29}/\mathrm{m}^{3}$
\cite{carrier_density} on top of a $s=20~\mathrm{nm}$ YIG film.
Such chirality is caused by constructive interference of the dipolar field of the spin waves and the Oersted field induced eddy currents in the superconductors.  

This superconducting gate in Fig.~\ref{magnon_gate}(a) generates an effective barrier that is dynamic, \textit{i.e.}, it depends on the magnon propagation direction and frequency. We may adopt concepts from mesoscopic electronics by structuring gates to trap spin waves, create one-dimensional magnon wave guides, or design magnonic point contacts. When the propagation is normal to the saturated magnetization, the effect of the barrier is chiral, \textit{viz.}, only the spin waves of particular propagation is affected. Such functionalities can be demonstrated by numerical calculations. Now consider placing a
microwave stripline with width $w_{s}=150$~nm and thickness $d_{s}=80$~nm centered at
the origin and biased by a uniform ac current of current density $I_{s}=10^{6}~\mathrm{A/cm^{2}}$  and frequency $\nu_{s}=2\pi
\times\omega_{s}=55$~GHz.
The pumping is chiral as reviewed in above Sec.~\ref{stripline_excitation}: when parallel to the external dc field and magnetization in the underlying YIG, the stripline launches the spin waves with positive $k_{y}^{\ast}$ into only half space as shown in Fig.~\ref{magnon_gate}(c) in
the form of a snapshot of the excited magnetization $\left\vert \mathbf{m}\right\vert $ \cite{Chiral_pumping_Yu,chiral_damping,stripline_poineering_1,stripline_poineering_2}.
When we place a superconducting gate made from an NbN film with thickness $d=40$~nm that covers the
YIG film from $y=w_{1}=3~\mathrm{\mu}$m to $w_{2}=100~\mathrm{\mu}$m, it totally reflects the coherent magnons excited by in stripline microwaves and propagating normal to the magnetization in one, but transmits magnons that approach from the opposite direction, as shown in Fig.~\ref{magnon_gate}(d). The total reflection relies on the induced ``barrier" Oersted field from the superconductor when the spin waves approach that only exists at its edge [Fig.~\ref{magnon_gate}(e)]. The physics of the normal metal is very different in that it only dampens the spin waves \cite{chiral_damping} without any reflection, as shown in Fig.~\ref{magnon_gate}(f).

The on-off ratio of the isolator is nearly unity over tens of GHz, in contrast to narrow-band resonant coupling effects \cite{magnon_trap,spin_wave_diode_1}. This setup is exceedingly simple and does not require spin-orbit interactions, in contrast to metal-based designs \cite{diode_PRX,DMI_circulator}.
    
Recently, the chiral or non-reciprocal interaction between the magnons in an in-plane saturated thin ferromagnetic film and Cooper pairs of the superconducting substrate via the magnetic dipolar coupling is also predicted by Kuznetsov and Fraerman \cite{Fraerman}. They found a non-reciprocal frequency shift in the form of $\Delta\omega({\bf k})\equiv [\omega({\bf k})-\omega(-{\bf k})]\propto ({\bf n}\times {\bf M}_s)\cdot {\bf k}$, i.e., in proportionality to the chirality index defined by the wave number ${\bf k}$, saturated magnetization ${\bf M}_s$, and surface normal ${\bf n}$ of the film. Silaev emphasized the importance of dipolar interaction between the Kittel magnons and Cooper pairs by demonstrating the shift of the ferromagnetic resonance of \textit{metallic} ferromagnets when sandwiched by the superconductors \cite{Silaev}. 
    
\subsubsection{Nonlinear magnon Doppler effect}
    
    \label{Spin_Doppler}
    
\begin{figure}[ptbh]
	\begin{centering}
		\includegraphics[width=0.99\textwidth]{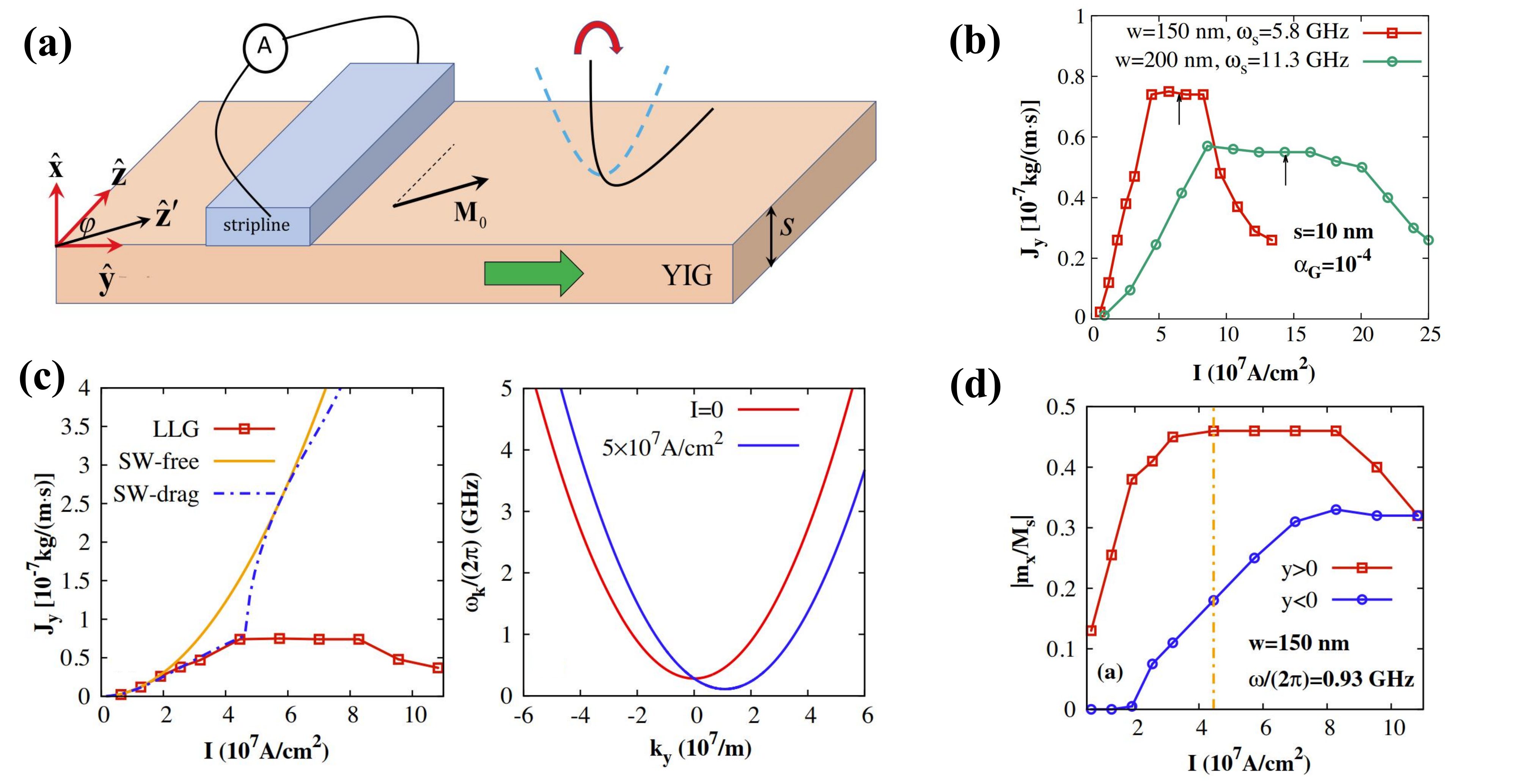}
		\par\end{centering}
	\caption{Spin-wave Doppler shift caused by the drag of a unidirectional spin current excited by a metal stripline along the $\hat{\mathbf{z}}
		$-direction (green arrow) in the 
		 thin YIG films with thickness $s\sim O(10)$~nm, \textit{i.e.}, a tilt of the parabolic magnon dispersion in the direction of the red arrow. Panel (b) shows a maximum spin current for thickness $s=10$ nm  from a numerical solution of  the LLG equation. Panel (c) shows that the mean-field theory results agree  with the numerical ones when the driving current $I$ is below the critical value. Panel (d) demonstrates the breaking of the chirality in the pumping by the nonlinearity. The figures are taken from Ref.~\cite{Doppler_Yu}.}
	\label{Doppler_effect}
\end{figure}

The theory discussed in this review focuses on the linear response regime with weakly excited magnetization, in which the lowest order Holstein-Primakoff expansion [Eq.~(\ref{Bogoliubov_a},\ref{Bogoliubov})] holds and magnons form an ensemble of independent harmonic oscillators. This picture breaks down at higher excitation power and temperature with a larger number of magnons. Many non-linear effects can be captured by the next terms in the expansion that corresponds to 3 and 4 magnon interactions. While non-linearities in excited magnets have a long history, a systematic study of chirality in interacting magnon systems appears to be lacking.
As discussed above (Sec.~\ref{stripline_excitation}), an AC current in a narrow stripline launches unidirectional spin waves into half of a magnetic film. When the amplitude increases with rising power, so does the interaction between the magnons. At a critical amplitude, the Kittel mode decays due to ``Suhl'' instabilities \cite{Suhl_1,Suhl_2}. Yu \textit{et al.} studied the nonlinear effects on a unidirectional spin current on the magnon transport in the configuration of Fig.~\ref{Doppler_effect}(a), predicting a tilt of magnon dispersion caused by an interaction-induced drag of a magnon current, \textit{i.e.}, a Doppler shift  \cite{Doppler_Yu}, previously reported for phonon \cite{Doppler_phonon} and  electrons \cite{Doppler_electron,instability_Doppler1,instability_Doppler2,Duine_Doppler}. Moreover, the interaction limits the maximal spin current that can be driven through a magnetic thin film, see Fig.~\ref{Doppler_effect}(b).

We define the magnetization spin current density from its conservation law as
\begin{equation}
	\tilde{\mathbf{j}}_{\delta}=\alpha_{\mathrm{ex}}\mu_{0}\gamma\mathbf{M}\times\nabla_{\delta}\mathbf{M}, \label{spin_current_density}%
\end{equation}
and the total spin current at the origin of a thin magnetic film of thickness $s$ [Fig.~\ref{Doppler_effect}(a)]
\begin{align}
{\mathbf{J}}_{y}(y=0)=-1/(2\omega_{M}\gamma\alpha_{\mathrm{ex}})\int_{-s}%
^{0}dx\tilde{\mathbf{j}}_{y}(x,y=0),
\end{align} 
where $\omega_{M}\equiv\mu_{0}\gamma
M_{s}$,
as a function of the applied AC electric current density $I(\omega)$. Figure~\ref{Doppler_effect}(b-d) summarizes the results of the mean-field theory reviewed below, which predicts a maximum spin current 
\begin{equation}
	\mathbf{J}_{y}^{\left(c\right)}=\hbar/(4\mathcal{V}_{0})\sqrt{\hbar\omega_{M}\alpha_{\mathrm{ex}}\mu_0\gamma H_{\rm app}}.
	\label{critical}
\end{equation} 
The nonlinearity also breaks the chiral pumping, as shown in Fig.~\ref{Doppler_effect}(d).

We illustrate the principle of the magnonic Doppler effect in terms of an in-plane magnetized YIG film with the surface normal along the $\hat{\mathbf{x}}$-direction and a static magnetic field $\mathbf{H}%
_{\mathrm{app}}$ applied along the $\hat{\mathbf{z}}^{\prime}$-direction, as shown in Fig.~\ref{Doppler_effect}(a). We adopt a Hamiltonian  in terms of spin field operators \textbf{S}, as before:
\begin{equation}
	\hat{H}=\mu_{0}\int\left[  \frac{\gamma\hbar}{2}\mathbf{S}(\mathbf{r}
	)\cdot\hat{\mathbf{H}}_{\mathrm{dip}}(\mathbf{r})+\frac{\gamma^{2}\hbar
		^{2}\alpha_{\mathrm{ex}}}{2}\nabla\mathbf{S}\cdot\nabla\mathbf{S}+\gamma
	\hbar\mathbf{S}\cdot\mathbf{H}_{\mathrm{ext}}\right]  d\mathbf{r},
\end{equation}
where $\mathbf{H}_{\mathrm{dip}}$ is the dipolar field and $\alpha_\mathrm{ex}$ is the exchange parameter.
In ultrathin films, the effect of the dipolar interaction on the magnetization dynamics may be disregarded. In previous Chapters, we replaced the spin Hamiltonian with that of non-interacting magnons, which is an approximation that we improve on here. In terms of a magnon field operator $\hat{\Theta}(\mathbf{r})$, the exact Holstein-Primakoff transformation reads 
\begin{subequations}
	\begin{align}
		&  \hat{S}_{x}(\mathbf{r})+i\hat{S}_{y^{\prime}}(\mathbf{r})=\hat{\Theta
		}^{\dagger}(\mathbf{r})\sqrt{2S-\hat{\Theta}^{\dagger}(\mathbf{r})\hat{\Theta
			}(\mathbf{r})},\\
		&  \hat{S}_{x}(\mathbf{r})-i\hat{S}_{y^{\prime}}(\mathbf{r})=\sqrt
		{2S-\hat{\Theta}^{\dagger}(\mathbf{r})\hat{\Theta}(\mathbf{r})}\hat{\Theta
		}(\mathbf{r}),\\
		&  \hat{S}_{z^{\prime}}(\mathbf{r})=-S+\hat{\Theta}^{\dagger}(\mathbf{r}%
		)\hat{\Theta}(\mathbf{r}).
	\end{align}
\end{subequations}
Here we focus on a thin magnetic film in which the wave interference leads to perpendicular standing spin  waves normal to
the interfaces (PSSWs). Assuming free boundary conditions, the magnon operator can be
expanded as
\begin{equation}
	\hat{\Theta}(\mathbf{r})=\sum_{l}\sqrt{\frac{2}{1+\delta_{l0}}}\frac{1}%
	{\sqrt{s}}\cos\left(  \frac{l\pi}{s}x\right)  \hat{\Psi}_{l}({\boldsymbol{\rho
	}}),
\end{equation}
where $l$ is the subband index and $s$ the film thickness. In terms of
$\hat{\Psi}_{l}({\boldsymbol{\rho}}),$ the in-plane magnon field operators for
subband $l$, the Zeeman and  linear exchange Hamiltonians read
\begin{subequations}
	\begin{align}
		\hat{H}_{\mathrm{Z}}&=\mu_{0}\gamma\hbar H_{\mathrm{app}}\sum_{l}\int\hat{\Psi}%
		_{l}^{\dagger}({\boldsymbol{\rho}})\hat{\Psi}_{l}({\boldsymbol{\rho}%
		})d{\boldsymbol{\rho}},\\
		\hat{H}_{\mathrm{ex}}^{\mathrm{L}}&=\mu_{0}\gamma\hbar M_{s}\alpha_{\mathrm{ex}}\sum_{l\geq1}\left(
		\frac{l\pi}{s}\right)  ^{2}\int\hat{\Psi}_{l}^{\dagger}({\boldsymbol{\rho}%
		})\hat{\Psi}_{l}({\boldsymbol{\rho}})d{\boldsymbol{\rho}}+\mu_{0}\gamma\hbar
		M_{s}\alpha_{\mathrm{ex}}\sum_{l}\int\nabla_{\boldsymbol{\rho}}\hat{\Psi}%
		_{l}^{\dagger}({\boldsymbol{\rho}})\cdot\nabla_{\boldsymbol{\rho}}\hat{\Psi
		}_{l}({\boldsymbol{\rho}})d{\boldsymbol{\rho}}.
	\end{align}
\end{subequations}
The subband edges of the magnon dispersion are therefore at
\begin{equation}
	E_{l}=\mu_{0}\gamma\hbar H_{\mathrm{app}}+\mu_{0}\gamma\hbar M_{s}%
	\alpha_{\mathrm{ex}}(l\pi/s)^{2}.
\end{equation}
In a YIG film with thickness below 10 nm and at temperatures $T\lesssim300$~K,
only the lowest three bands $l=\{0,1,2\}$ are significantly populated.

The leading non-linear terms of the exchange interaction read
\begin{align}
	\hat{H}_{\mathrm{ex}}^{\mathrm{NL}}  &=\sum_{l_{1}l_{2}l_{3}l_{4}}\mathcal{U}_{l_{1}l_{2}l_{3}l_{4}}\int\hat
	{\Psi}_{l_{1}}^{\dagger}({\boldsymbol{\rho}})\hat{\Psi}_{l_{2}}^{\dagger
	}({\boldsymbol{\rho}})\hat{\Psi}_{l_{3}}({\boldsymbol{\rho}})\hat{\Psi}%
	_{l_{4}}({\boldsymbol{\rho}})d{\boldsymbol{\rho}}\nonumber\\ &+\sum_{l_{1}l_{2}l_{3}l_{4}}\mathcal{V}_{l_{1}l_{2}l_{3}l_{4}}\int\hat
	{\Psi}_{l_{1}}^{\dagger}({\boldsymbol{\rho}})\hat{\Psi}_{l_{2}}^{\dagger
	}({\boldsymbol{\rho}})\nabla_{\boldsymbol{\rho}}\hat{\Psi}_{l_{3}%
	}({\boldsymbol{\rho}})\cdot\nabla_{\boldsymbol{\rho}}\hat{\Psi}_{l_{4}%
	}({\boldsymbol{\rho}})d{\boldsymbol{\rho}}+\mathrm{H.c.},
\end{align}
in which
\begin{subequations}
	\begin{align}
   \mathcal{U}_{l_{1}l_{2}l_{3}l_{4}}  &  =\frac{\mu_{0}\gamma^{2}\hbar^{2}\alpha_{\mathrm{ex}}}{4s}\frac{4}{\sqrt{(1+\delta_{l_{1}0})(1+\delta_{l_{2}0})(1+\delta_{l_{3}0})(1+\delta_{l_{4}0})}}\frac{l_{3}\pi}{s}\frac{l_{4}\pi}{s}\mathcal{A}_{l_{1}l_{2}l_{3}l_{4}},\\
   \mathcal{V}_{l_{1}l_{2}l_{3}l_{4}}  &  =\frac{\mu_{0}\gamma^{2}\hbar^{2}\alpha_{\mathrm{ex}}}{4s}\frac{4}{\sqrt{(1+\delta_{l_{1}0})(1+\delta_{l_{2}0})(1+\delta_{l_{3}0})(1+\delta_{l_{4}0})}}\mathcal{B}_{l_{1}l_{2}l_{3}l_{4}},
\label{scattering_terms}
	\end{align}
\end{subequations}
with form factors
\begin{subequations}
	\begin{align}
		\mathcal{A}_{l_{1}l_{2}l_{3}l_{4}}  &  =\frac{1}{s}\int_{-s}^{0}\cos\left(
		\frac{l_{1}\pi}{s}x\right)  \cos\left(  \frac{l_{2}\pi}{s}x\right)
		\sin\left(  \frac{l_{3}\pi}{s}x\right)  \sin\left(  \frac{l_{4}\pi}%
		{s}x\right)  dx,\\
		\mathcal{B}_{l_{1}l_{2}l_{3}l_{4}}  &  =\frac{1}{s}\int_{-s}^{0}\cos\left(
		\frac{l_{1}\pi}{s}x\right)  \cos\left(  \frac{l_{2}\pi}{s}x\right)
		\cos\left(  \frac{l_{3}\pi}{s}x\right)  \cos\left(  \frac{l_{4}\pi}%
		{s}x\right)  dx .
	\end{align}
\end{subequations}
Obviously, the interaction strength should increase with decreasing film thickness. Here we focus on the physical situation in which the microwaves excite the lowest band $l_{1}=0$
coherently to an amplitude $\langle\hat{\Psi}_{l=0}(\boldsymbol{\rho}%
)\rangle \ne 0,$ while
 higher bands are thermally populated. The matrix elements $\mathcal{A}_{0l_{2}l_{3}l_{4}}%
$ with $\{l_{3},l_{4}\}\neq0$ are, accordingly, governed by the selection rules
\begin{subequations}
\begin{align}
	\mathcal{A}_{0l_{2}l_{3}l_{4}}  &=\frac{1}{4}\left(  \delta_{l_{2}+l_{3},l_{4}}+\delta_{l_{2}+l_{4},l_{3}}-\delta_{l_{3}+l_{4},l_{2}}\right),\\
	\mathcal{B}_{0l_{2}l_{3}l_{4}}  &=\frac{1}{4}\left(  \delta_{l_{2}+l_{3},l_{4}}+\delta_{l_{2}+l_{4},l_{3}}+\delta_{l_{2}+l_{3}+l_{4},0}%
	+\delta_{l_{3}+l_{4},l_{2}}\right).
\end{align}
\end{subequations}

With commutator $[\hat{\Psi}_{l^{\prime}}(${$\boldsymbol{\rho}$}$^{\prime}),\hat{\Psi
}_{l}^{\dagger}(${$\boldsymbol{\rho}$}$)]=\delta_{ll^{\prime}}\delta
(${$\boldsymbol{\rho}$}$-${$\boldsymbol{\rho}$}$^{\prime})$ and the Heisenberg
equation of motion $i\hbar\partial_{t}\hat{\Psi}_{l^{\prime}}(\boldsymbol{\rho}^{\prime
})=[\hat{\Psi}_{l^{\prime}}(\boldsymbol{\rho}^{\prime}),\hat{H}],$ the coherent magnons in the lowest band ($l=0$) obey
\begin{align}
	i\hbar\frac{\partial\langle\hat{\Psi}_{l=0}(\boldsymbol{\rho})\rangle
	}{\partial t} &  =E_{l=0}\langle\hat{\Psi}_{l=0}(\boldsymbol{\rho}%
	)\rangle-\hbar\omega_{M}\alpha_{\mathrm{ex}}\nabla^{2}\langle\hat{\Psi}%
	_{l=0}(\boldsymbol{\rho})\rangle\nonumber\\
	&  +2\sum_{l_{2}l_{3}l_{4}}\mathcal{U}_{0l_{2}l_{3}l_{4}}\langle\hat{\Psi
	}_{l_{2}}^{\dagger}(\boldsymbol{\rho})\hat{\Psi}_{l_{3}}(\boldsymbol{\rho
	})\hat{\Psi}_{l_{4}}(\boldsymbol{\rho})\rangle+2\sum_{l_{1}l_{2}l_{3}%
	}\mathcal{U}_{l_{1}l_{2}l_{3}0}\langle\hat{\Psi}_{l_{1}}(\boldsymbol{\rho
	})\hat{\Psi}_{l_{2}}(\boldsymbol{\rho})\hat{\Psi}_{l_{3}}^{\dagger
	}(\boldsymbol{\rho})\rangle\nonumber\\
	&  +2\sum_{l_{2}l_{3}l_{4}}\mathcal{V}_{0l_{2}l_{3}l_{4}}\left(  \langle
	\hat{\Psi}_{l_{2}}^{\dagger}\nabla_{\boldsymbol{\rho}}\hat{\Psi}_{l_{3}}%
	\cdot\nabla_{\boldsymbol{\rho}}\hat{\Psi}_{l_{4}}\rangle-\nabla
	_{\boldsymbol{\rho}}\cdot\langle\hat{\Psi}_{l_{2}}\hat{\Psi}_{l_{3}}%
	\nabla_{\boldsymbol{\rho}}\hat{\Psi}_{l_{4}}^{\dagger}\rangle\right)
	,\label{nonlinear_EOM}
\end{align}
where $\omega_{H}=\mu_{0}\gamma H_{\mathrm{app}}$ and $\omega_{M}=\mu
_{0}\gamma M_{s}$, as defined before. The terms
involving $\mathcal{U}$ vanish in the mean-field approximation of the 3-magnon
amplitudes when $\langle\hat{\Psi}_{l\neq0}(\boldsymbol{\rho})\rangle=0.$ In the last term,  we recognize the magnon current density in subband $l^{\prime}$ (in units of
an angular momentum current J/m$^{2}$)
\begin{equation}
	\mathbf{J}_{l^{\prime}}({\boldsymbol{\rho}})=\frac{\hbar}{2i}\left(
	\langle\hat{\Psi}_{l^{\prime}}^{\dagger}({\boldsymbol{\rho}})\nabla
	_{\boldsymbol{\rho}}\hat{\Psi}_{l^{\prime}}({\boldsymbol{\rho}})\rangle
	-\langle\hat{\Psi}_{l^{\prime}}({\boldsymbol{\rho}})\nabla_{\boldsymbol{\rho}%
	}\hat{\Psi}_{l^{\prime}}^{\dagger}({\boldsymbol{\rho}})\rangle\right)
	,\label{momentum_current}%
\end{equation}
which when integrated leads to an expression in terms of magnon occupation numbers
\begin{equation}
	{\pmb {\cal J}}_{l^{\prime}}=\int\mathbf{J}_{l^{\prime}}({\boldsymbol{\rho}%
	})d{\boldsymbol{\rho}}=\frac{1}{2}\sum_{\mathbf{k}^{\prime}}\hbar
	\mathbf{k}^{\prime}\left(  \langle\hat{\Psi}_{l^{\prime}}^{\dagger}%
	(\mathbf{k}^{\prime})\hat{\Psi}_{l^{\prime}}(\mathbf{k}^{\prime}%
	)\rangle+\langle\hat{\Psi}_{l^{\prime}}(\mathbf{k}^{\prime})\hat{\Psi
	}_{l^{\prime}}^{\dagger}(\mathbf{k}^{\prime})\rangle\right)  .
\end{equation}
$\mathbf{J}$ is a spin current since (exchange) magnons carry spin $\hbar$.
The expressions are consistent with the magnon density current $\mathbf{\tilde
	{J}}_{l}$ defined by the Heisenberg equation of motion and the magnon conservation law
(in the absence of damping)
\begin{equation}
	\frac{\partial\hat{\rho}_{m}^{l}}{\partial t}=\frac{1}{i\hbar}[\hat{\rho}%
	_{m}^{l},\hat{H}_{\mathrm{ex}}^{\mathrm{L}}]=-\nabla\cdot\mathbf{\tilde{J}%
	}_{l},
\end{equation}
which leads to
\begin{equation}
	\left\langle \mathbf{\tilde{J}}_{l}({\boldsymbol{\rho}})\right\rangle
	=\omega_{M}\alpha_{\mathrm{ex}}\frac{1}{i}\left(  \langle\hat{\Psi}%
	_{l}^{\dagger}({\boldsymbol{\rho}})\nabla_{\boldsymbol{\rho}}\hat{\Psi}%
	_{l}({\boldsymbol{\rho}})\rangle-\langle\hat{\Psi}_{l}({\boldsymbol{\rho}%
	})\nabla_{\boldsymbol{\rho}}\hat{\Psi}_{l}^{\dagger}({\boldsymbol{\rho}%
	})\rangle\right),
\end{equation}
which equals $\mathbf{J}_{l}$ divided by the constant magnon mass.
The equation of motion of the coherent amplitude Eq.~(\ref{nonlinear_EOM}) therefore becomes 
\begin{align}
	i\hbar\frac{\partial\langle\hat{\Psi}_{0}(\boldsymbol{\rho})\rangle}{\partial
		t}=E_{l=0}\langle\hat{\Psi}_{0}(\boldsymbol{\rho})\rangle-\hbar\omega
	_{M}\alpha_{\mathrm{ex}}\nabla^{2}\langle\hat{\Psi}_{0}(\boldsymbol{\rho
	})\rangle +\frac{8i}{\hbar}\sum_{l^{\prime}}{\mathcal{V}}_{00l^{\prime}l^{\prime}%
	}\nabla_{\boldsymbol{\rho}}\langle\hat{\Psi}_{0}({\boldsymbol{\rho}}%
	)\rangle\cdot\mathbf{J}_{l^{\prime}}({\boldsymbol{\rho}}),
	\label{EOM2}%
\end{align}
driven by both incoherent and coherent magnons via the current density
$\mathbf{J}_{l^{\prime}}$.

The excited magnon current in the configuration of Fig.~\ref{Doppler_effect}(a) $\mathbf{J}_{y}(y>0)=\overline
{\mathbf{J}}_{y}\exp({-y/\delta})$ decays exponentially with distance from the
source on the scale of the decay length $\delta=2/\operatorname{Im}\kappa_{y}$,
 which as a  root of $(\omega_{s}-\mu_{0}\gamma H_{\mathrm{app}}-\omega_{M}%
\alpha_{\mathrm{ex}}\kappa_{y}^{2})^{2}+(\alpha_{G}\omega_{s})^{2}=0$ reads $\delta  \sim\sqrt{(\alpha_{\mathrm{ex}}\omega
	_{M})(\omega_{s}-\mu_{0}\gamma H_{\mathrm{app}})}/(\alpha_{G}\omega_{s})$.  Near the stripline, the magnon current in the lowest band obeys the integral equation,
\begin{equation}
	\overline{\mathbf{J}}_{y}=\frac{1}{\delta}  \frac{\mu^2_{0}{\gamma\hbar M_{s}s}}{2\hbar^2}\int\frac{dk_{y}}{2\pi}k_{y}\frac{\left\vert H_{x}(k_{y})-i H_{y}%
		(k_{y})\right\vert ^{2}}{(\omega_{s}-\tilde{\omega}_{k_{y}})^{2}+\alpha
		_{G}^{2}\omega_{s}^{2}}, \label{stripline_spin_current}%
\end{equation}
with Doppler-shifted magnon frequency
\begin{equation}
	\tilde{\omega}_{\mathbf{k}}=\mu_{0}\gamma H_{\mathrm{app}}+\omega_{M}%
	\alpha_{\mathrm{ex}}k^{2}-({8}/{\hbar^{2}}){\mathcal{V}}_{0}k_{y}%
	\overline{\mathbf{J}}_{y}, 
	\label{excitations_Doppler}%
\end{equation}
which poses a self-consistency problem. For weak driving, this description appears to correctly describe the excited spin current, see Fig.~\ref{Doppler_effect}(c), but deviates from the result for non-interacting magnons that scale as $|\mathbf{J}_{y}|\propto I^{2}$.
A negative frequency implies instability of the ferromagnetic ground state at a critical magnon current estimated by Eq.~(\ref{critical}). Thereby, by the back action of magnon currents on the magnetic order, the spin current limitations by nonlinearity are important design parameters in magnonic nano-devices.

Current-biased electric contacts made from heavy metals can generate gradients in the  magnon chemical potential or temperature by the injection of a magnon accumulation via the spin Hall effect or via Joule heating and spin Seebeck, respectively. We estimate the associated incoherent magnon currents by solving the linearized
Boltzmann equation in the relaxation time approximation. Assuming that the
drag term in the collision integral is small
\begin{equation}
	-\mathbf{v}_{\mathbf{k},l}\cdot\nabla T\frac{\partial f_{\mathbf{k},l}%
	}{\partial T}=-\frac{f_{\mathbf{k},l}-f_{\mathbf{k},l}^{(0)}}{\tau
		_{\mathbf{k},l}},
\end{equation}
where $\mathbf{v}_{\mathbf{k},l}=(1/\hbar)\partial E_{l}(\mathbf{k}%
)/\partial\mathbf{k}=2\omega_{M}\alpha_{\mathrm{ex}}\mathbf{k}$ is the magnon
group velocity, $f_{\mathbf{k},l}=\langle\hat{\Psi}_{l}^{\dagger}%
(\mathbf{k})\hat{\Psi}_{l}(\mathbf{k})\rangle$ is the magnon occupation of
the $l$-th subband, $f_{\mathbf{k},l}^{(0)}=1/\{\exp[E_{\mathbf{k},l}%
/(k_{B}T)]-1\}$ is the equilibrium Planck distribution at temperature $T$, and
$\tau_{\mathbf{k},l}$ is a magnon relaxation time. A uniform $\nabla
T=\mathcal{E}_{y}\hat{\mathbf{y}}$ then generates a magnon momentum current
\begin{equation}
	\mathbf{J}_{l}=\hbar\omega_{M}\alpha_{\mathrm{ex}}\mathcal{E}_{y}%
	\hat{\mathbf{y}}\int\frac{dk_{y}dk_{z}}{2\pi^{2}}k_{y}^{2}\tau_{\mathbf{k}%
		,l}\frac{\partial f_{\mathbf{k},l}^{(0)}}{\partial T}, \label{thermal_current}%
\end{equation}
which affects the coherent magnon amplitude by substitution into Eq.~(\ref{EOM2}). We predict that a temperature gradient of $4\,\mathrm{K/\mu m}$ at $T=300$~K generates the critical spin current [Eq.~(\ref{critical})].

\subsubsection{Chiral magnon-photon interaction and microwave diodes}
\label{Microwave_diodes}

\textbf{Chiral magnon-photon interaction}.---In Sec.~\ref{waveguide_fields}, we have pointed out  that  microwaves at special points or lines  in wave guides are circularly polarized with a sign that depends on their propagation direction. The excitations of magnetic particles placed onto these points and magnetized normal to the polarization plane couple preferentially to only one photon polarization and therefore wave vector \cite{waveguide_Yu_1,waveguide_Yu_2}. Here we discuss the consequences of this chiral magnon-photon interaction for cavities and wave guides loaded with more than one magnet.

We start from a Zeeman interaction for a magnetic particle at  $\mathbf{r}_0=(x_0,y_0,z_0)$  that is small compared to  the photon wavelength 
\begin{equation}
	\hat{H}_{\rm int}={\mu_0}\int {\bf H}({\bf r},t)\cdot{\bf M}({\bf r},t)d{\bf r} \approx \mu_0 {\bf H}({\bf r}_0,t)\cdot V_s {\bf M}(t),
\end{equation}
where $\mathbf{M}(t)$ is the uniform precession, \textit{i.e.}, the Kittel mode, and \(V_s\) is the particle volume. When  $\mathbf{M}_0 \parallel\hat{\bf y}$, the non-equilibrium magnetization can be expressed in terms of the magnon operator $\hat{\alpha}_{K}$,
\begin{equation}
	\hat{M}_{\beta=\{z,x\}}=-\sqrt{2M_s\gamma\hbar}\left(M_{\beta}^K\hat{\alpha}_K(t)+M_{\beta}^{K*}
	\hat{\alpha}^{\dagger}_K(t)\right),
\end{equation}
where the amplitudes $M_{z,x}^K$ are normalized as
\begin{equation}
	\int d{\bf r}\left(M_z^K({\bf r})M_x^{K*}({\bf r})-M_z^{K*}({\bf r})M_x^K({\bf r})\right)=-i/2.
\end{equation}
Since $M_x=iM_z$ this leads to $M_z^K={1}/({2\sqrt{V_s}})$ . Inserting the magnetic field operator Eq.~(\ref{HExp}), the interaction Hamiltonian becomes
\begin{equation}
	\hat{H}_{\rm int}=\hbar \sum_{k,\lambda}\big(g_{k,\lambda}\hat{p}_{k,\lambda}\hat{\alpha}^{\dagger}_K
	+h_{k,\lambda}\hat{p}_{k,\lambda}\hat{\alpha}_K+{\rm H.c.}\big).
\end{equation}
The rotating wave approximation corresponds to dropping the $h$-term while the coupling constant
\begin{equation}
	g_{k,\lambda}=\frac{\mu_0}{2}\sqrt{\frac{2\gamma M_sV_s}{\hbar L }}\left(i{\cal H}^{\lambda}_{k,x}(x_0,y_0)-{\cal H}^{\lambda}_{k_z,z}(x_0,y_0)\right)e^{ik_zz_0},
\end{equation}
where ${\cal H}$ denotes the magnetic field of the waveguide mode. Using  Eq.~(\ref{rec}) for the TE$_{10}$ mode in a rectangular waveguide
\begin{equation}
	g_k=\sqrt{\frac{\gamma M_s V_s}{\hbar Lab}}\frac{i}{\omega_k}\left(-k\sin\left(\frac{\pi x_0}{a}\right)+\frac{\pi}{a}\cos\left(\frac{\pi x_0}{a}\right)\right)e^{ikz_0}.
	\label{eqn:g_chiral}
\end{equation}
$|g_k|\ne |g_{-k}|$, so the coupling is non-reciprocal. Chiral coupling is realized at particular magnon frequencies $\omega_m$ and by placing the magnet at 
\begin{equation}
	x_0=\frac{a }{\pi}\arctan{ \left[ \frac{\pi } {a \sqrt{ \Big(\frac{ \omega_m}{c} \Big)^2-\Big( \frac{\pi}{a} \Big)^2 }   } \right]}.
\end{equation}
The total Hamiltonian is of Anderson-Fano type \cite{Fano} 
\begin{equation}
	\hat{H}=\hbar\omega_m \hat{m}^\dagger \hat{m}+\sum_{k}\hbar \omega_{k} \hat{p}_{k}^\dagger \hat{p}_{k}+\sum_{k} \hbar(g_{k}\hat{p}_{k}^\dagger \hat{m}+g^*_{k}\hat{m}^\dagger \hat{p}_{k}), 
\end{equation}
where $\omega_m$ is the Kittel mode frequency.

\textbf{Microwave diode}.---The chiral coupling with the magnets introduces a non-reciprocity of the microwave transmission that in contrast to an electronic diode is a linear effect. In terms of the input-output theory introduced in Sec.~\ref{dipolar_pumping}, the microwave scattering matrix of a wave guide with loaded one magnet that couples to the TE$_{10}$ mode above 
\begin{subequations}
\begin{align}
	&S_{21}(k)=1-i\frac{L}{v_{k}}\frac{|g_{k}|^2}{\omega_{k}-\omega_m(1-i\alpha_G)-\Sigma(\omega_{k})},\\
	&S_{11}(k)=-i\frac{L}{v_{k}}\frac{g_{k}g_{-k}^*}{\omega_{k}-\omega_m(1-i\alpha_G)-\Sigma(\omega_{k})},
\end{align}
\end{subequations}
where $v_k$ is the photon group velocity and
\begin{equation}
	\Sigma(\omega_k)=-i\frac{L}{2v_k}(|g_k|^2+|g_{-k}|^2)
	\label{eqn:self_energy_solo}
\end{equation}
is the magnon self-energy. $S_{11}(k)=0$ when the magnon-photon coupling is chiral ($g_k=0$ or $g_{-k}=0$) and the magnon does not reflect any photons. The transmission at resonance ($\omega_k=\omega_{m}$) then reads
\begin{align}
	S_{21}(k)=\left\{\begin{matrix}
		e^{ikL},\\
		\xi e^{ikL},
	\end{matrix}\right.\quad
	\begin{matrix}
		{\rm when}~~g_{k}=0,\\
		~{\rm when}~~g_{-k}=0,
	\end{matrix}
\end{align}
where the prefactor 
\begin{align}
	\xi=\frac{2\alpha_G\omega_{m}-|g_k|^2/v_k}{2\alpha_G\omega_{m}+|g_k|^2/v_k}
\end{align}
suppresses the amplitude of the transmitted waves because the damping $\alpha_G>0$ implies $|\xi|<1$. When $g_k=0$, the magnets do not interact with the microwave with momentum \textit{k} and acquire only  the propagation phase $kL$. On the other hand, when $g_{-k}=0$ the microwaves are absorbed and  re-emitted by the magnets, resulting in a double dissipative phase shift $\pi/2$. When the magnetic quality is high $2\alpha_G\omega_{m}\ll |g_k|^2/v_k$ and $\xi \rightarrow e^{i\pi}$ is a pure phase shift that directly reflects this process. For larger damping $2\alpha_G\omega_{m} \approx |g_k|^2/v_k$, the transmission vanishes, and the absorbed photon energy accumulate in the magnet \cite{magnon_trap}. Finally, when $2\alpha_G\omega_{m}\gg |g_k|^2/c_r$,   $\xi\rightarrow 1$ and damping renders the magnet invisible to the photons. The results hold for point-like magnets when their radius is much smaller than the wavelength of the microwaves and cross section of the wave guide. Otherwise, the microwave field is non-uniform over the magnet, and other than the Kittel mode can be excited. Such a situation can best be addressed by numerically solving the coupled LLG and Maxwell equations, but we expect that the qualitative features of point magnetic dipoles persist.

The chiral coupling between magnons and waveguide-photons also affects the reciprocal process of generating microwaves by magnetization dynamics. Any motion of the magnetization generates magnetic stray fields that can ideally escape only in one direction. This unidirectional photon emission  transfers energy and angular momentum from the magnet into the electromagnetic fields and is a form of radiative damping \cite{radiative_damping_1,radiative_damping_2,radiative_damping_exp}, and is closely related to the chiral pumping of spin waves \cite{waveguide_Yu_1}. Analogously, two magnetic spheres in a waveguide may trap photons, very similar to the trapping of spin waves by magnetic strip lines (refer to Sec.~\ref{Sec_magnon_trap}) \cite{magnon_trap}. 

\begin{figure}[ptbh]
	\begin{centering}
		\includegraphics[width=0.99\textwidth]{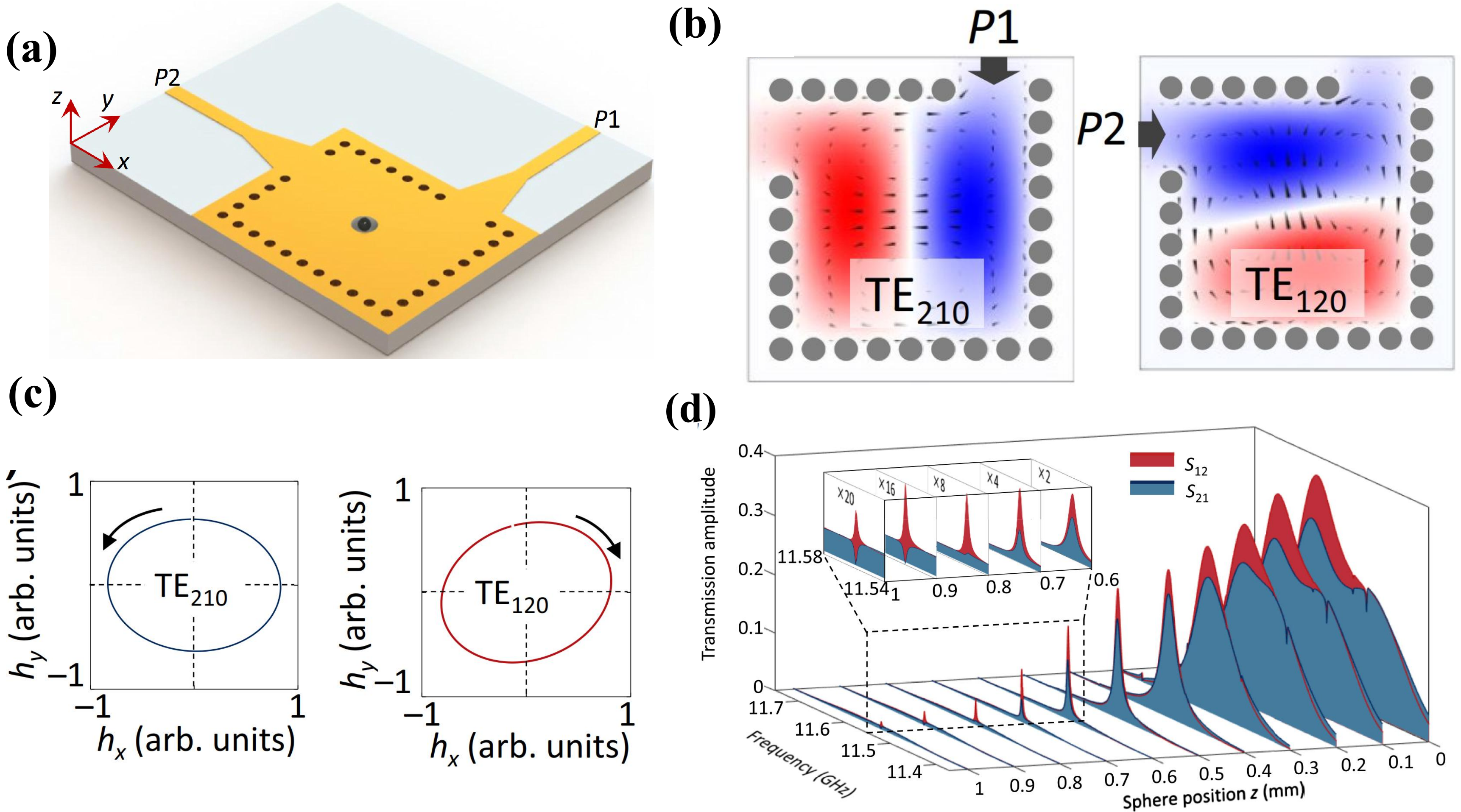}
		\par\end{centering}
	\caption{Observation of non-reciprocal microwave transmission by the chiral magnon-photon interaction. (a) illustrates the configuration in which a magnetic sphere is placed in the center of the planar cavity. When the cavity has a square shape, port $P_1$ is a feed of microwaves that at the magnet are right circularly polarized, whereas port $P_2$ at the same frequency sends microwaves of opposite polarity, as shown in (c) and (d). The chiral magnon-photon interaction then leads to non-reciprocal microwave transmission, as shown in (d), which emphasized the dependence om the perpendicular position of the YIG sphere relative to the cavity center $\{x,y,z\}=0$. The figures are taken from Ref.~\cite{Xufeng_exp}.}
	\label{Xufeng_exp}
\end{figure}

\textbf{Experimental evidence}.---Zhang \textit{et al}. exploited the principle that the magnon can only interact with the microwave photons of the same polarity to realize broadband non-reciprocity of microwave transmission by loading a specially designed microwave cavity with a YIG sphere \cite{Xufeng_exp}. The cavities are square or rectangular shaped slabs fabricated from a dielectric, as reproduced in Fig.~\ref{Xufeng_exp}(a). When unperturbed, the cavities support ${\rm TE}'_{210}$ and  ${\rm TE}'_{120}$ modes that at the center of the cavity $\{x=0,y=0,z\}$ are linearly polarized along $\hat{\bf y}$ and $\hat{\bf x}$. When fed by ports, $P_1$ and $P_2$, both modes are populated, but with different phases $\varphi_1=\pi/2$ and $\varphi_2=-\pi/2$, respectively, because their fields at the entrances differ. This leads to the excitation of right (left) circularly polarized magnetic fields, ${\rm TE}_{210}={\rm TE}'_{210}+e^{i\varphi_1}{\rm TE}'_{120}$ (${\rm TE}_{120}={\rm TE}'_{210}+e^{i\varphi_2}{\rm TE}'_{120}$), at the origin when fed by port $P_1$ ($P_2$), as shown in Fig.~\ref{Xufeng_exp}(b) and (c). A YIG sphere with magnetization along the $\hat{\bf z}$-direction at the center couples only to the right circularly polarized mode input from $P_1$, but not the one from $P_2$, causing non-reciprocity of the microwave transmission $S_{21} \ne S_{21}$, as shown in Fig.~\ref{Xufeng_exp}(d). The position $z$ of the YIG sphere in the dielectric affects the coupling strength by the spatial dependence of the microwave mode amplitude at the magnet. The non-reciprocity in the transmission is observed for a 0.5~GHz frequency band. It can be improved by loading the cavity with more or larger spheres that both enhance the interaction matrix elements \cite{circulating_polariton}.

Recently, Zhong \textit{et al.} also observed the chiral coupling effect between the Kittel magnons in the YIG magnetic sphere and the microwave photons in a cross-shaped cavity, where the field polarizations change as well at different positions \cite{Zhong}. They successfully displayed a functionality of nonreciprocal transmission using the
chiral condition in the magnon-photon coupling. Bourhill \textit{et al.} generated the circulating cavity magnon polaritons, which was predicted by Yu \textit{et al.} in terms of the chiral magnon-photon interaction \cite{circulating_polariton}, in a torus-shaped cavity that supports the degenerate clockwise and counterclockwise photon modes (refer to Sec.~\ref{waveguide_fields}) when loading a YIG sphere \cite{circulating_polariton_exp}. Via the chiral coupling preferably to one mode by choosing the polarization of magnons with proper magnetic field direction, the  degeneracy between clockwise and counterclockwise travelling microwaves is lifted due to their different coupling to the Kittel magnon. 
 
\subsubsection{Spin skin effect}  
\label{spin_skin_effect}

The exchange interaction between localized spins in condensed matter, whether symmetric or antisymmetric (DMI), acts only over interatomic distances between nearest neighbors that become much smaller in the next nearest neighbors. The magnetic dipole interaction is much weaker but generates long-range forces that decay algebraically with distance and importantly affect macroscopic magnets. Magnetic moments can also interact indirectly through non-magnetic atoms (superexchange) or a non-magnetic medium. Mobile electrons in metals mediate an oscillatory Ruderman-Kittel-Kasuya-Yosida (RKKY) \cite{RKKY_1,RKKY_2,RKKY_3} non-local exchange interaction that extends over nanometers. The dissipative coupling by non-equilibrium spin currents through metal spacers can synchronize  magnetic precessions over micrometers \cite{spin_pumping_RMP}. The exchange of spin waves synchronizes magnetic oscillators over even larger distances. 

Microwave photons in high-quality devices such as cavities and waveguides, as reviewed above, are well suited to communicate spin information because of their coherence over large distances and the strong interaction of spin ensembles such as ferromagnets with AC magnetic fields. Here we review the theoretical approaches and results for the chirally coupled magnets over large distances by microwaves and highlight the novel functionality of the spin (magnon) skin effect.

\textbf{Landau-Lifshitz-Gilbert (LLG) and Maxwell equations}.---The standard approach to formulate the coupling of magnets by microwaves is the LLG equation including the torques by the magnetic fields governed by the Maxwell equation \cite{spin_pumping_RMP}. This approach becomes tedious when considering multiple magnets, but we may illustrate the principle for just two magnetic particles (magnonic hydrogen molecule), labeled by 1 and 2 with coordinate $z_{1}>z_{2}$. They are tuned to interact with the TE$_{10}$ mode, of a rectangular microwave waveguide as in Fig.~\ref{waveguide_waves}(a). In Sec~\ref{dipolar_pumping} we formulated the magnetic field radiated from sphere 2 as experienced by sphere 1, $\tilde{\mathbf{H}}%
_{2\rightarrow 1}^{r}$, and from sphere 1 to sphere 2,  $\tilde{\mathbf{H}}_{1\rightarrow 2}^{r}$:
\begin{subequations} 
	\begin{align}
		\tilde{H}_{2\rightarrow 1,\delta }^{(r)}(\mathbf{r}_{1},t)& =-i\frac{\mu_{0}V_{s}}{v(k_{\omega })}\mathcal{H}_{k_{\omega },\delta }(\pmb{\rho}_{1})%
		\mathcal{H}_{k_{\omega },\eta }^{\ast }(\pmb{\rho}_{2})M_{2,\eta } e^{ik_{\omega }(z_{1}-z_{2})}, \\
		\tilde{H}_{1\rightarrow 2,\delta }^{(r)}(\mathbf{r}_{2},t)& =-i\frac{\mu
			_{0}V_{s}}{v(k_{\omega })}\mathcal{H}_{-k_{\omega },\delta }(\pmb{\rho}_{2})%
		\mathcal{H}_{-k_{\omega },\eta }^{\ast }(\pmb{\rho}_{1})M_{1,\eta } e^{ik_{\omega }(z_{1}-z_{2})},
	\end{align}
\end{subequations}
where $k_{\omega }$ is the photon momentum at frequency $\omega$,  $v(k)$ is the photon group velocity, and $\mathcal{H}$ is the magnetic field of the selected cavity mode, see Eq.~(\ref{rec}). The associated  radiative damping \cite{radiative_damping_exp,radiative_damping_1,radiative_damping_2,eddy_damping_3} reads
\begin{equation}
	\alpha _{\delta =\{x,z\}}^{\left( r\right) }(\omega)=\frac{\mu _{0}^{2}V_{s}}{%
		2\omega}\frac{\gamma M_{s}}{v(k_{\omega })}%
	\left( |\mathcal{H}_{k_{\omega },\delta }(\pmb{\rho})|^{2}+|\mathcal{H}%
	_{-k_{\omega },\delta }(\pmb{\rho})|^{2}\right) .
\end{equation}
Substituting these fields into the LLG equation for the two spheres \cite{waveguide_Yu_2}
\begin{subequations} 
	\begin{align}
		\frac{dM_{1,\alpha }}{dt}&=-\gamma \mu _{0}\varepsilon _{\alpha \beta
			\delta }M_{1,\beta }H_{1,\delta }^{\mathrm{eff}}+\gamma \mu _{0}\varepsilon
		_{\alpha \beta \delta }M_{1,\beta }\tilde{H}_{2\rightarrow 1,\delta }^{(r)}+\frac{\alpha _{G}+\alpha _{1,\delta }^{(r)}}{M_{1,s}}\varepsilon _{\alpha
			\beta \delta }M_{1,\beta }\frac{dM_{1,\delta }}{dt}, \\
		\frac{dM_{2,\alpha }}{dt}& =-\gamma \mu _{0}\varepsilon _{\alpha \beta
			\delta }M_{2,\beta }H_{2,\delta }^{\mathrm{eff}}+\gamma \mu _{0}\varepsilon
		_{\alpha \beta \delta }M_{2,\beta }\tilde{H}_{1\rightarrow 2,\delta }^{(r)}+\frac{\alpha _{G}+\alpha _{2,\delta }^{(r)}}{M_{2,s}}\varepsilon _{\alpha
			\beta \delta }M_{2,\beta }\frac{dM_{2,\delta }}{dt}.
	\end{align}
\end{subequations}
The in-phase and out-of-phase components of the radiated magnetic fields contribute \textit{field-like}
and \textit{damping-like} torques on the local magnetizations, respectively. 
Linearizing the coupled LLG equations and disregarding the small intrinsic
Gilbert damping $\alpha _{G}$, the collective mode frequencies solve
\begin{align}
	\left( \omega -\omega _{1}+i\omega _{1}\alpha _{1}^{(r)}\right) \left(
	\omega -\omega _{2}+i\omega _{2}\alpha _{2}^{(r)}\right)+\left( i{J}_{zx}-i{J}_{xz}+{J}_{xx}+{J}_{zz}\right) (i{P}_{zx}-i{P}_{xz}+{P}%
	_{xx}+{P}_{zz})/4 =0,
	\label{eqn:eigen_classical}
\end{align}  
where $\omega _{i}=\gamma \mu _{0}H_{\mathrm{eff},i}$, $\alpha
_{i}^{(r)}=(\alpha _{i,x}^{(r)}+\alpha _{i,z}^{(r)})/2$, and
\begin{subequations}
\begin{align}
	{J}_{\delta \eta }&=\frac{\gamma \mu _{0}^{2}V_{s}M_{s}}{v(k_{\omega })}%
	\mathcal{H}_{k_{\omega },\delta }(\pmb{\rho}_{1})\mathcal{H}_{k_{\omega
		},\eta }^{\ast }(\pmb{\rho}_{2})e^{ik_{\omega }(z_{1}-z_{2})}, \\
	{P}_{\delta \eta }&=\frac{\gamma \mu _{0}^{2}V_{s}M_{s}}{v(k_{\omega })}%
	\mathcal{H}_{-k_{\omega },\delta }(\pmb{\rho}_{2})\mathcal{H}_{-k_{\omega
		},\eta }^{\ast }(\pmb{\rho}_{1})e^{ik_{\omega }(z_{1}-z_{2})}.
\end{align}
\end{subequations}
When $\omega_1=\omega_2$, the collective modes are out-of-phase (optical) and in-phase (acoustic) with different imaginary parts (reciprocal lifetimes):
\begin{align}
	\omega_{\pm}=\omega_1-i\omega_1\alpha_1^r\pm \frac{i}{2}\sqrt{(iJ_{zz}-iJ_{xx}+J_{xx}+J_{zz})(iP_{zx}-iP_{xz}+P_{xx}+P_{zz})}.
\end{align}

\textbf{Master equation approach}.---The coherent and dissipative components of the coupling can be understood best by the
so-called master (or Lindblad) equation \cite{Molmer,Gardiner_chiral,chiral_emitter_light_5,chiral_optics}. The interaction between the magnons by exchanging photons can be expressed  in terms of a self-energy \(\Sigma\) by integrating over the photon degree of freedom. Turning to a chain of magnets in a cavity with equal distance \textit{d},  the self-energy due to the interaction between magnets with indices \textit{j} and \textit{l}  reads to leading order:
\begin{equation}
	\Sigma_{jl}(\omega)=\int \frac{dk}{2\pi }\frac{g_{j}(k)g_{l}^{\ast }(k)}{\omega-\omega
		_{k}+i0^{+}}=-i%
	\begin{cases}
		(\Gamma_{L}+\Gamma_{R})/2, & j=l\\
		\Gamma_{R}e^{ik_{\omega}(j-l)d}, & j>l\\
		\Gamma_{L}e^{ik_{\omega}(l-j)d}, & j<l
	\end{cases}
	, \label{Def:Sigma}%
\end{equation}
where $\Gamma_{R}=\tilde{g}^{2}(k_{\omega})/v(k_{\omega})$ and $\Gamma_{L}=\tilde{g}%
^{2}(-k_{\omega})/v(k_{\omega})$. 
\(\mathrm{Im} \Sigma\) contributes a dissipative \cite{input_output_Clerk,dissipative_Canming,dissipative_Tao,Canming_exp} and long-ranged coupling. The  coupling becomes non-reciprocal when $\Gamma
_{L}\neq\Gamma_{R}$. The phase $k_{0} (j-l)d$ is the phase accumulated by a photon when traveling from the $j$-th to the $l$-th magnet. \(| \Sigma |\)
does not depend on distance because we assume sufficiently
high quality of waveguide and magnets, but we will discuss the novel effect of dissipation \cite{non_Hermitian_skin_effect} at the end of this part. Coupling dissipative systems by traveling waves is a very general problem that recently attracts much attention \cite{Canming_review_1,Canming_review_2,Cavity_magnonics}. While the input-output theory (Sec.~\ref{dipolar_pumping}) is suited to solve these problems, it can also be 
treated by an effective Hamiltonian at the cost of losing hermiticity and real eigenvalues. For our problem
\begin{align}
	\hat{H}_{\rm eff}=\mathcal{M}^{\dagger}\tilde{H}_{\mathrm{eff}}\mathcal{M},
\end{align}
where the vector $\hat{\mathcal{M}}=\left(  \hat{m} 
_{1},\dots,\hat{m}_{N}\right)  ^{T}$ stores the magnon operators in magnets $\{1,\cdots,N\}$, and the matrix 
$\tilde{H}_{\mathrm{eff}}=\tilde{\omega}+\Sigma$ with elements
$\tilde{\omega}_{jl}=\omega_{m}\left(
1-i\alpha_{G}\right)  \delta_{jl}$.

For the magnonic hydrogen molecule introduced above, we introduce the coherent Hermitian and dissipative anti-Hermitian
Hamiltonians
\begin{subequations} 
	\begin{align}
		\hat{H}_{h}&=(\hat{H}_{\mathrm{eff}}+\hat{H}_{\mathrm{eff}%
		}^{\dagger })/2=\sum_{i=1,2}\omega _{i}\hat{m}_{i}^{\dagger }\hat{m}_{i}+i%
		\frac{\Sigma _{12}+\Sigma _{21}^{\ast }}{2}\hat{m}_{1}^{\dagger }\hat{m}_{2}+i\frac{\Sigma _{21}+\Sigma _{12}^{\ast }}{2}\hat{m}_{1}\hat{m}%
		_{2}^{\dagger }, \\
		\hat{H}_{ah}&= (\hat{H}_{\mathrm{eff}}-\hat{H}_{\mathrm{eff}}^{\dagger })/2=-i\sum_{i=1,2}\frac{\delta \omega _{m}}{2}\hat{m}%
		_{i}^{\dagger }\hat{m}_{i}+\frac{\Sigma _{12}-\Sigma _{21}^{\ast }}{2}\hat{m}%
		_{1}^{\dagger }\hat{m}_{2}+\frac{\Sigma _{21}-\Sigma _{12}^{\ast }}{2}\hat{m}_{1}^{\dagger }\hat{m}%
		_{2},
	\end{align}
\end{subequations}
where $\delta \omega _{m}=\Gamma _{R}+\Gamma _{L}+2\alpha _{G}\omega _{m}$, $%
\Sigma _{12}=-i\Gamma _{L}e^{ikd}$, and $\Sigma _{21}=-i\Gamma _{R}e^{ikd}$. The
master equation for the magnon density operator  $\hat{\rho}$ \cite%
{Molmer,Gardiner_chiral,chiral_emitter_light_5,chiral_optics} 
\begin{align}
	\partial _{t}\hat{\rho}=i\left[ \hat{\rho},\hat{H}_{h}\right] +\sum_{i}%
	\frac{\delta \omega _{m}}{2}\hat{\mathcal{L}}_{ii}\hat{\rho}+i\frac{\Sigma
		_{12}-\Sigma _{21}^{\ast }}{2}\hat{\mathcal{L}}_{12}\hat{\rho}+i\frac{\Sigma _{21}-\Sigma _{12}^{\ast }}{2}\hat{\mathcal{L}}_{21}\hat{%
		\rho},
\end{align}%
in which $\mathcal{L}_{ij}\hat{\rho}=2\hat{m}_{j}\hat{\rho}\hat{m}%
_{i}^{\dagger }-\hat{m}_{i}^{\dagger }\hat{m}_{j}\hat{\rho}-\hat{\rho}\hat{m}%
_{i}^{\dagger }\hat{m}_{j}$ is a relaxation operator (Lindblad
super-operator), while $\delta \omega _{m}$ and $i(\Sigma _{12}-\Sigma
_{21}^{\ast })/2$ are the self- and collective decay rates, respectively. For
perfect chiral coupling $\Sigma _{21}=0$ and at resonance, the master
equation in the frame rotating at frequency $\omega_m$ and $\hat{m}(t)=\tilde{m}e^{-i\omega _{%
		\mathrm{in}}t}$ leads to
\begin{equation}
	\frac{\partial }{\partial t}%
	\begin{pmatrix}
		\langle \tilde{m}_{1}\rangle \\ 
		\langle \tilde{m}_{2}\rangle%
	\end{pmatrix}%
	=%
	\begin{pmatrix}
		-\delta \omega _{m}/2 & -i\Sigma _{12} \\ 
		0 & -\delta \omega _{m}/2%
	\end{pmatrix}%
	\begin{pmatrix}
		\langle \tilde{m}_{1}\rangle \\ 
		\langle \tilde{m}_{2}\rangle%
	\end{pmatrix}%
	+%
	\begin{pmatrix}
		-iP \\ 
		-iP%
	\end{pmatrix}%
	,  \label{master1}
\end{equation}%
for the slowly varying envelopes $\tilde{m}_{1,2}$, where the average $\langle \hat{O}(t)\rangle =\langle \hat{O}\hat{\rho}%
(t)\rangle $.  The magnon numbers or accumulations obey the different equation
\begin{align}
	& \frac{\partial }{\partial t}%
	\begin{pmatrix}
		\langle \tilde{m}_{1}^{\dagger }\tilde{m}_{1}\rangle \\ 
		\langle \tilde{m}_{2}^{\dagger }\tilde{m}_{2}\rangle \\ 
		\langle \tilde{m}_{1}^{\dagger }\tilde{m}_{2}\rangle \\ 
		\langle \tilde{m}_{1}\tilde{m}_{2}^{\dagger }\rangle%
	\end{pmatrix}%
	=%
	\begin{pmatrix}
		iP & -iP & 0 & 0 \\ 
		0 & 0 & iP & -iP \\ 
		0 & -iP & iP & 0 \\ 
		-iP & 0 & 0 & iP%
	\end{pmatrix}%
	\begin{pmatrix}
		\langle \tilde{m}_{1}\rangle \\ 
		\langle \tilde{m}_{2}\rangle \\ 
		\langle \tilde{m}_{1}^{\dagger }\rangle \\ 
		\langle \tilde{m}_{2}^{\dagger }\rangle%
	\end{pmatrix}+
	\begin{pmatrix}
		-\delta \omega _{m} & 0 & -i\Sigma _{12} & i\Sigma _{12}^{\ast } \\ 
		0 & -\delta \omega _{m} & 0 & 0 \\ 
		0 & \Sigma _{12}^{\ast } & -\delta \omega _{m} & 0 \\ 
		0 & 0 & 0 & -\delta \omega _{m}%
	\end{pmatrix}%
	\begin{pmatrix}
		\langle \tilde{m}_{1}^{\dagger }\tilde{m}_{1}\rangle \\ 
		\langle \tilde{m}_{2}^{\dagger }\tilde{m}_{2}\rangle \\ 
		\langle \tilde{m}_{1}^{\dagger }\tilde{m}_{2}\rangle \\ 
		\langle \tilde{m}_{1}\tilde{m}_{2}^{\dagger }\rangle%
	\end{pmatrix}%
	.  \label{master2}
\end{align}%
$P$ drives the coherent amplitude according to Eq.~(\ref{master1}), while the dissipative coupling in Eq.~(\ref{master2}) collectively
reduces  the magnon numbers. The ratio between the magnon numbers on the two magnets
\begin{equation}
	\Lambda \equiv \left\vert \frac{\left\langle \hat{m}%
		_{1}^{\dagger}\hat{m}_1\right\rangle }{\left\langle \hat{m}_{2}^{\dagger}\hat{m}_2\right\rangle }\right\vert=\left\vert \frac{2\alpha _{G}\omega _{m}+\Gamma _{R}+\Gamma
		_{L}\left( 1-2e^{2ikd}\right) }{2\alpha _{G}\omega _{m}+\Gamma _{L}-\Gamma
		_{R}}\right\vert^2 .  \label{imbalance_waveguide_input}
\end{equation}
$\Lambda \gg 1$ also without chirality, \textit{i.e.}, when $%
\Gamma _{L}=\Gamma _{R}=\Gamma $: $\Lambda =\left\vert 1+2\Gamma \left(
1-e^{2ikd}\right) /\left( 2\alpha _{G}\omega _{m}\right) \right\vert^2 $ because the microwaves arrive only from the left. This geometrical imbalance depends strongly on the parameters, \textit{e.g.},  
$\Lambda \approx 5-4\cos (2kd)$ when $\Gamma _{R}\rightarrow 0$ and $\Gamma _{L}\gg \alpha _{G}\omega _{m}$ . When $kd=n\pi /2$ with $n$ being odd integer, $\Lambda =3$, and a ratio of the excited magnon numbers of $\Lambda \approx 9$. When $\Gamma _{L}=\Gamma
_{R}-2\alpha _{G}\omega _{m},$ $\left\vert \left\langle \hat{m}%
_{2}\right\rangle \right\vert =0$ and $\Lambda $ diverges: magnet $2$ cannot be excited because the input and emitted photons from the other magnet interfere destructively. 

\begin{figure}[ptbh]
	\begin{centering}
		\includegraphics[width=1\textwidth]{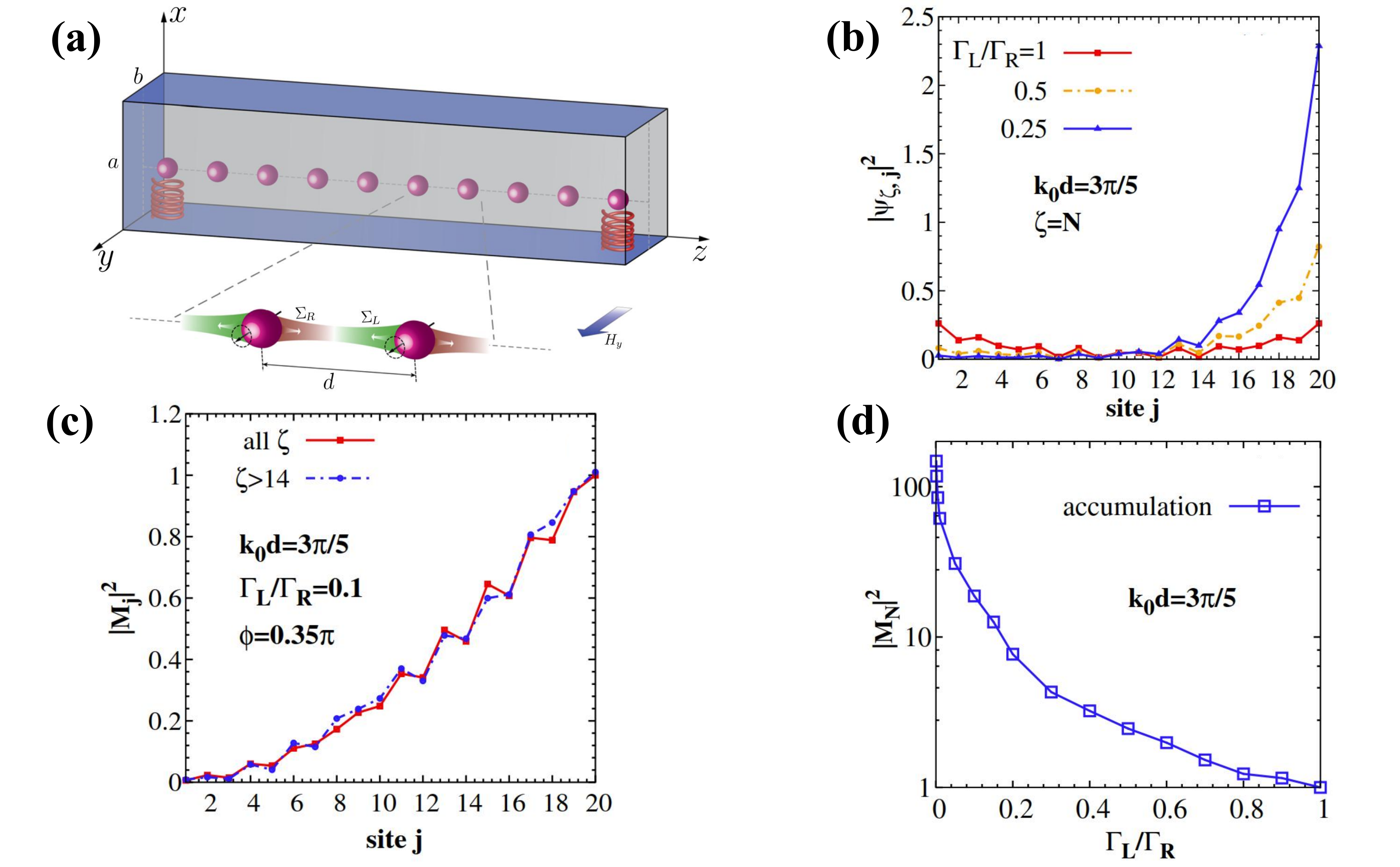}
		\par\end{centering}
	\caption{Magnon accumulation of chirally coupled magnets mediated by waveguide microwaves. (a) illustrates that a chain of magnets is placed in the microwave guide, between which the photons mediate an effective interaction. (b) plots the magnon accumulation $|\psi_{\zeta,j}|^2$ of the shortest-lived state with different chiralities. In these super-radiant states, the magnons accumulate at one boundary [(c)],  and increasingly with chirality [(d)].  The figures are taken from Ref.~\cite{waveguide_Yu_1}.}
	\label{Magnon_skin}
\end{figure}

\textbf{Non-Hermitian formalism}.---The coupling between magnets by photon exchange
in the waveguide gives rise to collective excitations. Solving the coupled Maxwell and LLG equations becomes tedious when the number of magnets increases \cite{waveguide_Yu_2}, such as for the configuration in Fig.~\ref{Magnon_skin}(a) in which a chain of $N$ magnets are placed in a microwave guide. On the other hand, by the non-Hermitian Hamiltonian $\hat{H}_{\mathrm{eff}}=\hbar \hat{\mathcal{M}}^{\dagger }\tilde{H}_{\mathrm{eff}}\hat{\mathcal{M}}$ with both coherent and dissipative couplings \cite{waveguide_Yu_1,waveguide_Yu_2} we may adopt  powerful techniques from quantum optics \cite{Molmer,subradiance1,Yuxiang_subradiance,subradiance3,subradiance4,subradiance5,Gardiner_chiral,input_output_Clerk,input_output_Cardiner}.

The right and left eigenvectors of $\tilde{H}_{\mathrm{eff}%
}$ are not the same. Let the right eigenvectors of $\tilde{H}_{\mathrm{eff}}$
be $\{\psi _{\zeta }\}$ with corresponding eigenvalues $\{\nu _{\zeta }\}$, where $\zeta \in \{1,\dots ,N\}$ label the collective modes. The right eigenvectors of $\tilde{H}_{\mathrm{eff}%
}^{\dagger }$ are $\{\phi _{\zeta }\}$ with corresponding eigenvalues $\{\nu
_{\zeta }^{\ast }\}$,  but $\phi _{\zeta }^{\dagger }$ is a left eigenvector. In the absence of degeneracies, \textit{i.e.}, $\forall _{\zeta \zeta
	^{\prime }}$ $\nu _{\zeta }\neq \nu _{\zeta ^{\prime }}$, the bi-orthonormality relation $\psi _{\zeta }^{\dagger }\phi _{\zeta ^{\prime }}=\delta
_{\zeta \zeta ^{\prime }}$. $\psi _{\zeta }^{\dagger }\phi _{\zeta}=1$ does not uniquely fix the individual normalization factors, but this has no physical consequences. 
The matrices $\mathcal{L}=\left( \phi _{1},\dots ,\phi _{N}\right) $
and $\mathcal{R}=\left( \psi _{1},\dots ,\psi _{N}\right) $ in terms of left
and right eigenvectors fulfil $\mathcal{R}^{\dagger }\mathcal{L}=%
\mathcal{L}^{\dagger }\mathcal{R}=I_{N}$, where $I_{N}$ is the $N\times N$
identity matrix and $\tilde{H}_{\rm eff}=\mathcal{R}\nu \mathcal{L}^{\dagger }$,
with matrix elements $\nu _{ij}=\left( \nu _{1},\dots ,\nu _{N}\right)
\delta _{ij}$.  
$\hat{\alpha}_{\zeta }=\phi _{\zeta }^{\dagger }\hat{\mathcal{M}}$ annihilates a quasiparticle  with \textquotedblleft wave function\textquotedblright\ $\psi _{\zeta }$.

We refer to the inhomogeneous distribution of the magnons in collective states, especially the conspicuous localization at the edges of a long array, as a ``magnon skin effect''.  
We illustrate it for a chain of $N$ identical magnets at
equal distance $z_{j+1}-z_{j}=d\ \left( 0<j<N\right) $ on a line in and parallel to a wave guide as in Fig.~\ref{Magnon_skin}(a). The magnets are tuned to the $\mathrm{TE}_{10}$ waveguide mode. Such a system has been realized already for $N=7$ but is in a closed cavity \cite{Magnon_dark_modes}. The traveling photons mediate the chiral interaction between Kittel modes of any two magnets as before, which leads to the Hamiltonian matrix 
\begin{equation}
	\tilde{H}_{\mathrm{eff}}=%
	\begin{pmatrix}
		\omega _{m}-i\alpha _{G}\omega _{m}-i\frac{\Gamma _{R}+\Gamma _{L}}{2} & 
		-i\Gamma _{L}e^{ik_0d} & -i\Gamma _{L}e^{2ik_0d} & \dots & -i\Gamma
		_{L}e^{(N-1)ik_0d} \\ 
		-i\Gamma _{R}e^{ik_0d} & \omega _{m}-i\alpha _{G}\omega _{m}-i\frac{\Gamma
			_{R}+\Gamma _{L}}{2} & -i\Gamma _{L}e^{ik_0d} & \dots & -i\Gamma
		_{L}e^{(N-2)ik_0d} \\ 
		-i\Gamma _{R}e^{2ik_0d} & -i\Gamma _{R}e^{ik_0d} & \omega _{m}-i\alpha
		_{G}\omega _{m}-i\frac{\Gamma _{R}+\Gamma _{L}}{2} & \dots & -i\Gamma
		_{L}e^{(N-3)k_0d} \\ 
		\vdots & \vdots & \vdots & \ddots & \vdots \\ 
		-i\Gamma _{R}e^{i(N-1)k_0d} & -i\Gamma _{R}e^{i(N-2)k_0d} & -i\Gamma
		_{R}e^{i(N-3)k_0d} & \dots & \omega _{m}-i\alpha _{G}\omega _{m}-i\frac{\Gamma
			_{R}+\Gamma _{L}}{2}%
	\end{pmatrix}%
	,  \label{H_matrix}
\end{equation}
where $k_0=\sqrt{{\omega _{m}^{2}}/{c^{2}}-\left({\pi }/{a}\right) ^{2}}$. $\Gamma_L$ ($\Gamma_R$) represents the coupling strength from the left to right (the right to left) with  $\Gamma_L \ne \Gamma_R$ . The microwaves emitted by magnet $j$ to the right in our scheme  are instantaneously felt by all magnets on the right with equal modulus but phase factor $e^{ik_0|z_{j}-z_{l}|},$ and analogously for the magnets to the left. The eigenvalues of  \(\tilde{H}_{\mathrm{eff}}\) from a band structure with generalized
Bloch states labelled $\zeta \in \{1,\dots ,N\}$ with right eigenvectors $%
\{\psi _{\zeta }\}$ and corresponding eigenvalues $\{\nu _{\zeta }\}$, $(\nu _{\zeta }-\tilde{H}_{\mathrm{eff}})\psi _{\zeta }=0$.
$\mathrm{Re} \, \nu _{\zeta }$ ($\mathrm{Im} \, \nu _{\zeta }$) is the resonance frequency (inverse lifetime) of the $\zeta $-th mode. The states with decay rates larger (smaller) than $\alpha_G\omega_m+(\Gamma_R+\Gamma_L)/2$ are called super-radiant (subradiant) \cite{Yuxiang_subradiance,waveguide_Yu_1,waveguide_Yu_2}. The most subradiant states are potentially useful to store information.

The rich features of the collective motion including both sub- and super-radiant states can be accessed analytically in some limits 
\cite{Yuxiang_subradiance,waveguide_Yu_1,waveguide_Yu_2}. To this end, we try a Bloch operator $\hat{\Psi}_{\kappa }=\frac{1}{\sqrt{N}}%
\sum_{j=1}^{N}e^{i{\kappa }z_{j}}\hat{m}_{j} $ with $z_{j}=(j-1)d$ and complex momentum $\kappa $ that obeys the equation of motion
\begin{equation}
	d\hat{\Psi}_{\kappa}/{dt}=-\frac{i}{\hbar}\left[ \hat{\Psi}_{\kappa },\tilde{H}_{\mathrm{eff}}\right]=-i\omega _{\kappa }\hat{%
		\Psi}_{\kappa }-\Gamma _{L}g_{\kappa }\hat{\Psi}_{k_0}+\Gamma _{R}h_{\kappa }%
	\hat{\Psi}_{-k_0}.  \label{EOM_Bloch_states}
\end{equation}
The solutions have complex dispersion relation 
\begin{equation}
	\omega _{\kappa }=-i\frac{\Gamma _{R}}{2}\frac{1+e^{i({\kappa }+k_0)d}}{1-e^{i({\kappa }+k_0)d}}+i\frac{\Gamma _{L}}{2}\frac{1+e^{i({\kappa }-k_0)d}}{1-e^{i({\kappa }-k_0)d}},  \label{dispersion}
\end{equation}%
and `leakage' 
\begin{subequations}
\begin{align}
	g_{\kappa }&=\frac{1}{1-e^{i({\kappa }-k_0)d}},\\
	h_{\kappa }&=\frac{e^{i({\kappa }+k_0)Nd}}{1-e^{i({\kappa }+k_0)d}}.
\end{align}
\end{subequations}
Equation~(\ref{EOM_Bloch_states}) is a closed equation for the unknown $\kappa $.
When the terms $g_{\kappa },h_{\kappa }$ in Eq.~(\ref{EOM_Bloch_states}) vanish, $\hat{\Psi}_{\kappa }$ is a proper solution. However, leakage and reflection at the edges mixes $\hat{\Psi}_{\kappa }$ with the plane waves $%
\hat{\Psi}_{k_0}\ $and $\hat{\Psi}_{-k_0}$, which renders the problem non-trivial. 
In general, the field operator $\hat{\alpha}$ should be a superposition of Bloch waves at the same frequency. For the simple chain, we expect degeneracy  
\begin{equation}
	\omega _{\kappa }=\omega _{\kappa ^{\prime }}
	\label{first}
\end{equation}
for not more than two states with different  $%
\kappa $ and $\kappa ^{\prime }$. Trying the superposition $\hat{\alpha}=g_{\kappa ^{\prime }}\hat{\Psi}_{\kappa }-g_{\kappa }%
\hat{\Psi}_{\kappa ^{\prime }}$ gives 
\begin{equation}
	{d\hat{\alpha}}/{dt}=-i\omega _{\kappa }\hat{\alpha}%
	+\Gamma _{R}\left( g_{\kappa ^{\prime }}h_{\kappa }-g_{\kappa }h_{\kappa
		^{\prime }}\right) \hat{\Psi}_{-k_0},
\end{equation}%
which solves the problem when
\begin{equation}
	g_{\kappa }h_{\kappa ^{\prime }}=g_{\kappa ^{\prime }}h_{\kappa }.
	\label{second}
\end{equation}%
Equations~(\ref{first}) and (\ref{second}) 
determine the complex unknowns $\kappa $ and $\kappa ^{\prime }$, which subsequently leads to 
\begin{subequations}
	\begin{align}
		\psi _{j}& \propto g_{\kappa ^{\prime }}e^{i\kappa z_{N-j}}-g_{\kappa
		}e^{i\kappa ^{\prime }z_{N-j}}, 
		\label{wavefunction_universal}\\
		\nu & ={\omega}_{m}(1-i\alpha_G)+\omega _{\kappa }. \label{solutions}
	\end{align}
\end{subequations}
Near the minima of $\omega _{\kappa }$, say around ${\kappa }={\kappa }%
_{\ast }$, the modes are sub-radiant with weak radiative damping Im$\, \omega _{\kappa }$. At the minimum of Eq.~(\ref{dispersion}) 
\begin{align}
	{\kappa }_{\ast }d& =\arcsin \frac{\Gamma _{R}-\Gamma _{L}}{\sqrt{\Gamma
			_{R}^{2}+\Gamma _{L}^{2}-2\Gamma _{R}\Gamma _{L}\cos (2k_0d)}}  -\arctan \frac{\Gamma _{R}-\Gamma _{L}}{(\Gamma _{R}+\Gamma _{L})\tan (k_0d)}.
\end{align}%
The $\arcsin $ is a two-valued function and hence we search for two extremal points in the first Brillouin zone $[-\pi /d,\pi /d]$. $\kappa _{\ast }$ and the corresponding $\kappa _{\ast }^{\prime }$ do not yet satisfy the eigenvalue equation Eq.~(\ref{second}). Trying $\kappa =\kappa _{\ast
}+\delta $ and $\kappa ^{\prime }=\kappa _{\ast }-\delta $, however, leads to the desired
eigenfunctions and frequencies
\begin{subequations}
	\begin{align}
		\psi _{\xi ,j}& \approx -2i\frac{e^{i\kappa _{\ast }z_{N-j}}}{1-e^{i(\kappa
				_{\ast }-k_0)d}}\sin (\delta _{\xi }z_{N-j}), \\
		\omega _{\xi }& =\omega _{{\kappa }_{\ast }}+\frac{\sin (k_0d)}{\cos ({\kappa }%
			_{\ast }d)-\cos (k_0d)}\frac{\Gamma _{R}(\delta _{\xi }d)^{2}/2}{1-\cos [(k_0+{%
				\kappa }_{\ast })d]},
	\end{align}
\end{subequations}
where $\xi =\{1,2,\cdots \}$,
which are symmetric even in the presence of chiral coupling, because sub-radiant modes do
not efficiently couple to the waveguide and are standing rather than propagating waves.

The super-radiant modes with $\mathrm{{Im}}\,\omega _{\kappa }\gg \Gamma _{R},\Gamma _{L}$
are near $\kappa \approx \pm k_0$, \textit{i.e.}, at complex momenta $\kappa =k_{0}+\eta $
and $\kappa ^{\prime }=-k_{0}+\eta ^{\prime }$ with small imaginary numbers $%
\eta $ and $\eta ^{\prime }$ that should be calculated numerically. $\mathrm{Im} \,\eta $ and $\mathrm{Im} \,\eta ^{\prime }$ are reciprocal skin depths of the edge states. Figure~\ref{Magnon_skin}(b) plots the magnon accumulation $|\psi_{\zeta,j}|^2$ of the  shortest-lived state. When $\Gamma_R=\Gamma_L$, it is symmetrically localized at both edges (red solid curve). With increasing chirality, the distribution becomes asymmetrically skewed to one boundary. When  excited by a phased array of microwave sources, almost all magnons concentrate at one boundary [Fig.~\ref{Magnon_skin}(c)] with large amplitudes when $\Gamma_L/\Gamma_R\rightarrow 0$ [Fig.~\ref{Magnon_skin}(d)]. Thus, the chirality much improves the excitation efficiency of spins.

Super-radiant and sub-radiant states play an important role in quantum optics for quantum communication and computing \cite{Yuxiang_subradiance,subradiance1,subradiance3,subradiance4,subradiance5}. There the magnetic moments are replaced by, for example, atomic electric dipoles that couple to electric rather than magnetic fields. Yao \textit{et al}. \cite{dissipative_Tao} pointed out that the observed level attraction between magnons and microwave photons in open cavities \cite{dissipative_Canming} is a consequence of the dissipative coupling communicated only by traveling waves. The magnonic skin effect addressed here is of no topological nature, but by tailoring the damping of the substrate we can achieve topological non-Hermitian skin effect as reviewed in the following, which is at the root of the sensitivity of the quasiparticle dispersion on the boundary conditions. When we replace the open by periodic boundary conditions we connect the ends of the wave guide to form a closed torus cavity \cite{circulating_polariton} that can be understood by the same formalism.  Full numerical simulation of the coupled LLG and Maxwell equations of a torus cavity and analytical analysis along the lines explained here agree well with each other \cite{circulating_polariton}. A closed system (with $\alpha_G=0$) is Hermitian since the microwaves cannot escape and the Bloch states are labeled by the real ``crystal momentum''. Comparing results for open and closed boundary conditions shed light on the concept of Bloch theorem and Brillouin zone in dissipative systems \cite{nonhermitian_review}.

 \textbf{Non-Hermitian skin effect}.---Closely related to the chiral coupling in optics and magnetism as reviewed here is the ``short-range asymmetric coupling"  in the non-Hermitian skin effect, which refers to the localization of bulk states at a boundary of a dynamic system \cite{Hatano_Nelson,GBZANDSE,GBZANDSE2,nonhermitian_review,SEBiorthogonal,SEhighorder,Review2,CBBC,nonBloch,bipolarSE,GreenSE,Biorthogonal}.  This effect has been observed in  systems with relatively short-range interactions \cite{SEquantum,toposensor}, \textit{e.g.}, in photonic lattice \cite{lightfunnel},  electric circuits \cite{SEcircuit}, and mechanical metamaterials \cite{SEmetamaterial}. The key features of the topology may be captured by a generalized Brillouin zone  \cite{GBZANDSE,GBZANDSE2} for wave vectors on the complex plane. In spite of the ubiquitous chirality in magnetic systems, their  non-Hermitian topology has rarely been addressed \cite{Non_Hermitian_Flebus_1,Non_Hermitian_Flebus_2}. We emphasize here the chiral interaction between the Kittel magnons of magnetic particles or wires  with
propagating spin waves in magnetic films \cite{Au_first,Chiral_pumping_grating,Chiral_pumping_Yu,magnon_trap,Haiming_exp_wire}, waveguide microwaves \cite{waveguide_Yu_2,waveguide_Yu_1,Xufeng_exp}, and surface acoustic waves (see below) \cite{phonon_Yu_2,Xu,phonon_Kei}.
These waves can propagate over long distances and thereby mediate the chiral interaction between magnets \cite{magnon_trap,Haiming_exp_wire,waveguide_Yu_2,waveguide_Yu_1,phonon_Yu_2,phonon_Yu_1}. Long-range chiral interactions might lead to a non-Hermitian skin effect similar to that known for  short-range chiral interaction in the Hatano-Nelson model \cite{Hatano_Nelson} because in both cases the energy tends to accumulate at one end. On the other hand,  the systems  reviewed above do not favor the coalescence
of bulk modes \cite{waveguide_Yu_1,phonon_Yu_2}. The predicted edge modes in magnetic systems  must still be confirmed by experiments.

\begin{figure}[ptbh]
	\begin{centering}
		\includegraphics[width=0.99\textwidth]{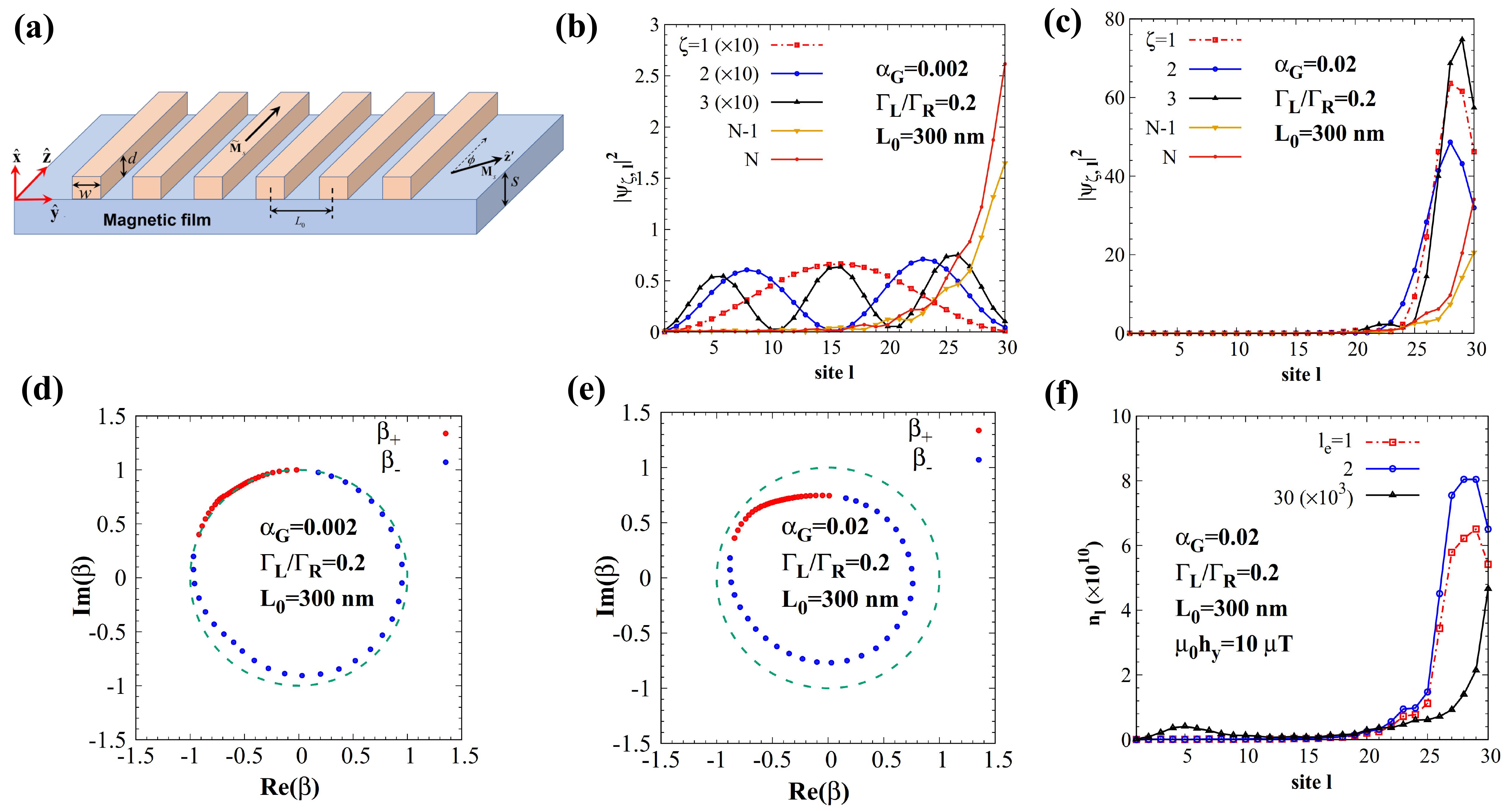}
		\par\end{centering}
	\caption{Non-Hermitian skin effect of the magnetic excitation in a periodic array of magnetic nanowires that are coupled chirally via spin waves of a thin magnetic substrate. (a) illustrates the coordinate systems for a film magnetization that rotates under an in-plane magnetic field and the geometric parameters. In (b) and (c), we calculate all the collective magnonic modes of wires with a small film Gilbert damping $\alpha_G=0.002$ [(b)] and a large one $\alpha_G=0.02$ [(c)]. Their complex momenta $\kappa$, parameterized by $\beta=e^{i\kappa L_0}$, are plotted in (d) and (e), respectively. (f) illustrates how the excited magnons accumulate at the right edge when a local stripline excites only the $l_e$-th wire. The figures are taken from Ref.~\cite{non_Hermitian_skin_effect}.}
	\label{non_Hermitian_topology}
\end{figure}

According to Yu and Zeng  \cite{non_Hermitian_skin_effect,magnon_defect_state}   the interference due to the propagation phase shift of the traveling waves can be detrimental to the skin effect of all the modes. They found a non-Hermitian topological phase in  dipolar-coupled
magnets with chiral coupling when the mediation traveling waves are dampened. Figure~\ref{non_Hermitian_topology}(a) illustrates the model of an array of magnetic wires  with spacing $L_0$ on top of a thin magnetic film, as discussed in Sec.~\ref{dipolar_pumping}. When  all the magnetizations are along the wires, the coupling between the Kittel mode in the wires and the spin waves in the films is chiral, see Sec.~\ref{dipolar_fields_1}. The spin waves in the film, therefore, mediate a chiral interaction (or asymmetric coupling) between the wire to the left and right when the damping of the spin waves of the film is sufficiently large (while that in the wires remains small) as in  Fig.~\ref{non_Hermitian_topology}(c), so \textit{all} the collective modes are localized at one edge. This is a non-Hermitian skin effect that vanishes when the damping tends to zero, as in Fig.~\ref{non_Hermitian_topology}(b) or the chirality is absent. 

The generalized Bloch wave function in Eq.~(\ref{EOM_Bloch_states}) leads to the magnon amplitude [Eq.~(\ref{wavefunction_universal})]
\[
\psi_j\propto (g_{\kappa'}e^{i\kappa (N-j)L_0} -g_{\kappa}e^{i\kappa' (N-j)L_0}).
\]
The distribution of a state over the wired is determined by $\beta_{\kappa}=e^{i\kappa L_0}$. When $|\beta_{\kappa}|>1$ ($|\beta_{\kappa}|<1$), the amplitude of 
$\psi_j$ decreases (increases) with increasing the sites from 1 to N, implying the
localization at the left (right) edge of the chain.  Figures~\ref{non_Hermitian_topology}(d) and (e) illustrate that $|\beta_{\kappa}|$ is close to the unit circle when the damping of the film is weak as in  Fig.~\ref{non_Hermitian_topology}(b), but strongly deviates otherwise, see Fig.~\ref{non_Hermitian_topology}(c). A nontrivial winding of the eigenfrequency emerges in the presence of both chirality and strong damping of the film spin waves. This non-Hermitian skin effect may operate in a non-local and non-reciprocal information processor since the excitation of the wire at one edge leads to a large amplitude at
the other edge, as shown in Fig.~\ref{non_Hermitian_topology}(e). Its sensitivity allows the
detection of weak microwaves that may be useful in classical and quantum information processing and metrology.

  \subsection{Magnon and surface phonon}
  
  \label{magnon_phonon}

Much of the above description of the magnon-photon interaction carries over to magnon-phonon systems, but many details are quite different as well. First, the coupling between magnetic and elastic degrees of freedom has a very different origin. We may distinguish three mechanisms: (i) The exchange coupling is proportional to the  overlap between neighboring localized atomic wave functions that is altered by lattice strains. (ii) Long-range modulation of the magnetization direction and sample shape costs magnetodipolar interaction energy. (iii) The orbital motion of an electron in a crystal  couples to its spin by the relativistic spin-orbit interaction.  Assuming small lattice displacements $\partial_j u_i$, where $\{i,j\}$ are Cartesian coordinates, the  elastic strain can be separated into  a symmetric $\varepsilon_{ij}=(\partial_j u_i+\partial_i u_j)/2$ and anti-symmetric  $\omega_{ij}=(\partial_j u_i-\partial_i u_j)/2$ strain tensor, with associated magneto-elastic and magneto-rotation energies, respectively.

Usually, phenomenological magnon-phonon interaction parameters can be derived by a Taylor expansion under proper  symmetry constraints.  Their number can be estimated by model Hamiltonians, first-principle calculation, or fitting experimental observables. The free-energy density
\begin{align}
	f_{\rm em}({\bf r})=\sum_{\alpha\beta}U_{\alpha\beta}\left[{\bf M}\right]\varepsilon_{\alpha\beta}+\sum_{\alpha\beta}V_{\alpha\beta}\left[{\bf M}\right]\omega_{\alpha\beta}
\end{align}
is invariant under time-reversal operation, so $U_{\alpha\beta} \left[{\bf M}\right]$ and $V_{\alpha\beta}\left[{\bf M}\right]$ are even in ${\bf M}$. This free energy should also be invariant under a combined global SO(3) rotation of ${\bf M}$ and the displacement fields ${\bf u}$. In the presence of a mirror symmetry plane, \(f_{\rm em}\)  is invariant with respect to reflection with respect to ${\bf x}_{\alpha}$-axis and ${\bf x}_{\beta}$-axis, leading to 
\begin{align}
	\{U,V\}_{\alpha\beta}(M_{\alpha},M_{\beta},M_{\gamma})=-\{U,V\}_{\alpha\beta}(-M_{\alpha},M_{\beta},M_{\gamma})=-\{U,V\}_{\alpha\beta}(M_{\alpha},-M_{\beta},M_{\gamma}).
\end{align}
We can therefore expand $U_{\alpha\beta}({\bf M})=B_{\alpha\beta}M_{\alpha}M_{\beta}/M_s^2$ and $V_{\alpha\beta}({\bf M})= C_{\alpha\beta} M_{\alpha}M_{\beta}/M_s^2$ to the leading order in the coefficients $B_{\alpha\beta}=B_{\beta\alpha}$ and $C_{\alpha\beta}=-C_{\beta\alpha}$, such that
\begin{align}
	f_{\rm em}({\bf r})=\frac{1}{M_s^2}\sum_{\alpha\beta}B_{\alpha\beta}M_{\alpha}M_{\beta}\varepsilon_{\alpha\beta}+\frac{1}{M_s^2}\sum_{\alpha\beta}C_{\alpha\beta}M_{\alpha}M_{\beta}\omega_{\alpha\beta},
\label{free_energy}
\end{align}
which is commonly used in the literature. In the following we will work with a quantized version to build the Hamiltonian appropriate for various phenomena.

\subsubsection{Chiral magnon-phonon interaction and phonon diodes: theory}
\label{SAW_diode_theory}

\begin{figure}[ptbh]
	\begin{centering}
		\includegraphics[width=0.99\textwidth]{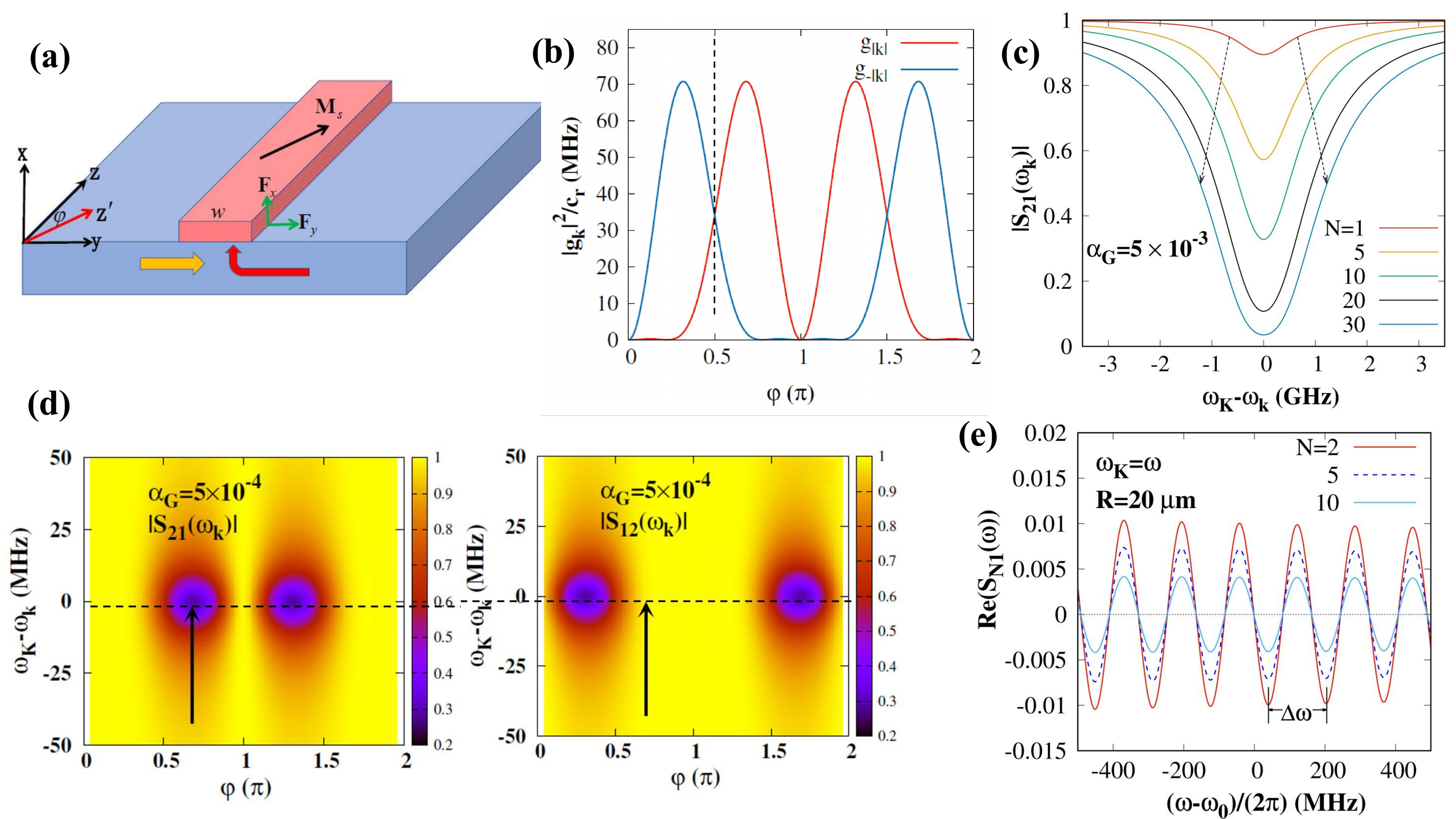}
		\par\end{centering}
	\caption{The SAW diode effect induced by magnetic wires on top of a dielectric substrate. (a) illustrates the coordinate systems for a wire magnetization that rotates under a saturating magnetic field. (b) The calculated non-reciprocal  ($|g_{k}|\ne |g_{-k}|)$ coupling Eq.~(\ref{eqn:coupling_strength}) between the Kittel magnon in a Ni wire and the SAW in gadolinium gallium garnet. The wire number and frequency dependence of the phonon transmission is discussed in (c) and (d).  In (c), the phonon transmission decreases with more magnetic wires. In (d) we show with different magnetization directions, the transmission is non-reciprocal with $|S_{21}(\omega_k)|\ne |S_{12}(\omega_k)|$, \textit{e.g.}, with the parameter indicated by the thick black arrow. (e) gives the microwave transmission between wires 1 and $N$. The interval $\Delta \omega$ between the dips of frequency remains unchanged with a different number of wires. The figures are taken from Ref.~\cite{phonon_Yu_2}.}
	\label{Yu_phonon_diode}
\end{figure}

\textbf{Magnetoelastic coupling}.---Magnetoelasticity addresses the coupling between the magnetization and lattice strains. Like magnetocrystalline anisotropy, it originates from the relativistic spin-orbit coupling. The magnetoelastic Hamiltonian depends on the crystal symmetry of the material \cite{Landau}. Here we illustrate the principles by an isotropic continuum model for Eq.~(\ref{free_energy}) that describes a wide class of materials well \cite{Kittel_old,parameters,Simon,Xu,DMI_phonon_exp}, \textit{viz.} \cite{Landau,Simon,Kittel_old},
\begin{equation}
	\hat{H}^{(a)}_c=\frac{1}{M_s^2}\int d{\bf r}\left(B_{\parallel}\sum_iM_i^2\varepsilon_{ii}+B_{\perp}\sum_{i\neq j}M_iM_j\varepsilon_{ij}\right),
	\label{magnetoelastic}
\end{equation}
where $B_{\parallel}$ and $B_{\perp}$ are the magnetoelastic constants, and $\varepsilon_{ij}=(\partial_j u_i+\partial_i u_j)/2$ denotes the strain tensor in terms of the displacement field $u_i({\bf r})$.
We are interested in the coupling of a magnetic wire with the phonons in a dielectric substrate in a transverse transport configuration illustrated in Fig.~\ref{Yu_phonon_diode}(a). The magnon-phonon coupling is then dominated by surface phonons with a larger amplitude at the magnetic interface and more efficient lateral propagation than the bulk phonons that we disregard in the following. We focus here on
the Rayleigh surface acoustic waves (SAWs) localized at the interface of an elastic medium to vacuum/air \cite{Kino1987,Viktorov1967}, as reviewed in Sec.~\ref{surface_acoustic_waves} for basic equations. The displacements generated by a SAW propagating perpendicular to the wire with momentum ${\bf k}\parallel\hat{\bf y}$ are of the form $(u_x(x,y),u_y(x,y))$ and cause strains $\varepsilon_{xx}$, $\varepsilon_{yy}$ and $\varepsilon_{xy}$. We restrict the treatment here to in-plane magnetizations \cite{Xu,Onose_exp,Nozaki_exp,DMI_phonon_exp}, saturated and controlled by a sufficiently strong magnetic field $-H_0\hat{\bf z}'$ with an angle $\varphi$ between ${\bf z}'$ with the wire. The equilibrium component $\sim -M_s\hat{\bf z}'$ is static, while the components $m_{x}{\bf x}+m_{y'}{\bf y}'$ describe magnetic excitations.  In this configuration, the Kittel modes in the wire are coupled to the SAWs propagating in the $\hat{\bf y}$-direction. The Hamiltonian Eq.~(\ref{magnetoelastic}) can be linearized to be 
\begin{align}
	\nonumber
	\hat{H}^{(a)}_c&=\frac{2\sin\varphi}{M_s}\int d{\bf r}\left(B_{\parallel}\cos\varphi m_{y'}\varepsilon_{yy}+B_{\perp} m_{x}\varepsilon_{xy}\right)\\
	&\simeq \frac{2\sin\varphi\cos\varphi}{M_s} {B_{\parallel}Ld}m_{y'}\left(u_y|_{\frac{w}{2}+y_l}-u_y|_{-\frac{w}{2}+y_l}\right)+\frac{\sin\varphi}{M_s}B_{\perp}Ldm_{x}\left(u_x|_{\frac{w}{2}+y_l}-u_x|_{-\frac{w}{2}+y_l}\right),
	\label{eqn:coupling}
\end{align}
where $y_l$ is the center coordinate of the $l$-th magnetic wire with length $L$.
Here we assume that the magnetic wire is so thin that the displacements of top and bottom surface are nearly the same such that the associated strains vanish.

\textbf{Magnetorotational coupling}.---\textcolor{blue}{Equation~(\ref{free_energy}) implies a magnetorotational coupling that is governed by the anti-symmetric strain tensor, which describes a local deviation of the anisotropy axes by a rotational motion to the 
lattice \cite{Maekawa}. The magnetorotational coupling depends on the magnetocrystalline and shape anisotropies that couple to the lattice strains by the antisymmetric strain tensor} $\omega_{ij}=(\partial_ju_i-\partial_iu_j)/2$ \cite{Maekawa,magnetorotation_1997}. For a uniaxial anisotropy along the wire $(\hat{\bf z})$ direction and SAWs propagating along the $\hat{\bf y}$-direction,  $\omega_{xz}$ and $\omega_{yz}$ and  magnon-phonon coupling vanish.
When, on the other hand, the easy axis is along the $\hat{\bf x}$-direction \cite{Xu}, the Hamiltonian
\begin{align}
\nonumber
	\hat{H}_c^{(b)}&=-\frac{2K_1}{M_s^2}\int d{\bf r}M_x({\bf r})\big[M_y\omega_{yx}({\bf r})+M_z\omega_{zx}({\bf r})\big]\\
	&\rightarrow\frac{\sin\varphi}{M_s}K_1L dm_{x}\left(u_x|_{\frac{w}{2}+y_l}-u_x|_{-\frac{w}{2}+y_l}\right),
	\label{magnetorotation_coupling}
\end{align}
where $K_1$ is the magnetic anisotropy constant. Comparing with Eq.~(\ref{eqn:coupling}), we conclude that a perpendicular anisotropy would shifts $B_{\perp}$ to $\tilde{B}_{\perp}=B_{\perp}+K_1$. When the easy axis is along the $\hat{\bf y}$-direction, $\tilde{B}_{\perp}=B_{\perp}-K_1$.

\textbf{Phonon diode}.---Let us apply a sufficiently large magnetic field $H_0$ such that the Kittel mode in the wire is circularly polarized. We can second-quantize the Hamiltonian by substituting the magnetization and displacement-field operators Eq.~(\ref{eqn:phonon_operator}) into Eqs.~(\ref{eqn:coupling}) and (\ref{magnetorotation_coupling}) such that
\begin{equation}
	\hat{H}_c=\hbar\sum_l\sum_{k}g_{l}(k)\hat{\beta}_l^{\dagger}\hat{b}_{k}+\mathrm{H.c.}.
	\label{Hamiltonian_magnon_phonon}
\end{equation}
The coupling constants
\begin{align}
	g_{k,l}=i\sin\varphi\sqrt{\frac{\gamma}{M_s\rho c_r}}\sqrt{\frac{d}{w}}\sin\left(\frac{kw}{2}\right)e^{iky_l}\xi_P\left(\tilde{B}_\perp-\cos\varphi B_\parallel\mathrm{sgn}(k)\frac{1+b^2}{a}\right)
	\label{eqn:coupling_strength}
\end{align}
depend on the sign of the phonon wave vector and are in general non-reciprocal $(|g_{|k|}|\ne|g_{-|k|})|$, as illustrated in Fig.~\ref{Yu_phonon_diode}(b) for Ni on gadolinium gallium garnet (GGG). The dimensionless parameters $a=\sqrt{1-(c_r/c_l)^2}$ and $b=\sqrt{1-\eta^2}$ also appear in Sec.~\ref{surface_acoustic_waves}. The couplings vanish when $\varphi=0$ and the non-reciprocity vanishes $(|g_{|k|}|=|g_{-|k|}|)$ when  $\varphi=\pi/2$ \cite{phonon_Yu_1}. The coupling is fully chiral $(g_{|k|}=0\ne|g_{-|k|})$ at the critical angle
\begin{align}
	\varphi_c=\arccos \left(\frac{\tilde{B}_{\perp}}{B_{\parallel}}\frac{a}{1+b^2} \right),
\end{align}
that depends only on material parameters.
Without the magnetoelastic coupling, the magnetorotation coupling alone cannot cause a non-reciprocity in the present  case of coupling between Kittel magnon and SAWs with uniaxial anisotropy. The general case with different anisotropies and the resonant spin waves in the extended magnet needs to be considered in future studies.

The chirality in the Hamiltonian implies chirality in the phonon transmission through the region under the magnetic wire. We address this \textit{phonon diode} effect in the linear and ballistic transport regime. The phonon transmission can be tuned not only by the choice of materials but also by device design. We address here an array of $N$ parallel magnetic nanowires (one-dimensional magnonic crystal). Assuming that their distance exceeds their widths we may disregard the inter-wire dipolar interaction. The scattering amplitude from state $|{k}\rangle$ to state $|{k'}\rangle$ is an element of the $T$-matrix \cite{scattering_PRB,scattering_PRE,Mahan}
\begin{align}
	T_{k'k}=\langle k'|\hat{T}|k\rangle=\delta_{kk'}+\frac{1}{\omega_{k}-\omega_{k'}+i\eta_{k'}}\sum_{ll'}g^*_{k',l}
	G_{N,ll'}(\omega_{k})g_{k,l'},
	\label{eqn:T_matrix}
\end{align}
where $\eta_{k'}$ is a phenomenological phonon lifetime broadening.
The summation over the wire index $\{l=1,2,\cdots,N\}$ represents the interference of multiply scattered phonons.  $G_{N}|_{ll'}$ is the Green function for the magnon propagation from  $l'$ to $l$ \cite{Mahan,Fetter,Haug,Abrikosov}:
\begin{equation}
	\left(G^{-1}_{N}(\omega)\right)_{ll'}=(\omega-\tilde{\omega}_{\rm K})\delta_{ll'}-\Sigma_{ll'}(\omega),
	\label{eqn:green_function}
\end{equation}
where $\tilde{\omega}_{\rm K}=\omega_{{\rm K}}-i\alpha_G\omega_{{\rm K}}$ with Gilbert constant $\alpha_G$, and 
\begin{align}
	\Sigma_{ll'}(\omega)=\sum_{k'} \frac{g_{k',l}g^*_{k',l'}}{\omega-\omega_{k'}+i\eta_{k'}}
\end{align}
 is a self-energy.   For high-quality elastic substrates  $\eta_{k}\rightarrow 0_+$ and
\begin{subequations}
	\begin{align}
		&\Sigma_{ll}(\omega)=-\frac{i}{2c_r}\left(|g_{k_*,l}|^2+|g_{-k_*,l}|^2\right)=-i\Gamma_l(\omega),\\
		&\Sigma_{l<l'}(\omega)=-\frac{i}{c_r}\tilde{g}_{k_*,l}\tilde{g}^*_{k_*,l'}e^{ik_*|y_l-y_{l'}|}=-i\Gamma_{R,ll'}(\omega),\\
		&\Sigma_{l>l'}(\omega)=-\frac{i}{c_r}\tilde{g}_{-k_*,l}\tilde{g}^*_{-k_*,l'}e^{ik_*|y_l-y_{l'}|}=-i\Gamma_{L,ll'}(\omega),
	\end{align}
\end{subequations}
where $k_*=\omega/c_r+i\eta_k/c_r\rightarrow \omega/c_r$ and $\tilde{g}_{k,l}=g_{k,l}e^{-iky_l}$. 

The diagonal elements $\Sigma_{ll}(\omega_{k})$ represent the sound attenuation by the magnonic crystal. The off-diagonal self-energies $\Sigma_{l\ne l'}(\omega_{k})$ describe the interaction between the magnetic wires by the exchange of surface phonons.
We can now write the magnon Green function $G_N(\omega)=\big(\omega-{\cal H}_N(\omega)\big)^{-1}$, where 
\begin{align}
	{\cal H}_N(\omega)&\equiv \tilde{\omega}_{\rm K}- i\begin{pmatrix}
		\Gamma_{1}(\omega)&\Gamma_{L,21}(\omega)&\cdots&\Gamma_{L,N1}(\omega)\\
		\Gamma_{R,12}(\omega)&\Gamma_{2}(\omega)&\cdots&\Gamma_{L,N2}(\omega)\\
		\cdots&\cdots&\cdots&\cdots\\
		\Gamma_{R,1N}(\omega)&\Gamma_{R,2N}(\omega)&\cdots&\Gamma_{N}(\omega)
	\end{pmatrix}
\end{align}
is  a non-Hermitian Hamiltonian matrix that is very similar to the dissipatively-coupled magnons for magnetic spheres in the waveguide as reviewed in Sec.~\ref{spin_skin_effect}.

We may rewrite the
${T}$-matrix Eq.~(\ref{eqn:T_matrix})
\begin{align}
	\begin{split}
		T_{k'k}&=\delta_{k'k}+\frac{1}{\omega_{k}-\omega_{k'}+i\eta}
		{\cal M}_{k'}^*
		G_N(\omega_{k})
		{\cal M}_{k}^T
	\end{split},
\end{align}
with ${\cal M}_{k}=\begin{pmatrix}g_{k,1},\cdots,g_{k,N}
\end{pmatrix}$.
The asymptotic scattering states for $k>0$ read
\begin{subequations}
	\begin{align}	
		&\lim_{y\rightarrow +\infty}\psi_t(y)=\sum_{k'}\langle y|k'\rangle T_{k'k},\\
		&\lim_{y\rightarrow -\infty}\psi_r(y)=\sum_{k'}\langle y|k'\rangle T_{k'k}.
	\end{align}
\end{subequations}
The elements of the phonon scattering matrix \(S\) therefore read \cite{scattering_PRB,scattering_PRE,Mahan}
\begin{subequations}
	\begin{align}
		&S_{21}(\omega_{k})=e^{ikD}
		\left(1-\frac{i}{c_r}
		{\cal M}_{k}^*
		G_N(\omega_{k})
		{\cal M}_{k}^T\right),\\
		&S_{11}(\omega_{k})=-\frac{i}{c_r}
		{\cal M}_{-k}^*
		G_N(\omega_{k}){\cal M}_{k}^T,
		\label{phonon_transmission}
	\end{align}
\end{subequations}
where the propagation distance $D$ governs  the  phase of the unperturbed SAW.
In the limit  $\alpha_G\rightarrow 0$ the phonon current is conserved and the scattering matrix  must be unitary, from which  $|S_{21}(k)|^2+|S_{11}(k)|^2=1$ follows. 
We can express the phonon scattering matrix by the collective magnon modes, \textit{i.e.}, the eigenvectors of the non-Hermitian Hamiltonian ${\cal H}_N(\omega_k)$, analogous to the procedure reviewed in Sec.~\ref{spin_skin_effect}. Let the right eigenvectors of ${\cal H}_N(\omega_k)$ be $\psi_{\zeta}$ with corresponding eigenvalue $\nu_{\zeta}$, ${\cal H}_N(\omega_k)\psi_{\zeta}=\nu_{\zeta}\psi_{\zeta}$, where $\zeta=\{1,\cdots,N\}$ labels the collective modes. We also define the right eigenvectors $\phi_{\zeta}$ of ${\cal H}^{\dagger}_N(\omega_k)$ with corresponding eigenvalue $\nu_{\zeta}^*$. The bi-linear expansion of the magnon (retarded) Green function then reads
\begin{align}
	G^r_N(\omega)=\sum_{\zeta}\psi_{\zeta}\phi_{\zeta}^{\dagger}\frac{1}{\omega-\nu_{\zeta}}.
	\label{retarded}
\end{align}
The net effect of many wires differs from adding those of a single wire due to interference. $|S_{21}(\omega_{k})|\ne |S_{12}(\omega_k)|$ implies the phonon diode and even isolator effect. Figure~\ref{Yu_phonon_diode}(c) and (d) illustrate the wire-number and frequency dependence of the non-reciprocal phonon transmission.

\textbf{Non-reciprocal microwave spectroscopy}.---Similar to the spin waves (Sec.~\ref{dipolar_pumping}), the phonon propagation can be also measured by microwave spectroscopy, in which the magnetic wires couple inductively to microwave striplines in the proximity of the magnetic wires \cite{Yu_Springer,chiral_damping,Haiming_exp_grating,Haiming_exp_wire,Chuanpu_NC,Dirk_transducer}. The magnetization dynamics pumps SAWs into the substrate in the source wire, while the such created SAWs excite the magnet in a distant detector.  We consider here $N$ parallel magnetic wires in which the microwaves drive the $i$-th wire while monitoring the non-local response of the $j$-th wire. In terms of the input-output theory introduced in Sec.~\ref{dipolar_pumping}, the microwave reflection at the $i$-th wire reads
\begin{align}
	\tilde{S}_{ii}(\omega)=1-i\kappa_p(G_N(\omega))_{ii}=1-i\kappa_p\sum_{\zeta=1}^N\frac{\left(\psi_{\zeta}(\omega)\right)_i\left(\phi_{\zeta}^{\dagger}(\omega)\right)_i}{\omega-\nu_{\zeta}},
\end{align}
which depends on a diagonal element of the magnon Green function, while the microwave signal transmission carried by the SAWs from the $i$-th to the $j$-th wire
\begin{align}
	\tilde{S}_{ji}(\omega)=-i\kappa_p(G_N(\omega))_{ji}=-i\kappa_p\sum_{\zeta=1}^N\frac{\left(\psi_{\zeta}(\omega)\right)_j\left(\phi_{\zeta}^{\dagger}(\omega)\right)_i}{\omega-\nu_{\zeta}},
\end{align}
is proportional to an off-diagonal element. In the chiral limit in which $g_{k}$ or $g_{-k}$ vanish, the microwave transmission is chiral such that either $S_{1N}$ and $S_{N1}$ are zero. The microwave transmission also depends on the number of wires, as illustrated in Fig.~\ref{Yu_phonon_diode}(e).

\begin{figure}[ptbh]
	\begin{centering}
		\includegraphics[width=1\textwidth]{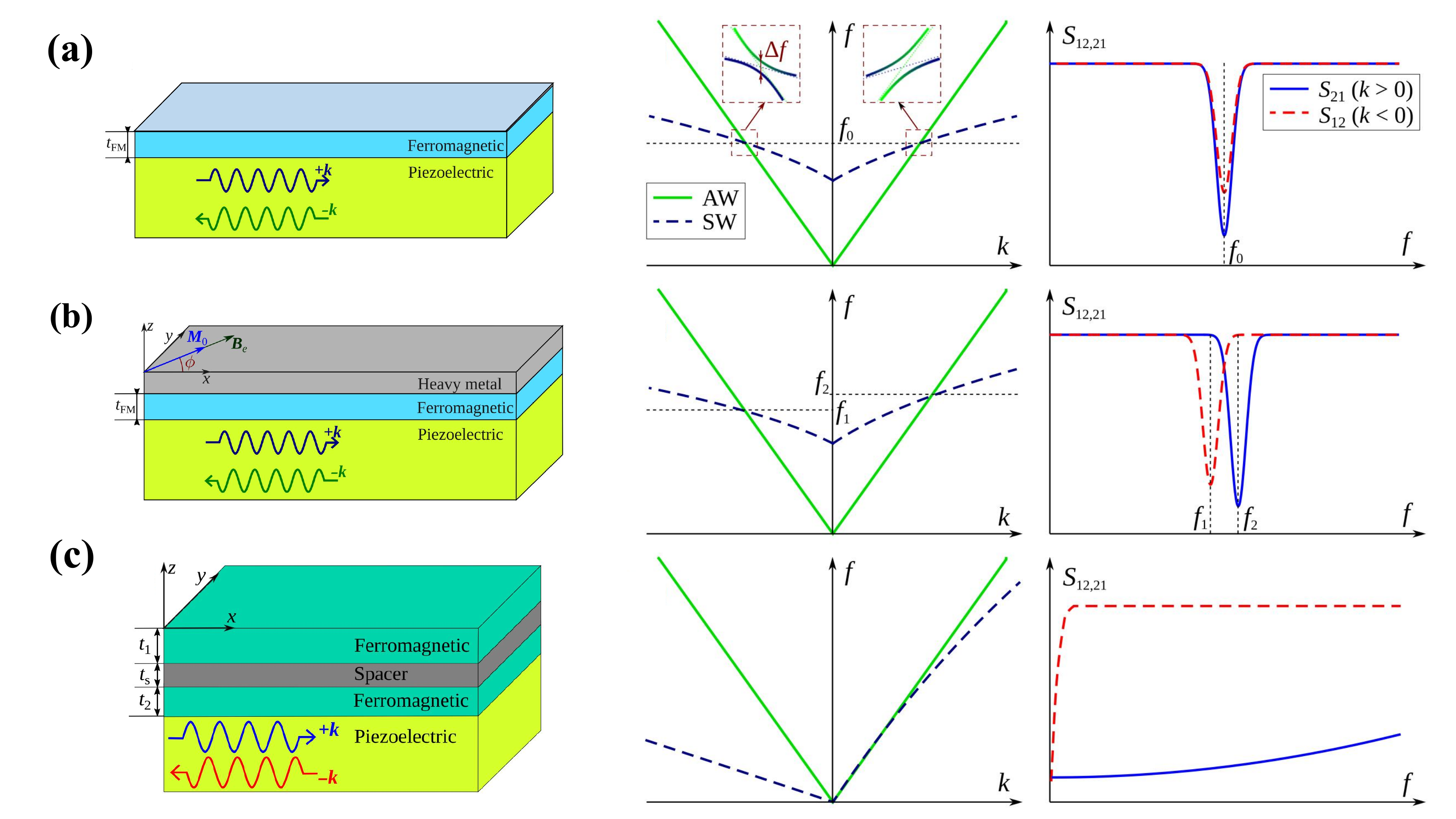}
		\par\end{centering}
	\caption{The SAW isolator induced by magnetic films on top of a piezoelectric substrate. In (a), the non-reciprocity is introduced by the pure magneto-elastic coupling, which induces different anti-crossing gaps around the resonant frequency $f_0$ at opposite momentum and thereby different SAW transmission around $f_0$. (b) employs the non-reciprocity of the spin waves induced by the interfacial DMI via the proximity of heavy metal, which induces anti-crossing gaps at two different resonant frequencies $f_1$ and $f_2$, respectively. This improves the performance of the isolator as shown by the SAW transmission. (c) proposes the realization of wide-band non-reciprocity via the coupling with a synthetic antiferromagnet that supports ``fast" spin waves propagating in one direction and ``slow" spin waves propagating in the other direction. The performance of the SAW is largely improved when the group velocity of the ``fast" spin waves matches the one of the SAWs. The figures are adopted from Refs.~\cite{Slavin_2018,Slavin_2019}.}
	\label{DMI_shift_phonon}
\end{figure}

\textbf{Phonon diode related to DMI}.---So far the generalized spin-orbit interaction of the surface acoustic waves is responsible for the chiral interaction with the magnet, where the time-reversal symmetry is broken. On the other hand, the relativistic spin-orbit coupling itself is also an important source for the phonon diode effect as reviewed here. This is reflected in its role in the non-reciprocity of spin-wave dispersion (Sec.~\ref{chiral_spin_waves}). In extended magnetic bilayers as shown in Fig.~\ref{DMI_shift_phonon}(a), the surface acoustic waves hybridize with the spin waves only when their frequency and momentum match. The dipolar interaction is not active but the relativistic magnetoelastic coupling is chiral, \textit{i.e.}, it differs for positive and negative resonant momenta \cite{magneto_elastic_1,magneto_elastic_2,magneto_elastic_3,magneto_elastic_4,magneto_elastic_5,phonon_Kei}. The different anti-crossing gaps  affect the propagation of surface acoustic waves in opposite directions [Fig.~\ref{DMI_shift_phonon}(a)].

Verba \textit{et al.} proposed to utilize the non-reciprocity of spin waves for such an isolator  device for acoustic waves that is transparent for waves in only one direction by modifying the simple bilayer in Fig.~\ref{DMI_shift_phonon}(a)  \cite{Slavin_2018,Slavin_2019}. As shown in Fig.~\ref{DMI_shift_phonon}(b), the interfacial DMI introduced via the heavy metal in proximity to the ferromagnetic film (refer to Sec.~\ref{DMI_spin_waves})  induces band gaps at different momentum magnitude and different resonant frequencies, say $f_1$ and $f_2$, for the opposite propagation direction \cite{Slavin_2018}. This can improve the performance of the isolator at these two resonant frequencies.  However, the frequency window for these devices in Fig.~\ref{DMI_shift_phonon}(a) and (b) is still narrow. To solve this problem, they further proposed to use the spin waves in the synthetic antiferromagnet \cite{Slavin_2019}, as illustrated in Fig.~\ref{DMI_shift_phonon}(c), in which the spin waves are non-reciprocal due to the chiral dipolar coupling (refer to Sec.~\ref{synthetic_antiferromagnet}) with different group velocities when propagating in opposite directions, \textit{i.e.}, ``slow" and ``fast" spin waves. When the group velocity for spin waves propagating in one direction matches the one of the surface acoustic waves, their coupling is strong in a large frequency window. This can induce the wide-band non-reciprocity of surface acoustic waves [Fig.~\ref{DMI_shift_phonon} (c)].

   \subsubsection{Phonon diodes by chiral magnon-phonon interaction: experimental evidence}

Non-reciprocal signal operation is an attractive feature in optical, acoustical, and microwave applications since it suppresses cross talk and power loss caused by back scattering \cite{nonreciprocity_plasma,optics_nonreciprocity_4}. Ideally, it leads to an \textit{isolator}, a two-port device that transmits waves or  power in one direction only \cite{what_is_not}, somewhat analogous to a \textit{perfect} diode in an electronic circuit. Magnetic elements can rectify sound potentially more efficiently than in conventional devices, such as high-quality dielectrics without piezoelectric effect but via pure magnetic effect. Here we briefly review several experiments that addressed the non-reciprocal propagation of SAWs on a piezoelectric substrate, usually emitted and detected by interdigital transducers (IDTs).  

An ac voltage biased IDT  emits SAWs via the strains induced in the piezoelectric layer, which in turn cause magnetostriction in an epitaxial ferromagnetic film, leading to a coupling between voltage (electric field) and magnetization. Excitation of the magnetization in a magnetostrictive layer induces strain and electric polarization in a piezoelectric material that can be picked up by an IDT. A magnetic element between the source and detectors modulates the phonon transmission amplitude as discussed in the previous Sec.~\ref{SAW_diode_theory}.   Table~\ref{table_chiral_SAW} summarizes a number of experimental realizations of non-reciprocal propagation of SAWs by proximity magnets, with different materials, proposed mechanisms, and relevant features.

  \begin{table}[ht]
	\caption{Experimental realizations of non-reciprocal propagation of Rayleigh SAWs by proximity magnets.} \label{table_chiral_SAW}
	\centering
	\begin{tabular}{ccc}
		\toprule
		\hspace{-2.5cm}Materials (Reference) & \hspace{-2.8cm} Proposed Mechanisms & \hspace{-0.3cm}Features \\
		\toprule
		\hspace{-2.5cm}\begin{minipage}[m]{.5\textwidth}
			\centering\vspace*{5pt}
			ZnO|Ga-doped YIG|GGG\\
			(Lewis and Patterson \cite{Lewis_1972})
			\vspace*{2pt}
		\end{minipage} &
		\hspace{-2.7cm}\begin{minipage}[m]{6.7cm}
			\begin{itemize}
			\item Magnetoelastic interaction and Damon-Eshbach chirality\\
			(qualitative argument)
			\end{itemize}
		\end{minipage} &
		\hspace{-0.5cm}\begin{minipage}[m]{5.5cm}
			\begin{itemize}
				\item Pioneering experimental efforts (1972)
				\item  SAW isolator functionality at 200~MHz.
				\item Thick magnetic film ($5~\mu$m)    		\end{itemize}
		\end{minipage}\\
		\toprule
		\hspace{-2.5cm}\begin{minipage}{.5\textwidth}
			\centering\vspace*{5pt}
			Ni|LiNiO$_3$\\
			(Sasaki \textit{et al.}  \cite{SAW_chiral_attenuation,Onose_exp})
			\vspace*{2pt}
		\end{minipage} &
		\hspace{-2.7cm}\begin{minipage}[m]{6.7cm}
			\begin{itemize}
				\item Magnetoelastic interaction 
				(interference of shear and longitudinal type magnetoelastic couplings) \cite{SAW_chiral_attenuation,Onose_exp} 
			\end{itemize}
		\end{minipage} &
		\hspace{-0.6cm}
		\begin{minipage}[m]{5.5cm}
			\begin{itemize}
				\item Ni thickness is 30~nm.
				\item Non-reciprocal SAW attenuation and transmission \cite{SAW_chiral_attenuation}
				\item Non-reciprocal phase velocity of SAWs \cite{SAW_chiral_attenuation}
				\item Non-reciprocal manipulation of the magnetization by SAW torque \cite{Onose_exp}
			\end{itemize}
		\end{minipage}\\
		\toprule
		\hspace{-2.5cm}\begin{minipage}{.5\textwidth}
			\centering\vspace*{5pt}
			(Si|Ni)|LiNiO$_3$\\
			(Tateno and Nozaki \cite{Nozaki_exp})
			\vspace*{2pt}
		\end{minipage} &
		\hspace{-2.7cm}\begin{minipage}[m]{6.7cm}
			\begin{itemize}
				\item Magnetoelastic interaction\\
				(enhanced shear component of the strain tensor) 
			\end{itemize}
		\end{minipage} &
		\hspace{-0.5cm}\begin{minipage}[m]{5.5cm}
			\begin{itemize}
				\item  Ni thickness is 20~nm, but the Si layer with thickness from tens to hundreds of nanometers can tune the amplitude of the shear strain in the Ni film.
				\item Enhanced non-reciprocal propagation of SAWs by Si layer 
			\end{itemize}
		\end{minipage}\\
		\toprule
		\hspace{-2.5cm}\begin{minipage}{.5\textwidth}
			\centering\vspace*{5pt}
			Ta|CoFeB|MgO|LiNiO$_3$\\
			(Xu \textit{et al.} \cite{Otani_exp})
			\vspace*{2pt}
		\end{minipage} &
		\hspace{-2.7cm}\begin{minipage}[m]{6.7cm}
			\begin{itemize}
				\item Magnetorotation coupling\\
				(non-reciprocity by the magnetoelastic coupling  is said to vanish when the magnetic film thickness is much smaller than the SAW wavelength.)
				\item DMI exists, but is believed to be too weak to be responsible for the SAW non-reciprocity. 
			\end{itemize}
		\end{minipage} &
		\hspace{-0.5cm}\begin{minipage}[m]{5.5cm}
			\begin{itemize}
				\item CoFeB thickness (1.6~nm) is too thin to be affected by shear strain of the SAW, in contrast to Refs.~\cite{Lewis_1972,SAW_chiral_attenuation,Onose_exp}.
				\item Observation of both non-reciprocal SAW attenuation and resonant magnetic field of the spin waves that match the SAW frequency.
				\item Non-reciprocal resonant magnetic field is attributed to the DMI with strength $D\sim 0.09~{\rm mJ/m^2}$.
			\end{itemize}
		\end{minipage}\\
		\toprule
		\hspace{-2.5cm}\begin{minipage}{.5\textwidth}
			\centering\vspace*{5pt}
			(CoFeB|Pt)|LiNiO$_3$\\
			(K\"u$\beta$ \textit{et al.} \cite{DMI_phonon_exp})
			\vspace*{2pt}
		\end{minipage} &
		\hspace{-2.7cm}\begin{minipage}[m]{6.7cm}
			\begin{itemize}
				\item Spin-wave non-reciprocity by DMI
				\item Magnetoelastic and magnetorotation couplings
			\end{itemize}
		\end{minipage} &
		\hspace{-0.5cm}\begin{minipage}[m]{5.5cm}
			\begin{itemize}
				\item  The SAW non-reciprocity increases with CoFeB thickness of 1.5 to 5~nm,.
				\item Measured DMI $D\sim -0.42~{\rm mJ/m^2}$ causes significant non-reciprocal resonant magnetic fields.
			\end{itemize}
		\end{minipage}\\
		\toprule
		\hspace{-2.5cm}\begin{minipage}{.5\textwidth}
			\centering\vspace*{5pt}
			(FeGaB|Al$_2$O$_3$|FeGaB)|LiNiO$_3$\\
			(Shah \textit{et al.} \cite{Page_exp,Page_exp_2})
			\vspace*{2pt}
		\end{minipage} &
		\hspace{-2.7cm}\begin{minipage}[m]{6.7cm}
			\begin{itemize}
				\item Spin-wave non-reciprocity in antiferromagnetically coupled magnetic bilayers
			\end{itemize}
		\end{minipage} &
		\hspace{-0.5cm}\begin{minipage}[m]{5.5cm}
			\begin{itemize}
				\item Record high isolation of SAWs
				\item Symmetry breaking by applied magnetic fields 
			\end{itemize}
		\end{minipage}\\
		\hline
	\end{tabular}
\end{table}

As early as 1972, Lewis and Patterson  successfully fabricated a 200-MHz SAW isolator  with an epitaxial layer of Ga-doped YIG (5~$\mu$m) on a GGG substrate \cite{Lewis_1972} covered by a thin piezoelectric ZnO  (1~$\mu$m)  layer. Now this phenomenon is well understood. The excitation of spin waves by the SAWs is observed  in a ferromagnetic thin-film–normal/metal (Co/Pt) bilayer \cite{acoustic_pumping_1,acoustic_pumping_2} more than ten years ago, but without resolving the  non-reciprocity. The inverse process, \textit{i.e.}, the propagation of the SAWs through the magnet, turns out to be very sensitive for detecting the non-reciprocity.

\begin{figure}[ptbh]
	\begin{centering}
		\includegraphics[width=1\textwidth]{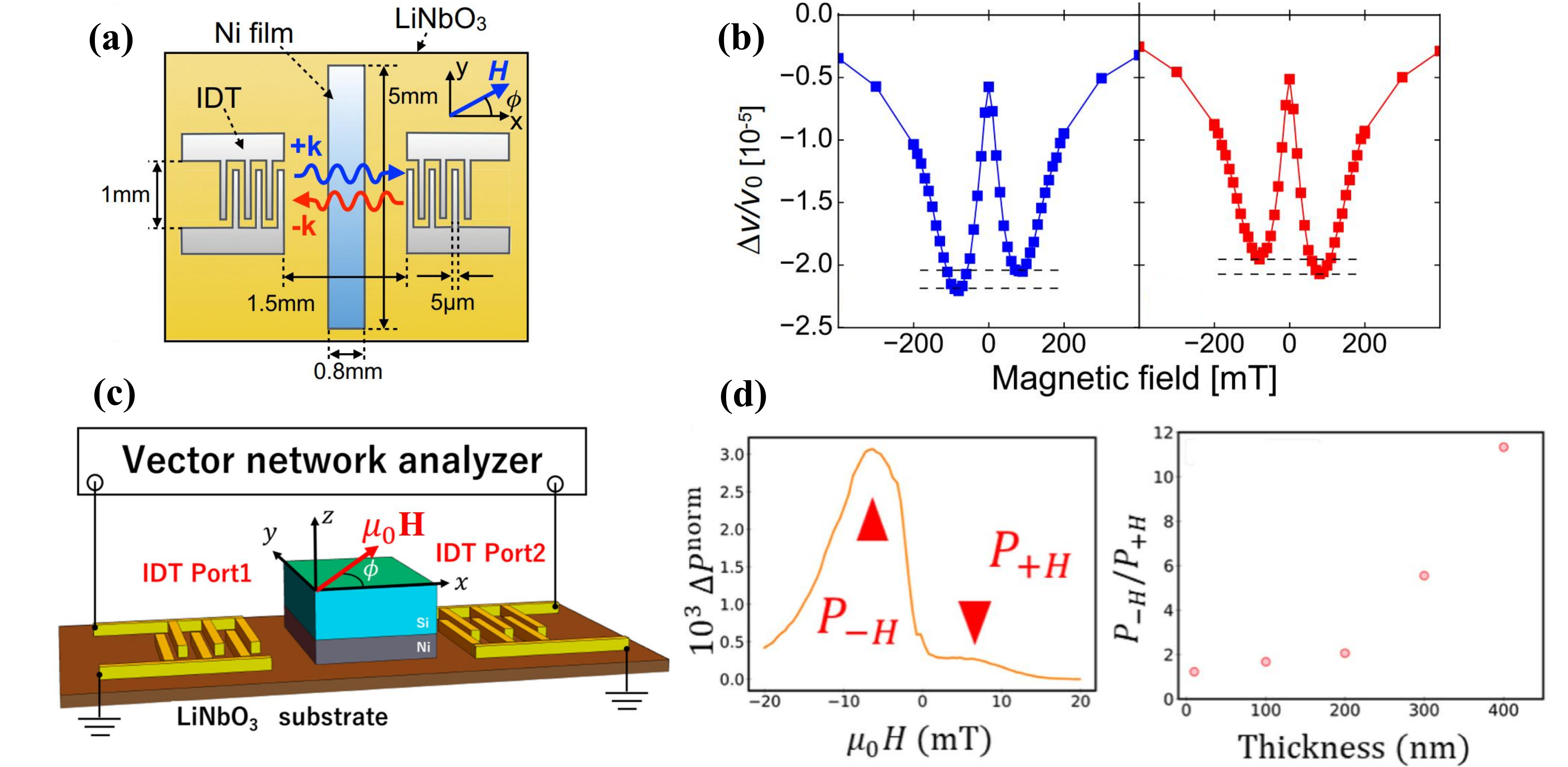}
		\par\end{centering}
	\caption{(a)Non-reciprocal SAW attenuation in a device of a Ni layer deposited on a LiNiO$_3$ substrate. (b) Detector voltage for positive and negative magnetic fields indicates different phase velocities of SAW propagating along $+k$ and $-k$ directions as a function of the magnetic field along $\phi=45^{\circ}$. (c) A Si layer is fabricated on the Ni film to tune the efficiency of non-reciprocity. (d) Measured non-reciprocity via the magnetic-field dependence of the absorption power of SAWs and $P_{-H}/P_{+H}$ as a function of Si thickness, where $P_{-H}$ and $P_{+H}$ are peak values in the negative and positive fields, respectively. The figures are taken from Refs.~\cite{SAW_chiral_attenuation} [(a) and (b)] and \cite{Nozaki_exp} [(c) and (d)].}
	\label{Nozaki_exp}
\end{figure}

Recently, Sasaki \textit{et al.} studied the propagation of SAWs in a Ni/LiNbO$_3$ hybrid device \cite{SAW_chiral_attenuation} in a device as shown in Fig.~\ref{Nozaki_exp}(a). 
The SAWs excited by the source IDT interact with the ferromagnet and are detected by a second IDT. The observable is then the transmission amplitude $S_{21}(\omega)$ from port 1 to 2.  Non-reciprocity is a $S_{12}(\omega)$ from port 2 to 1 with $|S_{12}(\omega)|\ne |S_{21}(\omega)|$ at fixed magnetic field, or an $|S_{12}(\omega)|$ that depends on the magnetization direction.  This is equivalent to amplitude  and phase velocity that depends on the sign of the wave vector, as shown in Fig.~\ref{Nozaki_exp}(b). The  used pulse excitation can measure both the phase and amplitude of a wave, with which the phase velocity is resolved. The authors attributed the observed non-reciprocity to the interference of shear-type and longitudinal-type magnetoelastic couplings \cite{SAW_chiral_attenuation}. It is noted that the longitudinal-type strain is diagonal such as $\epsilon_{xx}$ and $\epsilon_{yy}$, and the shear-type strain is off-diagonal such as $\epsilon_{xy}$. In the theory part Eq.~(\ref{eqn:coupling}), both have a contribution but can be canceled or enhanced depending on the magnetization direction Eq.~(\ref{eqn:coupling_strength}). But details between theory and observation are not compared because the device is not completely the same.

Figure~\ref{Nozaki_exp}(c) shows  a device with Si$/$Ni bilayer films deposited on LiNiO$_3$ substrate \cite{Nozaki_exp}. The Si layer turns out to tune the SAW transmission by changing the acoustic thickness of the magnetic (Ni) layer. An enhanced non-reciprocity is interpreted in terms of the shear strains that change sign with the propagation direction,  while the longitudinal strain remains the same, thereby locking the phonon polarization and momentum. The shear strain by the SAW increases with increasing Si spacer thickness, which improves the non-reciprocity by up to a factor of ten, as shown in Fig.~\ref{Nozaki_exp}(d).    
   
\begin{figure}[ptbh]
	\begin{centering}
		\includegraphics[width=1\textwidth]{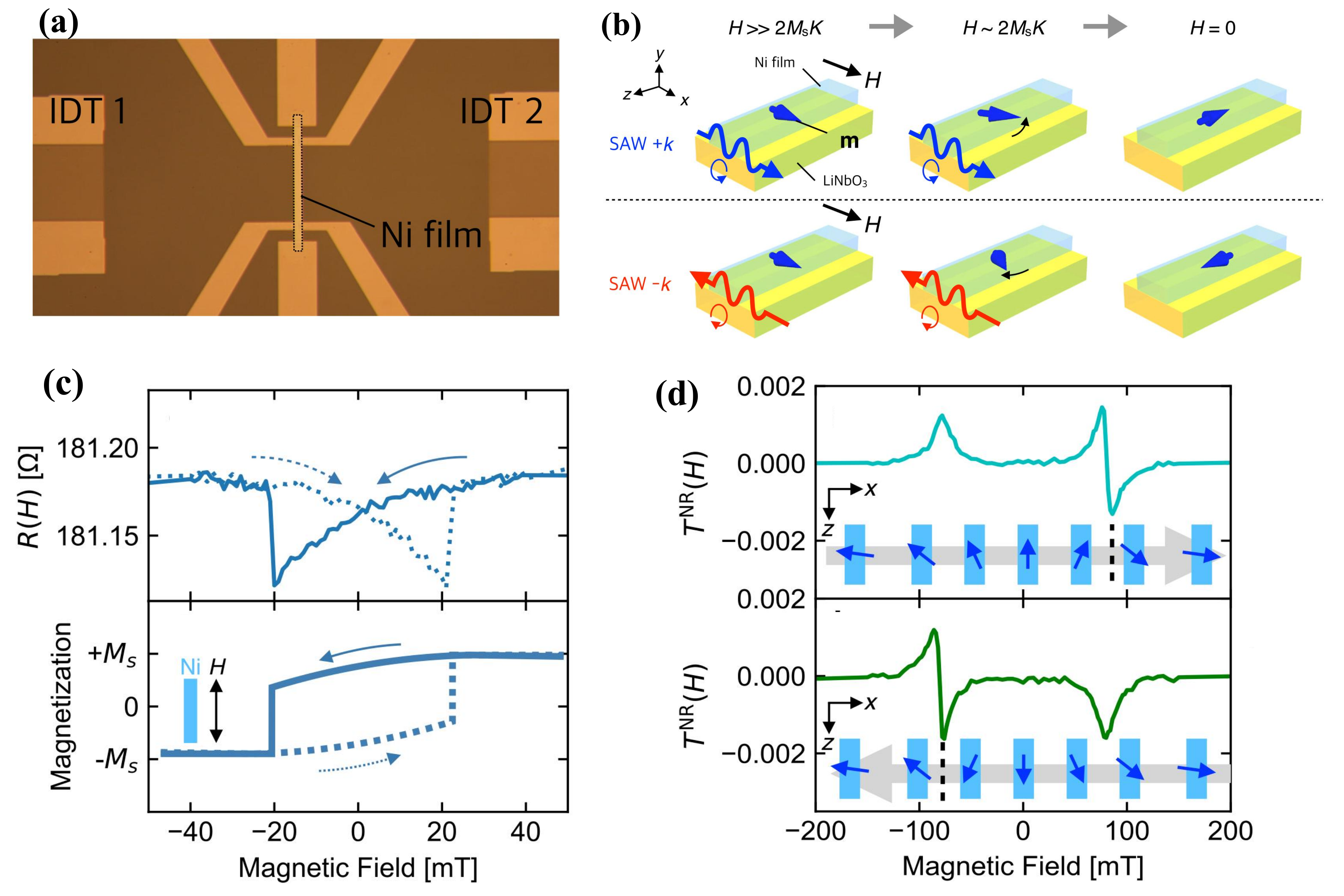}
		\par\end{centering}
	\caption{Magnetization switching in a Ni nanowire by SAWs. (a) is the configuration of the experimental setup of a Ni nanowire on a LiNiO$_3$ substrate. In (b), when the magnetization of Ni wire is parallel to the $\hat{\bf x}$-direction, the phonon transfers angular momentum to the magnetization. A SAW current controls the magnetization direction at $H=0$. Panels (c) and (d) show measurements of the magnetization direction ($+\hat{\bf z}$ or $-\hat{\bf z}$) after the switching via the magnetoresistance and the non-reciprocal SAW transmission. The figures are taken from Ref.~\cite{Onose_exp}.}
	\label{Ni_nanowire_phonon}
\end{figure}

A unidirectional SAW current can control the magnetization direction of a Ni wire by angular momentum transfer from the SAW to the magnetization~\cite{Onose_exp}. According to Fig.~\ref{SAW}, the rotation direction of SAWs is locked to the momentum that can be interpreted as a phonon angular momentum. 
Spin waves generate an anticlockwise precession of the magnetization around its  equilibrium direction. In Fig.~\ref{Ni_nanowire_phonon}(a), the excited SAWs provide a torque on the magnetization as illustrated in Fig.~\ref{Ni_nanowire_phonon}(b) for a magnetic field in the $\hat{\bf x}$-direction, \textit{i.e.}, perpendicular to the wire along $\hat{\bf z}$. The SAW coming from the left drives the magnetization, and when $H\rightarrow 0$ can even flip its direction. A SAW from the other side can flip it back. The direction of magnetization after the switch can be measured by the magnetoresistance, see Fig.~\ref{Ni_nanowire_phonon}(c) which depends on the magnitude of $m_z$.  The switching can be also observed in terms of the non-reciprocal transmission of SAWs in Fig.~\ref{Ni_nanowire_phonon}(d) by increasing the magnetic field along the $\hat{\bf x}$-direction. The authors interpreted their result by the change in angular momentum of the phonon by the magnetoelastic coupling, which  implies a torque on the magnetization that, if large enough, switches the magnetization. In Fig.~\ref{Ni_nanowire_phonon}(c), an applied magnetic field  anti-parallel to the magnetization  switches the magnetization direction accompanied by a jump of the resistance.  Figure~\ref{Ni_nanowire_phonon}(d) shows  field-dependent SAWs transmission spectra as a function of increasing the magnetic field after the switching event. 

So far, Ni ferromagnetic films or wires on piezoelectric substrates are common systems to measure the coupling between
 SAWs and spin waves. The attenuation of the SAWs relies on the relatively large Gilbert damping in Ni that moderates the dynamics but also reduces the excitation efficiency. Magnets with a smaller Gilbert damping may improve the efficiency of SAW attenuation.  Also, the other materials may profit from interactions, such as magnetorotation coupling and DMI. We are not aware of a systematic comparison of the materials listed in Table~\ref{table_chiral_SAW} and below.

\begin{figure}[ptbh]
	\begin{centering}
		\includegraphics[width=1.0\textwidth]{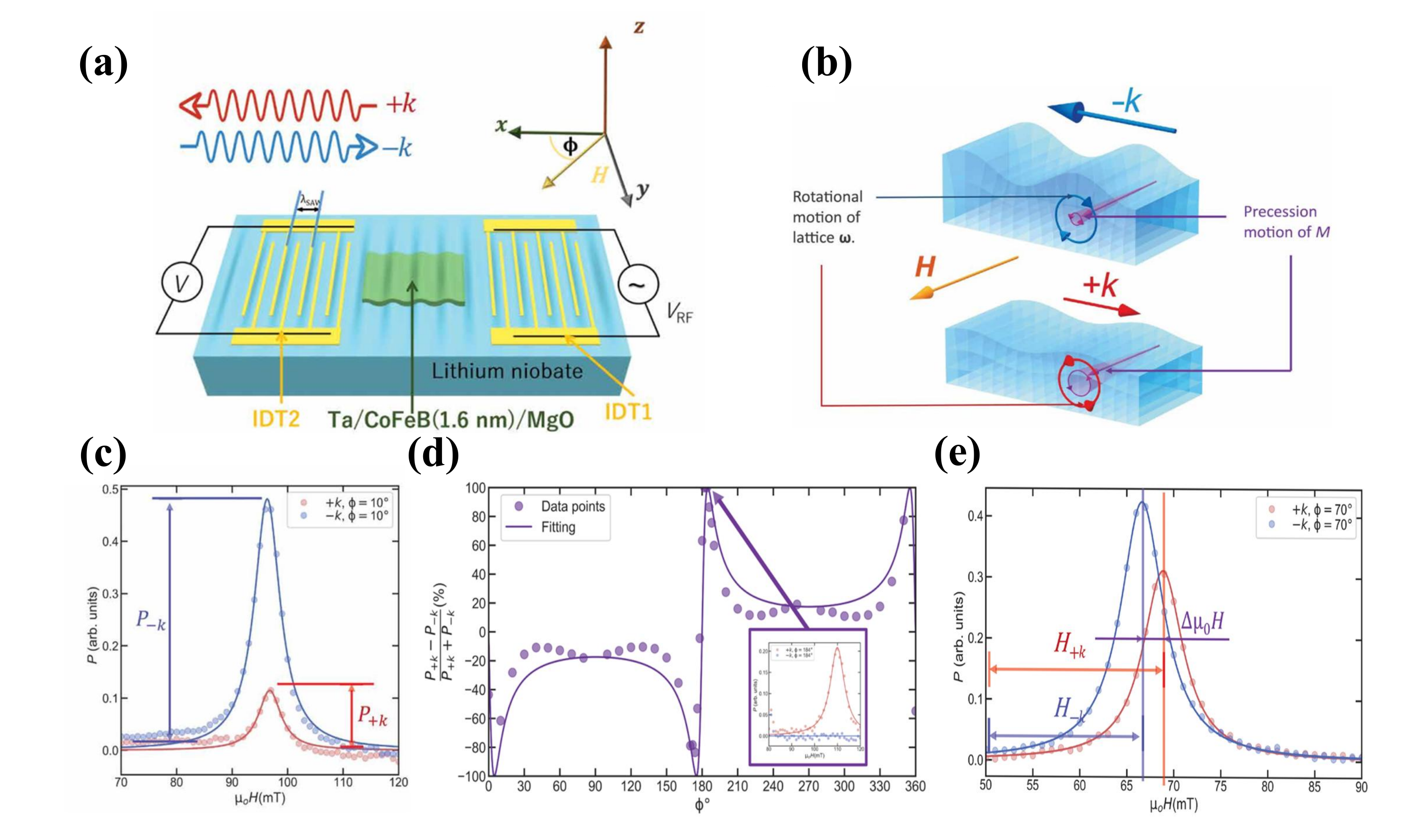}
		\par\end{centering}
	\caption{(a) Observation of non-reciprocal SAW attenuation by magnetorotation coupling in a Ta|CoFeB|MgO layer  on a LiNiO$_3$ substrate. Panel (b) illustrates the locking of the strain rotation direction and the propagating direction of SAWs, which is the key ingredient for the chiral interaction with a proximity magnet. In (c), the attenuation power of the SAWs with opposite linear momenta ($P_{+k}$ and $P_{-k}$) differs, here for a fixed magnetic field angle $\phi=10^{\degree}$. (d) shows the non-reciprocity degree $(P_{+k}-P_{-k})/(P_{+k}+P_{-k})$ as a function of $\phi$. Panel (e) shows the non-reciprocity of the spin-wave resonance field $H_{\rm k}^{\rm res}$ under reversal of the SAW wave vector from k to -k. The figures are taken from Ref.~\cite{Otani_exp}.}
	\label{Otani_exp}
\end{figure}

Xu \textit{et al.} \cite{Otani_exp} studied the non-reciprocal attenuation of SAWs by the proximity of a Ta|CoFeB(1.6~nm)|MgO layer as shown in Fig.~\ref{Otani_exp}(a). In contrast to the relatively thick Ni films (20-30~nm) in Refs.~\cite{Lewis_1972,SAW_chiral_attenuation,Onose_exp}, the magnetic film is too thin to feel a shear strain and magnetoelastic coupling from a SAW in the substrate. Nevertheless, SAWs  display a non-reciprocal damping that can be tuned by an in-plane magnetic field with angle  $\phi$ with respect to their propagation direction.  Figure~\ref{Otani_exp}(c) shows the attenuation power of the SAWs with opposite momenta ($P_{+k}$ and $P_{-k}$) for a fixed configuration $\phi=10^{\degree}$.  Figure~\ref{Otani_exp}(d)  demonstrates that the non-reciprocity $(P_{+k}-P_{-k})/(P_{+k}+P_{-k})$ depends on the magnetic field direction up 100$\%$ around $\phi=0^{}\degree$ or $180^{\degree}$, implying the SAWs are unidirectional and perfectly chiral. The authors argued that rather than the magnetoelastic coupling, the magnetorotation coupling by the perpendicular magnetic anisotropy causes the observed chirality \cite{Maekawa,magnetorotation_Soviet}. The non-reciprocal behavior of the resonance field $H_{\pm k}^{\rm res}$ under reversal of the SAW wave vector from $k$ to $-k$  in Fig.~\ref{Otani_exp}(e) is attributed to an interfacial DMI (refer to Sec.~\ref{DMI_spin_waves}).  $H_{+k}^{\rm res}-H_{-k}^{\rm res}$ can be fitted by a DMI coefficient $D=0.089\pm 0.011~{\rm mJ/m^2}$, which is too weak to cause the observed large SAW non-reciprocity, however.

\begin{figure}[ptbh]
	\begin{centering}
		\includegraphics[width=1\textwidth]{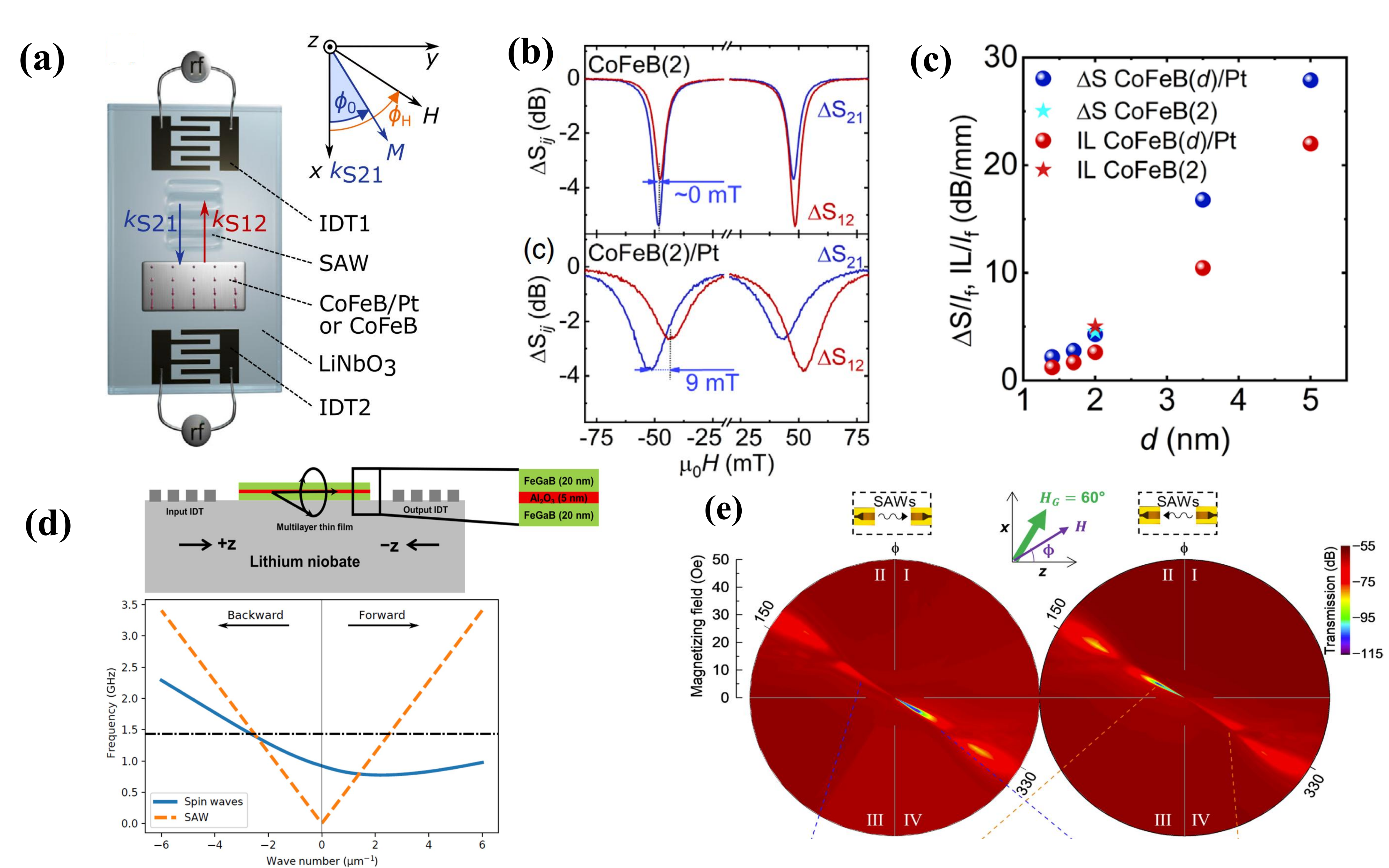}
		\par\end{centering}
	\caption{Observed non-reciprocity (``isolation") of SAWs. (a)-(c) illustrate the results of  (CoFeB|Pt)|LiNiO$_3$, while (c) and (d) are for a  tunnel junction  (FeGaB|Al$_2$O$_3$|FeGaB)|LiNiO$_3$  with anti-parallel magnetization that improves isolation. The figures are taken from Ref.~\cite{DMI_phonon_exp} [(a)-(c)] and Ref.~\cite{Page_exp} [(d) and (e)].}
	\label{Wixforth_exp}
\end{figure}

K\"u{\ss} \textit{et al.} measured the SAW non-reciprocity generated by a CoFeB ultrathin film on lithium niobate, with and without a Pt cover layer \cite{DMI_phonon_exp}, as shown in Fig.~\ref{Wixforth_exp}(b) for the configuration in Fig.~\ref{Wixforth_exp}(a). The non-reciprocity in the resonant magnetic fields is larger than that reported in Ref.~\cite{Otani_exp} because of 5 times  larger DMI $D\sim -0.42~{\rm mJ/m^2}$ in CoFeB|Pt bilayer compared to that in Ta|CoFeB \cite{Otani_exp}. With increasing CoFeB thickness from 1.5 to 5~nm, the SAW non-reciprocity increases, as shown in Fig.~\ref{Wixforth_exp}(c). A similar feature reported in Ref.~\cite{Nozaki_exp} when  covering a  Ni film with a Si layer was attributed to an increased shear strain and magnetoelastic coupling. In contrast to \cite{Otani_exp}, K\"u{\ss} \textit{et al}. concluded that the magnetoelastic and magnetorotation coupling are of the same order \cite{DMI_phonon_exp}. The DMI is stronger in Ref.~\cite{DMI_phonon_exp}, so it appears to contribute to the non-reciprocity for the thinner film. More dedicated experiments that systematically change  material and device parameters  appear necessary to fully understand the microscopic mechanisms.

The SAW non-reciprocity can also be tuned by the design of the magnetic devices. Shah \textit{et al}. reported  the transmission amplitudes of SAWs at $1435$~MHz on a LiNiO$_3$ substrate coated by ferromagnet|insulator|ferromagnet (FeGaB|Al$_2$O$_3$|FeGaB) spin valve \cite{Page_exp} as shown in Fig.~\ref{Wixforth_exp}(d).  Figure~\ref{Wixforth_exp}(e) shows a giant non-reciprocity (or ``isolation'') of  SAWs   as a function of magnitude (from $0$ to $50$~Oe) and direction $\phi$ of an applied field relative to the propagation direction. The resonance with the spin waves sharply and unidirectionally suppresses SAW propagation  only for special angles  $\phi=\{30^{\degree},330^{\degree}\}$ quite different from other devices \cite{Otani_exp,DMI_phonon_exp,Onose_exp,SAW_chiral_attenuation,Nozaki_exp}. Authors attributed the directionality to magnetic anisotropy induced by growing material under a $30^{\degree}$ magnetic field, and the isolation to the non-reciprocal dispersion of the spin waves in the antiferromagnetically coupled tunnel junctions, as shown in Fig.~\ref{Wixforth_exp}(d). It should be interesting to measure SAW transmission for parallel magnetized tunnel junctions. In the same hybrid magnetoelastic heterostructures, the non-reciprocity of phase accumulation of SAWs, as well as their non-reciprocal propagation losses, were later observed \cite{Page_exp_2}.

  \subsubsection{Chiral pumping of phonon}
  
  \label{Sec_chiral_phonon_pumping}

The excitation of spin waves by lattice vibrations \cite{acoustic_pumping_1,acoustic_pumping_2,acoustic_pumping_3,acoustic_pumping_4,acoustic_pumping_5} is well established, as discussed above. More recently, the inverse process, \textit{i.e.}, the pumping of phonons by magnetization dynamics has attracted attention  \cite{Simon,phonon_Yu_1,Klein_phonon,Kruglyak_phonon,Ruckriegel_phonon,Klein_phonon_PRX}. The non-reciprocal transmission of SAWs motivates the search for chiral or unidirectional pumping of SAWs, which can be useful for acoustic device applications \cite{acoustic_1,acoustic_2,acoustic_3,acoustic_4}. Chiral pumping of surface acoustic waves by proximity magnetization dynamics does not require piezoelectricity, allowing for excitation, manipulation, and detection of coherent
SAWs on arbitrary substrates. This functionality could operate in regimes out of reach of conventional acoustics. Here we review two mechanisms for unidirectional phonon pumping, \textit{i.e.}, by an interference effect and by a chiral magnon-phonon interaction.

\begin{figure}[ptbh]
	\begin{centering}
	\includegraphics[width=1\textwidth]{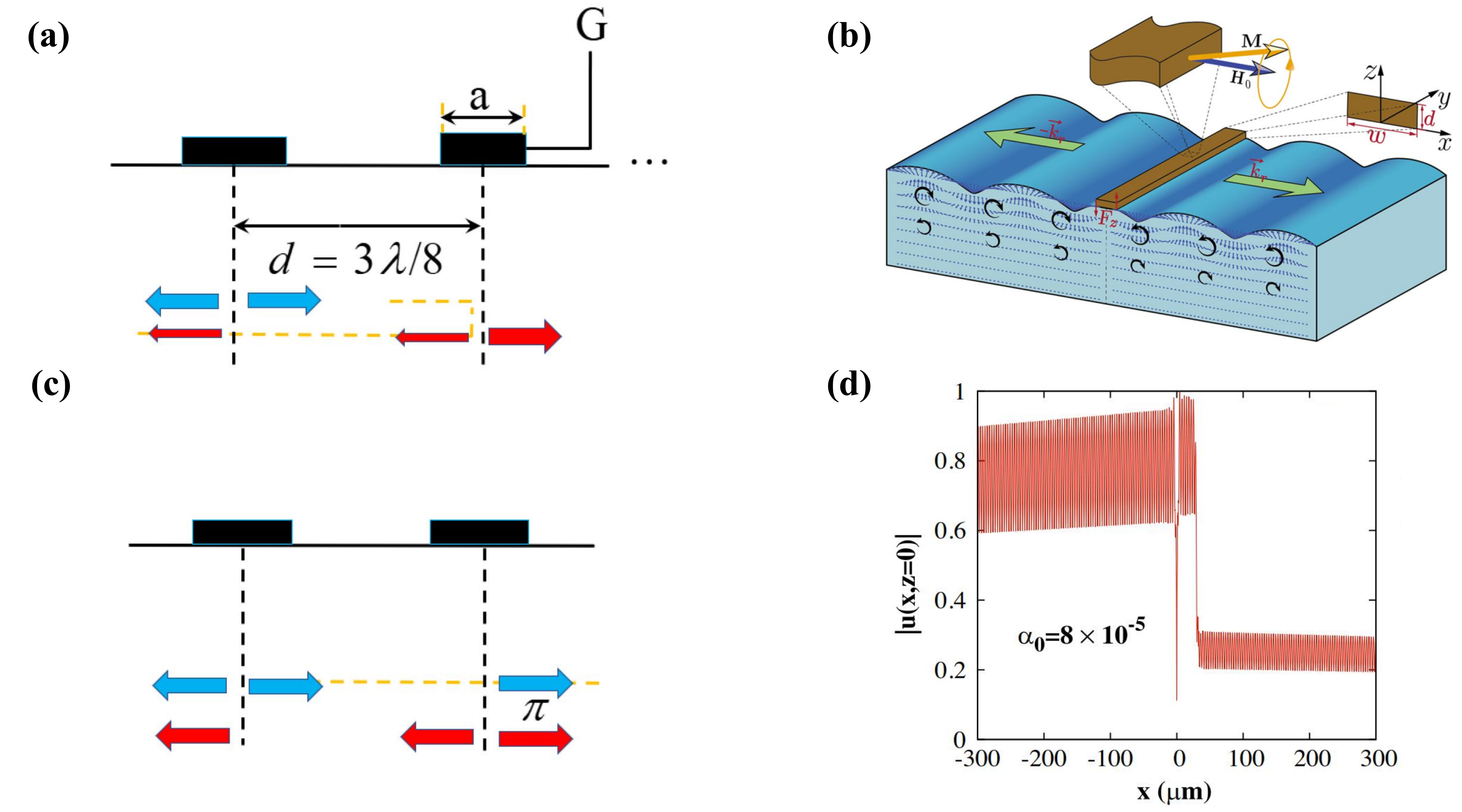}
	\par\end{centering}
	\caption{Comparison between electric [(a)] and magnetic [(c)] isolators for unidirectional pumping of SAWs. The blue arrows
represent the waves excited by the active left antenna,  while the red arrows are those reflected/transmitted [(a)] or re-emitted [(c)] waves. The unidirectionality in (a) is a purely geometric interference effect with a second grounded electrode (\(G\)). The chirality  
of the magnetic system in (c) is a purely dynamic effect. (b) and (d) illustrate the pumping of SAWs by one and two magnetic wires with magnetization perpendicular to the wire direction. The figures [(b) and (d)] are taken from Ref.~\cite{phonon_Yu_1}.}
	\label{Phonon_pumping}
\end{figure}

\textbf{Chiral phonon pumping by interference}.---A unidirectional pumping of surface acoustic waves can be realized in conventional (non-magnetic) devices  by an interference effect, as illustrated in Fig.~\ref{Phonon_pumping}(a). Here  properly spaced metal electrodes on a piezoelectric crystal reflect the SAWs by constructive interfering for sub-GHz frequencies. The active electrode generates the SAWs on the left, while a grounded electrode \(G\) centered at \textit{d} with width \(a
\) reflects incoming SAWs at the edges with amplitudes $r_{e}$ and $-r_{e}$, where in practice $\left\vert r_{e}\right\vert \ll1$ . The two reflected waves interfere as 
\begin{align}
	r_{e}e^{2ik(d-a/2)}-r_{e}e^{2ik(d+a/2)}=r_{e}(e^{-ika}-e^{ika})e^{2ikd}
	=-2ir_{e}\sin(ka)e^{2ikd}.
\end{align}
The grounded electrode $G$, therefore, causes a phase shift $-\pi/2$.  When located at a distance, \textit{e.g.}, $3/8$ of the SAW wavelength  $\lambda$ from the active electrode it generates a propagation phase shift $3\pi/2$ that for $r_{e}<0$, which leads to constructive interference at the active electrode. It causes destructive interference at the left side of the active electrode such that enhances the transmission to the right. Such enhancement is small with a small $|r_e|$. When fabricating an array of grounded electrodes, the transmission becomes unidirectional and acts as an excellent isolator. The SAW wavelength has to match the fixed electrode distance, so changing the frequency deteriorates performance. Highly unidirectional devices based on this principle can be realized by employing many electrodes, which found commercial applications \cite{acoustic_1,acoustic_2}.

  Zhang \textit{et al}. proposed a SAW isolator  with two thin magnetic wires on top of a high-quality dielectric \cite{phonon_Yu_1} that does not have to be piezoelectric.  When the magnetizations are perpendicular to the wire direction, as shown in Fig.~\ref{Phonon_pumping}(b), the Kittel magnon couples equally to the SAWs propagating in opposite directions since  Eq.~(\ref{eqn:coupling_strength}) for $\varphi=\pi/2$ and for a wire centered at $x=R_l$ becomes reciprocal, \textit{i.e.}, $g_{k}=i\sqrt{\gamma/(M_s\rho c_r)}\sqrt{{d}/{w}}\sin\left({kw}/{2}\right)\exp(ikR_l)\xi_P\tilde{B}_\perp$ and $|g_{k}|=|g_{-k}|$. Therefore, in the configuration shown in Fig.~\ref{Phonon_pumping}(b), the pumped amplitudes of the SAWs of opposite propagation are equal when excited by one magnetic wire, but oppositely polarized. A unidirectional pumping for this magnetization direction can be achieved by two magnetic wires centered at $R_1$ and $R_2$. In contrast to the interference effect in Fig.~\ref{Phonon_pumping}(a) that requires certain distances, here the destructive interference is 
caused by the \emph{dynamical} phase shift of $\pi$ remitted by the passive wire at the same  resonance frequency as the active one, see Fig.~\ref{Phonon_pumping}(c), independent of the width of the magnetic wire. In this case, SAWs are emitted only to the left of the double wire 
with a frequency that can be tuned by the applied magnetic field strength. In the coherent strong coupling regime, just two nanowires are
  sufficient to realize full chiral pumping. A static magnetic field can even switch the SAW beam completely off. With frequency or distance $k(R_2-R_1)=n\pi$ with $n\in\mathbb{Z}_{0}$, the waves to the left vanish, \textit{i.e.}, the phonons are trapped between the two wires to form an acoustic cavity. Figure~\ref{Phonon_pumping}(d) shows the displacement field at the GGG surface $\left\vert\mathbf{u}(x,z=0)\right\vert =\sqrt{u_{x}^{2}+u_{z}^{2}}|_{z=0}$ for an intrinsic Gilbert damping $\alpha_{0}=8\times10^{-5}$, YIG wires of thickness $d=200$~nm and width $w=2.5~\mathrm{\mu}$m centered at $R_{1}=0$ and $R_{2}=30~\mathrm{\mu}$m, and a (ordinary) Kittel frequency $\omega_{\mathrm{F}}=3$~GHz. The additional damping due to pumping SAWs into the substrate is then a significant $1.2\times10^{-4}$. 
The phonon chirality for these parameters is not perfect, and will be smaller for larger damping but is clearly measurable in real samples.  The device performance can be improved by an array of wires, magnetic materials with larger magnetoelasticity, or different substrates.  The mismatch of the acoustic parameters between the magnet and the substrate might influence the quantitative prediction, which should be addressed in future studies.

 \textbf{Chiral phonon pumping by chiral interaction}.---Since the magnon-phonon interaction  is intrinsically chiral, also a single wire can pump phonons unidirectionally. This becomes evident from solving the LLG coupled to the elastic equations \cite{phonon_Kei} and input-output theory as reviewed above (Sec.~\ref{dipolar_pumping}). We are not aware of the direct experimental evidence for the predicted chiral phonon pumping effect.
 
 Here we illustrate  the physical principles by the scattering theory of transport or Landauer-B\"uttiker formula \cite{Datta_scattering,RMP_scattering}. This approach allows a unified treatment of the pumping of magnon, photon, and phonon in terms of the Green function of many (randomly) coupled magnets in terms of their non-Hermitian Hamiltonian. 
 The sound velocity of long-wavelength  surface waves is constant, so energy and momentum currents are proportional to each other. The  left- and right-moving  currents $I_{L/R}$ in the $\pm \hat{\bf y}$-direction depend on the total injection rate ${\cal P}_{R/L}$ and sound velocity \(c_r\)  with
    \begin{align}
    	I_{R/L}={\cal P}_{R/L}c_r,
    \end{align}
    where ${\cal P}_{L/R}\equiv d\hat{N}_{L/R}/dt=(i/\hbar)[\hat{H},\hat{N}_{L/R}]$ from the Heisenberg equation of motion, in which $N_{L}$ and $N_R$ are the number of phonons injected with negative and positive velocity. Usually, the injection rate by one magnetic nanowire is not large, but we can enhance performance  with multiple sources such as $N$ parallel magnetic nanowires with equal FMR frequency.  When the wire  distance is sufficiently larger than the wire width, the interwire dipolar interaction is not important and the Hamiltonian $\hat{H}_c$ reduces to the sum of  Eq.~(\ref{Hamiltonian_magnon_phonon}) over the wires with index \(l\).
    
    The phonon injection rate reads   
    \begin{align}
    	{\cal P}_{R(L)}&=\frac{i}{\hbar}\left\langle \left[\hat{H}_c,\sum_l\hat{\beta}_l^{\dagger}\hat{\beta}_l\right]\right\rangle=-2{\rm Re}\left(\sum_l\sum_{k>0(<0)}g_{k,l}^*G^{<}_{lk}(t,t)\right),
    \end{align}
where we introduced the lesser Green functions $G^<_{lk}(t,t')\equiv -i\left\langle \hat{b}_{k}^{\dagger}(t')\hat{\beta}_l(t)\right\rangle$  and $G^<_{kl}(t,t')\equiv -i\left\langle \hat{\beta}^{\dagger}_l(t')\hat{b}_{k}(t)\right\rangle$. The time-ordered Green functions
    \begin{align}
    G^t_{lk}(t,t')=-i\theta(t-t')\langle \hat{\beta}_l(t)\hat{b}_{k}^{\dagger}(t')\rangle-i\theta(t'-t)\langle \hat{b}_{k}^{\dagger}(t')\hat{\beta}_l(t)\rangle
    \end{align}
evolve according to
    \begin{equation}
    	\Big(-i\frac{\partial}{\partial t'}-\omega_{k}\Big)G^t_{lk}(t,t')=\sum_{l'} g_{k,l'}{G}_{ll'}^t(t,t'),
    \end{equation}
    where the time-ordered Green function for the magnons reads
    \begin{align}
    {G}^t_{ll'}(t,t')=-i\theta(t-t')\langle \hat{\beta}_l(t)\hat{\beta}_{l'}^{\dagger}(t')\rangle-i\theta(t'-t)\langle\hat{\beta}_{l'}^{\dagger}(t')\hat{\beta}_l(t)\rangle.
    \end{align}
    In terms of the non-interacting phonon Green function $-i{\partial}/{\partial t'}-\omega_{k}\equiv [{\cal G}^t_{k}(t')]^{-1}$,
    \begin{equation}
    	G_{lk}^t(t,t')=\sum_{l'}\int dt_1 g_{k,l'}{G}_{ll'}^t(t,t_1){\cal G}_{k}^t(t_1,t').
    \end{equation}
It is convenient to introduce retarded \(G^r\) and advanced \(G^a\) Green functions (as explained in  textbooks \cite{Haug}), in terms of which the ``lesser" Green function in frequency space reads
    \begin{equation}
    	G_{lk}^<(\omega)=\sum_{l'}g_{k,l'}\big[{G}_{ll'}^r(\omega){\cal G}^<_{k}(\omega)+{G}_{ll'}^<(\omega){\cal G}_{k}^a(\omega)\big].
    \end{equation}
    and the phonons pumping rate becomes 
    \begin{align}
    	{\cal P}_{R(L)}=-2\sum_{k>0(<0)}\sum_{ll'}\int \frac{d\omega}{2\pi}{\rm Re}\left({g_{k,l}^*G^r_{ll'}(\omega)}g_{k,l'}{\cal G}^<_{k}(\omega)+g_{k,l}^*G^<_{ll'}(\omega)g_{k,l'}{\cal G}^a_{k}(\omega)\right).
    	\label{eqn:photon_pumping}
    \end{align}
Full chirality means that either of ${\cal P}_{R}$ and ${\cal P}_{L}$ should vanish, in which case the pumped phonon flows unidirectionally into half of the elastic substrate, making an analogy to the magnon and photons as reviewed. 
    
    The ``lesser'' and advanced non-interacting phonon Green functions read ${\cal G}_{k}^<(\omega)=2\pi if(\omega)\delta(\omega-\omega_{k})$, where $f(\omega)=1/(e^{\hbar\omega/(k_BT)}-1)$ is the Planck distribution at temperature $T$, and ${\cal G}_{k}^a(\omega)={1}/({\omega-\omega_{k}-i\eta_k})$, respectively.  Equation~(\ref{retarded}) is the retarded magnon Green function $G_N^r(\omega)$ and the spectral function ${\mathcal A}(\omega)=2i \textrm{Im} G^r_N(\omega)$. According to the fluctuation-dissipation theorem \cite{Haug}
    \begin{align}
    	G_{ll'}^<(\omega)=iF(\omega){\mathcal A}_{ll'}(\omega),
    \end{align}
where $F(\omega)$ is the non-equilibrium magnon distribution. The function
  $
    {\cal I}(\omega)\equiv \sum_{ll'}g_{k,l}^*G_{ll'}^<(\omega)g_{k,l'}$ obeys ${\cal I}^*(\omega)=-{\mathcal I}(\omega)$ and 
    \begin{align}
    	\nonumber
    	{\mathcal P}_{R(L)}&=\sum_{k>0(<0)}\sum_{ll'}{\rm Re}\left(ig_{k,l}^*G_{ll'}^r(\omega_k)g_{k,l'}\right)\left(F(\omega_k)-f(\omega_k)\right)-\sum_{k>0(<0)}\sum_{ll'}{\rm Re}\left(ig_{k,l}^*G_{ll'}^r(\omega_k)g_{k,l'}\right)f(\omega_k)\nonumber\\
    	&-\sum_{k>0(<0)}\sum_{ll'}{\rm Re}\left(ig_{k,l}^*G_{ll'}^{r*}(\omega_k)g_{k,l'}\right)F(\omega_k).
    	\label{Landuer_Buttiker}
    \end{align}
At thermal equilibrium, $F(\omega_k)=f(\omega_k)$ and ${\mathcal P}_{R(L)}$ vanishes. We recognize the passive phonon transmission in Eq.~(\ref{phonon_transmission}) with $\sum_{ll'}ig_{k,l}^*G_{ll'}^r(\omega_k)g_{k,l'}=c_r(1-S_{21}(\omega_k))$.

  We are interested in the scaling with the number of identical nanowires excited by a monochromatic microwave of frequency $\omega_{\rm K}$. With $F(\omega_k)\rightarrow f(\omega_k)+\delta f\delta(\omega_k-\omega_{\rm K})$ Eq.~(\ref{Landuer_Buttiker}) reduces to
    \begin{align}
    	{\mathcal P}_{R(L)}=({1}/{c_r}){\mathcal T}_{R(L)}\delta f,
    \end{align}
    where 
    \begin{subequations}
    \begin{align}
    	{\mathcal T}_R&=\frac{1}{2\pi}\sum_{ll'}{\rm Re}\left(ig_{k_*,l}^*(G_{ll'}^r(\omega_{\rm K})-G_{ll'}^{r*}(\omega_{\rm K}))g_{k_*,l'}\right),\\
    	{\mathcal T}_L&=\frac{1}{2\pi}\sum_{ll'}{\rm Re}\left(ig_{-k_*,l}^*(G_{ll'}^r(\omega_{\rm K})-G_{ll'}^{r*}(\omega_{\rm K}))g_{-k_*,l'}\right).
    \end{align}
\end{subequations}

For Ni nanowires on top of a GGG substrate and a magnetic field tuned to the critical angle at which ${\mathcal T}_L=0$, we arrive at an approximately linear scaling of ${\cal T}_R$ with the number of wires, which implies that a nanowire array of high magnetic quality \cite{phonon_Yu_2} can be a high-power and unidirectional source of SAWs. 
    
\begin{figure}[ptbh]
	\begin{centering}
		\includegraphics[width=0.98\textwidth]{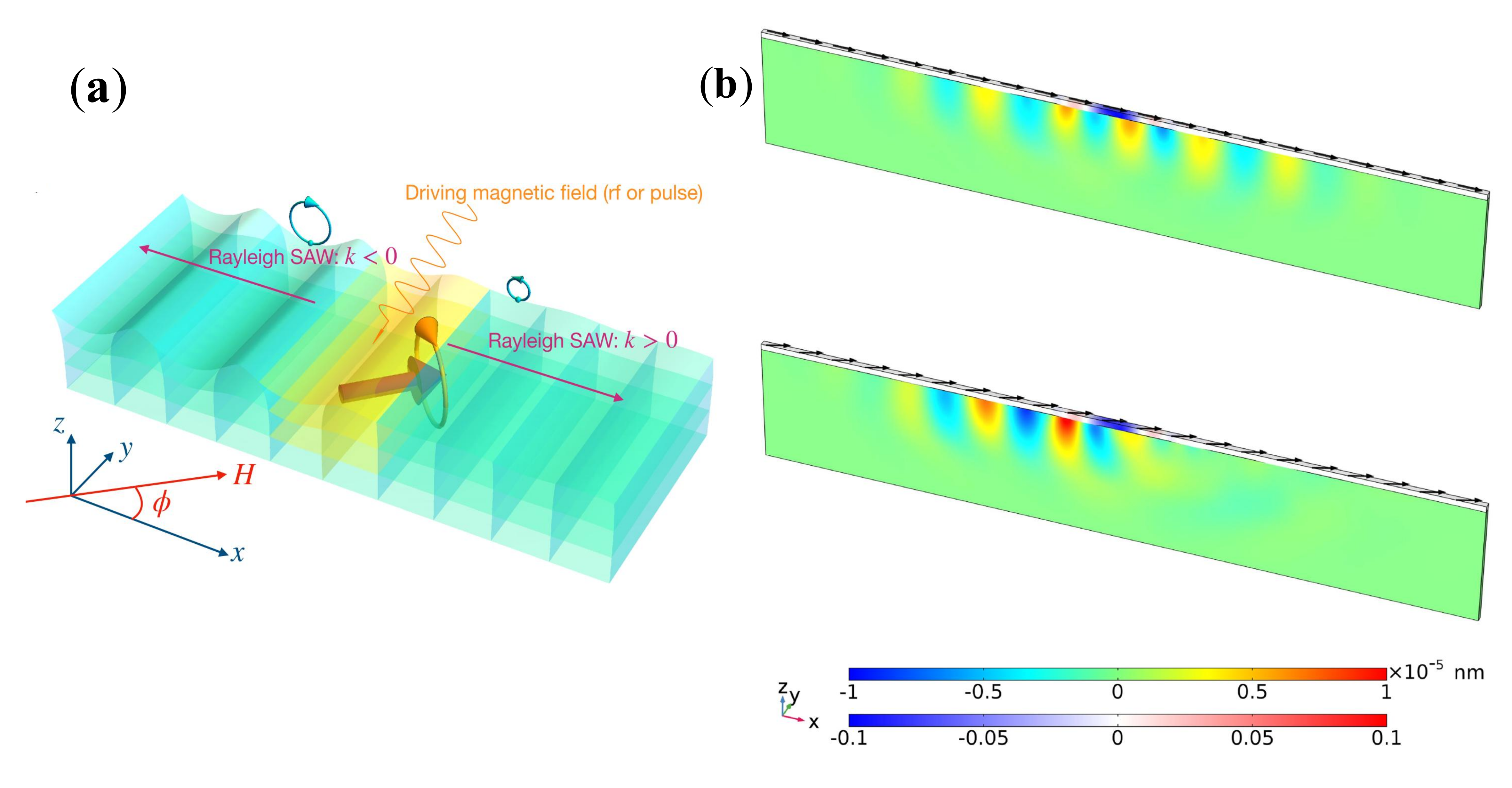}
		\par\end{centering}
	\caption{Numerical simulations of the magnon-phonon hybridization in a magnetic ﬁlm (YIG, 50~nm thick) on a dielectric substrate (GGG, 1~$\mu$m thick). (a) illustrates the configuration, in which the magnetization precession excites Rayleigh SAWs exclusively into one direction. (b) shows snapshots of the spin waves and SAW excited by a magnetic ﬁeld pulse at $t=1$~ns for the equilibrium  magnetization directions $\phi=0$ (upper panel) and $0.36\pi$ (lower panel) indicated by black arrows. The figures are taken from Ref.~\cite{phonon_Kei}.}
	\label{Phonon_pumping_simulation}
\end{figure}

  \textbf{Numerical simulation}.---Yamamoto \textit{et al.} numerically solved the coupled magnetization and surface acoustic wave dynamics 
formulated in \cite{phonon_Kei} for a magnetic wire with arbitrary thickness  \cite{phonon_Yu_1,phonon_Yu_2} in Fig.~\ref{Phonon_pumping_simulation}(a).  In Fig.~\ref{Phonon_pumping_simulation}(b) we show results for a 50~nm thick YIG wire on a $1~{\mu}$m thick GGG slab, while wire magnetization at an angle  $\phi$ relative to the  phonon propagation direction $\hat{\bf x}$. We conclude that i) when the wire magnetization is normal to the wire ($\phi=0$), the SAWs on both sides of the YIG film have the same amplitude (upper panel) as expected from the analytical results in Ref.~\cite{phonon_Yu_1}; ii), The symmetry is increasingly broken with  $\phi$ until at  $\phi=0.36\pi$ the excitation becomes fully unidirectional (lower panel), as also predicted \cite{phonon_Yu_2}.

Via the numerical simulation Cai \textit{et al.} found that the non-reciprocity can persist in the higher harmonic generation of SAWs by the magnetization dynamics \cite{Chengyuan_Cai}.

\section{Near-field spintronics: transverse spins as spin sources}
\label{Near_field_spintronics} 
The efficient transfer of spin information over large distances  and between different entities is a key objective in spintronics. As summarized in Tables~\ref{table_chiral_universality} and \ref{table_chiral_universality_continued}, the near fields of microwaves, spin waves, surface acoustic waves, and surface plasmon polaritons carry (transverse) angular moment normal to the wave vector. When they act as spin carriers in the medium, these transverse spins of evanescent waves can be unidirectionally excited because they are locked to the momentum or surface, as reviewed in above Sec.~\ref{Chiral_interaction}. On the other hand, they can act as ``spin" sources that pump the spin current of other carriers such as electrons. These effective transverse \textquotedblleft spins\textquotedblright\ can be transferred to electrons in spintronic devices with considerable efficiency. This holds the promise of ``near-field spintronics"  with a number of novel functionalities. In this section, we review the theoretical foundations and address the  experimental progress. 

   \subsection{Non-contact spin pumping by evanescent microwaves}
   \label{Sec_evanescent_pumping}
   
  As reviewed in Sec.~\ref{spin_density_waves}, in the vacuum, an electromagnetic field with electric and magnetic field components \textbf{E} and \textbf{H} at frequency $\omega$ carries a spin angular momentum density  \cite{Nori,Jackson,Nori_PRL_1,Nori_PRL_2}
  \begin{equation}
  	{\pmb {\mathcal{D}}}(\textbf{r})=\frac{1}{4\omega}\operatorname{Im}\left(  \varepsilon
  	_{0}\mathbf{E}^{\ast}\times\mathbf{E}+\mu_{0}\mathbf{H}^{\ast}\times
  	\mathbf{H}\right)  ,
  \end{equation}
  where $\mu_{0}$ $(\varepsilon_{0})$ is the vacuum permeability (permittivity). In a dispersive medium, the transverse spin is generally given by  Eq.~(\ref{transverse_spin_EM}) \cite{Nori,Nori_PRL_1,Nori_PRL_2}, to be addressed later.
In the microwave regime, the magnetic field component dominates. In evanescent fields,  ${\pmb {\mathcal{D}}}$  may be finite close to an interface or boundary even when it vanishes in the bulk  \cite{Jackson,nano_optics}. A photon transverse angular momentum is generated by excited ferromagnets, and at the special lines of microwave cavities or waveguides, to name a few. In analogy with the spin pumping by magnetization into metal contacts, we refer to the spin transfer from these near-fields to electrons as \textquotedblleft near-field spin pumping\textquotedblright. This process is contactless and avoids possible artifacts by
magnetic proximity. The reciprocal progress of near-field spin pumping is near-field spin transfer, that is, the transfer of spin angular momentum from electrons to other degrees of freedom via contactless interactions.
    
   \subsubsection{Near-field spin pumping by dipolar fields}

  Spin pumping refers to the generation of a spin current at the interface between an excited ferromagnet and an electric conductor, which is mediated by the exchange interaction between magnetic moments at the interface and the conduction electron spins \cite{spin_pumping1,spin_pumping_RMP}. This process is efficient when the magnet and conductor form a good electric contact, but is strongly suppressed by, \textit{e.g.},  surface contamination, and Schottky barriers. A good contact between the magnet and the conductor also favors proximity effects that may impede a straightforward interpretation of experimental results.
However, spins can also be pumped with high efficiency without an exchange interaction, but by the dipolar stray fields emitted by an excited ferromagnet   \cite{electron_spin_Yu}. These fields decay only slowly as a function of distance and may pump spins remotely, \textit{e.g.}, through tunnel junctions.  
   
\begin{figure}[ptbh]
	\begin{centering}
	\includegraphics[width=1\textwidth]{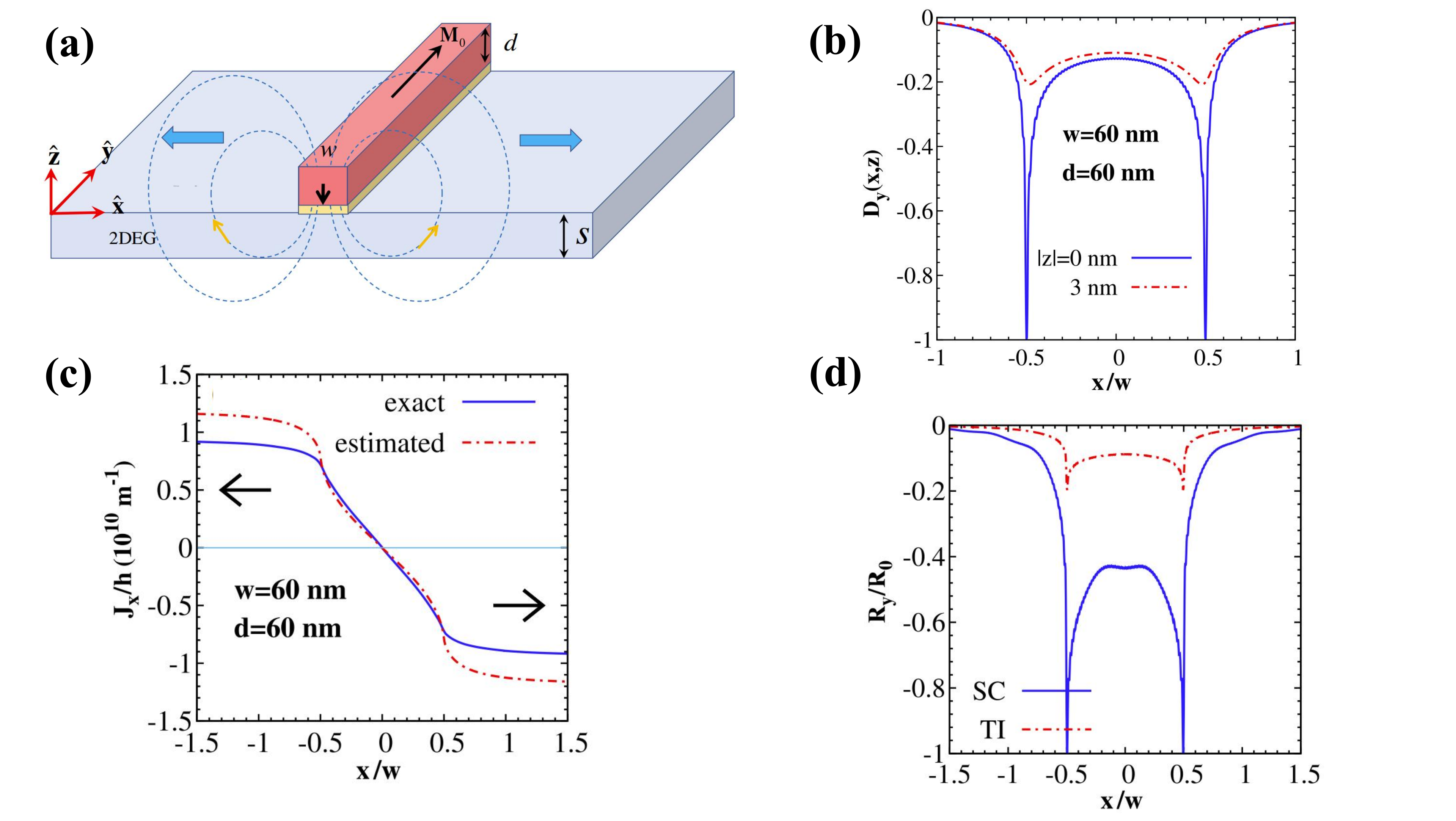}
	\par\end{centering}
	\caption{Near-field pumping of an electron spin current in a thin metal film by the stray field of precessing magnetization in a nearby wire. (a) shows the configuration with geometric parameters. (b) is the calculated DC component of the transverse spin generated by the stray field that pumps a constant spin current into both directions as in (c). The exact result is calculated by Eq.~(\ref{pumping_general}). The red dashed line labeled by ``estimated" assumes $k_x=\pi/(2w)$ in the spin susceptibility [Eq.~(\ref{pumping_simple})]. (d) shows the calculated spin injection rates into an electron gas with weak (SC) and strong (TI) spin-orbit couplings. The figures are taken from Ref.~\cite{electron_spin_Yu}.}
	\label{Evanescent_pumping}
\end{figure}

We illustrate the process for the configuration in Fig.~\ref{Evanescent_pumping}(a). Here we consider a magnetic nanowire of width $w$, thickness $d$, and equilibrium magnetization $\mathbf{M}_{s}$
pointing along the wire into the $\hat{\bf y}$-direction, deposited on top of a thin metal film with a thickness 
$s$. According to Eqs.~(\ref{near_dipolar_field}) and (\ref{transverse_spin_density}), the stray field emitted by the Kittel mode at the FMR frequency $\omega_{\mathrm{K}}$ \cite{Au_first,Dirk_transducer,Chuanpu_NC}, is circularly polarized in the $\hat{\bf x}$-$\hat{\bf z}$-plane (or spin along  $\hat{\bf y}$) near the edge of the magnetic wire. Figure~\ref{Evanescent_pumping}(b) shows the degree of transverse spin  of a magnetic wire with $d=w=60$~nm, as reviewed in Sec.~\ref{An_example}. In the following, we derive the spin current generated by this drive in a nonmagnetic electron gas  \cite{electron_spin_Yu}.

The theory in terms of the Green function is addressed in the effective one-dimensional case. The Zeeman interaction of an electron spin with the dynamic magnetic field $\mathbf{H}(x,t)$ 
   \begin{align}
   	\hat{H}_{\mathrm{Z}}\left(  t\right)  =\mu_{0}\gamma
   	_{e}\int\hat{\mathbf{s}}(x,t)\cdot\mathbf{H}(x,t)dx
   	\label{Zeeman_Hamiltonian}
   \end{align}
   perturbs the system Hamiltonian $\hat{H}_{0}=\hat{p}^2/(2m)$ of two-dimensional free electron gas and modifies the Heisenberg equation of motion for
   the electron spin density $\hat{\mathbf{s}}$ operator
   \begin{equation}
   	\frac{\partial\hat{\mathbf{s}}(x,t)}{\partial t}=\frac{i}{\hbar}\left[
   	\hat{H}_{0}+\hat{H}_{\mathrm{Z}},\hat{\mathbf{s}}(x,t)\right]  =\frac{i}%
   	{\hbar}\left[  \hat{H}_{0},\hat{\mathbf{s}}(x,t)\right]  -\mu_{0}\gamma
   	_{e}\hat{\mathbf{s}}(x,t)\times\mathbf{H}(x,t),
   \end{equation}
 where we use the commutator $[\hat{s}_{\beta}(x^{\prime},t),\hat{s}_{\alpha
   }(x,t)]=i\hbar\varepsilon_{\beta\alpha\delta}\hat{s}_{\delta}(x)\delta
   (x-x^{\prime})$. Here, $\gamma_{e}=-g_{e}\mu_{B}/\hbar$, while $\mu_{B}$ and $g_{e}$ are the Bohr magneton and  (effective) electron $g$-factor, respectively.
  We chose the Heisenberg representation in which the wave function does not depend on time, 
   \begin{equation}
   	\left\langle \frac{\partial\hat{\mathbf{s}}(x,t)}{\partial t}\right\rangle
   	=\frac{\partial\left\langle \hat{\mathbf{s}}(x,t)\right\rangle }{\partial t},
   \end{equation}
and the interaction representation for the perturbation \cite{Mahan,Vignale} with operators $\hat{A}%
   _{I}(t)=e^{iH_{0}t/\hbar}\hat{A}_{S}e^{-iH_{0}t/\hbar}$ evolving in time by $\hat{H}_{0}$. The spin-density rate
 of change then contains two terms%
   \begin{align}
   	\left\langle \frac{\partial\hat{\mathbf{s}}(x,t)}{\partial t}\right\rangle  &
   	=\frac{\partial\left\langle U^{\dagger}(t)\hat{\mathbf{s}}_{I}%
   		(x,t)U(t)\right\rangle }{\partial t}\nonumber\label{interpretation}\\
   	&  =\frac{i}{\hbar}\left\langle U^{\dagger}(t)\left[  \hat{H}_{\mathrm{Z}%
   	}(t),\hat{\mathbf{s}}_{I}(x,t)\right]  U(t)\right\rangle +\left\langle
   	U^{\dagger}(t)\frac{\partial\hat{\mathbf{s}}_{I}(x,t)}{\partial t}%
   	U(t)\right\rangle ,
   \end{align}
   where%
   \begin{equation}
   	U(t)=T_{t}\exp\left(  -\int_{-\infty}^{t}dt^{\prime}\hat{H}_{\mathrm{Z}}(t^{\prime
   	})\right)
   \end{equation}
   is the evolution operator governed by the perturbation, where \(T_t\) is the time-ordering operator.

   In the interaction representation the pump magnetic field does not appear in the
  operators, so for the free electron gas $\hat{H}_{0}^{I}=\hat{p}_{I}^{2}/(2m)$ and in the absence of spin decay processes we arrive at the spin-conservation that reads
   \begin{equation}
   	\frac{\partial\hat{\mathbf{s}}_{I}(x,t)}{\partial t}=\frac{i}{\hbar}[\hat
   	{H}_{0}^{I},\hat{\mathbf{s}}_{I}(x,t)]=-\frac{\partial}{\partial x}%
   	\hat{\pmb {\mathcal{J}}}_{I}(x,t), \label{local_spin}%
   \end{equation}
   where by the Fermion operator $\hat{f}_{\alpha}^{I}(x)$ (with ``hat" for not confusing with the distribution function) with spin $\alpha$ 
   \begin{equation}
   	\hat{\pmb {\mathcal{J}}}_{I}=\frac{\hbar^{2}}{4im}\hat{f}_{\alpha}^{I\dagger
   	}(x)\boldsymbol{\sigma}_{\alpha\alpha^{\prime}}\partial_{x}\hat{f}%
   	_{\alpha^{\prime}}^{I}(x)+\mathrm{H.c}. \label{spincurrent}%
   \end{equation}
   is the spin current density operator. The formal exact expectation value for the spin current is therefore%
   \begin{equation}
   	-\frac{\partial}{\partial x}{\pmb{ \mathcal{J}}}(x,t)=\frac{\partial
   		\left\langle U^{\dagger}(t)\hat{\mathbf{s}}_{I}(x,t)U(t)\right\rangle
   	}{\partial t}-\frac{i}{\hbar}\left\langle U^{\dagger}(t)\left[  \hat
   	{H}_{\mathrm{Z}}(t),\hat{\mathbf{s}}_{I}(x,t)\right]  U(t)\right\rangle .
   	\label{local_conservation}%
   \end{equation}
 Expanding the divergence of the spin current to
   first order in the microwave field $\mathbf{H}$:
   \begin{align}
   	\left\langle U^{\dagger}(t)\hat{s}_{\alpha}^{I}(x,t)U(t)\right\rangle  &
   	\rightarrow\left\langle \hat{s}_{\alpha}(x,t)\right\rangle _{l}=\frac{i}
   	{\hbar}\int_{0}^{t}dt^{\prime}\left\langle \left[  \hat{H}_{\mathrm{Z}}%
   	^{I}(t^{\prime}),\hat{s}_{\alpha}^{I}(x,t)\right]  \right\rangle \nonumber\\
   	&  =-\mu_{0}\gamma_{e}\int dx^{\prime}dt^{\prime}\chi_{\alpha\beta
   	}(x-x^{\prime},t-t^{\prime})H_{\beta}(x^{\prime},t^{\prime})+\mathcal{O}%
   	(\chi^{\left(  2\right)  }H^{2}),
   	\label{spin_density_excitation}
   \end{align}
    The first term is the spin susceptibility \cite{Mahan,Vignale}
   \begin{equation}
   	\chi_{\alpha\beta}(x-x^{\prime},t-t^{\prime})=i\Theta(t-t^{\prime
   	})\left\langle \left[  \hat{s}_{\alpha}^{I}(x,t),\hat{s}_{\beta}^{I}%
   	(x^{\prime},t^{\prime})\right]  \right\rangle 
   	\label{spin_susceptibility_2}
   \end{equation}
and we disregard the quadratic response function $\chi^{\left(  2\right)  }$. The second term in Eq. (\ref{local_conservation})
\begin{align}
   	&  -\frac{i}{\hbar}\left\langle U^{\dagger}(t)\left[  \hat{H}_{\mathrm{Z}%
   	}(t),\hat{s}_{\alpha}^{I}(x,t)\right]  U(t)\right\rangle \nonumber\\
   	&  \rightarrow\frac{\mu_{0}\gamma_{e}}{\hbar}i\varepsilon_{\beta\alpha\delta
   	}H_{\beta}(x,t)\int_{0}^{t}dt^{\prime}\left\langle \left[  \hat{H}%
   	_{\mathrm{Z}}(t^{\prime}),\hat{s}_{\delta}(x,t)\right]  \right\rangle =\mu
   	_{0}\gamma_{e}\varepsilon_{\beta\alpha\delta}H_{\beta}(x,t)\hat{s}_{\delta
   	}(x,t)|_{l}
   \end{align}
scales like $\sim\chi H^{2}$ because  the excited spin density ${\bf s}_l$ is proportional to the field as well  [Eq.~(\ref{spin_density_excitation})].  The spin current conservation law then reads 
   \begin{equation}
   	-\nabla\cdot{\pmb {\mathcal{J}}}(x,t)=\frac{\partial\left\langle
   		\mathbf{s}\right\rangle _{l}}{\partial t}+\mu_{0}\gamma_{e}\left\langle
   	\mathbf{s}\right\rangle _{l}\times\mathbf{H},
   	\label{spin_current_generation}
   \end{equation}
with two source terms on the right-hand side, \textit{i.e.}, a direct spin injection and its precession in the applied field, since the  dissipative component
   $\left\langle \mathbf{s}\right\rangle _{l}\nshortparallel\mathbf{H}$  feels an instantaneous torque.  The DC spin current generation is only contributed by the second nonlinear terms of ${\bf H}$.
   
   The spin-orbit coupling also introduces torques that can be included in the definition of the spin
   current \cite{torque_Niu} in order to enforce global spin conservation with spin injection rate 
   ${\pmb{ \mathcal{R}}}(t)=\left\langle U^{\dagger}(t)\partial_{t}%
   \hat{\mathbf{s}}_{I}(\boldsymbol{\rho},t)U(t)\right\rangle $ as the only source term.  The spin-injection rate beneath an evanescent field is a source term for the kinetic
spin Bloch equations \cite{Wu_review,Tao_diffusion}, from which we can calculate the spin transport with spin relaxation. We consider here two limiting cases of weak and strong spin-orbit coupling that correspond to spin current injection into a free two-dimensional electron gas and into the surface states of the topological insulator. The calculated results in Fig.~\ref{Evanescent_pumping}(c) and (d) reveal the high efficiency or the dipolar pumping process. The spin current flows with the same polarization and equal magnitude into opposite directions from the local source, as in the spin pumping by the exchange interaction.  This example proves that quite generally chiral excitation by dipolar radiation is not caused by a hidden symmetry but requires poles in the spin susceptibility generated by degenerate electron-hole pairs or the plasmon, magnon, and phonon excitations of a rigid ground state.

   \subsubsection{Conservation of photon transverse spin and electron spin}

  Here we address the conservation of the total transverse spin of the stray field and electron system, which is not obvious from inspection of the Hamiltonian Eq.~(\ref{Zeeman_Hamiltonian}) since the near-field spin pumping is intrinsically a nonlinear process in terms of quadratic ${\bf H}$. The free electron gas is a convenient system to address this conservation law. Its (Pauli) spin susceptibility is \cite{Mahan,Vignale},
  \begin{equation}
  	\chi(k,\omega)=\frac{\hbar^{2}}{2}\sum_{\mathbf{q}}\frac
  	{f(\xi_{\mathbf{q}})-f(\xi_{\mathbf{k}+\mathbf{q}})}{\hbar\omega+i0_{+}%
  		+\xi_{\mathbf{q}}-\xi_{\mathbf{k}+\mathbf{q}}}, \label{susceptibility_free}%
  \end{equation}
  where $\xi_{\mathbf{k}}=\hbar^{2}k^{2}/(2m^{\ast})-\mu$ is the electron energy
  with effective mass $m^{\ast},$ relative to the chemical potential $\mu$, and
  $f(\xi_{\mathbf{k}})=\{\exp[\xi_{\mathbf{k}}/(k_{B}T)]+1\}^{-1}$ is the
  Fermi-Dirac distribution at temperature $T$. In the microwave regime and not too narrow magnetic wires, $\left\vert k_{x}\right\vert <2k_{F},$ where $k_{F}$ is the 2DEG
  Fermi vector [in semiconductors $k_{F}=O\left(  \mathrm{nm}^{-1}\right)  $]
  and $\mathrm{Re}\chi(|k_{x}|,\omega\rightarrow0)=m^{\ast}/\left(  \pi\hbar
  ^{2}\right) $. The DC spin
  current derived above then reads \cite{electron_spin_Yu} 
  \begin{align}
  	{\pmb {\cal J}}_{x}^{\mathrm{DC}}(x)  &  =(\mu_{0}\gamma_{e})^{2}\int_{0}%
  	^{x}dx\sum_{k_{x}}e^{ik_{x}x}\left.  \partial_{\omega}%
  	\mathrm{\operatorname{Im}}\chi(|k_{x}|,\omega)\right\vert _{\omega
  		=0}\left\langle \dot{\mathbf{H}}(k_{x},t)\times\mathbf{H}%
  	(x,t)\right\rangle _{\mathrm{DC}},\label{pumping_general}%
  \end{align}
where we used the symmetry relations ${\pmb {\cal J}}_{x}^{\mathrm{D}C}%
  (k_{x})=-{\pmb {\cal J}}_{x}^{\mathrm{DC}}(-k_{x})$ in terms of the stray field Eq.~(\ref{near_dipolar_field}) of the Kittel mode of a magnetic nanowire and ${\pmb {\cal J}}%
  _{x}^{\mathrm{DC}}(x=0)=0$. In the low frequency limit  $\chi(|k_{x}|,\omega)\approx\chi(k_{\mathrm{ave}},\omega)$ with $k_{\mathrm{ave}}\sim\pi/(2w)$
  \begin{align}
  	{\pmb {\cal J}_{x}^{\mathrm{DC}}(x)}  &  \approx\left.  -2(\mu_{0}\gamma
  	_{e})^{2}\omega_{\mathrm{K}}\partial_{\omega}\mathrm{\operatorname{Im}}%
  	\chi\left(  |k_{x}|\rightarrow k_{\mathrm{ave}},\omega\right)  \right\vert
  	_{\omega=0}\int_{0}^{x}dx^{\prime}\mathrm{\operatorname{Im}}\left[
  	\tilde{\mathbf{H}}^{\ast}(x^{\prime})\times\tilde{\mathbf{H}}(x^{\prime
  	})\right]. \label{pumping_simple}%
  \end{align}
  Hence, the DC spin-current below the nanowire is (approximately) proportional
  to the transverse spin of the magnetic field, implying the transfer of the photon
  spin angular momentum to the electron spin with an efficiency governed by
  $\left.  \partial_{\omega}\mathrm{\operatorname{Im}}\chi\left(  |k_{x}%
  |\rightarrow k_{\mathrm{ave}},\omega\right)  \right\vert _{\omega=0}$. The
  spin current is polarized in the $-\hat{\bf y}$-direction, \textit{i.e.}, opposite to the
  magnetization direction of the nanowire.
  This implies the transfer of transverse spin of near fields to the electron spins, \textit{i.e.}, a conservation law.
  
  The experimental demonstration of the transfer of the transverse spin of the evanescent magnetic field to the electron spin needs a measurement of the transverse spin current of electrons. The future theoretical study should address parameter dependence such as the dependence of spin current flows on the magnetization direction. The numerical calculation predicted a transverse spin current density $10^{-13}~{\rm J/m}$ pumped into an InAs 2DEG \cite{electron_spin_Yu}, which is almost 2 orders of magnitude larger than what has been observed for the spin pumping by a Py slab into the graphene \cite{pumping_graphene}. We note that its signal should indeed be much smaller with a small $g$ factor. We therefore cannot exclude that the observations are caused by dipolar fields at the edge of Py and not exchange interactions at the interface.

   \subsubsection{Chiral spin pumping into carbon nanotube and Luttinger liquid}
   
  The near-field spin pumping turns out to be sensitive to the electron dispersion \cite{electron_spin_Yu}. The spin pumping current is chiral when the spin susceptibility displays
singularities that indicate collective states (\textit{e.g.}, magnons in Sec.~\ref{dipolar_pumping}). However, 1D systems with linear dispersion at the Fermi energy,
such as metallic carbon nanotubes, are an exception since spin pumping is chiral even without interactions.

  \textbf{Carbon nanotube}.--- Three-dimensional conventional conductors behave like non-interaction Fermi liquids, but when they are spatially confined, the electron-electron interaction can become important.  Here we illustrate the chiral spin pumping \cite{Chiral_pumping_Yu} into a correlated system, \textit{viz.}, a one-dimensional electron gas in a metallic nanotube with linear dispersion  (along the $\hat{\bf x}$-direction) perpendicular to a close-by magnetic nanowire (along the $\hat{\bf y}$-direction). We showed above that the spin density pumped by the Kittel magnon with frequency $\omega_{\mathrm{K}}$ can be written 
   \begin{equation}
   	\langle\hat{s}_{\alpha}(x,t)\rangle_{l}=-\mu_{0}\gamma_{e}\sum_{k}%
   	e^{ikx-i\omega_{\mathrm{K}}t}\chi(k,\omega_{\mathrm{K}})H_{\alpha}%
   	(k,\omega_{\mathrm{K}})+\mathrm{H.c.}, \label{excitation_linear}%
   \end{equation}
   where $H_{\alpha}(\left\vert k\right\vert ,\omega_{\mathrm{K}})=0$ and  $H_{\alpha}(-\left\vert k\right\vert ,\omega_{\mathrm{K}})\neq0$ [Eq.~(\ref{near_dipolar_field})] are chiral in
   momentum space \cite{Chiral_pumping_Yu,Chiral_pumping_grating}.
   The spin susceptibility  Eq. (\ref{susceptibility_free})   in one dimension is
   \begin{equation}
   	\chi_{F}(k,\omega)=\frac{\hbar^{2}}{2}\sum_{\tau}\sum_{q}\frac{n_{F}%
   		(\xi_{q,\tau})-n_{F}(\xi_{k+q,\tau})}{\hbar\omega+i0_{+}+\xi_{q,\tau}%
   		-\xi_{k+q,\tau}}, \label{ChiF}
   \end{equation}
   where $\tau$ labels the valley degree of freedom. For $n$-doped metallic carbon
   nanotube the dispersion close to the neutrality poin $\xi_{q,\tau}=\hbar v_{F}\left\vert q\right\vert -\mu$  is linear, where
   $v_{F}$ is the Fermi velocity. We are interested in the long
   wavelength response with $\left\vert k\right\vert \ll\mu/(\hbar v_{F})=k_{F}$,
   so $k+q$ is on the same branch as $q$ and
   \begin{equation}
   	\xi_{k+q,\tau}=\hbar v_{F}\left(  \left\vert q\right\vert +k\mathrm{sgn}%
   	(q)\right)  -\mu.
   \end{equation}
   A low temperatures $n_{F}(\xi_{q})=1-\Theta(\xi_{q})$ and
   \begin{align}
   	\chi_{F}(k,\omega)  &  =\frac{\hbar^{2}}{2}\sum_{\tau}\left(  \frac{\sum
   		_{q>0}\left(  \Theta(\xi_{k+q})-\Theta(\xi_{q})\right)  }{\hbar\omega
   		+i0_{+}-\hbar v_{F}k}+\frac{\sum_{q<0}\left(  \Theta(\xi_{k+q})-\Theta(\xi
   		_{q})\right)  }{\hbar\omega+i0_{+}+\hbar v_{F}k}\right) \nonumber\\
   	&  =\frac{\hbar kL}{2\pi}\left(  \frac{1}{\omega+i0_{+}- v_{F}k}+\frac
   	{1}{\omega+i0_{+}+ v_{F}k}\right)  , \label{carbon_nanotube}%
   \end{align}
where the valley degree of freedom contributes a factor of 2. The response
 becomes singular because the linear dispersion renders the excited electron
hole pairs degenerate. For positive frequency $\omega$, the two poles $k_{\pm
   }=\pm(\omega/(v_{F})+i0_{+})$ are in the upper and lower complex plane, respectively. When $x>0$ we close the contour in the upper complex plane, which leads to a vanishing spin density, while for $x<0$
   \begin{equation}
   	\langle\hat{s}_{\alpha}(x,t)\rangle_{l}=i\frac{\mu_{0}\gamma_{e}%
   		\omega_{\mathrm{K}}}{v_{F}^{2}}H_{\alpha}(k_{-},\omega_{\mathrm{K}}%
   	)e^{ik_{-}x-i\omega_{\mathrm{K}}t}+\mathrm{H.c.},~~~~x<0. \label{excited_spin}%
   \end{equation}
   Therefore, in contrast to the results for conventional conductors, the excited spin density lives in only half of the nanowire! Consequently, the DC  spin current
   \begin{align}
   	{\pmb {\mathcal {J}}}_{x}^{\mathrm{DC}}(x)  &  ={\pmb {\mathcal {J}}}%
   	_{x}^{\mathrm{DC}}\left(  x=\frac{w}{2}\right)  +\left\{
   	\begin{array}
   		[c]{c}%
   		-\mu_{0}\gamma_{e}\int_{\frac{w}{2}}^{x}dx^{\prime}\left.  \mathbf{s}%
   		_{l}(x^{\prime},t)\times\mathbf{H}(x^{\prime},t)\right\vert _{\mathrm{DC}}\\
   		0
   	\end{array}
   	~~\text{for }%
   	\begin{array}
   		[c]{c}%
   		x<0\\
   		x>0
   	\end{array}
   	\right. \nonumber\\
   	&  \rightarrow\left\{
   	\begin{array}
   		[c]{c}%
   		-\mu_{0}\gamma_{e}\int_{0}^{x}dx^{\prime}\left.  \mathbf{s}_{l}(x^{\prime
   		},t)\times\mathbf{H}(x^{\prime},t)\right\vert _{\mathrm{DC}}\\
   		0
   	\end{array}
   	~~\text{for }%
   	\begin{array}
   		[c]{c}%
   		x<0\\
   		x>0
   	\end{array}
   	\right.
   \end{align}
   flows only in the same half-space because $\pmb {\mathcal {J}}_{x}%
   ^{\mathrm{DC}}\left(  x=\frac{w}{2}\right)  \rightarrow0$ as we show now. The
   Fourier components of the spin current are
   \begin{equation}
   	{\pmb {\mathcal{J}}}_{x}(k)=-\frac{\mu_{0}\gamma_{e}}{ik}\sum_{k^{\prime}%
   	}\mathbf{s}_{l}(k^{\prime},t)\times\mathbf{H}(k-k^{\prime},t).
   \end{equation}
   Since the poles of $s_{l}(k,t)$ lie in the lower complex plane and for
   $x=w/2>0$,
   \begin{align}
   	{\pmb {\mathcal{J}}}_{x}\left(  x=\frac{w}{2},t\right)   &  =\sum_{k}%
   	e^{ikx}{\pmb{\mathcal{J}}}_{x}\left(  k,t\right)  =i\mu_{0}\gamma_{e}\sum_{k}\sum_{k^{\prime}}e^{i(k+k^{\prime})x}\frac
   	{1}{k+i0_{+}+k^{\prime}}\mathbf{s}_{l}(k^{\prime},t)\times\mathbf{H}(k,t)=0
   \end{align}
   by closing the integral path of $k^{\prime}$ in the upper complex plane. The mathematical origin for the chiral behavior are the singularities in the susceptibilities
   Eq.~(\ref{carbon_nanotube}) at low frequencies that are absent in higher dimensions.
   
\textbf{Luttinger liquid}.---Similar singularities in the response functions  render the instability of the ground state of the 1D electron gas when the interaction is switched on in favor of the Tomonaga-Luttinger liquid, but the results above still hold. The Hamiltonian in the presence of interactions becomes
   \cite{Mahan,Vignale}
   \begin{equation}
   	\hat{H}_{0}=\hbar v_{F}\sum_{k\sigma}\left\vert k\right\vert \hat{f}_{k\sigma
   	}^{\dagger}\hat{f}_{k\sigma}+\frac{1}{2L}\sum_{k}V_{k}\hat{\rho}(k)\hat{\rho
   	}(-k),
   \end{equation}
   where $V_{k}$ is the Fourier component of Coulomb potential in (quasi) one dimension,
   and the Fourier component of the density operator reads
   \begin{equation}
   	\hat{\rho}(k)=\sum_{q\sigma}f_{q-k/2,\sigma}^{\dagger}\hat{f}_{q+k/2,\sigma}.
   \end{equation}
   This model supports two kinds of Bosonic excitations \cite{Mahan}, \textit{i.e.}, charge
   and spin density waves with different group velocities. The spin density excitations have  a dispersion $\omega_{k}=\hbar
   v_{F}\left\vert k\right\vert $ and susceptibility (cf. Chap.~4 of Ref.~\cite{Mahan}) 
   \begin{equation}
   	\chi(k,\omega)=\frac{\hbar\left\vert k\right\vert L}{4\pi}\left(
   	\frac{1}{\omega-\omega
   		_{k}+i0_{+}}-\frac{1}{\omega+\omega_{k}+i0_{+}}\right)  .\label{susint}%
   \end{equation}
Only the first term in  Eq.~(\ref{susint}) contributes to the resonant response to the external
   microwaves with a positive $\omega,$ leading to  exactly the same response as
   the non-interacting case Eq.~(\ref{carbon_nanotube}) except for a factor of 2 for the valley degree of freedom. The Coulomb interaction dramatically modifies the ground state, but the spin-excitation channel is not affected by the spin pumping, so the chirality remains intact. 
   
   We conclude that chirality requires a singular susceptibility, which in
   general requires a collective excitation that is indicative of rigidity in the ground state. The
  one-dimensional system illustrates that this is not the same as long-range order. The long-range correlation in the Tomonaga-Luttinger liquid caused by the linear (or linearized) dispersion at the Fermi energy in 1D systems is sufficient to cause chiral spin pumping.

   \subsection{Surface plasmon polariton and electron spin}
  \label{SPP_spin_pumping}
 Metallic nanostructures are efficient antennas for high-frequency electromagnetic fields by virtue of their surface plasmons.  The plasmon polaritons confine light far beyond the wave length and thereby strongly enhance field amplitudes. Magneto-plasmonics addresses, for example, amplified magneto-optical effects \cite{nonreciprocity_plasma,SPP_magneto_optics}, surface plasmon resonance in ferromagnetic metals \cite{SPP_ferromagnet}, and the integration of plasmonics and spintronics \cite{plasmonics_spin_APL,plasmonics_spin_NC,plasmonics_spin_NJP,plasmonics_spin_PRB,plasmonics_spin_PRL}. Before addressing the spin-current generation by surface plasmon polariton, we address here  the inverse Faraday effect, \textit{i.e.}, the dynamics generation of magnetization by light.

 \subsubsection{Inverse Faraday effect in metals}
  In magneto-optics, the Faraday (Kerr) effect refers to the rotation of a linearly-polarized light plane  by a magnetic field or moment in transmission (reflection) \cite{Landau}. Pitaevskii predicted
\cite{inverse_Faraday_Pitaevskii} an inverse effect,
  	\textit{viz.}, the generation of a magnetic moment by circularly polarized light,
by the transfer of photon angular momentum to that of the electrons by a Raman scattering process.  Here we follow the derivation by Hertel \cite{inverse_Faraday_Hertel}  which sketches  a  simple physical picture, based on the charge conservation in an electron gas with density $n({\bf r},t)$ and velocity field ${\bf v}({\bf r},t)$ 
    \begin{align}
    	\partial n/\partial t+\nabla(n{\bf v})=0,
    	\label{continuity}
    \end{align}
and Ohm's Law 
\begin{align}
	{\bf j}\equiv en{\bf v}=\sigma {\bf E},
	\label{current}
\end{align}
where  $\sigma=ine^2/(m\omega)$ is the reactive dynamical conductivity of the free electron gas.
The electron gas responds to an electric field ${\bf E}({\bf r},t)=\pmb{\cal E}e^{i({\bf k}\cdot{\bf r}-\omega t)}$ dc and ac components:
\begin{subequations}
\begin{align}
	\label{density}
	n({\bf r},t)&\rightarrow \langle n({\bf r},t)\rangle +\delta n({\bf r},t),\\
	{\bf v}({\bf r},t)&\rightarrow \langle {\bf v}({\bf r},t)\rangle +\delta {\bf v}({\bf r},t).
	\label{velocity}
\end{align}
\end{subequations}
The dc response is a non-linear down-conversion process that follows from
substituting Eqs.~(\ref{density}) and (\ref{velocity}) into Eqs.~(\ref{continuity}) and (\ref{current}), \textit{i.e.},
\begin{subequations}
  \begin{align}
  	\delta n&=-\frac{i}{\omega e}\nabla(\sigma{\bf E}),\\
  	\delta{\bf v}&=\frac{\sigma}{\langle n \rangle e}{\bf E},
  \end{align}
  \end{subequations}
leading to
\begin{align}
	{\bf j}({\bf r},t)=\langle \delta n({\bf r},t)\delta {\bf v}({\bf r},t)\rangle \rightarrow -\frac{i}{4e\langle n\rangle \omega}\nabla\times (\sigma^*{\bf E}^*\times \sigma{\bf E}).
\end{align}
According to Maxwell's equation the current density ${\bf j}=\nabla \times {\bf M}$, which implies an induced orbital magnetization
\begin{align}
	{\bf M}=-\frac{|e|\varepsilon_0 \omega_{\rm P}^2}{4m \omega^3}{\rm Im}({\bf E}^*\times{\bf E}).
	\label{orbital_1}
\end{align}

This result can be derived from the angular momentum of electromagnetic waves \cite{Nori_angular_momentum} [see Eq.~(\ref{transverse_spin_EM}) and below]. It can be generalized to include dissipative media \cite{inverse_Faraday_quantum}. 
 In magnetic semiconductors,  off-resonant polarized light can shift the band edges by the spin-selective ac Stark effect, which is equivalent to  a strong effective magnetic field \cite{Inverse_Faraday_Brataas}. The orbital magnetization induced by the electric components of circularly polarized light is a non-linear and reactive effect that is believed to be a mechanism for ultrafast all-optical magnetization dynamics  \cite{ultrafast_1,ultrafast_2,ultrafast_RMP}. In the full theoretical treatment of light-induced magnetization reversal by the inverse Faraday effect, the spin-orbit interaction including spin-relaxation should be included.
 
\subsubsection{Spin current generation by surface plasmon polariton}
  \label{Sec_SPP_pumping}

\begin{figure}[ht]
	\begin{centering}
	\includegraphics[width=1\textwidth]{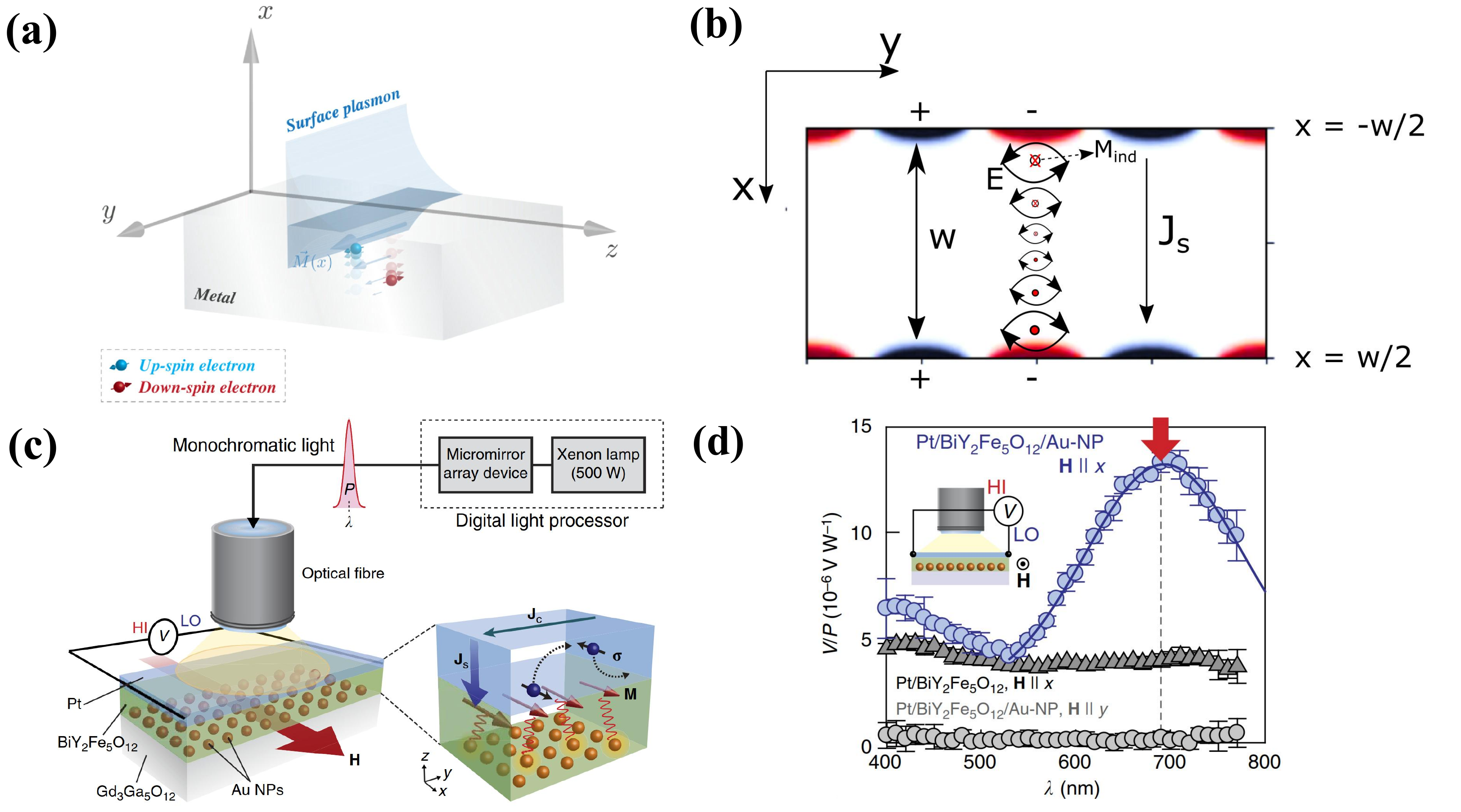}
	\par\end{centering}
	\caption{Generation of spin currents by surface plasmons. (a) A surface plasmon induces a constant magnetization profile whose gradient drives a spin current in the metal. (b)The edge plasmon of a graphene ribbon is the one-dimensional equivalent of the surface plasmon. It propagates along the edges and excites an out-of-plane spin polarization and current between the edges. (c) An experimental setup for a Pt|YIG  bilayer with embedded  Au nanoparticles. The spin pumping is found to be enhanced at the surface plasmon resonance in (d), providing evidence for the spin-current generation by surface plasmon resonance. The figures are taken from Refs.~\cite{plasmonics_spin_PRB} [(a)], \cite{plasmonics_spin_OOP} [(b)], and \cite{plasmonics_spin_NC} [(c) and (d)].}
	\label{SPP_pumping_figure}
\end{figure}

We now address the close connection of the inverse Faraday effect and the transverse spin of electromagnetic waves. The surface plasmon polariton, much like the surface acoustic wave, is an evanescent mode that can carry angular momentum (refer to Sec.~\ref{unification}). 
  Oue and Matsuo \cite{plasmonics_spin_PRB,plasmonics_spin_NJP} predicted angular momentum transfer from optical transverse spins captured by the surface plasmon polaritons to the orbital motion of conduction electrons, leading to a steady-state magnetization and electron spin current. When the magnetic field is linearly polarized, only the electric field contributes to the transverse spin angular momentum (SI unit)
  	\begin{align}
  		\nonumber
  		{\bf S}&=\frac{1}{4\omega}{\rm Im}\left(\varepsilon_0\frac{d(\omega \varepsilon_r)}{d\omega}{\bf E}^*\times {\bf E}+\mu_0\frac{d(\omega\mu_r)}{d\omega}{\bf H}^*\times {\bf H}\right)\\
  		&=\left( \frac{\varepsilon_0\varepsilon_r}{4\omega}+\frac{\varepsilon_0}{4}\frac{d\varepsilon_r}{d\omega}\right){\rm Im}\, {\bf E}^*\times {\bf E} \equiv  {\bf S}_{\rm EM}+{\bf S}_m.
  \end{align}
The first term  in the second line is purely electromagnetic, while the second term invokes the conduction electron dynamics that affects the dielectric constant. With gyromagnetic ratio $-|e|/(2m)$ and $\varepsilon_r=1-\omega_{\rm P}^2/\omega^2$, 
\begin{align}
	{\bf M}=-|e|/(2m){\bf S}_m
	=-\frac{|e|\varepsilon_0 \omega_{\rm P}^2}{4m \omega^3}{\rm Im}({\bf E}^*\times{\bf E}),
\end{align}
which agrees with Eq.~(\ref{orbital_1}). When the surface plasmon polariton propagates along the $\hat{\bf x}$-direction  as in Fig.~\ref{SPP_pumping_figure}(a),  the rotating electric field lies in the $x$-$z$ plane and generates an in-plane magnetization along $\hat{\bf y}$ that decays from the surface on the length scale $\lambda_{\rm SP}$, as also addressed in Sec.~\ref{Electric_field}.  

The magnetization is generated by a steady state orbital motion of the electrons and its stray field equals that of their Oersted field. Its gradient may drive an angular momentum current  $-(\hbar\sigma_0/m)\nabla M_y$ in a diffuse metal \cite{plasmonics_spin_PRB,plasmonics_spin_NJP},
with conductivity $\sigma_0$.
Oue \textit{et al.} \cite{plasmonics_spin_PRB,plasmonics_spin_NJP} treated the gradient as a source term in the spin diffusion equation. In terms of the spin accumulation $\delta\mu$ \cite{spin_diffusion_equation}
\begin{align}
\nabla^2\delta\mu=\frac{\delta\mu}{\lambda_s^2}+\frac{\hbar e}{m}\nabla^2M_y,
\end{align}
where $\lambda_s$ is the spin diffusion length. Here the connection between the magnetization generated by the rotating electric field couples to the electron spin by spin-flip relaxation that equilibrates the system in the presence of the Zeeman field in favor of a net spin polarization ~\cite{plasmonics_spin_PRB,plasmonics_spin_NJP}.

However, the gradient of a magnetic field should make a different effect from that of an electric field, with the latter drifting the system that is balanced by the momentum scattering. It is not really possible to bias the spin system by the gradient of magnetic field \textit{in the steady state} since the spin flip quickly balances the spin-polarization gradient. Future studies should also address how the electron spins couple to the orbital magnetization that is induced optically.

The diffusive DC spin current driven by this spin 
accumulation is then
\begin{align}
    {\bf J}_s=\frac{\sigma_0}{e}\nabla\delta\mu=-\frac{2(2\lambda_s/\lambda_{\rm SP})^2}{(2\lambda_s/\lambda_{\rm SP})^2-1}\frac{\hbar\sigma_0}{m}\nabla M_y.
\end{align}
The estimated magnitude of the spin current may achieve $10^5~{\rm A/m^2}$ \cite{plasmonics_spin_PRB}.

In Fig.~\ref{SPP_pumping_figure}(a) the excited transverse magnetization and spin polarization lie in the plane of the surface, while the spin current flows perpendicular to it, which renders it difficult to detect. Ukhtary \textit{et al}. \cite{plasmonics_spin_OOP}  proposed to employ the one-dimensional plasmons that propagate along the edges of graphene ribbons \cite{edge_plasmon}, that generate an out-of-plane spin polarization and current that flows between the edges as shown in Fig.~\ref{SPP_pumping_figure}(b).  The induced magnetization $\mu_0M_z(x)/d$, where  $d$ is the graphene thickness,  polarizes the spin of the electron in the direction of the magnetization. The spin accumulation is proportional to the difference between spin-up $n_{\uparrow}$ and spin-down $n_{\downarrow}$ electron densities \cite{spin_diffusion_equation}
\begin{align}
    \delta\mu_{\rm sou}(x)=\frac{2}{N_0}\left(n_{\uparrow}(x)-n_{\downarrow}(x)\right)=\frac{2\mu_0\mu_B}{d}M_z(x),
\end{align}
where $N_0$ is the density of states at the Fermi energy. The out-of-plane magnetization  drives a spin current between the edges.

\textbf{Experiments}.---References \cite{plasmonics_spin_NC,plasmonics_spin_PRL} reported experimental observation of spin current generation by surface plasmon resonance. The experiment \cite{plasmonics_spin_NC} employs Au nanoparticles embedded in films composed by the paramagnetic Pt and ferrimagnetic BiY$_2$Fe$_5$O$_{12}$ layers, as illustrated in Fig.~\ref{SPP_pumping_figure}(c), illuminated by unpolarized monochromatic light with wavelength $400$-$770$~nm. Surface plasmon resonance happens at light wavelength $690$~nm. The voltage at the two ends of the Pt layer comes from the inverse spin Hall effect and is measured in device Fig.~\ref{SPP_pumping_figure}(c). The voltage is found to be enhanced when illuminated by light at wavelength $690$~nm, as shown by the peak of the blue solid curve in Fig.~\ref{SPP_pumping_figure}(d). This enhancement of spin pumping was attributed to the excitation of magnons in BiY$_2$Fe$_5$O$_{12}$ by the near field of surface plasmon polaritons \cite{plasmonics_spin_NC}. The spin generation may be also explained by the spin Seebeck effect by enhanced heating at the resonance of surface plasmon polariton. However, recent measurements in Ref.~\cite{plasmonics_spin_PRL} reported the direct spin-current generation by surface plasmon resonance in the absence of the magnons.

 \subsection{Ac spin current generation by surface acoustic waves}
 
We derived in  Sec.~\ref{phonon_angular_momentum} that the volume elements in surface acoustic waves (SAWs) carry out a circular motion. This transverse phonon spin is locked to its momentum and can drive electronic spin dynamics \cite{spin_rotation_1,spin_rotation_2,spin_rotation_3,spin_rotation_Matsuo}, as briefly reviewed in this subsection.
  
 The close relation between electron angular momentum and mechanical angular momentum was revealed by Einstein and de Haas in 1915 by measuring the mechanical torque generated when reversing the magnetization of an iron cylinder. The reciprocal effect, \textit{i.e.}, the induced magnetization by mechanical rotation, was discovered by Barnett \cite{Barnett_1,Barnett_2}. A rigid rotation with angular velocity $\pmb{\omega}$ that in a rotating frame corresponds to an effective Barnett field $\pmb{\cal B}=\pmb{\omega}/\gamma$ that affects the magnetization by the Zeeman interaction $-{\bf M}\cdot\pmb{\cal B}$.  
 The Barnett field is thus not a real field, but a gauge field that arises when transforming to the rotation frame. It has no effect unless there is relaxation to the ground state. The Barnett effect can be employed to manipulate the magnetization of thin films and nanostructures \cite{Barnett_field_APL}.

\begin{figure}[ptbh]
	\begin{centering}
	\includegraphics[width=1.0\textwidth]{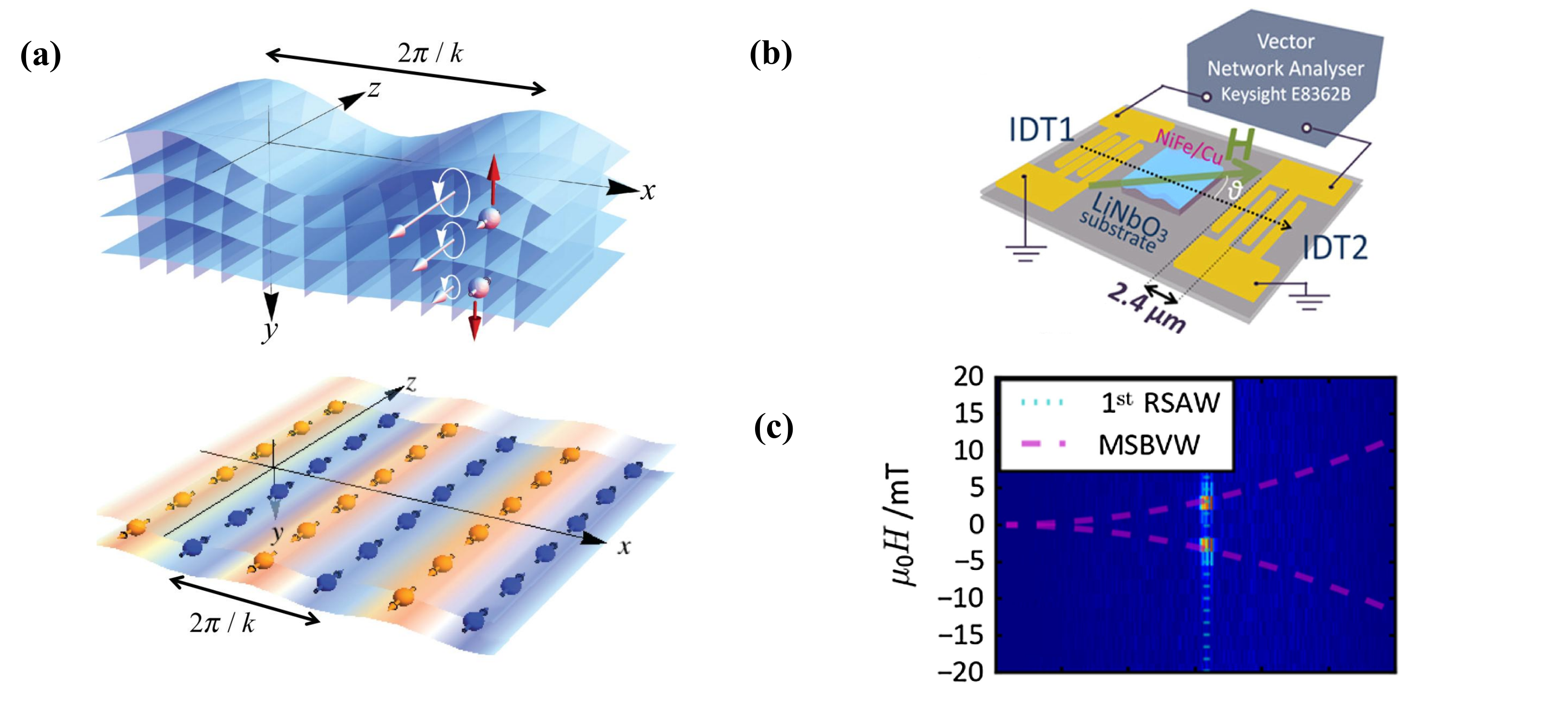}
	\par\end{centering}
	\caption{Generation of alternating spin currents in metal films by surface acoustic waves (SAWs). (a) sketches the concept that the Barnett spin-rotation coupling  generates an alternating spin current that can be measured in the configuration in panel (b) with a NiFe|Cu bilayer on top of a LiNbO$_3$ substrate. Here the IDTs excite and detect the SAWs.  (c) shows the absorption power of the ferromagnet as a function of an applied magnetic field and the SAW  frequency. The figures are taken from Refs.~\cite{mechanical_generation_1} [(a)], and \cite{mechanical_generation_exp} [(b) and (c)].}
	\label{Spin_rotation}
\end{figure}

Matsuo {\it et al}. studied the mechanical generation of ac spin currents in non-magnets by mechanical rotation and elastic deformations \cite{mechanical_generation_1,mechanical_generation_2,mechanical_generation_3,mechanical_generation_4,mechanical_generation_5} starting from the relativistic Dirac equation. The electron spin ${\bf s}$ couples with the Barnett-field density $\pmb{\cal B}({\bf r},t)$ via the spin-rotation coupling \cite{spin_rotation_1,spin_rotation_2,spin_rotation_3,spin_rotation_Matsuo} 
    \begin{align}
    	\hat{H}_{\rm SR}=-\int {\bf s}({\bf r})\cdot \pmb{\cal B}({\bf r},t) d{\bf r}.
\end{align}
A time-dependent effective magnetic field may thus parametrically pump an electron spin current. In the coordinate system Fig.~\ref{Spin_rotation}(a), the displacement field of an SAW  eigenmodes Eq.~(\ref{eqn:SAW_profile})  reads
 \begin{subequations}
 \begin{align}
     u_x&=i\frac{s}{|k|} u_0\left(\frac{2k^2}{k^2+s^2}e^{-qy}-e^{-sy}\right)e^{ikx-i\omega_k t},\\
     u_y&=-{\rm sgn}(k)u_0\left(\frac{2qs}{k^2+s^2}e^{-qy}-e^{-sy}\right)e^{ikx-i\omega_k t},
     \end{align}
     \end{subequations}
 where $u_0$ is the mode amplitude.
 The vorticity of fluid induces the spin-vorticiy coupling in metallic fluids such as mercury \cite{Mercury}. The rotational motion corresponds to an effective Barnett gauge field \cite{mechanical_generation_1,mechanical_generation_2,mechanical_generation_3,mechanical_generation_4,mechanical_generation_5}
 \begin{align}
     \pmb{\cal B}({\bf r},t)=\frac{1}{2}\nabla\times \dot{\bf u}=\frac{1}{2}\omega_k\eta^2|k|u_0e^{-sy}e^{ikx-i\omega_kt}\hat{\bf z}=\frac{1}{2c_t}\omega^2_ku_0e^{-sy}e^{ikx-i\omega_kt}\hat{\bf z},
     \label{Barnett_field}
 \end{align}
which is linearly polarized along the $\hat{\bf z}$-direction  and not locked to the propagation direction. 
In linear response, the induced spin polarization ${\bf s}_l({\bf r},t)$ of a free electron gas is parallel to the Barnett field which does not provide a torque on ${\bf s}_l({\bf r},t)$ or pump a dc spin current. On the other hand, the rate of change $\partial{\bf s}_l({\bf r},t)/\partial t$ is a source for an ac spin current. SAW phonon therefore cannot transfer their spins to conduction electrons, however, different from electromagnetic waves.

Matsuo {\it et al}. solved the spin diffusion equation for the spin accumulation $\delta\mu$ driven by the time-dependent Barnett gauge field Eq.~(\ref{Barnett_field}) \cite{mechanical_generation_1,mechanical_generation_2,spin_diffusion_equation}
\begin{align}
	\left(\frac{\partial }{\partial t}-D\nabla^2+\frac{1}{\tau_{\rm sf}}\right)\delta\mu=\hbar\frac{\partial {\cal B}}{\partial t}, 
\end{align} 
where $D$ is the spin diffusion constant and $\tau_{\rm sf}$ is the spin-flip time. According to Ohm's Law,  the ac spin current \(J_s\) is  polarized along the $\hat{\bf z}$-direction and proportional to the conductivity $\sigma_0$. In the strong scattering regime $\omega_k\tau_{\rm sf}\ll 1$ 
\begin{align}
	J_s=({\sigma_0}/{e})\nabla\delta\mu\approx \frac{\hbar \sigma_0}{2e}\tau_{\rm sf}\omega_k^4\frac{u_0}{c_t^2}\frac{\sqrt{1-\eta^2}}{\eta}e^{-sy}e^{ikx-i\omega_k t}.
\end{align}
It follows the spatiotemporal SAW profile, being a propagating wave in the plane and evanescent normal to it. The ac spin current is proportional to the spin-flip time, which is usually short for metals that have a large spin Hall angle. This hinders its observation 
by the inverse spin Hall effect that transforms it into an easily detectable transverse emf \cite{spin_Hall_effect,spin_Hall_1971,spin_Hall_Hirsh}, which can also detect an alternating spin current \cite{spin_back_flow,spin_back_flow_exp_1,spin_back_flow_exp_2}.

\textbf{Experiments}.---Instead, Kobayashi \textit{et al.} detected the ac spin current  via the spin transfer torque effect on a ferromagnetic contact \cite{mechanical_generation_exp}, as shown in Fig.~\ref{Spin_rotation}.  Here a NiFe|Cu bilayer is fabricated on a LiNbO$_3$ substrate in which SAWs are sent from IDT 1 to IDT 2. Matching the FMR and   SAWs frequencies opens a resonant dissipation channel between the phonons and the magnetic order that damps the SAW transmission amplitude. Figure~\ref{Spin_rotation}(c) shows the deduced absorption power of the ferromagnet as a function of frequency $f$, for a magnetic field parallel to the SAW propagation direction. The observed resonant absorption  is suppressed when rotating the field or inserting an insulating SiO$_2$ film, providing additional evidence that the Barnett effect can generate a spin current.

However, a direct magneto-mechanical effect cannot yet be excluded with certainty, since very similar effects might exists without the Cu spacer layer, as summarized in Table~\ref{table_chiral_SAW}. 
These experiments discovered non-reciprocal propagation of SAWs rendered by  magnetic cappings, which can be explained by magnetostriction.

\section{Summary and outlook}
\label{outlook}

To summarize, we have reviewed the universal physical mechanisms of the locking of the transverse spins of evanescent waves (or quasiparticles in their quanta) in spintronic nanostructures to their momentum and surface normal of their propagation plane. We have addressed the consequences of this generalized spin-orbit interaction that have been theoretically and experimentally explored in spintronics, magnetism, and magnonics.

\begin{figure}[ptbh]
	\begin{centering}
		\includegraphics[width=1\textwidth]{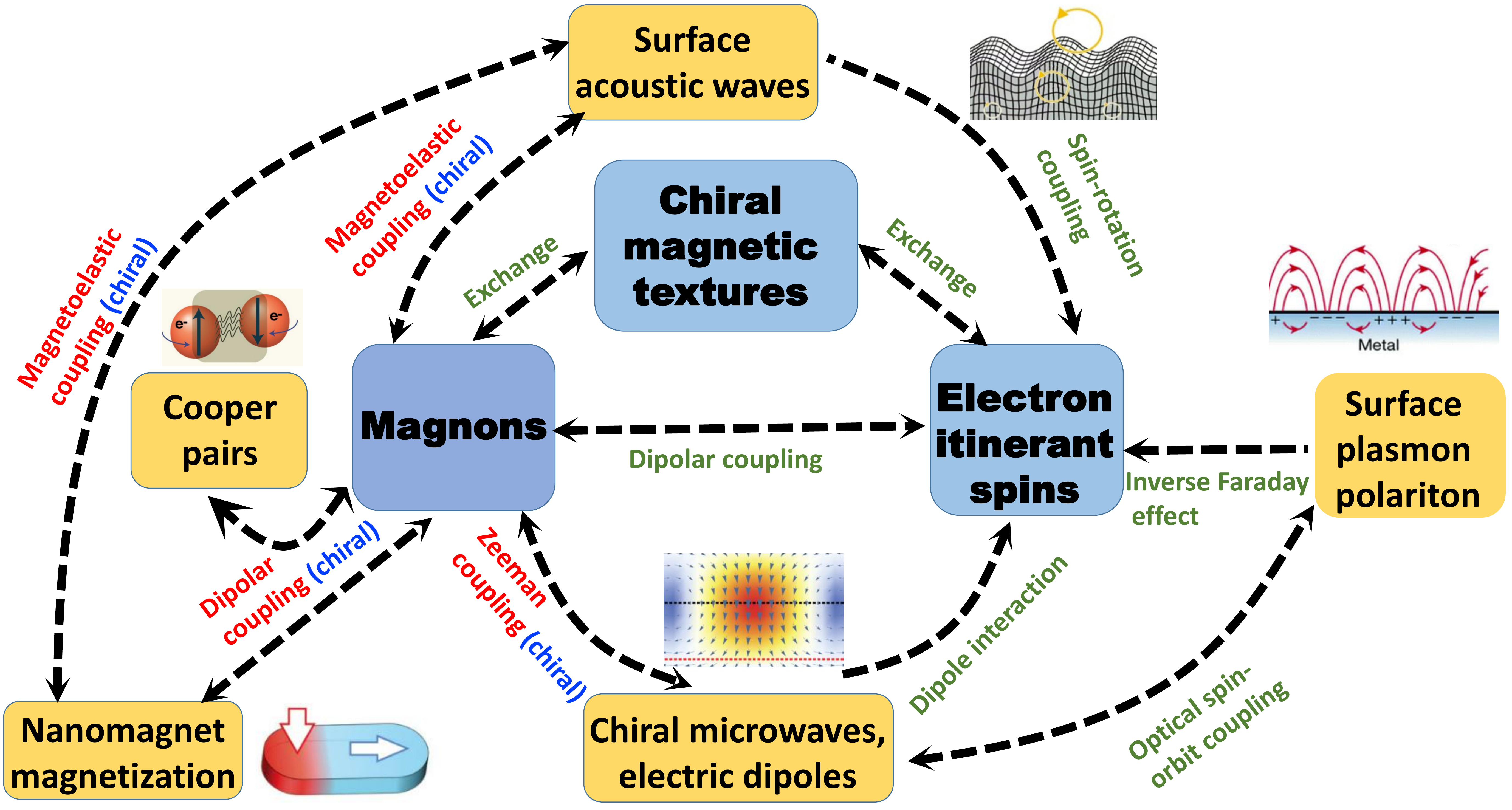}
		\par\end{centering}
	\caption{Excitations with generalized transverse ``spins" and chiral objects in spintronic nanostructures. Examples contain the itinerant magnons in the magnetic films, the confined magnons in the magnetic nanowires or spheres, the evanescent microwaves from the stripline, waveguide or cavity, the near optical fields of electric dipoles, the surface phonons, and the surface plasmon polaritons. The spins of electrons can be along any direction when itinerant, but can be fixed chirally in the chiral magnetic textures when they are local. The mutual interaction between these quasiparticles allows the transfer of (transverse) spin between each other. In particular, for the quasiparticles holding transverse spins, their interaction can be chiral, as highlighted by ``chiral" in blue, which facilitates the unidirectional, non-reciprocal, and sensitive control of spin transport and operation.}
	\label{chiral_spintronics}
\end{figure}

The electron's spin is an intrinsic degree of freedom originating from the relativistic effect. The transverse spin, on the other hand, is a generalization of the relativistic spin.  As summarized in Fig.~\ref{chiral_spintronics}, many of these transverse spins are even approximately conserved when transferring to the electron spin via the non-contact spin pumping and spin transfer processes, which thus may be useful family members of the ``spin" carriers in spintronics that may overcome the shortcomings of electron spin in the generation, transport, and detection. For example, magnons, phonons, and photons are important information carriers with very low dissipation in many materials and devices.

The relativistic effect also induces the spin-orbit interaction of electrons. It favors many chiral magnetic textures in magnetism and renders the static chiral interaction between nanomagnets possible, which was successfully exploited in the experiments for realizing logic operations. We have reviewed that by the evanescence of many waves, rather than the relativistic effect, the generalized transverse spin can be locked to its momentum (and surface normal). Thus the chirality acts as the generalized spin-orbit interaction. When these waves or quasiparticles interact with another object holding a definite transverse ``spin" due to the time-reversal symmetry breaking (typically, magnons as reviewed, or circularly polarized electric dipoles), the object favors to interact with the waves with the same spin, thus the pumped waves propagating in one direction via the generalized spin-orbit interaction. Figure~\ref{chiral_spintronics} summarizes various chiral couplings, as highlighted by ``chiral" in blue, between these quasiparticles that render the interaction non-reciprocal and even unidirectional. Many progresses in demonstrating the chiral interaction and novel spin states and functionalities in this aspect have been reviewed in this article.

This chiral interaction remains to be exploited for realizing functional spintronics devices (Sec.~\ref{devices}).  The present theoretical and experimental studies are performed favorably in one-dimensional structure and largely in the linear regime, but the inevitable additional dimension and nonlinearity in the practical devices  can enrich the spintronics functionality as well (Sec.~\ref{freedoms}). The generalization of the spin concept is not limited to the transverse case, but may also allow to include the mixed longitudinal spin, and, in a broader view, the orbital angular momentum of wave beams (Sec.~\ref{freedoms}). It has important implications for other materials such as low-dimensional, antiferromagnetic, synthetic, and even meta-materials (Sec.~\ref{materials}). The presented theory in this article is formulated in terms of quantum description via Hamiltonian and operators, which becomes classical when replacing operators with amplitudes, thus friendly to include the quantum effect such as quantum fluctuation, quantum noise, and quantum qubit when going to the low temperature, which we envision that chirality may bring unique functionalities in quantum sensing, communication, and metrology based on spintronics in the future (Sec.~\ref{quantum_regime}). We provide a detailed outlook for both static and dynamic chiralities below.

\subsection{Functional spintronic devices}

\label{devices}

\subsubsection{Wave isolators}
\label{wave_isolator} 

\begin{figure}[ht]
	\begin{centering}
		\includegraphics[width=1\textwidth]{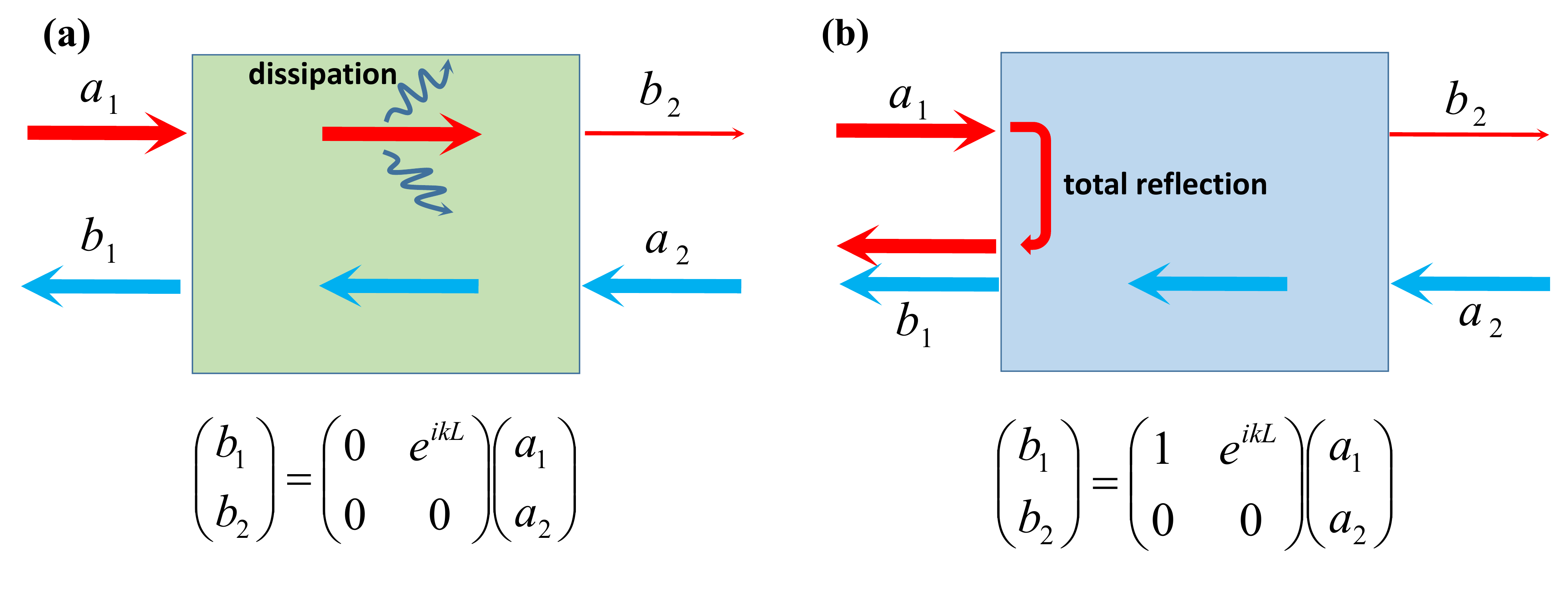}
		\par\end{centering}
	\caption{Wave isolators by chiral interactions with definitions in terms of the different scattering matrices. $a_{1,2}$ is the input amplitude and $b_{1,2}$ is the output amplitude of waves. In (a), the chiral interaction renders that only the waves propagating in one direction with input $a_1$ interact with the object (the green scattering region) that dampens the propagation unidirectionally, which is promising for the realization of an isolator. In (b), the gate (the gray region) only shifts the dispersion of waves propagating in one direction  and thus blocks the waves with input $a_1$ from this direction with a total reflection but allows the propagation via input $a_2$, which also leads to the broadband unidirectional functionality.}
	\label{isolators}
\end{figure}

Optical, microwave, acoustic, spin-wave, and electric-polarization isolators are those devices that block the propagating waves in one direction but allow them to pass in the opposite direction. As reviewed in this article, the non-reciprocal transmission of spin waves, microwaves, and acoustic waves \textit{etc.} are widely proposed when placing magnets in proximity to the magnetic films, microwave waveguides, and dielectric substrate via the chiral interaction mechanism \cite{Chiral_pumping_Yu,Chiral_pumping_grating,magnon_trap,Haiming_exp_grating,Haiming_exp_wire,Hanchen_damping,Au_first,DMI_circulator,bilayer_dipolar_1,bilayer_dipolar_2,Dirk_transducer,Chuanpu_NC,waveguide_Yu_1,waveguide_Yu_2,circulating_polariton,Xufeng_exp,Canming_exp,Teono_NV,Yu_Springer,Doppler_Yu,circulator_Tang,stripline_poineering_1,stripline_poineering_2,Yuxiang_subradiance,chiral_waveguide1,chiral_waveguide2,chiral_waveguide3,phonon_Yu_1,Xu,DMI_phonon_exp,Onose_exp,Nozaki_exp,phonon_Yu_2,Otani_exp,phonon_Kei,Page_exp,Page_exp_2,plasmonics_spin_NC,plasmonics_spin_APL,plasmonics_spin_PRB,plasmonics_spin_NJP,plasmonics_spin_PRL}. 
In particular, via this mechanism the non-reciprocal transmission of microwaves \cite{Xufeng_exp} and surface acoustic waves \cite{Lewis_1972,SAW_chiral_attenuation,Onose_exp,Nozaki_exp,Otani_exp,DMI_phonon_exp,Page_exp,Page_exp_2} have been experimentally observed. However, a simple non-reciprocal transmission of waves does not imply the realization of an isolator because an excellent isolator requires that the transmission in one direction is completely blocked. As a wave phenomenon, the phenomenology of an isolator is well captured by the scattering matrix in such as a two-terminal device in Fig.~\ref{isolators}, where $a_{1,2}$ is the input amplitude and $b_{1,2}$ is the output amplitude of waves.

To date, chiral interaction via the generalized spin-orbit interaction promises two schemes of wave isolators. Figure~\ref{isolators}(a) shows that due to the chiral interaction, the propagating waves with amplitude $a_1$ can interact with the object while $a_2$ does not, via which the dissipation of the object dampens $a_1$ with zero output $b_2$, but does not affect $a_2$. The associated scattering matrix in Fig.~\ref{isolators}(a) implies the unidirectional transmission $e^{ikL}$, where $k$ is the wave vector and $L$ is the length of the object. In this mechanism, the object itself suffers from the (chiral) radiative damping as well when in proximity to the propagating waves. An excellent isolator is thereby achieved only when balancing the radiative damping and the wave absorption \cite{phonon_Yu_2}. One can also use the gating effect to achieve an isolator as shown in Fig.~\ref{isolators}(b) that the gate only shifts the dispersion of waves of one direction and thus block the waves from this direction with a total reflection, such as the input $a_1$, but allows the propagation via input $a_2$, which also leads to the broadband unidirectional functionality without much complicated design \cite{chiral_gate,Slavin_2018,Slavin_2019}. This mechanism was proposed in gating the magnon by metal with interfacial DMI \cite{Slavin_2018,Slavin_2019} without chiral interaction and superconductor with chiral interaction \cite{chiral_gate}, but may be generalized in other devices as well in the future.

\subsubsection{Diodes}

A diode is a two-terminal (electronic) element that conducts momentum, particle, or energy current primarily in one direction with asymmetric conductance. That is, it has low resistance in one direction, but high resistance in the other. In electronics, this is achieved via the nonlinear modulation of the device. Obviously, this functionality is not defined via the scattering matrix like the isolators (Sec.~\ref{wave_isolator}), but the conductance for the current. This device does not require unidirectionality but achieves a high on-off ratio
may have this request.

Non-reciprocal wave transmission is a coherent process, but the waves in terms of wave packets behave like a particle as well, so they carry current with the group velocity. In this sense, the chiral waves reviewed in this article are candidates for the realization of various diodes. Indeed, the thermal incoherent transport in terms of these quasiparticles conducts thermal current that is governed by thermal fluctuation. An important difference to the electronics diodes occurs as well since the non-reciprocity holds in the linear response, seemingly acting like a Maxwell demon. The behavior of thermal noise itself under the chiral interaction, even at the equilibrium, is of fundamental interest, \textit{e.g.}, the accumulation of spin noise at one edge, which as well as holds the possible application such as the enhanced sensing efficiency with (quantum) noise \cite{quantum_noise}.

The control of thermal current in many spintronic nanostructures may have straightforward applications in thermal control in miniaturized devices since when the devices become smaller it is essential to steer thermal current or noise away from the hot spots towards sinks. As reviewed, the thermal transport in this respect is only studied in the context of the chiral spin Seebeck effect theoretically \cite{Chiral_pumping_Yu} and experimentally \cite{Luqiao_exp}. Thereby this largely unexplored area with transport in terms of various chiral quasiparticles and their interplay awaits further theoretical explorations and experimental observations in the future.

\subsubsection{Logic device}

\begin{figure}[ht]
	\begin{centering}
		\includegraphics[width=1.0\textwidth]{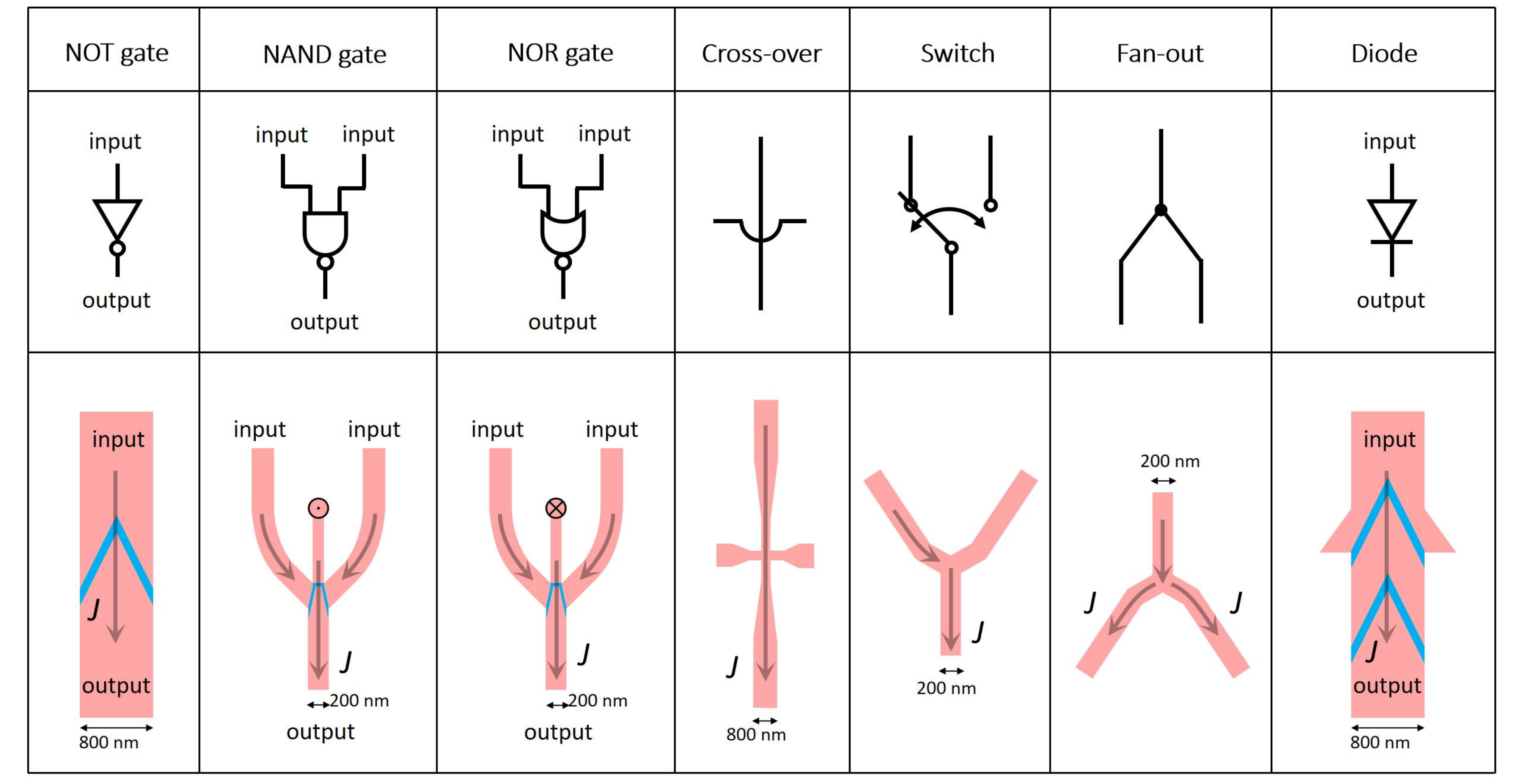}
		\par\end{centering}
	\caption{Complete set of domain-wall logic elements. \textit{i.e.}, NOT gate, NAND gate, NOR gate, cross-over, switch, fan-out, diode. Red and blue shaded regions indicate out-of-plane and in-plane anisotropy, respectively. The direction of the domain wall  flow is indicated by black arrows. The indicated dimensions are those of experimental devices. The figures are reproduced with permission from \cite{Ref_DMIChiral7}.}
	\label{DMIChiralCoupling3}
\end{figure}

The static chiral coupling between nanomagnets turns out to be useful for logic devices via the domain wall \cite{Ref_DMIChiral1,Ref_DMIChiral7}. Figure~\ref{DMIChiralCoupling3} shows  the complete set of domain-wall logic elements, \textit{i.e.}, NOT gate, NAND gate, NOR gate, cross-over, switch, fan-out, and diode in terms of the chirally coupled nanomagnets \cite{Ref_DMIChiral7}. The domain-wall logic scheme can be easily integrated with magnetic ``racetrack" memories  \cite{Ref_DMIDW8, Ref_RaceMemory2, Ref_DMIDW9}, allowing a logic-in-memory architecture, which saves time and power. However, several technical challenges prevent as yet industrial applications, such as intrinsic/extrinsic domain-wall pinning, efficient domain-wall writing, and non-destructive reading \cite{Ref_RaceMemory4}. Moreover, the demands on a domain-wall logic device such as tunable magnetic anisotropy and sizable DMI can be met only by several materials and requires additional fabrication steps.

 The spin-wave logic, or its extension to the devices with other waves, is an important application in magnonics \cite{spin_wave_computing}. The proposed isolators or diodes above may act as important elements in the spin-wave logic device. Importantly, the chiral interaction promises that the pumped waves propagate unidirectionally in only half space \cite{Chiral_pumping_Yu,Chiral_pumping_grating,magnon_trap,Haiming_exp_grating,Haiming_exp_wire,Hanchen_damping,Au_first,waveguide_Yu_1,waveguide_Yu_2,Teono_NV,Yu_Springer,Doppler_Yu,stripline_poineering_1,stripline_poineering_2,Yuxiang_subradiance,chiral_waveguide1,chiral_waveguide2,chiral_waveguide3,phonon_Yu_1,Xu,DMI_phonon_exp,Onose_exp,Nozaki_exp,phonon_Yu_2,Otani_exp,phonon_Kei,Page_exp,Page_exp_2,plasmonics_spin_NC,plasmonics_spin_APL,plasmonics_spin_PRB,plasmonics_spin_NJP,plasmonics_spin_PRL}, which is, therefore, a switch that is governed by the magnetization direction. The magnon gate by proximity superconductor becomes a perfect switch for the spin waves that are controlled by the temperature \cite{chiral_gate}. The future study may benefit from the dynamical chirality with the generalized spin-orbit interaction to realize various spin-wave gates similar to those by the static chiral interaction in Fig.~\ref{DMIChiralCoupling3}.

\subsection{Other degrees of freedom}

\label{freedoms}

\subsubsection{High spatial dimension}

Most chiral physics in this review is emphasized by the effective one-dimensional models since the chirality index $Z=
\hat{\bf k}\cdot(\hat{\pmb{{\bf n}}}\times \hat{\pmb{\sigma}})$ is a maximal integer $+1$ or $-1$, but in the other propagation directions, the waves are affected by the (non-integer) chirality index as well. Examples may vary from the Damon-Eshbach spin waves, magnetic fields of spin waves and short striplines to the electric fields of short dipoles wires and electric polarizations, where the waves' spins become longitudinal with $Z=0$ only when the waves are propagating along the magnetization or the wire directions. Interestingly, the non-reciprocal coupling between magnons of nanowires and films strongly depends on the propagation direction of magnons in the film, in which the chirality completely vanishes at particular angles \cite{Chiral_pumping_Yu,Chiral_pumping_grating,magnon_trap,Haiming_exp_grating,Haiming_exp_wire}. The chirality in the magnetoelastic coupling between proximity magnet and surface acoustic waves is strongly modulated by the magnetization direction \cite{phonon_Yu_1,phonon_Yu_2,phonon_Kei}. These provide flexible tunability of chirality of the coherent waves that can be switched on and off by changing the propagation direction of waves. The associated dependence of chirality on the spatial dimension via the generalized spin-orbit interaction thereby may be distinguished from the other non-reciprocal mechanisms in various experiments in the future.

In incoherent thermal transport, the quasiparticles of all  propagation directions play a role, but the chirality survives for a large fraction of them, thus easily manifesting in the transport measurement and providing steering on the generation and control of thermal current. These thermal dynamics are important design parameters in future spintronic nanostructures.

\subsubsection{Nonlinearity}

To date, the dynamical chirality in spintronics is focused more on the linear regime. Despite its great success in demonstrating the key mechanisms and easy observation, sometimes beyond the haven of linearity brings new functionality. The magnetization switching by spin current relies on a large spin flow. Also, the nonlinearity appears to be unavoidable in the miniaturized devices where the relative driving force becomes strong and causes a large current flow. Chiral interaction promises the unidirectional generation of the ``spin" current carried by various quasiparticles, which may modulate the ground state by their interaction with local order parameters, such as the Doppler shift of spin-wave dispersion under chiral pumping in thin magnetic films \cite{Doppler_Yu}. Another source of nonlinearity brought by the chirality is that the energy tends to accumulate at one edge when placing many magnets that are mediated by the chiral waves. It favors the nonlinearity even with a small excitation power \cite{waveguide_Yu_1,waveguide_Yu_2,non_Hermitian_skin_effect}. However, from the experimental viewpoint, the nonlinear spin dynamics in the presence of generalized spin-orbit interaction has not yet been explored much in the literature.

The near electric or magnetic fields, by virtue of their strong spatial localization at the sample surface, have enhanced strength that may serve as the sources for driving nonlinear dynamics in the spin nanostructures. Surface plasmon polaritons are successful examples that enhance many optical interactions \cite{SSP_1,SSP_textbook_1,SSP_textbook_2,SSP_review} such as the magneto-optic processes \cite{nonreciprocity_plasma}. The  interaction between the Damon-Eshbach modes and optical whispering gallery modes is strongly enhanced with their surface effects \cite{Sanchar_PRB,Sanchar_optimal,WGM1,WGM2,WGM3}. Future studies may benefit from this surface enhancement with other evanescent chiral waves.

\subsubsection{Chiral fluctuation beyond the quasiparticle scenario}

\textcolor{blue}{The quasiparticle concept for the low energy excitations of ordered phases are suitable theoretical instruments describing the dynamics of quantum materials  when the temperature is sufficiently lower than the critical ones. This review  focuses on the chirality of such low-energy excitations and their interactions, such as the magnons \cite{DE,Walker_sphere,Chiral_pumping_Yu,surface_roughness_Yu,Kei_topology,chiral_sensing}, microwave photons \cite{stripline_poineering_1,stripline_poineering_2,Yu_Springer,waveguide_Yu_1,waveguide_Yu_2,circulating_polariton,Xufeng_exp,Zhong}, optical photons \cite{chiral_optics,Nori,nano_optics}, phonons \cite{Kino1987,Viktorov1967,Xu,Onose_exp,Nozaki_exp,phonon_Yu_2,phonon_Yu_1,Otani_exp,Page_exp,Page_exp_2,phonon_Kei,DMI_phonon_exp,phonon_Yu_1,phonon_Kei,phonon_Yu_2,Chengyuan_Cai}, surface plasmons \cite{SSP_review,plasmonics_1,plasmonics_2,plasmonics_spin_NC,plasmonics_spin_APL,plasmonics_spin_PRB,plasmonics_spin_NJP,plasmonics_spin_PRL}, and electron spins \cite{electron_spin_Yu} (refer to Sec.~\ref{section3}). Their chirality is intrinsic by evanescence and characterized by the chirality index $Z$. The non-reciprocity of their excitation, transport, and detection in the linear response regime  emerges from this chirality (refer to Sec.~\ref{Chiral_interaction}).
The chirality affects transport in nonlinear regimes that can be formulated by a mean-field theory that accounts for the low-order interactions that arise between the quasiparticles \cite{Chengyuan_Cai,Doppler_Yu,Schwinger_boson,Lee_Lee}.}

\textcolor{blue}{However, at temperatures close to the critical ones, the fluctuations amplitudes  become comparable to the equilibrium order parameter. They become so short-lived that the quasiparticle description becomes inappropriate. In such a beyond-quasiparticle regime, the role of the chirality, as initially defined for the ground state and non-interacting quasiparticles with well-defined wave vectors,  raises theoretical challenges. The recently discovered enhancement of the thermal Hall effect around the ordering temperature in two-dimensional ferromagnets with perpendicular magnetic anisotropy emphasizes the important role of chiral fluctuation as governed by a spin chirality with three-spin index ${\bf S}_i\cdot({{\bf S}_j\times {\bf S}_k})$ \cite{Chiral_fluctuation_1,Chiral_fluctuation_2}. The numerical simulation by Monte Carlo and stochastic spin dynamics \cite{Chiral_fluctuation_Xiao} is beyond the scope of the present review.  }

\subsubsection{Orbital angular momentum}

The exploitation of the spin degree of freedom of electrons has led to the fast development of spintronics,  which focuses on the generation, detection, and manipulation of spin information. Recently, attention is attracted to the orbital degree of freedom of electrons as well, which is believed to be an important alternative information ``carrier" to spin. The associated branch is termed as ``orbitronics" \cite{Ref_Orbitronics1, Ref_Orbitronics2, Ref_Orbitronics3}. The orbital degree of freedom is often regarded as “frozen” due to the orbital quenching by the crystal fields. As a surprise, recent studies accumulated evidence that the dynamics and transport of the orbital information can be released under external stimuli such as an electric field in many materials. Similar to those in spintronics, orbital current can carry the orbital angular moment originating from orbital Hall effect \cite{Ref_OrbitalHall1, Ref_OrbitalHall2, Ref_OrbitalHall3} and orbital Rashba effect \cite{Ref_OrbitalRashba1, Ref_OrbitalRashba2, Ref_OrbitalRashba3, Ref_OrbitalRashba4, Ref_OrbitalRashba5}, and exert torques on the magnetization. Hence, the orbitronics may provide another route to manipulate and detect magnetization.

\begin{figure}[ht]
	\begin{centering}
		\includegraphics[width=0.9\textwidth]{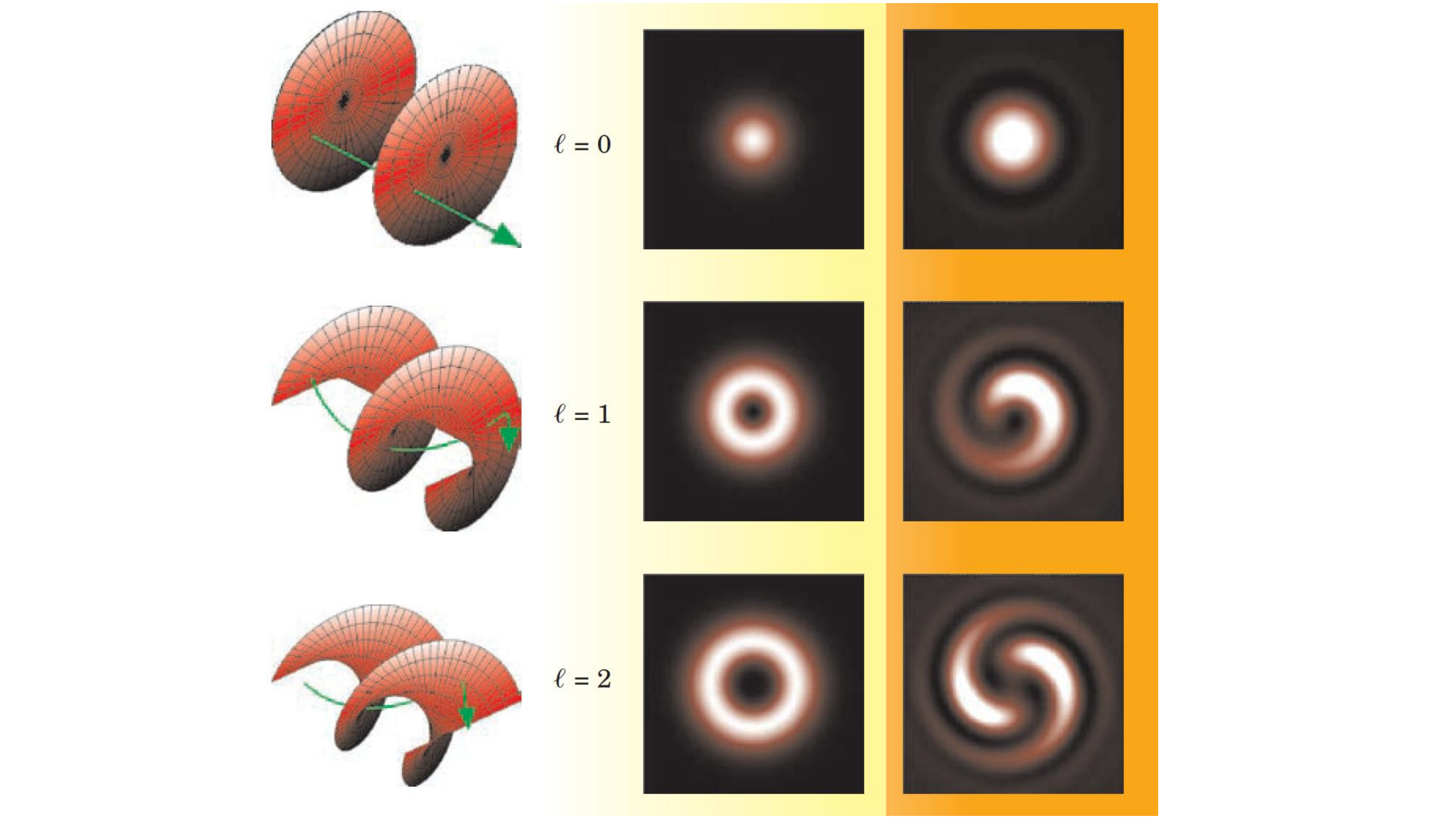}
		\par\end{centering}
	\caption{Orbital angular momentum of a
light beam with wavefront (left panel) and projection (right panel). The beams are labeled by the orbital angular-momentum quantum
number $l$. The associated beam’s orbital angular momentum is $L=l\hbar$. The figure is reproduced with permission from \cite{orbital_AM}.}
	\label{orbital_AM}
\end{figure}

This review focuses on the transverse spin of numerous evanescent waves as a generalization of the electron spin. Their locking to the momentum or surface normal of the waves' propagation plane is achieved by the integer chirality index. A close analogy of orbital angular momentum of waves to the one of electrons also exists, which has been well studied for decades in optics \cite{Allen}. Figure~\ref{orbital_AM} shows that the optical wave beams also carry the orbital angular momentum density that is indexed by an integer $l$, independent of the beam’s polarization \cite{orbital_AM}.  In this article, we have addressed the conservation of transverse spin transferring to the electron one in the context of near-field spin pumping in terms of microwaves \cite{electron_spin_Yu}. The optical spin of surface plasmon polaritons is also a source for the electron orbital magnetization by the inverse Faraday effect \cite{plasmonics_spin_NC,plasmonics_spin_APL,plasmonics_spin_PRB,plasmonics_spin_NJP,plasmonics_spin_PRL}. The counterpart, \textit{i.e.}, the mutual transfer between the orbital angular momentum of electromagnetic fields and electron's orbital one may be envisioned. The methodology and overview of this article may inspire similar prospects in the future. 

\subsubsection{Electric polarization}

The close analogy between magnetic dipoles in magnetism and electric dipoles in ferroelectricity originates from their duality, i.e. the associated electric ${\bf E}$ and magnetic
${\bf H}$ fields have identical forms, which have cross-fertilization of both fields for many years. Recently, their excitations are found to have many common features \cite{ferron_1,ferron_2,ferron_3}. The electric polarization can transport momentum and energy in ferroelectrics, similar to that in terms of magnons in a ferromagnet. Zhou \textit{et al}.  predicted that coherent polarization waves or ``ferrons" can be localized at the surface of ferroelectric insulators by interacting with their own electric stray fields \cite{surface_ferron}. Nevertheless, their properties turn out to be strikingly different from the magnetic counterpart, i.e. the surface ``Damon-Eshbach" magnons \cite{DE,Walker_sphere} in ferromagnets, although their generated stray electric fields are chiral. We envision the dynamical chirality with generalized spin-orbit interaction reviewed in this article should
stimulate similar activities in ferronics.

\subsection{Antiferromagnetic, two-dimensional, and synthetic materials}

\label{materials}

In this article, both the static and dynamical chiralities are mostly reviewed in the context of ferromagnets, dielectrics, conductors, and their hybridizations for the simplicity of illustrating the concepts. Rich functionalities of various spin textures and different alignments of nanomagnets are governed by the static chiral interaction. Universal viewpoints in terms of generalized spin-orbit interaction are addressed for the motion of quasiparticles in these materials. The saturation magnetization fixes the direction of the transverse spin of the magnon, so when being evanescent they propagate unidirectionally at the surface of a magnet. For the other evanescent waves, by the time-reversal symmetry, their transverse spins are allowed to be reversed such that they are allowed to propagate along opposite directions at one surface. The chiral interaction may rely on the breaking of time reversal symmetry that chooses the preferred direction. These implications should be broad that can be applied to other materials. For example, treating the chirality as spins also bring about the tunability of nematicity by the in-plane magnetic field in thin chiral superconductors \cite{Yu_nematicity}.

\subsubsection{Antiferromagnetic spintronics}

It is nowadays well appreciated that the antiferromagnet and ferrimagnet can host magnons of both left-handed and right-handed circular polarizations or ``chiralities"  \cite{Ref_AFMmagnon1, Ref_AFMmagnon2, Ref_AFMmagnon3} that are carried by the so-called $\alpha$ and $\beta$ modes, which provides an additional degree of freedom and tunability than that of ferromagnet. In particular, the magnon chirality can be manipulated via the modulation of the ratio of the sublattice magnetization \cite{Ref_FiMmagnon1, Ref_FiMmagnon2}. The manipulation of right-handed and left-handed chiralities in antiferromagnets and ferrimagnets enables the design of chirality-based magnonic devices such as field-effect transistor \cite{ Ref_AFMMagDevice1} and chirality-based logic devices \cite{ Ref_AFMMagDevice2, Ref_AFMMagDevice3}. 

Chirality already plays an important role in the magnonic excitations of the antiferromagnet. The Damon-Eshbach surface modes exist as well in the antiferromagnetic films \cite{DE_AFM_1,DE_AFM_2,DE_AFM_3,DE_AFM_4,DE_AFM_5,DE_AFM_6}. The dipolar interaction, although being canceled out at the equilibrium, turns out to act as a Rashba-like spin-orbit coupling between the $\alpha$ and $\beta$ magnon modes \cite{DE_AFM_7}, rendering an analogous spin dynamics to the electron.  Thanks to the time-reversal symmetry, spin-momentum locking of these modes are envisioned. The dipolar interaction between two magnetic layers, \textit{i.e.}, synthetic antiferromagnet, is chiral \cite{bilayer_dipolar_1,bilayer_dipolar_2,slow_wave,bilayer_2019}. Exploiting the accumulated experience of the chirality may bring inspiring functionalities in the control of the magnon dynamics of an antiferromagnet.    

\subsubsection{Two-dimensional magnets}

Previous works demonstrating the static and dynamic chiral interaction have mainly focused on magnetic nanostructures including nanomagnets, thin films, and nanowires, which are all very simple for clarifying the underlying mechanisms. For example, the Damon-Eshbach modes vanish in the ultrathin magnetic films, which after being ruled out can highlight the chiral dipolar interaction between nanowires and films. However, the continuum model appears to break down when the magnet is reduced to atomic thickness. On the other hand, reducing the thickness of a magnet not only downscales the size of spintronics devices but also brings physics to the new regime. The magnon conductivity in high-quality subnanometer YIG films turns out to reach record values because of the onset of two-dimensional (2D) diffusive magnon transport \cite{Xiangyang_record}.

But strictly speaking, these magnetic films are not ``true" 2D  magnets as there are still many dangling bonds and many subbands for magnons. The inevitable surface roughness at the interface plays an important role in the magnetic properties and quality. Recently, along with intensive studies of 2D nonmagnetic materials such as graphene \cite{Ref_2DGr} and monolayer transition-metal dichalcogenide \cite{Ref_2DTMD}, isolated monolayer 2D magnets without dangling bonds, \textit{e.g.}, Cr$_2$Ge$_2$Te$_6$ \cite{Ref_2DM1}, CrI$_3$ \cite{Ref_2DM2}, Fe$_3$GeTe$_2$ \cite{Ref_2DM3}, and CrCl$_3$ \cite{Ref_2DM4}) have been successfully prepared by mechanical exfoliation with Scotch tape, liquid exfoliation, and molecular-beam epitaxy (MBE). These materials provide an ideal platform to explore magnetic excitations in 2D limit \cite{Ref_2DSW1, Ref_2DSW2, Ref_2DSW3, Ref_2DSW4}. Due to the atomic thickness, 2D magnets are susceptible to external stimuli. Hence the chiral dynamic phenomena in 2D magnet-based systems can be effectively modulated by the means of the electric field, photon irradiation, pressure, chemical modification, and strain.

Similar to other 2D materials, 2D magnets can be stacked into high-quality heterostructures via the van der Waals force. The observation of strong DMI in 2D magnet stacks promises its potential in ultra-compact next-generation chiral spintronics \cite{Ref_2DDMI1, Ref_2DDMI2}. In particular, the relative twist in the bilayer heterostructure creates moir\'e patterns \cite{Ref_2DTwist2, Ref_2DTwist3} and potentially offers a versatile platform to study chirality-related physics as the twisting operation has the helicity.

\subsubsection{Synthetic and meta-materials}

Chiral waves of very different types, such as electromagnetic waves, spin waves, and acoustic waves, exist in many materials of very different properties. Synthetic and meta-materials enable to bring various functionalities together, such as the routing of various waves in one device.

Also, the static chirality is usually determined either by the structural chirality of the system or by DMI polarity, which is an intrinsic property of the materials. It usually keeps unchanged once constructed.  On the other hand, the capability of reversibly and locally modifying the chirality may offer another dimension to manipulate the chiral magnetism and enable the realization of multi-functional and reprogrammable devices. While the structural chirality with respect to the shape of the structure is difficult to change, it has been reported that the polarity of DMI can be reversibly tuned by modulating electronic density with electric gate \cite{Ref_NewDMI5}, by switching the polarization of multiferroics \cite{Ref_NewDMI6} and by chemisorption/desorption of functional atoms at the interface \cite{Ref_NewDMI7}.

\subsection{Quantum regime}

\label{quantum_regime}

Quantum information technology, harnessing the basic principles of quantum mechanics such as superpositions and entanglement of quantum states, is of great interest for precise measurement, information processing, and communication tasks \cite{Ref_QuantumInfo1, Ref_QuantumInfo2, Ref_QuantumInfo3, Ref_QuantumInfo4, Ref_QuantumInfo5}. In analogous to the classical bit that has the value of 0 and 1, the quantum information is encoded in the so-called qubit (quantum bit). Taking advantage of the unique quantum physics of superposition and entanglement, the simultaneous and coherent manipulation of qubits enables the accomplishment of unprecedented computation tasks beyond the capability of a classical supercomputer.

Although both spintronics and quantum information technology implement spins, these two fields developed relatively independently at the early stage. In recent years, the situation starts to change and their interdisciplinary has opened up a new avenue for both two fields \cite{Ref_QuantumSpin1, Ref_QuantumSpin2, Ref_QuantumSpin3,quantum_Mehrdad,quantum_Sanchar,Nakamura_science_1,Nakamura_science_2}. The sensing of magnon via superconducting qubit in the microwave cavity achieves quantum resolution, \textit{i.e.}, a single quanta \cite{Nakamura_science_1,Nakamura_science_2}. The quantum control of and communication with the quanta of surface acoustic waves has been achieved experimentally \cite{SAW_nature,SAW_science}. Microwave photon was intensively studied in the context of quantum electrodynamics \cite{QED}. These ingredients with chiral interaction may combine in the quantum regime.

The merit of chiral interaction between two objects is the lack of backaction when they are mediated by the chiral waves, thus rendering novel manipulation at the quantum level. The manipulation of microwave photons by the qubit is the key functionality in circuit quantum electrodynamics, in which the chiral interaction stabilizes the entanglement and quantum manipulation without backaction \cite{chiral_optics}. In the classical regime, the chiral interaction between an array of magnets favors the accumulation and greatly enhances the sensitivity of detection of microwaves \cite{waveguide_Yu_1,waveguide_Yu_2,non_Hermitian_skin_effect}.

This review intensively formulates the chiral coupling of various quasiparticles in terms of the quantum formalism, which may form a starting point to study quantum effects in the
spintronic nanostructures, such as spin-pumping-induced magnetic quantum noise
\cite{quantum_noise} at low temperatures, quantum squeezing, one-way quantum steering \cite{quantum_steering_chirality}, and entanglement
of the quasiparticles with microwaves via the magnet \cite{entanglement_Zou,Yalong}. The chiral interaction may improve the performance and sensitivity of quantum communication and quantum sensing with magnon, microwave photon, and surface phonon in several aspects.

\addcontentsline{toc}{section}{Declaration of competing interest}
\section*{Declaration of competing interest}
The authors declare no competing financial interests that could have appeared to influence the work reported in this paper.

\addcontentsline{toc}{section}{Acknowledgments}
\section*{Acknowledgments}
This work is financially supported by the National Natural Science Foundation of China (Grant No. 0214012051), the
startup grant of Huazhong University of Science and Technology (Grants No. 3004012185 and No. 3004012198),  the
startup grant of Peking University, as well as JSPS KAKENHI Grant No. 19H00645.
T.Y. started his research on chirality in spintronics with G.B. at the Kavli Institute of Nanoscience of the TU Delft five years ago,  acknowledging many useful discussions with his colleagues Sanchar Sharma, Xiang Zhang, Yuguang Chen, Yaroslav M. Blanter, and Toeno van der Sar. He continued these studies  at the Max Planck Institute for the Structure and Dynamics of Matter, thanking Michael A. Sentef, Dante M. Kennes, and Angel Rubio for their support. We thank Yu-Xiang Zhang, Weichao Yu, Kei Yamamoto, Bimu Yao, Jinwei Rao, Bowen Zeng, Jin Lan, Haiming Yu, Mehrdad Elyasi, Ji Zou, Xiangyang Wei, Akashdeep Kamra, Canming Hu, Ping Tang, Alejandro O. Leon, and Sergio Rezende for numerous inspiring discussions.

\end{document}